\documentclass{article}
\usepackage{breqn}
\usepackage{fancyhdr}
\usepackage{algorithm}
\usepackage{algpseudocode}
\usepackage{amsmath}
\usepackage{amsfonts}
\usepackage{titlesec}
\usepackage{xcolor}
\usepackage{graphicx}
\usepackage{caption}
\usepackage{subcaption}
\usepackage{url}
\usepackage{hyperref}
\usepackage{cite}
\usepackage{lastpage}
\usepackage{array}

\usepackage[left=3.3cm,right=3.3cm,top=3.9cm,bottom=3.9cm]{geometry}

\newcommand{\myline}{\noindent\rule{\linewidth}{1.5pt}\par}
\newcommand{\mytitle}{On the time complexity analysis of numerical percolation threshold estimation}
\newcommand{\myauthor}{Daniel García Solla}
\newcommand{\stirling}[2]{\genfrac{\{}{\}}{0pt}{}{#1}{#2}}
\makeatletter
\renewenvironment{abstract}{
    \if@twocolumn
      \section*{\abstractname}
    \else
      \begin{center}
        {\bfseries \Large\abstractname\vspace{\z@}}
      \end{center}
      \quotation
    \fi}
    {\if@twocolumn\else\endquotation\fi}

\makeatother

\title{
  \myline
  \textbf{\mytitle} 
  \\\vspace{0.2cm}
  \large\textit{}
  \myline}
\date{}

\author{
  \textbf{\myauthor} \\
  Universidad de Valladolid \\
  \texttt{daniel.garcia22@estudiantes.uva.es}  
}

\begin{document}

\maketitle
\begin{quote}
\begin{abstract}   
The main purpose of percolation theory is to model phase transitions in a variety of random systems, which is highly valuable in fields related to materials physics, biology, or otherwise unrelated areas like oil extraction or even quantum computing. Thus, one of the problems encountered is the calculation of the threshold at which such transition occurs, known as percolation threshold. Since there are no known closed forms to determine the threshold in an exact manner in systems with particular properties, it is decided to rely on numerical methods as the Monte Carlo approach, which provides a sufficiently accurate approximation to serve as a valid estimate in the projects or research where it is involved. However, in order to achieve an exact characterization of the threshold in two-dimensional systems with site percolation, in this work it is performed an analysis of the complexity, both temporal and spatial, of an algorithm that implements its computation from the aforementioned numerical method. Specifically, the conduction of an accurate analysis of the cost of such algorithm implies a deep enough knowledge about certain metrics regarding its duration, or work completed per iteration, which along with its formalization may contribute to the determination of the threshold based on these metrics. Nevertheless, as a result, various bounds are achieved for the best, average and worst cases of the execution on systems spanning several dimensions, revealing that in 1 and 2 the complexity is directly conditioned by the duration, although from 3 onwards no proof for this point has been found, notwithstanding the evidence suggesting its compliance. Furthermore, based on the average case, several methods are proposed that could be applied to characterize the threshold, although they have not been thoroughly explored beyond what is necessary for the complexity analysis.
\end{abstract}
\end{quote}
\vspace{1.5em}

\tableofcontents
\newpage

\section{Introduction}
As already mentioned in the abstract, percolation theory \cite{duminilcopin2017yearspercolation} has potential applications \cite{Ryazanov2024} in a large number of areas of scientific knowledge. Though, in this work we won't focus on its applicability, but rather on its theoretical development, in order to provide tools with which to contribute to its increasingly positive impact in these areas. Specifically, given the computational limitations we encounter when trying to estimate the percolation threshold, it is worth considering that a proper complexity analysis of the Monte Carlo algorithm involved is key to determine its effectiveness for its task, i.e., to estimate the threshold in the shortest feasible time. Then, with a reliable complexity bound, it will be possible to compare it with more advanced methods and select the most optimal one on the basis of the resulting data. Moreover, the method of analysis that will be proposed leads to several theoretically relevant results, including approaches that might be suitable to accurately characterize the threshold without the need for a numerical estimation. But, the most significant conclusion to be drawn from this work is the complexity bounds for systems of multiple dimensions. That is, for systems of 1 and 2 dimensions, we can leverage previous studies of other well-known examples such as the coupon collector's problem to demonstrate the validity of the approach, reaching bounds that are consistent with the experimental measurements. And, for systems of 3 or more, although no analysis has been performed for systems of 4 and above, it is uncertain whether the method provides a correct and exact outcome, since it cannot be demonstrated due to the complexity of the properties defining the system.

\section{Preliminaries}
In this section, we will start by defining the problem and the algorithm we will deal with in the rest of the article. Then, its functionality will be explained in detail and a preliminary analysis will be performed with standard techniques, resulting in complex equations with hardly any contribution to this problem and leading to the proposed analysis section.
\subsection{Problem definition}
\text The problem on which we will perform the time complexity analysis is stated as follows:
You are given a $n \times n$ matrix with its cells initialized in an empty state. At certain points in time, a function will be executed to insert an element in a cell whose position is indexed by the input parameters $i$ with respect to the matrix rows and $j$ for columns. The positions of these elements along the sequence of insertions are assumed to be random, i.e., each matrix cell has the same probability of having the invoked function to insert an element on it. Likewise, if an element already exists in such position, the insertion will be discarded, but the function call will be considered in a global counter. After a particular number of iterations, i.e., executions of the element insertion function, the system will contain enough of them in order to allow the existence of a path connecting the top and bottom matrix rows, which will cause the system to change its state. 
\\\\
The purpose of the following algorithm is to emulate this process and detect the point when such a path arises, also called the point where the system percolates, enabling us to study the system's status after the percolation event. 

\begin{algorithm}[H]
\caption{PercolationSystem2D Template}
\begin{algorithmic}[1]
\small
    \State \textbf{Class} \textsc{PercolationSystem2D}
        \State $\text{grid} \gets \text{None}$
        \State $\text{visited} \gets \text{None}$
        \State $\text{nei} \gets [[-1, -1], [-1, 0], [-1, 1],[0, -1],$\\$ [0, 1], [1, -1], [1, 0], [1, 1]]$
        \State $\text{percolate} \gets \text{False}$
        
        \Procedure{Initialize}{$n$}
            \State $\text{grid} \gets [[\text{False} \; \text{for} \; \_ \; \text{in} \; \text{range}(n)] \; \text{for} \; \_ \; \text{in} \; \text{range}(n)]$
            \State $\text{visited} \gets [[\text{False} \; \text{for} \; \_ \; \text{in} \; \text{range}(n)] \; \text{for} \; \_ \; \text{in} \; \text{range}(n)]$
            \State $\text{percolate} \gets \text{False}$
        \EndProcedure

        \Procedure{Insert}{$i, j$}
            \If{not $\text{grid}[i][j]$}
                \State $\text{grid}[i][j] \gets \text{True}$
                
                \text{//Clear visited matrix} $O(n^2)$
                
                \State $\text{percolate} \gets \text{helper}(i, j, 0)$
                
                \text{//Clear visited matrix} $O(n^2)$
                
                \State $\text{percolate} \gets \text{percolate}  \land  \text{helper}(i, j, \text{len}(\text{grid}) - 1)$
            \EndIf
        \EndProcedure  
        
        \Function{helper}{$i, j, \text{up}$}
            \If{$i == \text{up}$}
                \State \textbf{return True}
            \EndIf
            \State $\text{visited}[i][j] \gets \text{True}$
            
            \For{$k$ \textbf{in} $\text{nei}$}
                \State $\text{newI} \gets k[0] + i$
                \State $\text{newJ} \gets k[1] + j$
                \If{$(0 \leq \text{newI} < \text{len}(\text{grid})  \land  0 \leq \text{newJ} < \text{len}(\text{grid}[0])  \land $
                    \State $\text{grid}[\text{newI}][\text{newJ}]  \land  \lnot \text{visited}[\text{newI}][\text{newJ}]  \land $
                    \State $\text{helper}(\text{newI}, \text{newJ}, \text{up})$}
                    \State \textbf{return True}
                \EndIf
            \EndFor
            \State \textbf{return False}
        \EndFunction
        
        \Function{Percolate}{}
            \State \textbf{return} $\text{percolate}$
        \EndFunction 
\end{algorithmic}
\end{algorithm}

Above is shown our algorithm template that will be used to computationally describe the elemental operations for the prior process, so that we can initialize a new system with the $Initialize()$ function, execute the insertion attempt of an element with $Insert(i,j)$ \textit{(and its associated $helper()$ function)}, and check for percolation within the system using the Percolate() function.After defining the percolation system template, it is necessary to describe the algorithm responsible for finding the state transition point. 

\begin{algorithm}[H]
\caption{Percolation Simulation Process}
\begin{algorithmic}[1]
\State $\text{random.seed()}$
\State $\text{system} \gets \text{PercolationSystem2D()}$
\State $\text{time} \gets 0$
\State $\text{iterations} \gets 0$
\State $\text{system.Initialize}(n)$
\State $\text{time} \gets \text{get\_time()}$
\While{$\lnot \text{system.Percolate}()$}
    \State $\text{system.Insert}(\text{random.randint}(0, n - 1), \text{random.randint}(0, n - 1))$
    \State $\text{iterations} \gets \text{iterations} + 1$
\EndWhile
\State $\text{time} \gets \text{get\_time()} - \text{time}$
\State $\text{elements} \gets \sum\limits_{i} \sum\limits_{j} system.grid[i][j]$
\end{algorithmic}
\end{algorithm}

The previous process shows the simulation where elements are inserted into the system matrix as long as no percolation occurs. In particular, the $Insert()$ function is executed and the iteration counter is incremented as long as there is no path connecting the corresponding matrix rows.
\\\\
Finally, in order to specify the definition of path used in this context, the matrix will have the Moore's neighborhood property \cite{moore_neighborhood}. Thus, each cell will have a set of 8 neighbors composed of its horizontal, vertical, and diagonal adjacent cells whose location indices are described as $\{ (i-1, j-1), (i-1, j), (i-1, j+1), (i, j-1), (i, j+1), (i+1, j-1), (i+1, j), (i+1, j+1) \}$ with respect to a cell's location $(i, j)$. Given this information, we can define a valid path as a sequence of elements all adjacent to each other by the prior property that has at least one of those elements in the top matrix row and another in the opposite bottom row, establishing a connection between both. Note that instead of considering rows, it could also be defined with the left and right end matrix columns. However, both options are equally indifferent, since the process is invariant to this kind of transformation.

\subsection{Initial approach}

Before undertaking the analysis of the algorithm's runtime, and how this time grows as a function of the system size on which it operates, it is useful to examine the results provided for a given set of input parameters. In view of its complexity and the significant amount of states it can undergo throughout its execution, building an exhaustive trace up to an outcome is not the preferred option, as it would be tedious to construct and challenging to understand.
\\\\
Therefore, we will start by examining the terminal system state, and how the algorithm has reached that situation based on the operations defined previously. As can be seen, Figure 1 represents a matrix of size $20\times 20$ where a site-percolative process has been applied using the specified algorithm. Out of the $20\cdot 20=400$ constituent system cells, those colored in black are deemed empty, while the white ones contain a single element. That is, as the algorithm advances and inserts new elements, there will be cells that remain empty during the entire process, arriving at the terminal state without any element, and there will be others that will contain 1 element, discarding insertions that choose their position when the cell is already occupied.
\begin{figure}[H]
    \centering
    \includegraphics[width=5.6cm]{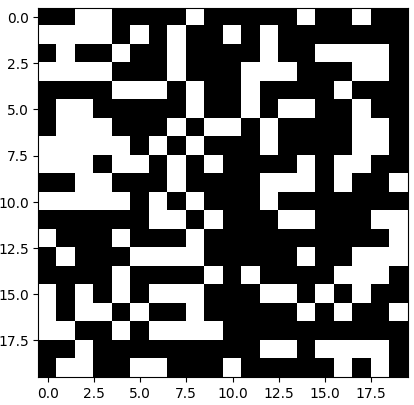}
    \caption{Resulting $20\times 20$ grid after simulation}\label{fig:grid20}
\end{figure}
This algorithm primarily aims to determine the number of iterations required to induce a state change within the system, a phenomenon known as phase transition in percolation theory. Essentially, this refers to the point at which the algorithm has inserted the necessary amount of elements into the system to produce a path connecting the top and bottom matrix rows. From that point on, any other elements added by successive iterations will be incapable of altering the system's state thereafter, since there will already be a path that has induced such phase transition before. As there is a correlation between the number of iterations that the algorithm takes to complete and the volume of elements present in the terminal state, in percolation theory the point at which this state is reached is usually defined as the percolation threshold $p_c$, represented by the ratio between the elements within the system and the total number of occupiable cells. That is, the occupancy ratio $p$ is derived by computing the quotient of the number of elements found at an iteration $i$ of the procedure by the maximum number of elements supported by the system, which in our case would be $n^2$ for executing the algorithm on a matrix. However, there are other categories of systems formed by polygonal structures such as hexagons, in which the neighborhood alters the algorithm properties including its runtime complexity, which is of special interest in this work, and the percolation threshold itself. Thus, upon the algorithm's completion, the point when $p=p_c$ satisfies is found, the percolation threshold reached in that simulation is returned, which in average with other thresholds acquired with a sufficiently large amount of simulations results in the intrinsic threshold for that system, distinct for each geometrical system composition {\it (and the restrictions applied to the existence of paths, if any)} \cite{Saberi_2015}.
\\\\
Returning to the particular example in Figure 1, the terminal state is reached in 191 iterations, yielding a total of 154 elements within the system, $400-154=246$ free cells, and an occupancy ratio $p=154/400=0.385$, resulting in a threshold $p_c=0.385$ for that particular simulation. However, a comparison of this particular value with other numerical approximations of $p_c$ for a square lattice reveals that it is comparatively smaller. The reason behind this result lies in the threshold of a specific simulation, which can fluctuate depending on the shape of the achieved path on the terminal state. For example, if the path is formed by the minimum number of elements necessary, in this case $n$, the final threshold will be $p_c=n/n^2=1/n$. On the other hand, in scenarios where the path is established after filling the whole matrix except the top and bottom rows, the threshold, very different from the previous one, would be $p_c=(n^2-2n)/n^2=1-2/n$. Therefore, to ascertain the actual threshold for the system, we run a considerably large number of simulations and average all the $p_c$ obtained\cite{Chen_2015}. Moreover, it should be noted that terminal states are not characterized by a singular path connecting the corresponding rows, in this example it is shown that there are many possibilities to trace a valid path, however, the calculation of the above $p_c$ for each simulation outcome is streamlined if we assume that the sequence of insertions performed by the algorithm leads to a unique path.
\\\\
To summarize, the process initiates with an entirely empty matrix. For each iteration the algorithm designates a uniformly random position within the system and verifies the presence of an element in that location. In the absence of an element, it inserts a new one in the cell indicated by that position and invokes the $helper()$ function, which checks for the existence of a valid path from the cell in which it is located to the row whose index is provided in one of its input parameters. This verification is performed recursively, that is, the function calls itself in the neighboring cells and according to their results it concludes whether the process has finished or not.

\begin{figure}[H]
    \centering
    \includegraphics[width=6cm]{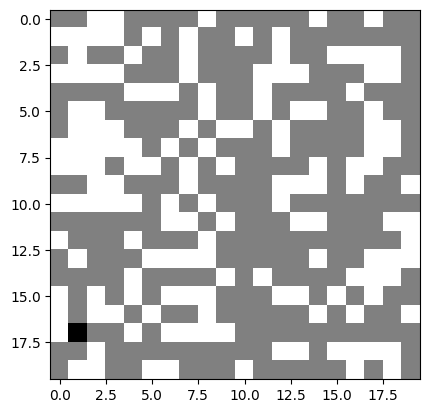}
    \caption{Resulting $20\times 20$ grid after simulation with the last inserted element highlighted}
    \label{fig:gridlast20}
\end{figure}

In Figure 2 the same terminal state is depicted as in Figure 1, albeit with the last added element to the system highlighted, in order to demarcate it from the rest. This is intended to elucidate alternative ways in which the algorithm could be implemented. First, as the iterations in the percolative process are executed and in each one of them there is an absence of a valid path, an element is inserted in a position that may or may not be adjacent to other already inserted elements. What this causes is a progressive change in the number of element clusters within the system over the iterations. In the realm of percolation theory, the study of these clusters, encompassing their formation and their union with others throughout the process to form larger clusters is of paramount significance, since it is the system geometry that delineates the neighborhood between elements, and therefore also the emergence and evolution of clusters along the process \cite{Grossman_1986}.
\\\\
The value of studying the system's clusters becomes apparent when the process reaches its terminal state. During any iteration from the initial to the penultimate one leading to $p=p_c$, there is an indeterminate number of clusters that are not of much relevance at the moment. Nonetheless, when the last iteration is executed and the process ends, an element is inserted adjacent to at least 2 others. Each of these neighboring elements must belong to a cluster that extend to either the uppermost or lowermost matrix row, accordingly. That is, if the process completion coincides with an element insertion, this implies that its adjacent elements constitute a cluster possessing at least 1 element in a row where valid paths start or end, specifically, one of the clusters must have that element in one row {\it (top one, for instance)} and the other in the opposite row {\it (bottom row)}. Hence, the insertion of such an element joins both clusters, thereby forming a larger cluster encompassing the valid path(s) that trigger the termination of the algorithm, typically denoted as a spanning cluster\cite{Cho_2013}.
\\\\
In light of this property, we can think of a data structure adept of effectively representing all the system's clusters at a given time, so that in subsequent iterations the elements verify the set to which their neighboring elements belong, if any, and perform union operations with those that can be joined to the corresponding cluster, or merge several clusters into a larger one. This data structure manifests as a disjoint set, which by applying the Union Find algorithm leads to far more computationally efficient implementations of this type of percolation detection algorithms \cite{Sedgewick_Wayne_2011}, although here we will only address the preceding implementation.

\begin{figure}[H]
    \centering
    \includegraphics[width=6cm]{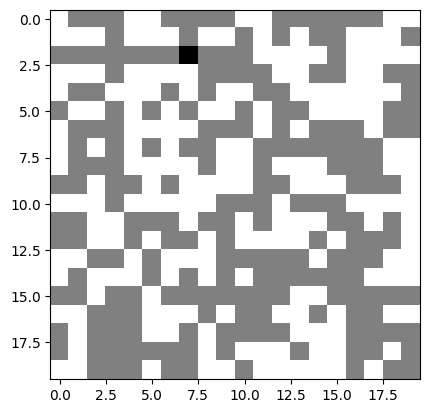}
    \caption{Resulting $20\times 20$ grid after simulation with the last inserted element highlighted and a different insertion sequence.}
    \label{fig:gridlast20_2}
\end{figure}
For instance, if we consider the terminal state illustrated in Figure 3 wherein an element is inserted at position $(2,7)$, it is seen how in that iteration the algorithm implemented by Union Find would perform several union operations. Initially, it would join the cluster of the inferior neighboring element with one of the 2 clusters where the upper neighboring elements belong. Subsequently , it would merge the resulting cluster with the one remaining from the upper row, and that result would be the final cluster containing one or more valid paths, i.e., the spanning cluster that causes the phase transition in the percolative process. Under this philosophy, the computational complexity of an insertion operation in the algorithm implemented with disjoint sets, on average, asymptotically approaches $O(\alpha (n))$. This portends that at every iteration the algorithm performs a nearly constant amount of work, since the inverse Ackermann function grows very slowly as $n$ tends to infinity \cite{InverseAckermann}. Conversely, the other recursive algorithm that we will analyze definitely performs more work, since the path existence verification depends on the number of elements that composing a cluster to which its neighboring elements belong.
\\\\
Before proceeding, it is imperative to acknowledge an important consideration regarding the way in which the elements are inserted. Theoretically, it is assumed that when the algorithm is tasked with selecting a grid cell to perform an insertion, it follows a uniform probability distribution that assigns all system cells the same probability of being selected for an insertion, for all the process iterations. This assumption remains valid solely within the theoretical analysis, since it is computationally impossible to guarantee pure randomness \cite{Zenil2011}. Instead, it will be necessary to consider the seed parameter used to specify the random generator which sequence will be generated in each process simulation. For example, the terminal state in Figure 1 was generated in a process where a random generator had previously been initialized with the seed value 42, giving a sequence of insertions that commences and ends as follows: $(3, 0),(8, 7),(7, 4),(3, 17)\cdots (13, 17),(0, 3),(2, 4),(17, 1)$. Meanwhile, in Figure 2 the random generator was initialized with seed 89, leaving a different sequence of insertions, and therefore a distinct simulation: $(2, 19),(8, 4),(11, 2),(4, 13)\cdots (12, 12),(19, 13),(8, 0),(2, 7)$
\\\\
These simulations are run in Python, so the usage of identical seeds in other languages does not assure the replication of sequences, unless the same generators are involved. Thus, in the case of wishing to calculate some system property that depends directly on the insertion sequence, such as the threshold $p_c$, it will be necessary to conduct multiple simulations with a seed of 0, which indicates a random sequence upon each initialization of the generator.

\subsubsection{Space complexity analysis}
After defining the algorithm's purpose and examining how it operates to reach it, we can proceed with its analysis. Primarily, the computational resources to be addressed are memory and time. Although here we will focus exclusively on the latter, it remains pertinent to have an idea of how much memory the algorithm consumes, since it is a constraining feature when deployed on certain systems.
\\\\
Regarding its space complexity, there are two fundamental points that define the memory needed to run a simulation of this percolative system depending on the parameters that determine its size. On the one hand, there is the matrix epitomizing the system where the elements will be inserted, which will have a generic size of $n\times n$. Thus, building the matrix computationally and storing it in memory incur a consumption of $n^2$ elementary memory units. Secondly, we must also account for the supplementary memory overhead for each insertion, since between any contiguous pair of iterations we can assume that the state of the former is not preserved, i.e., the algorithm executes the $helper()$ function again after resetting the register of cells visited by that procedure. As a result, the register of visited cells will require to be at least the same size as the system matrix, filling $n^2$ units of memory. Furthermore, whenever the algorithm executes the $helper()$ procedure, the recursive calls to that procedure must be buffered in a stack. Therefore, if we think about the worst-case scenario it may encounter, it becomes apparent that the maximum memory consumed by an insertion operation will never scale beyond $n^2$. For instance, if the matrix is populated with elements except for the rows where the paths start and end, and the elements are distributed to maximize the length of each recursive branch, at most as many recursive calls will have to be stored in the stack as the length of the branch with maximum length. This length is not trivial to find out, since it depends on how a cell's neighbors are explored, though it is clear it will not exceed $n^2-2n$ calls. Hence, by combining the memory required to represent the system with that demanded for each insertion operation, we arrive at a maximum of $3\cdot n^2 -2n$ total memory units for a given value of $n$, which translates into a resulting space complexity of $O(n^2)$.
\\\\
As a conclusion, the majority of implementations of this percolative process with recursive algorithms akin to the one described earlier will need at least the memory specified in the asymptotic bound $O(n^2)$, due to the requirement of storing the system's matrix. Nevertheless, the discrepancies among multiple algorithms can be substantial, contingent upon the factors delineated above, so considering the exact function that governs the actual memory usage instead of the asymptotic bound is key when studying the feasibility and efficiency of diverse implementations. In the case where more advanced data structures are used to obviate the need to store the whole system and to facilitate efficient insertion operations, the resulting bound may be reduced below $O(n^2)$.

\subsubsection{Runtime measurements}
Referring the time complexity, we proceed to evaluate the temporal resources requisite for the algorithm to complete its purpose. At first glance, when examining carefully the pseudo-code definition of the whole percolative process, a notable degree of complexity is noticed, hindering our ability to confidently estimate an average asymptotic bound for its runtime. As will be discussed later, this analytical complexity is more profound than it initially appears, especially when trying to determine the most accurate bound possible, analogous to what was already explained for the spatial complexity. For this reason, we should begin by applying other non-exact methods that will simplify the process of deriving such a bound. Fundamentally, the crux of the challenge in performing an exact analysis comes from the different "stages" in which the algorithm can be partitioned, as the count of elementary operations is not equivalent when the algorithm is inserting an element or checking if a cell´s state is free. So, as an approximation, we will measure the complete amount of time spent from the system matrix initialization until the last element is inserted and the algorithm detects the existence of one or more valid paths. Given the impracticality of exhaustively evaluating or replicating all the specific situations attainable by the process and measure its runtime, due to the vast quantity thereof, we shall instead measure the time required to run a sufficiently large number of simulations for different values of $n$ \cite{time_measurements_2D1}.

\begin{figure}[H]
    \centering
    \includegraphics[width=9cm,clip]{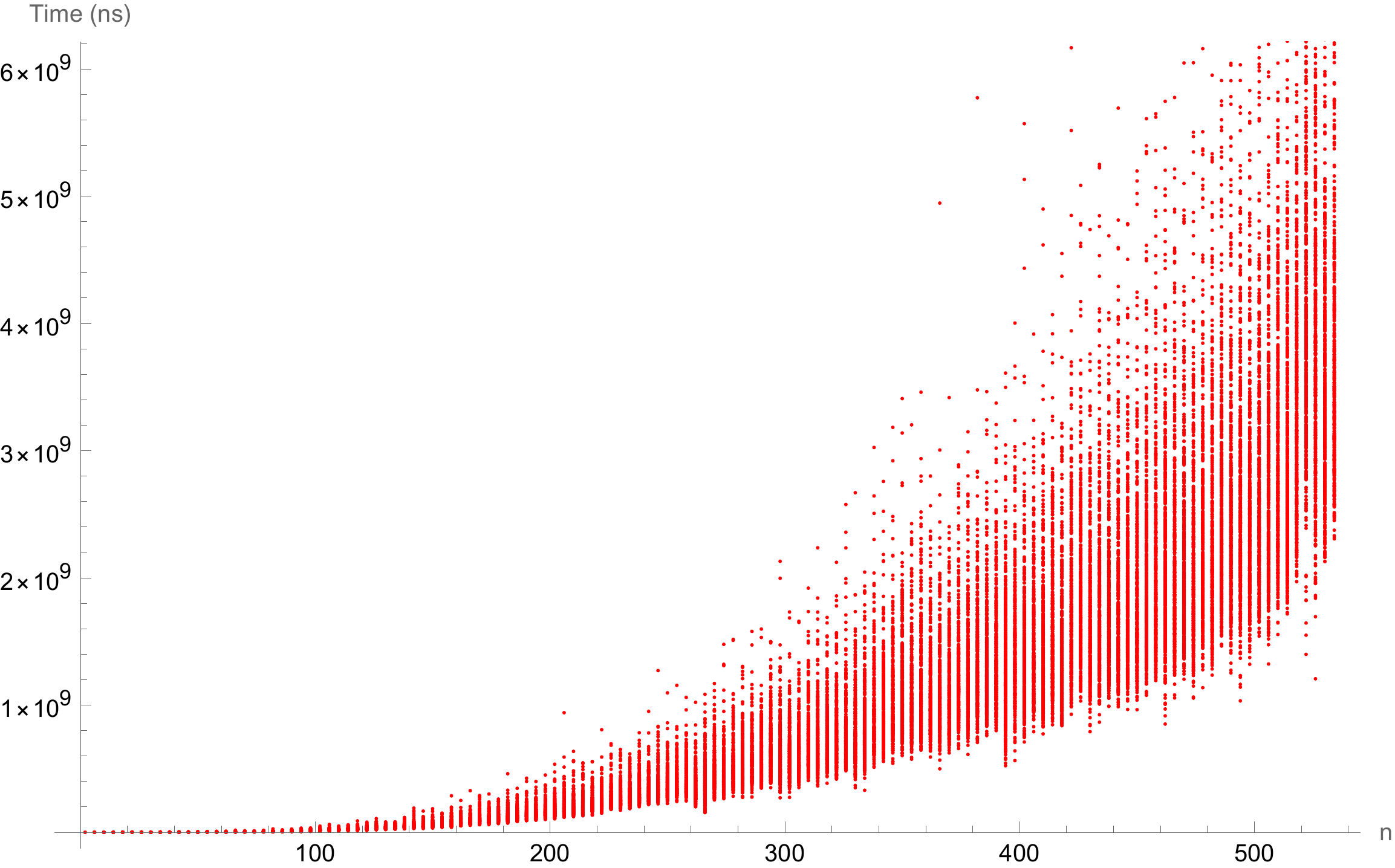}
    \caption{Sample time measurements in nanoseconds for $2\leq n \leq 534$}
    \label{fig:2Dtimes}
\end{figure}

In order to clearly identify how the runtime complexity grows as the system size increases, and to be capable of providing approximations towards modeling such growth, we undertake a sampling of empirical time measurements as shown in Figure 4. Primarily, a judicious range is selected for the matrix side value $n$, in this case starting at 2 and ending at 534. While a higher upper limit would be preferable in order to generate additional data points, simulations with large sizes are markedly time-consuming to compute. Thereby, in order to save computational resources, the axis step on which we represent the matrix side is not 1, but 4, i.e., the measurements are not performed on the sequence $2,3,4,5,\cdots 534$, but on $2,6,10,14,18,\cdots 534$. There is no major impact on the final outcome since the goal is to illustrate how the running time for each simulation behaves when the matrix size $n^2$ is large. Then if the step size is not set unusually high, the trend remains discernible without needing to extend the upper limit of $n$. In second place, for each value of $n$ not only one unique simulation is performed as the algorithm's state space is really large, which in conjunction with its definition causes a significant variability in its running time. In scenarios where a singular simulation is executed, some of them may follow sequences wherein elements are inserted in a manner conducive to obtaining a valid path with substantially fewer iterations relative to others, concomitantly reducing algorithmic overhead, depending on the amount of clusters to be formed. To address this issue, we choose to take $n$ measurements for each corresponding value of $n$. This choice, albeit arbitrary, hinges on the specific algorithm being analyzed. The primary rationale for applying this rule is the measurement dispersion when $n$ increases. As the state space becomes larger, maintaining a constant number of measurements would provide minimal informational value whenever $n$ grows, unless a high constant value is adopted, which would detrimentally impact the computation time.
\\\\
Lastly, there is another relevant detail that warrants attention when collecting the measurements: randomness. As seen before, the algorithm relies on a random number generator to designate the system cells where elements are inserted, and this generator must have a seed value as input parameter to correctly produce a specific sequence of $(i,j)$ pairs. Let's assume for a moment that every simulation corresponding to the data points measurements are performed with a random generator initialized with always the same seed value, that is, in every simulation run the generator is newly started with an identical seed value. Under this premise, it can be confirmed that all the generated sequences of $(i,j)$ positions would be identical for the same value of $n$. Though, this doesn't pose any issue for different sized matrices since the position pairs meet restrictions such as $0\leq i\leq n \land 0\leq j\leq n$, resulting in different particular values being returned despite using the same generator with the same initialization state, or in other words, the generation of a sequence of values depends on the range boundaries where each value resides. As a sample, Figure 5 exhibits a simulation with the same seed as in Figure 1, except with a different matrix size, yielding a significantly different result, noticeable both in the occupancy ratio $p=p_c=58/100$ and in the required number of iterations, 88.

\begin{figure}[H]
    \centering
    \includegraphics[width=4cm,clip]{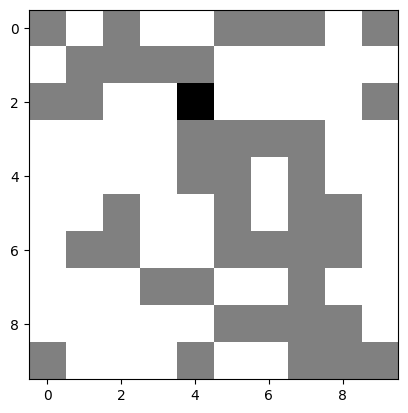}
    \caption{Resulting $10\times 10$ grid after a simulation with the same random generator as in Figure 1 {\it seed=42}}
    \label{fig:gridlast10}
\end{figure}

However, the problem is encountered in simulations of uniform matrix dimensions. For a certain $n$, if all the simulations are executed under the same seed value, the insertion sequence will be equal in all of them. And, the resultant time measurements would resemble those presented in Figure 4, notwithstanding their inaccuracy. The fundamental cause for the invalidity of the data points would be the inherent hardware variability. When the runtime of a procedure on the same hardware is measured multiple times, it is found not to be identical. This discrepancy arises due to the architecture itself and the state in which all its components are in. Such variability would correspond to a dispersion of the data points along the vertical for the value of $n$. Namely, the final plot with every measurement would reveal a similar growth as $n$ grows, but the dispersion of the data points for each matrix size would represent the time difference it has taken to run the same simulation with the same insertion sequence. If we carried out the analysis on that data, we would not be capturing information over the entire process state space, thereby making the inferred complexity bound deviate from the true growth rate.
\\\\
In order to address this issue, we can proceed in several ways. On the one hand, the easiest possibility involves leveraging a feature of the random generator whereby if we preclude it from being assigned a specific seed number at each simulation launch, it will automatically initialize itself with a “uniform” seed. Internally, it selects sufficiently random values so that each time a new instance of a generator is built, it can deliver a uniform random sequence relative to other generator instances. For example, using the current date and time or the process ID to determine the seed are some of the most commonly used techniques \cite{randomnumbersNaoiseHolohan}. In this manner, when running numerous simulations without a predefined seed, we will guarantee for each of them that its insertion sequence is uniformly random, successfully probing the entire sequence space. Alternatively, it is also possible to complete all the simulations with a unique random generator, initialized prior to the first simulation with an indifferently selected seed number. By doing so, treating all the insertion sequences collectively as a singular sequence would emulate the behavior of a “universal” random generator. Hence, by achieving a uniform complete sequence we can ensure as in the previous case that all simulations performed for a certain $n$ explore sufficiently well the algorithm's state space.

\subsubsection{State space and Sequence space}

Before estimating the process time complexity with the former data, the concepts of state space and sequence space must be correctly defined.
\\\\
At one side we have the state space, which refers to the set of possible states the system may be encountered at any iteration. Accordingly, if we have a matrix of size $n$ we can denote the state space as the set $S_{n}$. Overall, the set will be composed of matrices containing from 0 to $n^2$ elements, in any conceivable arrangement, in order to account for all the possibilities the algorithm has when selecting an element's insertion position. Thus, if we want to calculate how large the state space is to understand how long a simulation, or sequence of simulations, might last, we must find an expression for its cardinal.

\begin{align}    
    |S_{n}|=\sum _{k=0}^{n^2} \binom{n^2}{k}=2^{n^2}
\end{align}
\\\\
As can be seen above, the cardinality $|S_{n}|$ is achieved by summing the complete range of elements the system can contain $0\leq k \leq n^2$. For each amount of elements, we enumerate every permutation where they can be arranged within the system, which, with its structure being a square matrix, leaves a total of $\binom{n^2}{k}$ arrangements. And, by applying the sum of coefficients row property of the binomial theorem, it grants a cardinal of $2^{(n^2)}$ conceivable states for a matrix of side $n$. Finally, it should be noticed that regardless of its appearance, the cardinal function does not exhibit super-exponential growth \cite{99882}. The main reason is the interpretation of $n^2$ within this context, i.e., the value approaching infinity is $n$, therefore by resembling an exponential function of constant base and variable exponent with respect to $n$, even though the exponent growth is not linear, it is still typified an exponential function.
\\\\
Besides, we also have a sequence space \cite{Buchholz_2017} encapsulating the necessary information about the feasible process executions. Distinct from the previously discussed space, the purpose of this set is to store every possible insertion sequence that may manifest within a simulation. If we denote this set as $S'_{n}$, each element contained in it will comprise a sequence of position pairs $(i,j)$ representing all the positions randomly generated throughout an execution, sequentially ordered from the initial iteration to the terminal state. Then, to ascertain how many sequences a process may follow for a given matrix size, we must compute its cardinality, which prima facie could appear to be infinite. Under the assumption that the positions of an arbitrary $S'_{n}$ element are randomly selected, there exists a negligible possibility that 2 or more pairs $(i,j)$ will coincide in contiguous or proximate iterations, exemplified by the sequence: $(1,2),(6,2),(0,0),(0,0),(0,0),(4,1),\cdots$. Since the process persists until a path is constructed, there are infinite sequences wherein one of its positions is generated repeatedly an indeterminate number of times. Evidently, the more repetitions required, the less probable it is for that sequence to arise; nonetheless, it remains a possibility worth acknowledging. Thus, we could infer that the cardinality is infinite, $|S'_n|=\infty$.
\\\\
Although under these conditions the cardinal would be correct, it is impractical to hold a set with infinite elements, since it would be complex to formalize the exact enumeration and comparison with related sets. So, to compute a cardinal worthwhile for the complexity analysis, we must use the probability of each sequence's occurrence. To this end, we will calculate the total amount of existing sequences with a length constraint, which is what allows an infinite number of sequences to arise for a given $n$.
\begin{align}    
    |S'_{n}|=\prod _{k=0}^{I(n)} n^2=n^{2\cdot{I(n)}+1}
\end{align}

The limiting constraint for the number of sequences inside the set is found in the function $I(n)$, which returns the average number of iterations the algorithm needs to reach its terminal state. This function proceeds from the mean of a distribution that we will later see is not computable, owing to the lack of a closed form for an intermediate expression, but at the moment we will continue with it. If every existing sequence has a maximal length estimated as $I(n)$, enumerating all sequences involves finding out how many pairs $(i,j)$ can be generated at each iteration and in how many ways they can be arranged. Initially, any matrix cell can always be chosen to insert an element, so there are $n^2$ possibilities available, i.e., positional pairs. However, in subsequent iterations, pairs that have already appeared previously in the sequence may be selected again, since the insertion probability remains uniform for every matrix cell. And, since in this case we are ensuring the sequence is finite because of $I(n)$, the cases where there are multiple equal pairs merely represent situations where the algorithm abstains from inserting new elements, they do not suppose an impact on sequence lengths and set size as before. Hence, we will get another $n^2$ possibilities available for every iteration producing the sequence elements. As a consequence, if we have the same amount at each iteration, the product of the whole will be the cardinal of $S'_{n}$ we are seeking.
\\\\
Ultimately, even in the absence of an expression for $I(n)$, we can quickly realize it will require to equal at least $n$ for any input parameter. Specifically, for an $n$-sided matrix, a minimum of $n$ insertions is required to build a valid path, leaving a cardinal of $|S'_{n}|=n^{2\cdot n}$. Comparable to the cardinality of the state space, $|S'_{n}|$ is likewise regarded as exponential, although in this case it can be observed and verified that its growth is slower.
\begin{align}  
    L = \lim_{n \to \infty} \frac{n^{2n}}{2^{n^2}}  
\end{align}

\begin{align*}  
    \ln (L) = \lim_{n \to \infty} \ln ( \frac{n^{2n}}{2^{n^2}} ) = \lim_{n \to \infty} ( \ln(n^{2n}) - \ln(2^{n^2}) )
\end{align*}

\begin{align*}  
    \ln (L) = \lim_{n \to \infty} (2n \ln (n) - n^2 \ln (2)) = \lim_{n \to \infty} (n \ln (n) - n^2)
\end{align*}

\begin{align*}  
    \lim_{n \to \infty} \frac{ln(n)}{n}=0 \implies \lim_{n \to \infty} (n \ln (n) - n^2) = -\infty
\end{align*}

Given $C\cdot ln(n)<C\cdot n$ for $C>0$ and a sufficiently large $n$ \cite{488012}, also denoted as $O(log(n))\subset O(n)$, the ratio between $ln(n)$ and $n$ approaches 0, solving the original limit.

\begin{align*}  
    \ln (L) = -\infty \implies L = e^{-\infty} = 0    
\end{align*}

Consistently, when $n$ approaches infinity, the cardinal $|S'_{n}|$ grows slower than the state space one, which is also shown in Figure 6.

\begin{figure}[H]
    \centering
    \includegraphics[width=10cm,clip]{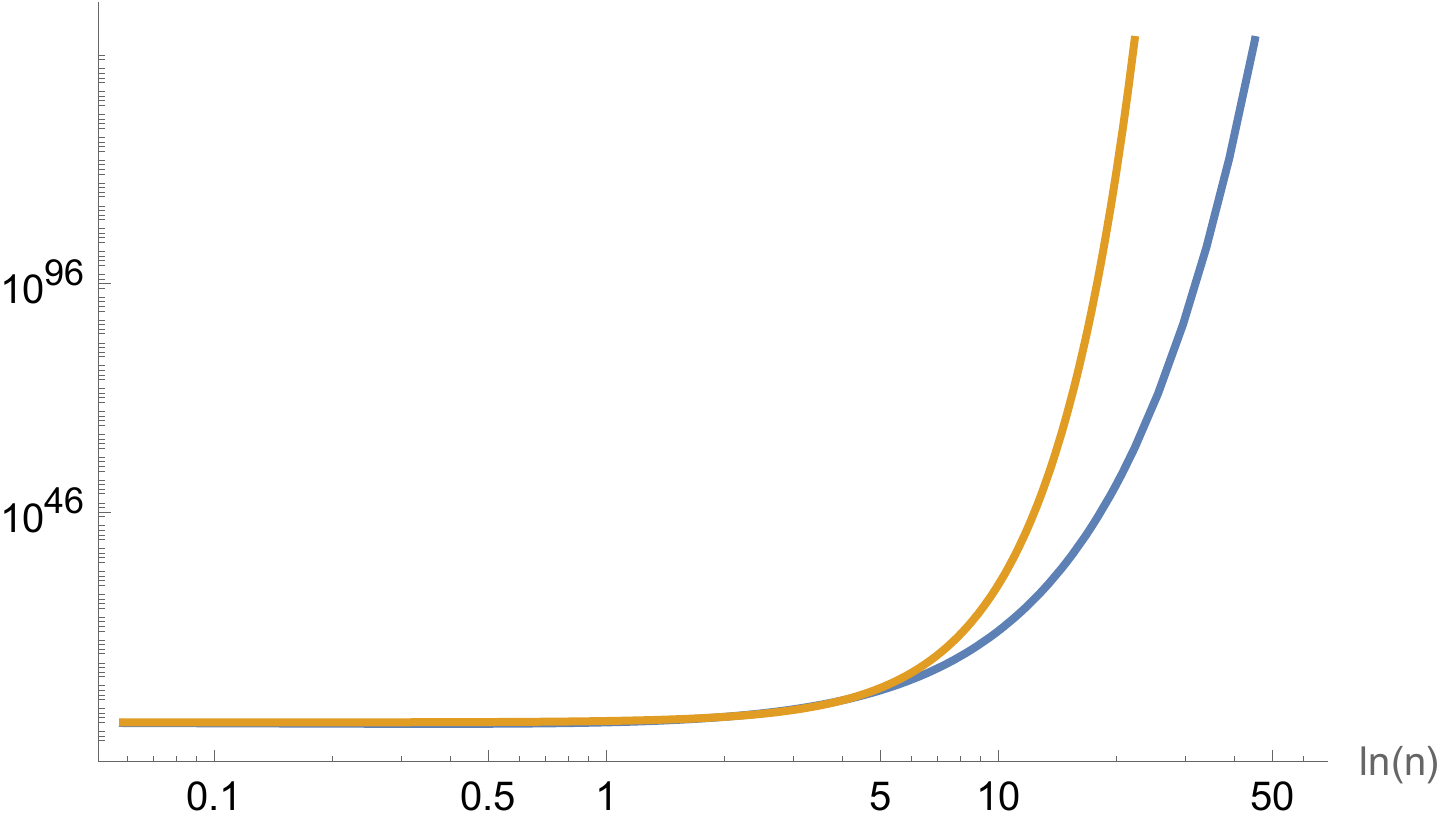}
    \caption{Logarithmic scaled plot displaying $|S'_{n}|$ and $|S_{n}|$ in blue and orange respectively.}
    \label{fig:cardinals}
\end{figure}

\subsubsection{Average time complexity estimation}
\label{subsec:Average-time-complexity-estimation}
Relying on the former sets, we continue to formalize the calculation of the algorithm's time complexity growth. At first, referring the data from Figure 4, it is apparent the existence of several bounds delimiting the data points spread for each group of simulations with the same $n$. These bounds, depending on their location, determine the algorithm's temporal growth in its worst and best case scenarios. On one side, the function that remains beneath all data points and remains in closest proximity, stands for the best case, which in our algorithm could be, for instance, sequences where elements are inserted in the most “efficient” way available, so the process terminal state is reached in few iterations. The worst case, in contrast, is the function bounding the data from above, involving sequences with a complementary “efficiency” to the preceding one. Nevertheless, these concrete situations are straightforward to analyze extensively and simple to provide an accurate expression for its growth, as we will discuss ahead. The difficulty stems from the average case, which proves to be a reliable metric for verifying and comparing the temporal complexity with alternative implementations. Graphically, it is visualized as the function composed by every point where most accumulation of final measurements is located in each $n$, therefore, if there were measurements for every valid simulation, it would be their mean.
\\\\
Henceforth, the key nature of all the specific details attended to during the measurement phase is established, as they are essential to accurately approximating the average case. To understand why, it is necessary to formalize its derivation, which, depending on how we use the above space sets, there are several approaches available. Still, the general procedure for finding the average time any algorithm needs to finish for a given input $n$ is invariant:

\begin{align}  
    T_{avg}(n) = \frac{\sum_{\forall s \in S_{n}} T(n, s)}{|S_n|}
\end{align}

Note that in this expression, $S_n$ doesn't refer to the state space, but rather to a set that depending on the algorithm's nature may contain all the different inputs it can receive, or structures describing valid executions, as in our case, the $(i,j)$ sequences within the sequence space. Then, if we are able to build it, this set will yield information about all the existing executions performable by the algorithm over an input of size $n$. Thereby, if we take each of its elements $s$ and compute the runtime for that case with the function $T(n,s)$, we will graphically retrieve a vertical at $x=n$ containing all the runtimes the algorithm can possibly require for that input size, where its maximum $\sup_{M \in T_{S_n}}$ corresponds to the worst case scenario and its infimum $\inf_{m \in T_{S_n}}$ to the best. With this, computing the average time $T_{avg}(n)$ is now reduced to computing the mean of all resulting runtimes, so they are aggregated and the total divided by the cardinal of the set $S_n$.
\\\\
As expected, exhaustively generating every instance of $S_n$ or sampling uniformly enough over the set such that complexity can be inferred from measurements is computationally expensive, as already stated. For this reason, running simulations with uniformly generated sequences, which is significantly more feasible, suffices to approximate the resulting runtimes set $T_{S_n}$ at whichever level of detail selected. Additionally, due to the rapid growth of $|S_n|$ when $n$ approaches infinity, the potential spread the measurements may suffer for a certain $n$ compels to run a number of simulations increasing with respect to $n$, ideally, at a similar rate as the cardinal. Since it is too costly to produce so many simulations, it is enough to keep it proportional to $n$, as mentioned above.\pagebreak

Within our particular algorithm, we must rely on the sequence space set $|S'_{n}|$ to find an expression for $T_{avg}(n)$, as it contains every conceivable scenario when running the process for a given $n$. This set's cardinal is already calculated in advance, although not in an exact closed form. Nevertheless, in addition to the cardinal, it is necessary to create a formula that, starting from a concrete sequence $s$, returns the time, as exact as possible, the algorithm will require to complete. 
\begin{align}  
    T_{avg}(n) = \frac{\sum_{\forall s \in S'_{n}} T(n, s)}{|S'_{n}|}
\end{align}
This task presents a significant challenge, yielding results of limited utility for the analysis. One, given the absence of a closed-form expression for $I(n)$, the sole conclusion we can draw for now about the denominator cardinal is that it must exceed a certain bound. Consequently, we establish a marked difference between the actual number of elements present within the set and the ones we are counting with the bound, so the final growth is likely to appear potentially skewed. Furthermore, even if we were able to provide an exact expression of $|S'_{n}|$ with a different approach, we would still encounter the problem of determining $T(n, s)$, which would return something equivalent to the amount of work performed by the algorithm in case the execution is defined by the insertion sequence $s$; there is no need to strictly return time units, since an instruction, basic operation, or elementary unit of work can be trivially translated by its unit equivalence to time. The reasons why $T(n, s)$ currently doesn't have a closed form will be discussed in forthcoming sections, but at this point we can attribute it to the system geometry and the modelling of each sequence $s$, i.e. the amount of them generating symmetries, adjacencies between intermediate clusters along the process, immediately influencing the runtime outcome, or the path count for a given process state, among others.
\\\\
Additionally, $T_{avg}(n)$ may be outlined through the state space, under specific constraints:
\begin{align*}  
    T_{avg}(n) = \frac{\sum_{\substack{\forall s \in S_{n} \\ i(s)=1}} T(n, s)}{|S_{n}|\cdot \frac{\sum_{\substack{\forall s \in S_{n} \\ i(s)=1}} 1}{|S_{n}|}} = \frac{\sum_{\substack{\forall s \in S_{n} \\ i(s)=1}} T(n, s)}{\sum_{\substack{\forall s \in S_{n} \\ i(s)=1}} 1}
\end{align*}
If we define $i(s)$ as a function that returns 1 if the matrix $s$ represents a terminal state and 0 otherwise, we can confine the state space $S_{n}$ to a set including exclusively every terminal state. Then, the average runtime computation approach would imply that in this context $T(n, s)$ does not receive an insertion sequence, but rather a terminal state, and returns the time needed to reach such a state. Therefore, average time is found by aggregating the time for all elements of the restricted $S_{n}$ set, namely, the execution times for all terminal states $(i(s)=1)$, divided by their quantity. In order to count how many there are, we multiply the state space cardinal by the ratio of terminal states among all the existing system states, leading to a simplification that leaves nothing but the actual number of terminal states in the denominator, which is meaningful considering that the average is computed with respect to every terminal state. And finally, as above, there isn't a straightforward approach to procure an exact expression for $T(n, s)$, note that in this case the formula has to return the runtime from a terminal state, so the variability within the sequences resulting in a particular state hinders the intermediate analysis for the same reasons, counting clusters along the process, adjacencies or symmetries between them, amongst others.
\\\\
Following an examination of the formal approach to average time complexity growth, and realizing that there isn't an attainable method to derive exact expressions for it {\it (for instance $i(s)$)}, we proceed to estimate such growth from the measurements provided in Figure 4. For this endeavor, we must first select a suitable model that will fit the data. But, it doesn't conform to any pre-established form, as the actual functional representation we aim to approximate is initially unknown. If we were dealing with the worst or best case scenario, as we will see ahead, it would have a concrete $n$-dependent form that we could parameterize as a model, yet due to the average case complexity we do not know beforehand its concrete substantive shape. However, for a particular $n$ we know the amount of elements that a system defined by that value will be able to hold. Moreover, the algorithm's workload, despite its dependence on concrete situations of its state and sequence spaces, will be ultimately a function of system size, as will those specific situations. Therefore, the model must inherently incorporate a dependence on $n$.

\begin{align}  
    T_{avg}(n) = n^x \quad \colon \quad x \in \mathbb{R}
\end{align}

Ultimately, it is settled to use a model with polynomial growth where its exponent $x$ will be fitted according to the data. One reason this form is preferred lies in its ability to bound all algorithm cases. For one thing, if $x$ is large enough, it will be able to upper bound the worst case with a function that is greater than any collected data point. Conversely, as we will discover when examining the best case, there exists a function of the form $n^x$ for which every data point lies above, with a correspondingly small exponent $x$. Then, if we manage to tune this parameter towards an intermediate value between the former ones, we will be able to bring the function growth closer to the actual average case growth rate, which is precisely the area where most data points are gathered. This process will be executed through a numerical method, ensuring an acceptable level of correlation and accuracy is achieved for the submitted data.

\begin{figure}[H]
    \centering
    \includegraphics[width=10cm,clip]{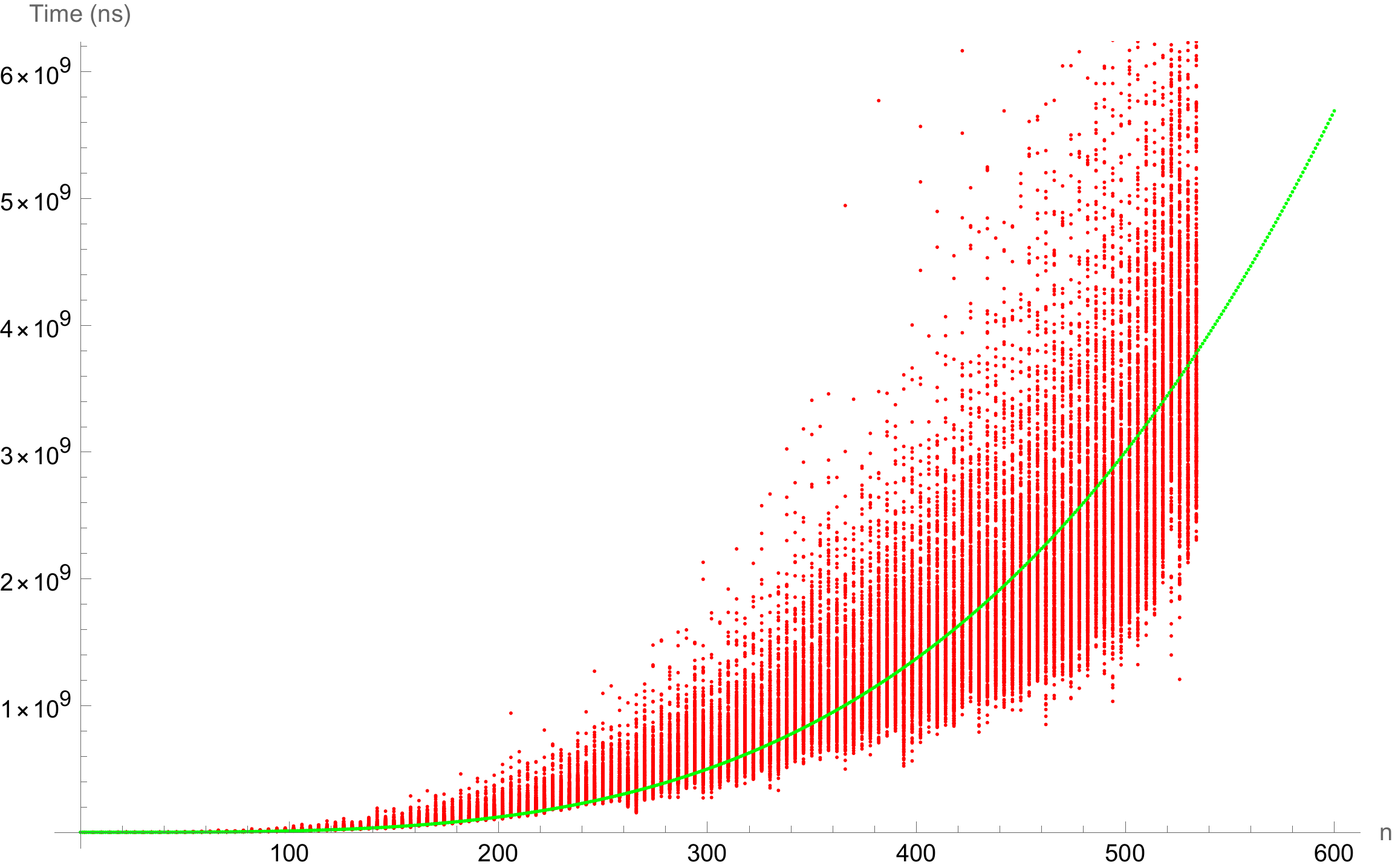}
    \caption{Time measurements from Figure 4 plotted in red and the resulting $T_{avg}(n)=n^{3.51}$ fitted model in green}
    \label{fig:2Dtimesfit}
\end{figure}

As shown in Figure 7, after fitting the $T_{avg}(n)$ model to the runtime measurements with Wolfram's $NonlinearModelFit[]$ function \cite{Wolfram_NonlinearModelFit_2008}, it is noticeable how it traverses through all the points where most measurements are found for each $n$. This occurs because the data points in each vertical follow a probability distribution determined by the algorithm's nature and its sequence space, such that the $n$ measurements will be located according to that distribution. For the worst case, it is observed that as $n$ increases it becomes more unlikely to occur, which is similarly true, though to a minor extent, for the best case. But, the intermediate range where the remaining points lie is the region over where the average case resides. Hence, if the fitted model grows consistently with this region, it is a fairly accurate indicator that the approximation is valid, as long as the model does not overfit or underfit nor suffers from any issues arising from the numerical method it proceeds, meaning it is capable of generalizing for $n$ values larger than those found in the data set used for the fit. 
\\\\
Resultantly, an approximate average time growth of $n^{3.51}$ is achieved, which computationally can be denoted as $\Theta (n^{3.51})$. Still, since it is an approximation, it is more suitable to use $O(n^{3.51})$ because the actual growth may not have exactly the same decimal places in its exponent, or may have a different but very similar form, predominantly with a slower growth. However, although we do not yet know its actual shape, we can check the goodness of fit with different metrics, as shown for example in the ANOVA table:

\[
\begin{array}{l|lll}
    \text{} & \text{DF} & \text{SS} & \text{MS} \\
   \hline
    \text{Model} & 1 & 1.16907\times 10^{23} & 1.16907\times 10^{23} \\
    \text{Error} & 35911 & 1.43512\times 10^{22} & 3.99632\times 10^{17} \\
    \text{Uncorrected Total} & 35912 & 1.31258\times 10^{23} & \text{} \\
    \text{Corrected Total} & 35911 & 5.71212\times 10^{22} & \text{} \\
\end{array}
\]

Overall, it is shown that a considerable part of data variability is explained by the model. Nevertheless, it is especially evident with the p-value and t-statistic:

\[
\begin{array}{l|llll}
    \text{Parameter} & \text{Estimate} & \text{Standard Error} & \text{T-Statistic} & \text{P-Value} \\
   \hline
    x & 3.51141 & 0.000299642 & 11718.7 & 0. \\
\end{array}
\]

In the p-value case, the result is nearly 0, to the extent that it is not representable with the established digit precision, thus we infer that it is close enough to 0 to satisfy $p<\alpha$ for almost any $\alpha$. In turn, the value of $t$ is meaningfully elevated, suggesting a strong rejection of the null hypothesis for the exponent.
\[
\begin{array}{l|lll}
    \text{Parameter} & \text{Estimate} & \text{Standard Error} & \text{Confidence Interval} \\
   \hline
    x & 3.51141 & 0.000299642 & \{3.51082,3.512\} \\
\end{array}
\]

Finally, we notice that the confidence interval is relatively narrow, resulting in an exponent value $x$ close to 3.511. Furthermore, the process returns a value for the adjusted $R^2$ of 0.890661, which by being notably close to 1 signals an adequate goodness-of-fit. Likewise, the value for the unadjusted $R^2$ is 0.890664, although at these fits we will focus on the former. In this case, it is inconsequential to account for the adjusted $R^2$ or the unadjusted one, since they are very close. But, in other models parameterized with additional variables, the disparity can be substantially larger. Consequently, it is convenient to penalize the inclusion of superfluous variables into the model, something that the adjusted $R^2$ aptly reflects. On such wise, by using this value we can reliably compare the goodness of fit for models with different forms. If we were to use the unadjusted $R^2$ metric, models that over-fit the data with parameters potentially irrelevant to the actual growth would have a high $R^2$ compared to simpler models that might better resemble the true solution \cite{Tatachar2021}.
\\\\
Even though we have constructed a function for $T_{avg}(n)$ that properly reflects the growth trend of runtime measurements, there are several nuances that can enhance the resultant model's performance. The foremost and most conspicuous aspect pertains to its form, and since we cannot determine it for the actual outcome, we must provisionally accept something similar to $n^x$. Yet, considering that measurements start in a range with $n\geq 2$ and time units are unnormalized, even if the algorithm runtime is low for the smallest $n$ values, time readings will be in nanoseconds, as seen in plots, resulting in potentially elevated data points at the range's onset. This suggests that the model doesn't necessarily have to traverse through the point $(0,0)$, or approach it closely when $n=0$, which is precisely a constraint that the previous model $n^x$ satisfies. The problem posed from this constraint is that the curve connecting all the points with a greater measurement density doesn't need to intersect such point, despite theoretically the function that truly models the average time should be 0 for a matrix of null size, causing a slight bias when fitting the data with a model that does comply with the constraint. In essence, since it is an approximation, the bias shouldn't influence the result meaningfully, although there are alternatives to prevent it.
\begin{align}  
    T_{avg}(n) = n^x+b
\end{align}
The most intuitive one is to incorporate an adjustable offset parameter $b$ to the model, thereby enabling it to determine the optimal vertical positioning for its predictions. By this means, when $n$ is small, the model will be able to better fit the high measurement values, and so for the remaining matrix sizes it will align as nearly as possible with the data points growth, since if the model height matches the dataset's one at the smallest $n$, the goodness of fit only depends on the growth rate determined by the exponent, which is what we are pursuing with this approach.
\\\\
The second detail we can work with to improve the model's performance is the fitting technique. Previously, by not using a linear model, it was imperative to apply a numerical method to find a suitable $x$ exponent. Among the options Wolfram offers us are Newton's method, some other founded on gradient optimization, or the Levenberg-Marquardt algorithm\cite{More1978}. These methods exhibit remarkably similar outcomes, though in certain situations they can get stuck at local minima or maxima, impeding the attainment of the global optimum solution. There exist mechanisms to overcome such issues, or solutions that require more numerical process iterations to ensure convergence. In our case, as there isn't a massive amount of data and the fitting model is rather simple, we don't need to be concerned about the convergence rate of such methods, but rather about the outcomes they yield. Therefore, to avoid potential pitfalls and secure an improved fit, as well as to count with an alternate fit to Figure 7 and, at the same time, generate additional insights about the growth we want to model, we will proceed with the simplest model available.
\begin{align}  
    T_{avg}(n) = m\cdot n+b
\end{align}

When ascribing $T_{avg}(n)$ a purely linear growth, we simplify the whole fitting process and prevent potentially falling into suboptimal solutions. This is attributed to the convexity of the cost function to be optimized \cite{2774106}, which guarantees that all local minima therein are at the same time global; therefore, the fitting method will only manage to reach that minimum, which proves to be the most optimal solution. However, we are causing a further inconvenience by adopting this method: we confine the growth of $T_{avg}(n)$ to a linear function, which serves as a lower bound for the algorithm's best case but does not upper bound its worst case. Moreover, as shown in measurements, the average case growth is patently nonlinear. To ameliorate this concern and estimate the average time from a linear model, a transformation is performed on data and matrix size axis such that the new line's slope is the exponent of the variable $n$, and the offset parameter $b$ retains the same role as in the original model.

\begin{align*}  
    T_{avg}(n) = b\cdot n^x 
\end{align*}
\begin{align*}  
    ln(T_{avg}(n)) = ln(b\cdot n^x) = ln(b) + ln(n^x)
\end{align*}
\begin{align}  
    ln(T_{avg}(n)) = ln(b) + x\cdot ln(n) \label{linear_model}
\end{align}

If we begin from the nonlinear model we originally had and include a constant $b$ as a multiplier, we can apply the function $ln()$ on both sides of the equality and leverage its properties to detach $b$ from the remaining model through an addition. Then, in the second addition the exponent $x$ can be extracted as a multiplier of the logarithm of $n$, resulting in the expression intended for model fitting, even though we need to undo the logarithm to get the actual prediction values.

\begin{align}
    T_{avg}(n) = e^{x\cdot ln(n) + ln(b)} \label{linear_model_actual}
\end{align}

As shown in the \textcolor{blue}{\ref{linear_model}} model expression, the left side of the equality resembles the dependent variable of a linear regression, whereas the other side corresponds to its slope, offset, and independent variable parameters. Respectively, the offset will also have undergone the logarithmic transformation, so it will match the term $ln(b)$. On the other hand, the slope of the new line becomes $x$, so the independent variable will be its multiplier $ln(n)$. Under this approach, we are required to transform all our runtime measurements as follows:
\begin{align*}
    t'=ln(t) \quad n'=ln(n) \quad b'=ln(b)
\end{align*}
\begin{align}
    t'=b'+x\cdot n'
\end{align}

For every time $t$, its new value for which the fit will be conducted will be its natural logarithm, primarily because $T_{avg}(n)$ is under the effect of the transformation in the model expression. Likewise, all matrix sides $n$ must be transformed for the same reason, since the new independent variable is redefined as $ln(n)$.

\begin{figure}[H]
    \centering
    \begin{subfigure}[b]{0.49\textwidth}
        \centering
        \includegraphics[width=\textwidth,clip]{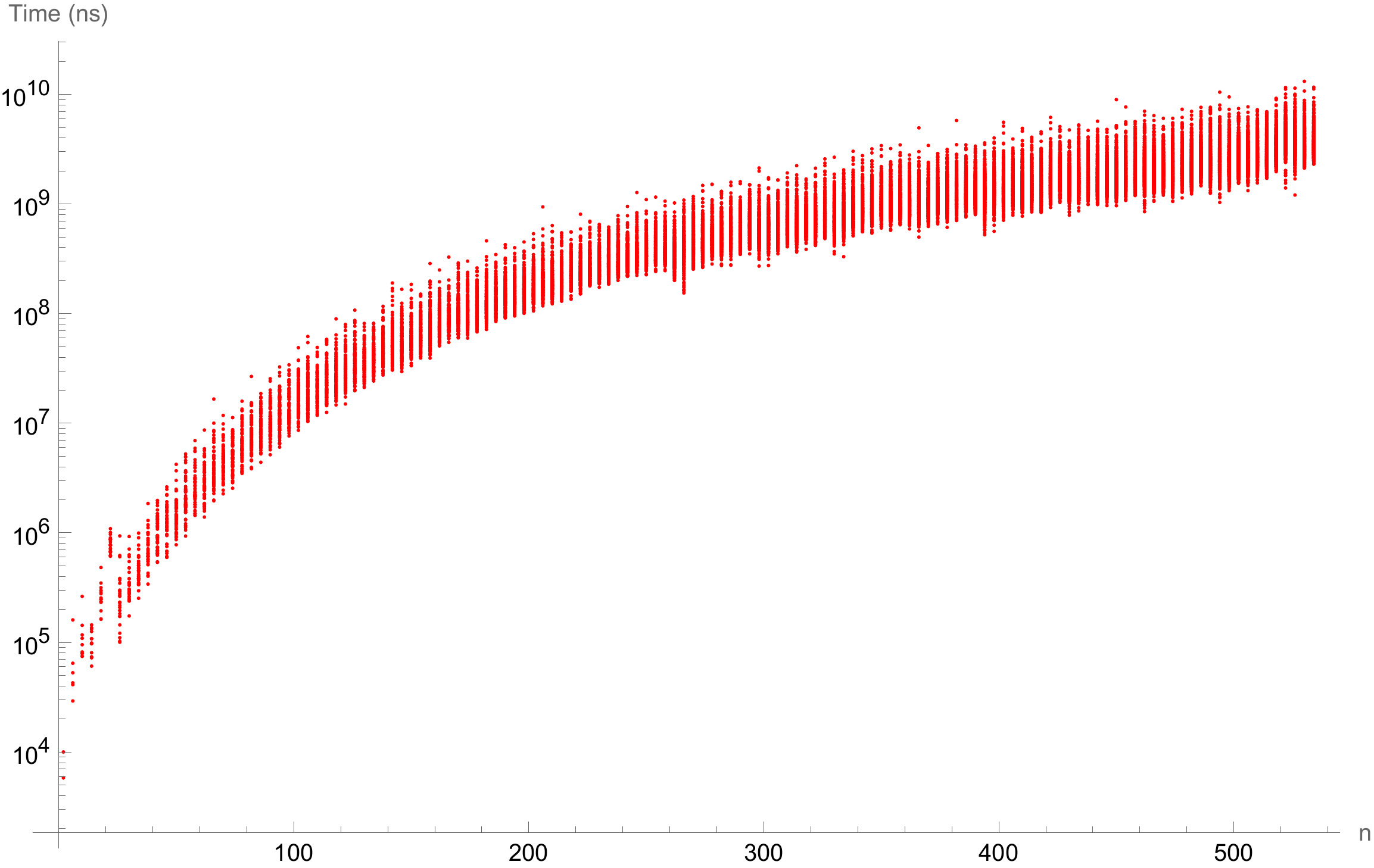}
        \caption{}
        \label{fig:2Dtimeslog}
    \end{subfigure}
    \hfill
    \begin{subfigure}[b]{0.49\textwidth}
        \centering
        \includegraphics[width=\textwidth,clip]{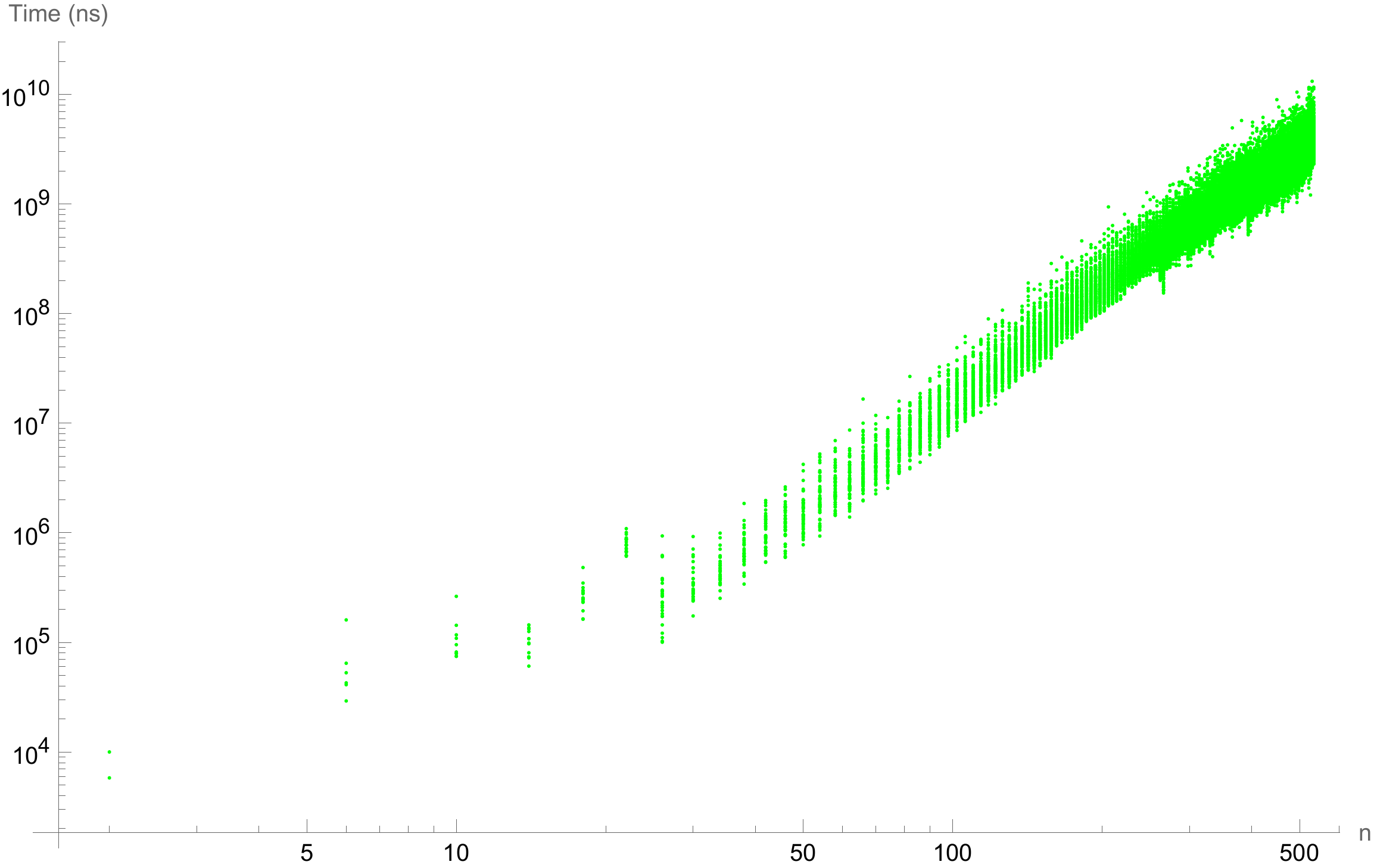}
        \caption{}
        \label{fig:2Dtimeslog2}
    \end{subfigure}
    \caption{Sample time measurements in nanoseconds from Figure 4. (a) Solely the transformation $t'=ln(t)$ has been applied upon runtimes (b) Similarly to the previous one, runtimes and also the $n$-axis are transformed, leading to the dataset where the linear model $t'=b'+x\cdot n'$ will be fitted.}
    \label{fig:time_comparison}
\end{figure}

Subsequent to the transformation, data transitions from having a growth visually similar to a function of the form $n^x$ to a linear one, as can be appreciated above. This allows us to apply the linear fit that will benefit from the efficient implementations of its multiple fitting methods and will faithfully model the growth of the original data as previously explained.

\begin{figure}[H]
    \centering
    \includegraphics[width=10cm,clip]{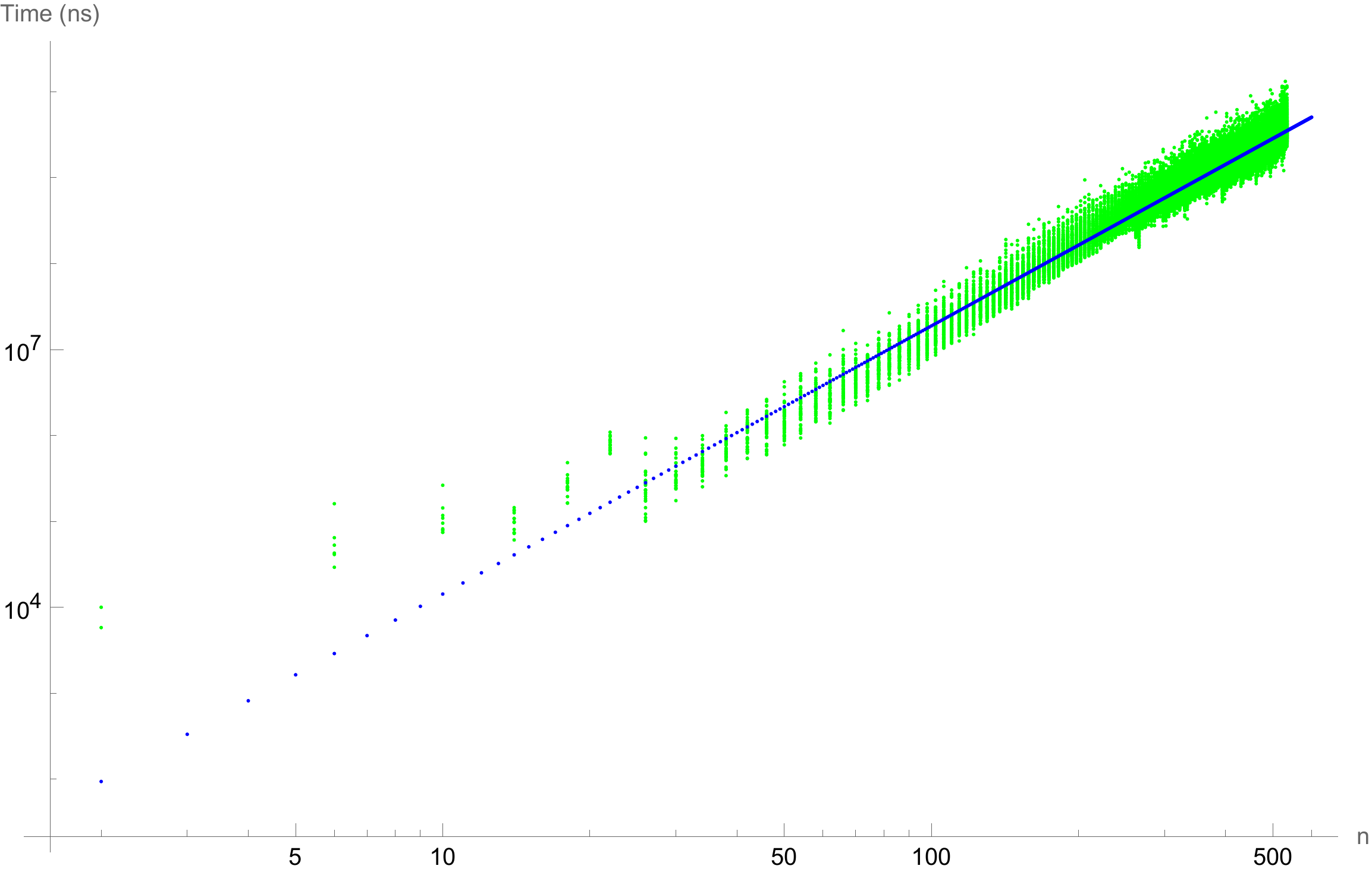}
    \caption{Transformed time measurements from Figure 8 (b) plotted in green and the resulting $T_{avg}(n) = e^{3.12\cdot ln(n) + 2.37}$ model in blue. {\it (Also referred to as $ln(T_{avg}(n)) = 3.12\cdot ln(n) + 2.37$ in its pure linear form.)}}
    \label{fig:2Dtimeslogfit}
\end{figure}

Figure 9 shows the outcome of fitting the linear model $ln(T_{avg}(n)) = b' + x\cdot n'$ to the transformed dataset. Here, the used fitting method stems from the same software as the previous one, although in this case the $LinearModelFit[]$ function \cite{Wolfram_LinearModelFit_2008} has been applied instead due to the model's nature. Therefore, the complexity growth derived from such a model can be denoted as $\Theta (n^{3.12})$, though more prudently $O (n^{3.12})$.

\[
\begin{array}{l|lllll}
    \text{} & \text{DF} & \text{SS} & \text{MS} & \text{F-Statistic}  \\
   \hline
    x & 1 & 87227.2 & 87227.2 & 833655.  \\
    \text{Error} & 35909 & 3757.24 & 0.104632 & \text{}  \\
    \text{Total} & 35910 & 90984.4 & \text{} & \text{}  \\
\end{array}
\]

Regarding the ANOVA table, we realize that modifying the model type leads to different metrics. In general, similar to the preceding fit, most of the data variability is explained by the model. But, now it is worth noting the new F-statistic metric of the slope $x$, whose high value signals a strong dependence between the independent variable $ln(n)$ and the data we fit the model to.

\[
\begin{array}{l|llll}
    \text{Parameter} & \text{Estimate} & \text{Standard Error} & \text{T-Statistic} & \text{P-Value} \\
   \hline
   ln(b) & 2.37858 & 0.0198432 & 119.868 & 0. \\
   x & 3.12065 & 0.00341785 & 913.047 & 0. \\
\end{array}
\]

For certainty, we can retrieve the p-value and the t-statistic for the slope and intercept parameters. Concerning the former, the fitting method returns values so small that they are not representable with the stated decimal precision, which, being so close to 0, proves that $p<<\alpha$. About the latter, its results are notably milder than before, despite being large enough to validate the model's usefulness, particularly for the independent variable, since the offset only provides the height correction so that at the smallest $n$ the model can “pivot” by varying the slope until it finds its most optimal value.

\[
\begin{array}{l|lll}
    \text{Parameter} & \text{Estimate} & \text{Standard Error} & \text{Confidence Interval} \\
   \hline
    ln(b) & 2.37858 & 0.0198432 & \{2.33968,2.41747\} \\
    x & 3.12065 & 0.00341785 & \{3.11396,3.12735\} \\
\end{array}
\]

When examining the confidence intervals for each parameter, we notice that both ranges exhibit a higher uncertainty in comparison to the exponent parameter of the initial fit. That phenomenon can be attributed to the inclusion of the offset in the model, which, since it is a varying parameter during the fit process,  introduces additional uncertainty on its own interval and propagates it to the remaining parameters, in this case the slope $x$; something further corroborated by the elevated standard error values observed.
\\\\
Lastly, the most suitable metric for comparing the model's goodness of fit with the prior one is its adjusted $R^2$. Here it is particularly necessary to use the adjusted one due to the amount of parameters, leading to $R_{adj}^2=0.958703$. The value's closeness to 1 indicates that the linear model under the logarithmic transformation better describes the data's features. Even though, if we account for aspects such as the algorithm's sequence space, the multitude of paths present at a certain state, and its difficulty to model them with closed-form expressions, it is plausible to presume that the actual function for the average runtime complexity growth is likely to display a more sophisticated form than the one used in the models of prior estimations.

\subsubsection{Approximation comparison}

\begin{figure}[H]
    \centering
    \includegraphics[width=10.5cm,clip]{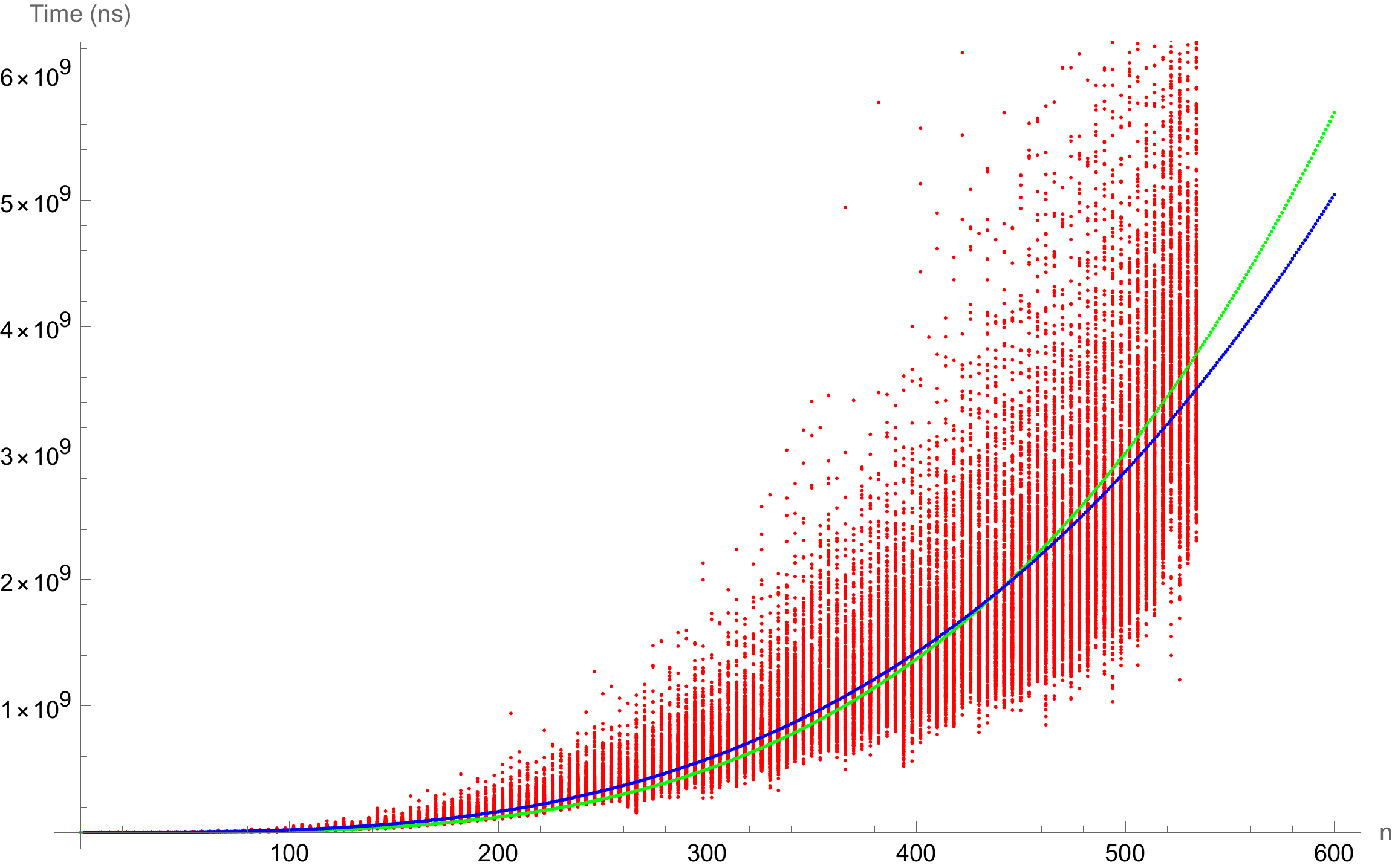}
    \caption{Time measurements}
    \label{fig:2Dtimesfitcomparison}
\end{figure}

In conclusion, it is prudent to furnish a broader comparison of both fits in order to elicit the maximum amount of information about the average runtime growth from data, since, being an approximation, we have no margin to deduce something as complex as its precise value.
\\
Initially, we have interpreted the fitted models primarily as the exact runtime growth via the notation $\Theta (n^{3.12})$. Nevertheless, we also need to account for the confidence intervals and the uncertainty inherent in such a procedure. Moreover, if we were to rely on this notation, the actual time growth would require to match a function with exactly that growth, which is highly improbable since any deviation in the exponent's decimal places would invalidate its application. Alternatively, we adopt the perspective of framing the final model as an upper bound for the actual time growth through the notation $O(n^{3.12})$, which is seemingly more pragmatic as it allows the margin between the best and estimated average cases for the true solution to lie. With all this, irrespective of whether we approximate with the original polynomial model or the linearly transformed counterpart, what actually matters is the exponent determinant for the estimate growth, which is shown in asymptotic notation:
\begin{align}
    T_{avg}(n) = O(e^{x\cdot ln(n) + ln(b)}) = O(e^{ln(n^x)}\cdot e^{ln(b)}) = O(b\cdot n^x)
\end{align}
\begin{align*}
    b=e^{b'}=e^{2.37858}\approx 10.78957 \quad x\approx 3.12
\end{align*}
\begin{align}
    O(b\cdot n^x)=O(10.78957\cdot n^{3.12})=O(n^{3.12})
\end{align}

In the first $n^x$ model we fitted the unique variable parameter was the exponent, no further simplification is available when applying asymptotic notation. Conversely, if we start from the linear model inversely transformed to obtain its true prediction values and simplify it in a form equivalent to $b\cdot n^x$, we will realize that the only difference with the preceding one is the constant $b$. This constant emanates from the inverse transformation of the linear model's offset parameter, which, as illustrated above, has a value of approximately 10.78957. On substituting it in the model and finding its asymptotic growth, we can simplify it by removing $b$ from the expression, since when evaluating the model growth with respect to the variable $n$, the only affecting factor when nearing infinity is the exponent form this variable is raised to. That is, despite the presence of the constant $b$, a model with a higher exponent will grow faster when $n$ is close enough to infinity, consequently, when expressing the growth with asymptotic notation, the offset that previously served to increase the goodness of fit must now be disregarded.
\\\\
Finally, as evidenced in Figure 10 and corroborated by the numerical results of both fits, a significant divergence between them is observed. In the original, the growth is characterized by $O(n^{3.51})$, while in the linear one it remained quite a bit lower, at $O(n^{3.12})$. Statistically, the best fit for the current measurements is the second one, since it is able to handle the height problem of the model described initially. However, the data may present slight, or not so small, variations depending on how they are collected, both in terms of hardware, the proportion of measurements taken for each $n$ and related criteria. This, besides altering the resulting fit, has an implication on the influence that the data points height can bring, which is what grants the linear model a decent accuracy. Otherwise, we could normalize the dataset so that the data points corresponding to the smallest $n$ have a value close to 0, correcting with such a linear transformation the remaining dataset points and simplifying the fitting process, since in this case a model of the form $n^x$ would not require any height rectification. Such transformation will not be conducted as the results achieved up to this stage are adequate to draw an idea of how the average time grows. If we had to offer a single result as a definitive approximation, we could choose the linear model for the advantages it demonstrates. Yet, the contribution of the original model remains noteworthy.
\\\\
Not coincidentally, the linear model with the corrected height demonstrates a diminished growth than its alternative. That is due to the other model being uniquely allowed to vary its exponent, which will adjust its growth by pivoting from the point $(0,0)$, while the linear model will vary the height of that point to find an optimal pivot in order to determine the best exponent available, which is the only feature remaining in the asymptotic bound. By altering the pivot point $(0,y)$, the linear model will need a smaller exponent to keep the error metrics minimized. In contrast, the only recourse for the other model is to reach suboptimal minima for these metrics caused by the initial values of $n$, which will be impossible to fit properly. Therefore, as these values invariably exceed the pivot point of the model, the exponent will tend to be larger after fitting, which we can verify in the achieved models and in the upcoming fits we will perform on various systems.
\\\\
To conclude, there isn't a straightforward way for us to draw a valid asymptotic growth from the two previously discussed in the approximations, and the limitation of not knowing the exact form of the average time function doesn't help either. Nevertheless, before we performed these approximations, the sole information we had about this function was that its growth was bounded by the algorithm's best and worst cases. At present, we have two functions whereby we can shrink the previous bounds. That is, even if the $n^x$ model results in a growth that we know is greater than it ought to, we can set an upper bound for the growth of the average time equal to its fit result, $O(n^{3.51})$. And, on the other hand, we get a fairly smaller growth in the linear model, but despite being a better fit, leaves us without additional insights into whether the target function is actually larger or smaller. As a workaround, we could establish a lower bound with its fit result $\Omega (n^{3.12})$, and estimate the real growth as an intermediate function between this bound and the aforementioned one. Although this is not entirely accurate on the lower bound's side, we know that the average time will be far from the upper bound and potentially close to the lower one, even marginally below it.
\subsubsection{Insertion complexity analysis}
After estimating the average runtime for the algorithm over the entire percolative process, we produced some expressions that could be relevant in case we are required to benchmark or contrast fundamentally different algorithms. Nonetheless, they are still approximations, even if they are accurate according to the fit with which they were achieved, they are not so regarding the algorithm itself. Primarily, as explained previously, this is caused by their form, which is the most noteworthy limitation. Up to this point we have disregarded the possibility that it is feasible to derive such a shape, but here it is evident that with approximations we cannot advance in this matter, so it is necessary to proceed with a comprehensive algorithm analysis. With comprehensive we must think of thorough, namely, if we manage to formalize and properly model the entire percolative process, or at least as many of its fundamental components as possible, we will have certainty about its time complexity, especially in the edge cases such as the worst or best, as well as solutions that despite being partially complete for the average case, as happened for the cardinal of the sequence space, will lead to approaches whereby the analysis can be completed with other techniques discussed later.
\\\\
At the outset, we have to evaluate the conceivable ways to set up the foundation for our analysis. For example, if our objective were to assess performance on a particular hardware architecture, we would have to decompose the algorithm into elementary operations that correspond to the instruction set of that architecture, which entails several phases. One of these phases involves the identification of these operations, which focuses on considering all the high-level functions or procedures of our algorithm and tracing them, while simultaneously decomposing all the statements they contain, recursively if necessary, with the objective of reaching atomic statements at their most granular level, being these potential candidates for the elementary operations sought. Following this breakdown, there might emerge multiple, potentially similar operations warranting further analysis. Therefore, it would be decided whether to select only one of them, the most significant one for the magnitudes that need to be measured, or treat them all as a single operation, applying a transformation to the values they contribute to these magnitudes to normalize them and avoid distinctions between them. The second phase consists of finding an expression that returns the number of elementary operations performed by the algorithm on average for an input of a given size, in our case the matrix size $n$. At this point, it is worth noting that this function could be formed by combining several more precise subexpressions that compute the number of operations performed in a specific execution. Depending on the algorithm, such execution will be defined by the input or, in our case, by the insertion sequence followed. This is often not easy to calculate, unless the algorithm is sufficiently simple or a property exists that we know and can harness to simplify the process and reach closed form expressions, which is the main problem in this regard. Should such an expression be deduced, knowing the resources consumed by a single operation, the desired function can be calculated to model the growth of these resources as the input size increases. 
\\\\
Notwithstanding, the function obtained in this manner would depend on the preliminary setup where a specific architecture has been selected, exempting others, or lacking precision. To remedy this, it is necessary to abstract the fundamental operations reached by the previous approach to a higher level, enabling generalization across all hardware platforms capable of executing our algorithm. Apart from this primary objective, it is also advisable that the time complexity analysis is based on purely theoretical foundations. That is, although time resources are strongly dependent on implementation details, both physical and software, isolating the theoretical definition and analysis of the algorithm from those aspects influenced by such implementation grants a precise formal evaluation of efficiency in relation to the algorithm's duty, and simplifies comparison with other algorithms that tackle the task through different data structures or paradigms.
\\\\
In our specific case, we will begin by delving as deeply as possible into basic operations, distinguishing those that appear in the theoretical definition of the algorithm and represent a unitary amount of work from those that result from breaking down some of its statements into machine instructions. Initially, we have the percolation process consisting of an indeterminate number of iterations. In these iterations, the algorithm performs tasks such as using the random number generator, checking if a cell is free, if not, checking for the existence of a path, etc. Thus, we can distinguish between the complexity due to the set of iterations that comprise the process and the work performed on each one. In this way, we will denote the work done in an iteration as Insertion Complexity, due to the possibility that an element may be inserted into the system. Thus, since an insertion is a function that composes the insertion sequence of an execution, we can express the total complexity in terms of operations that decompose each insertion. To identify these operations, we must persist to dissect it into simpler procedures, even considering situations that may signify the final elementary operation through which the entire complexity of the algorithm is expressed. However, the key point is to isolate each insertion so that its analysis is independent of its sequence. Accordingly, we can express the Insertion Complexity in terms of parameters that define a system state but not the way it has been reached in the insertion sequence. To illustrate, with $n$ and the number of elements existing within the system, we can define a set of states. In each of them, the algorithm's performance will fluctuate depending on how they are arranged. However, by varying or increasing the parameters defining this set, the complexity will grow in a certain manner, which is what will be used to complete the complexity of the entire process.

\begin{algorithm}[H]
\caption{Insertion operation}
\begin{algorithmic}[1]
\small
        \Procedure{Insert}{$i, j$}
            \If{not $\text{grid}[i][j]$}
                \State $\text{grid}[i][j] \gets \text{True}$
                
                \text{//Reset visited matrix}
                
                \State $\text{percolate} \gets \text{helper}(i, j, 0)$
                
                \text{//Reset visited matrix}
                
                \State $\text{percolate} \gets \text{percolate}  \land  \text{helper}(i, j, \text{len}(\text{grid}) - 1)$
            \EndIf
        \EndProcedure  
        
        \Function{helper}{$i, j, \text{up}$}
            \If{$i == \text{up}$}
                \State \textbf{return True}
            \EndIf
            \State $\text{visited}[i][j] \gets \text{True}$
            
            \For{$k$ \textbf{in} $\text{nei}$}
                \State $\text{newI} \gets k[0] + i$
                \State $\text{newJ} \gets k[1] + j$
                \If{$(0 \leq \text{newI} < \text{len}(\text{grid})  \land  0 \leq \text{newJ} < \text{len}(\text{grid}[0])  \land $
                    \State $\text{grid}[\text{newI}][\text{newJ}]  \land  \lnot \text{visited}[\text{newI}][\text{newJ}]  \land $
                    \State $\text{helper}(\text{newI}, \text{newJ}, \text{up})$}
                    \State \textbf{return True}
                \EndIf
            \EndFor
            \State \textbf{return False}
        \EndFunction
\end{algorithmic}
\end{algorithm}

Of the entire algorithm, the statements used to get an element inserted are presented above, which is the part we will focus on now. In short, there are three fundamental aspects through which we can perform the analysis. First, it is checked whether the cell at position $(i,j)$ has an element or not, determining whether the call to the insert function does any additional work. In case it is empty, an element is inserted into that cell and it is verified if from it, through its adjacent cells, a path of elements can be formed to the top and bottom system rows with two calls to $helper()$. However, for these calls to work correctly, the register of visited cells must be reset ahead of each invocation.
\\\\
On the one hand, we might expect that resetting this register affects the runtime complexity growth, since in this case, it is a matrix the same size as the system on which the state of each cell is stored, and as the size of the system increases, the algorithm is further delayed in performing this operation. Still, whether this occurs greatly depends on the particular implementation. For example, the simplest idea is to have an auxiliary matrix over which all its cells are traversed in each insertion, setting them in the correct state as corresponds. This would ensure a constant memory overhead, since the entire matrix is stored throughout the procedure, and a temporal growth of $\Theta (n^2)$ {\it (for that specific operation)} if we consider an elementary operation to be a state change in one of its cells.
\\\\
Apart from this implementation, we can decide to sacrifice conceptual simplicity to make this reset more efficient. As an alternative, we can leverage the insight that, during most process iterations, the number of unvisited elements surpasses the count of visited ones, since the proportion of elements that can potentially form a fully traversable cluster and the rest of the system's matrix cells, containing elements or not, is low. Thus, if we store the positions traversed by the function that checks for the existence of a path in a linear data structure, we can later access it to return only those cells to their initial state. In this scenario, both the memory consumed and the time complexity of the reset will depend on the average size of the cluster over which the insertion is executed, something that is not easy to formalize but can be guaranteed to not exceed the total number of elements in the system, therefore resulting in $O(n^2)$. As a downside, we don't know the exact length that this linear structure will have, and to avoid unnecessary resizing—since implementing it in a linked way would produce an evident overhead in the resources of each elementary operation—the best option is to initialize it as long as the total number of elements, always reaching $\Theta(n^2)$ in terms of memory but avoiding an increase in runtime, which is what we care about. But, if we go further and try to reduce the reset time to a constant amount for any system, we would need to sacrifice more memory and even depend on certain low-level instructions being executable as theoretically needed. One possibility for achieving this is to always have a reset register in a specific location in memory and another where the algorithm operates. Once an insertion is completed, the register used by the algorithm must be reset in some future iteration, possibly the next one or afterwards. In the meantime, the empty matrix should be copied and pasted over the utilized register through a low-level elementary operation. If this could be accomplished, the memory consumption would be double $2 \cdot n^2$, but its complexity as the system size increases would remain equivalent to $O(n^2)$, unlike the time complexity, stabilizing at a constant bound $O(1)$. Moreover, with this approach, we would save a reset operation between both executions of $helper()$, since the second will always traverse the same cluster as the first one, so by alternating both registers, the state changes of the initially used register can be reversed.
\\\\
Regardless of the way in which the management of these visited elements is carried out, particularly their reset, which is what can slow down the beginning of the subsequent functions the most, we can disregard its effect on the overall time complexity. The main reason is the difference between the elementary operation on which we would base the analysis if we finally considered it and what we will obtain in this way, which would greatly complicate its formalization. But also, due to the great variability in the work required to execute that specific procedure, caused by the compilation and execution circumstances of the algorithm. To show this variability and to see to what extent the algorithm's performance changes when executing this operation under certain conditions, we will measure the time consumed by this reset of the visited cells' register. Subsequently, we will use the same routine as before to fit a model to the collected data, and thus accurately calculate its time complexity, given the known actual model form in this instance.
\\\\
In reference to measurements, they will be carried out in the range of $2\leq n\leq 620$ with a step of 2 and recording $n$ measurements for each value of $n$. In each instance, a binary matrix will be generated uniformly at random, and then the time required to set all its elements to a specific value will be measured. Concerning the fitting model, in this case we know it will be of the form $n^x$ due to the nature of the process. However, we will also fit the logarithmically transformed data with a linear model to avoid the issue of measurement height, as we did not normalize them this occasion either. Lastly, as a significant point, we initialize a new random generator with a different seed for each measurement, avoiding biases that hardware might introduce to the resulting data and ensuring uniformity in the distribution of elements across the matrices. And, naturally, the same programming language with the identical environment is used as in the previous measurements for complete process simulations.

\begin{figure}[H]
    \centering
    \includegraphics[width=10cm,clip]{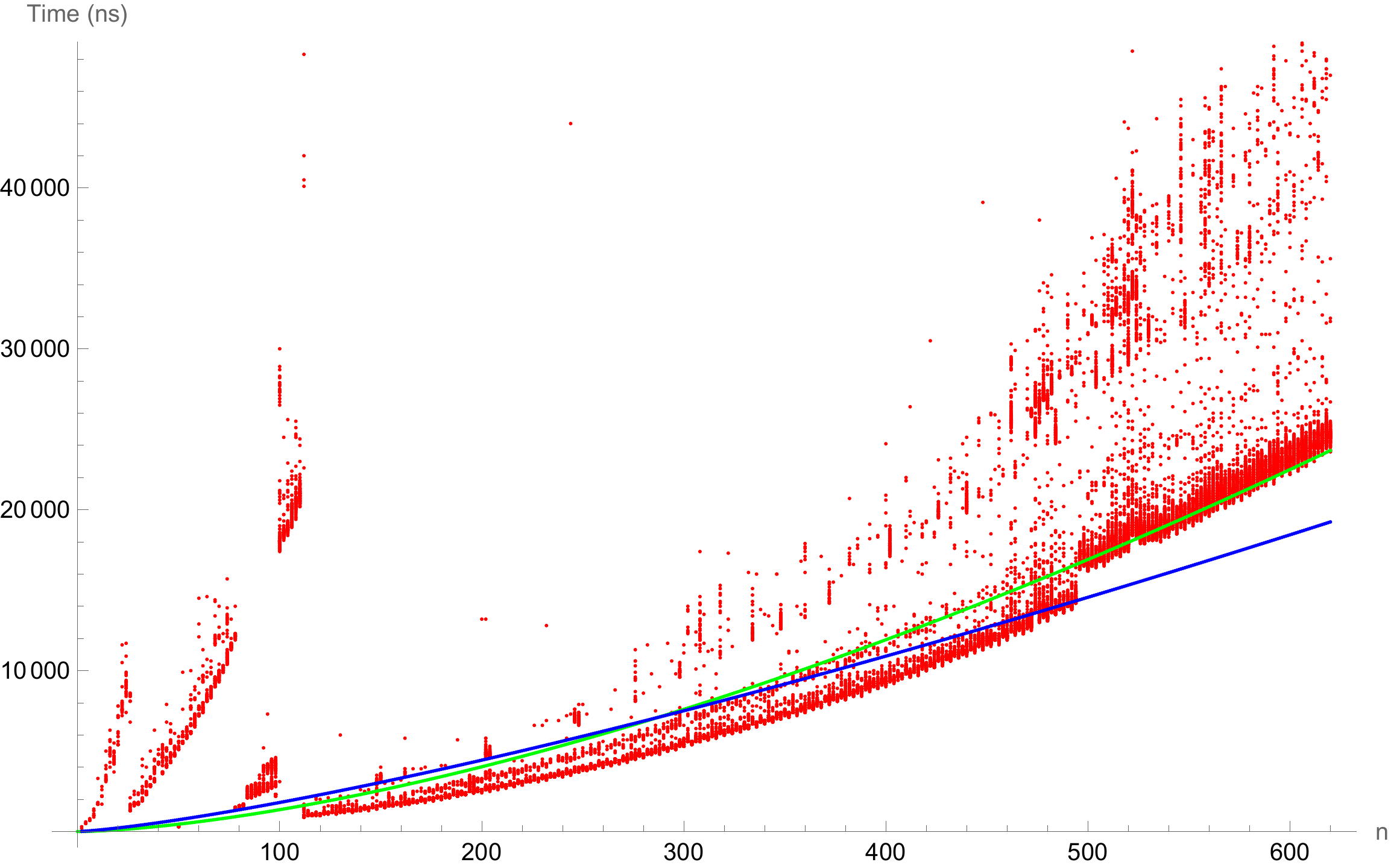}
    \caption{Runtime measurements of the register of visited cells reset procedure from a uniformly random initial state. In green is plotted the $n^{1.57}$ model and in blue the one inferred from the linear fit $e^{1.53 + 1.29\cdot ln(n)}$ }
    \label{fig:2Dmatrixfit}
\end{figure}

Given the results in Figure 11 \cite{time_measurements_matrix1}, the linear model estimates the complexity of the clean operation as $O(n^{1.29})$, while the other fit suggests a higher growth rate, $O(n^{1.57})$. As for the latter, it can be seen from the bottom table that the fit is of commendable quality, revealing a reduced uncertainty in its confidence interval for the exponent $x$.

\[
\begin{array}{l|lllll}
    \text{Parameter} & \text{Estimate} & \text{Standard Error} & \text{t-Statistic} & \text{P-Value} & \text{Confidence Interval} \\
   \hline
    x & 1.56654 & 0.0001406 & 11141.8 & 0. & \{1.56626,1.56681\} \\
\end{array}
\]

On the contrary, the linear model presents a slight additional difficulty in fitting the data. By being able to vary the height of the entire curve, the fitting method uses this variability in height to reduce the error versus the variability in the exponent. As a result, it provides a less accurate and more uncertain outcome, which can be verified in the t-statistic and the confidence intervals for both parameters, especially that of the offset $ln(b)$. Furthermore, we observe that the p-values provide virtually no additional information.

\[
\begin{array}{l|lllll}
 \text{Parameter} & \text{Estimate} & \text{Standard Error} & \text{t-Statistic} & \text{P-Value} & \text{Confidence Interval} \\
 \hline
 ln(b) & 1.52689 & 0.0176543 & 86.4884 & 0. & \{1.49229, 1.5615\} \\
 x & 1.29678 & 0.00296685 & 437.09 & 0. & \{1.29096, 1.30259\} \\
\end{array}
\]

Due to the simplicity of the analyzed procedure, it will suffice to assess the adjusted $R^2$ to compare the results of both models, which in the case of the linear one is equivalent to 0.665455, while in the other model $R_{adj}^2=0.930861$. In line with the above, the uncertainty of the linear model translates to a lower adjusted $R^2$, indicating a worse generalization for large values of $n$ outside the measurement range.
\\\\
In this experiment, the reset operation time was measured using a matrix that required state changes in some of its entries, as it was randomly initialized and it is rather unlikely for its initial state to be homogeneous. Subsequently, it is advisable to examine whether the growth of this time varies significantly when the initial conditions change, as internal optimizations, of which we may be unaware, can be introduced. Thus, we will replicate the measurements precisely as previously conducted, albeit with an initial state where the matrix is homogeneous and each of its entries has the same value as the one it will be assigned throughout the reset. Such matrix could be initialized with the opposite value; however, by proceeding as described, we can observe the duration of the initial iterations of the algorithm, during which we know that the number of traversable elements is substantially smaller than the register size, making this the situation where the reset would perform the least amount of useful work.

\begin{figure}[H]
    \centering
    \includegraphics[width=10cm,clip]{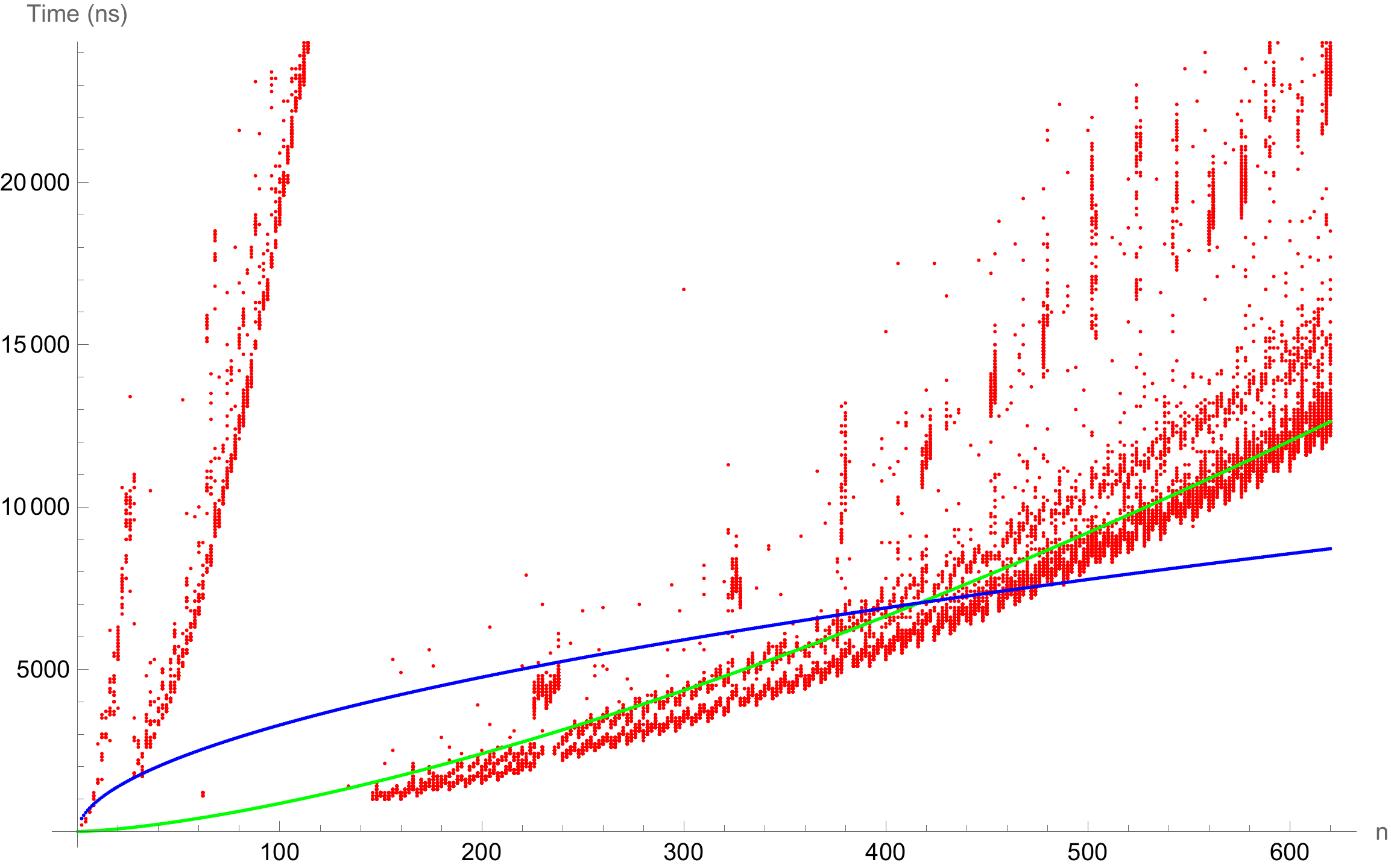}
    \caption{Runtime measurements of the register of visited cells reset procedure from a homogeneous initial state. In green is plotted the $n^{1.47}$ model and in blue the one derived from the linear fit $e^{5.63 + 0.54\cdot ln(n)}$ }
    \label{fig:2Dmatrixfit2}
\end{figure}

In this scenario whose results are shown in Figure 12 \cite{time_measurements_matrix2}, the linear model approximates temporal growth as $O(n^{0.54})$, and the polynomial as $O(x^{1.47})$. With this, and examining the appearance of the obtained data and the form of both functions, it is enough to calculate their corresponding $R_{adj}^2$ to check their goodness of fit. On one hand, the linear model returns $R_{adj}^2=0.163168$. This result, so close to 0, indicates, as is observable graphically, a poor generalization of the dataset's behavior. Although the form of the model is adequate, the objective pursued by the fitting method causes the error to be minimized by bringing the function closer to the data points with a small $n$, missing the opportunity to correctly model the trend that the other model does capture. For its part, that model achieves $R_{adj}^2=0.635656$, which is not as high a value as in previous fits, but considering the measurements of this procedure, we can assure that it is a decent value.
\\\\
Theoretically, we know with certainty that each reset of the visited matrix takes $n^2$ operations, leading to a complexity of $\Theta(n^2)$. However, after conducting experimental measurements of its execution time, it is revealed that the actual growth does not reach such magnitude, but remains below with functions like $O(n^{1.57})$. To understand why this difference occurs, we need to inspect the data points for small values of $n$ in both data sets. In the first one, the measurements seem to grow at a rate of $n^2$ until $n$ reaches a certain threshold, after which the rest of the runtimes grow according to the approximation. Meanwhile, in the second set, something very similar happens; the measurements grow a bit faster than $n^2$ up to a point marginally further from $n=0$ than in the previous case, subsequently conforming the growth rate determined by the fitted model. Definitely, this phenomenon is typically induced by a factor external to the algorithm, probably hardware, or even related to its compiler. In this case, using Java, there are many systems that can provoke this, from JIT to garbage collector \cite{VanPuttenKennedy2022}. As for the hardware, it can stem from the specific CPU technology and the particular memory hierarchy \cite{MeyerSandersSibeyn2003}. Although, the precise origin of these effects is beyond the scope of this discourse, since the important matter is to show the influence of these factors on the time complexity and thus be able to perform a higher-level theoretical analysis, which will grant us accuracy. Finally, it is worth noting that in the second experiment this growth transition is more pronounced, and it is not a coincidence that it arises when the matrix is initialized in a homogeneous state.
\\\\
After this interlude, we proceed to detail why resetting the visited register can be disregarded in the complexity analysis of an insertion operation. Firstly, if we consider that the algorithm runs on a Turing machine\cite{AroraBarak2009}, we can leverage its unbounded (but not infinite) amount of memory to achieve a constant time complexity for this operation. Thus, if each time a register with all its entries in the unvisited status is needed, an additional amount of memory, which has not been modified at any previous moment and whose size is equivalent to that register, is allocated, we can use that memory area for the next insertion register, or allocate two consecutive memory areas with those characteristics for the two resets that must be performed in each insertion. In this manner, the time spent to reset the register depends on the duration of a memory allocation, which theoretically can be assumed to be constant $O(1)$. The sole drawback of this method for achieving a matrix reset is its implementation in a real system. Theoretically, the amount of memory required by multiple insertions throughout the execution of our algorithm is finite, making it feasible on a Turing machine but incongruously large for a physical one. For it to be viable, a garbage collector must reset the unused memory to the correct state after each allocation. That is, with each insertion, 1 or 2 blocks of size equivalent to the visited register are allocated, and the memory reserved in previous iterations is erased so that these areas can be successfully reallocated in future insertions. Therefore, the garbage collector's operational speed must at least match the pace of new matrix segment allocations. Otherwise, the complexity would hinge on the speed of the garbage collector, which, if it were a constant slightly greater than that required for allocation, would ensure constant complexity. However, if it depends on the size of the block to be deallocated, the complexity would surely exceed $O(1)$.
\\\\
In light of this, given that the previous option may present viability issues, it is also convenient to contemplate the implementations that are more likely to appear within our algorithm. As for the simplest one, wherein the entire register is traversed changing the status of every element individually, it has already been established that theoretically its complexity is $\Theta (n^2)$, as long as the elementary operation being considered is a state change in one of its entries. Even so, experimentally we see that there exists a marked difference compared to the theoretical approach, most likely caused by compiler optimizations which we do not know for certain at this point. Revisiting to the complexity of an insertion operation, since in all its executions it is necessary to reset the register, at least $n^2$ elementary state change operations are produced, plus the remaining work required by the path existence verification.
\begin{align*}
    T_{ins}(n)=n^2+w \quad \colon \quad w\in O(n^2) \implies T_{ins}(n)= \Theta(n^2)
\end{align*}
At first glance, without delving into the details of the verification process, we might conclude that the average runtime of an insertion, denoted henceforth as $T_{ins}(n)$, grows exactly in proportion to the state changes required by a reset. Despite occurring more than once per insertion, its asymptotic growth does not differ from that shown above. This assertion is partially correct, since the state change in a register cell has a cost, and the elementary operation considered in the $helper()$ function for subsequent verification has its own cost. Specifically, we will define an elementary operation in this verification as the execution of $helper(i,j,up)$ on an entry of the system matrix located at position $(i,j)$, with the $up$ parameter holding no significant relevance. We will later elaborate on this decision and explain the advantages it offers in the final analysis. However, according to the algorithm's definition, this operation includes a state change in the visited register, along with additional statements that, collectively, guarantee the complete operation will consume more resources than a single state change in the register.
\\\\
Therefore, ensuring that $T_{ins}(n)$ grows in the form $\Theta (n^2)$ would imply that a state change in the register, as an elementary operation, demands an asymptotically similar time to an execution of $helper()$ on a cell. Thus, given that both operations are theoretically incomparable, the bound for the temporal growth of an insertion becomes inconsistent with respect to the elementary operations on which its analysis is based. The unique relationship between both operations, that could enable the construction of a correct and more specific bound, is the state change in the register that executes inside $helper()$. If we assumed that all the extra work of this function is not accounted for in the analysis, the prior problem would disappear. However, by strictly accounting for all the workload, the only conclusion we can draw from this analysis is $T_{ins}(n) = O(n^2)$. That is, assuming that the elementary operation of an insertion is an execution of $helper()$, a maximum of $n^2$ of them will be needed to complete any insertion, also adding some extra work to clear the visited register, which is omitted by this bound since traversing all its entries does not exceed it.
\\\\
Up to this point, considering the reset of the register as an iteration over all its elements does not provide sufficient information to decide whether or not to take it into account for obtaining an accurate bound of $T_{ins}(n)$. In the first analyzed implementation, it was theoretically possible to ignore this reset, although with certain limitations for its ultimate execution. Therefore, another option is to examine the implementation of this procedure using a linear data structure to store the register entries that need to be reset for the next iteration or subsequent check. In this scenario, we can start with the two elementary operations that previously caused inconsistency: one is the execution of $helper()$, and the other is a state change in the register. Here, despite needing to access a linear data structure to mark a cell as visited or not visited, the elementary operation should be treated in the same way as before; the only difference would be if exact time measures were required, which would likely be higher. In the first instance, each insertion performs a certain number of elementary operations that depend on the system's state during the iteration in which they are executed. Specifically, if we focus on the number of $helper()$ executions, which, as we will see below, is what determines how the work of an insertion grows, we will see that it depends on the number of elements in the system's matrix during that iteration, and more importantly, on their distribution. In this context, it is assumed that the work of an insertion is non-zero, since when the algorithm attempts to insert an element into an occupied cell, it always performs constant work. In the case where a complete insertion is required, it will be verified whether the system percolates through an initial call to $helper()$ on the corresponding cell. Following this, a depth-first search \cite{TarjanZwick2024} is conducted through neighboring cells containing elements to check whether any of the traversed cells are in the rows of the matrix that enable the formation of a path. During this traversal, each processed element will proceed from an execution of $helper()$, counting as an elementary operation towards the total workload. Therefore, to determine the total number of operations in a specific insertion, it is necessary to know how many neighboring cells with an element each pair $(i,j)$ visited by the algorithm has, which in total corresponds to the number of elements in the cluster to which the newly inserted element initially belongs.
\\\\
Given the large number of possible cluster combinations that can be encountered in any iteration of the algorithm, and since we only need the size of the cluster on which the insertion is performed, we will denote the average cluster size in a matrix of side $n$ with $k$ elements distributed within it as $c(n,k)$. Thus, we know that for each iteration, there are $c(n,k)$ executions of $helper()$, during which each one involves a state change in the visited register. Additionally, in this register, only $c(n,k)$ entries need to be reverted to their unvisited state, as with the current implementation we know exactly which ones they were, avoiding the traversal of the remaining $n^2 - c(n,k)$ register entries \textit{(also considered elementary operations)}.
\begin{align*}
    T_{ins}(n)=4\cdot c(n,k)+2\cdot c(n,k) \implies T_{ins}(n)= \Theta(c(n,k))
\end{align*}

Overall, the algorithm checks if the system percolates with two calls to $helper()$, one for each row where valid paths start or end. Moreover, before these executions, the visited register is reset once per invocation, resulting in a total of $c(n,k)$ elementary operations per call to $helper()$ and the same amount of state changes per reset. Considering that each execution of $helper()$ also produces a state change, the final expression reflects a duplication of state changes, as shown above in the term multiplied by 4. However, even though both elementary operations are not theoretically comparable, we know that in the case of state changes, the same number of occurrences applies to both the procedure of resetting the matrix and the depth-first traversal that verifies the existence of a path. That is, the insertion is performed in a state defined by $n$ and $k$, both of which are invariant during the procedure. As the size of the matrix increases, $c(n,k)$ tends to be larger depending on the number of elements, which is guaranteed to be constant for each insertion. In this way, when the parameter $n$ that defines the runtime of an insertion tends to infinity, in each insertion the number of state changes in the visited register will grow in the same manner as indicated by $c(n,k)$, regardless of the number of times that amount of state changes needs to be carried out.
\begin{align*}
    x\cdot c(n,k) = \Theta(c(n,k)) \quad \forall x>0 \quad \colon \quad x\in \mathbb{R}
\end{align*}
Thus, knowing that $c(n,k)$ coincides with the number of elementary operations counted by the executions of $helper()$, the concluding bound for $T_{ins}(n)$ is exactly the average cluster size. In essence, this bound depends solely on the depth-first search conducted at each insertion, not on the amount of resets of the register that may be performed, since as shown above, for any positive value of $x$ representing the number of resets completed in an insertion, the number of elementary operations caused by this implementation grows precisely as indicated by $c(n,k)$, and this work can be disregarded due to the composition of each operation provided by $helper()$.
\\\\
To conclude, the practical implementation of the register reset influences the information we have to ensure whether its effect on the complexity $T_{ins}(n)$ can be omitted or not. Yet, theoretically, the latter assures that its asymptotic behavior is indifferent to this procedure, so from this point on, we will analyze the complexity of an insertion as the time it takes for the algorithm to perform the depth traversal along the elements of the corresponding cluster. If an exact account of the temporal resources spent is necessary, the overall runtime could be expressed as the one spent by a state change, represented by $t$, and the time required by an execution of $helper()$ \textit{(without counting the inner state change)} on a system cell, denoted by $t'$, as follows:
\begin{align}
    T_{ins}(n)=4\cdot t\cdot c(n,k)+2\cdot t'\cdot c(n,k)
\end{align}
Now, we are aware of the direct correlation between the complexity of an insertion and the average size of a cluster of elements $c(n,k)$, so if we had an expression for this feature, we would automatically know its exact growth, specifically for insertions in systems with $k$ elements. Unfortunately, deriving a closed-form of $c(n,k)$ for all positive $n$ and $0 \leq k \leq n^2$ is not a viable option, as we will realize later, so it is convenient to focus the analysis on alternative properties that present the same growth starting from different, and possibly simpler, approaches. For this purpose, if we look at the definition of the function $helper()$, we will see that it follows a common pattern for carrying out a depth-first search, with a base case where it is detected if the traversal has reached the row with a certain index and a loop in which all neighboring cells are traversed to invoke the function recursively on them. With this, we can attempt to calculate the total number of executions of $helper()$ throughout the traversal, which is equivalent to the number of elements comprising the verified cluster. Initially, we can proceed to count it using the branching factor of the graph representing the cluster of elements as follows\cite{MIT_SearchComplexity_2003}:
\begin{align*}
    T_{ins}(n) = 1+b+b^2+b^3+\cdots +b^m=\frac{b^m-1}{b-1}\implies T_{ins}(n)=O(b^m)
\end{align*}
The initial invocation, being performed on one element, counts as one elementary operation. Subsequently, it generates $b$ calls to the traversable neighboring cells, and for each of those, it initiates another $b$ invocations until reaching the boundaries of the matrix or the cluster of elements, which in this case is modeled as the maximum depth of the traversal $m$. In this way, we would arrive at a total count of elementary operations that grows at the same rate as $b^m$, although this approach has a drawback. In this model, it is presumed that for every traversed cell, there exists a constant number $b$ of traversable neighboring cells. This assumption may not hold in any arbitrary cluster, as if it is very dense, there will be cells with part of their vicinity restricted by visited cells. One option to address this issue would be to turn $b$ into a function dependent on the traversal depth reached at a particular cell. Yet, in this case, it would also need to account for the geometry of the cluster, making it an unfeasible option to formalize exhaustively. Additionally, another drawback of this analysis is the exact calculation of the maximum depth $m$, which, like the other parameter, would require introducing a dependency on the structure of the cluster and, furthermore, on the particular cell where the cluster traversal starts.
\\\\
Therefore, owing the inability to proceed from the branching factor, an alternative method must be devised to obtain a precise value for the workload of such a traversal. To this end, we will consider the runtime required by $helper()$ to execute on a cell as the sum of the times taken by the recursive calls on its traversable neighboring cells, and so forth, until the execution time of the traversal over the entire cluster is accumulated. Here, it is valuable to underline that the premise that the entire cluster is always traversed emanates from the recursion's stopping condition. In every iteration except the final one, a full traversal is ensured, since the cluster does not reach the rows that allow path formation. Conversely, in the final iteration of the algorithm, the appearance of the spanning cluster with that insertion assures the fulfillment of the stopping condition. As a result, depending on the order of neighboring cells chosen for the recursive calls and the geometry of the cluster, fewer elements of the entire cluster may be traversed. But, since this situation invariably occurs in the final insertion and not in any other, we can also disregard this effect on the overall complexity because the number of iterations wherein it occurs is, in all cases, constant and equal to 1. Therefore, as the system size increases significantly, the contribution of the last insertion to the total workload remains constant relative to the rate at which the number of iterations the algorithm requires to complete grows. Accordingly, regarding the complexity of an insertion, if we set aside the counting of elementary operations and denote the insertion time as $T(i,j)$, we can express it as a function of the runtime for its neighboring cells in accordance with the traversal's definition:
\begin{align}
    T(i,j) = T(i,j-1) + T(i,j+1) + T(i-1,j-1) + T(i-1,j+1) \\
           + T(i-1,j) + T(i+1,j) + T(i+1,j-1) + T(i+1,j+1)\notag
\end{align}
As can be seen, it is a recursive definition \cite{Izadkhah2022} with which we can attempt to derive a closed-form expression for $T(i,j)$, thus accounting for the time required for an insertion into the cell $(i,j)$. This would allow us to later determine the average across the entire system as a function of $n$, and thereby obtain an asymptotic bound as it grows. However, in order for the recurrence to yield a solution, it is necessary to introduce constraints that model situations where the recursive case executes fewer recursive calls than those specified in its definition. For instance, if any of the position indices exceeds the system's bounds, the time contributed by that execution is 0:
\begin{align}
    T(i,j)=0 \quad \text{if } i=0\lor j=0 \lor i>n\lor j>n \lor V_{\lambda}(i,j)=1
\end{align}
Furthermore, there are cases where some neighboring cells have been previously traversed, causing $T(i,j)$ for those cells to be 0. Therefore, the formalization of whether the cells can be visited or not in a particular iteration of the traversal, which will be denoted as $\lambda$ since it differs from the iterations executed in our algorithm, will be achieved through a function $V_{\lambda}(i,j)$ that receives the pair of integers corresponding to the position of a cell and returns 1 or 0 depending on whether they are visited or not visited, respectively.

\begin{align}
    & V_{\lambda}(i,j): \{(i,j) \in (0,1,2,\cdots n)\times (0,1,2,\cdots n)\} \to \{0,1\}
\\
    & V_{\lambda}(i,j)=\begin{cases}
    0 & \text{if $(i,j)$ cell is unvisited} \\
    1 & \text{if $(i,j)$ cell is visited}
    \end{cases} \notag
\end{align}
With this definition, each time a $helper()$ execution occurs over a cell, +1 iterations will be counted in the traversal, resulting in a different function for each pair of values $\lambda$ and $\lambda +1$. Additionally, if it becomes necessary to build a set of all visited elements to discover some common alternative pattern, it could be obtained as shown below.
\begin{align}
    \mathbb{V}_{\lambda}=\{(i,j) \in (0,1,2,\cdots n)\times (0,1,2,\cdots n) \enspace / \enspace V_{\lambda}(i,j)=1 \}
\end{align}
Consequently, having a distinct function for each iteration of the traversal \(\lambda\) in order to represent the visited cells complicates the possible solution of \(T(i,j)\). Prior to applying the constraints to discover such solution, there arises a problem which will prevent specifying expressions for the function and the set of visited cells. On one hand, the most evident issue is that \(V_{\lambda}(i,j)\) depends on the shape of the cluster to be traversed, which in turn depends on the number of elements \(k\) present in the system at that moment. Simply the absence of knowledge concerning the average cluster size \(c(n,k)\), or the total number of clusters that can be formed with that amount of elements, considering all possibilities in terms of their sizes, makes it impossible to construct an expression for such function. Moreover, even if we had a way to know all of this and precisely model the visited cells of the system, for instance with a trigonometric function whose domain consists of the matrix cells, we would still need a generating function that, depending on $\lambda$, returns the correct $V_{\lambda}(i,j)$ for each iteration of the traversal. Such a function might be obtained, but regarding the potential variability in the order in which neighboring cells are traversed, the function would likewise depend on this order, being infeasible to formalize altogether such a number of possibilities. All this precludes the usability of the constraint that models the information about visited cells. Nevertheless, we can propose another approach to model it in order to start working with the recurrence and determine its solvability.

\begin{align}
    V(i,j)=\frac{T(i',j')-T(i,j)}{n^2}
\end{align}
So far, the specific iteration in which the traversal was found determined the visited function. However, considering that transitions from unvisited to visited occur when moving from one cell to its neighboring cell, notwithstanding the presence of multiple formerly visited neighbors, we can dispense with the index $\lambda$ and characterize the state of a cell based on the subsequent ones in its traversal. Thus, $V(i,j)$ can be defined as the ratio of the execution time difference between the neighboring cell to be traversed in the iteration $\lambda + 1$ and the current one to the total number of cells in the system. This may not be entirely accurate regarding $V(i,j)$, since execution time is used to represent the state of a cell, yet by including it into the recurrence, we might be able to exploit some property to derive an expression.
\begin{align}
    T(i,j) = V(i,j)\cdot (T(i,j-1) + T(i,j+1) + T(i-1,j-1) + T(i-1,j+1) \\
           + T(i-1,j) + T(i+1,j) + T(i+1,j-1) + T(i+1,j+1)) \notag
\end{align}

\begin{align*}
    T(i,j) = \frac{T(i',j') - T(i,j)}{n^2} \cdot ( T(i,j-1) + T(i,j+1) + T(i-1,j-1) + T(i-1,j+1)\\
    + T(i-1,j) + T(i+1,j) + T(i+1,j-1) + T(i+1,j+1) )
\end{align*}
The way to include the term in $T(i,j)$ is by multiplying it with the cumulative runtimes of all the neighboring cells. Consequently, the proportion will affect the workload required to resolve the call to $helper()$ in $(i,j)$, being resolved successively according to each subcall it generates.

\begin{align*}
    T(i,j) \cdot n^2 = (T(i',j') - T(i,j)) \cdot ( T(i,j-1) + T(i,j+1) + T(i-1,j-1) + T(i-1,j+1)\\+ T(i-1,j) + T(i+1,j) + T(i+1,j-1) + T(i+1,j+1) )
\end{align*}

\begin{align*}
     T(i,j) \cdot n^2 = T(i',j') ( T(i,j-1) + T(i,j+1) + T(i-1,j-1) + T(i-1,j+1) \\ + T(i-1,j) + T(i+1,j) + T(i+1,j-1) + T(i+1,j+1) ) \\
     - T(i,j) ( T(i,j-1) + T(i,j+1) + T(i-1,j-1) + T(i-1,j+1) \\+ T(i-1,j) + T(i+1,j) + T(i+1,j-1) + T(i+1,j+1) )
\end{align*}
At this juncture, it is convenient to combine all the possible terms in the cell that is being processed $(i,j)$, and the one that will be evaluated in the next iteration $(i',j')$. Since we do not know which one it will be, for now it is left as shown below. Nevertheless, practically, it could be expanded by multiplying all terms by $T(i',j')$, substituting each pair $(i',j')$ with the corresponding one.
\begin{align*}
    T(i,j) ( n^2 + T(i,j-1) + T(i,j+1) + T(i-1,j-1) + T(i-1,j+1)\\ + T(i-1,j) + T(i+1,j) + T(i+1,j-1) + T(i+1,j+1)) \\= T(i',j') ( T(i,j-1) + T(i,j+1) + T(i-1,j-1) + T(i-1,j+1) \\+ T(i-1,j) + T(i+1,j) + T(i+1,j-1) + T(i+1,j+1))
\end{align*}
Specifically, each term containing the coordinates $(i',j')$, when multiplied by another term representing a known cell, can be explicitly regarded as the position pair that uniquely identifies that cell to be processed in the succeeding traversal iteration:
\begin{align*}
    T(i,j) ( n^2 + T(i,j-1) + T(i,j+1) + T(i-1,j-1) + T(i-1,j+1)\\ + T(i-1,j) + T(i+1,j) + T(i+1,j-1) + T(i+1,j+1)) \\= T(i,j-1)^2 + T(i,j+1)^2 + T(i-1,j-1)^2 + T(i-1,j+1)^2 \\+ T(i-1,j)^2 + T(i+1,j)^2 + T(i+1,j-1)^2 + T(i+1,j+1)^2
\end{align*}

After applying the substitution for each pair of corresponding position indices, we can provide a definitive expression for the runtime $T(i,j)$:

\noindent
\parbox{\textwidth}{ \scriptsize
\begin{flalign}
    T(i,j) &= (1-\frac{E(\alpha)}{n^2})\cdot \frac{T(i,j-1)^2 + T(i,j+1)^2 + T(i-1,j-1)^2 + T(i-1,j+1)^2 + T(i-1,j)^2 + T(i+1,j)^2 + T(i+1,j-1)^2 + T(i+1,j+1)^2}{n^2 + T(i,j-1) + T(i,j+1) + T(i-1,j-1) + T(i-1,j+1) + T(i-1,j) + T(i+1,j) + T(i+1,j-1) + T(i+1,j+1)}
\end{flalign}
}
In this final step, apart from isolating the term corresponding to the recursive case in the formula, an additional multiplier representing the probability that the given time is accounted for is introduced. According to the procedure's definition, an insertion is performed if the target cell for element insertion is empty, implying no workload is done for every already filled cell. To include this effect in the recurrence, $E(\alpha)$ is denoted as the expected number of elements in the system at iteration $\alpha$ of the algorithm. Consequently, the ratio of the number of elements to the total system size, which equals the probability that a cell is already occupied when an element is inserted, can be computed. As we will see later, $E(\alpha)$ indeed has an expression we can work with; however, for now, it will remain as indicated above since the final recurrence does not yield a similarly manageable solution. Regarding the numerator, all terms are squared, and substituting an expression as large as $T(i,j)$ with its corresponding indices and attempting to operate with not just one but the eight terms that model the neighborhood used by our system, renders it practically impossible to identify a pattern that correctly expresses $T(i,j)$ in terms of its parameters. Even when substituting concrete values in the denominator, we would achieve a formula with the same characteristics, which does not contribute to our purpose. Additionally, we must simultaneously consider the base cases in which the formula should return 0. This presents an added difficulty since expressing the condition of whether any of the indices has reached the system's bounds would imply adding another multiplier for each term, probably similar to a Kronecker delta function \cite{Xing2019,Sadd2005}, further complicating the solution. This is not convenient because the goal is to obtain an asymptotic bound, and if the solution is too elaborate, the process for finding such a bound also becomes intricate or even impossible, depending on the resulting function.
\\\\
In conclusion, basing the analysis of an insertion operation on the previous recurrence is of comparable difficulty to the approach grounded on the average cluster size $c(n,k)$. So, it is necessary to continue the analysis by exploring other methods to quantify this work, avoiding dependence on the geometry of our system. Nonetheless, before finalizing, it becomes apparent how the recurrence methodology may be effective under the presumption of a constant and well-known branching factor. \cite{125421}.
\begin{align}
    T(b,d)=b+b\cdot T(b,d-1)\enspace ,\enspace T(b,1)=b
\end{align}
\begin{align*}
    T(b,d)=b+b^2+b^3+\cdots +b^{d-1}\cdot T(b,1) \implies T(b,m)=O(b^m)
\end{align*}
Albeit, as evidenced in its early formulation, $b$ is not invariant throughout the insertion process, and similarly, the maximum traversal path length $m$ remains indeterminate. 
\subsubsection{Insertion complexity as an elementary operation}
Upon discovering the various methodologies for computing the exact runtime required for an insertion within the system, we realize that they are impractical for our purposes. The dearth of closed-form expressions and the vast size of the state space make it onerous to determine the exact workload our algorithm necessitates in a state defined by parameters such as the number of elements or, more crucially, the insertion sequence, as it is the latter that distinguishes it from all other existing configurations for $k$ elements. As an alternative, it is possible to resort to much simpler yet significantly more imprecise and misleading approximations regarding the total complexity they would entail. Therefore, in order to attain an exact asymptotic bound to meet the requirement of being as precise as possible in our analysis, it is essential to forgo these alternatives and propose a theoretically sound approach that brings us closer to the sought-bound. To this end, we can abstract the runtime of an insertion and instead focus on the aggregated work exerted across all the iterations incurred by the percolation process. Hence, we shall skip delving into the particulars of what takes place within each iteration or insertion, should it occur, and simply examine the sequence of steps concluding in the terminal state of the system.
\begin{align}
    T_{avg}(n) = \sum _{i=0}^{I(n)} (1-\frac{E(i)}{n^2})\cdot T_{ins}(n) \label{complexity_elemental_insertion}
\end{align}
To begin with, we can define the average runtime of our algorithm as the sum of the time taken by each iteration composing it. As noticed, from the iteration with index 0 to the last one, determined by the function $I(n)$, the execution time is obtained by multiplying the time an insertion would demand on the system by the probability of its occurrence. If we had a way to model, for each sequence of insertions, which ones eventually ensue and in which iteration they do so, each term of the above sum would yield an exact result with respect to the definition of the algorithm. Regrettably, this would once again lead us to the same situations in which conducting an exhaustive model in such a vast space of possibilities precludes reaching the desired solution. Consequently, it is preferable to use the expected number of elements $E(i)$ to faithfully represent the fact that not every iteration accounts for the time of an insertion.
\\\\
The underlying approach to the aforementioned statement is based on the selection of the basic operations that constitute $T_{avg}(n)$. In this case, since the complexity of the entire process cannot be derived from an extensive analysis of a particular insertion, such insertion is deemed an elementary operation. This way, the runtime can be expressed as a function of the time taken by an insertion, given that it is now an elementary operation. Moreover, this choice simplifies the analysis in relation to the system's geometry, because if we replace the matrix by another object \cite{Parviainen2004,GrunbaumShephard1987,GawlinskiStanley1999,Melchert2013} with a different neighborhood of cells \cite{Breckling2011,Zaitsev2017,Packard1985} or a different total number of elements, the formula would undergo minimal changes; mainly due to the contribution of the elementary operation. Its temporal magnitude is not genuinely relevant when assessing its growth, so any value proportional to this scale, including the unit that accounts for a single basic operation, would suffice in the above specification. Consequently, choosing such an elementary operation inherently entails the assumption that all insertions demand the same amount of time within the same system. Still, this is not accurate as established beforehand. In the initial iterations of any simulation, given that the system commences in a homogeneous state devoid of any elements, the insertions, at least the initial one, are guaranteed to take a constant time equivalent to an elementary operation of $helper()$, as observed earlier. In subsequent iterations, as the number of inserted elements increases, the formation of larger or smaller clusters can alter the workload they need, but on average, it is evident they will utilize more than one elementary operation. Thus, as a simulation advances, we can perceive that each insertion increasingly takes longer to complete, thereby contradicting the premise of this analysis.
\subsubsection{Average insertion runtime}
\label{subsubsec:AverageInsertionRuntime}
To mitigate the impact of this problem on the resulting complexity, an initial option worth exploring involves obtaining an expression for the average work required for a single insertion. In principle, we do not have exact knowledge about the runtime of an insertion in a system with $k$ elements, but we know that whenever an insertion occurs, there will be a quantity between 0 and $n^2$ of them. Hence, we can express the average as follows:
\begin{align}
    T_{ins}(n) = \frac{\sum _{k=0}^{n^2} c(n,k)}{n^2+1}
\end{align}
Formally, each insertion requires an amount of work proportional to $c(n,k)$. Therefore, by summing all the terms where $k$ traverses the possible quantities of elements that an insertion may encounter, we obtain the total number of elementary operations {\it ($helper()$ executions)}. Dividing this by the number of summands results in the average we seek. However, we again encounter the problem of not precisely knowing the average cluster size that an insertion will come across. The difference now is that we can use the known bounds of this value to provide an approximate formula, which can be subsequently substituted into the average runtime of an insertion. First of all, the range of values with respect to $k$ and $n$ that we are interested in is $0 \leq k \leq n^2$, as this includes all possible cluster sizes. For a specific $k$, they are reduced to $0 \leq c(n,k) \leq k$. This represents the possibilities where the system might be empty and no cluster exists, or there may be one cluster of the same size as $k$. In between, there can be one or multiple clusters in random areas of the matrix with a size indicated by the previous range. In summary, this can be translated into an upper and lower bound for the desired function. On the one hand, when $k \geq 1$, the minimum size for any existing cluster is assured to be 1, since any element with at least one neighboring cell occupied will form a cluster of larger size. Likewise, for any valid value of $k$, the largest cluster that can be formed with such a number of elements is of size $k$, as can be trivially deduced, although the probability of such a situation occurring is low. Therefore, even without knowing the exact number, we know that it will be bounded above by the function $y = k$ and below by $x = 1$. Moreover, if we consider the simplest possible case in which there is only one element inside the system, we realize that the average cluster size does not differ from that quantity, since the only possibility is that the single element is regarded as a cluster located in any of the $n^2$ cells of the matrix. This case, unlike all others except $k=0$, provides an additional constraint to the function we seek. That is, when $k=0$ the absence of clusters is guaranteed, so $c(n,0) = 0$. Likewise, if $k=1$ the only value that $c(n,1)$ can return is 1, so graphically the function will pass intersect the point $(1,1)$. Complementarily, there is another case where the function returns a value matching the number of elements, which is when $k=n^2$. In this scenario, the only possibility is that the entire matrix is filled with elements constituting a single cluster, so its size will be $n^2$.
\begin{figure}[H]
    \centering
    \includegraphics[width=10cm,clip]{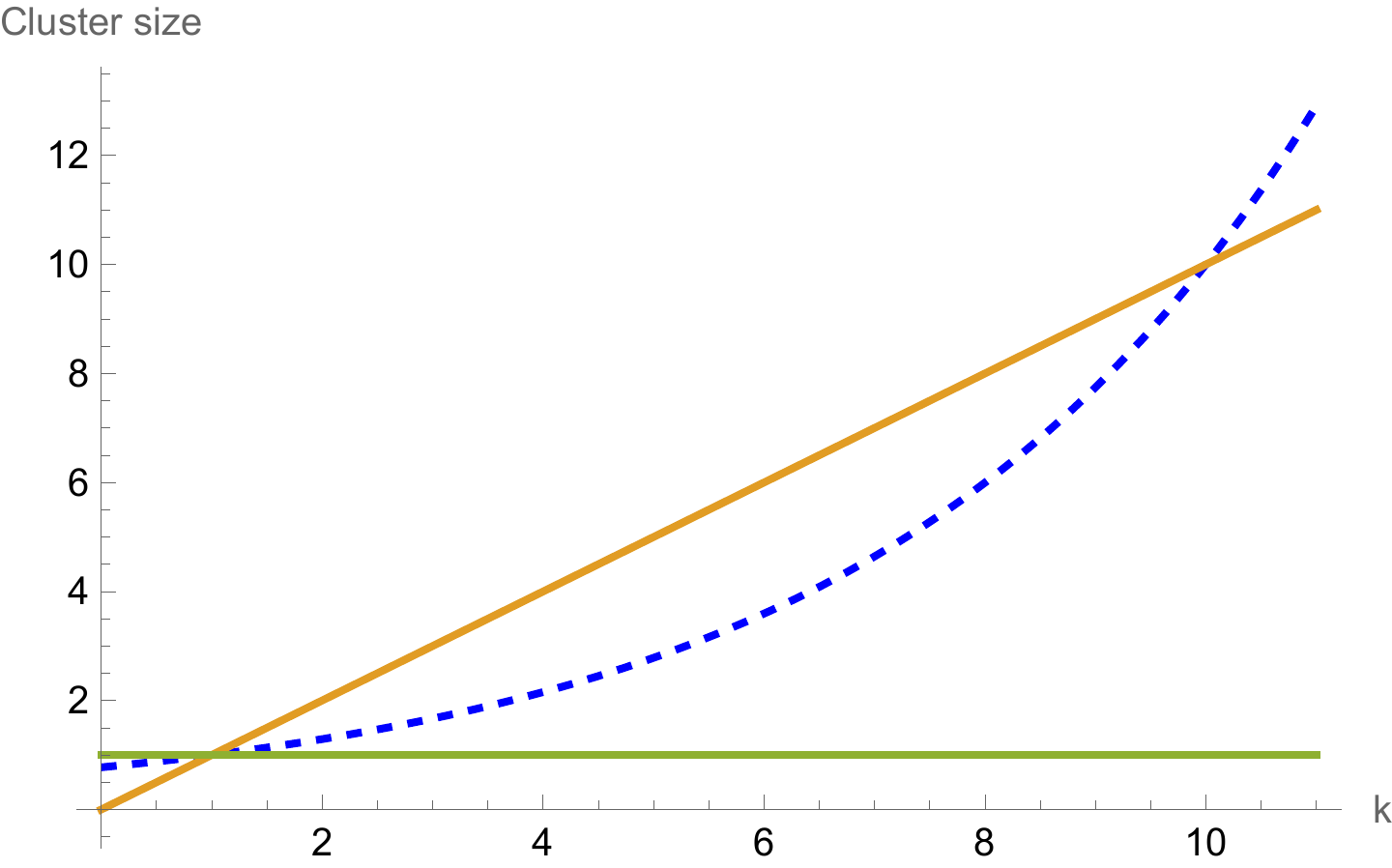}
    \caption{Constraints $y=k$ and $y=1$ are plotted in orange and green respectively. In blue with dashed line, the function $e^{\frac{(k-1)}{n-1}\cdot \log (n)}$}
    \label{fig:ClusterSizeInitial}
\end{figure}

In Figure 13, both constraints are plotted alongside a function that approximates $c(n,k)$. The function is chosen arbitrarily but meets certain requirements to graphically represent $c(n,k)$ with the highest fidelity available. The first condition is that it intersects the point $(1,1)$, which we know the function we aim to approximate must also satisfy. Additionally, regarding its growth, the function must be injective and monotonically increasing within the specified range. This stems from the nature of the percolation process, whereby each iteration has a determined number of elements already inserted. Thus, for any subsequent iteration, this number must be greater than or at least equal to the previous one. As it progressively increases, the probability that some of these elements form a cluster rises, leading to an increase in the average cluster size metric for the entire system. Consequently, as presented above, the average size must be low and close to 1 when the system is almost empty, since among all possible states, there are few forming clusters. In contrast, and in a complementary manner, when $k$ is sufficiently large, the state space restricted by this number of elements contains many more states featuring clusters than in the previous case. This is because, with such a large amount of elements, the probability that a set of occupied cells are neighbors increases.
\\\\
The main task at this point is to determine how such a probability increases within the described range. To accomplish this, we shall propose a function whose form fits all the constraints currently identified.
\begin{align}
    c(n,k) = \frac{a(n)}{a(n)-k+b(n)}
\end{align}
As observed in Figure 13, we need to identify a function analogous to $\frac{1}{-k}$ when $k<0$. With it, we model the required growth dynamics and satisfy the injectivity condition over the interval from $k=-\infty$ to $k=0$. However, in its current formulation, it would fail to be applicable within the desired range of $k$ values. Moreover, it lacks dependence on $n$, which would vary its range proportionally. To rectify this issue and adhere to the remaining constraints, two transformations are necessary. The first involves including an offset along the abscissa axis to reposition the entire function rightward, ensuring proper alignment. This can be effectively accomplished by incorporating a term in the denominator, denoted $b(n)$, that shifts the value of $k$ evaluated as a function of the system size. Secondly, introducing only the offset would not provide the necessary curvature for the function to intersect all prescribed fixed points of edge cases. Consequently, another transformation is required to establish a degree of freedom in this aspect to subsequently resolve its form. Above, the correction is effectuated by adding another term, similar to the preceding offset, to the denominator. The distinction lies in the application of this identical $a(n)$ as a multiplicative factor for the entire function. To grasp why this form works as expected, we must examine the asymptote that determines the upper limit for the range of $k$. Without any correction, this asymptote was originally vertical when neared from left, compelling the function to approach it with a certain growth rate. Upon the inclusion of $a(n)$, presuming it provides strictly positive values, this asymptote transitions from the point $k=0$ to $k=a(n)$, which, in conjunction with the scaling applied to all points, exerts a direct influence on the curvature. This phenomenon can also be observed by evaluating the curvature metric \cite{Coolidge1952} for the function both prior to and subsequent to the transformation:
\begin{align}
    \kappa_{c(n,k)} = \frac{\frac{\partial ^2 c(n,k)}{\partial k^2}}{\left(1+\left(\frac{\partial c(n,k)}{\partial k}\right)^2\right)^{\frac{3}{2}}}
\end{align}
In the case where $c_0(n,k)=\frac{1}{b(n)-k}$, its curvature can be calculated as follows:
\begin{align}
    \frac{\partial \left( \frac{1}{b(n) - k} \right)}{\partial k} = \frac{1}{(b(n)-k)^2} \quad  \quad \frac{\partial ^2 \left( \frac{1}{b(n) - k} \right)}{\partial k^2} = \frac{\partial \left( \frac{1}{(b(n)-k)^2} \right)}{\partial k} = \frac{2}{(b(n)-k)^3}
\end{align}

\begin{align}
    \kappa_{c_0(n,k)} = \frac{\frac{2}{(b(n)-k)^3}}{\left(1+\left(\frac{1}{(b(n)-k)^2}\right)^2\right)^{\frac{3}{2}}} = \frac{2}{\left(1+\frac{1}{(b(n)-k)^4}\right)^{\frac{3}{2}}\cdot (b(n)-k)^3}
\end{align}

In contrast, when both transformations are applied $c_1(n,k) = \frac{a(n)}{a(n)-k+b(n)}$:

\begin{align}
    \frac{\partial \left( \frac{a(n)}{a(n)-k+b(n)} \right)}{\partial k} = \frac{a(n)}{(a(n)-k+b(n))^2} \quad  \quad \frac{\partial ^2 \left( \frac{a(n)}{a(n)-k+b(n)} \right)}{\partial k^2} = \frac{\partial \left( \frac{a(n)}{(a(n)-k+b(n))^2} \right)}{\partial k} = \frac{2\cdot a(n)}{(a(n)-k+b(n))^3}
\end{align}

\begin{align}
    \kappa_{c_1(n,k)} = \frac{\frac{2\cdot a(n)}{(a(n)-k+b(n))^3}}{\left(1+\left(\frac{a(n)}{(a(n)-k+b(n))^2}\right)^2\right)^{\frac{3}{2}}} = \frac{2\cdot a(n)}{\left(1+\left(\frac{a(n)}{(a(n)-k+b(n))^2}\right)^2\right)^{\frac{3}{2}}\cdot (a(n)-k+b(n))^3}
\end{align}
Now, to demonstrate that the previous transformations are correct, we have an expression $\kappa_{c_1(n,k)}$ for its curvature, which allows us to determine if any of the terms $a(n)$ or $b(n)$ are involved in a variation of such curvature. Regarding the influence that $b(n)$ may wield, since this is where it is most apparent, we know that its sole purpose is to translate the function $c_1(n,k)$ along the x-axis according to its output value. Therefore, by uniformly shifting the entire function, it should not modify any property of the function related to its growth or curvature with respect to itself. That is, although the points at which it grows or ceases to do so are shifted, their proportion must be maintained.
\begin{align}
    \frac{a(n)}{a(n)-k+(b(n)+\alpha)} = c_1(n,k-\alpha) = \frac{a(n)}{a(n)-(k-\alpha)+b(n)}
\end{align}
Firstly, demonstrating that $b(n)$ effectively translates the function is trivial, since adding a real quantity $\alpha$ to it is equivalent to subtracting it from $k$, resulting in a translation of $\alpha$ units towards the side indicated by the sign of the offset. Additionally, we can verify that this property equally applies to its curvature metric, which returns a faithfully representative amount of the function's curvature at each $k$. Consequently, we shall simultaneously prove that this curvature remains invariant under translations induced by $b(n)$, implying that this term does not affect the curvature, and indicating that the remaining term $a(n)$ is responsible for managing this property.
\begin{align}
    \kappa_{c_1(n,k)}(k+\alpha,a(n),b(n)) = \kappa_{c_1(n,k)}(k,a(n),b(n)+\beta)
\end{align}
If we parameterize the metric $\kappa_{c_1(n,k)}$ based on the number of elements and the value of the transformation terms, we can characterize the preceding invariance by including two constants $\alpha$ and $\beta$. That is, the curvature of $c_1(n,k)$ at the point $k+\alpha$ must be equivalent to that at the point $k$ with its value of $b(n)$ displaced by a different amount $\beta$. By carrying out this approach, the relationships we obtain between the two constants will dictate the correctness of the premise that $b(n)$ uniquely translates the function.
\begin{align}
    \frac{2\cdot a(n)}{(a(n) - (k +\alpha) + b(n))^3 \left(1 + \frac{a(n)^2}{(a(n) - (k +\alpha) + b(n))^4}\right)^{3/2}} = \frac{2\cdot a(n)}{(a(n) - k + b(n) + \beta)^3 \left(1 + \frac{a(n)^2}{(a(n) - k + b(n) + \beta)^4}\right)^{3/2}} \label{offsetinvarianceoriginal}
\end{align}

\begin{align}
    \frac{1}{(a(n) - k -\alpha + b(n))^3 \left(1 + \frac{a(n)^2}{(a(n) - k -\alpha + b(n))^4}\right)^{3/2}} = \frac{1}{(a(n) - k + b(n) + \beta)^3 \left(1 + \frac{a(n)^2}{(a(n) - k + b(n) + \beta)^4}\right)^{3/2}}
\end{align}

\begin{align}
    (a(n) - k -\alpha + b(n))^3 \left(1 + \frac{a(n)^2}{(a(n) - k -\alpha + b(n))^4}\right)^{3/2} = (a(n) - k + b(n) + \beta)^3 \left(1 + \frac{a(n)^2}{(a(n) - k + b(n) + \beta)^4}\right)^{3/2}
\end{align}

\begin{align}
    &(a(n) - k -\alpha + b(n))^3 \left(\frac{(a(n) - k -\alpha + b(n))^4+a(n)^2}{(a(n) - k -\alpha + b(n))^4}\right)^{3/2} =\\&= (a(n) - k + b(n) + \beta)^3 \left( \frac{(a(n) - k + b(n) + \beta)^4+a(n)^2}{(a(n) - k + b(n) + \beta)^4}\right)^{3/2} \notag
\end{align}

\begin{align}
    &(a(n) - k -\alpha + b(n))^3 \left(\frac{((a(n) - k -\alpha + b(n))^4+a(n)^2)^{3/2}}{(a(n) - k -\alpha + b(n))^6}\right) =\\&= (a(n) - k + b(n) + \beta)^3 \left( \frac{((a(n) - k + b(n) + \beta)^4+a(n)^2)^{3/2}}{(a(n) - k + b(n) + \beta)^6}\right) \notag
\end{align}

\begin{align}
    \frac{((a(n) - k -\alpha + b(n))^4+a(n)^2)^{3/2}}{(a(n) - k -\alpha + b(n))^3} =  \frac{((a(n) - k + b(n) + \beta)^4+a(n)^2)^{3/2}}{(a(n) - k + b(n) + \beta)^3}
\end{align}

\begin{align}
    \left(\frac{\sqrt{(a(n) - k -\alpha + b(n))^4+a(n)^2}}{a(n) - k -\alpha + b(n)}\right)^3 =  \left(\frac{\sqrt{(a(n) - k + b(n) + \beta)^4+a(n)^2}}{a(n) - k + b(n) + \beta}\right)^3
\end{align}
Given our focus on real solutions and acknowledging that the radicand within the square root function remains invariably positive, coupled with the condition that both denominators must be non-zero, we arrive at the following equality:
\begin{align}
    \frac{(a(n) - k -\alpha + b(n))^4+a(n)^2}{(a(n) - k -\alpha + b(n))^2} =  \pm \frac{(a(n) - k + b(n) + \beta)^4+a(n)^2}{(a(n) - k + b(n) + \beta)^2} \label{solutions}
\end{align}

Considering the values of $\alpha$ and $\beta$ that cause any of the denominators to become zero are:
\begin{align}
    a(n) - k -\alpha + b(n)=0 \implies \alpha = a(n) - k + b(n)
\end{align}
\begin{align}
    a(n) - k + b(n) + \beta=0 \implies \beta=-(a(n) - k + b(n)) \label{denominatorbeta}
\end{align}

Given the previous sign alternation, we need to determine all potential solutions to the equation. So, in the case where both sides are positive, we proceed as follows:
\begin{align}
    \frac{(a(n) - k -\alpha + b(n))^4+a(n)^2}{(a(n) - k -\alpha + b(n))^2} = \frac{(a(n) - k + b(n) + \beta)^4+a(n)^2}{(a(n) - k + b(n) + \beta)^2}
\end{align}

\begin{align}
    &(a(n) - k -\alpha + b(n))^4\cdot (a(n) - k + b(n) + \beta)^2 + a(n)^2\cdot (a(n) - k + b(n) + \beta)^2 =\\&= (a(n) - k + b(n) + \beta)^4\cdot (a(n) - k -\alpha + b(n))^2+a(n)^2\cdot (a(n) - k -\alpha + b(n))^2 \notag
\end{align}

\begin{align}
    &(a(n) - k -\alpha + b(n))^4\cdot (a(n) - k + b(n) + \beta)^2 + a(n)^2\cdot (a(n) - k + b(n) + \beta)^2 \\&- (a(n) - k -\alpha + b(n))^2\cdot ((a(n) - k + b(n) + \beta)^4+a(n)^2)= 0 \notag
\end{align}

\begin{align}
    &(a(n) - k -\alpha + b(n))^4\cdot (a(n) - k + b(n) + \beta)^2 + a(n)^2\cdot (a(n) - k + b(n) + \beta)^2 \\&+ (a(n) - k -\alpha + b(n))^2\cdot (-(a(n) - k + b(n) + \beta)^4-a(n)^2)= 0 \notag
\end{align}
To achieve an effective simplification, we apply the substitution $(a(n) - k -\alpha + b(n))^2=u$
\begin{align}
    u^2\cdot (a(n) - k + b(n) + \beta)^2 + a(n)^2\cdot (a(n) - k + b(n) + \beta)^2 + u\cdot (-(a(n) - k + b(n) + \beta)^4-a(n)^2)= 0 
\end{align}

\begin{align}
    ((a(n) - k + b(n) + \beta)^2-u)(a(n)^2-u\cdot (a(n) - k + b(n) + \beta)^2)=0
\end{align}
Thus, by reversing the substitution:
\begin{align}
    ((a(n) - k + b(n) + \beta)^2-(a(n) - k -\alpha + b(n))^2)\cdot (a(n)^2-(a(n) - k + b(n) + \beta)^2(a(n) - k -\alpha + b(n))^2)=0
\end{align}
For the equality to be valid, one of the multiplicands must be 0, hence, for one of them:
\begin{align}
    ((a(n) - k + b(n) + \beta)^2-(a(n) - k -\alpha + b(n))^2)=0
\end{align}
\begin{align}
    a(n) - k + b(n) + \beta = \pm(a(n) - k -\alpha + b(n))
\end{align}

\begin{align}
    a(n) - k + b(n) + \beta = a(n) - k -\alpha + b(n) \implies \alpha=-\beta
\end{align}

\begin{align}
    a(n) - k + b(n) + \beta = -(a(n) - k -\alpha + b(n)) \implies \alpha = 2\cdot (a(n) - k + b(n)) + \beta
\end{align}

Once these solutions have been reached, we proceed to check the remaining multiplicand:
\begin{align}
    (a(n)^2-(a(n) - k + b(n) + \beta)^2(a(n) - k -\alpha + b(n))^2)=0
\end{align}

\begin{align}
    a(n)^2 = (a(n) - k + b(n) + \beta)^2(a(n) - k -\alpha + b(n))^2
\end{align}

\begin{align}
    a(n) - k + b(n) + \beta = \pm \frac{a(n)}{a(n) - k -\alpha + b(n)}
\end{align}

\begin{align}
    a(n) - k -\alpha + b(n) = \pm \frac{a(n)}{a(n) - k + b(n) + \beta} \implies \alpha  = a(n) - k + b(n) \pm \frac{a(n)}{a(n) - k + b(n) + \beta}
\end{align}

Then, all solutions for $\alpha$ that we attain by operating with \textcolor{blue}{\ref{solutions}} with both sides in positive are:
\begin{align}
\alpha = \begin{cases}
        -\beta\\
        2\cdot (a(n) - k + b(n)) + \beta\\
        a(n) - k + b(n) + \frac{a(n)}{a(n) - k + b(n) + \beta} \\
        a(n) - k + b(n) - \frac{a(n)}{a(n) - k + b(n) + \beta}
    \end{cases}
\end{align}

Henceforth, we must verify the existence of additional solutions when the signs of \textcolor{blue}{\ref{solutions}} are inverse:
\begin{align}
    \frac{(a(n) - k -\alpha + b(n))^4+a(n)^2}{(a(n) - k -\alpha + b(n))^2} = -\frac{(a(n) - k + b(n) + \beta)^4+a(n)^2}{(a(n) - k + b(n) + \beta)^2}
\end{align}

\begin{align}
    &(a(n) - k -\alpha + b(n))^4\cdot (a(n) - k + b(n) + \beta)^2 + a(n)^2\cdot (a(n) - k + b(n) + \beta)^2 =\\&= -(a(n) - k + b(n) + \beta)^4\cdot (a(n) - k -\alpha + b(n))^2-a(n)^2\cdot (a(n) - k -\alpha + b(n))^2 \notag
\end{align}

\begin{align}
    &(a(n) - k -\alpha + b(n))^4\cdot (a(n) - k + b(n) + \beta)^2 + a(n)^2\cdot (a(n) - k + b(n) + \beta)^2 \\&+(a(n) - k -\alpha + b(n))^2((a(n) - k + b(n) + \beta)^4+a(n)^2)= 0 \notag
\end{align}
By applying and reversing the same substitution as before, the subsequent factored form is produced:
\begin{align}
    ((a(n) - k + b(n) + \beta)^2+(a(n) - k -\alpha + b(n))^2)\cdot (a(n)^2+(a(n) - k + b(n) + \beta)^2(a(n) - k -\alpha + b(n))^2)=0
\end{align}
The solutions, as previously derived, are found by equating each factor to zero. So, for the initial one:
\begin{align}
    (a(n) - k + b(n) + \beta)^2+(a(n) - k -\alpha + b(n))^2=0
\end{align}

\begin{align}
    (a(n) - k + b(n) + \beta)^2=-(a(n) - k -\alpha + b(n))^2
\end{align}

\begin{align}
    a(n) - k + b(n) + \beta=\pm(a(n) - k -\alpha + b(n))\cdot i \implies \alpha = \begin{cases}
        a(n) - k + b(n) + (a(n) + b(n) + \beta - k)\cdot i\\
        a(n) - k + b(n) + (-a(n) - b(n) - \beta + k)\cdot i
    \end{cases}
\end{align}
For the remaining multiplier:
\begin{align}
    a(n)^2+(a(n) - k + b(n) + \beta)^2(a(n) - k -\alpha + b(n))^2=0
\end{align}

\begin{align}
    (a(n) - k -\alpha + b(n))^2=\frac{-a(n)^2}{(a(n) - k + b(n) + \beta)^2}
\end{align}

\begin{align}
    a(n) - k -\alpha + b(n)=\pm \frac{a(n)}{a(n) - k + b(n) + \beta}\cdot i
\end{align}
Conversely, the results deduced by manipulating \textcolor{blue}{\ref{solutions}} with one term positive and the other with its sign inverted are the following:
\begin{align}
\alpha = \begin{cases}
        a(n) - k + b(n) + (a(n) + b(n) + \beta - k)\cdot i\\
        a(n) - k + b(n) + (-a(n) - b(n) - \beta + k)\cdot i\\
        a(n) - k + b(n) + \frac{a(n)}{a(n) - k + b(n) + \beta}\cdot i\\
        a(n) - k + b(n) - \frac{a(n)}{a(n) - k + b(n) + \beta}\cdot i
    \end{cases}
\end{align}
Once we have obtained all the potential solutions for $\alpha$ and $\beta$, we must check if they satisfy the original equality \textcolor{blue}{\ref{offsetinvarianceoriginal}}. On one hand, this process turns out to be straightforward for the first case where $\alpha = -\beta$, resulting in a tautology that validates the solution. In other cases, such as $\alpha = 2 \cdot (a(n) - k + b(n)) + \beta$, the condition $\beta \neq k - b(n) \land a(n)=0$ emerges, which contradicts the constraint $a(n)>0$, so it is discarded. For the remaining 2 cases where $\alpha$ does not depend, apparently, on any complex value, we obtain solutions of the form $\beta \neq -a(n) - b(n) + k$ with $a(n)>0$ and $a(n)<0$ respectively. Thus, the one that requires negative values for $a(n)$ is immediately excluded, and the other one could be considered, since the only value where $\beta$ is undefined is the same as in the restriction \textcolor{blue}{\ref{denominatorbeta}}, where the denominator of the expression cannot equal 0. Regarding the last 4 cases, they are ignored for having a complex part. It is true that substituting them into the original equality may or may not be suitable, but having initially defined the constants $\alpha$ and $\beta$ as real parameters representing an offset in the curvature metric, it is pointless to use complex quantities for such translation, since the only quantity translated would be its real part.
\\\\
In conclusion, some of the previous solutions, especially $\alpha = -\beta$, demonstrate the invariance in the curvature of the function $c_1(n,k)$ when its offset $b(n)$ is modified. That is, for any real value $\beta$ that increases or decreases the value of the parameter $b(n)$, there will always exist another constant $\alpha$ with the opposite sign that shifts the curvature metric to the same location as it was positioned with the previous displacement, without changing its shape. In this way, it is ensured that any change in the parameter $b(n)$ does not affect the curvature of the function, but rather its translation along the x-axis, as the curvature metric remains invariant under such translation. Subsequently, after proving that one of the parameters used to construct the shape of the function $c(n,k)$ is exclusively responsible for adjusting its horizontal translation, it remains to be determined whether the other parameter $a(n)$ exerts any influence on the curvature. Such determination is key because, otherwise, the function could not be adjusted to the constraints of the average cluster size in our problem.

\begin{align}
    \kappa_{c_1(n,k)}(k+\alpha,a(n),b(n)) = \kappa_{c_1(n,k)}(k,a(n)+\beta,b(n))
\end{align}
For this purpose, we will use an analogous approach to that utilized for checking $b(n)$, with the difference that constants will be incorporated to different parameters. Fundamentally, to show that $a(n)$ only affects the curvature and does not produce any transformation that $b(n)$ might generate, we must prove that the curvature metric is not invariant under translations of this parameter. That is, if $a(n)$ did not influence the curvature, a change of $\beta$ in it would cause a transformation that maintains the curvature metric in the same form, but potentially with a different position with respect to $k$. Therefore, that difference in position is modeled as a real constant $\alpha$ added to the variable $k$. Thus, if we found a solution like $\alpha = \pm \beta$ or some function depending of $\alpha$, it would indicate that $a(n)$ produces the same effect as $b(n)$, so the curvature metric would remain unchanged with variations in $a(n)$, implying that it does not affect the curvature. Also, it could cancel the effect of $b(n)$ if this were the case, since both values may be equal. Consequently, by substituting the corresponding constant values into both expressions, we obtain:

\begin{align}
    \frac{2\cdot a(n)}{(a(n) - (k +\alpha) + b(n))^3 \left(1 + \frac{a(n)^2}{(a(n) - (k +\alpha) + b(n))^4}\right)^{3/2}} = \frac{2\cdot (a(n)+\beta)}{(a(n) + \beta - k + b(n))^3 \left(1 + \frac{(a(n)+\beta)^2}{(a(n) + \beta - k + b(n))^4}\right)^{3/2}}
\end{align}

\begin{align}
    &a(n)\cdot (a(n) + \beta - k + b(n))^3 \left(1 + \frac{(a(n)+\beta)^2}{(a(n) + \beta - k + b(n))^4}\right)^{3/2} =\\&= (a(n)+\beta)\cdot (a(n) - (k +\alpha) + b(n))^3 \left(1 + \frac{a(n)^2}{(a(n) - (k +\alpha) + b(n))^4}\right)^{3/2} \notag
\end{align}

\begin{align}
    \frac{(a(n) + \beta - k + b(n))^3 \left(1 + \frac{(a(n)+\beta)^2}{(a(n) + \beta - k + b(n))^4}\right)^{3/2}}{(a(n) - k -\alpha + b(n))^3 \left(1 + \frac{a(n)^2}{(a(n) - k -\alpha + b(n))^4}\right)^{3/2}} = \frac{a(n)+\beta}{a(n)}
\end{align}

\begin{align}
    \frac{(a(n) + \beta - k + b(n)) \left(1 + \frac{(a(n)+\beta)^2}{(a(n) + \beta - k + b(n))^4}\right)^{1/2}}{(a(n) - k -\alpha + b(n)) \left(1 + \frac{a(n)^2}{(a(n) - k -\alpha + b(n))^4}\right)^{1/2}} = \left(\frac{a(n)+\beta}{a(n)}\right)^{1/3}
\end{align}

\begin{align}
    \frac{(a(n) + \beta - k + b(n))^2 \left(1 + \frac{(a(n)+\beta)^2}{(a(n) + \beta - k + b(n))^4}\right)}{(a(n) - k -\alpha + b(n))^2 \left(1 + \frac{a(n)^2}{(a(n) - k -\alpha + b(n))^4}\right)} = \pm \left(\frac{a(n)+\beta}{a(n)}\right)^{2/3}
\end{align}

\begin{align}
    \frac{(a(n) + \beta - k + b(n))^2 \left(1 + \frac{(a(n)+\beta)^2}{(a(n) + \beta - k + b(n))^4}\right)}{\pm \left(\frac{a(n)+\beta}{a(n)}\right)^{2/3}} = (a(n) - k -\alpha + b(n))^2 \left(1 + \frac{a(n)^2}{(a(n) - k -\alpha + b(n))^4}\right) \label{usubstitution}
\end{align}
We proceed to substitute the left-hand side of \textcolor{blue}{\ref{usubstitution}} with a new variable $u$:
\begin{align}
    u = (a(n) - k -\alpha + b(n))^2 \left(1 + \frac{a(n)^2}{(a(n) - k -\alpha + b(n))^4}\right)
\end{align}
By solving for $\alpha$ as a function of the remaining variables, we conclude with the following solutions:
\begin{align}
    \alpha = \begin{cases}
        a(n)+b(n)-k+ \frac{\sqrt{u+ \sqrt{u^2-4 a(n)^2}}}{\sqrt{2}}\\
        a(n)+b(n)-k+ \frac{\sqrt{u- \sqrt{u^2-4 a(n)^2}}}{\sqrt{2}}\\
        a(n)+b(n)-k- \frac{\sqrt{u+ \sqrt{u^2-4 a(n)^2}}}{\sqrt{2}}\\
        a(n)+b(n)-k- \frac{\sqrt{u- \sqrt{u^2-4 a(n)^2}}}{\sqrt{2}}
    \end{cases}
\end{align}
For each of the above solutions, it is necessary to consider the $\pm$ sign in the denominator of $u$, yielding more than those shown. Thus, following the same procedure used for verifying the curvature invariance in $b(n)$, we derive restrictions that exclude the solutions, either because they have a complex part or because they contradict the parameter constraints of our problem. Hence, the sole method to satisfy the initial equality is for both constants to be equal and equivalent to zero, $\alpha = \pm \beta = 0$. Consequently, this unique solution implies that the only manner for two constants to modify $a(n)$, translate the function $c(n,k)$ horizontally with respect to $k$, and do so without altering the curvature, is for both to be null, denoting a zero displacement on the x-axis and in the variation of $a(n)$.
\\\\
In summary, a potential form for the average cluster size $c(n,k)$ has been proposed, involving different terms that can be adjusted to fit its properties. On one hand, $b(n)$ has been observed to induce a horizontal shift in the function without affecting its curvature, allowing it to relocate all the points in the necessary space so that some of them meet the other constraints of the function. Additionally, it has been verified that $a(n)$ is not capable of performing the same transformation as the other term, given its effect on the curvature. This can be seen not only in this sense, as $a(n)$ could perform a vertical displacement, or even on a complex axis inherent to the metric, without varying the curvature. The main reason lies in the numerator of the curvature metric, which, being $2 \cdot a(n)$, or $2 \cdot a(n) + \beta$ in the case it changes, directly impacts the metric's value. Therefore, verifying that the term does not shift the function in the same way as $b(n)$ is sufficient to establish that it controls the curvature as a function of its value.
\\\\
At this juncture, we know the restrictions that $c(n,k)$ must satisfy for it to operate correctly in the formula for the average insertion time, we know what each of its terms does, but there still isn't a reliable mean to determine the form of these terms. That is, $a(n)$ must return a positive real value, but how it does so as a function of $n$ is still undetermined. Therefore, we will proceed to apply these restrictions to the form we have for the average cluster size, which can provide information in this regard:
\begin{align}
    c(n,1) = \frac{a(n)}{a(n)-1+b(n)} = 1 \quad \quad c(n,n^2) = \frac{a(n)}{a(n)-n^2+b(n)} = n^2
\end{align}
The first restriction to be checked is the one that forces the curve intersect the point $(1,1)$. By substituting these values into the formula, we obtain the following:
\begin{align*}
    \frac{a(n)}{a(n)-1+b(n)} = 1
\end{align*}
\begin{align*}
    a(n) = a(n)-1+b(n) \implies b(n)=1
\end{align*}
Given this condition, it results in a constant value for $b(n)$. This might seem unusual since the term depends on $n$ and hence could potentially yield a more elaborate function. However, the offset depends on how the curvature reaches the reference points of the edge cases and the shape of the function itself. So, obtaining a constant value can actually be beneficial for future calculations. Moreover, the constraint that the function must intersect the point $(n^2, n^2)$ is missing. Specifically, in the case where the matrix is full, the average size of a cluster can only equal the system's size, although practically this situation cannot occur because if there are more elements than $n^2 - n$, at least one of them will be in one of the rows that form paths in the system, simultaneously bordering cells located in a cluster that reaches the row at the other extreme. Nevertheless, considering this case is theoretically valid and beneficial for simplifying the process of determining a candidate function for $c(n,k)$, as it serves its purpose for the corresponding input values.
\begin{align*}
    \frac{a(n)}{a(n)-n^2+b(n)} = n^2
\end{align*}
\begin{align*}
    a(n) =  n^2\cdot a(n)-n^4+n^2\cdot b(n)
\end{align*}
\begin{align*}
    a(n) = \frac{n^2\cdot (b(n)-n^2)}{(1-n^2)} = \frac{n^2\cdot (1-n^2)}{(1-n^2)} \implies a(n)=n^2
\end{align*}
After introducing the condition of the point where it returns $n^2$ when there are the same number of elements, we ascertain that the term $a(n)$ must be equivalent to this quantity. In this case, the term is indeed a function of $n$, as its very definition indicates. Furthermore, since $a(n)$ is required to be positive, we can see that the expression guarantees this condition, given that we are working with strictly positive values of $n$, and even if we were to assign it negative values, it would still return their square in positive, which is what our function demands.
\begin{align*}
    c(n,k)=\frac{n^2}{n^2-k+1}
\end{align*}
Ultimately, by substituting the prior solutions into the original form of $c(n,k)$, we obtain the upper function, which serves as a preliminary estimate of the average cluster size based on its most basic constraints.
\begin{figure}[H]
    \centering
    \includegraphics[width=10cm,clip]{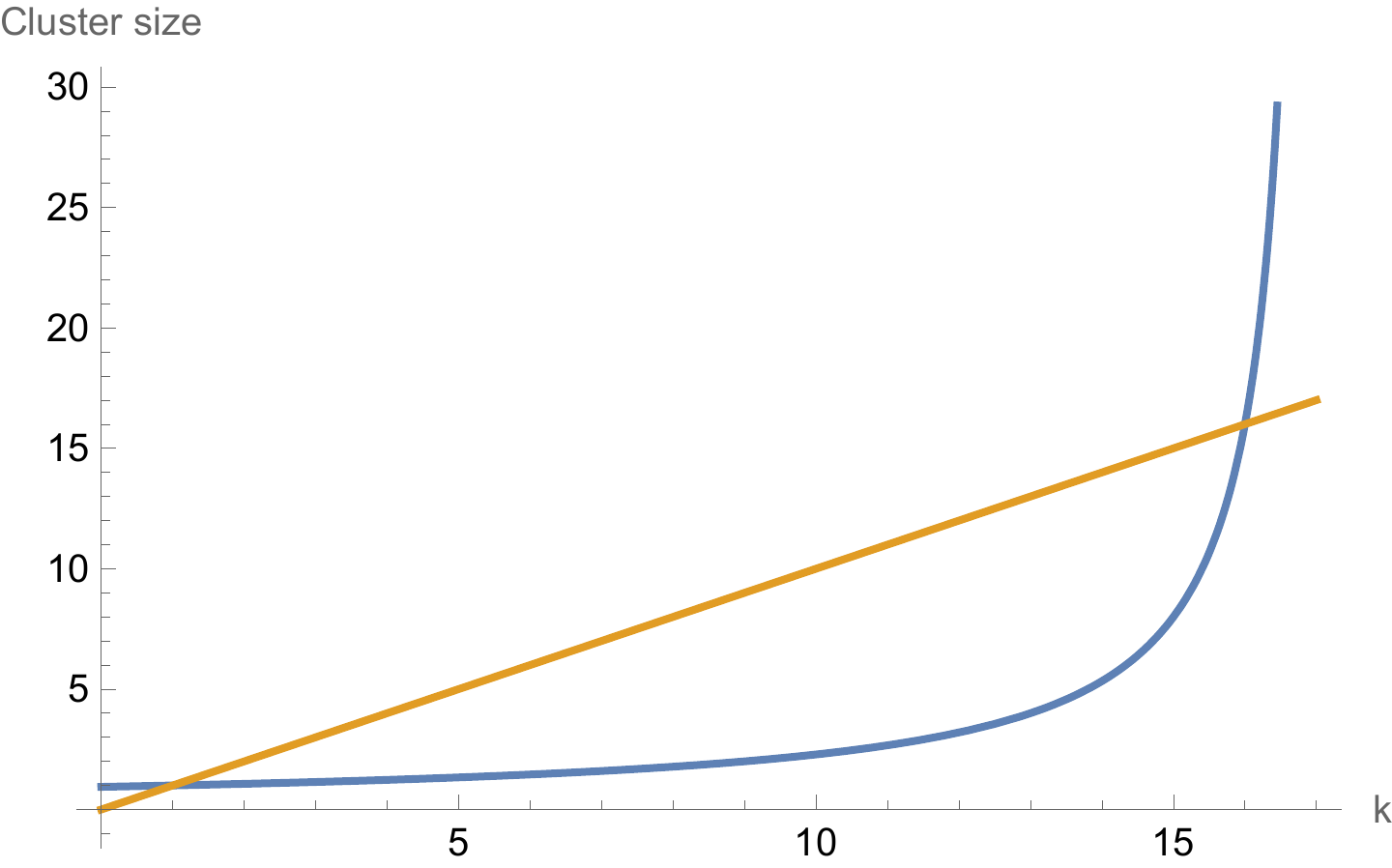}
    \caption{Plot of the average cluster size function in blue along with its worst case $c(n,k)=k$ in orange for a matrix of size $4\times 4$}
    \label{fig:ClusterSizeRestrictions}
\end{figure}
Above, the resulting function is displayed alongside the case where the cluster size is the maximum possible, demonstrating that all imposed restrictions are satisfactorily met. Additionally, to verify that the form of the function is suitable, a particular worst-case constraint can be applied to determine if the result of $a(n)$ matches the plotted function. In other words, the scenario where the cluster size coincides with $k$.
\begin{align}
    c(n,k) = \frac{a(n)}{a(n)-k+b(n)} = k
\end{align}
Assuming that $b(n)$ consistently holds the constant value as dictated by the condition where the system contains only one element, we can reformulate and proceed as follows:
\begin{align}
    \frac{a(n)}{a(n)-k+1} = k
\end{align}
\begin{align*}
    a(n) = k\cdot a(n)-k^2+k = k\cdot (a(n)+1)-k^2
\end{align*}
\begin{align*}
    k=\frac{-(a(n)+1)\pm \sqrt{(a(n)+1)^2-4\cdot a(n)}}{-2}=\frac{a(n)+1\pm \sqrt{(a(n)+1)^2-4\cdot a(n)}}{2}
\end{align*}
\begin{align*}
    k=\frac{a(n)+1\pm \sqrt{a(n)^2+2\cdot a(n)+1-4\cdot a(n)}}{2}=\frac{a(n)+1\pm \sqrt{(a(n)-1)^2}}{2}
\end{align*}

Therefore, several solutions to the quadratic equation are gathered:
\begin{align}
    k=\frac{a(n)+1\pm |a(n)-1|}{2}=\begin{cases}
        \frac{a(n)+1+a(n)-1}{2}=a(n)\\
        \frac{a(n)+1-a(n)+1}{2}=1
    \end{cases}
\end{align}
Among all the options, the one that articulates $a(n)$ in terms of $k$ is unique, thus the expression for $c(n,k)$ is stated as follows:
\begin{align}
    c(n,k) = \frac{k}{k-k+1} = k
\end{align}
Hence, it is demonstrated that by enforcing the constraint on the function such that, in each iteration, there exists a cluster whose size is commensurate with the total number of elements, the expected outcome is attained. Thus, it is shown how the curvature of this function can be adapted to suit the various scenarios the algorithm may encounter, thereby modeling each one adequately. Concerning the horizontal intercept, it remains constrained by the condition that all cases intersect at the point $(1,1)$, which hinders its variability.
\\\\
We now have at our disposal the complete set of functions $c(n,k)$ requisite for the modeling of the specified algorithm cases. However, the average time one we will use to compute the average insertion runtime and subsequently the total process time complexity, although expressed in a closed form, cannot be guaranteed to be equal to the actual function that models this magnitude. So, given all the known conditions it must met, we can still propose a method to obtain a function approximately equivalent to it. To this end, we begin with the assumption that the optimal scenario for the algorithm, in terms of the average cluster size, is one where all clusters have an equivalent size of 1 regardless of the number of elements within the system or specific state. Conversely, the worst-case scenario is characterized by a cluster of size $k$, implying that the average case must reside somewhere between these two bounds. Furthermore, we count with a rough understanding of the expected graphical form of this function based on the algorithm's behavior, empirically, and additional constraints. With all this information, even in the absence of precise knowledge about the function's placement between the bounds marked by the aforementioned cases, it is still possible to compute an average of all the functions contemplated within this range.
\begin{align}
    c(n,k,\xi) = \left(\frac{k-1}{n^2-1}\right)^\xi (n^2-1)+1
\end{align}
Initially, it is imperative to identify a function like the above one, so it enables us to represent all the possibilities for $c(n,k)$ to reside in that range \cite{DiewertWales1989}. Specifically, we seek a function whose curvature can vary according to a parameter $\xi$, spanning from $c(n,k) = k$ to asymptotically nearing the best case $c(n,k) = 1$. Nevertheless, it should be acknowledged that this optimal case will never truly be reached since, for that to occur, the parameter $\xi$ would need to be infinite. The choice of this function is one among many to achieve the desired curvature and behavior in relation to the parameter $\xi$, and as such, it may differ from the result already obtained for the average case. It might even deviate from the actual function we are attempting to find. However, asymptotically, the mean will still serve for the rest of the analysis, in case it exists.
\begin{figure}[H]
    \centering
    \includegraphics[width=10cm,clip]{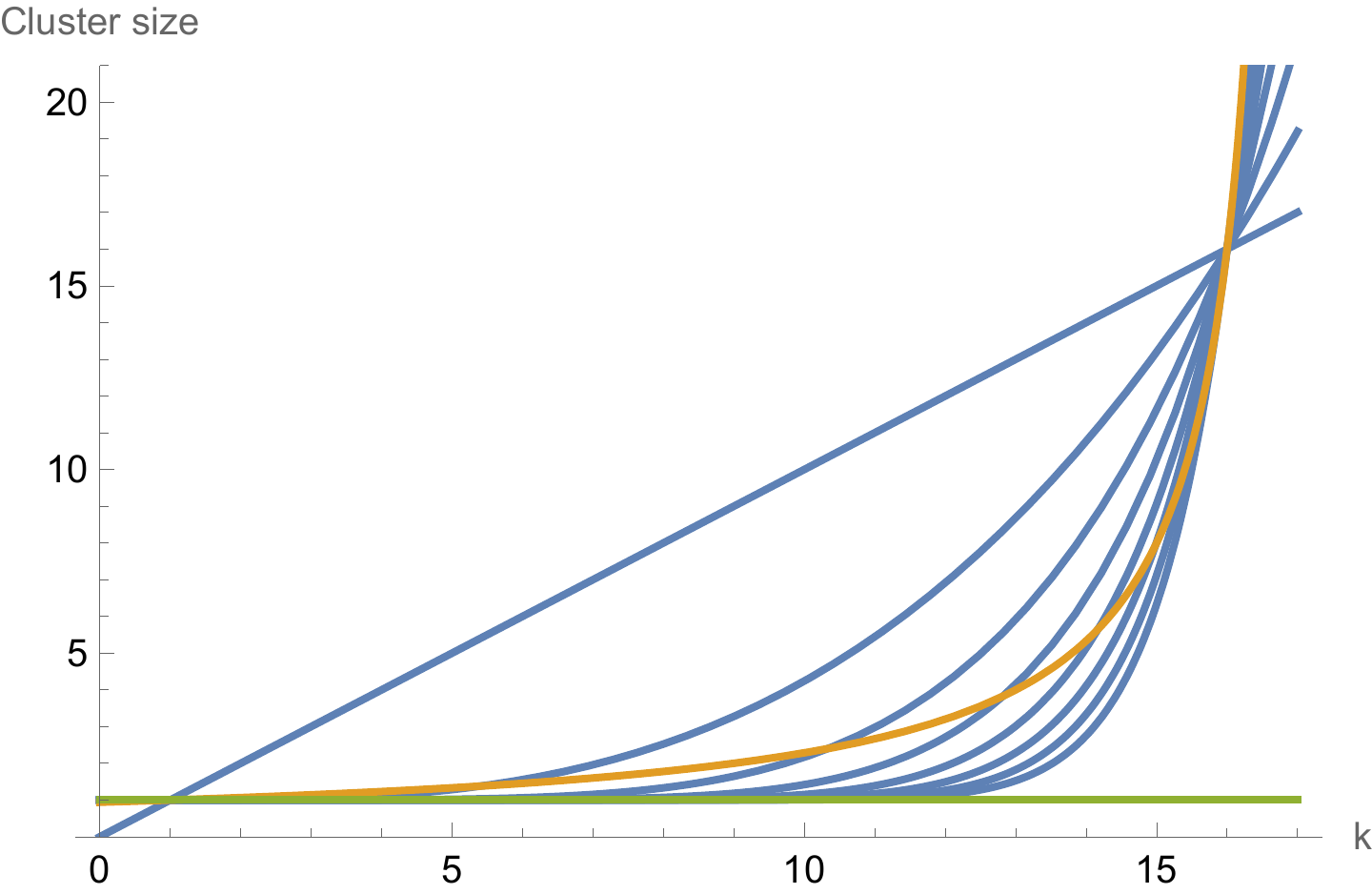}
    \caption{Best case cluster size $c(n,k)=1$ plotted in green, the one derived from its constraints in orange, and in blue the set of $c(n,k,\xi)$ functions where $1\leq \xi \leq n^2$ with a step of value 2 and $n=4$}
    \label{fig:ClusterSizeCurvatures}
\end{figure}
In Figure 15, the previously defined function is displayed alongside several functions $c(n,k,\xi)$ with specific values for the parameter $\xi$. Despite their similarity in curvature, they are not exactly identical. Consequently, it remains undetermined whether the parameterized function or the original function more accurately represent the average cluster size; thus, both may be deemed applicable. In this context, it is known that the actual function will be situated between $c(n,k,1)$ and $c(n,k,\infty)$. The former corresponds to the initial case, while the latter, when the curvature approaches infinity, approximates the best-case scenario. Therefore, by averaging all intermediate functions \cite{Wong2006}, we can achieve an asymptotic approximation that may be useful for calculating the average insertion time.

\begin{align}
    \overline{c}(n,k) = \overline{c}(n,k,\xi) = \lim_{U \to \infty} \frac{1}{U-1} \int_{1}^{U} c(n,k,\xi) \, d\xi
\end{align}
To compute the average \cite{Dougherty2016} of all the previously plotted functions, we must account for the full spectrum of curvatures spanning from the worst to the best scenarios, which correspond to the interval $\xi \in [1, \infty)$. Then, by integrating $\xi$ over this range and dividing the result by its length, we obtain the approximation $\overline{c}(n,k)$ for the average cluster size \cite{Stewart2012}:
\begin{align}
    \lim_{U \to \infty} \frac{1}{U-1} \int_{1}^{U} \left(\frac{k-1}{n^2-1}\right)^\xi (n^2-1)+1 \, d\xi
\end{align}

\begin{align}
    \lim_{U \to \infty} \left( \frac{1}{U-1} \int_{1}^{U} \left(\frac{k-1}{n^2-1}\right)^\xi (n^2-1)  \, d\xi + \frac{1}{U-1} \int_{1}^{U} 1 \, d\xi \right)
\end{align}

\begin{align}
    \lim_{U \to \infty} \left( \frac{1}{U-1} \int_{1}^{U} \left(\frac{k-1}{n^2-1}\right)^\xi (n^2-1) \, d\xi + \frac{1}{U-1} \left[ \xi \right]_{1}^{U} \right)
\end{align}
\begin{align}
    \lim_{U \to \infty} \left( \frac{(n^2-1)}{U-1} \int_{1}^{U} \left(\frac{k-1}{n^2-1}\right)^\xi \, d\xi + 1 \right)
\end{align}

\begin{align}
    \lim_{U \to \infty} \left( \frac{(n^2-1)}{U-1} \left[ \frac{\left(\frac{k-1}{n^2-1}\right)^\xi}{ln\left( \frac{k-1}{n^2-1} \right)}\right]_{1}^{U}  + 1 \right)
\end{align}

\begin{align}
    \lim_{U \to \infty} \frac{(n^2-1)}{U-1} \left( \frac{\left(\frac{k-1}{n^2-1}\right)^U}{ln\left( \frac{k-1}{n^2-1} \right)} - \frac{\frac{k-1}{n^2-1}}{ln\left( \frac{k-1}{n^2-1} \right)}\right) + 1
\end{align}

\begin{align}
    \lim_{U \to \infty} \frac{(n^2-1)}{U-1} \left( \frac{\left(\frac{k-1}{n^2-1}\right)^U - \left(\frac{k-1}{n^2-1}\right)}{ln\left( \frac{k-1}{n^2-1} \right)}\right)  + 1 
\end{align}
After having removed the inner integral, we proceed to evaluate the limit:
\begin{align}
    \lim_{U \to \infty} \left(\frac{\left(\frac{k-1}{n^2-1}\right)^U - \left(\frac{k-1}{n^2-1}\right)}{U-1}\right)  \frac{(n^2-1)}{ln\left( \frac{k-1}{n^2-1} \right)}  + 1 
\end{align}
First, we extract all components from the limit that remain constant with respect to $U$, retaining only the variable portion. Subsequently, any term that persists invariant but cannot be easily extracted is negligible when the limit's variable approaches infinity.
\begin{align}
    \lim_{U \to \infty} \left(\frac{\left(\frac{k-1}{n^2-1}\right)^U }{U-1}\right)  \frac{(n^2-1)}{ln\left( \frac{k-1}{n^2-1} \right)}  + 1 
\end{align}
\begin{align}
    \lim_{U \to \infty} \left(\frac{U}{U-1}\right) \cdot \lim_{U \to \infty} \left(\frac{\left(\frac{k-1}{n^2-1}\right)^U }{U}\right)  \frac{(n^2-1)}{ln\left( \frac{k-1}{n^2-1} \right)}  + 1 
\end{align}
By evaluating the limit multiplying on the left side, it simplifies as follows:
\begin{align}
    \lim_{U \to \infty} \left(\frac{\left(\frac{k-1}{n^2-1}\right)^U }{U}\right)  \frac{(n^2-1)}{ln\left( \frac{k-1}{n^2-1} \right)}  + 1 
\end{align}
\begin{align}
    \lim_{U \to \infty} \left(\frac{k-1}{n^2-1}\right)^U = \begin{cases}
\infty & \text{if } \frac{k-1}{n^2-1} > 1 \\
0 & \text{if } \frac{k-1}{n^2-1} < 1 \\
1 & \text{if } \frac{k-1}{n^2-1} = 1
\end{cases}
\end{align}
With these solutions for the limit of the previous numerator, the following results for the entire limit are attained:
\begin{align}
    \lim_{U \to \infty} \left(\frac{\left(\frac{k-1}{n^2-1}\right)^U }{U}\right)  \frac{(n^2-1)}{ln\left( \frac{k-1}{n^2-1} \right)}  + 1 = \begin{cases}
\infty & \text{if } \frac{k-1}{n^2-1} > 1 \\
1 & \text{if } \frac{k-1}{n^2-1} < 1 \\
1 & \text{if } \frac{k-1}{n^2-1} = 1
\end{cases}
\end{align}
According to the outcomes, when the the quantity of elements within the system is less than or equal to the maximum $n^2$, the mean value converges to 1 due to the last additive term. Nevertheless, the result does not converge to an average function that fits the expected form of $c(n,k)$. Furthermore, in the remaining case, it returns infinity, ensuring that it does not converge to any real value, nor does it return any form usable as average \cite{Heilman2015}.
\\\\
In summary, by obtaining the average of all functions with different curvature between both edge cases of the algorithm \cite{GrayVogt2019}, a constant value equal to the best scenario is yielded, predominantly due to the algebraic characteristics of the parameterized function involving $\xi$, which does not seem suitable for the apparent form of $c(n,k)$. So, to increase the likelihood of achieving convergence to a function of $n,k$ parameters in the antecedent procedure, the incorporation of a decay term as delineated below is advisable:
\begin{align}
    \overline{c}(n,k,\xi) = \lim_{U \to \infty} \frac{1}{U-1} \int_{1}^{U} \lambda^{-U} \cdot c(n,k,\xi) \, d\xi
\end{align}
With this approach, we could ensure that the limit as $U$ tends to infinity does not lead to the nullification of all terms containing the desired parameters within the solution. Although, this may also deliver a convergent solution when the number of elements $k$ surpasses its maximum, a case that would not be reasonable to consider, as the number of elements cannot exceed the amount of system's cells. Furthermore, beyond the necessity of computing a suitable base $\lambda$ for this purpose, it is possible that the form of the function we want to obtain may change, which would render it ineffective. Consequently, lacking a reliable mean regarding the curvature of the average cluster size, the only remaining recourse is to compute the average insertion time using the sole concrete expression of $c(n,k)$ at our disposal.
\begin{align}
T_{ins}(n) = \frac{\sum_{k=0}^{n^2} \frac{n^2}{n^2 - k + 1}}{n^2+1} 
&= \frac{\sum_{k=0}^{n^2} \frac{n^2}{(n^2 + 1) - k}}{n^2+1} \\ \notag
&= \frac{n^2 \sum_{k=0}^{n^2} \frac{1}{(n^2 + 1) - k}}{n^2+1} \\ \notag
&= \frac{n^2 \sum_{j=1}^{n^2 + 1} \frac{1}{j}}{n^2+1} \quad (\text{Let } j = (n^2 + 1) - k)\\ \notag
&= \frac{n^2 \cdot \mathit{H}_{n^2 + 1}}{n^2+1} \implies T_{ins}(n) = O(log(n))
\end{align}
Asymptotically, since there are equivalent terms in both numerator and denominator, the rate of increase is governed by the harmonic number, resembling $O(\log(n))$ as upper bound.

\subsubsection{Expected iterations for process termination} \label{Expected_iterations_for_process_termination}
In advancing the analysis of the average runtime associated with the percolation process, several considerations must be addressed prior to concluding with the specific insertion runtime. Primarily, an insertion operation will require an amount of elementary operations directly dependent on $c(n,k)$. This implies that it will vary with respect to the iterations, which are determinative of $k$. Nonetheless, the problem we face is that the only closed form of $c(n, k)$ available so far is not guaranteed to return the actual value of the quantity it models, preventing us from using it in calculations involving the average insertion time for a system defined by $n$, or even in the expression for the complete process runtime.
\\\\
Hence, to proceed with the last stage of the analysis, there are several alternatives we can use as a foundation for the ensuing discourse. On one hand, the simplest option given the present information is to utilize $c(n,k)$, knowing that it may not have the theoretically valid form. Due to its simplicity, subsequent calculations would be streamlined, and, we could even rely on the average insertion time, which also does not have an overly elaborate expression. However, the discrepancy between an analysis based on this approach and the actual complexity bounds may be substantial. On the other hand, we can ignore the work performed by each insertion and assume it to be constant throughout the process. In this way, each insertion would be accounted for, asymptotically, equivalent as if we used the average insertion time. The mean execution time for an insertion from when the system is empty until it is full is a constant value, which may be greater or reduced depending on the system's size. Therefore, treating the execution time of an insertion as constant aligns with this approach. Despite the potential error relative to the precise analysis regarding all possible quantities of elements, asymptotically, the behavior of this difference remains a priori unknown. Thus, even if the ultimate complexity of the algorithm is left dependent on the time an insertion might take, sufficient information can be extracted to model much of its performance as we scale up $n$.
\\\\
Now, to complete the rest of the complexity, assuming that an insertion constitutes an elementary operation, we must focus on the iterations during which these insertions are executed, and how many of them are necessary to terminate the process. As previously discussed in \textcolor{blue}{\ref{complexity_elemental_insertion}}, each iteration of the algorithm entails an insertion conducted with a probability contingent upon the expected number of elements present in that iteration; this formula will be computed later. Additionally, we had defined $I(n)$ as the function that returns the average number of iterations required for a process to conclude, factoring in all its variability. To understand the source of this function, we shall initially define the random variable $\mathbf{I}_n$ as the exact number of iterations required for a process completion in a system of size determined by $n$.
\begin{align*}
    I(n) = \mathbb{E}[\mathbf{I}_n]
\end{align*}
With this, presuming comprehensive knowledge of the specified variable, we can derive an expression for $I(n)$ from its expected value. That is, for all the possibilities considered in the random variable, its mean, which analogous to the average number of iterations a process may entail, constitutes the requisite value our function captures. However, discovering the precise attributes that define this variable proves to be a complex endeavor.
\begin{align*}
    \mathbf{I}_n \sim \mathit{P}_n
\end{align*}
This variable will adhere a probability distribution $\mathit{P}_n$, which is currently unknown. To calculate it, we could pursue an approach where we thoroughly analyze the state space of the problem and derive a function $P(\mathbf{I}_n=i)$ to characterize the distribution, subsequently computing its mean. But, this task confronts several daunting challenges. Regarding the state space, calculating its cardinality is relatively unproblematic and indeed beneficial. Contrarily, in order to determine whether a process at iteration $i$ has ended or not, we need to model the condition in which a particular state $s \in S_n$ contains a valid path. More precisely, we must count how many states in $S_n$ with $k$ elements in their matrix contain a path that triggers percolation transition. Initially, we need restrictions on the state space to identify all states with a certain number of elements and then further restrict to those presenting a path. This is where the principal difficulty in exactly computing $I(n)$ stems, as modeling the later constraint to determine whether a state has a path or not is rather unfriendly. Therefore, as an alternative, we may opt for empirically approximating the unknown probability distribution.
\\\\
For this purpose, a dataset is assembled with simulations and their relevant information, enabling a reasonable approximation. The dataset will consist of entries detailing the system size employed in the simulation, the number of iterations taken to complete, and the quantity of elements in their terminal state. Additionally, all simulations will have their own uniformly initialized random number generator to ensure maximal randomness and uniformity in the results, as bias induced by the generator could affect the shape of the expressions obtained or even invalidate the procedure. Moreover, it is advisable to select a wide range of matrix sizes. However, since in this case only the number of iterations required for process completion is being approximated, it is preferable to use the largest sizes feasible within the available computational resources.

\begin{tabular}{|c|>{\raggedright\arraybackslash}p{5cm}|>{\raggedright\arraybackslash}p{5cm}|c|}
    \hline
    \textbf{Matrix size} & \textbf{Iterations} & \textbf{Elements} & \textbf{Simulations} \\
    \hline
    (3, 3) & [6, 5, 7, ..., 9, 6, 4] & [5, 5, 4, ..., 6, 4, 4] & 100000 \\
    (4, 4) & [13, 10, 10, ..., 13, 8, 13] & [8, 7, 8, ..., 8, 7, 10] & 100000 \\
    (5, 5) & [11, 22, 7, ..., 17, 8, 25] & [10, 11, 7, ..., 13, 7, 16] & 1100000 \\
    (6, 6) & [27, 10, 16, ..., 25, 15, 28] & [20, 10, 14, ..., 20, 14, 22] & 1100000 \\
    (7, 7) & [59, 25, 33, ..., 19, 17, 29] & [33, 22, 24, ..., 19, 15, 26] & 1100000 \\
    (8, 8) & [18, 62, 40, ..., 49, 63, 29] & [15, 38, 29, ..., 36, 37, 25] & 1100000 \\
    (9, 9) & [53, 33, 25, ..., 47, 45, 83] & [38, 28, 22, ..., 32, 36, 48] & 1100000 \\
    (10, 10) & [126, 87, 88, ..., 139, 135, 84] & [95, 69, 73, ..., 93, 98, 74] & 1000000 \\
    (11, 11) & [73, 65, 84, ..., 106, 69, 84] & [55, 49, 59, ..., 65, 54, 64] & 1100000 \\
    (12, 12) & [118, 74, 59, ..., 63, 88, 118] & [78, 61, 50, ..., 52, 68, 72] & 100000 \\
    (13, 13) & [104, 86, 56, ..., 86, 100, 76] & [76, 71, 49, ..., 69, 76, 62] & 1100000 \\
    (14, 14) & [144, 138, 120, ..., 120, 115, 89] & [102, 96, 88, ..., 94, 92, 74] & 1100000 \\
    (15, 15) & [97, 89, 111, ..., 166, 122, 81] & [75, 71, 92, ..., 115, 99, 67] & 1100000 \\
    (16, 16) & [112, 124, 136, ..., 126, 200, 173] & [86, 103, 107, ..., 100, 130, 121] & 100000 \\
    (17, 17) & [153, 180, 95, ..., 119, 174, 104] & [116, 126, 87, ..., 91, 134, 84] & 100000 \\
    (18, 18) & [205, 189, 175, ..., 198, 136, 180] & [151, 141, 140, ..., 156, 114, 137] & 100000 \\
    (19, 19) & [258, 145, 218, ..., 196, 145, 195] & [181, 121, 158, ..., 154, 122, 141] & 100000 \\
    (20, 20) & [168, 263, 203, ..., 202, 189, 152] & [136, 192, 171, ..., 159, 157, 121] & 100000 \\
    (21, 21) & [237, 268, 271, ..., 261, 194, 177] & [186, 193, 201, ..., 185, 154, 155] & 100000 \\
    \hline
\end{tabular}
\vspace{1em}

In the preceding table, a preliminary description of the set of simulations performed for the dataset is provided \cite{dataset}. As discerned, for each pair $(n,n)$ of possible sizes, a number of simulations is conducted, ranging from 100000 to 1100000 in certain instances. This variance is inconsequential, as the approximations computed above the first number are sufficiently robust, though it is always advisable to have extra data available just in case. For each simulation, the iterations and corresponding elements are recorded at its end, storing both quantities in arrays with consistent indexing. That is, both magnitudes are stored in order as if the arrays were stacks. Cumulatively, the dataset holds 207,950,010 entries, each associated with a distinct simulation. These entries do not all represent systems of size $(n,n)$; some simulations have been performed with other configurations, which we will analyze later, including non-square matrix sizes such as $(n,1)$ or $(n,10)$.
\\\\
Given this data, to derive the probability distribution of $\mathbf{I}_n$, we shall utilize the iteration array for a certain system size. With the array, we will enumerate all the iterations and arrange them into a structure conducive to visualizing and subsequently fitting the underlying distribution displayed by the data. As for its visualization, we know that each number of iterations will span from 0 to $\infty$, with an indeterminate number of repeated values, as can be seen in some areas of the dataset. Consequently, we will construct a histogram from the recurrence of iteration counts for each array. That is, for each system size, we will tally the frequency of each iteration count's occurrence within the iteration array for that size. In this manner, by counting all the contained values, we will be able to plot the frequencies of occurrence in a chart that depicts the density function of the probability distribution we seek to approximate. Furthermore, we can normalize the frequencies relative to their total count to facilitate the subsequent fitting process.
\begin{figure}[H]
    \centering
    \includegraphics[width=8cm,clip]{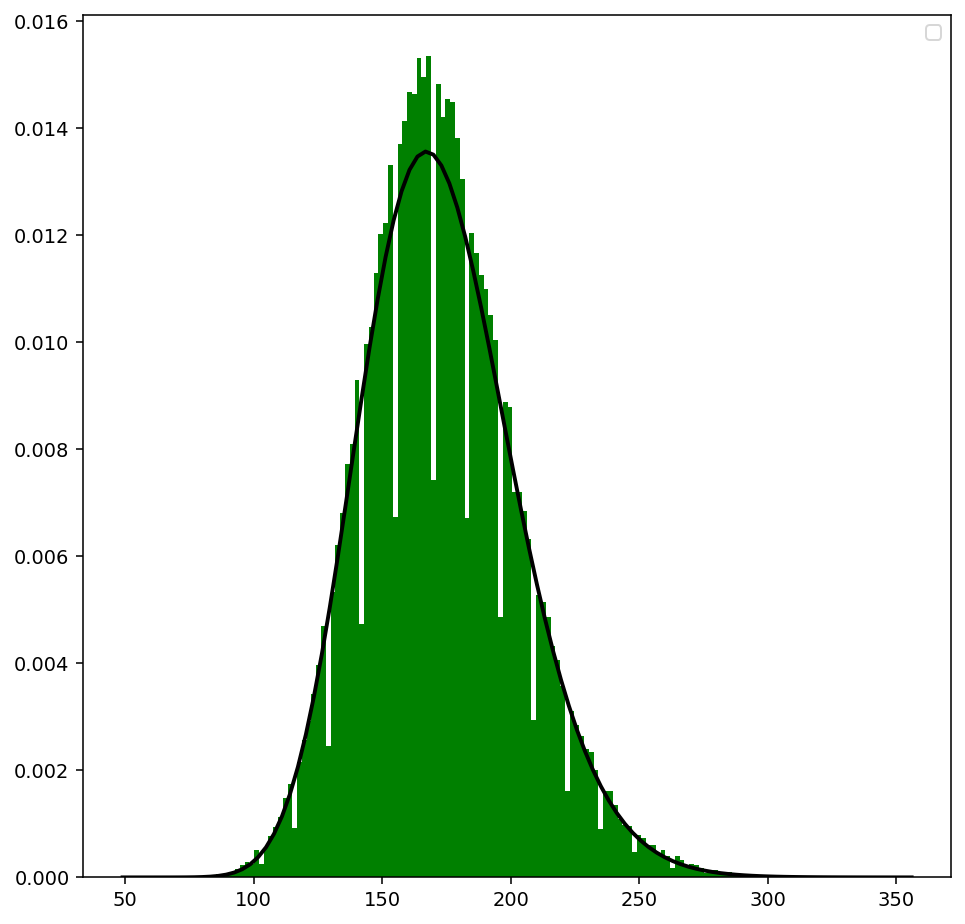}
    \caption{Histogram of the iterations array for a system with a square matrix of dimension $n=18$. The black curve represents a fitted Gamma distribution model. The x-axis denotes the number of iterations, while the y-axis shows their normalized frequencies.}
    \label{fig:IterationsDistribution}
\end{figure}
Figure 16 displays the histogram corresponding to the probability density function of the distribution $\mathit{P}_{18}$. At first glance, it lacks a definitive and recognizable form that would enable us to categorize the type of distribution unequivocally. However, the shape bears a resemblance to a Gamma distribution \cite{Stacy1962,Hosch2024,Theodoridis2020}, so assuming it follows such a distribution could be a suitable starting point, notwithstanding the absence of conclusive evidence affirming that $\mathit{P}_{18}$ conforms to this specific distribution.

\begin{align}
    p_n(i, \alpha(n), \beta(n)) = \frac{\beta(n)^{\alpha(n)} i^{\alpha(n) - 1} e^{-\beta(n) i}}{\Gamma(\alpha(n))}, \quad i \geq 0, \alpha(n) > 0, \beta(n) > 0
\end{align}
Denoting $i$ as any valid iteration count, $p_n(i, \alpha(n), \beta(n))$ constitutes the model employed to fit the histogram. That is, the standard Gamma density function will be parameterized to find the most optimal values for its parameters, thereby yielding the following cumulative distribution function \cite{Papoulis1965}:
\begin{align}
    \mathit{P}_n(i, \alpha(n), \beta(n)) = \int_{0}^{i}  \mathit{P}_n(t, \alpha(n), \beta(n))\, dt = \int_{0}^{i} \frac{\beta(n)^{\alpha(n)} t^{\alpha(n) - 1} e^{-\beta(n) t}}{\Gamma(\alpha(n))} \, dt = \frac{\gamma(\alpha(n), \beta(n) \cdot i)}{\Gamma(\alpha(n))}
\end{align}
With this, we shall proceed with the histogram fitting, for which the $gamma.fit()$ function from the SciPy \cite{SciPy2020} library in Python will be applied. This function defaults to the Maximum Likelihood Estimation method \cite{PanFang2002,Richards2018,Kitchin1994}, ensuring an acceptable outcome. Consequently, after fitting the histogram frequencies, an estimated shape parameter of $\alpha(18)=27.205474564768814$ and an estimated scale parameter of $\theta(18)=5.726297951242216$ are obtained. Thus, given that the aforementioned parameters \cite{KudryavtsevShestakov2022} are sufficient to define the mean and standard deviation of a Gamma distribution, we can estimate the value of $I(n)$ for its pertinent $n$:
\begin{align*}
    I(n) = \mathbb{E}[\mathbf{I}_n] = \frac{\alpha(n)}{\beta(n)} = \alpha(n) \cdot \theta(n) \implies I(18)\approx 155.78
\end{align*}
Since the fit provides us with $\theta (n)$, it can be straightforwardly converted to its rate parameter $\beta$ with the equivalence $\beta(n)=\frac{1}{\theta(n)}$. Given the graph and the specific value for $I(18)$, it is revealed that the returned value is coherently situated near the peak frequency on the histogram, which serves as a stalwart indication of the fit's veracity.

\begin{figure}[H]
    \centering
    \includegraphics[width=8cm,clip]{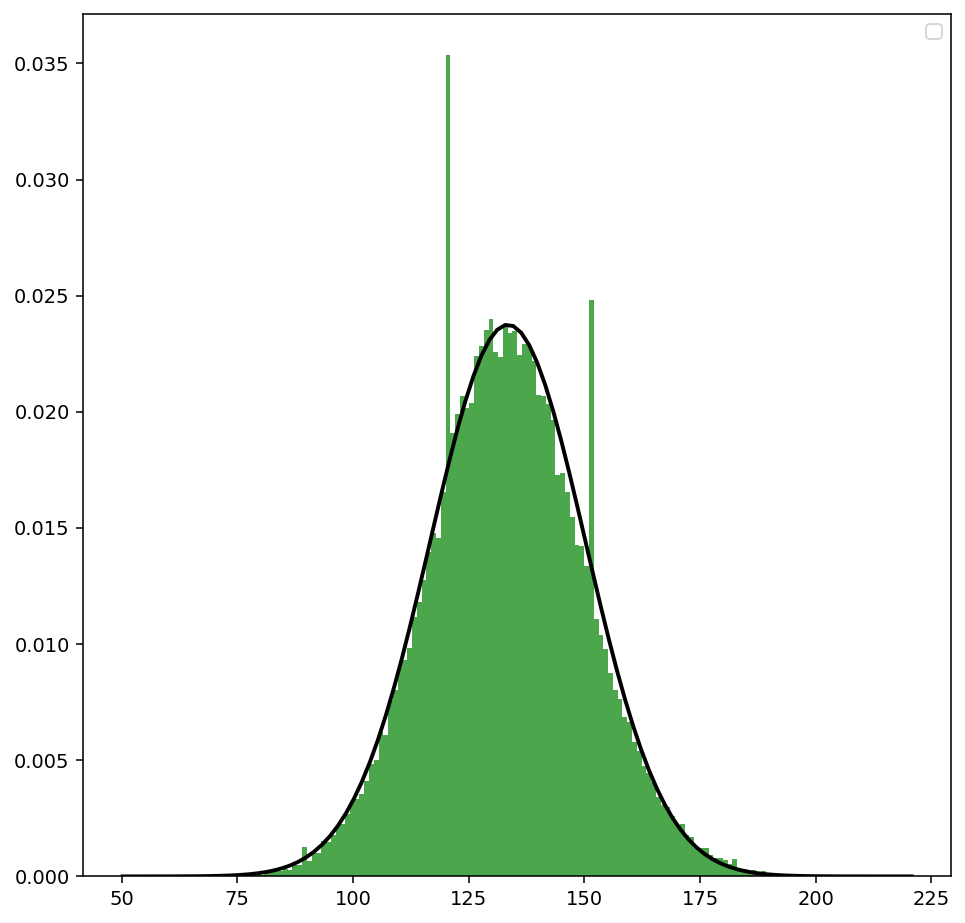}
    \caption{Histogram of the number of elements at terminal state array for a system with a square matrix of dimension $n=18$. The black curve overlaid represents a fitted Normal distribution model. The horizontal axis indicates the number of elements, and y-axis shows their normalized frequencies.}
    \label{fig:ElementsDistribution}
\end{figure}
In addition, considering that for an identical matrix size, we hold the quantity of elements in the terminal state for every simulation, we can apply an analogous procedure to fit the corresponding probability distribution to this magnitude. Hence, if we denote $\mathbf{E}_n$ as the random variable representing the number of elements in these states for a certain $n$, we can assert from the shape of the histogram in Figure 17 that it conforms to a normal distribution \cite{Bryc1995}, although a binomial distribution \cite{Edwards1960} is also a plausible characterization.

\begin{align}
    p_n(j, \mu(n), \sigma(n)) = \frac{1}{\sigma(n) \sqrt{2 \pi}} e^{-\frac{(j - \mu(n))^2}{2\sigma(n)^2}}, \quad -\infty < j < \infty
\end{align}
To facilitate the fitting procedure, a normal distribution featured by the density function $p_n(j; \mu(n), \sigma(n))$ will be adopted, where $j$ denotes a quantity of elements. Thus, the cumulative distribution function adhered to by the random variable can be derived as follows \cite{Soch2024,Harris2014}:
\begin{align}
    \mathit{P}_n(j, \mu(n), \sigma(n)) &= \int_{-\infty}^{j} f(t, \mu(n), \sigma(n)) \, dt \\ \notag
    &= \int_{-\infty}^{j} \frac{1}{\sigma(n) \sqrt{2 \pi}} e^{-\frac{(t - \mu(n))^2}{2\sigma(n)^2}} \, dt \\ \notag
    &= \frac{1}{2} \left[ 1 + \frac{2}{\sqrt{\pi}} \int_{0}^{\frac{j - \mu(n)}{\sigma(n) \sqrt{2}}} e^{-t^2} \, dt \right]\\ \notag
    &= \frac{1}{2} \left[ 1 + \text{erf}\left( \frac{j - \mu(n)}{\sigma(n) \sqrt{2}} \right) \right]
\end{align}

In this way, by using the same library as before, with the difference that the fitting function $norm.fit()$ now incorporates a normal distribution model, outputs a mean of $\mu(18)=133.4678$ and a standard deviation of $\sigma(18)=16.795075562795187$. Herein, since one of the parameters defining the distribution is its mean, the average of elements in a terminal state is computed by:
\begin{align}
   \mathbb{E}[\mathbf{E}_n] = \mu(n) \implies \mu(18)= 133.4678
\end{align}
In summary, once the probability distributions are fitted to the histograms, we can realize that the only value this procedure has lies in the resolution, or rather estimation, of $I(n)$ and $\mathbb{E}[\mathbf{E}_n]$ for a specific case. That is, modeling the complexity of the algorithm for all $n$ implies having precise information about each density function $p_n(i, \alpha(n), \beta(n))$, or something equivalent with which the exact shape of the probability distributions can be known. For now, we will not focus on the relationship between both distributions of $\mathbf{I}_n$ and the corresponding number of elements, since later with an expression for $E(i)$ it will be shown how the latter is deductible from the former. Consequently, even narrowing our task to defining the first distribution, we would need to somehow ensure that for different values of $n$ the fit with a Gamma will be valid, a situation that is currently unknown. Additionally, for almost any form we choose to fit or model the respective histogram, we will have a variable number of parameters, depending on how elaborate its density function is. Each of them has a specific value for every matrix size, so they will be functions dependent on that magnitude. Therefore, for solving the probability distribution we would either need to know the exact dependence of every parameter with respect to $n$, or to perform multiple fits like the previous one and thereafter another one intended to discover such dependence. In any case, it would be impractical to proceed. On the one hand, without even information about the distribution of the random variable, it is counterproductive to try to find a dependency between the parameters of a similar distribution, such as the Gamma in this instance, and $n$. Secondly, reiterating the previous fit for all matrix sizes in the dataset would not provide enough generalization capability for a subsequent model, also unknown beforehand, to fit the parameters of the distribution. The main reason is the lack of measurements, since the dataset spans system sizes ranging from $(3,3)$ to something less than $(50,50)$, which is insufficient given the apparent complexity of the expressions the parameters might present. Even knowing their form, it may be impossible to provide an exact result with such a fit, since there exist the possibility that the outcome involves irrational quantities that cannot be easily inferred. Ergo, in conclusion, to faithfully grasp all the properties of $I(n)$ and its contribution to the complexity of the algorithm, as well as the clusters of elements, it is indispensable to transcend and be thorough in its deduction.

\subsection{Additional approaches}
In addition to the standard method for analyzing the algorithm reviewed so far, which will subsequently serve as foundation for the definitive methodology, there exist numerous techniques we can conceive of as basis for the analysis. Hence, here are presented the ones that are most likely to lead to a successful outcome, although we will not delve into the details of each one.
\\\\
To commence, given that formally the system is conformed of a matrix, which computationally would correspond to a rank-2 tensor, it is possible to leverage the properties of this data structure, as well as the algorithms associated with it, to infer information about the percolation process \cite{Meng2023}. For example, we know that the process ends when there is a sufficient number of elements in the system that allows the appearance of a path, with the disadvantage that we must account for the specific arrangement of those elements to precisely determine its existence. Accordingly, we could hypothesize that by computing an average of every attainable arrangement and evaluating a metric according to the resulting mean, it would be possible to determine the existence of paths, and thus the approximate iteration where the terminal state is reached. Since this average originates from many system states, it may alternatively be estimated by analyzing individual states with an appropriate algorithm.
\begin{figure}[H]
    \centering
    \includegraphics[width=10cm,clip]{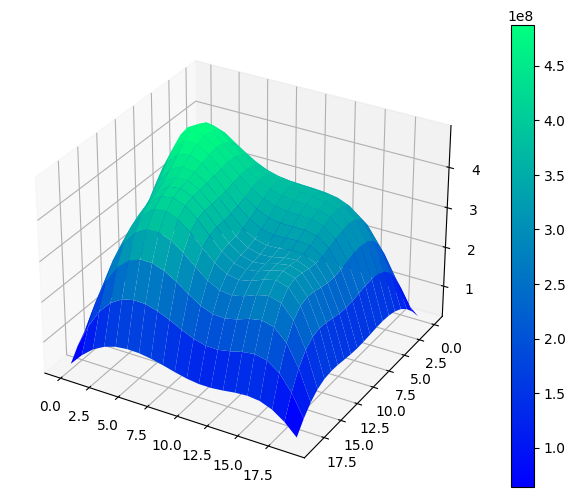}
    \caption{3D plot of a $20\times 20$ matrix resulting of a 10-stacked convolution with filter $\{[1, 1, 1], [1, 0, 1], [1, 1, 1]\}$ applied to the terminal state of Figure 1}
    \label{fig:convolution}
\end{figure}
If we take the terminal state from Figure 1 and apply a convolution multiple times with a kernel that captures information about the neighborhood for each cell, we will achieve a result analogous to that illustrated in Figure 18. For its construction, a convolution was performed between the filter $\{[1, 1, 1], [1, 0, 1], [1, 1, 1]\}$ and the terminal state. Subsequently, the process was repeated with the resulting matrix instead of the original terminal state, 10 times in total. As can be seen in its three-dimensional representation, the generated shape apparently depends on the distribution and number of elements from which the convolution sequence was initiated.
\begin{figure}[H]
    \centering
    \includegraphics[width=15cm,clip]{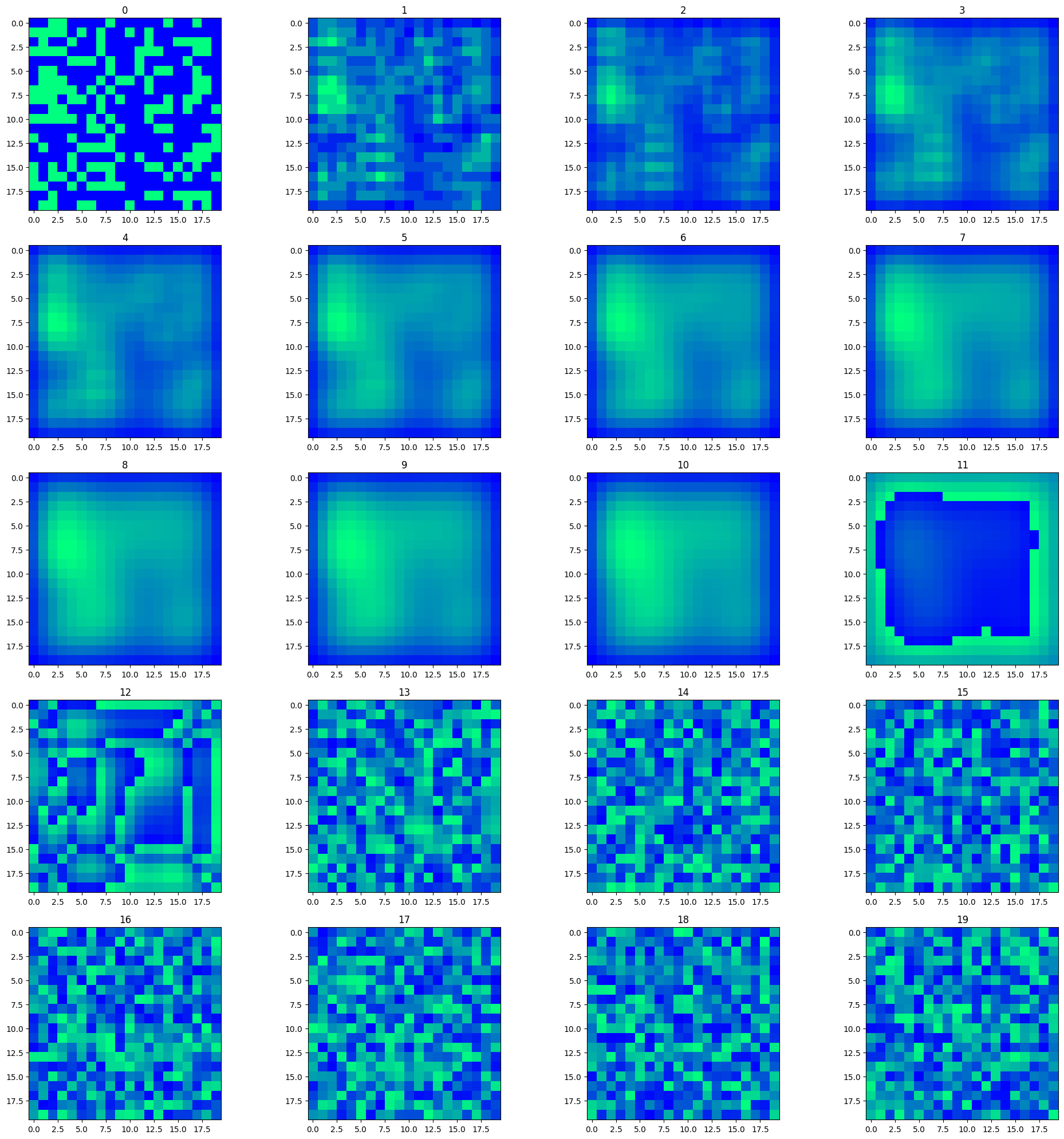}
    \caption{Convolution stack from Figure 18, index is shown above each matrix}
    \label{fig:convolutions}
\end{figure}
However, according to the intermediate results where the filter is iteratively applied, from iteration 15 onward the content of the matrix collapses and degrades into noise. Hence, only those preceding this iteration have significance, where we can identify a pattern that may be useful for the analysis. On the one hand, the filter is not randomly selected, but rather aims to measure the number of elements that exist in the vicinity of each cell, and since this algorithm is founded on Moore neighborhood, a $3\times3$ matrix is sufficient. As for the interpretation of the outcome, we observe that there is a concentration of cells with higher value in areas with a greater amount of elements in the initial state. Since the point of highest concentration can be located anywhere in the system, and the unique condition sought is whether a path can arise from it, or from other matrix points resulting from the stacked convolution, the analysis may continue by attempting to detect such a path or a necessary condition for its existence. To achieve this, there is the possibility to choose a more complex filter or a chain of different filters for each index within the stack, selected such that the final matrix presents some specific pattern or easily identifiable mark. The primary issue that could arise here is the choice of the number of filters used or the length of the stack, as at some intermediate step the system may collapse all its cells into noise. However, the advantage of this solution is the invariance, although not in all cases, to translations and rotations of the elements in the system, which depending on the selected filter can be easily ignored. In this way, by obtaining a stack of filters and a function that determines its length, we could study the relationship between the terminal states and the existence of paths within them, and other system states. Additionally, being able to simultaneously apply the properties of the Fourier transform \cite{WeissteinFFT,Podlozhnyuk2007} to optimize, and even extract further data from the system modeled in this context as a signal, it is plausible to arrive at an expression for $I(n)$.
\\\\
And, regarding the average cluster size, there exists another alternative we can explore. Specifically, it involves finding a model that takes as input any state $s\in S_n$ with $k$ elements, and through a mapping in a latent space \cite{Mayet2023,AspertiTonelli2023} comprised of all elements of $S_n$, it returns the desired value of the average cluster size. To this end, by training a convolutional model that can subsequently be formalized in the simplest way available, an embeddings model to process each state of the system, or a combination of both, the aforementioned mapping could be accomplished, despite the potential difficulty in formalizing and understanding the result of such systems.

\section{Methodology}
\label{Methodology}
In this section, the method proposed in this work will be introduced to achieve the most accurate analysis available of the algorithm's runtime for each of its cases. Thus, all the tools comprising it will be formalized, detailing their origin, confirming their validity and coherence with empirical results, and studying various alternatives that may provide relevant information for our purpose, or contribute to the theoretical resolution of the problem addressed by the algorithm in the context of percolation theory. Previously, some of them have already been presented, although most have not been completely resolved due to the approach followed. Therefore, we will now proceed with one that enables their resolution, acknowledging that not all of them currently have a closed form, nor is it known whether they exist, yet. Once the notation to be used has been clarified and all the required intermediate expressions have been explained, they will be applied throughout the procedure in the subsequent sections. That is to say, in this section, the formalization of the analysis is only detailed, while in the others this content will be leveraged to obtain concrete results, which may be exact temporal growth in the best-case scenarios, or functions that bound such growth with a loose margin. In case the latter happens, we will have to settle with restrictions for the resulting growth, even knowing that the approach is valid. Finally, one of the algorithm's cases will be specifically analyzed, which, as we will see, is invariant with respect to the system's aspect ratio, so it will be treated independently from the rest. This will provide us with a reference point to validate other bounds, particularly the average case one. If we can parameterize it in terms of the specific case, we can verify its equivalence. However, given the properties that the sequence of insertions must follow in both cases, this might not be achievable.
\subsection{Average cluster size estimation}
\label{AverageClusterSizeEstimation}
To begin with, the fundamental characteristic that defines the runtime of each insertion into the system is the average cluster size, so we will start the analysis by providing an estimated expression for this magnitude. The reason for not calculating an exact expression is the lack of viable approaches to achieve it. As we will see later, there are methods that guarantee modeling this quantity, but their nature suggests a high degree of difficulty in this respect, while not ensuring that what they offer is an utilizable formula. That is, even if an output congruent with our requirements is achieved, it is important that this result can be correctly interpreted in asymptotic terms. In this way, if the cluster size derived from a summation whose limits are impractical to evaluate, or even if the sum itself does not have a generating function, an alternative representation using simpler functions, or convergence to a function of the polynomial, exponential, etc., type, it would be practically impossible to find an asymptotic bound for this magnitude, complicating the analysis. Furthermore, the complexity analysis has more phases, so finishing this one with a formula devoid of a clear growth would compromise the viability of the rest of the process.
\\\\
Herewith, it is necessary to find a faithful estimate for $c(n,k)$ with respect to its representational significance, unlike the solutions we obtained previously, which merely aimed to meet the imposed restrictions. For this purpose, it is key to define what a cluster of elements is and its relationship with the insertion algorithm when it verifies the existence of a path within the system. At first, a cluster can be defined as a set of cells, each in a position $(i,j)$, that contain an element and satisfy a specific condition. Such condition states that for any pair of cells $(a,b)$ and $(c,d)$ from the set, there must exist a path between them formed exclusively by cells contained in the set. This ensures that all the individual components of a cluster are connected to each other, always depending on its neighborhood properties. Thus, among all the definitions with which this concept can be constructed, it is plausible that this one may guide us towards a decent approximation of its average size. Particularly, it is from the condition imposed on the cells of a cluster that we can extract a similarity to what the nodes of a tree must satisfy, with the difference that in this case, it permits the existence of more than one path between each pair of position tuples. What this difference represents is that, by modeling the system with a graph whose edges correspond to neighborhoods of adjacent cells and the cluster as one of its subgraphs with all its vertices associated to cells containing an element, structurally, there may exist areas of the cluster with a higher concentration of adjacent elements compared to other sparse regions, varying the number of existing paths connecting its cells. However, underlyingly, it is one of its Spanning Trees \cite{ShrockWu2000,Manna1992} that determines the geometry of the cluster, which we will use to obtain the sought approximation. For this, given that the complexity of an insertion in our algorithm is characterized by the size of a cluster traversed with a depth-first search, we will try to leverage the properties of this traversal to encompass all combinations of Spanning Trees placeable over the system \cite{Babalievski1998} within a function comprising a part of the average cluster size estimation. So, we continue analyzing the operation of a depth-first traversal as convenient:

\begin{figure}[H]
    \centering
    \includegraphics[width=14.5cm,clip]{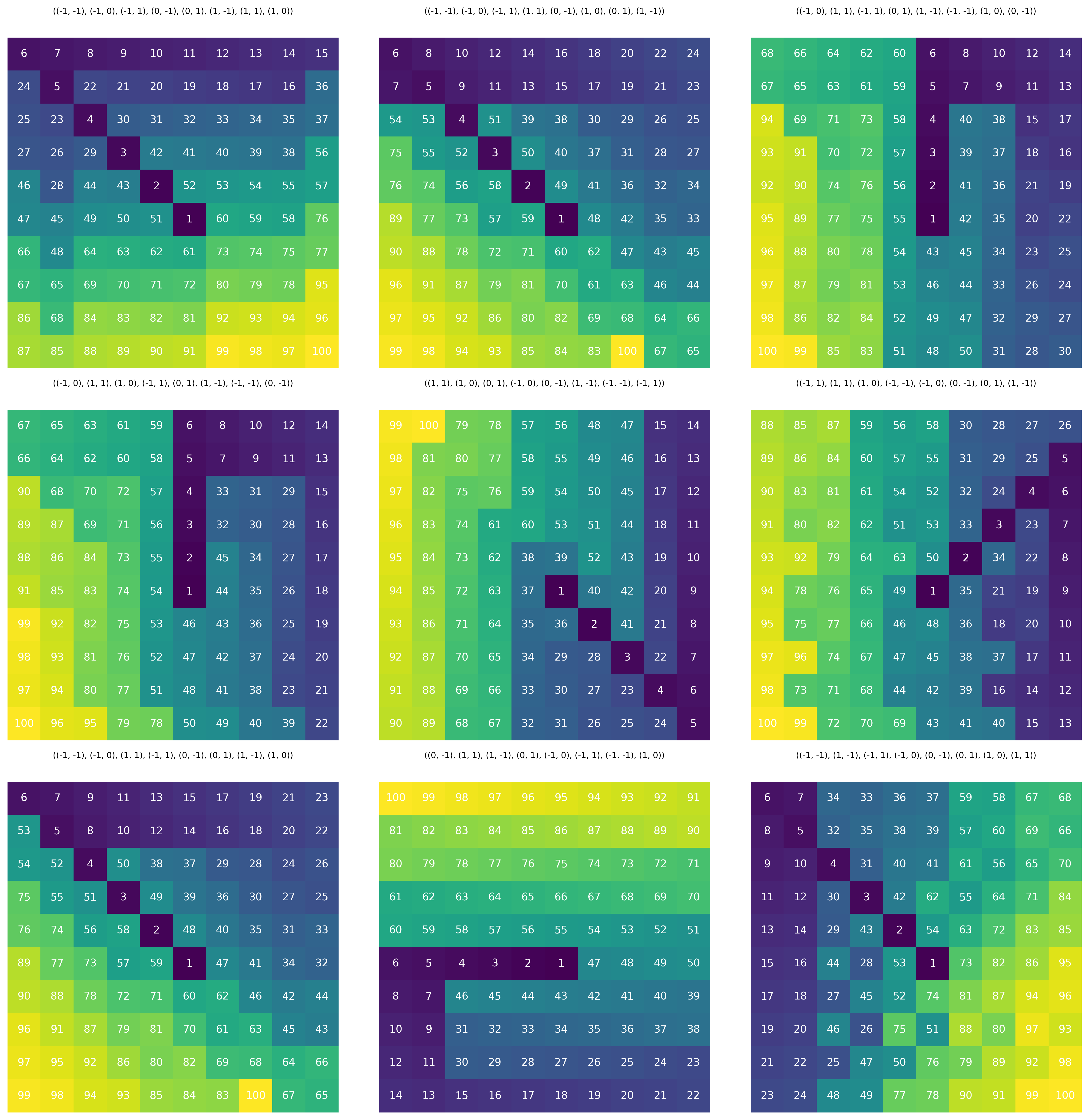}
    \caption{Depth-first search with no specific stopping condition performed at $(5,5)$ 0-indexed matrix cell. Traversing order is shown above every subplot.}
    \label{fig:dfs_count_grid}
\end{figure}
In Figure 20 it is shown the result of executing a program that performs the same traversal as our algorithm on a cluster. Specifically, it is a depth-first search with no specific stopping condition, that is, a traversal over the entire matrix whose only stopping condition is not to overstep the matrix bounds or to navigate through cells that have already been visited. To depict the traversal steps and the cells it transits in each of them, every cell is numbered with the 1-indexed step number in which the traversal went through it. In this way, starting in a cell located as close to the center as possible to leverage any potential symmetries that may arise, an ordered sequence of cells is obtained according to the assigned index. Also, this plot highlights the factors that influence the generation of this sequence, which we must contemplate for the main purpose of this section. On one hand, the order in which the next neighboring cell to visit is chosen influences the direction that the traversal follows. Out of a total of $8!=40320$ possible orders, only 9 are showcased above, where the patterns followed by each one are illustrated. Likewise, the starting position of the traversal modifies the pattern, and not invariantly under translations within the matrix, since it has bounds over which the depth-first traversal moves if it encounters any. Nonetheless, these bounds can be shifted over an infinite space, which is where the matrix would be located if it were infinite, so the initial cell that provides the most information in case of any translation over such space is the one at the center of the matrix.
\subsubsection{Traversal reach probability scalar field}
\label{subsubsec:traversalReachProbabilityScalarField}
To contextualize this in relation to the element clusters of our problem, we first need to devise a method to avoid dependency on their geometry when computing their size, as well as avoiding having to manage particular cases arising from their combinations or translations over the system. The approach we will follow consists of using the occupancy ratio of the system's matrix in a certain iteration to, from it, build a function that will map all the cells of the system to a real value according to their position. Thus, we will later be able to calculate certain global metrics that will lead us toward a reliable estimation of $c(n,k)$. The process to arrive at this function assumes that the probability of a cluster of size $k$ existing varies with the iteration in which the algorithm is found, as it is this magnitude that dictates how many elements will exist at each moment. Previously, it has been studied what happens as the algorithm progresses and inserts elements, which for now can be summarized in the input parameter $k$ of the average size. Therefore, assuming that in a certain iteration we know that there is that exact number of elements, the first thing we will be able to know is the minimum and maximum size for a possible cluster, whose respective probabilities of appearance depend on the specific system state. To avoid including the state in this phase of the estimation, we proceed from the definition of an insertion operation. First, select the cell on which to insert, which for simplicity we can assume is $(\lfloor \frac{n}{2} \rfloor, \lfloor \frac{n}{2} \rfloor)$. After this, it traverses the entire cluster, so its size will be determined by the steps required for the traversal and how many cells it explores.
\\\\
Given the known quantity of elements existing in that iteration, we can calculate for each cell the probability of it being occupied, resulting in $\frac{k}{n^2}$ if the matrix is square or $\frac{k}{n\cdot m}$ if its aspect ratio is not 1. With this, and along with the specific way of traversing the cells in the depth-first search, we can estimate the probability of a certain number of cells being traversed in a traversal step, that is, the traversal splitting into $0\cdots 8$ branches. So, in the first one all adjacent cells are examined as indicated by the neighborhood used and the established order of exploration, but only those that are occupied will contribute to the creation of more branches in the traversal. A priori, we do not know exactly how many of them meet this condition, nor in what position they are, but we do know that the occupancy probability on them remains $\frac{k}{n^2}$. Therefore, in the initial cell we assume that the probability of being traversed is 1, since $helper()$ always starts from it. And subsequently, for each level of adjacent cells, the probability of them being traversed is $\left(\frac{k}{n^2}\right)^s$, where $s$ denotes the current traversal step. The origin of this expression lies in the post-traversal analysis of cells neighboring the initial one. In that case, the next level of neighborhood will have as the probability of being traversable the ratio of elements in the system relative to its overall size. But, from the beginning of the traversal, the probability that the previous step has occurred, which enables the potential subsequent traversal towards the next level of neighborhood, is $\frac{k}{n^2}\cdot \frac{k}{n^2}$.

\begin{figure}[H]
    \centering
    \begin{subfigure}[b]{0.49\textwidth}
        \centering
        \includegraphics[width=\textwidth,clip]{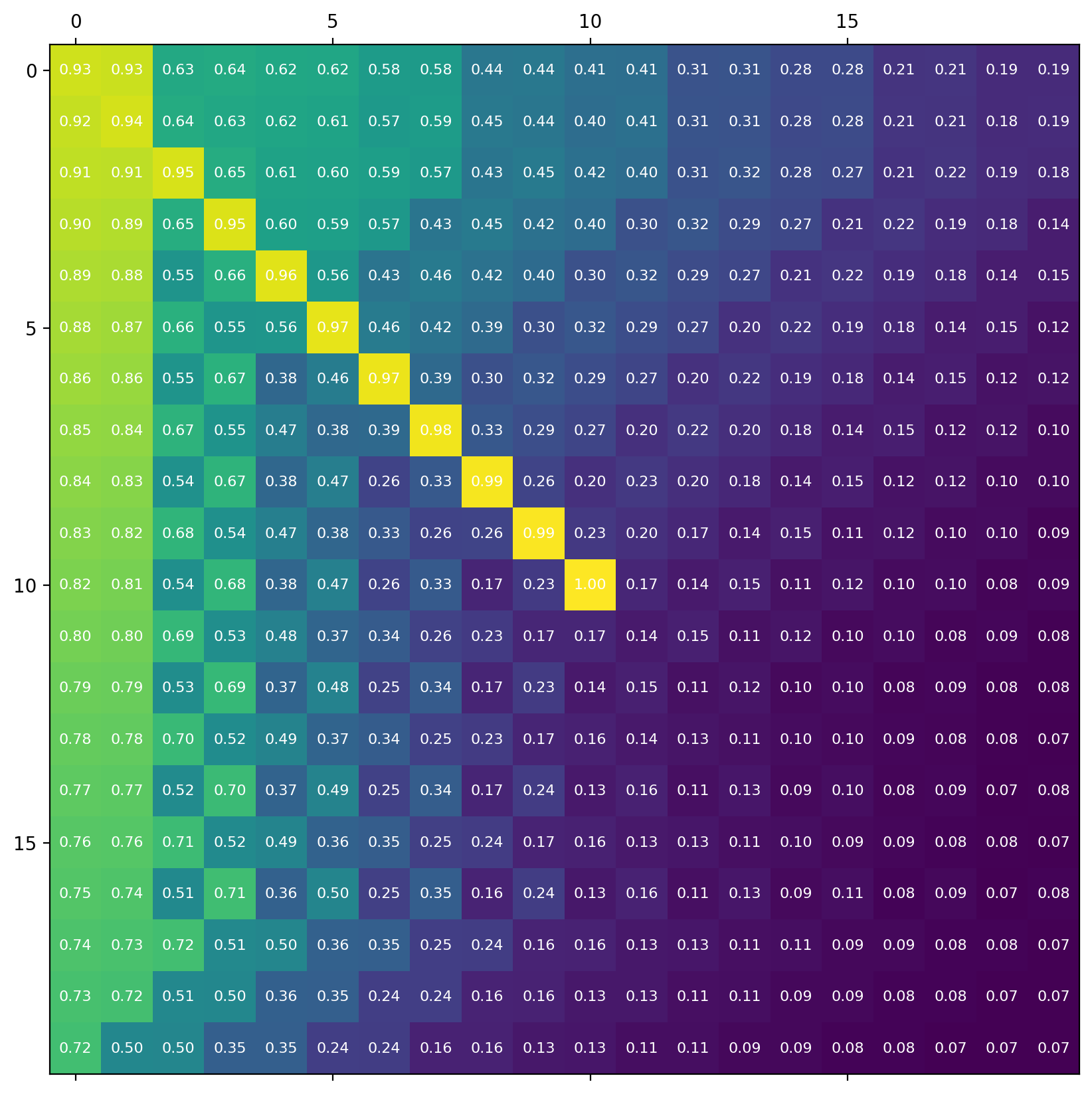}
        \caption{}
        \label{fig:dfs_probability}
    \end{subfigure}
    \hfill
    \begin{subfigure}[b]{0.49\textwidth}
        \centering
        \includegraphics[width=\textwidth,clip]{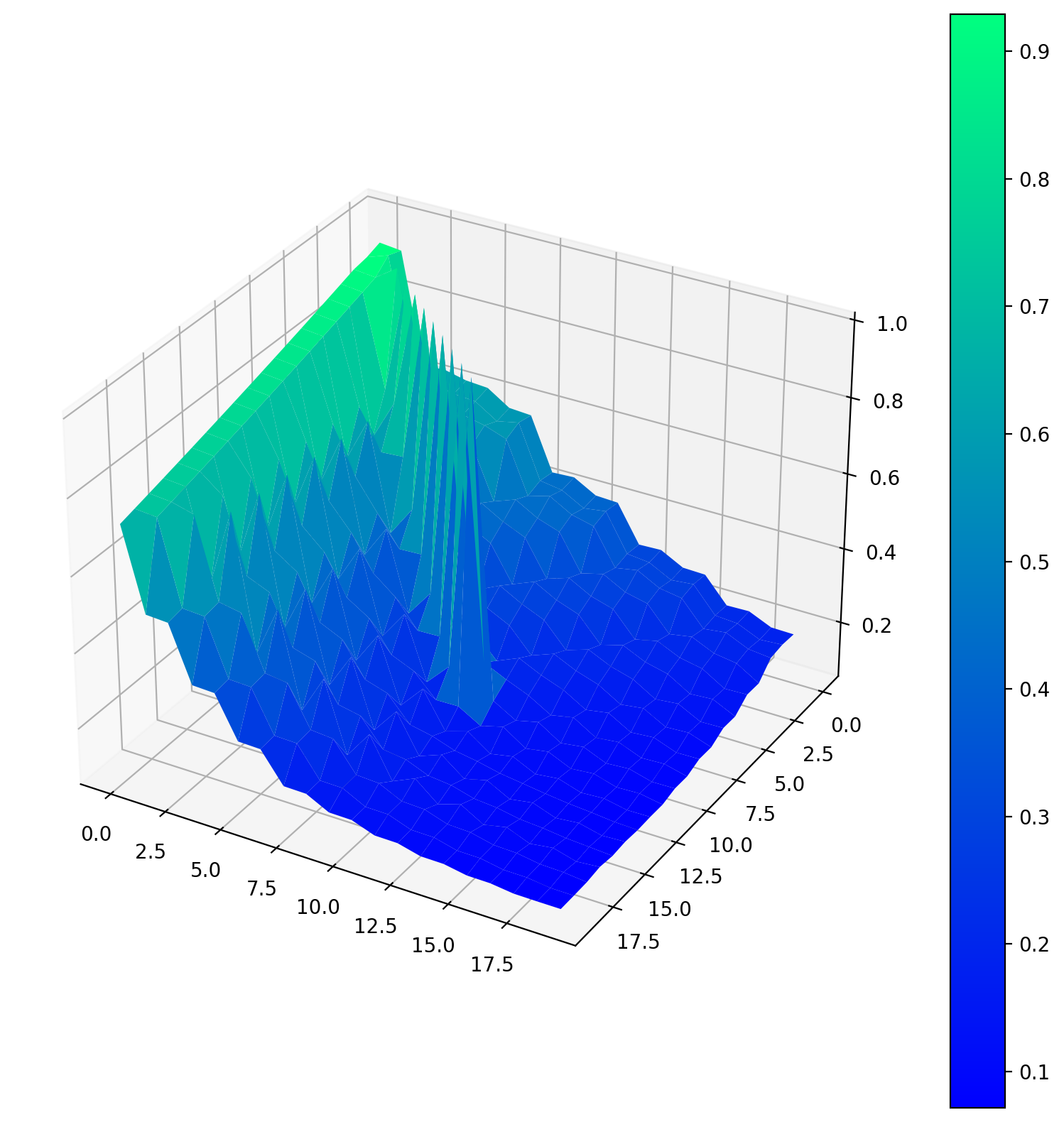}
        \caption{}
        \label{fig:dfs_probability3D}
    \end{subfigure}
    \caption{(a) DFS traversal from the last subplot within Figure 20 displaying the probability of every cell of being traversed at step $s$, with $\frac{k}{n^2}=0.993$. (b) 3D plot of the matrix values displayed in plot (a)}
    \label{fig:dfs_cluster}
\end{figure}

To visualize the relationship between the matrix occupancy ratio and the elements traversed by $helper()$, the same traversal as the last one in Figure 20 is shown above. The difference is that instead of displaying the step in each cell, the probability that such cell is traversed is illustrated under the assumption that there are $k$ elements in the system in unknown but uniformly distributed positions.
\begin{align}
   Pr[(i,j)\text{ is traversed}]=\left(\frac{k}{n^2}\right)^s \enspace \colon \enspace 1\leq s\leq n^2
\end{align}
As noticeable, the sequence of traversed cells mirrors the one from the previous experience since it is determined by the order of cells to be explored, which has remained the same. However, regarding the values of their cells, they decrease with step $s$, because the probability that a cell in the sequence will be visited by the traversal diminishes with the length of steps required to reach it from the initial cell. Fundamentally, the main reason for this phenomenon is the uncertainty that prevents us from ensuring the positions of the $k$ elements on the system. Thus, merely knowing the quantity that exists, a particular traversal like the one from Figure 21 only has the margin to determine whether, assuming $s$ steps have been successfully traversed, the next cell will be traversable or not, which corresponds to its probability of being traversable in step $s+1$. Nevertheless, our goal is to obtain a function that returns this probability for all cells in the system, regardless of the path followed or the starting cell. Therefore, with respect to the starting cell, we can ensure that its translations within the matrix affect the specific value that will emanate from each position $(i,j)$. Still, by placing the system on an infinite space interpretable as an infinitely large matrix, the only cell that will ensure symmetry concerning the matrix's bounds will be the center one. If we select any other position, notwithstanding some of its distances to the edges might be smaller or larger than the others, the contribution of all executable traversals from that cell will lead to a result equivalent to a translation of the one generated from the center cell. On the other hand, concerning all possible traversals produced from the order of exploration of neighboring cells, since there is a small quantity, we can employ a technique that involves their complete enumeration. Specifically, we will calculate the average of all traversals to determine the result the function should return for a given occupancy ratio $p$ \cite{Grimmett2018}.
\begin{figure}[H]
    \centering
    \begin{subfigure}[b]{0.49\textwidth}
        \centering
        \includegraphics[width=\textwidth,clip]{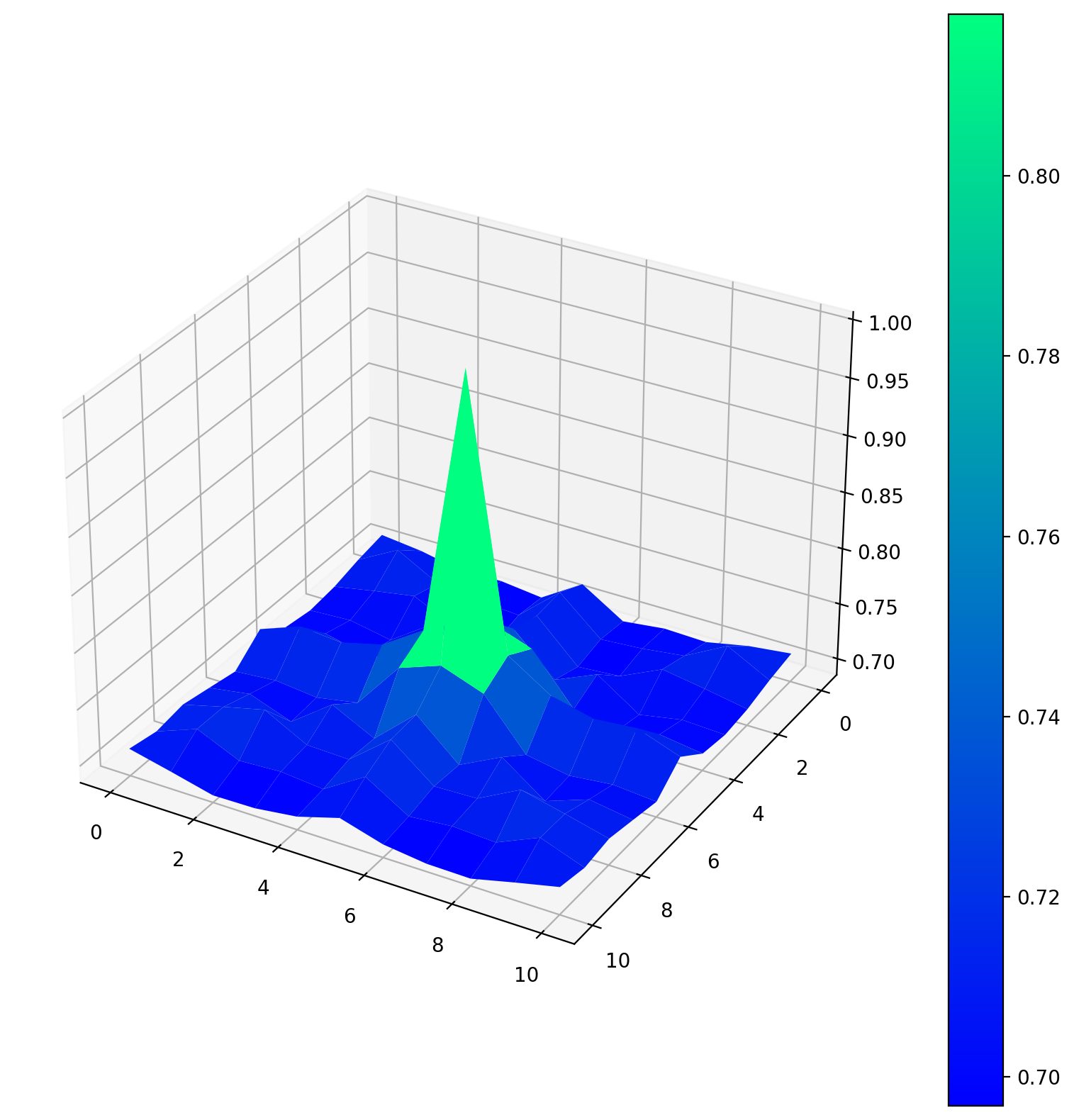}
        \caption{}
        \label{fig:dfs_average3D}
    \end{subfigure}
    \hfill
    \begin{subfigure}[b]{0.49\textwidth}
        \centering
        \includegraphics[width=\textwidth,clip]{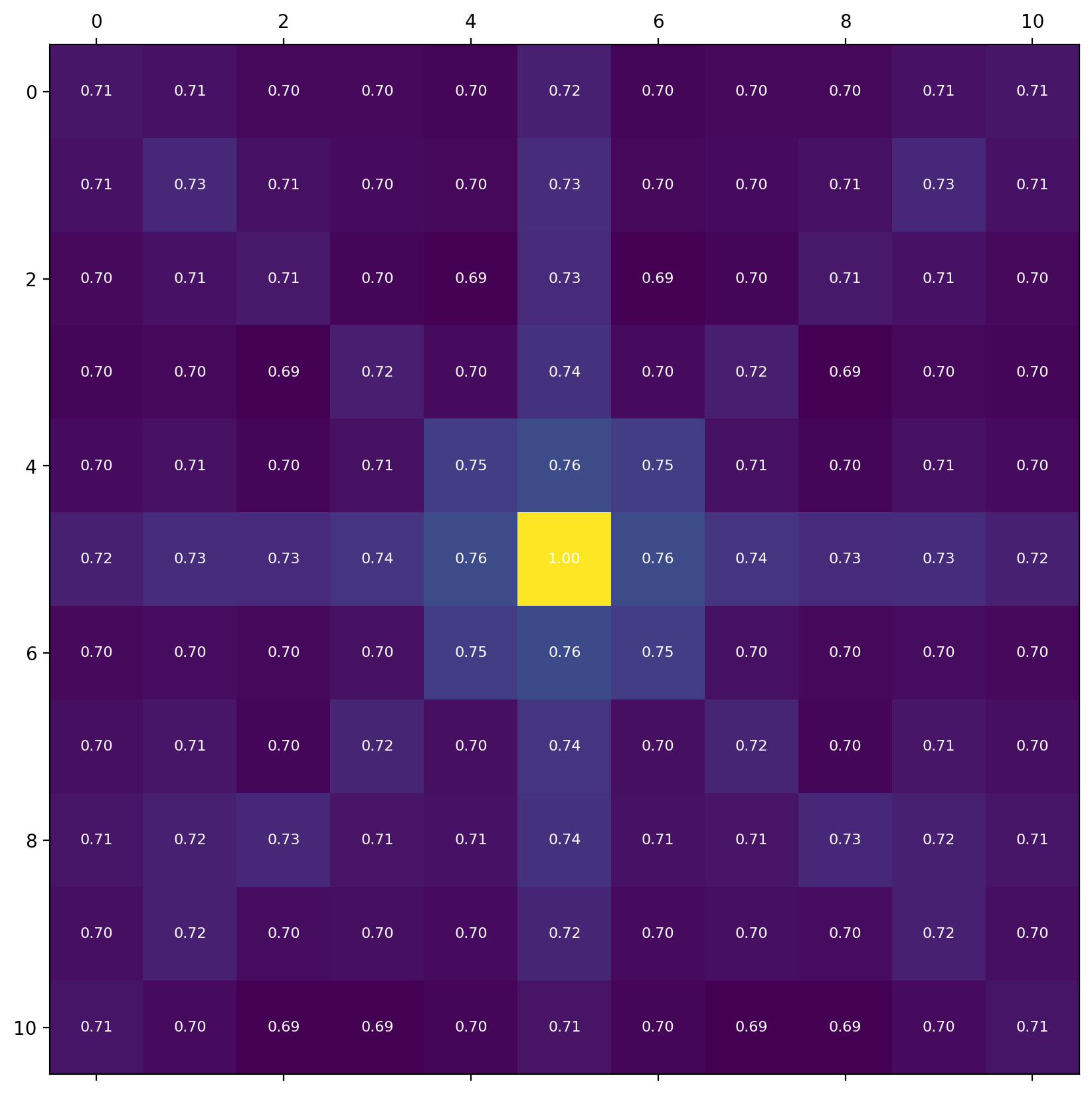}
        \caption{}
        \label{fig:dfs_average}
    \end{subfigure}
    \caption{(a) 3D plot displaying the average of the 40320 different traversals over a square matrix of size $n=11$, starting at cell $(5,5)$ and using an occupancy ratio of $p=0.993$. (b) matrix containing the 3D plot values of (a)}
    \label{fig:dfs_average_dual}
\end{figure}
As seen in Figure 20, we can generate each of the 40320 existing traversals over a system with well-defined parameters. In this example, considering a matrix of size $11\times11$ and an occupancy ratio of $p=0.993$, which equates to approximately $k=p\cdot n^2\approx120$ elements, all generated traversals will return a scalar field with the probabilities that each cell is traversed concerning the assigned exploration order.
\begin{align}
   \Phi(i,j,\xi) = \left(\frac{k}{n^2}\right)^{s_{\xi}(i,j)} \enspace\colon\enspace 0<\xi\leq40320
\end{align}
Thus, if we denote $\Phi(i, j, \xi)$ as the scalar field that assigns each cell $(i, j)$ the probability of being traversed at step $s_{\xi}(i, j)$ of the traversal with index $\xi$, we can obtain the mean $f(i, j)$, as presented in Figure 22, of all scalar fields returned by $\Phi(i, j, \xi)$:
\begin{align}
   f(i,j)=\overline{\Phi}(i,j,\xi) = \frac{1}{8!}\cdot\sum_{\xi=1}^{8!} \Phi(i,j,\xi) = \frac{1}{8!}\cdot\sum_{\xi=1}^{8!} \left(\frac{k}{n^2}\right)^{s_{\xi}(i,j)} 
\end{align}

\begin{figure}[H]
    \centering
    \begin{subfigure}[b]{0.49\textwidth}
        \centering
        \includegraphics[width=\textwidth,clip]{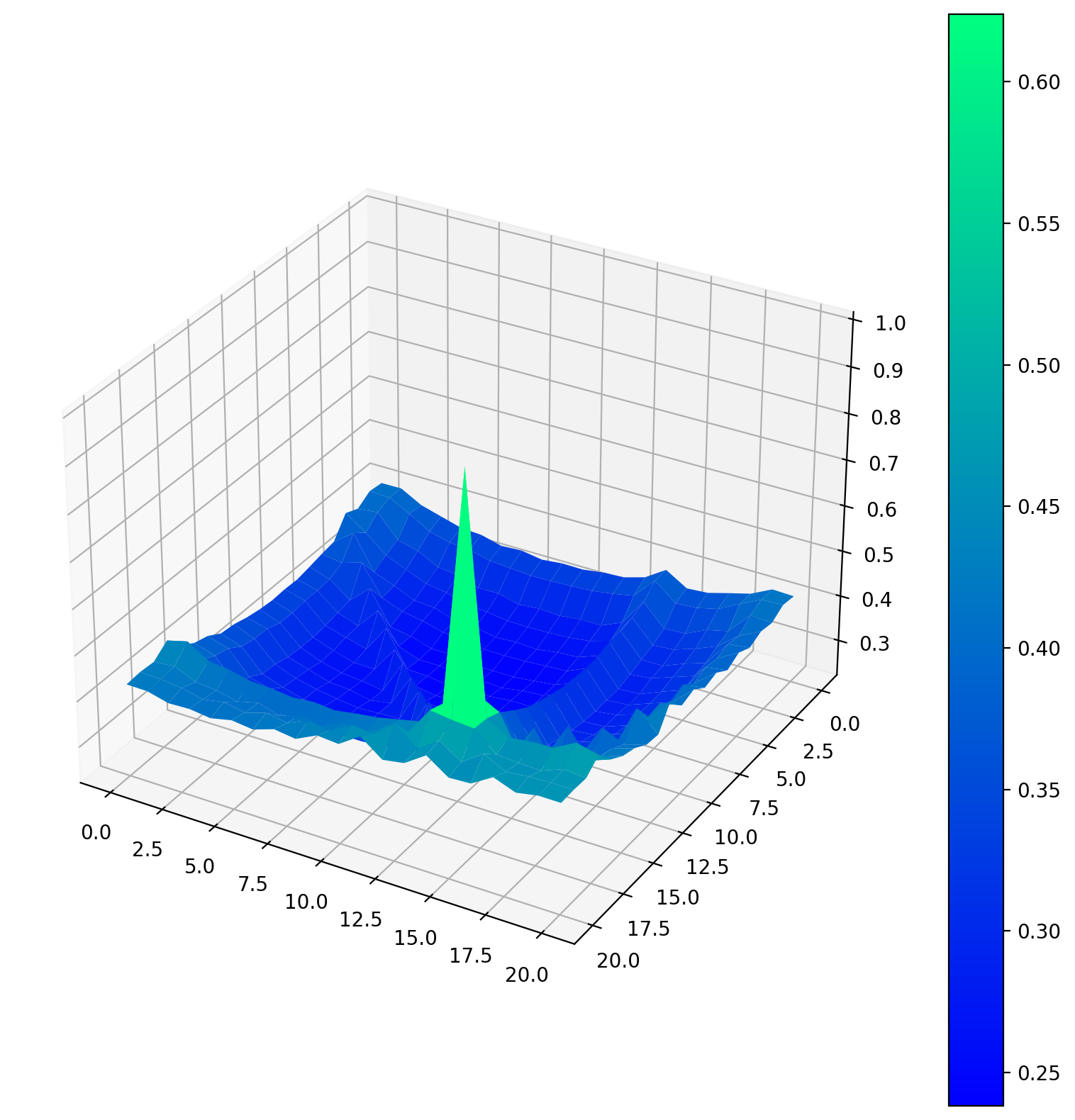}
        \caption{}
        \label{fig:dfs_average3D2}
    \end{subfigure}
    \hfill
    \begin{subfigure}[b]{0.49\textwidth}
        \centering
        \includegraphics[width=\textwidth,clip]{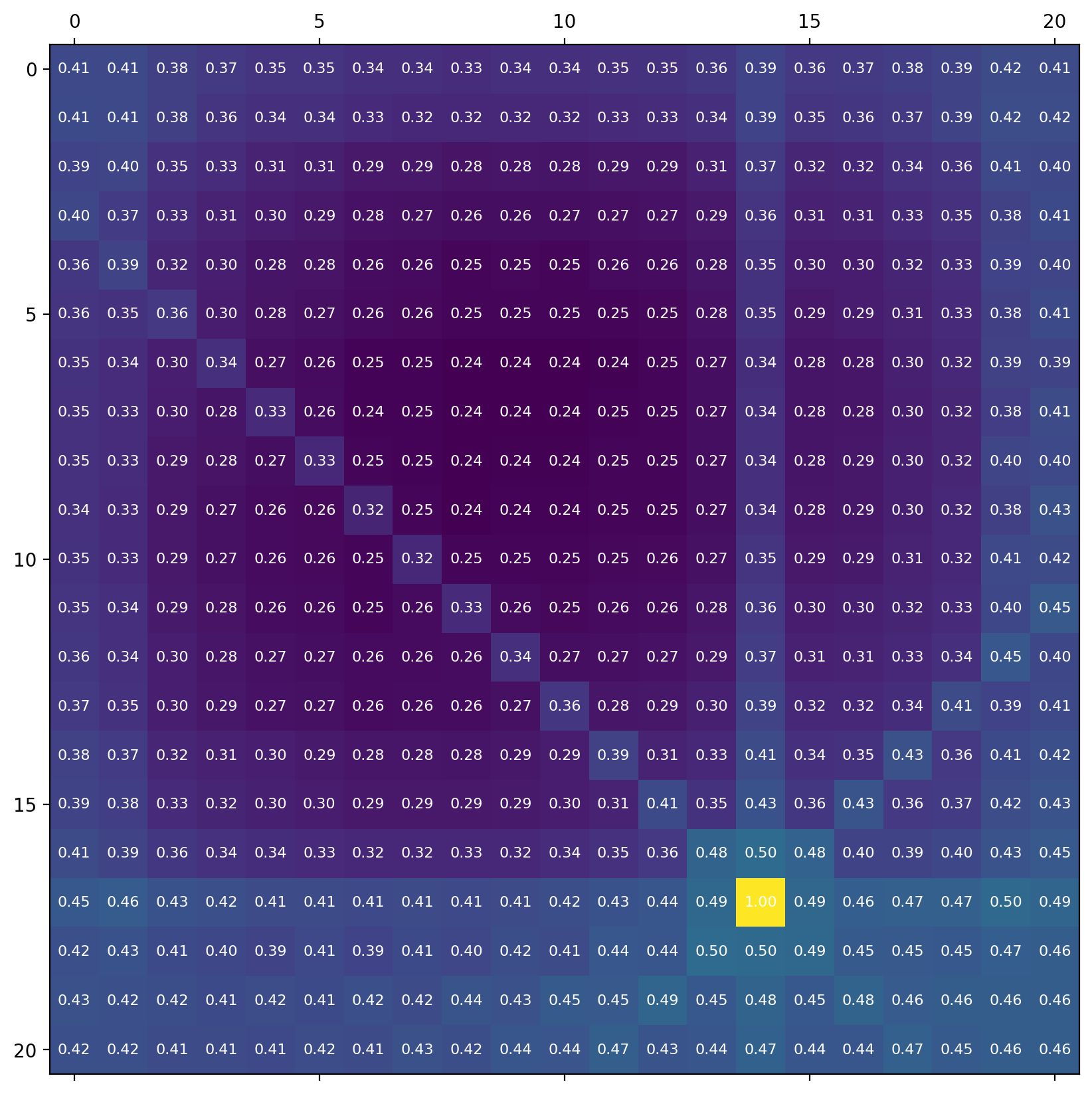}
        \caption{}
        \label{fig:dfs_average2}
    \end{subfigure}
    \caption{(a) 3D plot displaying the average of the 40320 different traversals over a square matrix of size $n=21$, starting at cell $(17,14)$ and using an occupancy ratio of $p=0.993$. (b) matrix containing the 3D plot values of (a)}
    \label{fig:dfs_average_dual2}
\end{figure}

As an important note, for all traversals, it is assumed that the starting cell is at the center with position $(\lfloor \frac{n}{2} \rfloor, \lfloor \frac{n}{2} \rfloor)$, since any other choice would result in a field equivalent to that of Figure 22 under the effect of a translation. Thus, once the average of all the scalar fields has been computed, denoted as a function of the occupancy ratio $p_{k,n}=\frac{k}{n^2}$ as $f(i,j,p_{k,n})$, we need to establish the procedure to estimate $c(n,k)$ with the current knowledge. Primarily, the information contained in the image of the field $f(i,j,p_{k,n})$ determines, probabilistically, the size of the expected cluster in which an element will be inserted. That is, from the probability of a traversal having a specific length, the expected size of the cluster being processed can be inferred, given that the sum of the lengths of all branches of the traversal equals the cluster size. In summary, what we achieve with this field is the ability to estimate the size of a cluster without needing to consider its geometry, which is a significant advantage. On the other hand, the use of this field manages to avoid the translation effect of the cluster across the matrix, since if the average of Figure 22 is repeated with any other starting cell, it will be seen that the resulting shape is identical, but translated. This invariance is mainly explained by the specific formula with which the field is constructed and the probability of each cell computed. Therefore, when proceeding with its construction, we are faced with 2 alternatives. The first consists of finding expressions for each $s_{\xi}(i,j)$, and then substituting them within the summation of the average of the scalar fields. Despite being relatively feasible, the final result would need a correction discussed below, so it is discarded.
\\\\
As for the alternative approach, it is founded on utilizing the occupancy ratio $p_{k,n}$ as the base of a continuous exponential function. And, to reach the conclusion that $f(i,j,p_{k,n})$ adheres to this form, it is imperative to analyze the decrease in probability from the center of the matrix to the cells at the edges. Initially, our approach dictated that, starting from any cell, the probability that there is an element in adjacent cells to it, and subsequently in the neighbors of these cells, was $p_{k,n}^s$. However, traversing the cells this way is conceptually more similar to a breadth-first traversal than the one used in our algorithm, given that the probability propagation occurs simultaneously in every neighboring cell. Hence, if we had used this traversal in our analysis, there would only emerge 1 scalar field, which in this case would result in the solution field, avoiding the process of computing the average of all those obtained with the depth-first traversal, but providing a result that does not capture sufficient information about the neighborhood of each cell. That is, the number of lines in the scalar field formed by cells with a higher probability than others, or equal in the case of BFS \cite{Siklossy1973}, which divide the area of the whole field into sections representing the possible traversal directions, differ in both types of traversal. On the one hand, in Figures 22 and 23, 4 lines of cells with higher probabilities are observed, corresponding to each of the 8 advance directions of the Moore neighborhood. In contrast, a breadth-first traversal would present only 2 diagonal lines with values similar to those of the remaining cells, failing to correctly segment the area.
\begin{figure}[H]
    \centering
    \includegraphics[width=10cm,clip]{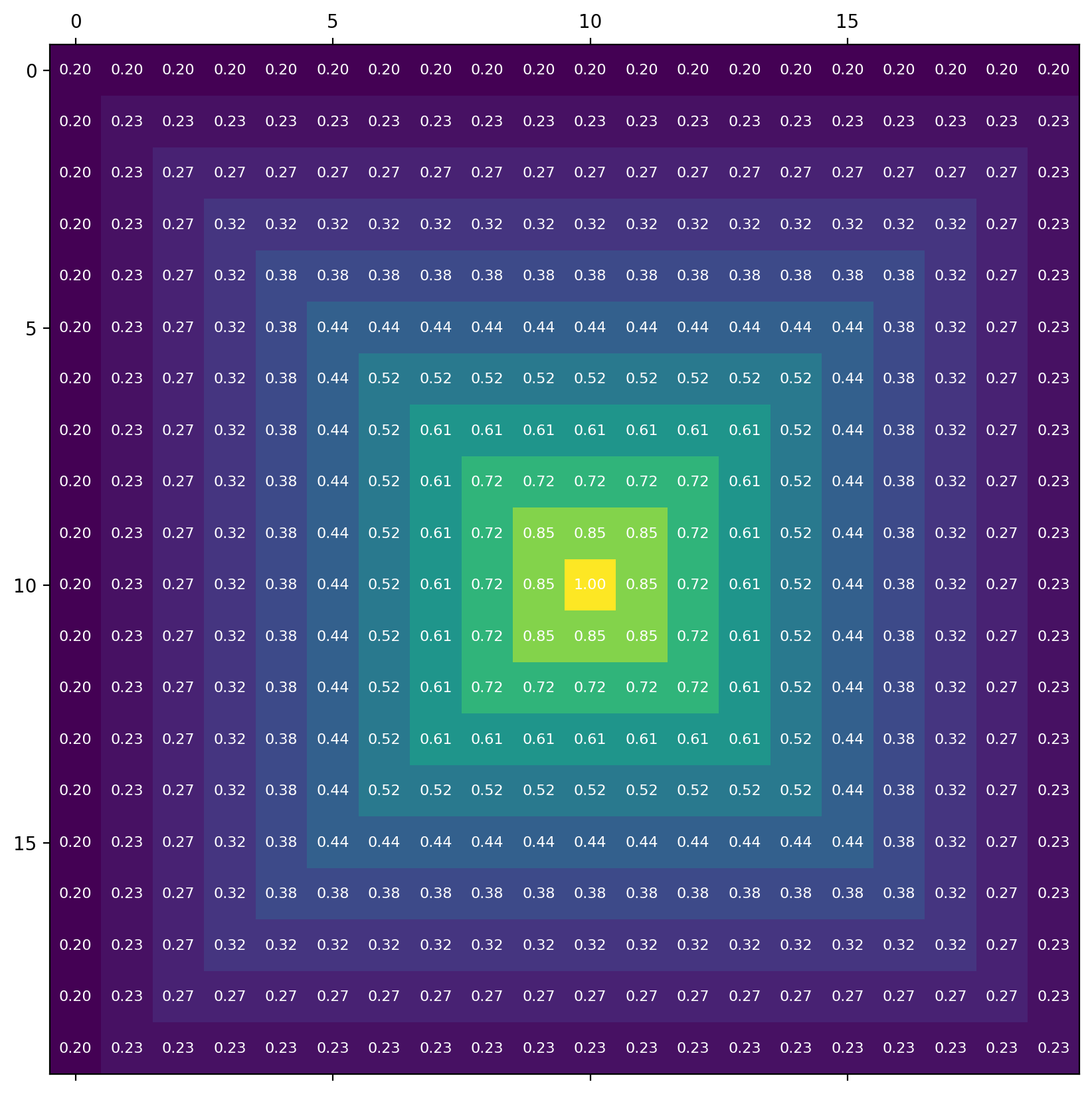}
    \caption{Breadth-first search with no specific stopping condition performed at (10,10) 0-indexed matrix cell. }
    \label{fig:bfs}
\end{figure}
Additionally, the last traversal presents an advantage that will lead to the exponential expression of $f(i,j,p_{k,n})$. While the average of the depth-first traversals accumulates probability at the matrix bounds, the breadth-first one correctly represents the progressive decrease of probability from the initial cell. That is, the accumulation of probability occurs around the position where the traversal starts, which adequately models the notion that the variation in probability propagates through the cells in the advance directions indicated by the system's neighborhood. Therefore, with this we can establish an initial expression for $f(i,j,p_{k,n})$ in the case that our algorithm uses a breadth-first traversal. From this expression, we can later apply a correction that incorporates the neighborhood properties the depth-first traversal captures:
\begin{align}
   f(x,y,p_{k,n}) = p_{k,n}^{|x|+|y|}
\end{align}
By denoting the coordinate axes on which the positions of the matrix cells are located as $x$ and $y$, we arrive at a scalar field like the following:
\begin{figure}[H]
    \centering
    \includegraphics[width=10cm,clip]{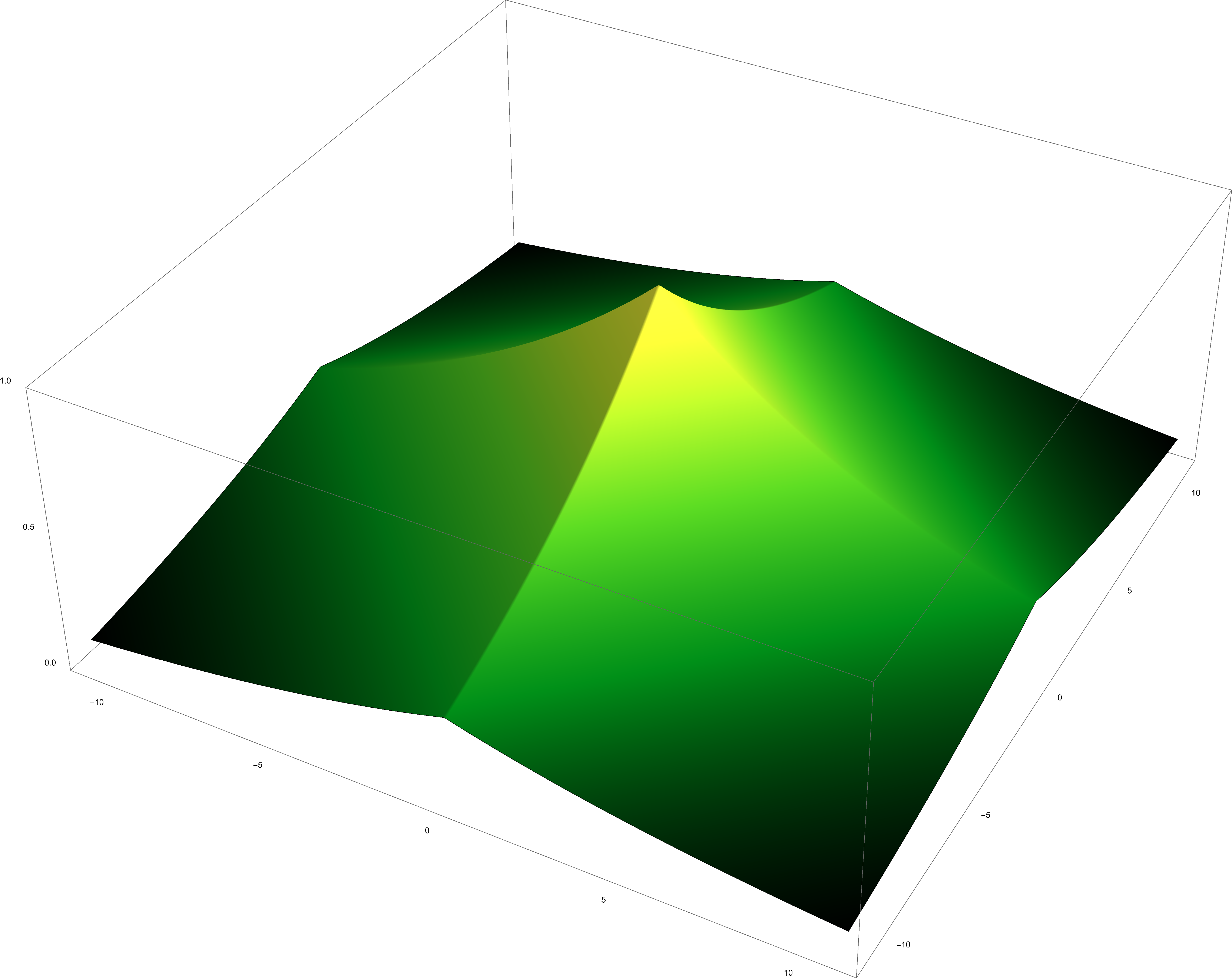}
    \caption{$f(x,y,p_{k,n}) = p_{k,n}^{|x|+|y|}$ probability scalar field corresponding with a BFS traversal plotted for the ranges $-11\leq x,y \leq 11,0\leq z\leq1$ and $p_{k,n}=0.9$}
    \label{fig:ClusterScalarFieldBFS}
\end{figure}
The resemblance to the field in Figure 24 arises from the exponentiation of the occupancy ratio. In a breadth-first search traversal, only the advance directions parallel to the coordinate axes are present, which in the formula are found in the exponent as $|x|$ for the $x=0$ direction and $|y|$ for the perpendicular $y=0$ one. Additionally, it is easily verifiable how the shape of the resulting function is invariant under translations, since $f(x+\alpha,y+\beta,p_{k,n})$ results in the same function shifted by an amount $(\alpha,\beta)$ with respect to its center, which can also be empirically observed by varying the initial cell of the traversal. Fundamentally, this invariance stems from the absence of matrix boundaries, which allows probability to expand over the infinite space where the matrix can increase its size indefinitely. However, the scalar field of the breadth-first traversal lacks a correction to model the remaining neighborhood directions. Specifically, the expression shown earlier needs to account for the directions corresponding to the diagonal neighboring cells. Therefore, to divide the scalar field into more regions according to the actual neighborhood, we need to include in the exponent the lines $y=x$ and $y=-x$.
\begin{align}
   f(x,y,p_{k,n}) = p_{k,n}^{\left(|x|+|y|+|x+y|+|x-y|\right)}
\end{align}
\begin{figure}[H]
    \centering
    \includegraphics[width=10cm,clip]{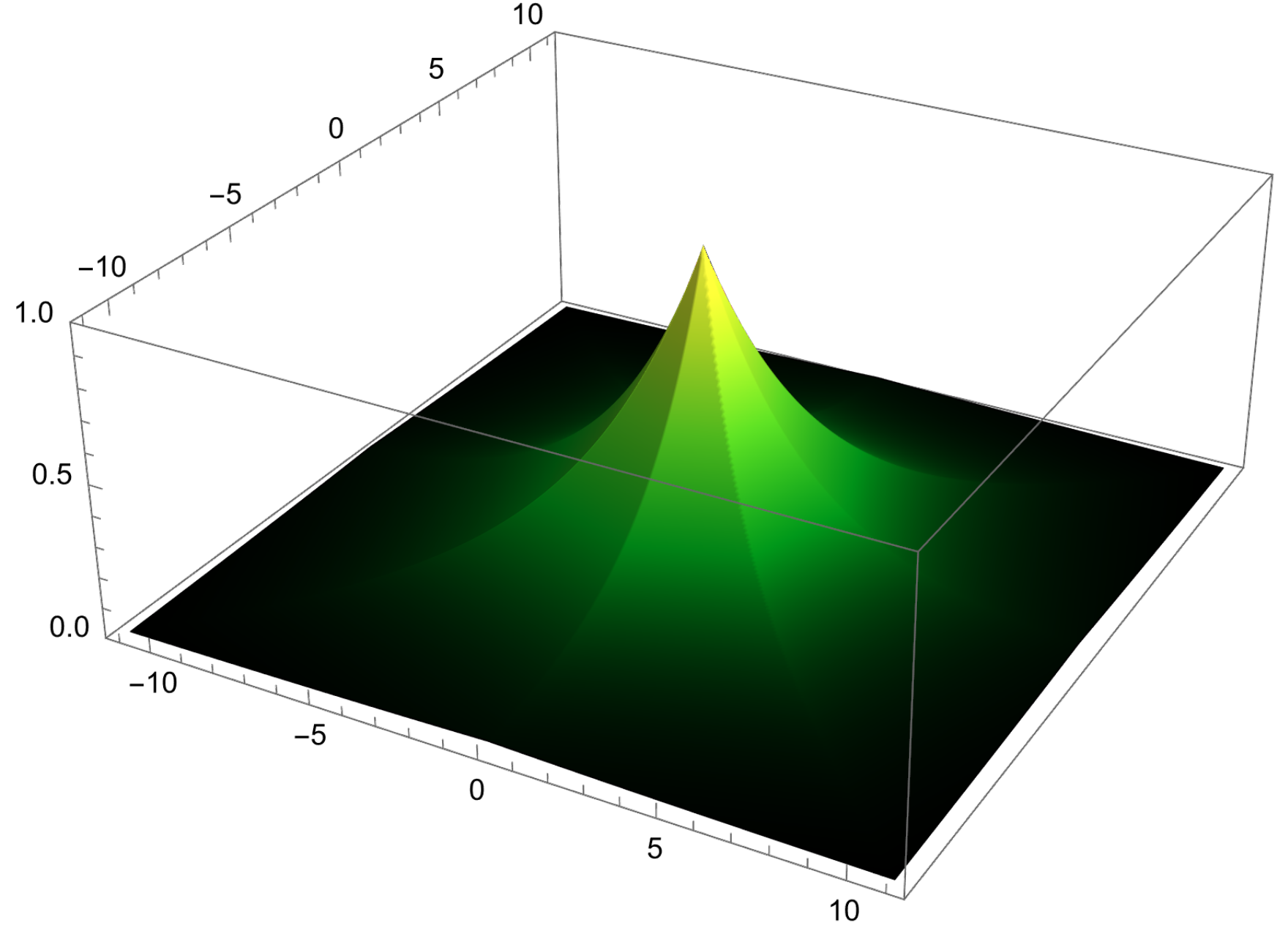}
    \caption{$f(x,y,p_{k,n}) = p_{k,n}^{\left(|x|+|y|+|x+y|+|x-y|\right)}$ probability scalar field corresponding with a DFS traversal plotted for the ranges $-11\leq x,y \leq 11,0\leq z\leq1$ and $p_{k,n}=0.9$}
    \label{fig:ClusterScalarField}
\end{figure}
After including the correction, there is an increase in the number of regions corresponding to the advance directions, whose final amount matches the number of neighboring cells used by the Moore neighborhood property. Thus, we obtain a scalar field $f(x,y,p_{k,n})$ like the one shown above with which we determine, for a point in space $(x,y)$ where a cell would be located, the probability that a depth-first search (DFS) traversal starting at the coordinate origin $(0,0)$ will reach that cell. Regarding its shape, it is recognizable how the probability accumulates near the center, as was the case in the breadth-first search (BFS). Nonetheless, if we look at empirical results of scalar fields obtained from a DFS, we will notice that the probability accumulates far from the center, albeit in a similar proportion. This may seem to contradict the final shape depicted in Figure 26. However, the explanation for this phenomenon resides in the behavior of both traversals when they reach the edges of the matrix. On the one hand, the BFS propagates the probability evenly until it reaches the bounds, while the DFS moves through them or even changes direction according to the order of cells to explore. Therefore, considering that the system's matrix has an unbounded size in this context, allowing the traversal to expand the probability in an infinite space, we can ensure that its bounds will not affect the constitution of the scalar field. Additionally, the probabilities near the edges when a DFS traversal is applied in a bounded matrix \textit{(since, in an infinite matrix, the DFS would not end)} are complementarily proportional to those near the center in a BFS \cite{Kozen1992,EverittHutter2015,Zhang1999,Maxwell2015,ZhangKorf1993}.
\subsubsection{Continuous sum of scalar field probabilities}
\label{subsubsec:ContinuousSumScalarFieldProbabilities}
Ultimately, after having a specific expression for the probability scalar field \cite{Apostol2000}, we can infer from it the average cluster size with respect to the occupancy ratio. That is, relating each probability to the contribution of its underlying cell to the length of a traversal, and by accounting for the executions of $helper()$ as an elementary operation in the complexity of an insertion, their sum will be a representative magnitude of the average cluster size processed in that traversal. Specifically, given the scalar field is infinite, all the values returned by each point in the set $\mathbb{R}^2$ will be summed. Still, several considerations must be made before estimating $c(n,k)$ with this method. The first relates to the normalization of probabilities and the scale of the final magnitude. With regards to the scalar field, its range is $[0,1]$, so all the probabilities it provides are correctly normalized. Yet, by performing the sum, it is evident that values will be produced that exceed the maximum of the previous range. This is not an inconvenience, as regardless of whether we normalize the result or not, it must be comparable to the range $[0,p_{n,k}\cdot n^2]$, since that is where the possible values for a cluster size will reside. Therefore, if a transformation of the final sum's range is needed, it will be performed with the aim of matching the valid range of cluster sizes, not probabilities. On the other hand, there are two options for conducting the sum over the entire $\mathbb{R}^2$ depending on the context, through integration of the continuous scalar field or a double summation. Although both alternatives represent the same magnitude, the discrete nature of the problem indicates that the sum must also be performed discreetly. Nevertheless, in this case, we will proceed with both methods to acquire more estimates for subsequently evaluate the validity of this approach.
\begin{align}
   c(n,k)= \int_{-\infty}^{\infty} \int_{-\infty}^{\infty} f(x,y,p_{k,n}) \, dx \, dy = \int_{-\infty}^{\infty} \int_{-\infty}^{\infty} p_{k,n}^{\left(|x|+|y|+|x+y|+|x-y|\right)} \, dx \, dy
\end{align}
For the continuous sum, $c(n,k)$ will be estimated as the double integral of the scalar field function $f(x,y,p_{k,n})$ with the neighborhood corrections included, from $-\infty$ to $\infty$ along both axes. But, given the amount of absolute values present in the exponent, it is necessary to prove certain symmetry properties before resolving the integral in order to simplify the process.
\begin{align}
   f(x,y,p_{k,n})=f(-x,y,p_{k,n})\\\notag
   f(x,y,p_{k,n})=f(x,-y,p_{k,n})\\\notag
   f(x,-y,p_{k,n})=f(-x,y,p_{k,n})\\\notag
   f(-x,y,p_{k,n})=f(x,-y,p_{k,n})\notag
\end{align}
Previously, we used the directions stemming from the neighborhood of a cell to construct the scalar field, resulting in 4 lines that divided $\mathbb{R}^2$ into 8 equal zones. With this, and weighing its shape when graphed, we can identify several symmetries of the field with respect to each of the lines \cite{Thomas2014}. On the one hand, there are those located on the axes $x=0$ and $y=0$, where the field will be symmetric with respect to the plane perpendicular to $\mathbb{R}^2$ that they form if the value returned at a point $(x,y)$ is equal to that of another point $(-x,y)$, or similarly with the remaining axis. Additionally, symmetries are also proposed with respect to planes situated on the lines that model the neighboring diagonal cells, $y=x$ and $y=-x$. For these latter symmetries to hold, the probability at a point $(x,-y)$ must be equivalent to that of another point in the form $(-x,y)$, or vice versa.

\begin{align}
   p_{k,n}^{|x|+|y|+|x+y|+|x-y|} = p_{k,n}^{|-x|+|y|+|-x+y|+|-x-y|}\\\notag
   p_{k,n}^{|x|+|y|+|x+y|+|x-y|} = p_{k,n}^{|x|+|-y|+|x-y|+|x+y|}\\\notag
   p_{k,n}^{|x|+|-y|+|x-y|+|x+y|} = p_{k,n}^{|-x|+|y|+|-x+y|+|-x-y|}\\\notag
   p_{k,n}^{|-x|+|y|+|-x+y|+|-x-y|} = p_{k,n}^{|x|+|-y|+|x-y|+|x+y|}\notag \label{symmetries}
\end{align}
Following the substitution of the point forms for each of the symmetries, all the previous equalities hold under the following conditions:
\begin{align}
   |\pm x|=|\pm x| \implies \sqrt{(\pm x)^2}=\sqrt{(\pm x)^2} \implies \sqrt{x^2}=\sqrt{x^2}
\end{align}
Initially, the base of the exponentiation is ignored by applying logarithms on both functions. Thus, we are left with the exponent terms, among which are some like $|x|$ and $|-x|$ on opposite sides of the equality. However, all those of this form can be cancelled in the same way.
\begin{align}
   |\pm x\pm y|=|\pm x\pm y| \implies \begin{cases}
       \sqrt{(+x+y)^2}=\sqrt{(-x-y)^2}\\
       \sqrt{(+x-y)^2}=\sqrt{(-x+y)^2}
   \end{cases} \implies \begin{cases}
       \sqrt{x^2-2xy+y^2}=\sqrt{x^2-2xy+y^2}\\
       \sqrt{x^2+2xy+y^2}=\sqrt{x^2+2xy+y^2}
   \end{cases}
\end{align}
Furthermore, the remaining terms are also equivalent, since their signs are complementary, and as observed above, as long as this condition is met, the absolute values will be equal \cite{HuHuang2010}. Consequently, it has been proven that the scalar field is symmetric with respect to the 4 planes defined by the corresponding lines, which allows us to simplify the integral sum of their probabilities as follows:

\begin{align}
   c(n,k)=  4 \left(\int_{0}^{\infty} \int_{0}^{x} p_{k,n}^{x+y+x+y+x-y} \, dy \, dx + \int_{0}^{\infty} \int_{0}^{y} p_{k,n}^{x+y+x+y-x+y} \, dx \, dy \right)
\end{align}
For each of the 4 sectors divided by $x=0$ and $y=0$, the field is integrated from the beginning of one of the axes to the line $y=x$, subsequently integrating the output function to infinity. For simplicity, this sum is carried out in the positive sector, so in both integrals the absolute value $|x-y|$ is decomposed into $x-y$ and $y-x$ depending on the side it is on relative to the line $y=x$.

\begin{align}
   c(n,k)=  4 \left(\int_{0}^{\infty} \int_{0}^{x} p_{k,n}^{3x+y} \, dy \, dx + \int_{0}^{\infty} \int_{0}^{y} p_{k,n}^{x+3y} \, dx \, dy \right)
\end{align}

\begin{align}
   c(n,k)=  4 \left(\int_{0}^{\infty} \left . \frac{p_{k,n}^{3x+y}}{ln(p_{k,n})} \right|_{y=0}^{y=x} \, dx + \int_{0}^{\infty} \left . \frac{p_{k,n}^{x+3y}}{ln(p_{k,n})} \right|_{x=0}^{x=y} \, dy \right)
\end{align}

\begin{align}
   c(n,k)=  4 \left(\int_{0}^{\infty} \frac{p_{k,n}^{4x}-p_{k,n}^{3x}}{ln(p_{k,n})} \, dx + \int_{0}^{\infty} \frac{p_{k,n}^{4y}-p_{k,n}^{3y}}{ln(p_{k,n})} \, dy \right)
\end{align}
Given that the region is divided into 2 symmetric subregions, the total integral is computable by integrating over a single one of the 8 subregions:

\begin{align}
   c(n,k) = 8 \int_{0}^{\infty} \int_{0}^{x} p_{k,n}^{x+y+x+y+x-y} \, dy \, dx = 8 \int_{0}^{\infty} \frac{p_{k,n}^{4x}-p_{k,n}^{3x}}{ln(p_{k,n})} \, dx
\end{align}
Before continuing with the bounds, it is convenient to separate the resolution of its antiderivative:
\begin{align}
   \int \frac{p_{k,n}^{4x}-p_{k,n}^{3x}}{ln(p_{k,n})} \, dx = \int \frac{p_{k,n}^{4x}}{ln(p_{k,n})} \, dx - \int \frac{p_{k,n}^{3x}}{ln(p_{k,n})} \, dx
\end{align}

\begin{align}
   \int \frac{p_{k,n}^{4x}}{ln(p_{k,n})} \, dx - \int \frac{p_{k,n}^{3x}}{ln(p_{k,n})} \, dx = \frac{p_{k,n}^{4x}}{4\cdot ln^2(p_{k,n})} - \frac{p_{k,n}^{3x}}{3\cdot ln^2(p_{k,n})} + C_1 - C_2
\end{align}
Finally, evaluating the antiderivative in its corresponding bounds, we obtain:
\begin{align}
   c(n,k) = 8 \left(\left . \frac{p_{k,n}^{4x}}{4\cdot ln^2(p_{k,n})} - \frac{p_{k,n}^{3x}}{3\cdot ln^2(p_{k,n})} \right|_{x=0}^{x=\infty} \right)
\end{align}

\begin{align}
   c(n,k) = 8 \left(\lim_{x \to \infty} \left(\frac{3p_{k,n}^{4x} - 4p_{k,n}^{3x}}{12 \cdot \ln^2(p_{k,n})}\right) + \frac{1}{12 \cdot \ln^2(p_{k,n})}\right)
\end{align}
To avoid complicating the evaluation of the limit, we can leverage the restriction $0\leq p_{k,n}\leq1$, since the occupancy ratio depends on the number of elements, which is never less than 0 nor greater than the number of cells in the system matrix. Therefore, the numerator of the limit tends to 0, yielding a final expression for the desired estimation:
\begin{align}
   \hat{c}_0(n,k) = 8 \left(\frac{1}{12 \cdot \ln^2(p_{k,n})}\right) = \frac{2}{3 \cdot \ln^2(p_{k,n})}
\end{align}
To discern the various expressions we have for the same magnitude, the estimates will be denoted as $\hat{c}_i(n,k)$.
\begin{figure}[H]
    \centering
    \begin{subfigure}[b]{0.49\textwidth}
        \centering
        \includegraphics[width=\textwidth,clip]{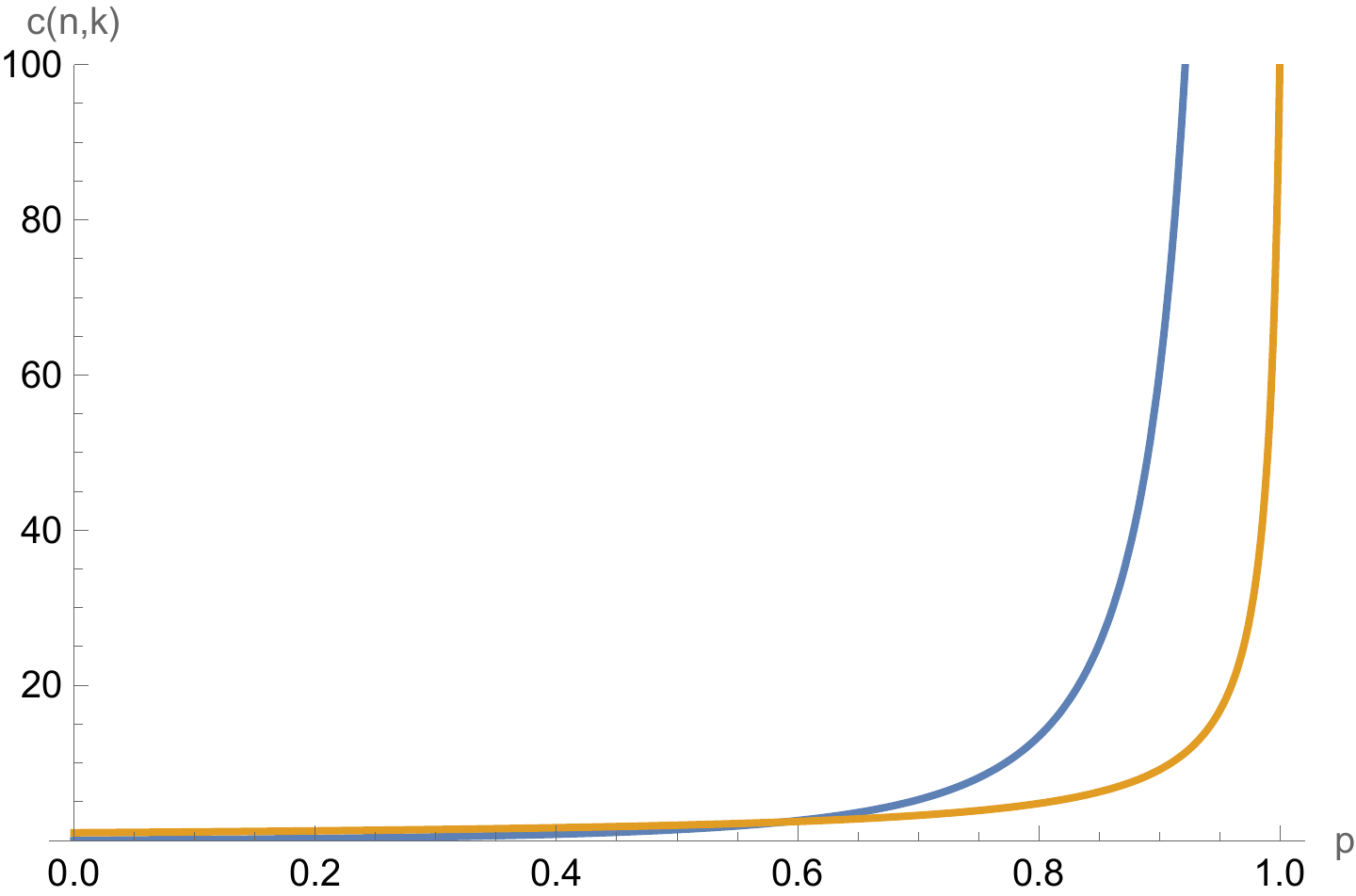}
        \caption{}
        \label{fig:ClusterSizeEstimation}
    \end{subfigure}
    \hfill
    \begin{subfigure}[b]{0.49\textwidth}
        \centering
        \includegraphics[width=\textwidth,clip]{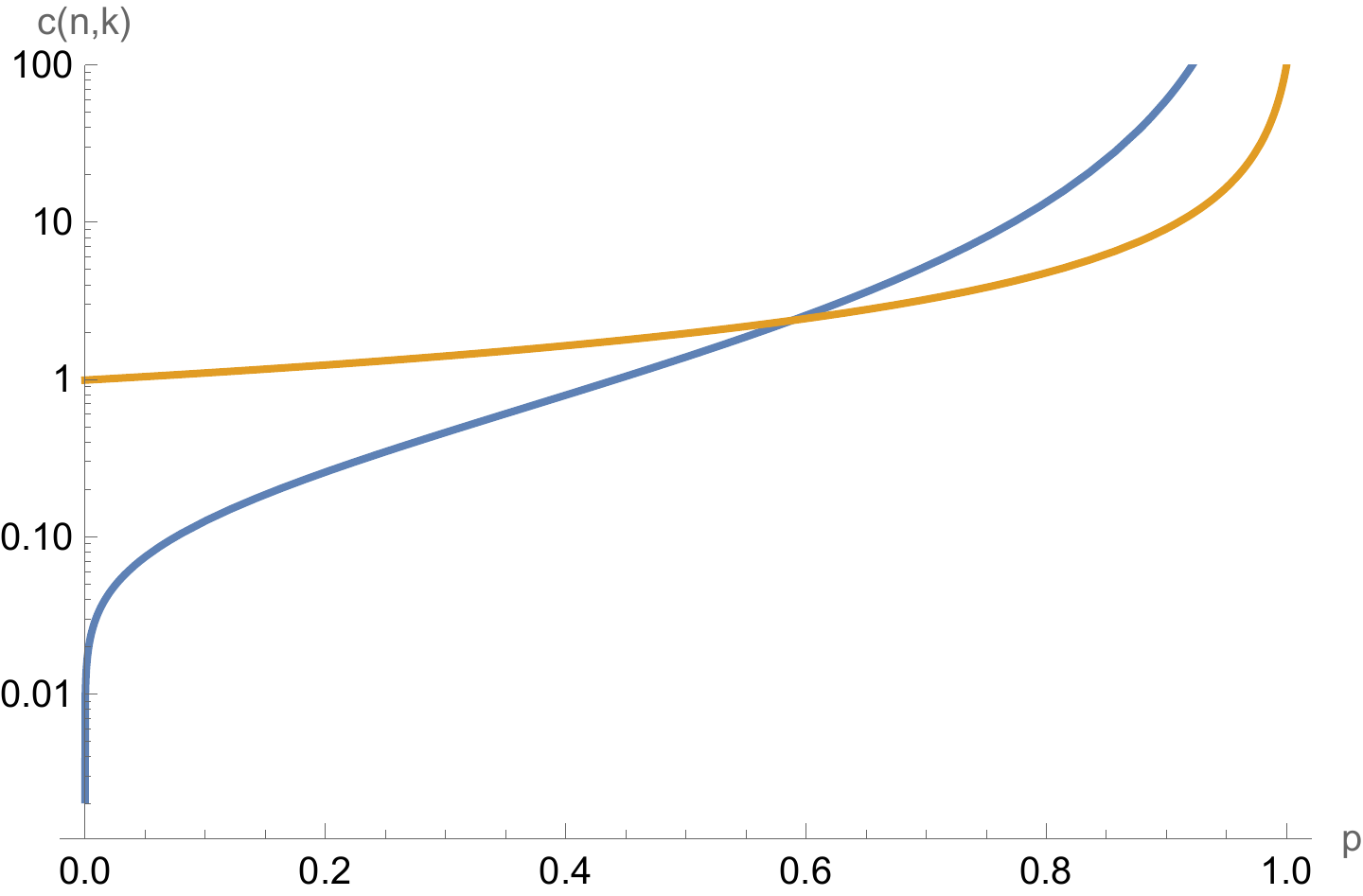}
        \caption{}
        \label{fig:ClusterSizeEstimationLog}
    \end{subfigure}
    \caption{(a) $c(n,k)$ plotted in orange when the occupancy ratio varies between 0 and 1, substituting on the expression the corresponding $k$ for a system of size $n=10$. In blue, the first estimation of the average cluster size $\hat{c}_0(n,k)$ is shown. (b) Plot (a) under a logarithmic transformation of its image.}
    \label{fig:ClusterSizeEstimationDual}
\end{figure}
Even though the function we originally obtained for $c(n,k)$ is not guaranteed to be correct for this magnitude, it serves as a good reference to compare with more advanced estimations from the scalar field, mainly due to its form and how it adheres to the imposed constraints. Thus, after constructing an expression for $\hat{c}_0(n,k)$ based on the continuous sum of probabilities, we can see that they are very similar for low occupancy ratio values, which is a good sign. On the other hand, the estimation diverges from $c(n,k)$ when the ratio tends to 0, more noticeably when both are in logarithmic scale. Additionally, the estimation does not reach the maximum value of $n^2$ when $p_{k,n}$ tends to its supremum 1, but rather tends to infinity at a similar rate as $c(n,k)$ approaches that value. 
\subsubsection{Discrete sum of scalar field probabilities}
But, before extensively comparing both functions and analyzing their correctness, we can construct another estimator $\hat{c}_1(n,k)$ based on the discrete sum of probabilities proposed at the beginning.
\begin{align}
   \hat{c}_1(n,k) = \sum _{x=-\infty }^{\infty } \sum _{y=-\infty }^{\infty } f(x,y,p_{k,n}) = \sum _{x=-\infty }^{\infty } \sum _{y=-\infty }^{\infty } p_{k,n}^{| y-x| +| x+y| +| x| +| y| }
\end{align}
Thus, in case we only need to account for the discrete values that any pair of positions $(x, y)$ can assume in the sum, a double summation is applied from $-\infty$ to $\infty$ on both axes. To resolve this, given that the scalar field is the same as in the continuous estimator, the sums are simplifiable by considering the distinct symmetric regions of that field.

\begin{align}
   \hat{c}_1(n,k) = 8\cdot \sum _{x=2}^{\infty } \sum _{y=1}^{x-1} p_{k,n}^{x+x+x+y+y-y} + 4\cdot \left(\sum _{x=1}^{\infty } p_{k,n}^{3 x} + \sum _{x=1}^{\infty } p_{k,n}^{4 x}\right) + 1
\end{align}
On the one hand, the 8 subregions it is divided into with respect to its symmetry axes will be calculated by multiplying the contribution of one of those situated in the positive quadrant by the corresponding value. As it is a discrete sum, it is convenient to exclude in the sum of the subregion the value of the point $(0,0)$, along with the points of the form $(x,y)$ or $(x,0)$, as their contributions would later need to be deducted from the total sum. In this way, by knowing how many lines compose the symmetry, we can sum them independently of the subregions, together with the point $(0,0)$ on which the field returns 1, to the final value of the estimator. As for the lines $x=0$ and $y=0$, since $f(x,0,p_{k,n})=p_{k,n}^{3|x|}$ and both are equal, 4 times the segment that composes half the length of one of them is summed, to prevent the central point from being included. Similarly, the remaining lines in the form $y=\pm x$ are summed in the same way, with the only difference being that the field over the points of that form equals $p_{k,n}^{4|x|}$.

\begin{align}
   \sum _{x=1}^{\infty } p_{k,n}^{3 x} = \frac{1}{1-p_{k,n}^3}-1 = \frac{p_{k,n}^3}{1-p_{k,n}^3}
\end{align}
\begin{align}
   \sum _{x=1}^{\infty } p_{k,n}^{4 x} = \frac{1}{1-p_{k,n}^4}-1 = \frac{p_{k,n}^4}{1-p_{k,n}^4}
\end{align}
Initially, the easiest part to solve are the segments of the symmetry lines, being both terms geometric sums.

\begin{align}
   \sum _{x=2}^{\infty } \sum _{y=1}^{x-1} p_{k,n}^{3x+y} &= \sum _{x=2}^{\infty } p_{k,n}^{3x}\cdot\sum _{y=1}^{x-1} p_{k,n}^{y} = \\ \notag
   &= \sum _{x=2}^{\infty } p_{k,n}^{3x}\cdot\left(p_{k,n}\frac{1-p_{k,n}^{x-1}}{1-p_{k,n}}\right) =\\ \notag
   &= \sum _{x=2}^{\infty } \frac{p_{k,n}^{4 x} - p_{k,n}^{3 x + 1}}{p_{k,n}-1} = \sum _{x=2}^{\infty } \frac{p_{k,n}^{3 x} \left(p_{k,n}^x-p_{k,n}\right)}{p_{k,n}-1}
\end{align}
Regarding the remaining term of each subregion, the first step is to eliminate the inner summation and express its result as a function of $x$ to subsequently solve the other sum. Thus, after converting the first one into a geometric series and providing a closed form as simplified as possible for its value, we can solve the remaining summation by breaking it down into several geometric sums.
\begin{align}
   \sum _{x=2}^{\infty } \frac{p_{k,n}^{3 x} \left(p_{k,n}^x-p_{k,n}\right)}{p_{k,n}-1} &= \sum _{x=2}^{\infty } \frac{p_{k,n}^{4 x}-p_{k,n}^{3x+1}}{p_{k,n}-1} = \\ \notag
   &= \frac{1}{p_{k,n}-1} \left(\sum _{x=2}^{\infty } p_{k,n}^{4 x} - \sum _{x=2}^{\infty } p_{k,n}^{3x+1}\right) = \\ \notag
   &= \frac{1}{p_{k,n}-1} \left(\frac{p_{k,n}^8}{p_{k,n}^4-1} - \sum _{x=2}^{\infty } p_{k,n}^{3x+1}\right) = \\ \notag
   &= \frac{1}{p_{k,n}-1} \left(\frac{p_{k,n}^8}{1-p_{k,n}^4} - \frac{p_{k,n}^7}{1 -p_{k,n}^3}\right) = \frac{p_{k,n}^7}{\left(p_{k,n}^4-p_{k,n}-1\right) p_{k,n}^3+1}
\end{align}
Ultimately, with the expressions obtained for each of the terms, the final expression for the estimator results in:

\begin{align}
   \hat{c}_1(n,k) = 8\cdot \left(\frac{p_{k,n}^7}{\left(p_{k,n}^4-p_{k,n}-1\right) p_{k,n}^3+1}\right) + 4\cdot (\frac{p_{k,n}^3}{1-p_{k,n}^3} + \frac{p_{k,n}^4}{1-p_{k,n}^4}) + 1
\end{align}
And, after some simplifications, it is rewritable as:
\begin{align}
   \hat{c}_1(n,k) = \frac{p_{k,n} \left(p_{k,n}^5-p_{k,n}^4+p_{k,n}^3+2 p_{k,n}^2+p_{k,n}-1\right)+1}{(p_{k,n}-1)^2 \left(p_{k,n}^2+1\right) \left(p_{k,n}^2+p_{k,n}+1\right)}
\end{align}
\subsubsection{Estimators comparison}
\begin{figure}[H]
    \centering
    \begin{subfigure}[b]{0.49\textwidth}
        \centering
        \includegraphics[width=\textwidth,clip]{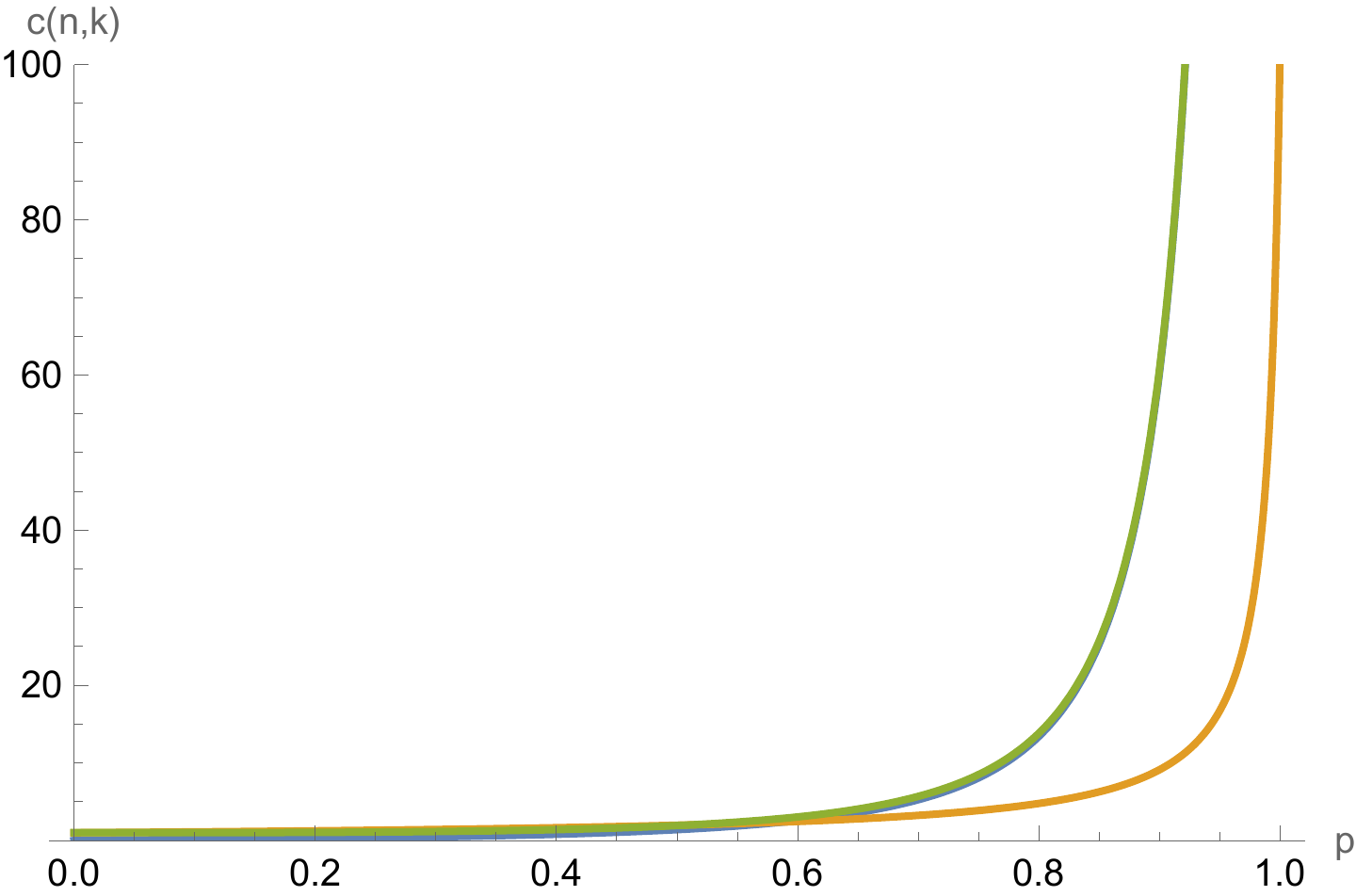}
        \caption{}
        \label{fig:ClusterSizeEstimations}
    \end{subfigure}
    \hfill
    \begin{subfigure}[b]{0.49\textwidth}
        \centering
        \includegraphics[width=\textwidth,clip]{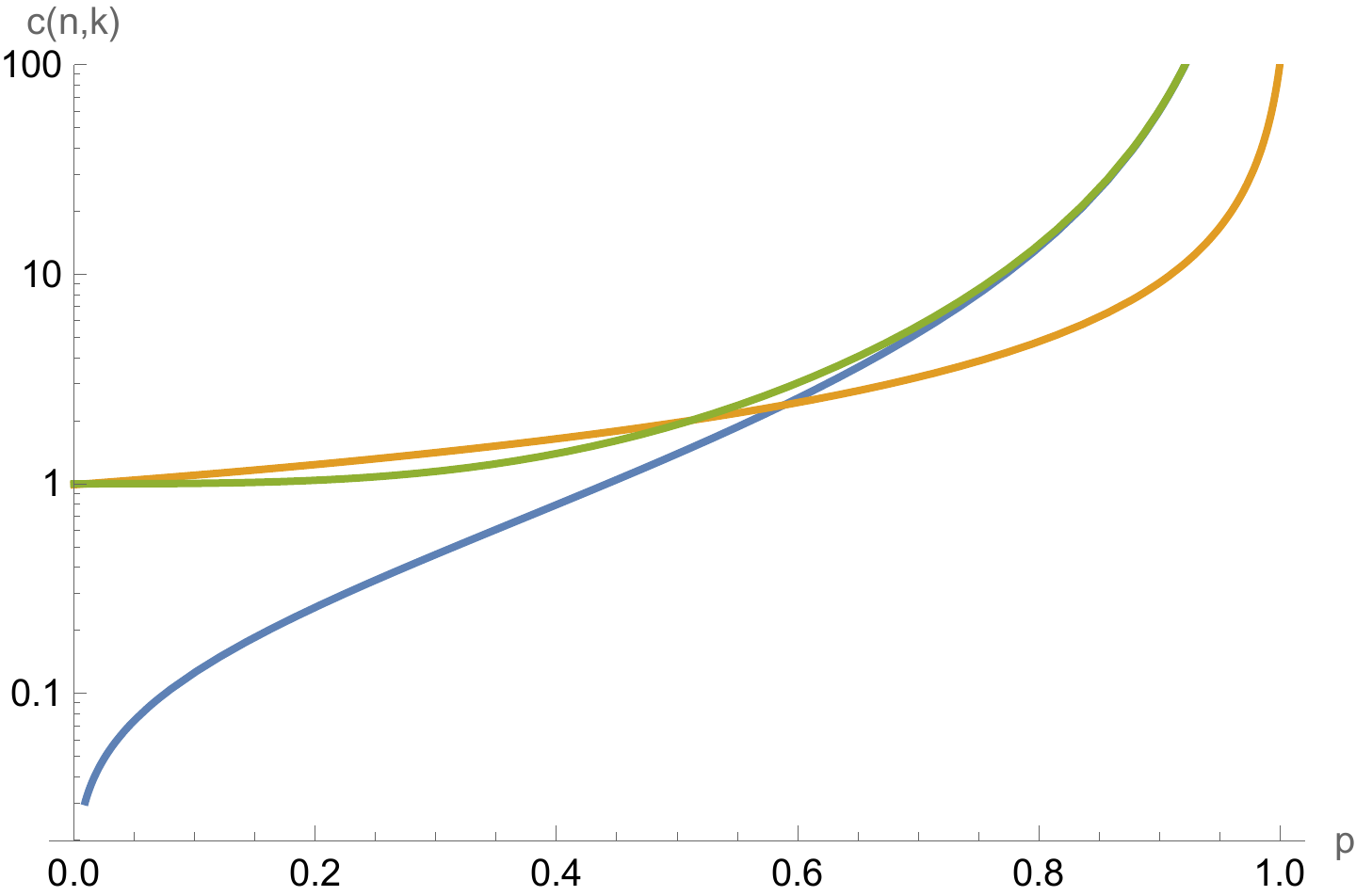}
        \caption{}
        \label{fig:ClusterSizeEstimationsLog}
    \end{subfigure}
    \caption{(a) $c(n,k)$ plotted in orange, $\hat{c}_0(n,k)$ in blue, and $\hat{c}_1(n,k)$ in green when the occupancy ratio varies between 0 and 1. (b) Plot (a) under a logarithmic transformation of its image.}
    \label{fig:ClusterSizeEstimationsDual}
\end{figure}
Both estimators appear to be similar to $c(n,k)$, though they exhibit certain differences that determine their validity. Concerning the continuous approximation, it is noted in the logarithmic plot how its image covers the range $[0,\infty)$, which is easily verifiable through the limits as the occupancy ratio reaches its extremes. In this way, as the ratio approaches 0, the estimator diverges from the other 2 functions and tends to 0. Although this may not seem significant in the normal scale plot, this difference indeed may affect the estimator's correctness when calculating the complexity of the entire percolative process. The main reason for this behavior is the amount of integrated volume in the scalar field, which, as a continuous quantity, has the possibility of reaching very small values as the occupancy ratio decreases, causing the estimator to fail at modeling the constraint for the average cluster size when $k=1$. On the other hand, the estimator derived from the discrete sum does not present this problem, since even if the number of elements on the system is small, the central point of the scalar field with value 1 will always be accounted for in the sum. Lastly, despite both estimators differ at low values of $p_{k,n}$, the opposite holds for the rest of the domain. Specifically, from a certain threshold onwards, they show an equivalent asymptotic growth, noticeable in their ratio when the number of elements tends to its maximum value, that is, the domain's upper limit:
\begin{align}
   &\lim_{p_{k,n}\to1^{-}} \frac{\displaystyle\frac{p_{k,n} \left(p_{k,n}^5-p_{k,n}^4+p_{k,n}^3+2 p_{k,n}^2+p_{k,n}-1\right)+1}{(p_{k,n}-1)^2 \left(p_{k,n}^2+1\right) \left(p_{k,n}^2+p_{k,n}+1\right)}}{\displaystyle\frac{2}{3 \cdot \ln^2(p_{k,n})}} =\\ \notag
    &= \lim_{p_{k,n}\to1^{-}} \frac{3\ln^2(p_{k,n})\left(p_{k,n} \left(p_{k,n}^5-p_{k,n}^4+p_{k,n}^3+2 p_{k,n}^2+p_{k,n}-1\right)+1\right)}{2(p_{k,n}-1)^2 \left(p_{k,n}^2+1\right) \left(p_{k,n}^2+p_{k,n}+1\right)} = \\ \notag
    &= \frac{3}{2}\cdot\lim_{p_{k,n}\to1^{-}} \left( \frac{p_{k,n} \left(p_{k,n}^5-p_{k,n}^4+p_{k,n}^3+2 p_{k,n}^2+p_{k,n}-1\right)+1}{\left(p_{k,n}^2+1\right) \left(p_{k,n}^2+p_{k,n}+1\right)}\right) \cdot \lim_{p_{k,n}\to1^{-}} \frac{\ln^2(p_{k,n})}{(p_{k,n}-1)^2} = \\ \notag
    &= \frac{3}{2}\cdot\frac{2}{3} \cdot \lim_{p_{k,n}\to1^{-}} \frac{\ln^2(p_{k,n})}{(p_{k,n}-1)^2} = \lim_{p_{k,n}\to1^{-}} \frac{\ln^2(p_{k,n})}{(p_{k,n}-1)^2} = \lim_{p_{k,n}\to1^{-}} \left(\frac{\frac{1}{p_{k,n}}}{1}\right)^2 = 1
\end{align}
The ratio of both estimators as the number of elements increases to its maximum value tends to 1, so their growth is asymptotically equivalent \cite{Wolfram2018}. Consequently, given that both have the same asymptotic behavior for high values of the occupancy ratio \cite{Kato1968,Chakravarthy2020}, the choice between the two estimators for complexity analysis will depend on the influence of their difference in the initial iterations of the algorithm, even though, primarily, it is preferable to consider the discrete sum of probabilities, since it fully meets all the imposed constraints.
\subsubsection{Image range correction}
\label{subsubsec:ImageRangeCorrection}
Nevertheless, as evidenced in the graphs and limits, both estimators return increasingly larger values up to infinity as they approach $p_{k,n}=1$, which represents a significant difference from the expected actual value. To correct this issue, the values produced by both estimators must be normalized, since initially we did not apply any transformation or normalization to the probabilities we summed from the field. Specifically, it is required that when the number of elements $k$ equals the maximum number of them that fit in the system, the estimators return $n^2$ instead of infinity. Thus, of all the transformations to correct the range covered by the image, we will select one that complicates the resulting estimators expressions as least as possible.
\begin{align}
   \hat{c}_0(n,k) = n^2\left(\frac{\displaystyle\frac{2}{3 \cdot \ln^2(p_{k,n})}}{\displaystyle \frac{2}{3 \cdot \ln^2(p_{k,n})}+n^2}\right) = \frac{2 n^2}{\left(3 ln^2(p_{k,n})\right) \left(n^2+\frac{2}{3 ln^2(p_{k,n})}\right)} = \frac{2 n^2}{3 n^2 ln^2(p_{k,n})+2}
\end{align}
Given the relative simplicity of the initial expression of the first estimator, it is applied first in that expression to enhance its clarity and discernibility. In summary, the idea behind the transformation is based on preventing the function from reaching infinite at the upper bound of its domain, achieved by dividing itself by the entire original expression \cite{Zayed1996}. This normalizes the value at that point to 1, which can then be multiplied by any desired amount. However, at all domain points, it would return the same value, so it is necessary to add the same amount determined by the multiplier as an offset to the denominator. This approach ensures the function retains its original properties, especially its growth in proportion to its image, which will now be upper-bounded by $n^2$ \cite{MacLane1986}.
\begin{align}
   \hat{c}_1(n,k) &= n^2\left(\frac{\displaystyle \frac{p_{k,n} \left(p_{k,n}^5-p_{k,n}^4+p_{k,n}^3+2 p_{k,n}^2+p_{k,n}-1\right)+1}{(p_{k,n}-1)^2 \left(p_{k,n}^2+1\right) \left(p_{k,n}^2+p_{k,n}+1\right)}}{\displaystyle \frac{p_{k,n} \left(p_{k,n}^5-p_{k,n}^4+p_{k,n}^3+2 p_{k,n}^2+p_{k,n}-1\right)+1}{(p_{k,n}-1)^2 \left(p_{k,n}^2+1\right) \left(p_{k,n}^2+p_{k,n}+1\right)} + n^2}\right) = \\ \notag
   &=\frac{n^2 \left(p_{k,n} \left(p_{k,n}^5-p_{k,n}^4+p_{k,n}^3+2 p_{k,n}^2+p_{k,n}-1\right)+1\right)}{\left((p_{k,n}-1)^2 \left(p_{k,n}^2+1\right) \left(p_{k,n}^2+p_{k,n}+1\right)\right) \left(n^2+\frac{p_{k,n} \left(p_{k,n}^5-p_{k,n}^4+p_{k,n}^3+2 p_{k,n}^2+p_{k,n}-1\right)+1}{(p_{k,n}-1)^2 \left(p_{k,n}^2+1\right) \left(p_{k,n}^2+p_{k,n}+1\right)}\right)} = \\ \notag
   &=\frac{n^2 \left(p_{k,n} \left(p_{k,n}^5-p_{k,n}^4+p_{k,n}^3+2 p_{k,n}^2+p_{k,n}-1\right)+1\right)}{n^2 \left(p_{k,n}^2+1\right) \left(p_{k,n}^2+p_{k,n}+1\right) (p_{k,n}-1)^2+p_{k,n} \left(p_{k,n}^5-p_{k,n}^4+p_{k,n}^3+2 p_{k,n}^2+p_{k,n}-1\right)+1} 
\end{align}

\begin{figure}[H]
    \centering
    \begin{subfigure}[b]{0.42\textwidth}
        \centering
        \includegraphics[width=\textwidth,clip]{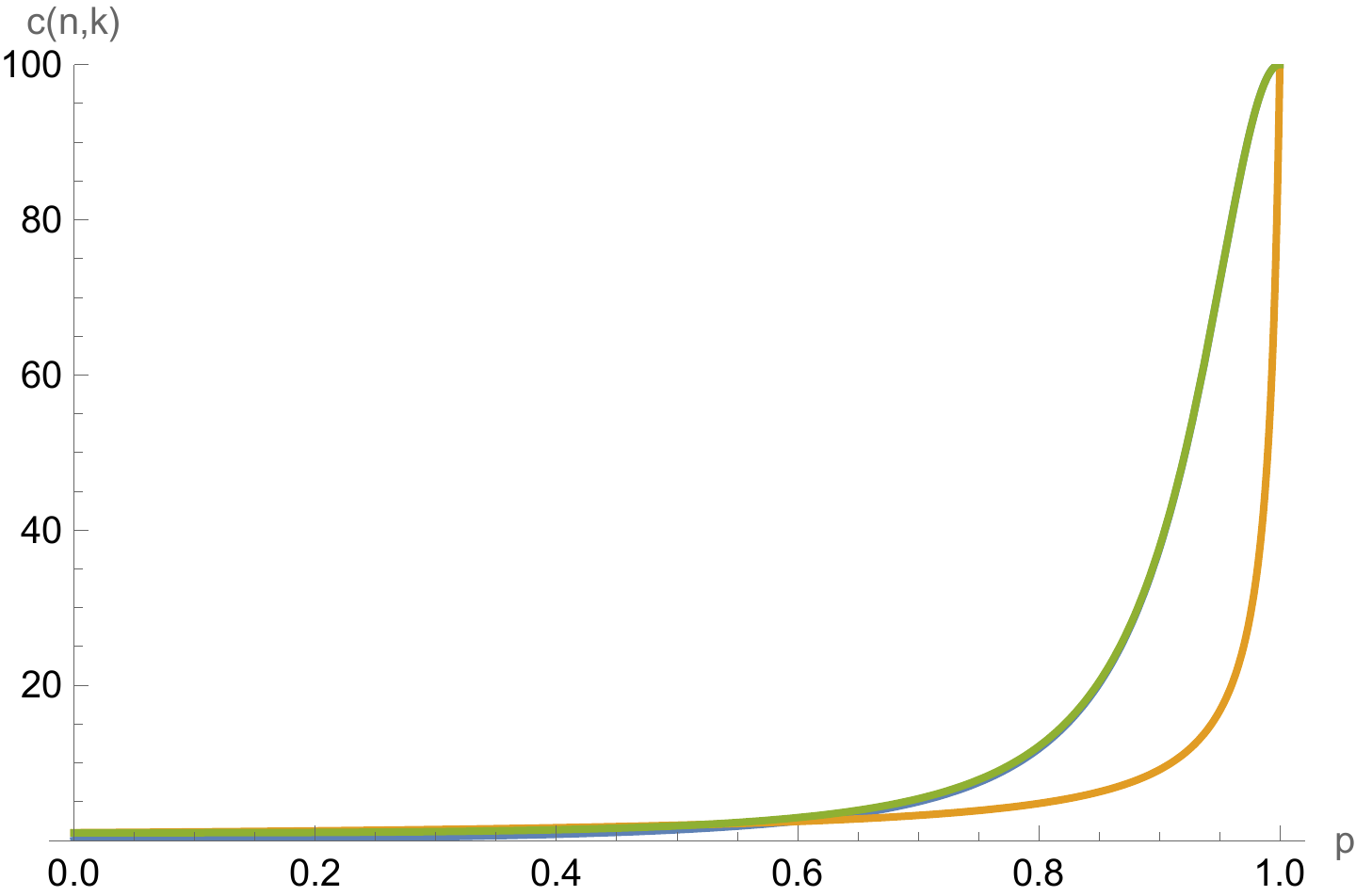}
        \caption{}
        \label{fig:ClusterSizeEstimationsFinal}
    \end{subfigure}
    \hfill
    \begin{subfigure}[b]{0.42\textwidth}
        \centering
        \includegraphics[width=\textwidth,clip]{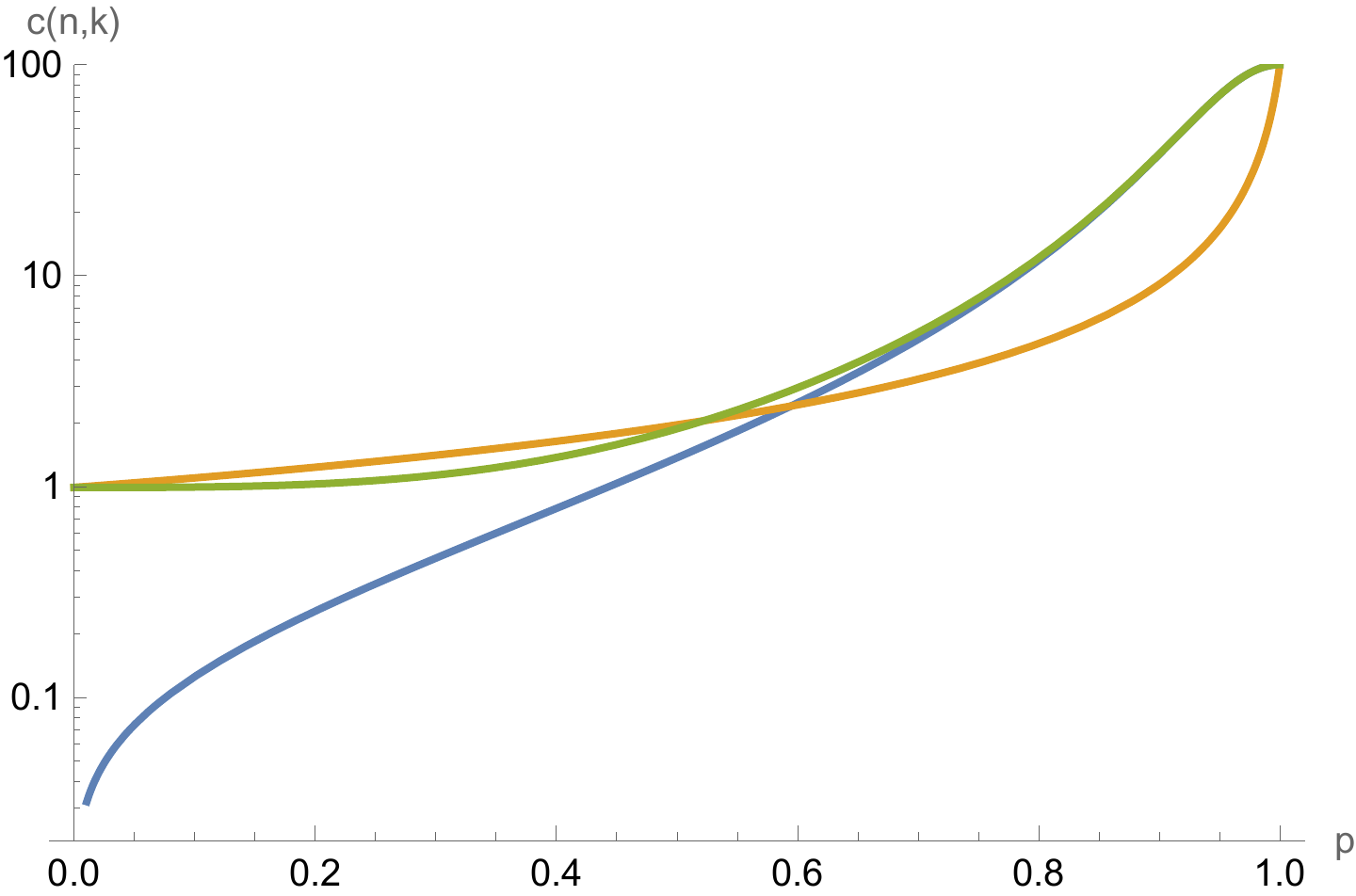}
        \caption{}
        \label{fig:ClusterSizeEstimationsLogFinal}
    \end{subfigure}
    \caption{(a) (b) Plots from Figure 28 after applying the image range correction to both estimators.}
    \label{fig:ClusterSizeEstimationsFinalDual}
\end{figure}
Finally, after applying the transformation to both estimators, Figure 29 demonstrates how they reach the correct value for the average cluster size when the system matrix fills up. With regard to their growth, the difference between the continuous estimator and its expected value for occupancy ratios close to 0 persists. Nevertheless, their equivalent asymptotic growth towards the other end of the domain also persists, which this time appears slightly different from the growth of $c(n,k)$, which is not significant because with large matrix sizes the difference progressively diminishes. In conclusion, from now on we can use either estimator or the previous formula $c(n,k)$ as the average cluster size in the complexity analysis. It is true that for now none of them have been proven to model this magnitude correctly, since the approach of $c(n,k)$ was based on meeting certain restrictions and the estimators stem from a probabilistic approach. Even so, as we proceed with the details of the rest of the analysis, it is likely that we will encounter situations that force us to discard one of them.

\subsection{Expected elements metric}

It has been previously mentioned that the number of elements in a particular iteration of the system has is formulable as a closed form, or at least the expected quantity of them. This would be of great help for the remaining analysis, both in expressions involving the occupancy ratio as well as the estimators for average cluster size and the computation of the probability of an insertion occurring. So, despite we can't consider all the possible states that the percolative process may undergo and perform the complexity analysis through its study or formalization, we can infer a fairly useful metric for our purpose without requiring to conduct exhaustive considerations. Such metric is the expected number of elements within the matrix at a certain iteration $i$, also denoted by $E(i)$. Thus, although the metric might not be entirely accurate, it sufficiently captures the information present in the possible system states.
\\\\
With a rigorous calculation of $E(i)$, we can ensure when performing the analysis that, for all active iterations of the process, there is an exact number of elements in the system equivalent to $E(i)$. Moreover, by not returning discrete values, contrary to what might be expected, it provides a considerable theoretical advantage, since the analysis will be performed upon the effect of the iterations over the system, not specifically upon the discrete number of elements on it. Thereby, we can also arrive at some probability distributions that will help us understand the structure of the problem by approaching certain properties that would be complex to compute in a discrete and extensive way.
\\\\
To calculate $E(i)$, we must analyze what happens at each process iteration from the beginning at $i=0$ until the number of iterations is sufficiently high for the process to stop. When it starts, the matrix is completely empty, so after the algorithm executes the insertion attempt, it surely inserts an element on a random grid position. This information is valuable for the construction of the $E(i)$ general case, because now we know that $E(0)=0$ and $E(1)=1$ occurs on every algorithm execution. The problem arises when $i>1$, where the probability of being able to insert an element in the matrix, or the opposite, in the case of having chosen a position held by an element inserted in a previous iteration, appears.
\\\\
The most immediate approach involves considering the difference between the expected number of elements in one iteration $i$ and the next $i+1$. Naively, the amount of elements in $i+1$ will be at least equal to the elements in $i$ ($E(i)$), plus a quantity that we don't necessarily know but can estimate. So, the first attempt for $E(i)$ will have the following form:
\begin{equation}
    E(i)=E(i-1)+Q
\end{equation}
Now, we have narrowed down the problem to the Q term, i.e., to calculate how much $E(i)$ increases each iteration. When an insertion is executed, the position is considered to be uniformly random. Therefore, for a certain state $E(i)$, the element will be inserted only if the chosen position corresponds to a valid matrix cell. Hence, we can ensure that the probability of adding an element to the system as a result of the insertion can be calculated as follows:
\begin{equation}
    P(i)=1-\frac{E(i-1)}{n^2}
\end{equation}
$P(i)$ is mainly formed by the complement to the probability of the selected position being occupied, with $E(i-1)$ representing the expected elements at the previous iteration and $n^2$ the total system elements, since the matrix is square shaped. Now, as this probability returns a value between 0 and 1, it can be analogized to the $Q$ value we want to calculate, because the $Q$ increment must also be between 0 and 1, corresponding to the 0 or 1 elements that can be inserted each iteration. 
\begin{align}
    E(i)=E(i-1)+(1-\frac{E(i-1)}{n^2})
\end{align}
To summarize, $E(i)$ equals to the number of elements in the previous iteration $E(i-1)$, which is also an expected quantity, plus the probability of inserting an element in the current iteration. Then, to obtain a valid expression for $E(i)$, we use the base cases $E(0)=0,E(1)=1$ and work out the previous recurrence relation:
\begin{align}
    E(i)=\frac{n^2E(i-1)+n^2-E(i-1)}{n^2}
\end{align}
Rewriting the prior result for convenience, it remains:
\begin{align}
    E(i)=1+\frac{n^2-1}{n^2}\cdot E(i-1)
\end{align}
Note that now the only recursive term is isolated in $E(i-1)$, which simplifies the resolution process. Hence, to detect the solution pattern, it is beneficial to unfold $E(i)$ for some values of $i$:
\begin{align*}    
    E(2)=1+\frac{n^2-1}{n^2}\cdot E(1)=1+\frac{n^2-1}{n^2}
\end{align*}
Since $E(1)=1$, we can skip the term.
\begin{align*}    
    E(3)=1+\frac{n^2-1}{n^2}\cdot (1+\frac{n^2-1}{n^2})\\
    =1+(\frac{n^2-1}{n^2})+(\frac{n^2-1}{n^2})^2
\end{align*}
At this point, a pattern emerges for the upcoming solution:
\begin{align*}    
    E(4)=1+\frac{n^2-1}{n^2}\cdot (1+(\frac{n^2-1}{n^2})+(\frac{n^2-1}{n^2})^2)\\
    =1+(\frac{n^2-1}{n^2})+(\frac{n^2-1}{n^2})^2+(\frac{n^2-1}{n^2})^3
\end{align*}
Then, the expanded general term will resemble like this:

\begin{align*}    
    E(i)=(\frac{n^2-1}{n^2})^0+(\frac{n^2-1}{n^2})^1+\cdots +(\frac{n^2-1}{n^2})^{i-1}E(1)
\end{align*}
As observed, each term of $E(i)$ is a geometric series sum of a ratio that does not vary with iterations, since the problem matrix maintains its dimensions throughout the percolation process. Also, note that the last term must be multiplied by $E(1)$, but ignored since it´s always $1$.

\begin{align*}    
    E(i)=\sum_{i=0}^{i-1} (\frac{n^2-1}{n^2})^i=\frac{1-(\frac{n^2-1}{n^2})^i}{1-\frac{n^2-1}{n^2}}
\end{align*}
After applying the definition of a geometric series sum we obtain the above expression, and by simplifying it leads to the final form \cite{Tufto2021,Alex2011}:
\begin{align}    
    E(i)=n^2(1-(1-\frac{1}{n^2})^i)
\end{align}
Upon deriving the complete expression, it is worth checking its correctness, and especially its fidelity to what we want to model. For this purpose, we shall consider a different starting point. This time, for $i>1$ the formula is posed considering the probability of whether or not a new element is inserted in a certain iteration. So, denoting as $P(i-1)$ the matrix occupancy ratio $\frac{E(i-1)}{n^2}$, we can calculate the expected number of elements it will contain in the next iteration as follows:

\begin{align}    
    E(i)=(1-P(i-1)) (E(i-1)+1)+P(i-1)E(i-1)
\end{align}
After substituting the actual value of $P(i-1)$ inside the expression and simplifying it, we arrive at the following outcome:

\begin{align}    
    E(i)=\frac{(n^2-E(i-1))(E(i-1)+1)+E(i-1)^2}{n^2}=\frac{n^2E(i-1)+n^2-E(i-1)}{n^2}
\end{align}
As shown \cite{Filip2018}, it´s equivalent to an intermediate step of $E(i)$ calculation with the original method, which implies its resolution will also be identical. But, in addition to this verification, many others may be performed, especially empirical ones, which show how the formula for the expected number of elements fit its actual magnitude. Mainly, although they do not guarantee that the expected amount is the closest to the actual one, nor that the expression for the expectation is unique or correct, experimentally checking the values it returns in certain situations serves as a guide to discard it if the results differ significantly. For this purpose, in this case we will use as an empirical measure the histograms from the simulations dataset in the previous section \textcolor{blue}{\ref{Expected_iterations_for_process_termination}}, with which the probability distributions for the number of elements needed for the algorithm to finish can be elucidated. Given that the dataset entries also keep information about the elements at the terminal state of all simulations, their corresponding number of iterations together with the formula for the expected number of elements serve to obtain a distribution equivalent to the histogram with the actual measured quantities. Thus, depending on the similarity between the two, it will be determined how accurate the metric is for the elements of each algorithm's iteration.

\begin{figure}[H]
    \centering
    \begin{subfigure}[b]{0.49\textwidth}
        \centering
        \includegraphics[width=\textwidth,clip]{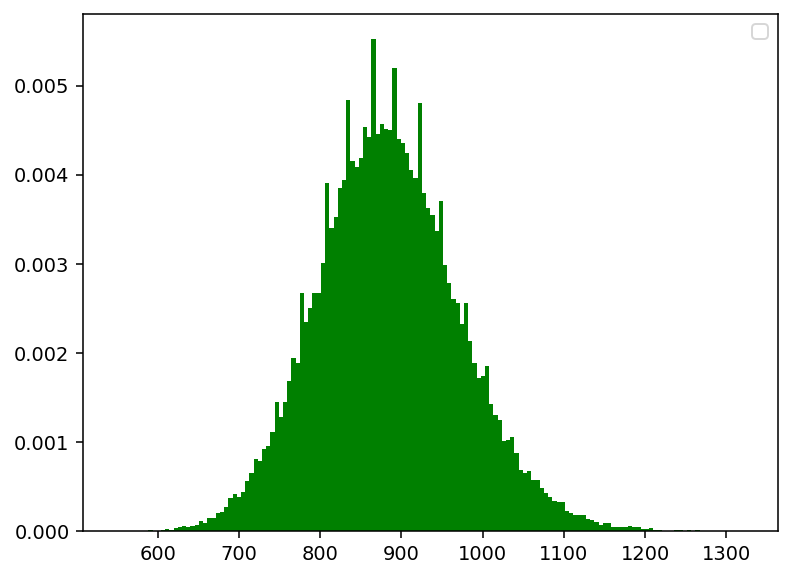}
        \caption{}
        \label{fig:iterations_distribution4141}
    \end{subfigure}
    \hfill
    \begin{subfigure}[b]{0.49\textwidth}
        \centering
        \includegraphics[width=\textwidth,clip]{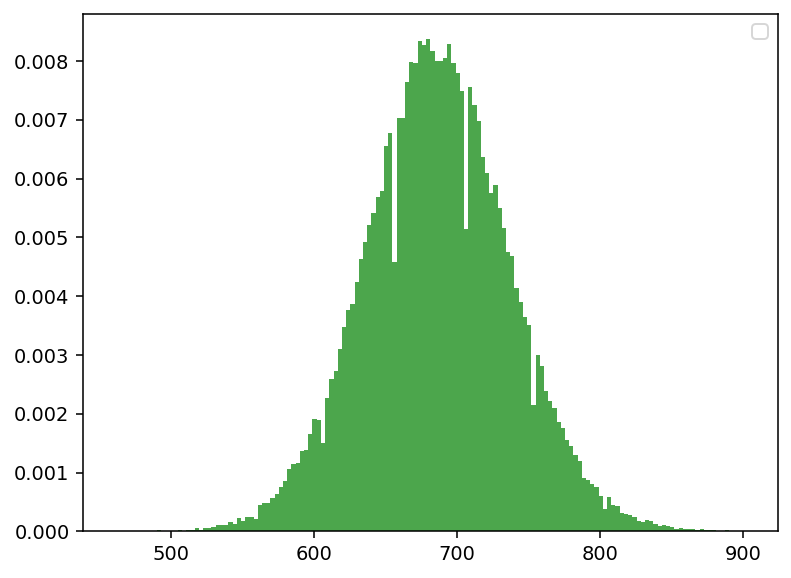}
        \caption{}
        \label{fig:elements_distribution4141}
    \end{subfigure}
    \caption{(a) Histogram of the iterations array for a system with a square matrix of dimension $n=41$. (b) Correspondingly, the histogram of the number of elements at terminal state array for a system with a square matrix of the same dimension.}
    \label{fig:41distributionDual}
\end{figure}
At first, to ensure a certain accuracy during the process, simulations performed for a matrix size near the upper limit of the dataset are selected. With this, we will acquire a well-defined shape of the histograms, which will allow us to better perceive the potential deficiencies of the metric for the number of elements. Regarding the simulation data in a matrix of size $n=41$, it is appreciable how the histogram of the iterations, which shows the distribution of the random variable $\mathbf{I}_n$, is similar to a gamma distribution, while the histogram corresponding to the elements at the terminal state is more akin to a binomial distribution, or normal if the matrix is sufficiently large. Thus, by denoting $\mathbf{E}_n$ as the random variable that, analogous to the previous one, determines the exact number of elements present in the terminal state of a matrix with side $n$, we can represent all the measurement information of the dataset in sets for both random variables. In this way, each simulation will be characterized by the pair $(i_\lambda,e_\lambda)$ for $0<\lambda\leq100000$, with both sets forming the entry of the dataset for the respective matrix size.

\begin{align*}    
    \mathbf{I}_n\approx\{i_1, i_2, i_3, \dots, i_{100000}\} \quad \mathbf{E}_n\approx\{e_1, e_2, e_3, \dots, e_{100000}\}
\end{align*}
After formalizing the sequences of values constituting the histograms, it is possible to examine if a transformation in any number of iterations $i_\lambda$ using the previous expression $E(i)$ results in a number of elements close to the one at the terminal state. That is, even without knowing exactly if the distribution that $\mathbf{I}_n$ follows is genuinely a gamma distribution, and notwithstanding the parameters that define it, we have measurements that capture the shape to which its true distribution tends, and therefore, we can apply certain functions to them that will result in a transformation of the distribution's shape.
\begin{align*}    
    \mathbf{\hat{E}}_n\approx\{E(i_1), E(i_2), E(i_3), \dots, E(i_{100000})\}
\end{align*}
Thus, by defining the set $\mathbf{\hat{E}}_n$ as the collection of all measurements originating from the set of iterations where all have been transformed into an expected amount of elements, we can plot it in the form of a histogram and contrast it with the set of measurements of the original element count.
\begin{figure}[H]
    \centering
    \includegraphics[width=10cm,clip]{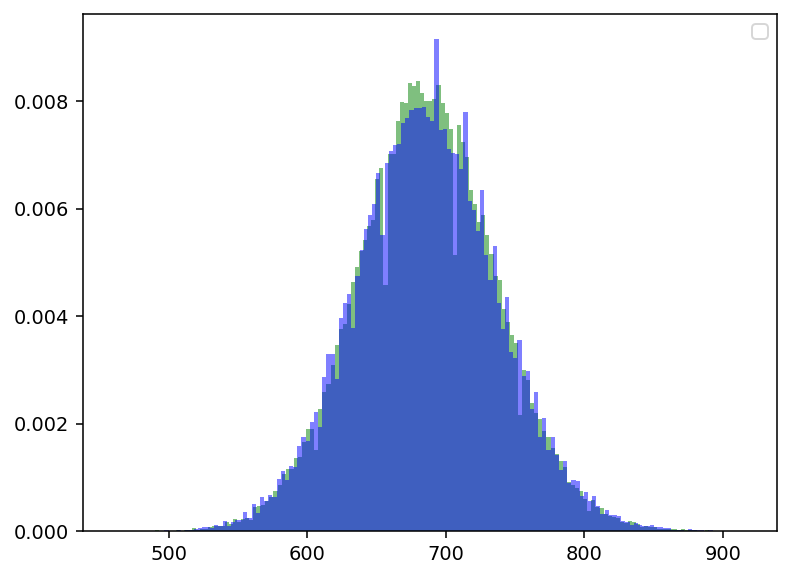}
    \caption{Histogram of the iterations array for a system with a square matrix of dimension $n=41$ transformed by the function $E(i)$ plotted in blue, while in green it is plotted the (b) histogram from Figure 30}
    \label{fig:estimated_elements_distribution}
\end{figure}
The foundation of this process lies in the relationship between the values of the pairs $(i_\lambda,e_\lambda)$ contained in the dataset. If one represents the iterations in which the terminal state has been reached, and the other the count of elements in that state, it is plausible to expect that $E(i_\lambda)$ will return a value as close as possible to the measured one. Consequently, after constructing the set of transformed iterations and arranging it in the form of a histogram, a clear similarity is observed compared to the actual measurements, which can also be verifiable with the rest of the matrix sizes in the dataset, although for small ones the difference may be slightly greater.
\\\\
In conclusion, we can now use $E(i)$ as a metric to predict how many elements are present in a specific iteration. Regarding its correctness, it empirically fits well to such magnitude throughout a process execution, showing strong evidence that the values it returns are sufficiently precise, especially for large system sizes. However, the optimal method to demonstrate its validity is to continue with the analysis and occasionally evaluate some conditions it must fulfill, or results it must provide in comparison with alternative metrics. Finally, regarding its precision, it remains uncertain whether the difference between the metric and the actual magnitude is significant enough to consider it an unreliable measure throughout the execution of the algorithm. Although, given the verifications conducted so far, it is suitable for our purpose.
\subsection{Probabilistic approach for expected number of elements}
Apart from the metric $E(i)$ that will serve to determine the expected number of elements per iteration, it is worth highlight that its use in the construction of certain probability distributions may be erroneous, given its continuous nature. Therefore, another additional metric is proposed which, instead of returning an expected quantity per iteration, represents the probability that a certain number of elements exists in that iteration. That is, to estimate how many cells have been occupied up to a certain point in the execution of the algorithm, the probability of that metric can be multiplied by the size of the system, in case the probability determines the existence of such a quantity of elements. To this end, there are various ways to arrive at an expression that adequately models the probability, although in this case, one will be followed that is suitable for the potential uses of the metric in probability distributions. Specifically, the initial approach consists of that, given a sequence of $i$ insertion attempts, which corresponds to the situation in which the algorithm has executed that amount of iterations, the probability that there exist a quantity $k \leq i$ of elements is equivalent to the probability that the set with all position pairs of the insertion sequence has a cardinal of $k$. In this manner, if we denote the metric as $H(i,k)$, we can reach an expression like the following \cite{Lavrov2024}:
\begin{align}    
    H(i,k)=\frac{(n^2)!}{(n^2)^i (n^2-k)!} \stirling{i}{k}
\end{align}
The metric represents the probability that a sequence of $i$ insertions has $k$ distinct position pairs, which correspond to the cells where an element has been inserted, because any other appearance of any of those pairs coincides with an insertion on an already occupied cell. Concerning its computation, it involves the ratio of sequences with $k$ distinct position pairs and the total count of existing sequences of size $i$. The latter quantity is easy to obtain since on a square matrix with side $n$ there are $n^{2i}$ possible sequences, given that a cell is selectable in multiple insertions, even if they do not occur. And, the number of them capable of forming a set with $k$ distinct position pairs stems from the number of partitions into $k$ non-zero subsets constructible on the insertion sequence \cite{SedgewickFlajolet2009}. In other words, the different ways to distribute the $i$ position pairs over $k$ sets such that all have some pair. After obtaining such count, it is multiplied by the possible specific cells that can compose the set of $k$ elements, which corresponds to the descending factorial of the first $k$ factors of $n^2$.
\begin{figure}[H]
    \centering
    \includegraphics[width=10cm,clip]{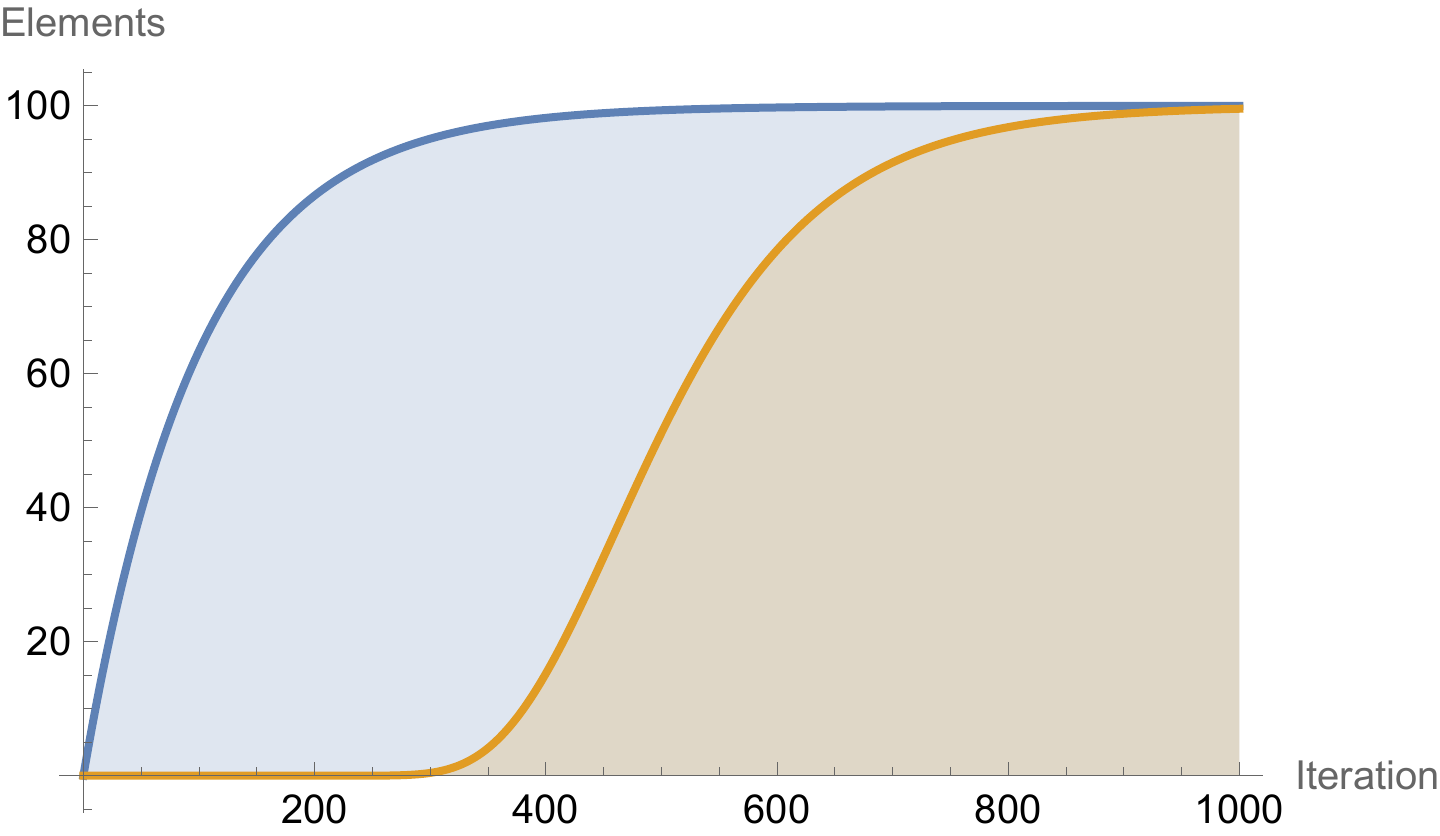}
    \caption{Expected number of elements metric $E(i)$ plotted in blue and the probability that in an iteration $i$ the system contains $n^2$ elements, given by the metric $n^2H(i,n^2)$ displayed in orange, for a square matrix of side $n=10$}
    \label{fig:ExpectedElementsAndProbability}
\end{figure}
The main distinction between the obtained expression and the metric $E(i)$ is appreciable graphically for an arbitrary matrix size. First, the expected number of elements returns increasingly higher values until it stabilizes near the size of the system in early iterations, whereas the new metric returns a probability of 0, that is, 0 expected elements in early iterations, until it grows and stabilizes at the same value as the previous metric. This is due to the pigeonhole principle, as the probability that there exist $n^2$ distinct elements in a system that has not undergone $n^2$ iterations must be 0, rendering it impossible for the event to occur before that iteration. With this, we can ensure that $H(i,n^2)n^2$ is not a good estimate for the elements in each iteration of the process, because in early iterations it is verifiable that for a sufficiently large matrix the probability of inserting an element is high, so multiple insertions will occur and their total quantity will increase considerably until stabilizing later; that is, the larger the system, the more elements will be inserted in early iterations, as most of the cells are empty.
\\\\
Contrarily, $E(i)$ seems to adhere correctly to the expected behavior, thus enabling the verification of its expression's correctness through the probabilities returned by the metric $H(i,k)$. In this way, considering that $H(i,k)$ returns the probability that in a certain iteration there are exactly $k$ elements, we can denote $\hat{E}(i)$ as another estimator of the expected count of elements in iteration $i$ of a process. And, to construct its expression, it is sufficient to apply the same criteria as in the expectation of a probability distribution, in which for each iteration from the first to the specified one, the expected number of elements contributes to the total measure as indicated by the metric.
\begin{align}
    \hat{E}(i)=\sum _{k=0}^i H(i,k)k = \sum _{k=0}^i \frac{(n^2)!k}{(n^2)^i (n^2-k)!} \stirling{i}{k}
\end{align}
Thus, depending on the similarity between the resolution of the previous sum to the original metric $E(i)$, we will be able to know how accurate both are. However, the Stirling number of the second kind \cite{RennieDobson1969} in the numerator complicates the process, so we will first proceed to evaluate the metric over a certain range of iterations.
\begin{align}
    &\hat{E}(0)=0\\ 
    &\hat{E}(1)=1\\ 
    &\hat{E}(2)=2-\frac{1}{n^2}\\ 
    &\hat{E}(3)=\frac{1}{n^4}-\frac{3}{n^2}+3\\ 
    &\hat{E}(4)=\frac{-6 n^4+4 n^2-1}{n^6}+4\\ 
    &\hat{E}(5)=\frac{1-5 \left(2 n^6-2 n^4+n^2\right)}{n^8}+5\\ 
    &\hat{E}(6)=\frac{\left(2 n^2-1\right) \left(n^4-n^2+1\right) \left(3 n^4-3 n^2+1\right)}{n^{10}}\\ 
    &\hat{E}(7)=\frac{7 \left(n^2-1\right) \left(n^5-n^3+n\right)^2+1}{n^{12}}\\
    &\hat{E}(8)=\frac{\left(2 n^2-1\right) \left(2 n^4-2 n^2+1\right) \left(2 n^8-4 n^6+6 n^4-4 n^2+1\right)}{n^{14}}\\
    &\hat{E}(9)=\frac{\left(3 n^4-3 n^2+1\right) \left(3 \left(n^{10}-3 n^8+6 n^6-7 n^4+5 n^2-2\right) n^2+1\right)}{n^{16}}
\end{align}
\begin{align}  
    &\hat{E}(10)=\frac{\left(2 n^2-1\right) \left(n^8-2 n^6+4 n^4-3 n^2+1\right) \left(5 \left(n^6-2 n^4+2 n^2-1\right) n^2+1\right)}{n^{18}} 
\end{align}
After completing these intermediate evaluations, we can try to find a function that generates all of them from the iteration number, perhaps by formulating a recursion or by observing patterns among them. Nevertheless, this procedure can be automated with Wolfram's $FindSequenceFunction[]$ feature \cite{Wolfram_FindSequenceFunction_2008}. Therefore, after applying it with the previous functions and setting the variable $i$ as generator, we obtain:

\begin{align}
    \hat{E}(i)=n^2\left(1-\left(1-\frac{1}{n^2}\right)^i\right)
\end{align}
As notable, the resulting expression for the new metric is equivalent to $E(i)$, so it is convenient to perform intermediate evaluations for that metric and check that the same functions are indeed produced.

\begin{align}
    &E(0)=0\\ 
    &E(1)=1\\ 
    &E(2)=2-\frac{1}{n^2}\\ 
    &E(3)=\frac{1}{n^4}-\frac{3}{n^2}+3\\ 
    &E(4)=\frac{-6 n^4+4 n^2-1}{n^6}+4\\ 
    &E(5)=\frac{1-5 \left(2 n^6-2 n^4+n^2\right)}{n^8}+5\\
    &E(6)=\frac{\left(2 n^2-1\right) \left(n^4-n^2+1\right) \left(3 n^4-3 n^2+1\right)}{n^{10}}\\
    &\vdots\\
    &E(i)=n^2\left(1-\left(1-\frac{1}{n^2}\right)^i\right)
\end{align}
Ultimately, all intermediate evaluations of $E(i)$ coincide with those of $\hat{E}(i)$, resulting in the same expression for their sequence. This congruence strongly suggests that both approaches previously constructed have a common basis. Therefore, in conclusion, both $E(i)$ and its analogue $H(i,k)$ fulfill their purpose, notwithstanding the differing contexts in which they are applicable. That is, while the former is ideal for accurately estimating the elements in a certain iteration and thus using it in expressions involved in the process analysis, the latter calculates the probability of the existence of a certain number of elements, proving useful in probability distributions.
\subsection{Expected iterations for process termination}
As an example of usage of both metrics in the analysis, which facilitates the verification of their correctness depending on the context where they are applied, we will proceed to formalize the previous function $I(n)$ that returns the expected number of iterations involved in a process over a system of size $n^2$. So far, our knowledge about this function is that it originates from the expectation of a random variable that follows an unknown probability distribution, although with a shape similar to a Gamma, according to experimental measurements. However, if we try to find an expression for this distribution as accurately as possible, without assuming any apparent form, it becomes imperative to introduce a new function $p(n,k)$ returning the number of existing terminal states for a system of size $n^2$ and $k$ elements.

\begin{align}
    Pr[\mathbf{I}_n\leq i]=\frac{p(n,E(i))}{\displaystyle\binom{n^2}{E(i)}}
\end{align}
To calculate the probability that the process finishes at most in a certain iteration $i$, we first need to know how many elements are expected in that iteration, determined by the metric $E(i)$. With this, the probability is defined as the ratio between all the terminal states of the system with $E(i)$ elements on it and the total count of possible states, whether terminal or not. On one hand, it has been seen before that the amount of states when the system has exactly $k$ elements is given by the binomial number $\binom{n^2}{k}$, resulting in the denominator of the above expression $\binom{n^2}{E(i)}$ in the case of not having an exact number of elements. On the other hand, in the numerator, we must account for how many of the existing states with that quantity of elements are terminal, that is, states that contain at least one path formed by elements that connect the opposing rows of the matrix. For now, such quantity will be represented with the function $p(n,k)$, since obtaining its expression may not be available, as we will see shortly.
\\\\
However, by employing the metric $E(i)$ for the previous probability, we are allowing both the numerator and the denominator to be evaluated with continuous values, which could create a discrepancy with respect to the desired result. Specifically, all occurrences of $E(i)$ in the probability expression actually denote a discrete number of elements, which allows for the generation of all possible states and the enumeration of the terminal ones. In this way, the use of the metric causes the amount of elements returned for two contiguous iterations not to reach a discrete value unless truncation operations are applied. That is, for the expected number of elements to increase by 1 unit, several iterations must elapse, especially when the occupation ratio is high, so in all of them, values that cannot be obtained in the percolation process are being considered, since at all times there is a discrete number of elements, although in certain situations it is necessary to estimate it continuously. Therefore, to avoid potential issues that may arise in this regard, we can rewrite the original probability with the metric $H(i,k)$, which, despite returning continuous values, affords the capability to compute the probability similarly to an expectation.

\begin{align}
    Pr[\mathbf{I}_n\leq i]=\sum _{j=0}^i \frac{p(n,j)}{\displaystyle\binom{n^2}{j}}H(i,j)
\end{align}
In this case, we start from the original expression by substituting the value of $E(i)$ with a fixed and discrete quantity $j$. Hence, given that the index of the iteration over which the probability is calculated does not vary, we can multiply the entire probability by the metric $H(i,j)$ evaluated at the index determined by the iteration and the number of elements $j$ that are assumed to exist in the system, resulting in the sum when the elements vary between 0 and the iteration index to the desired probability. Still, this alternative may present problems when $j>n^2$, causing its value to be indeterminate. Although,  due to the absence of an explicit expression for the function that quantifies the terminal states, it is not known if any term of the numerator would solve this problem by canceling the denominator's value resulting from the gamma function decomposition of the binomial number. Furthermore, if the function $p(n,j)$ returns 0 terminal states for invalid values of the number of elements, the solution would involve considering only the sum in the range $0\leq j\leq n^2$. In any case, with this approach, $I(n)$ can be specified as follows:

\begin{align}
     I(n)=\mathbb{E}[\mathbf{I}_n] =\sum _{i'=0}^\infty \left(Pr[\mathbf{I}_n\leq i']-Pr[\mathbf{I}_n\leq i'-1]\right)i'
\end{align}
The expectation of the random variable $\mathbf{I}_n$, which represents the function that determines the average length of a percolation process, emerges, from the previous approach, via the sum of the probabilities that a process ends exactly at iteration $i'$, each multiplied by its corresponding index. However, since the exact probability of such event is not available, it is formulated as the difference between the probability that a process has ended at a certain iteration and the same probability in the previous one, leading to:

\begin{align}
     I(n)= \sum _{i'=0}^\infty \left(\sum _{j=0}^{i'} \frac{p(n,j)}{\displaystyle\binom{n^2}{j}}H(i',j) - \sum _{j=0}^{i'-1} \frac{p(n,j)}{\displaystyle\binom{n^2}{j}}H(i'-1,j)\right)i'
\end{align}
Furthermore, to simplify the expression, we may also use the complement of the probability that the event corresponding to the termination of a process occurs, that is, the probability that the algorithm needs more than a certain number of iterations to terminate:
\begin{align}
     I(n)=\sum _{i'=0}^\infty Pr[\mathbf{I}_n> i'] = \sum _{i'=0}^\infty \left(1-\sum _{j=0}^{i'} \frac{p(n,j)}{\displaystyle\binom{n^2}{j}}H(i',j)\right)
\end{align}
As our random variable is constrained to positive integers, precluding any process from attaining a negative iteration count, the use of the cumulative distribution function instead of the probability of the exact event $\mathbf{I}_n=i'$ is equivalent. In conclusion, the only magnitude to be determined is the number of terminal states given by $p(n,k)$, which, unlike others such as the average cluster size, proves to be less tractable and straightforward. Later on, an expression for this function will be attempted in order to have $I(n)$ ready for use in the analysis of the average time of the percolation process. Moreover, when considering systems with an aspect ratio different from 1, especially edge cases such as matrices of dimension $(n,1)$, we can easily extract expressions to estimate the potential form of the case where the system's matrix is square. 

\subsection{Best case analysis}
\label{subsec:BestCaseAnalysis}
To begin the complete analysis of the algorithm, it is useful to consider a case that, regardless of the aspect ratio of the system's matrix, or even the dimension of the object used as system, will always result in the same time complexity. Such scenario is the best one, because the goal that ends the percolation process in the shortest possible time shares the same properties in systems composed of objects of different dimensions. For example, in matrices of sizes $(n,1)$, $(n,n)$, as well as in 3-dimensional systems $(n,x,y)$ formed by subsystems of matrices $(x,y)$, the shortest path between both zones of the system where these paths begin or end will be of length $n$. Therefore, with this information, we can ensure that the time complexity of this case can never drop under $\Theta(n)$, given that at least $n$ iterations will be needed to insert the same number of elements that form the minimum path, also assuming that in each iteration the workload to check if the system percolates is constant. However, to calculate the exact bound for this case, it is necessary to define the restrictions that a sequence of insertions must follow to result in the best-case complexity, that is, to generate a path of length $n$ within the system so that in each iteration the work of the depth-first search of the function $helper()$ is minimized.

\begin{align}
     T_{best}(n) = \frac{n}{2}+2\cdot\frac{n}{4}+6\cdot\frac{n}{8}+14\cdot\frac{n}{16}+30\cdot\frac{n}{32}+\cdots+n
\end{align}
At the outset, all position pairs of the insertion sequence will be different, indicating that no insertion has ever taken place on an already occupied cell, which would avert an increase in execution time and deviate from the best case. Thus, for each of the $n$ insertions, the sequence must select the position pairs in such a way as to minimize their adjacency during the process. In this way, considering the final path as a straight row of cells, at the start of the process, the first insertions will be guaranteed a constant time complexity with respect to the average cluster size, since only 1 call to $helper()$ will be made. To calculate how many calls possess this guarantee, it is enough to find the maximum number of elements that can be placed on the row where the final path will lie, with the condition that none of them are adjacent. As seen in the first term of the above formula, there are only $n/2$ elements, as by placing the first in one of the extreme cells of the row, the minimum distance that must separate the next inserted element is of one cell. Thus, placing 1 element in each even cell of the row and leaving the rest of the spaces free achieves this quantity. Post the first $n/2$ iterations, any inserted element will need to perform a depth-first search over a cluster of size 3, although the formula only counts the elements that are not the inserted one. Therefore, following the same philosophy, of the $n/2$ free spaces in the row, the next elements will be inserted alternately in these spaces so that if the row is traversed from one side to the other, the insertion will occur in the first found space and the next encountered space will be left free, ensuring a minimum cluster size in each insertion. Hence, of the original $n/2$ free cells, there are only $n/4$ left to occupy, continuing the same process until reaching the point where the last insertion does $n$ work, which indicates that it has had to check a cluster the same size as the final path.
\\\\
Consequently, the expression for the complexity $T_{best}(n)$ will contain a series of terms that correspond to the workload for each insertion sequence's segment with an equivalent cluster size. At first glance, it might be considered the largest growing term of $T_{best}(n)$ as the asymptotic growth of the best case, but, given the variable number of terms until reaching $n$, it is necessary to generate them all through a summation. For each one, we can decompose it into a function $g(x)$ that returns the number of iterations during which a traversal of the same size is performed, and another function $q(x)$ that computes the size of such cluster.

\begin{align}
     g(x)=\frac{n}{2^x}
\end{align}
The simplest to obtain is $g(x)$, as the first $n/2$ cells will traverse a cluster size of 1. If this amount were 0 during these iterations, they would not be counted as useful work of the algorithm, which is why in the next $n/4$, the inserted element in the cluster size is disregarded. Subsequently, after repeating the process and occupying half of the free $n/2$ cells, it repeats again in the following $n/8$, so it can be clearly observed how the function that determines this number of iterations for each cluster size is equivalent to the length of the final path divided by powers of 2. On the other hand, $q(x)$ is defined recursively from the pattern that forms the cluster size. That is, if in a segment of the insertion sequence there is a constant cluster size $g(\lambda-1)$, in the next segment the size will be $2(g(\lambda-1)+1)$, which is equivalent to the elements located on both sides of the inserted element, plus the inserted element itself, multiplied by the 2 sides towards which the traversal is performed.

\begin{align}
     q(2)=2,q(3)=6\quad q(x)=2(q(x-1)+1) \enspace \colon\enspace x>2
\end{align}
For simplicity, the sequence begins with the second term, where the cluster size is 2, that is, 1 element on each side of the inserted one, which itself is excluded. Consequently, by solving the recurrence relation, its expression as a function of the term's position within $T_{best}(n)$ is obtained:

\begin{align}
     q_h(x)=2q_h(x-1)\implies q_h(x)=A\cdot 2^x
\end{align}
\begin{align}
     q_p(x)=2(q_p(x)+1) \implies q_p(x)=-2
\end{align}

\begin{align}
     q(x)=q_h(x)+q_p(x) \implies q(x)=2(q_h(x-1)+q_p(x-1)+1)=2(A\cdot 2^{x-1}-1)
\end{align}

\begin{align}
\begin{cases}
    q(2)=2=2(A\cdot 2^{2-1}-1)\\
    q(3)=6=2(A\cdot 2^{3-1}-1)
\end{cases}\implies A=1 \implies \boxed{q(x)=2^x-2}
\end{align}

The sequence produced by $q(x)$ is also catalogued in \href{https://oeis.org/A095121}{A095121}, whose formula stems from solving its recurrence. Therefore, with the iteration count $g(x)$ and the cluster size $q(x)$ for each segment of the insertion sequence, we can construct the best case time complexity by adding its first term $n/2$ and the sum of the product of the previous functions.

\begin{align}
     T_{best}(n) = \frac{n}{2}+\sum_{x=2}^{\lambda} q(x)g(x) = \frac{n}{2}+\sum_{x=2}^{\lambda} \frac{(2^x-2)n}{2^x}
\end{align}
For now, the upper limit of the sum will be left as a function of $\lambda$, so that, once an expression for the full complexity has been achieved, it becomes easier to operate.

\begin{align}
     T_{best}(n) &= \frac{n}{2}+\sum_{x=2}^{\lambda} \frac{(2^x-2)n}{2^x}=\frac{n}{2}+n\sum_{x=2}^{\lambda} \frac{2^x-2}{2^x}=\\\notag
     &=\frac{n}{2}+n\left(\sum_{x=2}^{\lambda} 1 - \sum_{x=2}^{\lambda} +2^{1-x}\right)=\\\notag
     &=\frac{n}{2}+n\left((\lambda-1) - 2\sum_{x=2}^{\lambda} (\frac{1}{2})^x\right)=\frac{n}{2}+n\left((\lambda-1) - 1 + \frac{1}{2^{\lambda-1}}\right)= n \left(\lambda+2^{1-\lambda}-\frac{3}{2}\right)
\end{align}
Then, to find the number of segments $\lambda$ into which the insertion sequence should be partitioned for the process to reach a terminal state, we start from the premise that the minimum path has length $n$, and that in the last insertion the cluster size traversed is at least half of its length. Thus, we can set up an inequality to obtain the index of the last term of the sum in the expression $T_{best}(n)$, which is equivalent to the number of halving operations required for the sequence to reach unity, reflecting this idea in its outcome:

\begin{align}
     q(x)\leq \frac{n}{2} \quad 2^x-2\leq \frac{n}{2} \quad 2^x\leq \frac{n}{2}+2 \quad x\leq \log_2(\frac{n}{2}+2) \enspace\colon\enspace n>-4
\end{align}
Disregarding the constants and multipliers, asymptotically there are $\log_2(n)$ terms in each sum of the runtime complexity, which makes sense, since the insertion sequence of length $n$ is halved until it reaches one \cite{Knuth1997}, at which point the cluster size is equal to at least half the length of the minimum path.

\begin{align}
     q(x)\leq n \quad 2^x-2\leq n \quad 2^x\leq n+2 \quad x\leq \log_2(n+2) \enspace\colon\enspace n>-2
\end{align}
However, posing the inequality such that the cluster size attains the full path length yields an asymptotically equivalent outcome:

\begin{align}
     \lim_{n\to\infty} \frac{\log_2(\frac{n}{2}+2)}{\log_2(n+2)}=\lim_{n\to\infty} \frac{\log_2(1+\frac{4}{n})+\log_2(\frac{n}{2})}{\log_2(1+\frac{2}{n})+\log_2(n)}=\lim_{n\to\infty} \frac{\log_2(\frac{n}{2})}{\log_2(n)}=\lim_{n\to\infty} \frac{\frac{1}{n}}{\frac{1}{n}}=1
\end{align}
Hence, knowing that the number of terms $\lambda$ in the sum is of the form $\log_2(n)$, we arrive at the final time complexity for the best case:

\begin{align}
     T_{best}(n) = n \left(\log_2(n)+2^{1-\log_2(n)}-\frac{3}{2}\right) = n \left(\log_2(n)+\frac{2}{n}-\frac{3}{2}\right) \implies \boxed{T_{best}(n) = O(n\log(n))}
\end{align}
Additionally, it is convenient to verify that the asymptotic growth of the number of terms does not influence the length of the final path. To this end, the first $\lambda$ terms are summed, in which only the count of elements inserted in each segment of the insertion sequence is found. Subsequently, it is checked whether the outcome is consistent with the length of the minimum path.

\begin{align}
     \sum_{x=1}^{\lambda} g(x) = \sum_{x=1}^{\lambda} \frac{n}{2^x}=n\sum_{x=1}^{\lambda} \left(\frac{1}{2}\right)^x=n\left(1-\left(\frac{1}{2}\right)^\lambda\right)
\end{align}
Thus, replacing $\lambda$ with the asymptotic growth of the number of terms $\log_2(n)$, we reach:

\begin{align}
     \sum_{x=1}^{\log_2(n)} g(x) = n\left(1-\left(\frac{1}{2}\right)^{\log_2(n)}\right)=n-1
\end{align}
Despite not returning the exact length of the shortest path, which accounts for all insertions effected during the process, the result grows in the same manner as $n$, so it could be interpreted that the remaining unit corresponds to the last element of the insertion sequence, which may not be counted. Similarly, it can also be due to the approach's disregard for the particular cases determined by the parity of $n$, as asymptotically the growth would remain the same in all of them.

\begin{align}
     c(n,k)=2^{\displaystyle\left\lfloor \log _2\left(\frac{2 n^2}{n^2-k}\right)\right\rfloor }-2
\end{align}
Ultimately, from $q(x)$, we can deduce the average cluster size during the best case of the algorithm, as detailed below:

\begin{figure}[H]
    \centering
    \includegraphics[width=8.5cm,clip]{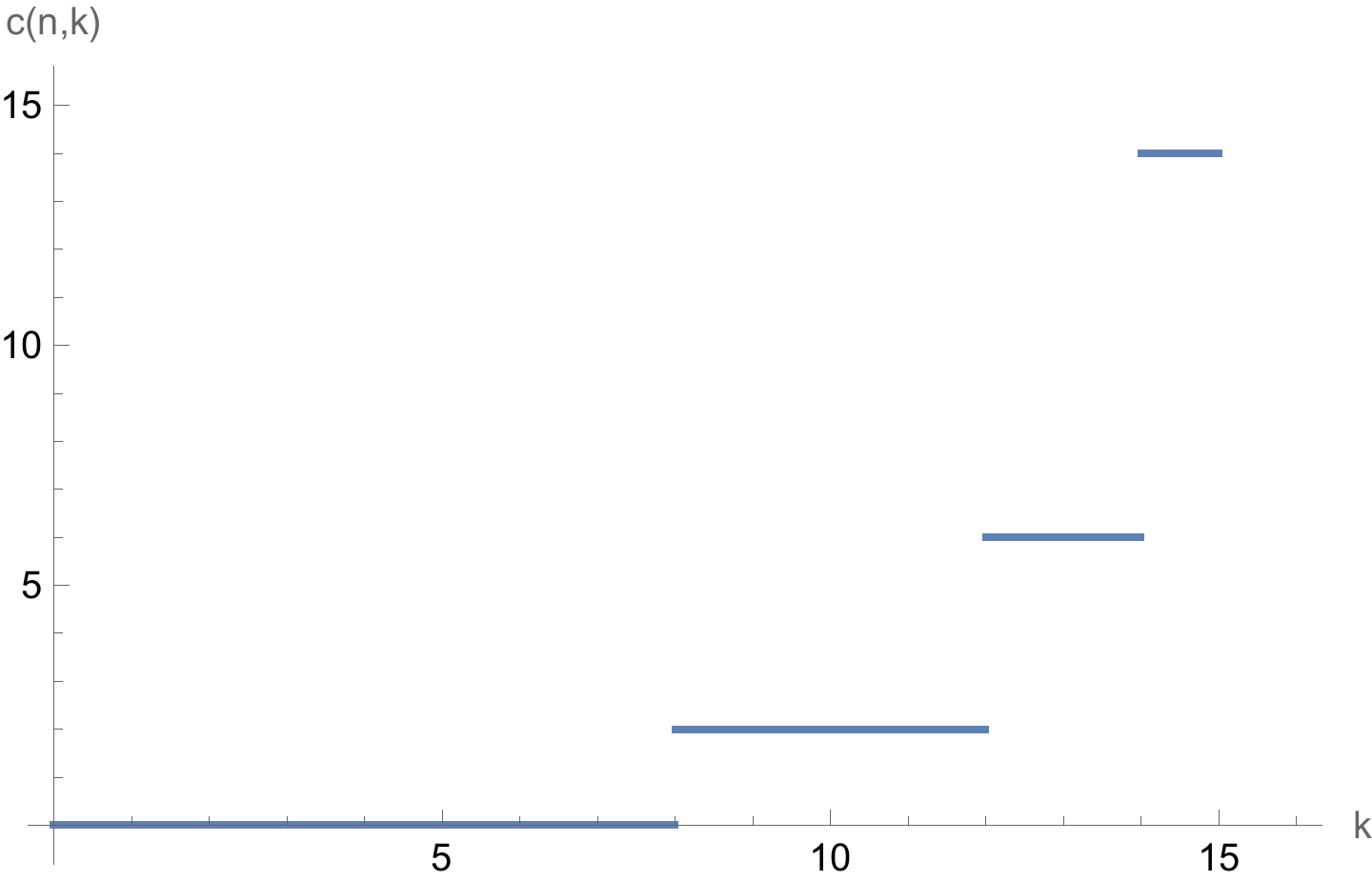}
    \caption{Best case $c(n,k)$ for $n=16$}
    \label{fig:BestCaseCluster}
\end{figure}
As seen, the initial iterations during which the number of items inserted is equivalent to half of the minimum path length have a null cluster size. Although in $T_{best}()$ this is counted as 1, it results in the same asymptotic growth. Hence, the sole difference lies in the accuracy with which execution time is measured. Subsequently, the next half of the previous iterations correspond to a size $q(2)=2$, and so on until the completion of the solution path. In summary, the best case of the algorithm has a time complexity of $O(n\log(n))$, applicable across systems of different dimensions, as will be seen next. This is because the minimum path and the properties an insertion sequence must present to result in this case are independent of the overall system size. Finally, it is worth noting that the probability of the algorithm empirically falling into this case is remarkably low, as achieving an insertion sequence without insertions over already occupied cells is rare, unless the system is large enough.

\section{1-dimensional case analysis}
\label{1-dimensional-case-analysis}
Regarding the dimensions a percolative system may present, in this section, we will begin with a specific one-dimensional case that will simplify the analysis process, allowing for the verification of each result's coherence in higher dimensions. Thus, by completely solving a basic case in terms of dimensionality, it becomes possible to discern the validity of the approximations required for magnitudes that are complicated to evaluate exactly, which will be crucial in those higher dimensions. Consequently, we initially proceed with the analysis of systems with matrices of size $(n,1)$, which are likewise referred to as 1-dimensional systems \cite{Bureeva2010}.

\begin{figure}[H]
    \centering
    \includegraphics[width=4.7cm,clip]{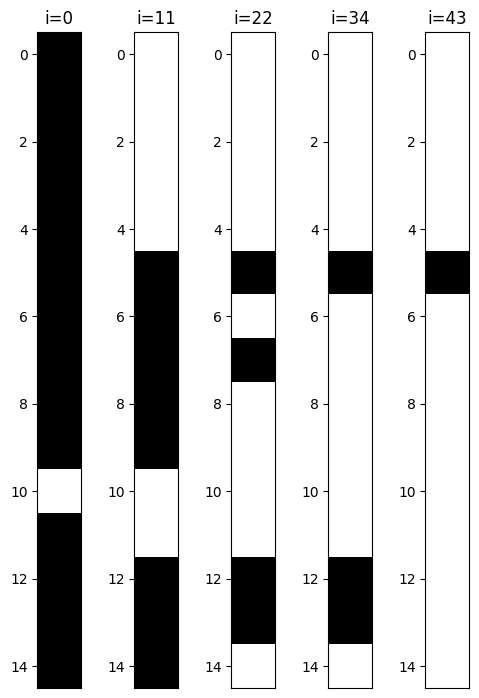}
    \caption{Intermediate states of a percolative process undertaken on a 1D system of size $n=100$}
    \label{fig:1D}
\end{figure}
Above are shown some intermediate states of a 1-dimensional system during the execution of a process \cite{MaltheSorenssen2024}. It is the same mechanism as in 2 dimensions, with the difference that the only existing path between the opposite elements of the system is of length $n$. There is also the possibility of considering systems of size $(1,n)$, although the process would always end in the first iteration since in that case all paths would have a constant length, which is not useful for the analysis. Hence, initially, it could be hypothesized that the average complexity of a one-dimensional system coincides with the best case of a two-dimensional one where a path of length $n$ is formed, although perhaps without the restriction that an insertion occurs in every iteration. To check the extent to which this is true, the approaches from preceding sections will be used to obtain expressions that calculate the execution time of the algorithm on such systems in different situations, both for the average case and edge cases.
\\\\
But, before commencing, it is pertinent to mention the similarity of this process on one-dimensional systems with the coupon collector's problem, as it is a widely researched field that offers a substantial amount of results with which we can guarantee the correctness of the following analysis. Specifically, if we consider the size of the one-dimensional system as the target number of coupons to be collected during a process, and each iteration as the extraction/obtaining of a coupon, a bijection is established between our algorithm and the problem, so all the metrics already calculated will serve to verify if they coincide with the expressions derived in the analysis of our algorithm from the previous approaches, thus demonstrating the validity of the methodology proposed, at least for this dimension.

\subsection{Average cluster size}
The initial metric for commencing the analysis is the average cluster size, key for quantifying the work of each insertion in a process. In one dimension, given the simplicity of the system and its state space, it is possible to build an exact expression for such magnitude, although the analysis will also be conducted with approximations calculated from the probabilistic estimators in section \textcolor{blue}{\ref{AverageClusterSizeEstimation}} to check the fidelity of these approximations to the actual magnitude.

\subsubsection{Probabilistic approximation}
Regarding the estimators for $c(n,k)$, they can be inferred from the sum of probabilities of the scalar field $(x,0,p_{k,n})$ mentioned earlier, although in this case not in its entire domain, but rather only in the values that represent the depth traversal of a one-dimensional system.

\begin{align}
   \hat{c}_0(n,k)= \int_{-\infty}^{\infty} f(x,0,p_{k,n}) \, dx = \int_{-\infty}^{\infty} f(0,y,p_{k,n}) \, dy = \int_{-\infty}^{\infty} p_{k,n}^{3|x|} \, dx
\end{align}
For example, in the continuous estimator, the integration is performed with respect to a coordinate axis that coincides with one of the symmetry axes of the field. In this way, the sum of the probabilities that a traversal reaches each point in the domain of the integrated line, starting at $x=0$, is computed. Apart from these axes, integration could be effected with respect to the diagonals $y=\pm x$, although they yield higher probabilities due to contributions from the diagonal neighboring cells of Moore's neighborhood, so they are discarded.

\begin{align}
   \hat{c}_0(n,k) = 2\int_{0}^{\infty} p_{k,n}^{3x} \, dx = \left . \frac{2\cdot p_{k,n}^{3x}}{3\cdot ln(p_{k,n})} \right|_{x=0}^{x=\infty} = -\frac{2}{3\cdot ln(p_{k,n})}
\end{align}
After simplifying the integration by splitting the sum into two symmetric regions, an estimator is obtained to which a correction must be applied in the range of its image for it to meet the constraints of $c(n,k)$, that is, normalize its result as was previously done with the estimators in the two-dimensional case.
\begin{align}
   \hat{c}_0(n,k) = n\left(\frac{\displaystyle-\frac{2}{3\cdot ln(p_{k,n})}}{\displaystyle -\frac{2}{3\cdot ln(p_{k,n})}+n}\right) = \frac{2\cdot n}{2-3\cdot n\cdot ln(p_{k,n})}
\end{align}
Likewise, in addition to $\hat{c}_0(n,k)$, it is convenient to compute the discrete sum, so it is posed along the same axis of symmetry:
\begin{align}
   \hat{c}_1(n,k) = \sum _{x=-\infty}^{\infty} f(x,0,p_{k,n}) = \sum _{y=-\infty}^{\infty} f(0,y,p_{k,n}) 
\end{align}
Here, to facilitate its resolution, it is necessary to add the point $x=0$ independently from the rest, allowing the initial operation to be decomposed into simpler geometric sums \cite{Christensen2002}:
\begin{align}
   \hat{c}_1(n,k) = 1+2\sum _{x=1}^{\infty} p_{k,n}^{3 x} = 1+2\left(\frac{1}{1-p_{k,n}^3}-1\right) = \frac{1+p_{k,n}^3}{1-p_{k,n}^3}
\end{align}
And, Analogous to the continuous case, the same transformation is applied to normalize the range of its image, resulting in the ultimate expression for the discrete estimator:

\begin{align}
   \hat{c}_1(n,k) = n\left(\frac{\displaystyle\frac{1+p_{k,n}^3}{1-p_{k,n}^3}}{\displaystyle \frac{1+p_{k,n}^3}{1-p_{k,n}^3}+n}\right) = \frac{n \left(p_{k,n}^3+1\right)}{-n\cdot p_{k,n}^3+n+p_{k,n}^3+1}
\end{align}

\subsubsection{Actual cluster size}
\label{subsubsec:ActualClusterSize1D}
In a system of size $(n,1)$, it becomes apparent that its state space when it contains $0 \leq k \leq n$ elements is considerably smaller than in other situations. Consequently, calculating an exact form for $c(n,k)$ is not as complex as in systems of 2 or more dimensions. In this way, we can attempt to obtain this expression through the fundamental ratio of an average, that is, by considering the ratio between the sum of the sizes of all clusters in the system states and the number of clusters present in those states.

\begin{align}    
    c(n,k)=\frac{s(n,k)}{N(n,k)}
\end{align}
If we denote $s(n,k)$ as the sum of the cluster sizes and $N(n,k)$ as the total count of clusters in all system states with $k$ elements, their ratio will correspond to the average cluster size $c(n,k)$. Specifically, the magnitudes of both functions span the entire state space with a certain number of elements as constraint. That is, for each state, there exists a number of clusters present in the system, which are counted together with the quantities of the remaining states in $N(n,k)$. In turn, each of these clusters has a size, so the sum of all of them in a specific state, together with the same sum carried out in the rest of states, is accounted for in the global $s(n,k)$.
\begin{align}    
    s(n,k)=k\binom{n}{k}
\end{align}
In total, there exist $\binom{n}{k}$ states in a one-dimensional system with $k$ elements, and since we know that each state has exactly that amount of elements, although we do not know a priori how many clusters they form, it can be inferred that the sum of all their sizes is equivalent to $k$. Ergo, by accounting for $k$ units of cluster size for each of the $\binom{n}{k}$ states, we arrive at the formula for $s(n,k)$. On the other hand, the total amount of clusters in these states must be counted, and as this is a non-trivial operation, expressions will first be computed for specific cases where the number of elements is small. For example, for $k=1$ there are only $n$ clusters of size 1, so dividing by $\binom{n}{1}=n$ results in an average size of 1, which represents one of the simplest situations to consider for this magnitude. The problem for the generalization of $N(c,k)$ lies in larger quantities of elements, increasing the state space and therefore also the combinations of clusters therein, that is, the partitions into which an amount $k$ is dividable to form several cluster sizes to be accounted for, and the number of them located in boundary situations such as the edges of the system, which encumbers the enumeration.
\begin{align}    
    c(n,1)&=1\\ 
    c(n,2)&=\frac{\displaystyle2 \binom{n}{2}}{\displaystyle2 \left(\binom{n}{2}-(n-1)\right)+(n-1)}\\
    c(n,3)&=\frac{\displaystyle3 \binom{n}{3}}{\displaystyle3 \left(\binom{n}{3}-(n-2)-((n-4) (n-3)+2 (n-3))\right)+(n-2)+2 ((n-4) (n-3)+2 (n-3))}
\end{align}

\begin{align}
    &c(n,4)=4 \binom{n}{4}\left( 2 \left(\binom{n-3}{2}+(n-5) (n-4)+2 (n-4)\right)+\frac{4}{24} (n-6) (n-5) (n-4) (n-3)+(n-3)+ \right. \notag\\& 
    \left. 3 \left(-(\binom{n-3}{2}+(n-5) (n-4)+2 (n-4))+\binom{n}{4}-\frac{1}{24} (n-6) (n-5) (n-4) (n-3)-(n-3)\right)\right)^{-1}
\end{align}
Nonetheless, for the first four quantities of elements, we compute the expression with each of its partitions. Thus, when $k=2$, the number of clusters of size 2 depends on the shifts that can be effectuated within the system $(n-1)$, which, when added to the rest of the combinations of the pair of clusters of size 1, produces $c(n,2)$. With $k=3$, we would also need to account for the situation in which there exist a cluster of size 2 and another of size 1, although the process is analogous, applied to the next expression where $k=4$, with diverse cases like 3 clusters of sizes 2-1-1, or 2 clusters of size 2. Accordingly, once the complete expressions for $c(n,k)$ have been deduced in the previous cases, there is the possibility of using some Wolfram primitive to find a generating function that generalizes in terms of $k$ and obtains an expression for all valid values of system size and number of elements. However, the lack of information prevents these methods from working, and even searching for a pattern at first glance is complicated, so the formulas are simplified.

\begin{align}    
    c(n,1)&=\frac{n}{n-1+1}\\
    c(n,2)&=\frac{n}{n-2+1}\\
    c(n,3)&=\frac{n}{n-3+1}\\
    c(n,4)&=\frac{n}{n-4+1}\\\notag
    &\vdots
\end{align}
After expanding all the polynomials and transforming the quantities derived from binomial numbers into Gamma functions, every expression can be rewritten in its simplest form as shown above. Through this step, we can infer a possible generalization that serves to model $c(n,k)$ exactly. A priori, it cannot be assured that the inferred expression works for all required parameter values, yet, by proposing the calculation of the number of clusters in all states recursively, its validity is demonstrated \cite{Chaunier2024}. In summary, the proof consists of considering a system of arbitrary size $n$ and analyzing what happens with the number of clusters when one of the cells at the ends is removed/added. Since it is a one-dimensional system, such analysis is simple enough for the proposed recursion to yield a solution, which can be substituted in place of $N(n, k)$ in the previous definition based on the ratio between the size and count of clusters, resulting in the concluding expression for $c(n, k)$.

\begin{align}
    c(n,k)=\frac{n}{n-k+1}
\end{align}

\begin{figure}[H]
    \centering
    \includegraphics[width=10cm,clip]{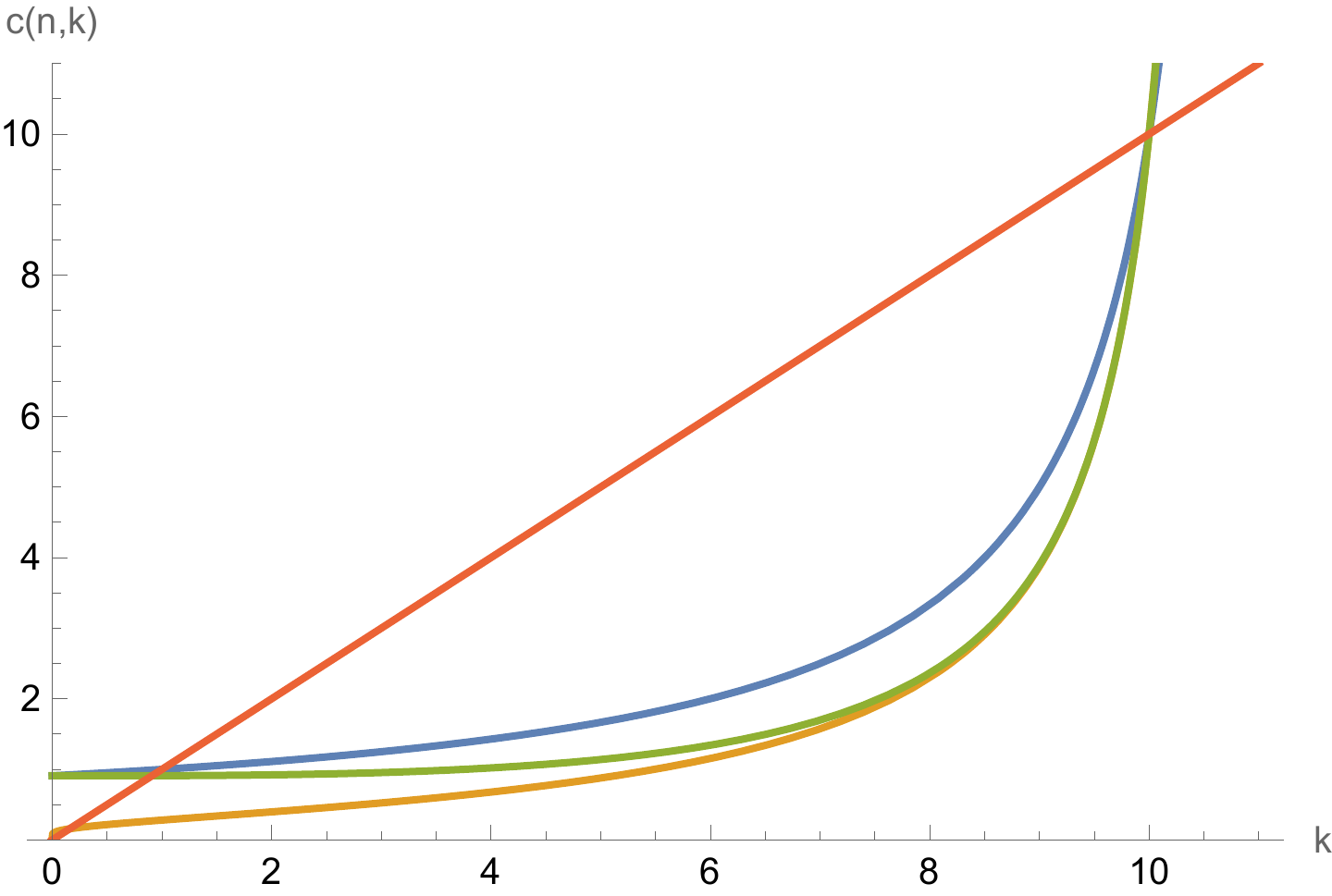}
    \caption{Actual $c(n,k)$ function plotted in blue, its estimators $\hat{c}_0(n,k)$ and $\hat{c}_1(n,k)$ displayed in orange and green respectively, and the edge case $c(n,k)=k$ in red, for a 1D system of size $n=10$}
    \label{fig:Clusters1D}
\end{figure}
Finally, it is worth highlighting the congruity between the actual metric $c(n,k)$ for one-dimensional systems and the one previously obtained for square matrices from the fundamental constraints of the magnitude it models, both graphically and algebraically. Likewise, regarding its estimators, it is noticeable in Figure 35 how the discrete estimator fits notably better to $c(n,k)$ compared to its continuous variant. This is mainly due to the difference in low values of $k$, as in the rest both estimators are asymptotically equivalent. Consequently, for the purposes of average case analysis, the discretionary selection should favor $\hat{c}_1(n,k)$.

\subsection{Expected iterations for process termination}
For this analysis, we are not only required to know the clusters of elements formed in each system state, but we also need to ascertain when the process will end, that is, the count of iterations during which the insertions occur. Hence, through the probability distribution expressions for $\mathbf{I}_n$ and the exact number of terminal states for each number of elements in the system, which in this case is trivial due to its dimensionality, we can deduce $I(n)$ for processes in 1D systems.

\begin{align}
    Pr[\mathbf{I}_n\leq i]=\sum _{j=0}^i \frac{p(n,j)}{\displaystyle\binom{n^2}{j}}H(i,j) = \sum _{j=0}^i \frac{\displaystyle\delta _{j,n}}{\displaystyle\binom{n}{j}} \frac{n!}{n^i (n-j)!} \stirling{i}{j}
\end{align}
Initially, the cumulative distribution function is defined from the previously constructed metric $H(n,k)$ and $p(n,k)$, which returns the number of terminal states. With respect to the latter, it can be readily deduced that among all doable quantities of elements, the only one that manages to generate the existing terminal state in 1D systems is $k=n$, since the sole possibility of forming a path that connects both system ends is when it is entirely occupied. Then, for any quantity of elements, the function $p(n,k)$ must return 0, except when $k=n$, in which case it must account for the unique terminal state. Therefore, the most forthright approach to model this sequence is with the Kronecker delta function, as illustrated above.

\begin{align}
    Pr[\mathbf{I}_n\leq i] = \frac{1}{\displaystyle\binom{n}{n}} \frac{n!}{n^i (n-n)!} \stirling{i}{n} = \frac{n!}{n^i} \stirling{i}{n}
\end{align}
\begin{figure}[H]
    \centering
    \includegraphics[width=10cm,clip]{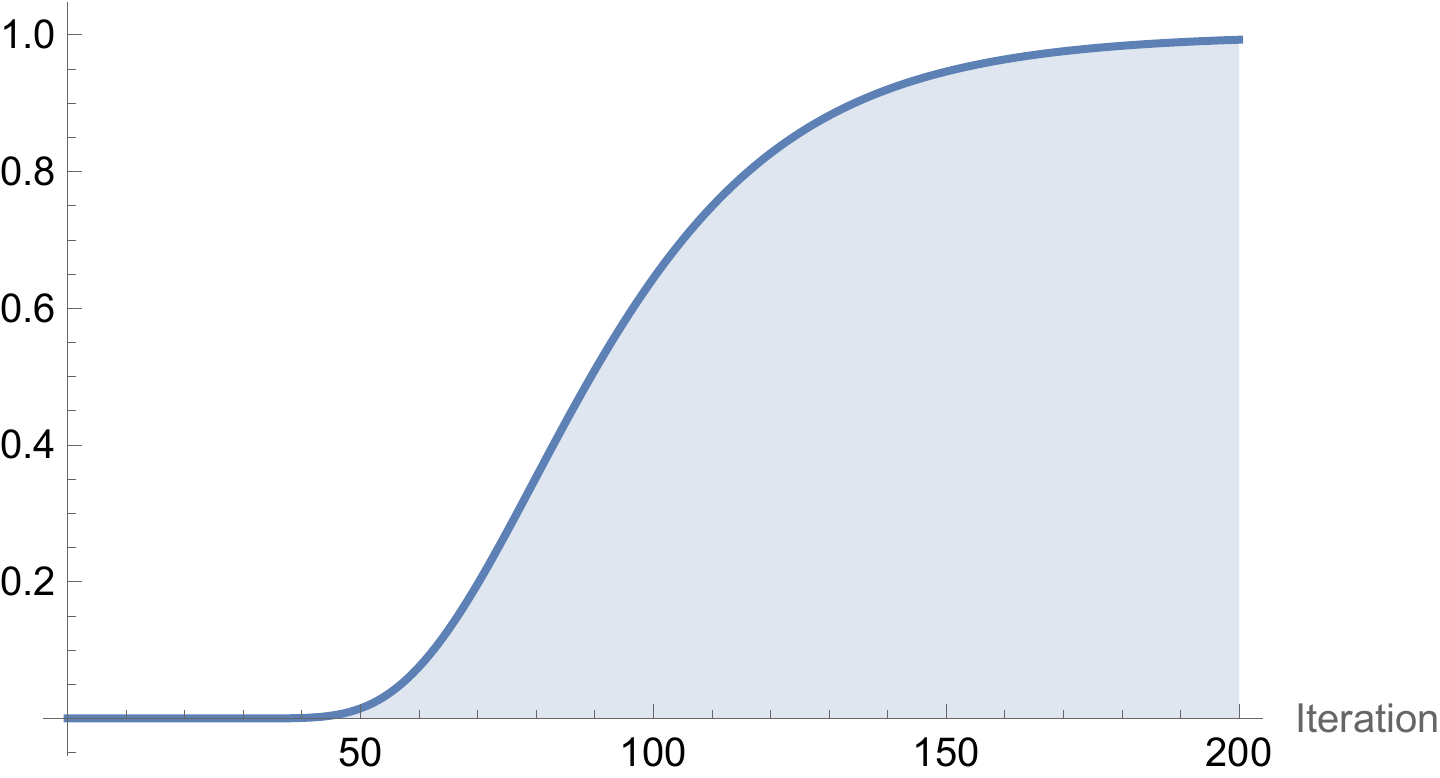}
    \caption{$Pr[\mathbf{I}_n\leq i]$ plotted for a 1D system of size $n=25$}
    \label{fig:1DCDF}
\end{figure}
After determining the probability distribution that the random variable $\mathbf{I}_n$ adheres to, it is possible to deduce the probability that this variable equals a specific iteration, leading to its density function:

\begin{align}
    Pr[\mathbf{I}_n= i] = Pr[\mathbf{I}_n\leq i]-Pr[\mathbf{I}_n\leq i-1] = \frac{n!}{n^i} \stirling{i}{n}-\frac{n!}{n^{i-1}} \stirling{i-1}{n} = \frac{n!}{n^i} \stirling{i-1}{n-1}
\end{align}
\begin{figure}[H]
    \centering
    \includegraphics[width=10cm,clip]{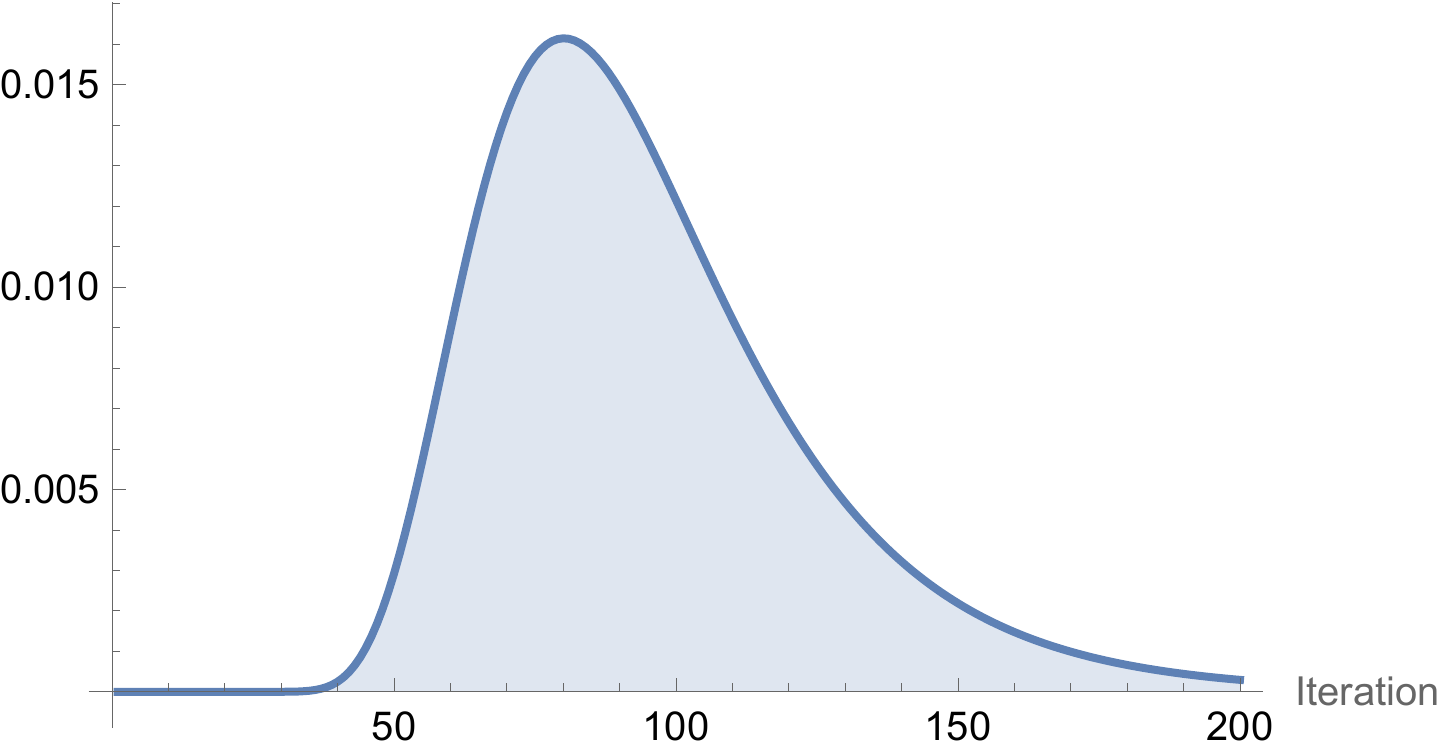}
    \caption{$Pr[\mathbf{I}_n= i]$ density function plotted for a 1D system of size $n=25$}
    \label{fig:1DPDF}
\end{figure}
The functions shown above determine the probability distribution of the variable that models the iterations required for a percolative process in 1 dimension to terminate, both its cumulative distribution function and the density one. Though, to experimentally verify its validity, its shape is contrasted with the histogram of iterations from the simulations dataset built previously, so that the comparison elucidates the potential disparity between the theoretically generated distribution and the one resulting from a considerable number of simulations.

\begin{figure}[H]
    \centering
    \includegraphics[width=10cm,clip]{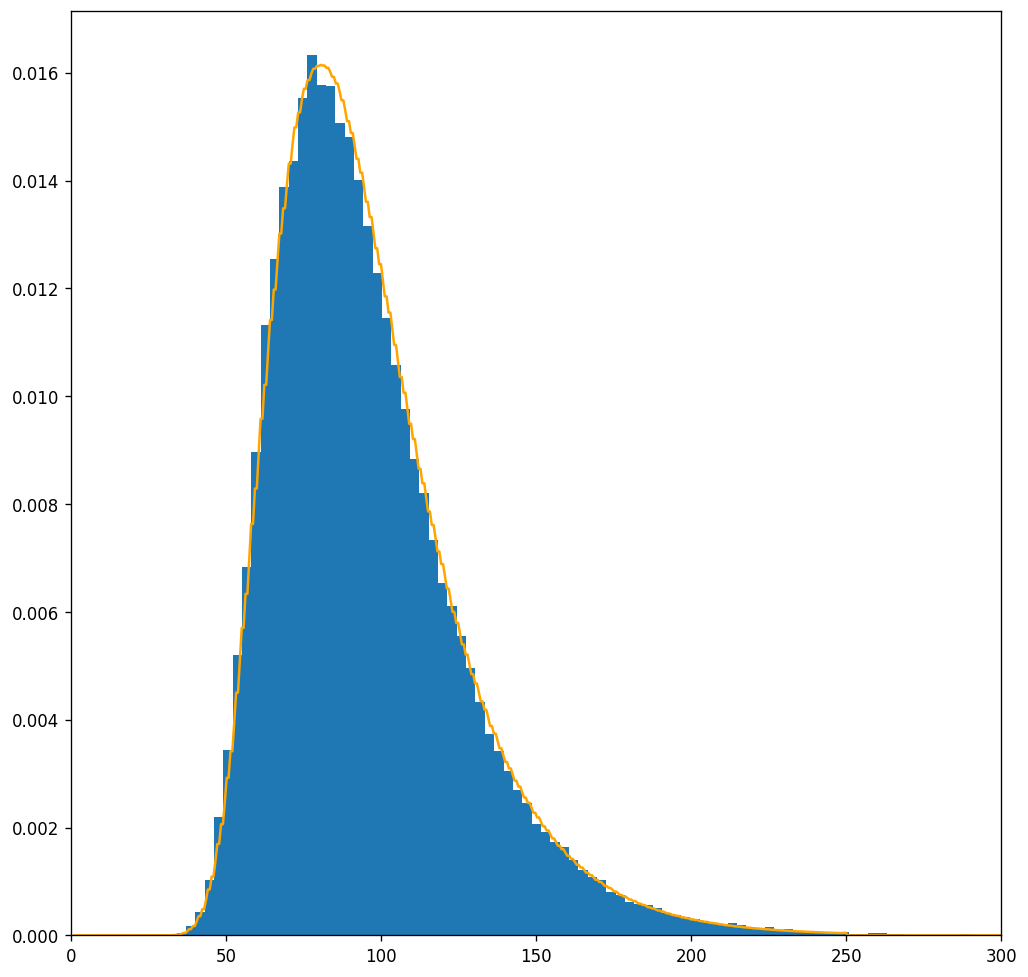}
    \caption{$Pr[\mathbf{I}_n= i]$ density function plotted for a 1D system of size $n=25$ in orange and the histogram of iterations from simulations dataset depicted in blue.}
    \label{fig:1DIterationsDensity}
\end{figure}
For the case of a medium-sized system, it is evident that the distribution adequately characterizes the empirical results of the histogram formed by data from 100,000 simulations, which indicates that theoretically the resulting distribution will not present any issues regarding the correctness of the following metrics. So, with the variable $\mathbf{I}_n$ completely defined by its distribution, we can proceed to calculate its expectation, which will allow us in the average case to know how many iterations a 1D process needs on average to reach its terminal state.
\begin{align}
     I(n)=\sum _{i=0}^\infty Pr[\mathbf{I}_n> i] = \sum _{i=0}^\infty \left(1-Pr[\mathbf{I}_n\leq i]\right) = \sum _{i=0}^\infty \left(1-\frac{n!}{n^{i}} \stirling{i}{n}\right)
\end{align}
Of all the existing methods, the simplest involves using the complement of the cumulative distribution function, so after solving the sum, we arrive at the following \cite{Joriki2016,Pandey2019}:
\begin{align}
     I(n)=n\cdot H_n
\end{align}
Additionally, if we denote $I(n,k)$ as the average number of iterations required for the process to insert $k$ elements, we can leverage a result from the coupon collector's problem to obtain its exact expression \cite{Riedel2017}:

\begin{align}
     I(n,k)=n(H_n-H_{n-k})
\end{align}
\begin{figure}[H]
    \centering
    \includegraphics[width=10cm,clip]{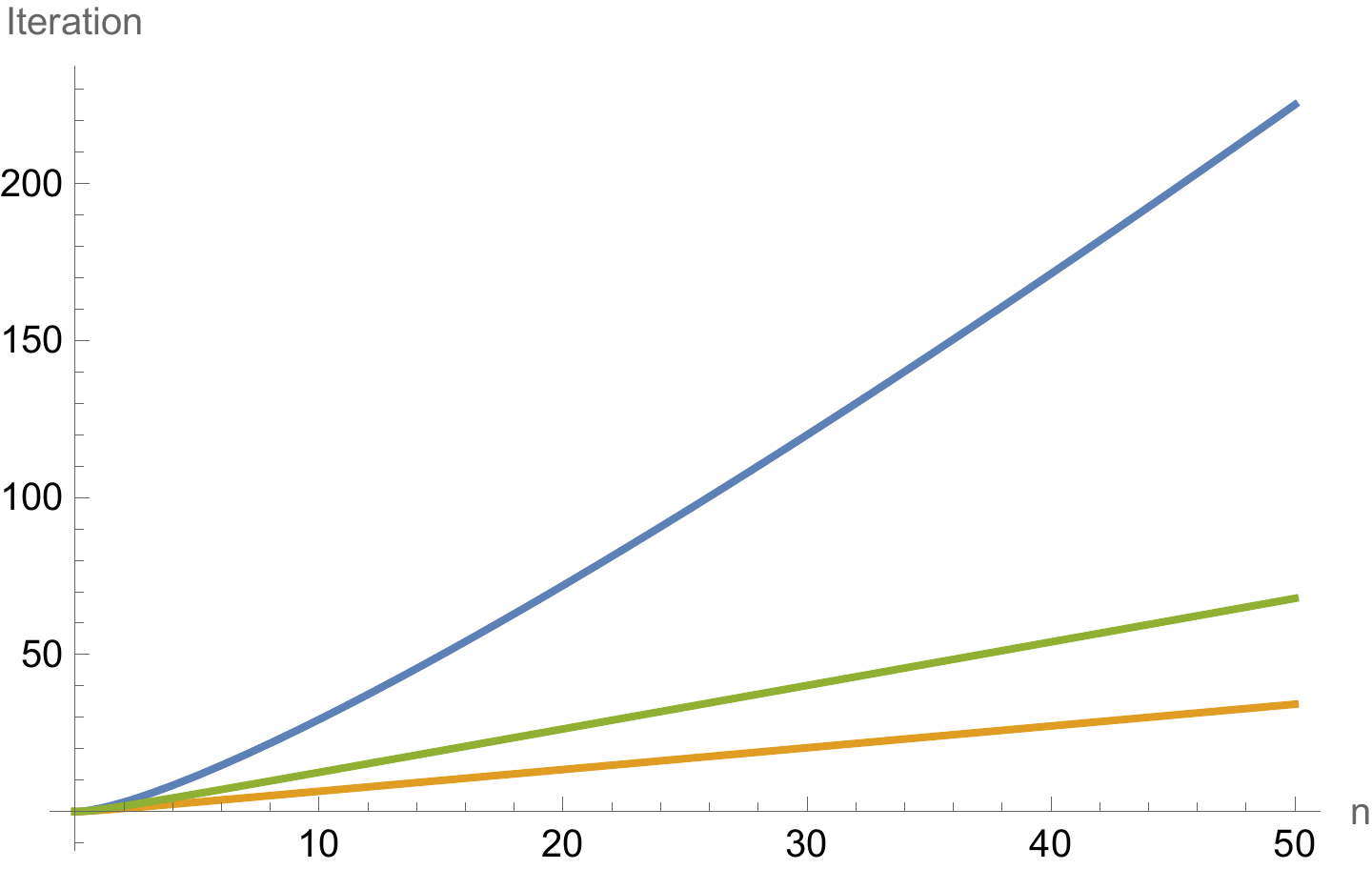}
    \caption{$I(n,n)$ plotted in blue, $I(n,n/2)$ in orange and $I(n,\frac{3}{4}n)$ in green.}
    \label{fig:1DIterations}
\end{figure}

Lastly, in the analysis, $I(n)$ or $I(n,k)$ will be employed interchangeably, depending on the conditions imposed on the process. However, it should be noted that for the average time complexity, the entire system must be filled with elements, so $I(n)=I(n,n)$ will hold. Additionally, regarding the formalization of all cases reachable by the process if it could finish before filling the system, above $I(n,k)$ is shown for different quantities of elements. With that, it is possible to represent situations where, on average, the process ends before achieving $n$ effective insertions, which in one dimension is not useful but in higher dimensions is key for addressing cases that differ from the average, both towards the best-case and the worst-case in execution time \cite{Kan2005}.
\subsection{Worst case analysis}
Regarding the latter, by using the aforementioned metrics for the average cluster size and the average iteration count determining the duration of the process, its time complexity is stated as follows:

\begin{align}
    T_{worst}(n) = \sum _{i=0}^{I(n,n)} \left(1-\frac{E(i)}{n}\right)\cdot c(n,k)=\sum _{i=0}^{I(n)} \left(1-\frac{E(i)}{n}\right)\cdot E(i)
\end{align}
Essentially, the worst-case runtime is the cumulative workload performed across all iterations. In each of them, the work is proportional to the cluster size indicated by $c(n,k)$. However, under worst-case conditions, it is assumed that the cluster always has the same size as the number of elements inserted up to that iteration, resulting in an average size of $E(i)$ units. Additionally, the probability of an insertion occurring must be included, which in this case is equivalent to the probability that an insertion encounters an unoccupied cell. To this end, the work of each insertion is multiplied by the complement of the occupancy ratio calculated from the metric $E(i)$, as $H(i,k)$ alone does not serve as an estimator of the element count at a given moment.
\begin{align}
    E(i)=n\left(1-\left(1-\frac{1}{n}\right)^i\right)
\end{align}
For this analysis, both in the worst and in other cases \cite{KrityakierneThanatipanonda2023,AlistarhDavies2021}, the $E(i)$ metric must be modified so that the maximum number of elements is restricted to $n$, instead of $n^2$ as it used to be in 2-dimensional systems. Consequently, it is sufficient to replace all corresponding $n^2$ terms with the maximum count of cells in a 1D system, which is $n$.

\begin{align}
    T_{worst}(n) = \int _{0}^{I(n)} \left(1-\frac{E(i)}{n}\right)\cdot E(i) \, di
\end{align}
For simplicity, we will first calculate the sum of the complexity $T_{worst}(n)$ in its integral form, with the intention of subsequently verifying its equivalence to the discrete sum. This is justified by the observation that, although both quantities are different up to a certain $n$, when $n$ grows, their asymptotic behavior converges, so it is indifferent to sum it in a continuous or discrete manner. Nonetheless, both will be performed just in case they differ under specific conditions, both in this analysis and in higher dimensions.
\begin{align}
    \int \left(1-\frac{E(i)}{n}\right)\cdot E(i) \, di &= \int E(i) \, di - \frac{1}{n}\int E(i)^2 \, di=\\ \notag
    &= \int n\left(1-\left(1-\frac{1}{n}\right)^i\right) \, di - \frac{1}{n}\int n^2\left(1-\left(1-\frac{1}{n}\right)^i\right)^2 \, di =\\ \notag
    &=n\left(i-\int \left(1-\frac{1}{n}\right)^i \, di\right) - n\int \left(1-\left(1-\frac{1}{n}\right)^i\right)^2 \, di =\\ \notag
    &=n\left(i-\frac{\displaystyle\left(1-\frac{1}{n}\right)^i}{\displaystyle ln\left(1-\frac{1}{n}\right)}\right) - n\int \left(1-\left(1-\frac{1}{n}\right)^i\right)^2 \, di=\\\notag
    &=n\left(i-\frac{\displaystyle\left(1-\frac{1}{n}\right)^i}{\displaystyle ln\left(1-\frac{1}{n}\right)}\right) - n\int 1 -2\left(1-\frac{1}{n}\right)^i +\left(1-\frac{1}{n}\right)^{2i} \, di =\notag \\ \notag
    &=n\left(i-\frac{\displaystyle\left(1-\frac{1}{n}\right)^i}{\displaystyle ln\left(1-\frac{1}{n}\right)}\right) - n\left(i - \frac{\displaystyle 2\left(1-\frac{1}{n}\right)^i}{\displaystyle ln\left(1-\frac{1}{n}\right)} + \frac{\displaystyle\left(1-\frac{1}{n}\right)^{2i}}{\displaystyle 2\cdot ln\left(1-\frac{1}{n}\right)}\right) =\\ \notag
    &=n\left(-\frac{\displaystyle\left(1-\frac{1}{n}\right)^i}{\displaystyle ln\left(1-\frac{1}{n}\right)} + \frac{\displaystyle 2\left(1-\frac{1}{n}\right)^i}{\displaystyle ln\left(1-\frac{1}{n}\right)} - \frac{\displaystyle\left(1-\frac{1}{n}\right)^{2i}}{\displaystyle 2\cdot ln\left(1-\frac{1}{n}\right)}\right)=
\end{align}
\begin{align}    
    =-\frac{(n-1)^i n^{1-2 i} ((n-1)^i-2 n^i)}{2 (ln(n-1)-ln(n))}+C \notag
\end{align}
After determining its antiderivative, it is substituted into the integral expression of $T_{worst}(n)$ and evaluated at its corresponding limits:

\begin{align}
    T_{worst}(n) = \left . -\frac{(n-1)^i n^{1-2 i} ((n-1)^i-2 n^i)}{2 (ln(n-1)-ln(n))} \right|_{i=0}^{i=I(n)} = -\frac{ n \left(\left(\frac{n-1}{n}\right)^{I(n)}-1\right)^2}{2\cdot ln\left(\frac{n-1}{n}\right)}
\end{align}
Now, the resulting function could be considered an exact bound of the worst-case time complexity. However, it does not provide much information, given that it is not a simple form, despite being fully simplified. Thus, the first step toward finding a simpler form that represents its asymptotic growth is to determine that of $I(n,k)$ as the size of the system increases:

\begin{align}
    \lim_{n\to\infty} I(n,k)&=\lim_{n\to\infty} n(H_n-H_{n-k})=\\ \notag
    &=\lim_{n\to\infty} n(\log(n)-\log(n-k))=\\ \notag
    &=\lim_{n\to\infty} \frac{\displaystyle\log(n)-\log(n-k)}{\frac{1}{n}}=\\ \notag
    &=\lim_{n\to\infty} \frac{\frac{1}{n}-\frac{1}{n-k}}{\frac{-1}{n^2}}=\lim_{n\to\infty} \frac{k n}{n-k} = k\lim_{n\to\infty} \frac{n}{n-k} = k \quad \colon \quad \boxed{n>0\enspace\land\enspace 0\leq k<n}
\end{align}
Since the harmonic number is infinitesimally comparable to the logarithm for $n\to\infty$ \cite{SondowWeisstein,user17762_2013,AbramowitzStegun1968,Mayank2013}, it can be deduced that the iteration count required to reach $k$ elements in the system depends exclusively on that quantity of elements, as long as it is not equal to the maximum $k=n$, in which case its asymptotic growth would be equivalent to $I(n,n)=n H_n$. Therefore, in order to ascertain the growth of the original expression $T_{worst}(n)$, it is advisable to replace the number of iterations $I(n)$ by $I(n,k)$ when evaluating its limit as $n\to\infty$, as this allows us to know the relationship between the number of elements the process must reach and its time complexity.

\begin{align}
    \lim_{n\to\infty} T_{worst}(n) &= \lim_{n\to\infty} -\frac{ n \left(\left(\frac{n-1}{n}\right)^{I(n,k)}-1\right)^2}{2\cdot ln\left(\frac{n-1}{n}\right)} =\\ \notag
    &=\lim_{n\to\infty} -\frac{ n \left(1-\frac{I(n,k)}{n}-1\right)^2}{\frac{-2}{n}}=\\ \notag
    &=\lim_{n\to\infty} \frac{ n \left(-\frac{I(n,k)}{n}\right)^2}{\frac{2}{n}} = \lim_{n\to\infty} \frac{ n^2 \frac{I(n,k)^2}{n^2}}{2} = \boxed{\lim_{n\to\infty} \frac{I(n,k)^2}{2}}
\end{align}
At this point, we could perform the substitution $I(n,k) \sim k$ and then analyze its limit as $k \to n$, since in 1 dimension all the system cells must be filled for the process to terminate. Still, as the substitution can only be performed when $k$ is smaller than the maximum element count, it is necessary to check whether the resulting form of this limit actually has the same asymptotic growth as the original expression $T_{worst}(n)$:

\begin{align}
    \lim_{n\to\infty} \frac{T_{worst}(n)}{n^2} &= \lim_{n\to\infty} -\frac{ n \left(\left(\frac{n-1}{n}\right)^{n H_n}-1\right)^2}{2\cdot ln\left(\frac{n-1}{n}\right)n^2}  =\\ \notag
    &=\lim_{n\to\infty} -\frac{\left(\left(\frac{n-1}{n}\right)^{n H_n}-1\right)^2}{2\cdot ln\left(\frac{n-1}{n}\right)n} = \\ \notag
    &=\lim_{n\to\infty} -\frac{\left(\left(\frac{n-1}{n}\right)^{n H_n}-1\right)^2}{\frac{-2}{n}n} = \\ \notag
    &=\lim_{n\to\infty} \frac{\left(\left(1-\frac{1}{n}\right)^{n H_n}-1\right)^2}{2} = \frac{1}{2} \quad (O(n)\subset O(n\log(n)))
\end{align}
If we assume that the complexity grows equally as $n^2$, the outcome of the upper limit is a real number, thereby suggesting that both functions are asymptotically equivalent. Moreover, $\frac{1}{2}$ corresponds to the constant term that has not been included in the growth $n^2$, originating from $\frac{I(n,k)^2}{2}$.

\begin{align}
    \lim_{n\to\infty} \frac{T_{worst}(n)}{\left(n H_n\right)^2} = \lim_{n\to\infty} -\frac{ n \left(\left(\frac{n-1}{n}\right)^{n H_n}-1\right)^2}{2\cdot ln\left(\frac{n-1}{n}\right)\left(n H_n\right)^2} = \lim_{n\to\infty} \frac{ \left(\left(\frac{n-1}{n}\right)^{n H_n}-1\right)^2}{2\left( H_n\right)^2} = \lim_{n\to\infty} \frac{1}{2\left( H_n\right)^2} = 0
\end{align}
Yet, in the case $k = n$, it is also necessary to verify that the original function does not exhibit identical growth as $\left(n H_n\right)^2$, as seen above. This is because $\left(n H_n\right)^2$ is the actual asymptotic growth of $\frac{I(n,n)^2}{2}$ in the edge case of ending a process with the system full of elements, producing an indetermination between selecting the number of elements as a definitive complexity bound or the iterations needed to reach them. Hence, whenever the asymptotic growth of $T_{worst}(n)$ is evaluated in this special case, it is essential to check which of the two bounds is valid. In summary, by positing the sum of the time complexity in its integral form, it results in a growth of the order $O(n^2)$, although it still needs to be verified if the discrete approach yields the same bound.

\begin{align}
    T_{worst}(n) &=\sum _{i=0}^{I(n)} \left(1-\frac{E(i)}{n}\right)\cdot E(i) =\\\notag
    &=\sum _{i=0}^{I(n)} n\left(1-\left(1-\frac{1}{n}\right)^i\right)-\frac{n^2\left(1-\left(1-\frac{1}{n}\right)^i\right)^2}{n} =\\ \notag
    &=\sum _{i=0}^{I(n)} n\left(1-\left(1-\frac{1}{n}\right)^i\right)-n\left(1-\left(1-\frac{1}{n}\right)^i\right)^2 =\\ \notag
    &=\frac{\displaystyle(n-1) n^{1-2 I(n)} \left(n^{I(n)}-(n-1)^{I(n)}\right) \left(n^{I(n)+1}-(n-1)^{I(n)+1}\right)}{\displaystyle 2 n-1}
\end{align}

In a similar vein, the growth of the complexity derived from the discrete sum is not discernible at first glance, so its limit when $n\to\infty$ is computed:

\begin{align}
    & \lim_{n\to\infty} \frac{\displaystyle(n-1) n^{1-2 I(n,k)} \left(n^{I(n,k)}-(n-1)^{I(n,k)}\right) \left(n^{I(n,k)+1}-(n-1)^{I(n,k)+1}\right)}{\displaystyle 2 n-1} =\\ \notag
    &=\lim_{n\to\infty} \frac{\displaystyle n\cdot n^{1-2 I(n,k)} \left(n^{I(n,k)}-n^{I(n,k)}\left(1-\frac{I(n,k)}{n}\right)\right) \left(n^{I(n,k)+1}-n^{I(n,k)+1}\left(1-\frac{I(n,k)+1}{n}\right)\right)}{\displaystyle 2n} =\\ \notag
    &=\lim_{n\to\infty} \frac{\displaystyle n^{1-2 I(n,k)} \left(\frac{n^{I(n,k)}I(n,k)}{n}\right) \left(\frac{n^{I(n,k)+1}(I(n,k)+1)}{n}\right)}{\displaystyle 2} =\\ \notag
    &=\lim_{n\to\infty} \frac{\displaystyle n^{1-2 I(n,k)} \left(n^{I(n,k)}I(n,k)\right) \left(n^{I(n,k)+1}(I(n,k)+1)\right)}{\displaystyle 2n^2} = \lim_{n\to\infty} \frac{I(n,k)\left(I(n,k)+1\right)}{2} \sim \boxed{\lim_{n\to\infty} \frac{I(n,k)^2}{2}}
\end{align}
In this case, the terms $I(n,k)$ found in the simplification of the limit expression cause it to differ from the growth of the previous continuous sum, notwithstanding their asymptotic equivalence.

\begin{align}
   & \lim_{n\to\infty} \frac{\displaystyle(n-1) n^{1-2 I(n)} \left(n^{I(n)}-(n-1)^{I(n)}\right) \left(n^{I(n)+1}-(n-1)^{I(n)+1}\right)}{\displaystyle (2 n-1)n^2} =\\\notag
   &=\lim_{n\to\infty} \frac{(n-1) n \left(\left(\frac{n-1}{n}\right)^{n H_n}-1\right) \left(n \left(\frac{n-1}{n}\right)^{n H_n+1}-n\right)}{(2 n-1)n^2}=\\\notag
   &=\lim_{n\to\infty} \frac{(n-1) n \left(-1\right) \left(-n\right)}{(2 n-1)n^2}\sim \lim_{n\to\infty} \frac{n^3}{2n^3}=\frac{1}{2}
\end{align}
Therefore, following the same procedure as before, it is confirmed that the bound $O(n^2)$ results in the same growth as $T_{worst}(n)$, which is consistent the prior outcomes. Likewise, we must verify that the discrete sum does not grow at the same rate as $\left(n H_n\right)^2$, stemming from the same case $k=n$ where $I(n,k)$ induces an indetermination in the analysis.

\begin{align}
   & \lim_{n\to\infty} \frac{\displaystyle(n-1) n^{1-2 I(n)} \left(n^{I(n)}-(n-1)^{I(n)}\right) \left(n^{I(n)+1}-(n-1)^{I(n)+1}\right)}{\displaystyle (2 n-1)\left(n H_n\right)^2} =\\\notag
   &=\lim_{n\to\infty} \frac{(n-1) n \left(\left(\frac{n-1}{n}\right)^{n H_n}-1\right) \left(n \left(\frac{n-1}{n}\right)^{n H_n+1}-n\right)}{(2 n-1)\left(n H_n\right)^2}=\\\notag
   &=\lim_{n\to\infty} \frac{(n-1) n \left(-1\right) \left(e^{-\gamma }-n\right)}{(2 n-1)\left(n H_n\right)^2}\sim \lim_{n\to\infty} \frac{n^3}{2n^3\left(H_n\right)^2}=\lim_{n\to\infty} \frac{1}{2\left(H_n\right)^2}=0
\end{align}

Consequently, the outcome is consistent with the continuous sum, demonstrating that $\left(n H_n\right)^2$ grows faster than the time complexity when $n$ increases, so it is excluded as an asymptotic bound.

\begin{figure}[H]
    \centering
    \begin{subfigure}[b]{0.49\textwidth}
        \centering
        \includegraphics[width=\textwidth,clip]{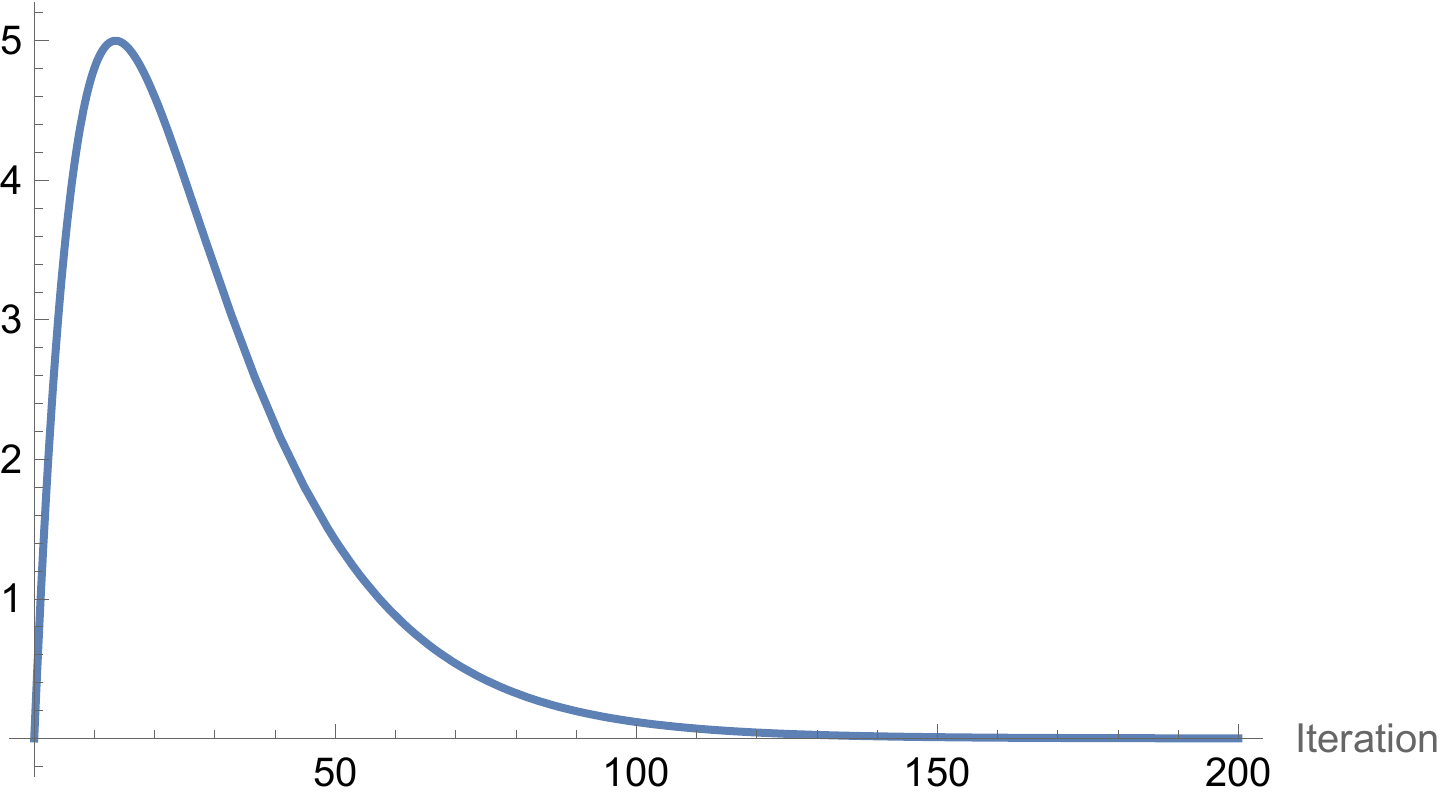}
        \caption{}
        \label{fig:1DWorstCaseInsertion}
    \end{subfigure}
    \hfill
    \begin{subfigure}[b]{0.49\textwidth}
        \centering
        \includegraphics[width=\textwidth,clip]{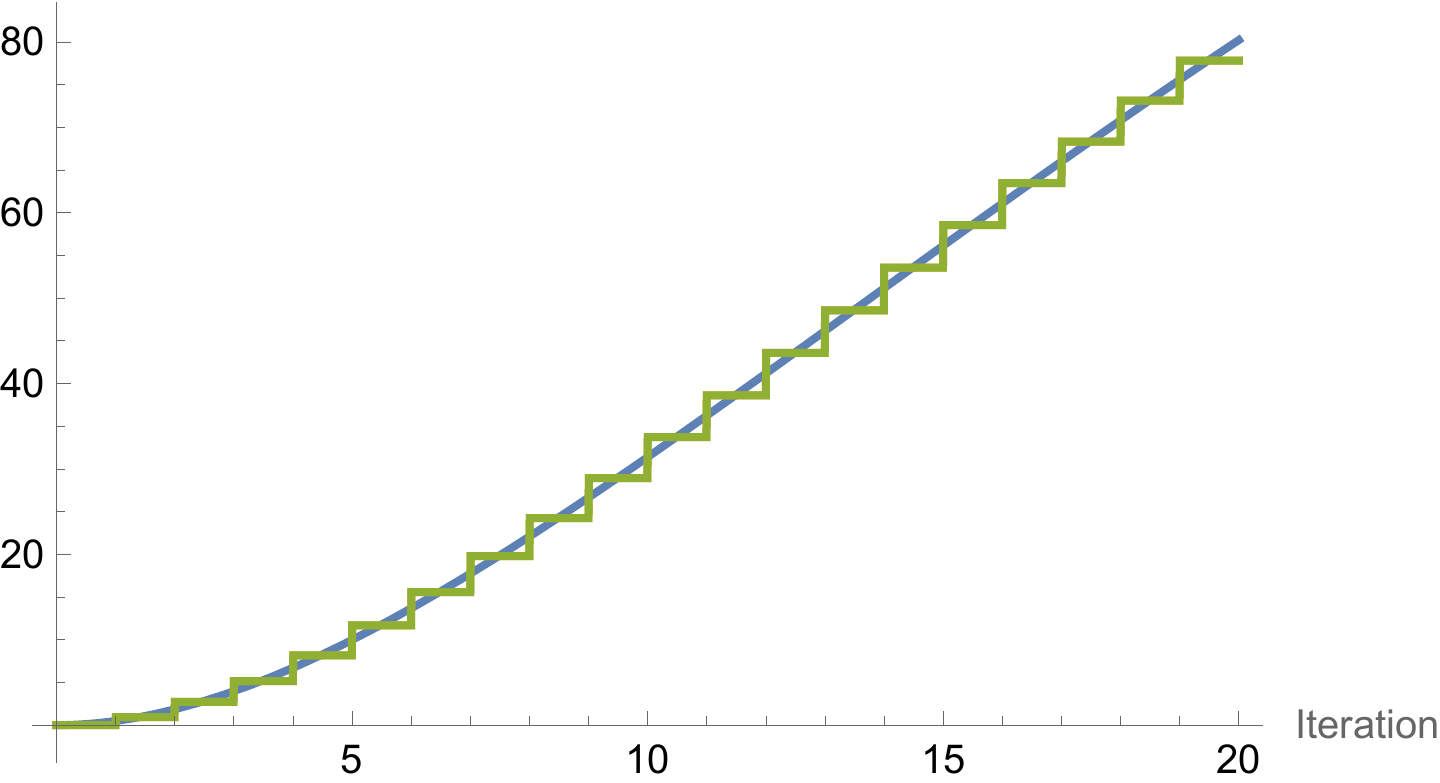}
        \caption{}
        \label{fig:1DWorstCaseSums}
    \end{subfigure}
    \caption{(a) $\left(1-\frac{E(i)}{n}\right)\cdot E(i)$ insertion complexity plotted for a 1D system of size $n=20$ (b) $T_{worst}(n)$ in its integral formulation plotted in blue for a 1D system of size $n=20$, and in green its discrete variant.}
    \label{fig:1DWorstCaseDual}
\end{figure}
In conclusion, it has been demonstrated that, for the worst case, it is indifferent to use the discrete or continuous sum of the insertion complexity throughout all iterations of a process. Ultimately, its complexity adheres to an asymptotic bound as the following:

\begin{align}
   T_{worst}(n)\sim \frac{n^2}{2}=O(n^2)
\end{align}
The basis of this bound lies in the number of elements traversed in all insertions, and how many of these are needed to complete the process. On one hand, each insertion traverses an increasing quantity of elements proportional to the amount of insertions performed, and since $n$ are needed to fill the system, its complexity asymptotically reaches $n^2$ elementary operations.

\subsection{Average case analysis}
\label{1D-Average-case-analysis}
At this point, we know that the average time complexity of a process lies between $O(n \log(n))$ and $O(n^2)$, corresponding to the best and worst cases, respectively. Therefore, to find the exact bound between these two, we proceed analogously to the worst-case analysis. Specifically, the function $T_{avg}(n)$ is posed as the sum of the work performed at each iteration, with the difference that in this case, the formula $c(n,k)$ is used to represent it, as the number of elementary operations is proportional to the average cluster size \cite{StaufferAharony1992}.

\begin{align}
    T_{avg}(n) = \sum _{i=0}^{I(n,n)} \left(1-\frac{E(i)}{n}\right)\cdot c(n,E(i))=\sum _{i=0}^{I(n)} \left(1-\frac{E(i)}{n}\right)\cdot \frac{n}{n-E(i)+1}
\end{align}
As for its parameter $k$, it is replaced by the metric $E(i)$, which is the best estimate of the element count per iteration. However, it will later be verified that the resulting complexity does not vary if we start from the sum of the probabilities that each quantity of elements exists up to $min(i,n)$. Similarly, in this case, it is also necessary to fill the system to complete the process, so the sum is computed up to $I(n)$.
\begin{align}
    T_{avg}(n) = \int _{0}^{I(n)} \left(1-\frac{E(i)}{n}\right)\cdot \frac{n}{n-E(i)+1} \, di
\end{align}

Initially, for simplicity, the continuous sum will be computed first, subsequently extracting an upper bound that can be compared with the result of the discrete sum, as currently there is no guarantee that both are asymptotically equal as $n\to\infty$.

\begin{align}
    \int \left(1-\frac{E(i)}{n}\right)\cdot \frac{n}{n-E(i)+1} \, di &= \int \left(1-\frac{n\left(1-\left(1-\frac{1}{n}\right)^i\right)}{n}\right)\cdot \frac{n}{n-n\left(1-\left(1-\frac{1}{n}\right)^i\right)+1} \, di\\ \notag
    &=\int \frac{n\left(1-\frac{1}{n}\right)^i}{n-n\left(1-\left(1-\frac{1}{n}\right)^i\right)+1} \, di=\\ \notag
    &=\frac{1}{ln \left(1-\frac{1}{n}\right)}\int \frac{1}{u} \, du= \quad \left[u=n-n\left(1-\left(1-\frac{1}{n}\right)^i\right)+1\right]\\ \notag
    &=\frac{ln \left(u\right)}{ln \left(1-\frac{1}{n}\right)}=\frac{ln \left(n \left(1-\frac{1}{n}\right)^i+1\right)}{ln \left(1-\frac{1}{n}\right)}+C
\end{align}
After solving the antiderivative of the continuous sum, it is substituted into its original expression and evaluated within the specified bounds:

\begin{align}
    T_{avg}(n) = \left . \frac{ln \left(n \left(1-\frac{1}{n}\right)^i+1\right)}{ln \left(1-\frac{1}{n}\right)} \right|_{i=0}^{i=I(n)} = \frac{ln \left(n \left(1-\frac{1}{n}\right)^{I(n)}+1\right)-ln (n+1)}{ln \left(1-\frac{1}{n}\right)}
\end{align}
As it happened in the worst case, the final formula for the average time complexity does not present a clear asymptotic bound to categorize the growth rate at first glance, even though it is itself an exact bound of its growth. Therefore, for this purpose, $I(n)$ is replaced by $I(n,k)$ and the relation between the amount of elements at terminal state and the complexity is examined when $n\to\infty$.

\begin{align}
    \lim_{n\to\infty} T_{avg}(n)&=\lim_{n\to\infty} \frac{ln \left(n \left(1-\frac{1}{n}\right)^{I(n,k)}+1\right)-ln (n+1)}{ln \left(1-\frac{1}{n}\right)} =\\ \notag
    &= \lim_{n\to\infty} \frac{ln \left(n \left(1-\frac{I(n,k)}{n}\right)+1\right)-ln (n+1)}{-\frac{1}{n}}= \\ \notag
    &=\lim_{n\to\infty} \frac{ln \left(n-I(n,k)\right)-ln (n)}{-\frac{1}{n}}=\\ \notag
    &=\lim_{n\to\infty} \frac{\displaystyle\frac{1}{n-I(n,k)}-\frac{1}{n}}{\displaystyle\frac{1}{n^2}}=\\\notag
    &=\lim_{n\to\infty} I(n,k)\frac{n}{n-I(n,k)}=\boxed{\lim_{n\to\infty} I(n,k)} \quad [k=o(n)]
\end{align}

For sufficiently large system sizes, the upper limit suggests that the time complexity is asymptotically equivalent to the number of iterations required for the process to end, which in turn is comparable to the element count found in its terminal state. Nevertheless, this only holds when the system does not reach its full capacity, that is, $k \ne n$, so it is key to check whether $O(n)$, derived from the limit when the number of elements approaches its maximum, or $O(n\log(n))$ from the actual number of iterations that the process lasts, is the actual bound that grows at the same rate as $T_{avg}(n)$.

\begin{align}
    \lim_{n\to\infty} \frac{T_{avg}(n)}{n}&=\lim_{n\to\infty} \frac{ln \left(n \left(1-\frac{1}{n}\right)^{n H_n}+1\right)-ln (n+1)}{ln \left(1-\frac{1}{n}\right)n} =\\ \notag
    &= \lim_{n\to\infty} \frac{ln \left(e^{-\gamma }+1\right)-ln (n+1)}{-\frac{n}{n}} = \\ \notag
    &= \lim_{n\to\infty} -ln \left(e^{-\gamma }+1\right)+ln (n+1) \sim \lim_{n\to\infty} ln (n+1) = \infty
\end{align}
When choosing $O(n)$, it is observed that the time complexity grows faster, so unlike the worst case, it is discarded.

\begin{align}
    \lim_{n\to\infty} \frac{T_{avg}(n)}{n H_n}&=\lim_{n\to\infty} \frac{ln \left(n \left(1-\frac{1}{n}\right)^{n H_n}+1\right)-ln (n+1)}{ln \left(1-\frac{1}{n}\right)n H_n} =\\ \notag
    &= \lim_{n\to\infty} \frac{ln \left(e^{-\gamma }+1\right)-ln (n+1)}{-\frac{n H_n}{n}} = \\ \notag
    &= \lim_{n\to\infty} \frac{ln \left(e^{-\gamma }+1\right)-ln (n+1)}{-H_n} = \\ \notag
    &= \lim_{n\to\infty} \frac{ln \left(e^{-\gamma }+1\right)}{-H_n} - \frac{ln (n+1)}{-H_n} = \\ \notag
    &= \lim_{n\to\infty} \frac{ln (n+1)}{H_n} =\lim_{n\to\infty} \frac{ln (1+\frac{1}{n}) + ln(n)}{H_n}= \lim_{n\to\infty} \frac{ln(n)}{H_n}=1
\end{align}
On the other hand \cite{Alzer2011,XuZhangZhao2023}, by verifying the bound $O(nH_n) \sim O(n\log(n))$, it yields a real ratio equal to 1 with respect to the growth of the time complexity. It can be concluded that, on average, the algorithm performs a work proportional to the number of iterations required to fill the $n$ cells of the one-dimensional system. Still, to ensure that the resulting bound is valid, it is advisable to verify the outcome provided by the discrete complexity approach, since in the worst-case scenario of the algorithm, both were equal, but here, there are no guarantees that the same will hold. 

\begin{align}
    T_{avg}(n) &= \sum _{i=0}^{I(n)} \left(1-\frac{E(i)}{n}\right)\cdot \frac{n}{n-E(i)+1} =\\ \notag
    &= \sum _{i=0}^{I(n)} \left(1-\frac{n\left(1-\left(1-\frac{1}{n}\right)^i\right)}{n}\right)\cdot \frac{n}{n-n\left(1-\left(1-\frac{1}{n}\right)^i\right)+1} =\\ \notag
    &= \sum _{i=0}^{I(n)} \frac{n\left(1-\frac{1}{n}\right)^i}{n-n\left(1-\left(1-\frac{1}{n}\right)^i\right)+1}=
\end{align}
\begin{align}
    =\frac{\displaystyle\psi _{\frac{n-1}{n}}^{(0)}\left(I(n)-\frac{\log \left(-\frac{1}{n}\right)}{ln \left(\frac{n-1}{n}\right)}+1\right)-\psi _{\frac{n-1}{n}}^{(0)}\left(-\frac{ln \left(-\frac{1}{n}\right)}{ ln \left(\frac{n-1}{n}\right)}\right)}{\displaystyle ln \left(\frac{n-1}{n}\right)} \notag
\end{align}
By applying the $Sum[]$ function in Wolfram \cite{WolframSum}, we obtain the above expression for the average complexity of a simulation. Hence, the same process is repeated as for the continuous approach.

\begin{align}
    \lim_{n\to\infty} \frac{ln \left(I(n,k)-\frac{ln \left(-\frac{1}{n}\right)}{ln \left(\frac{n-1}{n}\right)}+1\right)}{ln \left(n \left(\frac{n-1}{n}\right)^{I(n,k)}+1\right)}=1 \quad;\quad \lim_{n\to\infty} \frac{ln \left(-\frac{ln \left(-\frac{1}{n}\right)}{ln \left(\frac{n-1}{n}\right)}\right)}{ln (n+1)}=1
\end{align}
But, in this case, the expression for $T_{avg}(n)$ contains Digamma functions \cite{SrivastavaChoi2012}, which significantly complicate the resolution of its limit for increasingly large system sizes. Cautiously, the parameter $q$ characterizing both Digammas can be substituted by 1, since at infinity it tends to 1 from the left and none of the functions need to be differentiated. Furthermore, the asymptotic growth of each can be proposed as the logarithm of its evaluation point, as shown above. Specifically, by substituting them with their apparent asymptotic growth and taking the limit when $n \to \infty$ of their ratio concerning the numerator terms of the complexity resulting from the continuous approach, it is easily verifiable that its result implies they are asymptotically equivalent. Nonetheless, this does not imply that each of them can be substituted by the equivalent asymptotic growth of the Digamma function, as its domain is complex. Consequently, now the limit of $T_{avg}(n)$ will not be calculated directly, but rather the limit of each term in the summation. In this way, if a limit exists for all summands, the term representing the work of an iteration is replaceable by its limit, and then the sum resolved.

\begin{align}
    \lim_{n\to\infty} \left(1-\frac{E(i)}{n}\right)\cdot \frac{n}{n-E(i)+1} &= \lim_{n\to\infty} \frac{n\left(1-\frac{1}{n}\right)^i}{n-n\left(1-\left(1-\frac{1}{n}\right)^i\right)+1}= \\ \notag
    &= \lim_{n\to\infty} \frac{n\left(1-\frac{1}{n}\right)^i}{n\left(1-\frac{1}{n}\right)^i+1}= \\ \notag
    &= \lim_{n\to\infty} \frac{n\cdot e^{-i/n}}{n\cdot e^{-i/n}+1}= \\ \notag
    &= \left(e^{\lim_{n\to\infty} \frac{-i}{n}}\right)\lim_{n\to\infty} \frac{n}{n\cdot e^{-i/n}+1}=\\ \notag
    &= \left(e^{\lim_{n\to\infty} \frac{-i}{n}}\right)\lim_{n\to\infty} e^{i/n}= \quad \left[n\cdot e^{-i/n}+1\sim n\cdot e^{-i/n}\colon n\to\infty\right]\\ \notag
    &= \left(e^{\lim_{n\to\infty} \frac{-i}{n}}\right)\left(e^{\lim_{n\to\infty} \frac{i}{n}}\right)=\frac{e^{\lim_{n\to\infty} \frac{i}{n}}}{e^{\lim_{n\to\infty} \frac{i}{n}}}=1\quad [i=o(nH_n)]
\end{align}
\begin{align}
    \lim_{n\to\infty} \frac{n\cdot e^{-i/n}}{n\cdot e^{-i/n}+1}=\lim_{n\to\infty} \frac{n\cdot e^{-H_n}}{n\cdot e^{-H_n}+1}=\lim_{n\to\infty} \frac{\frac{n}{n\cdot e^\gamma}}{\frac{n}{n\cdot e^\gamma}+1}=\frac{1}{1+e^\gamma} \quad [i=\Theta(nH_n)]
\end{align}

When evaluating the limit of the complexity of each iteration as $n\to\infty$, with the insertion probability included, it is concluded that for any asymptotic growth of $i$ with respect to $n$, it converges to 1, save for the edge case where the iterations reach the terminal state. Thus, each term of the average complexity of the complete process is interchangeable by the value of the limit, leading to the bound deduced from the continuous approach.

\begin{align}
    \lim_{n\to\infty} T_{avg}(n) &= \lim_{n\to\infty} \sum _{i=0}^{I(n)} \left(1-\frac{E(i)}{n}\right)\cdot \frac{n}{n-E(i)+1} = \\ \notag
    &= \lim_{n\to\infty} \sum _{i=0}^{I(n)} 1 \sim \lim_{n\to\infty} nH_n + 1 
\end{align}

Despite the complexity limit has an additional term +1, the asymptotic bound corresponds to the number of iterations in which the terminal state is reached. This occurs because the sum being considered is the upper with respect to the integral, so if the limit were solved with the lower sum, the additional term would likely be -1, representing a smaller workload than the algorithm actually performs.

\begin{align}
    \lim_{n\to\infty} \frac{T_{avg}(n)}{nH_n}=\lim_{n\to\infty} \frac{nH_n+1}{nH_n}=1 \implies \boxed{T_{avg}(n)=\Theta(n\log(n))}
\end{align}
After confirming that asymptotically the additional term does not influence the final bound, although it does affect the exact work the algorithm performs, which would be valuable if an exact execution time accounting were needed, it is advisable to visualize the growth of all the sums, both the upper and lower as well as the integral approach.

\begin{figure}[H]
    \centering
    \begin{subfigure}[b]{0.49\textwidth}
        \centering
        \includegraphics[width=\textwidth,clip]{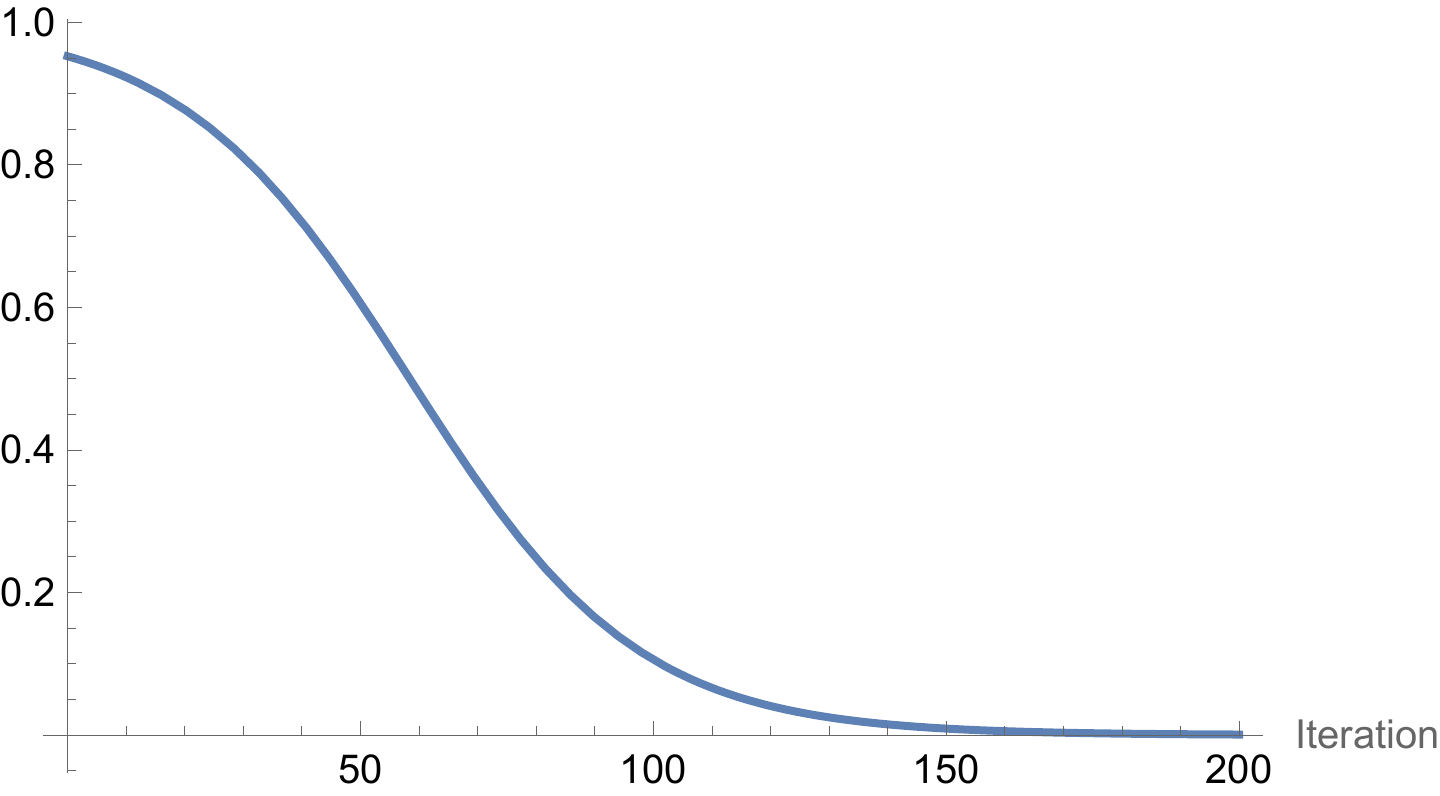}
        \caption{}
        \label{fig:1DAverageCaseInsertion}
    \end{subfigure}
    \hfill
    \begin{subfigure}[b]{0.49\textwidth}
        \centering
        \includegraphics[width=\textwidth,clip]{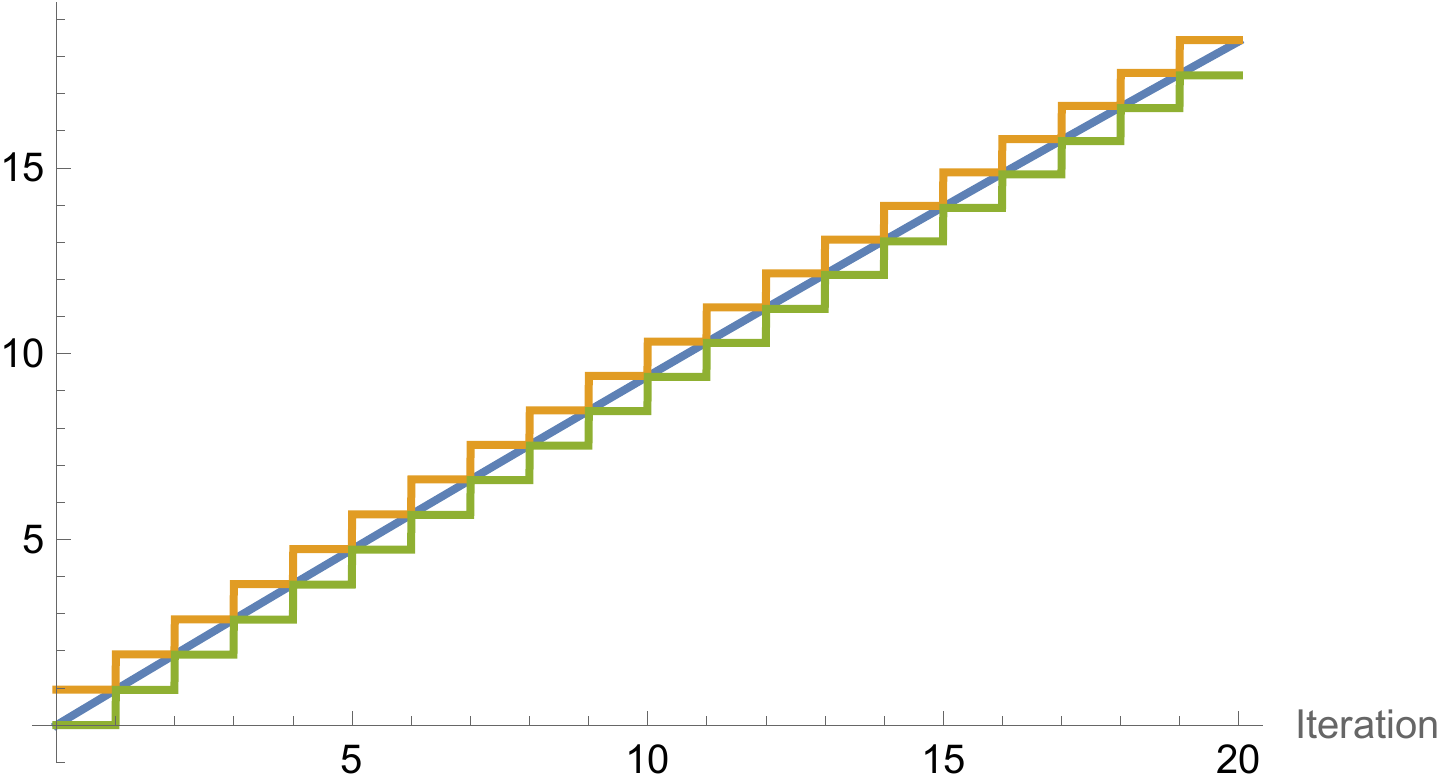}
        \caption{}
        \label{fig:1DAverageCaseSums}
    \end{subfigure}
    \caption{(a) $\left(1-\frac{E(i)}{n}\right)\cdot c(n,E(i))$ insertion complexity plotted for a 1D system of size $n=20$ (b) $T_{avg}(n)$ in its integral formulation plotted in blue for a 1D system of size $n=20$, and in green/orange its discrete lower and upper sum variants, respectively.}
    \label{fig:1DAverageCaseDual}
\end{figure}
As seen, the upper and lower discrete sums appear to be symmetrical with respect to the continuous sum, so either one would result in an equivalent asymptotic bound for the average process time complexity.

\begin{align}
    T_{avg}(n) = \sum _{i=0}^{I(n)} \sum _{j=0}^{i} \left(1-\frac{j}{n}\right)\cdot c(n,j)\cdot H(i,j)=\sum _{i=0}^{I(n)} \sum _{j=0}^{i} \left(1-\frac{j}{n}\right)\cdot \frac{n}{n-j+1}\frac{n!}{n^i (n-j)!} \stirling{i}{j}
\end{align}
Additionally, the execution time can be constructed without relying on the metric $E(i)$, probabilistically. As seen above, the expected work of an iteration is expressed as the sum of the probabilities that certain work is performed for varying amounts of elements ranging from 0 to the index of the outer summation, representing all executed by the process on average. In each of the iteration work probabilities, the metric $H(i,k)$ adapted for 1-dimensional systems is employed, implying its $n^2$ terms are replaced by $n$.

\begin{align}
    T_{iter}(n,i) = \sum _{j=0}^{i} \left(1-\frac{j}{n}\right)\cdot \frac{n}{n-j+1}\frac{n!}{n^i (n-j)!} \stirling{i}{j}
\end{align}
Mainly, it is advisable to verify that the asymptotic growth of this approach is equivalent to the previous ones, since in this way it is confirmed that both metrics for the expected number of elements and its probabilistic variant $H(i,k)$ fulfill their purpose in the analysis, at least in one dimension. Therefore, the sum of the execution time for an iteration is solved, which will subsequently serve to find the complexity of the entire process. Nonetheless, given the nature of the expression $T_{iter}(n,i)$, the $FindSequenceFunction[]$ function from Wolfram is used, to which a sequence of complexity evaluations of an iteration for different values of $i$ will be provided as input, such as the following:

\begin{align}
    T_{iter}(n,0) &= \frac{n}{n+1}\\
    T_{iter}(n,1) &= \frac{n-1}{n}\\
    T_{iter}(n,2) &= \frac{(n-1) n-1}{n^2}\\
    T_{iter}(n,3) &= \frac{(n-2) n (n+1)-1}{n^3}\\
    T_{iter}(n,4) &= \frac{n (n ((n-1) n-3)-3)-1}{n^4}\\
    T_{iter}(n,5) &= \frac{\left(n^2+n+1\right) ((n-3) n (n+1)-1)}{n^5}\\
    T_{iter}(n,6) &= \frac{n (n (n (n ((n-1) n-5)-10)-10)-5)-1}{n^6}\\
    T_{iter}(n,7) &= \frac{n (n+1) (n (n (n ((n-2) n-4)-11)-9)-6)-1}{n^7}\\
    T_{iter}(n,8) &= \frac{n (n (n (n (n (n ((n-1) n-7)-21)-35)-35)-21)-7)-1}{n^8}\\\notag
    \vdots\\
    T_{iter}(n,i) &=1-\frac{n \left(1+\frac{1}{n}\right)^i}{(n+1)^2}
\end{align}
Ultimately, with an expression for the sequence of evaluations, which corresponds with the amount of work of a particular insertion, we proceed to substitute it into the average complexity of the percolation process:

\begin{align}
    T_{avg}(n) &= \sum _{i=0}^{I(n)} T_{iter}(n,i) =\\\notag
    &= \sum _{i=0}^{I(n)} 1-\frac{n \left(1+\frac{1}{n}\right)^i}{(n+1)^2}=\\\notag
    &=-\frac{n^2 \left(\frac{1}{n}+1\right)^{I(n)}-n^2 I(n)-2 n^2+n \left(\frac{1}{n}+1\right)^{I(n)}-2 n I(n)-2 n-I(n)-1}{(n+1)^2}
\end{align}

As in previous occasions, the exact asymptotic bound is not immediately apparent, so the limit is calculated as $n \to \infty$:

\begin{align}
    \lim_{n\to\infty} T_{avg}(n)&=\lim_{n\to\infty} \frac{-n^2 \left(\frac{1}{n}+1\right)^{I(n,k)}+n^2 I(n,k)+2 n^2-n \left(\frac{1}{n}+1\right)^{I(n,k)}+2 n I(n,k)+2 n+I(n,k)+1}{(n+1)^2}=\\\notag
    &=\lim_{n\to\infty} \frac{-n^2 \left(\frac{1}{n}+1\right)^{I(n,k)}}{(n+1)^2} + \frac{n^2 I(n,k)}{(n+1)^2} -\frac{n \left(\frac{1}{n}+1\right)^{I(n,k)}}{(n+1)^2}+\frac{2 n^2+2 n I(n,k)+2 n+I(n,k)+1}{(n+1)^2}=\\ \notag
    &=\lim_{n\to\infty} \frac{-n^2 \left(\frac{1}{n}+1\right)^{I(n,k)}}{(n+1)^2} + I(n,k) -\frac{n \left(\frac{1}{n}+1\right)^{I(n,k)}}{(n+1)^2}+2=\\\notag  
    &=\lim_{n\to\infty} -1-\frac{I(n,k)}{n} + I(n,k) -\frac{1+\frac{I(n,k)}{n}}{n}+2=\notag\\ \notag
    &=\lim_{n\to\infty} -\frac{I(n,k)}{n} + I(n,k) -\frac{1}{n}-\frac{I(n,k)}{n^2}+1=\lim_{n\to\infty} I(n,k)+1-\frac{I(n,k)}{n}\sim \boxed{\lim_{n\to\infty} I(n,k)+1}
\end{align}
Asymptotically, given the properties of the function $I(n,k)$, it can be concluded that the result of the limit is equal to the initial outcome obtained from integrating the iteration complexity. Hence, both methodologies initially seem to yield the same bound for the complexity of the process. Still, as previously noted, it is necessary to verify whether this bound equals exactly the function $I(n)$ or the number of elements $n$ stemming from the limit of $I(n,k)$ when $k\to n$.

\begin{align}
    \lim_{n\to\infty} \frac{T_{avg}(n)}{n}&=\lim_{n\to\infty} \frac{-n^2 \left(\frac{1}{n}+1\right)^{I(n,n)}+n^2 I(n,n)+2 n^2-n \left(\frac{1}{n}+1\right)^{I(n,n)}+2 n I(n,n)+2 n+I(n,n)+1}{n(n+1)^2}=\\\notag
    &=\lim_{n\to\infty} \frac{-n^2 \left(\frac{1}{n}+1\right)^{n H_n}}{n(n+1)^2} + \frac{n^3 H_n}{n(n+1)^2} -\frac{n \left(\frac{1}{n}+1\right)^{n H_n}}{n(n+1)^2}+\frac{2 n^2+2 n^2 H_n+2 n+n H_n+1}{n(n+1)^2}=\\ \notag
    &=\lim_{n\to\infty} \frac{-n^2 \left(\frac{1}{n}+1\right)^{n H_n}}{n(n+1)^2} + \frac{n^3 H_n}{n(n+1)^2} -\frac{n \left(\frac{1}{n}+1\right)^{n H_n}}{n(n+1)^2}=\\ \notag
    &=\lim_{n\to\infty} \frac{-n^2 \left(\frac{1}{n}+1\right)^{n H_n}}{n(n+1)^2} + H_n -\frac{n \left(\frac{1}{n}+1\right)^{n H_n}}{n(n+1)^2}
\end{align}
\begin{align}
    &=\lim_{n\to\infty} \frac{-e^{H_n}}{n} + H_n -\frac{e^{H_n}}{n^2}=\notag\\ \notag
    &=\lim_{n\to\infty} H_n -\frac{(n+1) e^{H_n}}{n^2}=\lim_{n\to\infty} H_n -e^{\gamma }=\infty
\end{align}

For the number of elements $n$, the comparison with the growth of $T_{avg}(n)$ produces a result contingent upon $H_n$, approaching infinity as the system increases in size. Thus, it is inferable that the actual bound for the complexity based on the number of elements can be constructed by multiplying its value by $H_n$, coinciding with $I(n)$:

\begin{align}
    \lim_{n\to\infty} \frac{T_{avg}(n)}{n H_n}&=\lim_{n\to\infty} \frac{-n^2 \left(\frac{1}{n}+1\right)^{I(n,n)}+n^2 I(n,n)+2 n^2-n \left(\frac{1}{n}+1\right)^{I(n,n)}+2 n I(n,n)+2 n+I(n,n)+1}{n H_n (n+1)^2}=\\\notag
    &=\lim_{n\to\infty} \frac{-n^2 \left(\frac{1}{n}+1\right)^{n H_n}}{n H_n (n+1)^2} + \frac{n^3 H_n}{n H_n (n+1)^2} -\frac{n \left(\frac{1}{n}+1\right)^{n H_n}}{n H_n (n+1)^2}+\frac{2 n^2+2 n^2 H_n+2 n+n H_n+1}{n H_n (n+1)^2}=\\ \notag
    &=\lim_{n\to\infty} \frac{-n^2 \left(\frac{1}{n}+1\right)^{n H_n}}{n H_n (n+1)^2} + 1 -\frac{n \left(\frac{1}{n}+1\right)^{n H_n}}{n H_n (n+1)^2}=\\ \notag
    &=\lim_{n\to\infty} \frac{-e^{H_n}}{n H_n} + 1 -\frac{e^{H_n}}{n^2 H_n}=\\ \notag
    &=\lim_{n\to\infty} 1 -\frac{(n+1) e^{H_n}}{n^2 H_n}=\lim_{n\to\infty} 1 -\frac{e^{\gamma }}{H_n}=1 \implies \boxed{T_{avg}(n)=\Theta(n\log(n))}
\end{align}
As expected, the previous limit demonstrates that the bound for the time complexity $T_{avg}(n)$ proposed probabilistically results in the same as the approach with the metric $E(i)$. Hence, at least in one-dimensional systems, we know that the use of either metric is indifferent in estimating the number of elements for a given iteration \cite{Jockovic2011}.

\subsubsection{Continuous average cluster size estimator test}
Currently, the complexity has been computed from the actual formula for the average cluster size. However, in higher dimensions, an exact formula for such magnitude may not exist, or even if it does, we could lack access to it. So, it is advisable to test if the estimators we have generate a consistent bound using the same prior approach.

\begin{align}
    T_{avg}(n) = \sum _{i=0}^{I(n,n)} \left(1-\frac{E(i)}{n}\right)\cdot \hat{c}_0(n,E(i))=\sum _{i=0}^{I(n)} \left(1-\frac{E(i)}{n}\right)\cdot \frac{2 n}{2-3 n \ln (p_{E(i),n})}
\end{align}
Starting with the continuous estimator, we will use the metric $E(i)$ for simplicity in terms that require knowing the count of elements for the corresponding iteration, such as the system's occupancy ratio.

\begin{align}
    T_{avg}(n) = \int _{0}^{I(n)} \left(1-\frac{E(i)}{n}\right)\cdot \frac{2 n}{2-3 n \ln (\frac{E(i)}{n})} \, di
\end{align}

As formerly, the antiderivative is solved first prior to evaluating it at the sum limits:

\begin{align}
    \int \left(1-\frac{E(i)}{n}\right)\cdot \frac{2 n}{2-3 n \ln (\frac{E(i)}{n})} \, di &= \int \frac{2 n \left(1-\frac{1}{n}\right)^i}{2-3 n \ln \left(1-\left(1-\frac{1}{n}\right)^i\right)} \, di=\\\notag
    &= -2n\int \frac{\left(1-\frac{1}{n}\right)^i}{3 n \ln \left(1-\left(1-\frac{1}{n}\right)^i\right)-2} \, di=\\\notag
    &= \frac{2e^{\frac{2}{3n}}}{3\ln(1-\frac{1}{n})}\int \frac{e^{\frac{u}{3n}}}{u} \, du= \quad [u=3 n \ln \left(1-\left(1-\frac{1}{n}\right)^i\right)-2]\\\notag
    &= \frac{2e^{\frac{2}{3n}}}{3\ln(1-\frac{1}{n})}\int \frac{e^{v}}{v} \, dv= \quad [v=\frac{u}{3n}]\\\notag
    &= \frac{2e^{\frac{2}{3n}}}{3\ln(1-\frac{1}{n})} \text{Ei}\left(v\right)=\\\notag
    &= \frac{2e^{\frac{2}{3n}}}{3\ln(1-\frac{1}{n})} \text{Ei}\left(\frac{u}{3n}\right)=\\\notag
    &= \frac{2e^{\frac{2}{3n}}}{3\ln(1-\frac{1}{n})} \text{Ei}\left(\frac{3 n \ln \left(1-\left(1-\frac{1}{n}\right)^i\right)-2}{3n}\right)=\\\notag
    &=\frac{2 e^{\frac{2}{3 n}} \text{Ei}\left(\ln \left(1-\left(\frac{n-1}{n}\right)^i\right)-\frac{2}{3 n}\right)}{3 \ln \left(\frac{n-1}{n}\right)}+C
\end{align}

Consequently, the complexity formulation based on the approach of the continuous sum of each insertion workload produces:

\begin{align}
    T_{avg}(n) = \left . \frac{2 e^{\frac{2}{3 n}} \text{Ei}\left(\ln \left(1-\left(\frac{n-1}{n}\right)^i\right)-\frac{2}{3 n}\right)}{3 \ln \left(\frac{n-1}{n}\right)} \right|_{i=0}^{i=I(n)} = \frac{2 e^{\frac{2}{3 n}} \text{Ei}\left(\ln \left(1-\left(\frac{n-1}{n}\right)^{I(n)}\right)-\frac{2}{3 n}\right)}{3 \ln \left(\frac{n-1}{n}\right)}
\end{align}
Given the shape of the continuous estimator for the average cluster size, unlike its exact expression, it leads to a time complexity that includes an $\text{Ei}(z)$ function with a complex domain \cite{Masina2019}. This poses a drawback compared to alternative approaches, since in this case, not only is the asymptotic bound of the complexity not apparent, but when computing its limit, it will be necessary to consider different substitutions of the previous function for certain cases of its accumulation point. For example, if the argument of $\text{Ei}(z)$ approaches 0, the function must be substituted with a particular infinitesimal. Additionally, if the argument has multiple asymptotic growths, tending to $-\infty$ or $\infty$, the limit will require the substitution of other infinitesimals that correctly adapt to the behavior of $\text{Ei}(z)$ at the evaluation point $n\to\infty$. So, initially, the limit of the entire expression is calculated for a complexity that depends on the iterations necessary to achieve a certain number of elements $k$:

\begin{align}
    \lim_{n\to\infty} T_{avg}(n)&=\lim_{n\to\infty} \frac{2 e^{\frac{2}{3 n}} \text{Ei}\left(\ln \left(1-\left(\frac{n-1}{n}\right)^{I(n,k)}\right)-\frac{2}{3 n}\right)}{3 \ln \left(\frac{n-1}{n}\right)}=\\\notag
    &=\lim_{n\to\infty} \frac{n2 e^{\frac{2}{3 n}} \text{Ei}\left(\ln \left(\frac{I(n,k)}{n}\right)-\frac{2}{3 n}\right)}{-3}=\\\notag
    &=\lim_{n\to\infty} \frac{n2 e^{\frac{2}{3 n}} \left(\frac{e^{\ln \left(\frac{I(n,k)}{n}\right)-\frac{2}{3 n}}}{\ln \left(\frac{I(n,k)}{n}\right)-\frac{2}{3 n}}\right)}{-3}=\\\notag
    &=\lim_{n\to\infty} \frac{2 I(n,k) n}{2-3 n \ln \left(\frac{I(n,k)}{n}\right)}\sim \lim_{n\to\infty} \frac{-2 I(n,k)}{3\ln \left(\frac{I(n,k)}{n}\right)} \quad [k=o(n)]
\end{align}
First, by substituting the argument with its infinitesimal at $-\infty$, the previous asymptotic growth for $T_{avg}(n)$ is obtained. Nevertheless, now the bound is only valid for values of $I(n,k)$ that do not grow faster than $O(n)$, so the upper limit does not provide useful information about the desired growth for $T_{avg}(n)$. Thus, as it is known that the complexity is equivalent to $I(n)$ through the analysis with the actual $c(n,k)$, we proceed to evaluate the same limit directly with $I(n)$, instead of its equivalent for the quantity of elements:

\begin{align}
    \lim_{n\to\infty} T_{avg}(n)&=\lim_{n\to\infty} \frac{2 e^{\frac{2}{3 n}} \text{Ei}\left(\ln \left(1-\left(\frac{n-1}{n}\right)^{I(n,n)}\right)-\frac{2}{3 n}\right)}{3 \ln \left(\frac{n-1}{n}\right)}=\\\notag
    &=\lim_{n\to\infty} \frac{2 e^{\frac{2}{3 n}} \text{Ei}\left(\ln \left(1-\left(\frac{n-1}{n}\right)^{n H_n}\right)-\frac{2}{3 n}\right)}{-\frac{3}{n}}=\\\notag
    &=\lim_{n\to\infty} \frac{n2 e^{\frac{2}{3 n}} \text{Ei}\left(\ln \left(1-\frac{1}{e^{\gamma}n}\right)-\frac{2}{3 n}\right)}{-3}=\\\notag
    &=\lim_{n\to\infty} -\frac{2n}{3} e^{\frac{2}{3 n}} \ln\left(\ln \left(1-\frac{1}{e^{\gamma}n}\right)-\frac{2}{3 n}\right)=\\\notag
    &=\lim_{n\to\infty} -\frac{2n}{3} e^{\frac{2}{3 n}} \ln\left(-\frac{1}{e^{\gamma}n}-\frac{2}{3 n}\right)=\\\notag
    &=\lim_{n\to\infty} -\frac{2n}{3} e^{\frac{2}{3 n}} \ln\left( -\frac{2+3 e^{-\gamma }}{3 n} \right)=\\\notag
    &=\lim_{n\to\infty} -\frac{2n}{3} e^{\frac{2}{3 n}} \left(-\ln (n)-\gamma +i \pi +\ln \left(1+\frac{2 e^{\gamma }}{3}\right)\right)=\boxed{\lim_{n\to\infty} \frac{2}{3} e^{\frac{2}{3 n}} \left(n\ln(n)\right)}
\end{align}
Presently, by including another infinitesimal, the result bears a clearer resemblance to the actual complexity of the algorithm, so we proceed to ascertain their equivalence:
\begin{align}
    \lim_{n\to\infty} \frac{T_{avg}(n)}{nH_n}=\lim_{n\to\infty} \frac{2 e^{\frac{2}{3 n}} \left(n\ln(n)\right)}{3nH_n} =\lim_{n\to\infty} \frac{2 e^{\frac{2}{3 n}}}{3}=\frac{2}{3}
\end{align}

Since the estimator does not match the $c(n,k)$ actual function, the exact asymptotic bound of its growth differs by a constant, reflected in the result of the previous limit. However, upon convergence, it can be concluded that the continuous estimator provides a resulting complexity with an asymptotic growth comparable to that derived from $c(n,k)$. Nonetheless, thus far, only the integral of the complexity of each iteration in $T_{avg}(n)$ has been calculated, so it is mandatory to verify that its discrete variant results in the same bound or an equivalent one when $n \to \infty$.

\begin{align}
    T_{avg}(n) =\sum _{i=0}^{I(n)} \left(1-\frac{E(i)}{n}\right)\cdot \frac{2 n}{2-3 n \ln (\frac{E(i)}{n})}=\sum _{i=0}^{I(n)} \frac{2 n \left(1-\frac{1}{n}\right)^i}{2-3 n \ln \left(1-\left(1-\frac{1}{n}\right)^i\right)}
\end{align}

Originally, the definition of $T_{avg}(n)$ was established through a discrete sum over the iterations of the process, so obtaining the asymptotic bound from that definition ensures its correctness, compared to the results of the continuous sum. To begin with, since the nature of the summands complicates the summation process, we proceed directly with the limit as $n\to\infty$ of each individual summand. In this way, by examining their convergence for different asymptotic growths of the iteration index $i$, it becomes feasible to replace the convergence values of the limit, if they exist, as the work of each iteration in the sum of $T_{avg}(n)$, with its resolution being the asymptotic bound sought.

\begin{align}
    \lim_{n\to\infty} \frac{2 n \left(1-\frac{1}{n}\right)^i}{2-3 n \ln \left(1-\left(1-\frac{1}{n}\right)^i\right)} &= \lim_{n\to\infty} \frac{2 n \left(1-\frac{1}{n}\right)^i}{2-3 n \ln \left(1-\left(1-\frac{1}{n}\right)^i\right)} =\\ \notag
    &= \lim_{n\to\infty} \frac{2 n e^{-i/n}}{2-3 n \ln \left(1-e^{-i/n}\right)} =\\ \notag
    &= \lim_{n\to\infty} \frac{2 n (1-\frac{i}{n})}{2-3 n \ln \left(1-(1-\frac{i}{n})\right)} =\\ \notag
    &= \lim_{n\to\infty} \frac{2 n}{2-3 n \ln \left(\frac{i}{n}\right)} =\\ \notag
    &= \lim_{n\to\infty} -\frac{2}{3}\frac{n}{n \ln \left(\frac{i}{n}\right)} =\lim_{n\to\infty} \frac{-2}{3 \ln \left(\frac{i}{n}\right)}=0 \quad [O(i)\subset O(n)]
\end{align}

The iterations of a process cover a range that spans from 0 to $I(n)$. Within the first segment of this range, determined by all the values of $i$ arising from growth functions smaller than $O(n)$, it is demonstrable that the limit of the expression determining the work of an insertion converges to 0. As it is formed from the continuous estimator, and due to its behavior for small values of the system's occupancy ratio, the limit's outcome remains consistent. 

\begin{align}
    \lim_{n\to\infty} \frac{2 n e^{-i/n}}{2-3 n \ln \left(1-e^{-i/n}\right)}
    &= \lim_{n\to\infty} \frac{2 n}{e(2-3 n \ln \left(1-\frac{1}{e}\right))} =\\ \notag
    &= \lim_{n\to\infty} \frac{2}{e(-3 \ln \left(1-\frac{1}{e}\right))} =\frac{2}{3e - 3e \ln(e-1)} \quad [i= \Theta(n)]
\end{align}

Conversely, when $i$ grows at the same rate as the system size, the limit converges to a real value. Therefore, at that point, the work of an insertion in the sum of $T_{avg}(n)$ can be substituted by such value.

\begin{align}
    \lim_{n\to\infty} \frac{2 n e^{-i/n}}{2-3 n \ln \left(1-e^{-i/n}\right)}
    &= \lim_{n\to\infty} \frac{2 n e^{-nH_n/n}}{2-3 n \ln \left(1-e^{-nH_n/n}\right)} =\\ \notag
    &= \lim_{n\to\infty} \frac{2 n}{n e^\gamma(2-3 n \ln \left(1-\frac{1}{ne^\gamma}\right))} =\\ \notag
    &= \lim_{n\to\infty} \frac{2}{e^\gamma(2+\frac{3}{e^\gamma})} =\frac{1}{\frac{3}{2}+e^\gamma} \quad [i= \Theta(nH_n)]
\end{align}
On the other hand, an examination of the convergence at the maximum value reached by $i$ also yields a real value, which suggests that, among all the values covered by the asymptotic growths of $i$ from the previous limits, the work of an insertion similarly converges to one or several real values that are eligible for substitution into the sum.

\begin{align}
    \lim_{n\to\infty} \frac{2 n e^{-n\ln(\ln(n))/n}}{2-3 n \ln \left(1-e^{-n\ln(\ln(n))/n}\right)}
    =\frac{2}{3}\enspace;\enspace \lim_{n\to\infty} \frac{2 n e^{-n\ln(\sqrt{n})/n}}{2-3 n \ln \left(1-e^{-n\ln(\sqrt{n})/n}\right)}
    =\frac{2}{3}
\end{align}
Initially, it is not known whether in that range for $i$ the limit converges to a single value or a function of $i$. Nevertheless, based on the above results, we may infer that the value is potentially unique, although it is not proven for all functions with a growth bounded by $o(n)$ and $O(nH_n)$. Therefore, it is assumed that the value to which it converges in this range is $\frac{2}{3}$, and depending on the outcome of the sum, it will subsequently be concluded whether the convergence in this range is uniform with respect to the assumption, or if, conversely \cite{Cardstdani2024}, there are points where it converges to values different from $\frac{2}{3}$.

\begin{align}
    \lim_{n\to\infty} T_{avg}(n) &= \lim_{n\to\infty} \sum _{i=0}^{I(n)} \frac{2 n \left(1-\frac{1}{n}\right)^i}{2-3 n \ln \left(1-\left(1-\frac{1}{n}\right)^i\right)} = \\ \notag
    &=\lim_{n\to\infty} \sum _{i=0}^{n-1} 0 + \frac{2}{3e - 3e \ln(e-1)} + \left(\sum _{i=n+1}^{nH_n-1} \frac{2}{3}\right) + \frac{1}{\frac{3}{2}+e^\gamma} = \\ \notag
    &=\lim_{n\to\infty} \left(\sum _{i=n+1}^{nH_n-1} \frac{2}{3}\right) + \frac{2}{3e - 3e \ln(e-1)} + \frac{1}{\frac{3}{2}+e^\gamma} = \\ \notag
    &=\lim_{n\to\infty} \frac{2}{3}\left(-1 - n + n H_n\right) + \frac{2}{3e - 3e \ln(e-1)} + \frac{1}{\frac{3}{2}+e^\gamma}
\end{align}

With the convergence values of the limit for different asymptotic growths of $i$, the sum of the workload at each iteration can be segmented based on the previous growth. On one hand, as long as $i$ grows slower than $n$, the limit value of the summands converges to 0, so the sum of all of them will be 0. On the other hand, if the growth is higher than that threshold but lower than $O(nH_n)$, the limit returns $\frac{2}{3}$, so in this case the summands will be replaced by that value. And, in specific scenarios where the growth is exactly equal to one of the previous thresholds, the constants to which the limit converges are included, which asymptotically can be ignored, but for the correctness of the $T_{avg}(n)$ formulation its inclusion is required.

\begin{align}
    \lim_{n\to\infty} \frac{T_{avg}(n)}{n H_n} &= \lim_{n\to\infty} \frac{\frac{2}{3}\left(-1 - n + n H_n\right) + \frac{2}{3e - 3e \ln(e-1)} + \frac{1}{\frac{3}{2}+e^\gamma}}{n H_n} = \\ \notag
    &= \lim_{n\to\infty} \frac{\frac{2}{3}\left(-1 - n + n H_n\right) }{n H_n} = \lim_{n\to\infty} \frac{2}{3}\left(-\frac{1}{nH_n}-\frac{1}{H_n}+1\right)=\frac{2}{3}
\end{align}

After observing the final form of $T_{avg}(n)$, it becomes evident that the term with the greatest growth in the expression is $nH_n$. Consequently, it is necessary to verify that the growth bound is exactly equivalent to $I(n)$, as established earlier. Furthermore, the ratio between the complexity and $I(n)$ converges, so the average number of iterations results in the bound on the time complexity of the process.

\begin{figure}[H]
    \centering
    \includegraphics[width=10cm,clip]{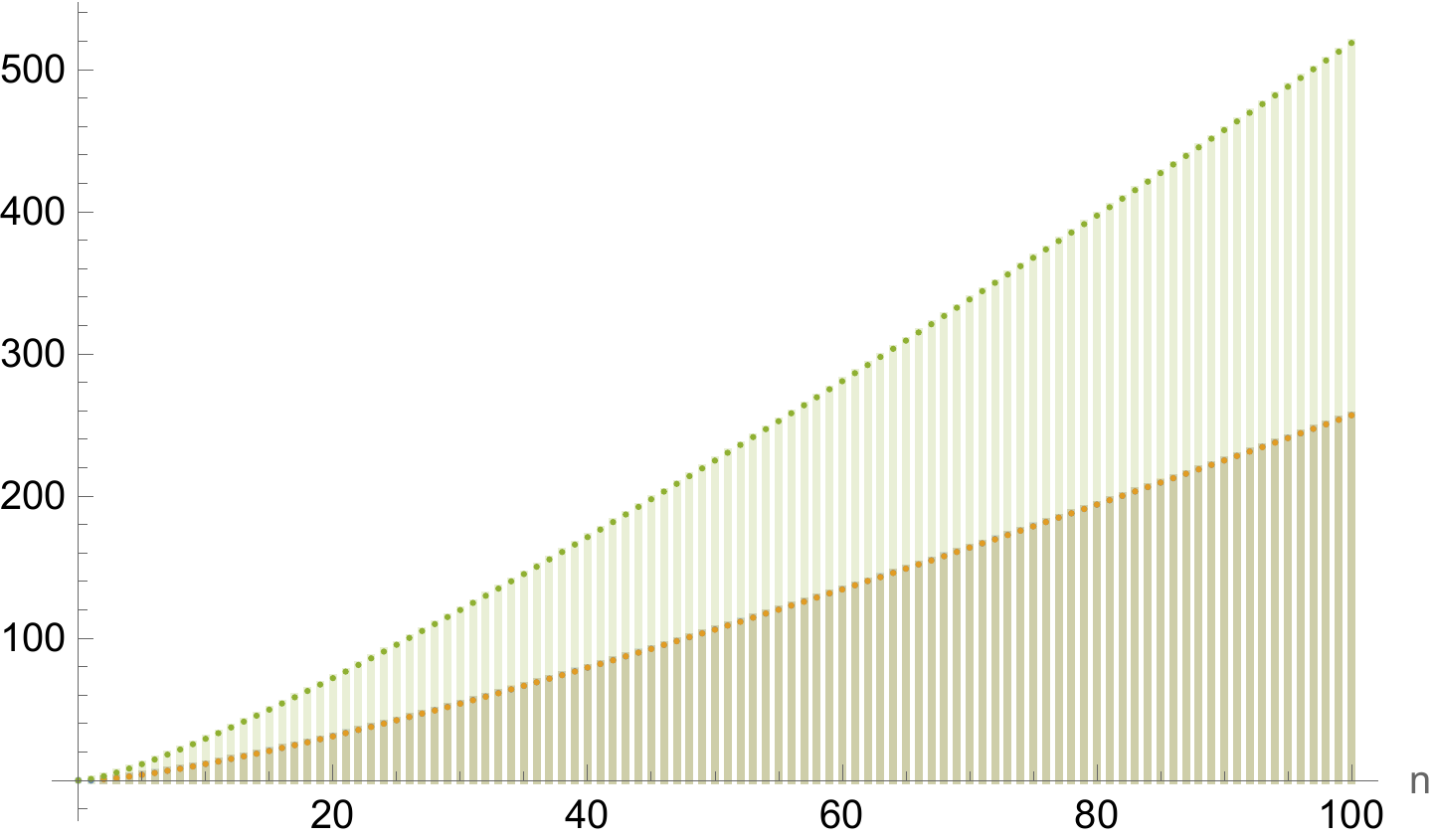}
    \caption{Integral and discrete sum formulation of $T_{avg}(n)$ plotted in blue and orange. In green, $nH_n$ function for $0\leq n \leq 100$. }
    \label{fig:1DContinuousComplexity}
\end{figure}
Graphically, the relationship between the average number of iterations a process takes and the complexity obtained from the continuous cluster size estimator, both from the original approach and the integral, exhibits linearity. That is, given their asymptotic equivalence, where the number of iterations serves as a valid bound for $T_{avg}(n)$, there exists a constant which, when multiplied by $I(n)$, results in the function that determines the time complexity, and vice versa. Furthermore, depending on the values returned by both functions when $n$ is near the lower bound of its domain, an offset parameter could also be considered to support the aforementioned constant, establishing a linear relationship between both functions, heeding that such a transformation is asymptotically invariant. Additionally, it is worth noting that in the upper graph, both the complexity arising from the discrete sum and the integral are calculated numerically with a precision of 20 digits, yet, they appear identical, which is consistent with the results of the previous analysis. But, to visualize more clearly the asymptotic equivalence as $n$ grows, a logarithmic transformation that "normalizes" the range of all functions is convenient. In this way, the effect of the linear difference between them is mitigated, preventing significant deviations within the displayed domain.

\begin{figure}[H]
    \centering
    \includegraphics[width=10cm,clip]{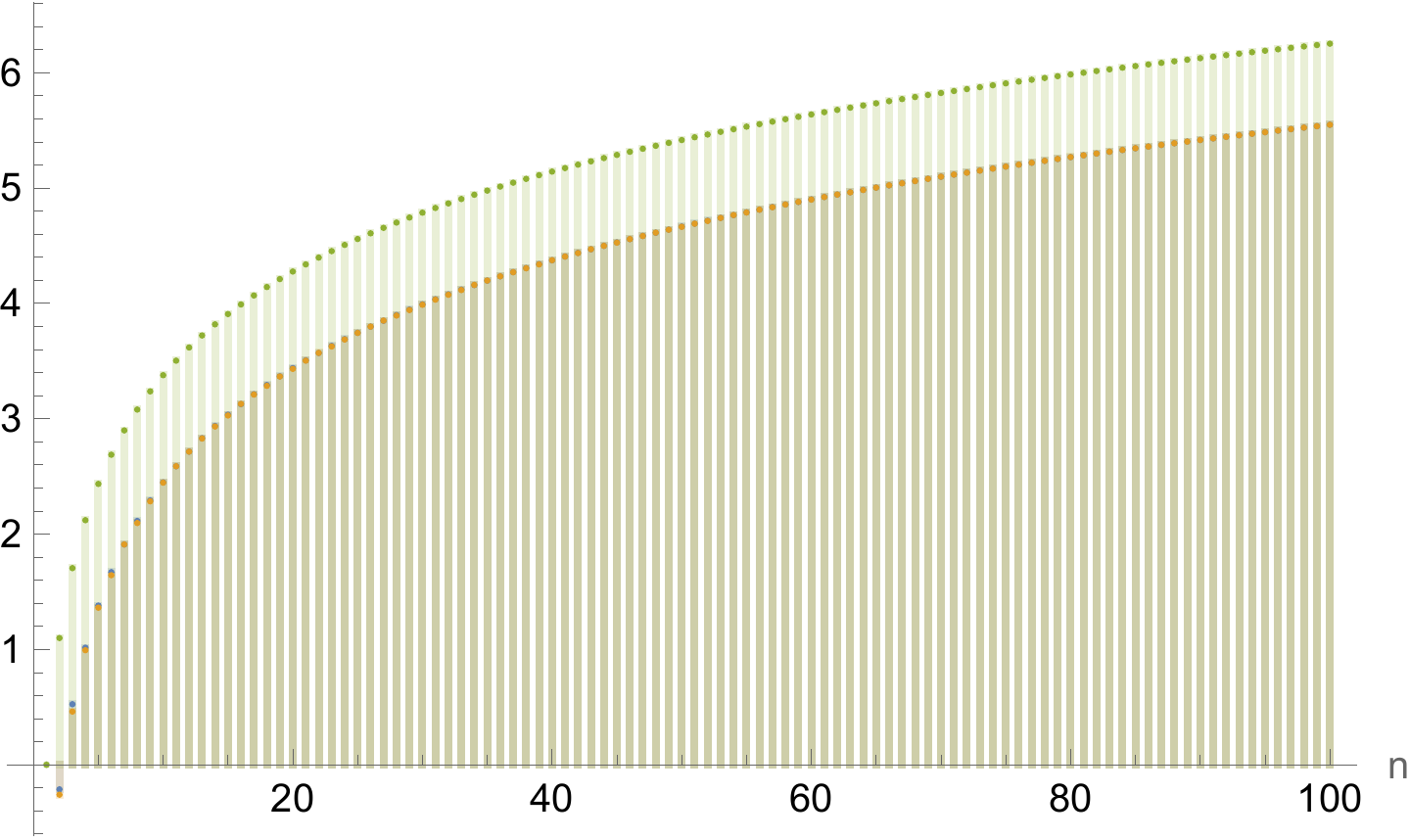}
    \caption{Plot from Figure 42 under a logarithmic transformation of the function's range.}
    \label{fig:1DContinuousComplexityLog}
\end{figure}
It is evident that, on a logarithmic scale, the process complexity and the number of iterations increase analogously. This is a good indicator, along with previous analysis, that the continuous estimator of the average cluster size fulfills its intended purpose effectively, at least in 1-dimensional systems \cite{BarakPellegBerend2022,Grimmett1999}.
\subsubsection{Discrete average cluster size estimator test}
On the other hand, besides performing the analysis with the actual formulation of $c(n,k)$ and its continuous estimator, it is advisable to verify that the results obtained from its discrete estimator do not differ from those already acquired. Hence, with an asymptotic bound derived from the previous procedures, it is expected that the discrete estimator will achieve the same growth. However, given the differences between the two estimators, this may not necessarily occur, so the analysis is repeated with $\hat{c}_1(n,k)$:
\begin{align}
    T_{avg}(n) = \sum _{i=0}^{I(n,n)} \left(1-\frac{E(i)}{n}\right)\cdot \hat{c}_1(n,E(i))=\sum _{i=0}^{I(n)} \left(1-\frac{E(i)}{n}\right)\cdot \frac{n \left(p_{E(i),n}^3+1\right)}{-n\cdot p_{E(i),n}^3+n+p_{E(i),n}^3+1}
\end{align}
As before, the complexity expression follows the same philosophy, aggregating for each iteration of the process the expected work, equivalent to the number of elementary operations attributed to the algorithm's execution. In this case, for simplicity and by using an estimator, the metric $E(i)$ is selected as the straightforward approach. And, regarding the average cluster size, this metric is substituted in the estimator $\hat{c}_1(n,k)$, yielding an expression as a function of the system's occupancy ratio within the iteration determined by the summation index.
\begin{align}
    T_{avg}(n) = \int _{0}^{I(n)} \left(1-\frac{E(i)}{n}\right)\cdot \frac{n \left(p_{E(i),n}^3+1\right)}{-n\cdot p_{E(i),n}^3+n+p_{E(i),n}^3+1} \, di
\end{align}
Initially, prior to addressing the sum proposed in the definition, the continuous sum is computed first, which will provide us with a preliminary result that potentially could be asymptotically equal to the one we are looking for, although this equivalence is not verified.

\begin{align}
    &\int \left(1-\frac{E(i)}{n}\right)\cdot \frac{n \left(p_{E(i),n}^3+1\right)}{-n\cdot p_{E(i),n}^3+n+p_{E(i),n}^3+1} \, di =\\ \notag
    &=\int \frac{\left(1 + \left(1 - \left(1 - \frac{1}{n}\right)^i\right)^3\right) \left(1 - \frac{1}{n}\right)^i n}{1 + \left(1 - \left(1 - \frac{1}{n}\right)^i\right)^3 + n - \left(1 - \left(1 - \frac{1}{n}\right)^i\right)^3 n} \, di = \\\notag
    &=\frac{n}{3 (n-1)^{4/3} (n+1)^{2/3} \ln \left(\frac{n-1}{n}\right)}(-3 (n+1)^{2/3} \sqrt[3]{n-1} \left(\frac{n-1}{n}\right)^i \\\notag
    &+2 n \ln \left(\sqrt[3]{n-1} \left(\left(\frac{n-1}{n}\right)^i-1\right)+\sqrt[3]{n+1}\right) \\ \notag
    &-n \ln \left(-\left(\sqrt[3]{n-1} \sqrt[3]{n+1} \left(\left(\frac{n-1}{n}\right)^i-1\right)\right)+(n-1)^{2/3} \left(\left(\frac{n-1}{n}\right)^i-1\right)^2+(n+1)^{2/3}\right) \\ \notag
    &+2 \sqrt{3} n \tan ^{-1}\left(\frac{2 \sqrt[3]{\frac{n-1}{n+1}} \left(\left(\frac{n-1}{n}\right)^i-1\right)-1}{\sqrt{3}}\right)+3 (n+1)^{2/3} \sqrt[3]{n-1}) + C    
\end{align}
After substituting the occupancy ratio with the value of the metric $E(i)$ and attempting to find its antiderivative in order to subsequently substitute its formulation within the evaluation limits, it is noticeable that in this case it requires a considerable amount of steps and transformations for its derivation. Consequently, using the $Integrate[]$ function from Wolfram \cite{WolframIntegrate}, the antiderivative is automatically computed, revealing a closed form that is, however, more intricate compared to other cases.

\begin{align}
    &\int \frac{\left(1 + \left(1 - \left(1 - \frac{1}{n}\right)^i\right)^3\right) \left(1 - \frac{1}{n}\right)^i n}{1 + \left(1 - \left(1 - \frac{1}{n}\right)^i\right)^3 + n - \left(1 - \left(1 - \frac{1}{n}\right)^i\right)^3 n} \, di = \\\notag
    &=   n \left(\frac{2 \left(\sum_{\left\{Z:\>Z \left(3n - 3\right) + Z^{3} \left(n - 1\right) + Z^{2} \left(3 - 3n\right) + 2 = 0\right\}} \frac{\ln\left(\left|\left(1 - \frac{1}{n}\right)^{i} - Z\right|\right) \, n}{Z^{2} \left(3n - 3\right) + 3n + Z \left(6 - 6n\right) - 3}\right)}{\ln\left(1 - \frac{1}{n}\right) \left(n - 1\right)} - \frac{\left(1 - \frac{1}{n}\right)^{i}}{\ln\left(1 - \frac{1}{n}\right) \left(n - 1\right)}\right) + C
\end{align}

Moreover, leveraging the capabilities offered by Wolfram's engine, we can rewrite its solution by summing over the roots of a polynomial \cite{WolframRootSum}. At first glance, it does not seem worthwhile to proceed with this option, as the next step involves evaluating it at the integration limits, which becomes complicated if any of those limits are in the summands of the expression. Thus, although it may be useful under certain circumstances, it is preferable, for this analysis, to use the initial closed form, since we will later need to find an asymptotic bound for the evaluation result. This implies that the simpler $T_{avg}(n)$ is algebraically, the shorter the ensuing procedure will be.

\begin{align}
    T_{avg}(n) &= \int _{0}^{I(n)} \frac{\left(1 + \left(1 - \left(1 - \frac{1}{n}\right)^i\right)^3\right) \left(1 - \frac{1}{n}\right)^i n}{1 + \left(1 - \left(1 - \frac{1}{n}\right)^i\right)^3 + n - \left(1 - \left(1 - \frac{1}{n}\right)^i\right)^3 n} \, di=\\\notag
    &=\frac{n}{3 (n-1)^{4/3} (n+1)^{2/3} \ln \left(\frac{n-1}{n}\right)}(-3 (n+1)^{2/3} \sqrt[3]{n-1} \left(\frac{n-1}{n}\right)^i \\\notag
    &+2 n \ln \left(\sqrt[3]{n-1} \left(\left(\frac{n-1}{n}\right)^i-1\right)+\sqrt[3]{n+1}\right) \\ \notag
    &-n \ln \left(-\left(\sqrt[3]{n-1} \sqrt[3]{n+1} \left(\left(\frac{n-1}{n}\right)^i-1\right)\right)+(n-1)^{2/3} \left(\left(\frac{n-1}{n}\right)^i-1\right)^2+(n+1)^{2/3}\right) \\ \notag
    &+2 \sqrt{3} n \tan ^{-1}\left(\frac{2 \sqrt[3]{\frac{n-1}{n+1}} \left(\left(\frac{n-1}{n}\right)^i-1\right)-1}{\sqrt{3}}\right)+3 (n+1)^{2/3} \sqrt[3]{n-1}) \left . \right|_{i=0}^{i=I(n)}=\\\notag
    &=\frac{n}{9 (n-1)^{4/3} (n+1)^{2/3} \ln \left(\frac{n-1}{n}\right)} (-9 (n+1)^{2/3} n^{-I(n)} (n-1)^{I(n)+\frac{1}{3}} \\ \notag
    &+n (6 \ln \left(\sqrt[3]{n-1} \left(\left(\frac{n-1}{n}\right)^{I(n)}-1\right)+\sqrt[3]{n+1}\right) \\\notag
    &-3 \ln \left(-\left(\sqrt[3]{n-1} \sqrt[3]{n+1} \left(\left(\frac{n-1}{n}\right)^{I(n)}-1\right)\right)+(n-1)^{2/3} \left(\left(\frac{n-1}{n}\right)^{I(n)}-1\right)^2+(n+1)^{2/3}\right) \\ \notag
    &+6 \sqrt{3} \tan ^{-1}\left(\frac{2 \sqrt[3]{\frac{n-1}{n+1}} \left(\left(\frac{n-1}{n}\right)^{I(n)}-1\right)-1}{\sqrt{3}}\right)+\pi  \sqrt{3})+9 (n+1)^{2/3} \sqrt[3]{n-1})
\end{align}

Evaluating the antiderivative within its limits of integration yields the upper formula, whose asymptotic growth could be inferred by the term of highest growth were it not for the one with an arctan function. Additionally, it is not a small number of terms that need to be compared asymptotically to determine which one denotes the final bound. Hence, analogously to previous procedures, the limit of the whole expression is calculated as $n \to \infty$, substituting $I(n)$ with its counterpart $I(n,k)$ in order to express the complexity as a function of the number of iterations the process takes to fill a certain quantity of elements. Subsequently, the growth of $T_{avg}(n)$ can be inferred from it.

\begin{align}
    \lim_{n\to\infty} T_{avg}(n)=\lim_{n\to\infty} \int _{0}^{I(n,k)} \frac{\left(1 + \left(1 - \left(1 - \frac{1}{n}\right)^i\right)^3\right) \left(1 - \frac{1}{n}\right)^i n}{1 + \left(1 - \left(1 - \frac{1}{n}\right)^i\right)^3 + n - \left(1 - \left(1 - \frac{1}{n}\right)^i\right)^3 n} \, di \sim I(n,k) \quad [k=o(n)]
\end{align}

Given its intricacy, the limit of the complexity will be determined using the symbolic calculation tools provided by Wolfram. Specifically, applying the $Limit[]$ function \cite{WolframLimit}, it is concluded that, similar to what originally occurred with the complexity approach based on $c(n,k)$, the asymptotic growth depends solely on the iterations it takes for the process to reach a threshold of elements, where this can vary from 0 to the size of the system. However, as this threshold approaches $n$, it cannot be guaranteed that the limit value will be the same as the exact number of iterations needed to reach such amount of elements in the system, so both possibilities are checked, where one converges and the other diverges, respectively:

\begin{align}
    \lim_{n\to\infty} \frac{T_{avg}(n)}{n H_n}=\lim_{n\to\infty} \frac{\int _{0}^{I(n,k)} \frac{\left(1 + \left(1 - \left(1 - \frac{1}{n}\right)^i\right)^3\right) \left(1 - \frac{1}{n}\right)^i n}{1 + \left(1 - \left(1 - \frac{1}{n}\right)^i\right)^3 + n - \left(1 - \left(1 - \frac{1}{n}\right)^i\right)^3 n} \, di}{n H_n} = \frac{2}{3}
\end{align}
In the case of $I(n)$, which coincides with the previous bounds, the limit produces a real value equal to that obtained in the analysis using the continuous estimator, allowing us to conclude that the asymptotic growth of the time complexity is equivalent to the average duration of a process, which is an indicator of the reliability of the discrete estimator.
\begin{align}
    \lim_{n\to\infty} \frac{T_{avg}(n)}{n}=\lim_{n\to\infty} \frac{\int _{0}^{I(n,k)} \frac{\left(1 + \left(1 - \left(1 - \frac{1}{n}\right)^i\right)^3\right) \left(1 - \frac{1}{n}\right)^i n}{1 + \left(1 - \left(1 - \frac{1}{n}\right)^i\right)^3 + n - \left(1 - \left(1 - \frac{1}{n}\right)^i\right)^3 n} \, di}{n} = \infty
\end{align}
Conversely, by assessing whether the complexity grows at the same rate as the bound from the limit when $k\to n$ of the asymptotically equivalent expression to $T_{avg}(n)$, the limit diverges to infinity, which can be inferred without performing this computation since if the exact bound is $nH_n$, to obtain a growth function $O(n)$, the bound must be divided by a logarithmic growth quantity, which at infinity causes its ratio to the time complexity to diverge. It is pertinent to mention that the previous operations were performed automatically with Wolfram, due to the length of the expressions. This is also reinforced by its simplicity, despite the length, since to infer an asymptotic bound, it is sufficient to select one of the terms in which it can be expanded, whose growth must be the greatest. Finally, after verifying that the complexity based on the integral sum of the work of all iterations of a process results in the same bound $I(n)$ as with the continuous estimator or the actual formula $c(n,k)$, it can be inferred, for now, that the discrete estimator successfully models the metric of the average cluster size in such a way that they are asymptotically interchangeable in the analysis. Nevertheless, it remains to be verified that the original definition of complexity with the discrete sum results in the same bound, so the previous process is repeated with:

\begin{align}
    T_{avg}(n) =\sum _{i=0}^{I(n)} \left(1-\frac{E(i)}{n}\right)\cdot \frac{n \left(p_{E(i),n}^3+1\right)}{-n\cdot p_{E(i),n}^3+n+p_{E(i),n}^3+1}=\\\notag
    =\sum _{i=0}^{I(n)} \frac{\left(1 + \left(1 - \left(1 - \frac{1}{n}\right)^i\right)^3\right) \left(1 - \frac{1}{n}\right)^i n}{1 + \left(1 - \left(1 - \frac{1}{n}\right)^i\right)^3 + n - \left(1 - \left(1 - \frac{1}{n}\right)^i\right)^3 n}
\end{align}
For this resolution, as with the previous integral, Wolfram will be used to compute the final result, given the length of the intermediate procedure and the nature of the functions it includes. Still, for simplicity, the sum will be decomposed based on the terms in which the work of an insertion can be expanded:

\begin{align}
    T_{avg}(n) &=\sum _{i=0}^{I(n)} \frac{\left(1 + \left(1 - \left(1 - \frac{1}{n}\right)^i\right)^3\right) \left(1 - \frac{1}{n}\right)^i n}{1 + \left(1 - \left(1 - \frac{1}{n}\right)^i\right)^3 + n - \left(1 - \left(1 - \frac{1}{n}\right)^i\right)^3 n}=\\\notag
    &=\sum _{i=0}^{I(n)} \left(\frac{2 n \left(1-\frac{1}{n}\right)^i}{3 n \left(1-\frac{1}{n}\right)^i-3 \left(1-\frac{1}{n}\right)^i-3 n \left(1-\frac{1}{n}\right)^{2 i}+3 \left(1-\frac{1}{n}\right)^{2 i}+n \left(1-\frac{1}{n}\right)^{3 i}-\left(1-\frac{1}{n}\right)^{3 i}+2}\right) \\\notag
    &-\sum _{i=0}^{I(n)} \left(\frac{3 n \left(1-\frac{1}{n}\right)^{2 i}}{3 n \left(1-\frac{1}{n}\right)^i-3 \left(1-\frac{1}{n}\right)^i-3 n \left(1-\frac{1}{n}\right)^{2 i}+3 \left(1-\frac{1}{n}\right)^{2 i}+n \left(1-\frac{1}{n}\right)^{3 i}-\left(1-\frac{1}{n}\right)^{3 i}+2}\right) \\\notag
    &+\sum _{i=0}^{I(n)} \left(\frac{3 n \left(1-\frac{1}{n}\right)^{3 i}}{3 n \left(1-\frac{1}{n}\right)^i-3 \left(1-\frac{1}{n}\right)^i-3 n \left(1-\frac{1}{n}\right)^{2 i}+3 \left(1-\frac{1}{n}\right)^{2 i}+n \left(1-\frac{1}{n}\right)^{3 i}-\left(1-\frac{1}{n}\right)^{3 i}+2}\right) \\\notag
    &-\sum _{i=0}^{I(n)} \left(\frac{n \left(1-\frac{1}{n}\right)^{4 i}}{3 n \left(1-\frac{1}{n}\right)^i-3 \left(1-\frac{1}{n}\right)^i-3 n \left(1-\frac{1}{n}\right)^{2 i}+3 \left(1-\frac{1}{n}\right)^{2 i}+n \left(1-\frac{1}{n}\right)^{3 i}-\left(1-\frac{1}{n}\right)^{3 i}+2}\right)=\\\notag
    &=\frac{n}{3 (n-1)^2 (n+1) \ln \left(\frac{n-1}{n}\right)} (3 \left(n^2-1\right) n^{-n H_n} \left((n-1)^{n H_n+1}-n^{n H_n+1}\right) (\ln (n-1)-\ln (n)) \\\notag
    &+n \sqrt[3]{-(n-1)^2 (n+1)} (-2 \psi _{\frac{n-1}{n}}^{(0)}\left(n H_n+\frac{\ln \left(\frac{\sqrt[3]{-(n-1)^2 (n+1)}}{n-1}+1\right)}{\ln (n)-\ln (n-1)}+1\right) \\ \notag
    &+i \left(\sqrt{3}+i\right) (\psi _{\frac{n-1}{n}}^{(0)}\left(\frac{\ln \left(\frac{i \left(i+\sqrt{3}\right) \sqrt[3]{-(n-1)^2 (n+1)}}{2 (n-1)}+1\right)}{\ln (n)-\ln (n-1)}\right) \\ \notag 
    &-\psi _{\frac{n-1}{n}}^{(0)}\left(n H_n+\frac{\ln \left(\frac{i \left(i+\sqrt{3}\right) \sqrt[3]{-(n-1)^2 (n+1)}}{2 (n-1)}+1\right)}{\ln (n)-\ln (n-1)}+1\right)) \\ \notag 
    &+2 \psi _{\frac{n-1}{n}}^{(0)}\left(\frac{\ln \left(\frac{\sqrt[3]{-(n-1)^2 (n+1)}}{n-1}+1\right)}{\ln (n)-\ln (n-1)}\right)\\\notag
    &+\left(1+i \sqrt{3}\right) \psi _{\frac{n-1}{n}}^{(0)}\left(n H_n+\frac{\ln \left(1-\frac{i \left(-i+\sqrt{3}\right) \sqrt[3]{-(n-1)^2 (n+1)}}{2 (n-1)}\right)}{\ln (n)-\ln (n-1)}+1\right) 
\end{align}
\begin{align}
    &+\left(-1-i \sqrt{3}\right) \psi _{\frac{n-1}{n}}^{(0)}\left(\frac{\ln \left(1-\frac{i \left(-i+\sqrt{3}\right) \sqrt[3]{-(n-1)^2 (n+1)}}{2 (n-1)}\right)}{\ln (n)-\ln (n-1)}\right)))
\end{align}
Alternatively, there exists a more succinct expression, as shown below, however, since it is a sum of roots of a polynomial, it proves impractical for our purpose:

\begin{align}
    T_{avg}(n) &=\sum _{i=0}^{I(n)} \frac{\left(1 + \left(1 - \left(1 - \frac{1}{n}\right)^i\right)^3\right) \left(1 - \frac{1}{n}\right)^i n}{1 + \left(1 - \left(1 - \frac{1}{n}\right)^i\right)^3 + n - \left(1 - \left(1 - \frac{1}{n}\right)^i\right)^3 n}=\\\notag
    &=n \left( 2n \sum_{\{\omega : n \omega^3 - 3 n \omega^2 + 3 n \omega - \omega^3 + 3 \omega^2 - 3 \omega + 2 = 0\}} 
\left( -\omega \psi^{(0)}\left( \frac{n-1}{n} \right) 
\left( n H_n - \frac{\log(\omega)}{\log\left(\frac{n-1}{n}\right)} + 1 \right) \right. \right.\\ \notag
&\left. + \psi^{(0)}\left( \frac{n-1}{n} \right) 
\left( n H_n - \frac{\log(\omega)}{\log\left(\frac{n-1}{n}\right)} + 1 \right) 
+ n \omega H_n \log \left(\frac{n-1}{n}\right) + \omega \log \left(\frac{n-1}{n}\right) \right) \\ \notag
&- 2n \sum_{\{\omega : n \omega^3 - 3 n \omega^2 + 3 n \omega - \omega^3 + 3 \omega^2 - 3 \omega + 2 = 0\}} 
\left( \psi^{(0)}\left( \frac{n-1}{n} \right) 
\left( \frac{\log(\omega)}{\log\left(\frac{n-1}{n}\right)} \right) \right. \\ \notag
&\left. - \omega \psi^{(0)}\left( \frac{n-1}{n} \right) 
\left( \frac{\log(\omega)}{\log\left(\frac{n-1}{n}\right)} \right) + 3n^2 
\left( \frac{n-1}{n} \right) n H_n \log \left(\frac{n-1}{n}\right) 
- 6n^2 H_n \log \left(\frac{n-1}{n}\right) \right. \\ \notag
&\left. - 3 \left( \frac{n-1}{n} \right) n H_n \log \left(\frac{n-1}{n}\right) \right) 
+ 3n^2 \log \left( \frac{n-1}{n} \right) - 9n \log \left( \frac{n-1}{n} \right) \left. \right)  \\ \notag
&\bigg/ \left( 3(n-1)(n+1) \log \left( \frac{n-1}{n} \right) \right)
\end{align}

Considering these solutions for the time complexity of the process in its fundamental definition, it is possible to attempt to numerically compute the limit as the system size approaches infinity, or even use functions that computationally obtain a solution automatically with appropriate algorithms, taking into account that in this context the value of $I(n,n)$ has been directly substituted into the resulting formulas. Nevertheless, the first formulation of the discrete sum features sophisticated terms, such as Digamma functions and arguments complex enough to be challenging to simplify and evaluate, even using computational symbolic calculation methods. On the other hand, its alternative involves sums over roots of polynomials, which can also be complex. As such, this alternative is not a closed form that can be easily evaluated for the limit as $n\to\infty$, not even computationally. Therefore, the limit of the work of an arbitrary insertion at the same accumulation point is determined next. In parallel with previous approaches, if the limit of all summands converges to a value, it can be replaced into the summation, leading to the solution of the original limit sought. Additionally, if there are multiple convergence values depending on the asymptotic growth of the iteration index, the extant quantities of the different existing growths can be decomposed, wherein each summand corresponds to its respective convergence value.

\begin{align}
    &\lim_{n\to\infty} \frac{\left(1 + \left(1 - \left(1 - \frac{1}{n}\right)^i\right)^3\right) \left(1 - \frac{1}{n}\right)^i n}{1 + \left(1 - \left(1 - \frac{1}{n}\right)^i\right)^3 + n - \left(1 - \left(1 - \frac{1}{n}\right)^i\right)^3 n}=\\\notag
    &=\lim_{n\to\infty} \frac{\left(1 + \left(1 - e^{-i/n}\right)^3\right) e^{-i/n} n}{1 + \left(1 - e^{-i/n}\right)^3 + n - \left(1 - e^{-i/n}\right)^3 n}=\\\notag
    &=\lim_{n\to\infty} \frac{2 n e^{-\frac{i}{n}}}{-n \left(1-e^{-\frac{i}{n}}\right)^3+\left(1-e^{-\frac{i}{n}}\right)^3+n+1}-\frac{3 n e^{-\frac{2 i}{n}}}{-n \left(1-e^{-\frac{i}{n}}\right)^3+\left(1-e^{-\frac{i}{n}}\right)^3+n+1}\\\notag
    &+\frac{3 n e^{-\frac{3 i}{n}}}{-n \left(1-e^{-\frac{i}{n}}\right)^3+\left(1-e^{-\frac{i}{n}}\right)^3+n+1}-\frac{n e^{-\frac{4 i}{n}}}{-n \left(1-e^{-\frac{i}{n}}\right)^3+\left(1-e^{-\frac{i}{n}}\right)^3+n+1}=\\\notag
    &=\lim_{n\to\infty} \frac{n e^{-\frac{i}{n}} \left(3 e^{i/n}-3 e^{\frac{2 i}{n}}+2 e^{\frac{3 i}{n}}-1\right)}{-3 n e^{i/n}+3 n e^{\frac{2 i}{n}}+3 e^{i/n}-3 e^{\frac{2 i}{n}}+2 e^{\frac{3 i}{n}}+n-1}=\\\notag
    &=\lim_{n\to\infty} \frac{\displaystyle n e^{-\frac{i}{n}} \left(2 e^{i/n}-1\right) \left(-e^{i/n}+e^{\frac{2 i}{n}}+1\right)}{\displaystyle-3 n e^{i/n}+3 n e^{\frac{2 i}{n}}+3 e^{i/n}-3 e^{\frac{2 i}{n}}+2 e^{\frac{3 i}{n}}+n-1}=\\\notag
    &=\begin{cases}
    \lim_{n\to\infty} \frac{n}{n+1}=1 & \text{if } i=o(n)\\
    \lim_{n\to\infty} \frac{\left(\left(1-\frac{1}{e}\right)^3+1\right) n}{e \left(\left(1-\frac{1}{e}\right)^3 (-n)+n+\left(1-\frac{1}{e}\right)^3+1\right)}=\frac{\displaystyle (2 e-1) (e (e-1)+1)}{\displaystyle e (3 e (e-1)+1)} & \text{if } i= \Theta(n)\\
    \lim_{n\to\infty} \frac{\displaystyle n e^{-H_n} \left(2 e^{H_n}-1\right) \left(-e^{H_n}+e^{2 H_n}+1\right)}{\displaystyle-3 n e^{H_n}+3 n e^{2 H_n}+3 e^{H_n}-3 e^{2 H_n}+2 e^{3 H_n}+n-1}=\frac{\displaystyle 2}{\displaystyle 3 + 2e^{\gamma}} & \text{if } i=\Theta(nH_n)
    \end{cases}
\end{align}
First, the analysis begins with the cases where the iteration index encompasses all values from 0 to $n$, also referred to as $i=O(n)$, covering the exact growth value or situated at the supremum of the sum interval $\Theta(nH_n)$. These are the easiest to resolve since, due to the substitution of infinitesimals in the first step, the trend of exponential growths can be simplified. Thus, in the case where $i$ grows slower than $n$, all terms will be equivalent to 1 in the limit, resulting in a convergence value of 1 for the respective range of $i$ values. Similarly, when the iteration index matches the exact growth rate of $n$, all its appearances can be replaced by a parametrization like $c\cdot n$ (with $c$ being a constant) to represent any function with equivalent growth, and proceed from there. However, since we have just considered all values of $i$ with growth less than the exact bound to be verified, it is not necessary to consider constant values less than 1, because they have already been counted infinitesimally in the previous case. Likewise, the same applies to constants greater than 1, so calculating the limit when $i=n$ results in a real convergence value, different from 1, and irrational. Lastly, the special case corresponding to the last iteration, equal to the supremum of the range $I(n)$, returns a value with similar properties to the previous case. In summary, the infinitesimal initially introduced can be simplified by acknowledging the similarity between the harmonic number and the logarithm, enabling all exponential terms to be split into the sum of $n$ and a constant term.

\begin{align}
    &\lim_{n\to\infty} \frac{\left(1 + \left(1 - \left(1 - \frac{1}{n}\right)^i\right)^3\right) \left(1 - \frac{1}{n}\right)^i n}{1 + \left(1 - \left(1 - \frac{1}{n}\right)^i\right)^3 + n - \left(1 - \left(1 - \frac{1}{n}\right)^i\right)^3 n}=\\\notag
    &=\lim_{n\to\infty} \frac{n e^{-\ln ^k(n)} \left(\left(1-e^{-\ln ^k(n)}\right)^3+1\right)}{-n \left(1-e^{-\ln ^k(n)}\right)^3+\left(1-e^{-\ln ^k(n)}\right)^3+n+1}\\\notag
    &=\lim_{n\to\infty} -\frac{n e^{-4 \ln ^k(n)}}{-n \left(1-e^{-\ln ^k(n)}\right)^3+\left(1-e^{-\ln ^k(n)}\right)^3+n+1}+\frac{3 n e^{-3 \ln ^k(n)}}{-n \left(1-e^{-\ln ^k(n)}\right)^3+\left(1-e^{-\ln ^k(n)}\right)^3+n+1}\\\notag
    &-\frac{3 n e^{-2 \ln ^k(n)}}{-n \left(1-e^{-\ln ^k(n)}\right)^3+\left(1-e^{-\ln ^k(n)}\right)^3+n+1}+\frac{2 n e^{-\ln ^k(n)}}{-n \left(1-e^{-\ln ^k(n)}\right)^3+\left(1-e^{-\ln ^k(n)}\right)^3+n+1}=\\\notag
    &=\lim_{n\to\infty} \frac{n e^{-\ln ^k(n)} \left(2 e^{\ln ^k(n)}-1\right) \left(-e^{\ln ^k(n)}+e^{2 \ln ^k(n)}+1\right)}{-3 n e^{\ln ^k(n)}+3 n e^{2 \ln ^k(n)}+3 e^{\ln ^k(n)}-3 e^{2 \ln ^k(n)}+2 e^{3 \ln ^k(n)}+n-1}=\frac{2}{3}
\end{align}
Furthermore, the range that remains to be examined is computed separately, as observed above. In this last case, the iteration index generates all values whose growth is between $n$ and $nH_n$, or in other words, its asymptotic bound is in the set $O(nH_n)-O(n)$. To verify the convergence of the limit in all functions within this set, it is necessary to define a parameterization from which these functions are derived. The simplest one, and the one used here, is $n\cdot\log^k(n)$ for real values of $k$ between 0 and 1. In this way, when $k$ reaches its infimum, the resulting growth is equivalent to the one checked previously $\Theta(n)$, and conversely, in the supremum it reaches $\Theta(nH_n)$. Given its continuity, for all values $k\in(0,1)$ the parameterization manages to generate growths in the desired set, allowing for its substitution into the infinitesimals of the expression. Unlike the case $i=\Theta(nH_n)$, this time the resulting logarithm in the exponential terms cannot be simplified, but the complete formula can be expanded and the $Assuming[]$ \cite{WolframAssuming} function in Wolfram can be utilized to computationally confirm that for the values of $k$ in its respective range the limit converges to $\frac{2}{3}$.

\begin{align}
    \lim_{n\to\infty} T_{avg}(n) &= \lim_{n\to\infty} \frac{\left(1 + \left(1 - \left(1 - \frac{1}{n}\right)^i\right)^3\right) \left(1 - \frac{1}{n}\right)^i n}{1 + \left(1 - \left(1 - \frac{1}{n}\right)^i\right)^3 + n - \left(1 - \left(1 - \frac{1}{n}\right)^i\right)^3 n} = \\ \notag
    &=\lim_{n\to\infty} \left(\sum _{i=0}^{n-1} 1\right) + \frac{(2 e-1) (e (e-1)+1)}{e (3 e (e-1)+1)} + \left(\sum _{i=n+1}^{nH_n-1} \frac{2}{3}\right) + \frac{1}{\frac{3}{2}+e^\gamma} = \\ \notag
    &=\lim_{n\to\infty} \left(\sum _{i=n+1}^{nH_n-1} \frac{2}{3}\right) + \left(\sum _{i=0}^{n-1} 1\right) + \frac{(2 e-1) (e (e-1)+1)}{e (3 e (e-1)+1)} + \frac{1}{\frac{3}{2}+e^\gamma} = \\ \notag
    &=\lim_{n\to\infty} \frac{2}{3} \left(n H_n-n-1\right) + n + \frac{(2 e-1) (e (e-1)+1)}{e (3 e (e-1)+1)} + \frac{1}{\frac{3}{2}+e^\gamma}    
\end{align}

At this juncture, knowing that for all values of $i$ that the sum of $T_{avg}(n)$ iterates through, its limit converges to distinct values, the sum can be decomposed based on the intervals that correspond to these convergence values, subsequently obtaining the limit value of the complexity that could not originally be computed by solving the initial sum. First, the constants of the $\Theta(n)$ and $\Theta(nH_n)$ cases are directly replaced, which do not alter the asymptotic growth of the complexity due to their nature. As for the remaining cases, the resolution of their sums is considerably simplified, producing an expression that asymptotically reveals an equivalence the bounds derived from prior analyses with the continuous estimator and the exact formula $c(n,k)$, so it is verified that the ratio between the complexity and the predicted bound $I(n)$ converges to a real value at infinity:

\begin{align}
    \lim_{n\to\infty} \frac{T_{avg}(n)}{nH_n}&=\lim_{n\to\infty} \frac{\frac{2}{3} \left(n H_n-n-1\right) +n+ \frac{(2 e-1) (e (e-1)+1)}{e (3 e (e-1)+1)} + \frac{1}{\frac{3}{2}+e^\gamma}}{nH_n}=\\\notag
    &=\lim_{n\to\infty} \frac{\frac{2}{3} \left(n H_n-n-1\right)+n}{nH_n}=\lim_{n\to\infty} \frac{2}{3}(1-\frac{1}{H_n}-\frac{1}{nH_n})+\frac{1}{H_n}=\frac{2}{3}
\end{align}
As expected, for sufficiently large system sizes, the ratio converges to $\frac{2}{3}$, so it can be concluded that the complexity evaluated from the discrete estimator has the same asymptotic bound as the other cases, thereby suggesting that the methodology works as intended, at least in this type of systems.

\begin{figure}[H]
    \centering
    \begin{subfigure}[b]{0.49\textwidth}
        \centering
        \includegraphics[width=\textwidth,clip]{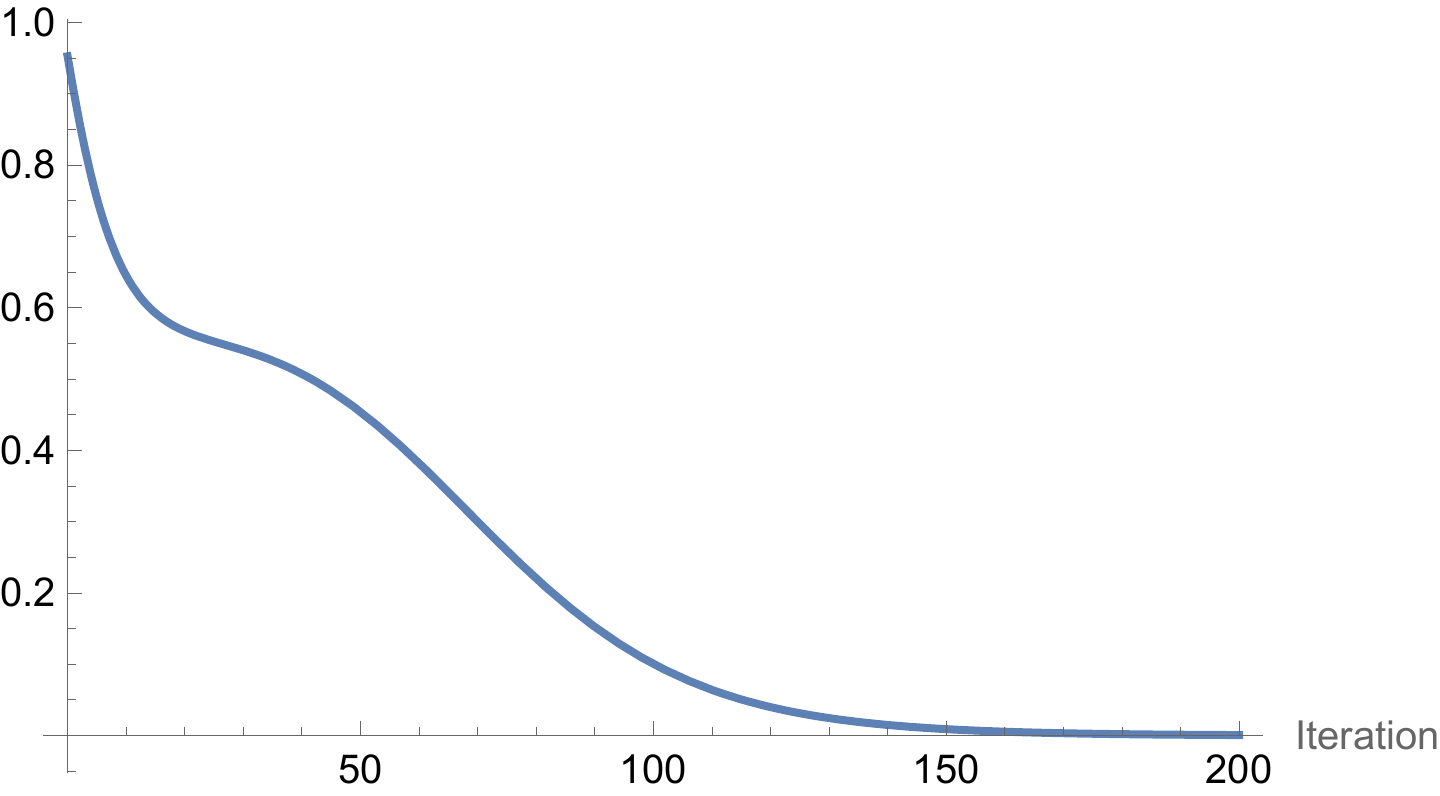}
        \caption{}
        \label{fig:1DDiscreteEstimatorComplexity}
    \end{subfigure}
    \hfill
    \begin{subfigure}[b]{0.49\textwidth}
        \centering
        \includegraphics[width=\textwidth,clip]{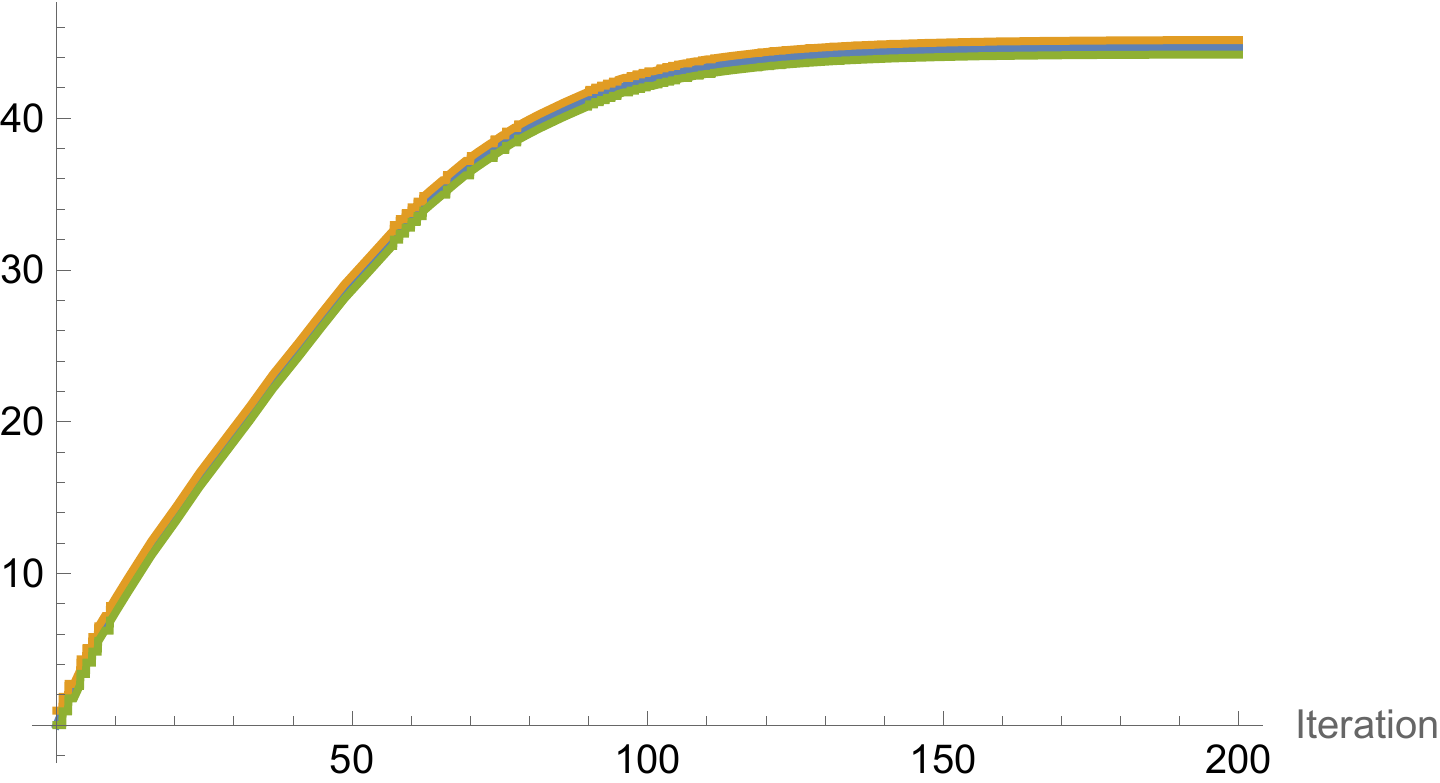}
        \caption{}
        \label{fig:1DDiscreteEstimatorSum1}
    \end{subfigure}
    \caption{(a) $\left(1-\frac{E(i)}{n}\right)\cdot \hat{c}_1(n,E(i))$ insertion complexity plotted for a 1D system of size $n=20$ (b) $T_{avg}(n)$ in its integral formulation plotted in blue for a 1D system of size $n=20$, and in green/orange its discrete lower and upper sum variants, respectively.}
    \label{fig:1DDiscreteEstimatorDual}
\end{figure}

Finally, to graphically elucidate why the cumulative work of each insertion results in the same asymptotic bound regardless of whether it is computed in a continuous or discrete context, we plot the functions that represent the insertion cost, the integral sum of such cost, and the discrete upper and lower sums. On one hand, in the work of each insertion from the discrete estimator, a difference is observed in the decay rate in the first iterations compared to the graph obtained in the initial analysis with the exact expression $c(n,k)$. Despite this, the decrease to 0 as the system size increases is proportional, which is inferred from the same bound obtained for both analyses. On the other hand, the integral sum of such work is noted to coincide with the discrete sums, both upper and lower, of equivalent magnitude. By default, the discrete sum from the original definition is the upper one, although the its alternative is obtained by subtracting the first term, or by evaluating the function at a smaller $n$.

\begin{figure}[H]
    \centering
    \includegraphics[width=10cm,clip]{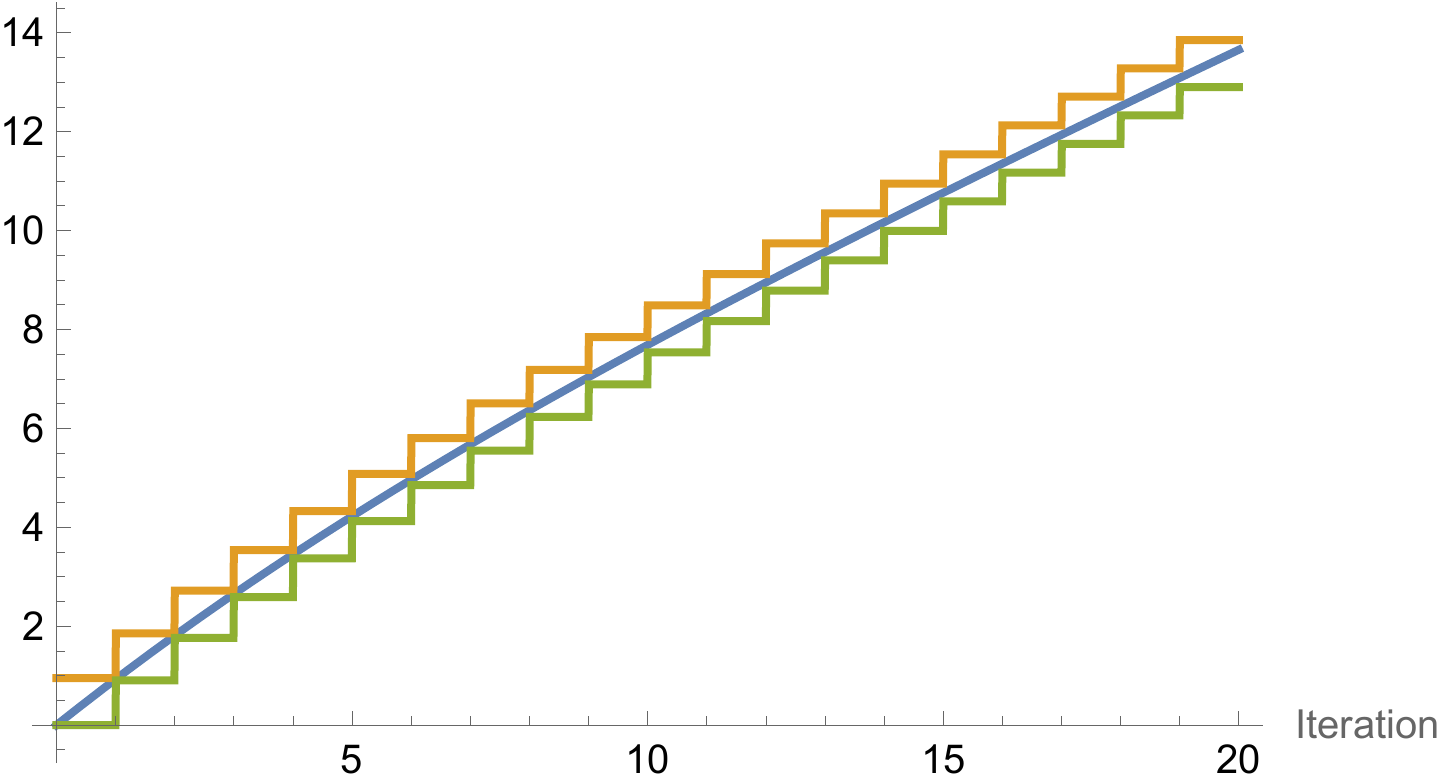}
    \caption{$T_{avg}(n)$ plotted in blue along with its discrete upper and lower sums in orange and green, respectively, for a system of size $n=20$}
    \label{fig:1DDiscreteEstimatorSum2}
\end{figure}

To enhance the observation of this coincidence, despite its inexactitude across all real domain values, the graph is augmented by evaluating the sums in a reduced number of iterations. Above, it is apparent how the amount of area summed by the three, though not identical, grows proportionally with the number of iterations, which leads to the coincidence of the asymptotic bound obtained in both the integral and the discrete sum.

\subsection{Empirical time measurements}
Prior to concluding this section, we proceed to measure the execution time of a percolation process in 1-dimensional systems, which will subsequently be fitted to a model based on the asymptotic bound derived in the theoretical analysis. In this way, we can observe the growth of the execution time cost as the system increases in size, contrasting the previous results.

\begin{figure}[H]
    \centering
    \includegraphics[width=10cm,clip]{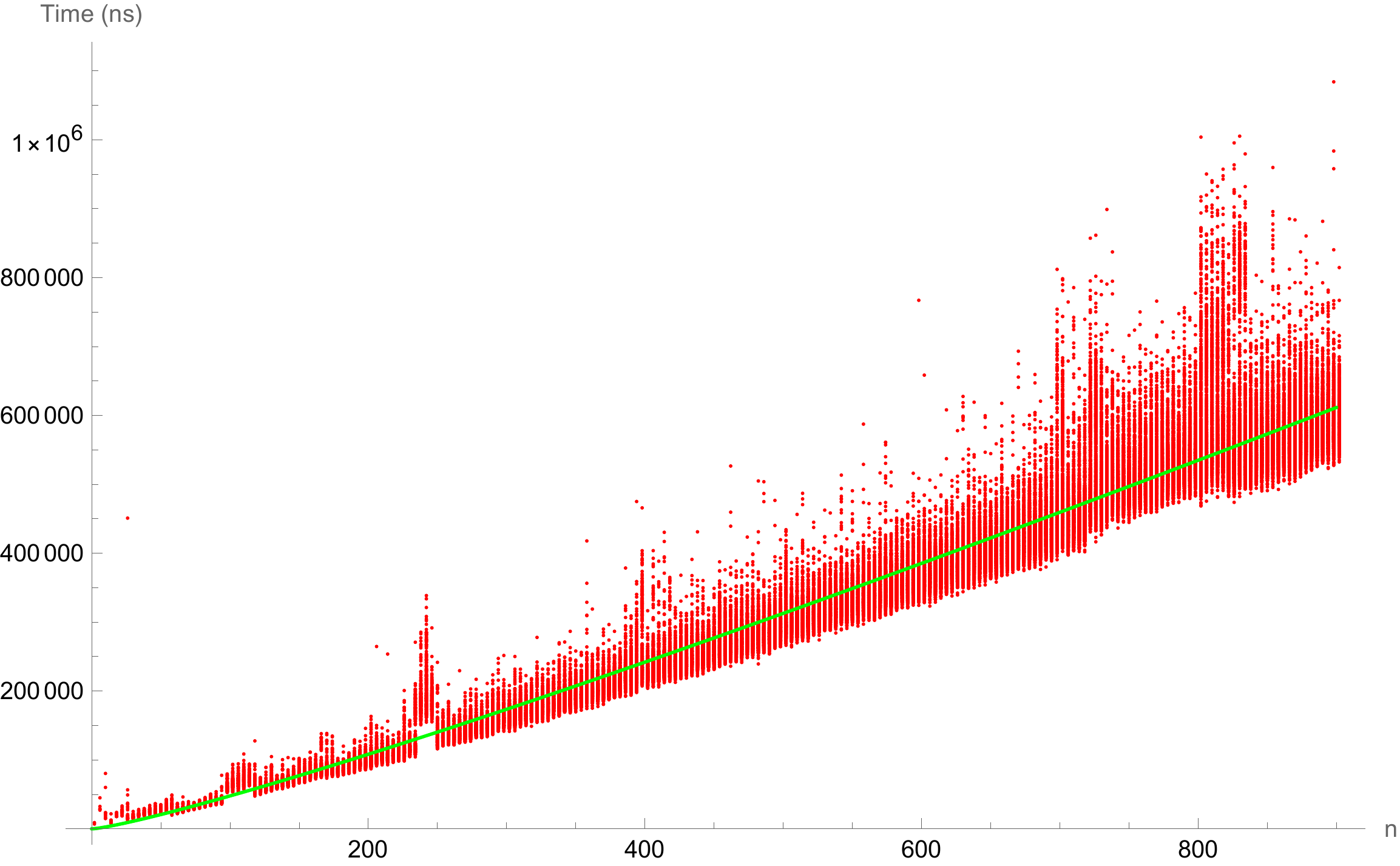}
    \caption{Runtime measurements of a percolative process over a 1D system plotted in red. In green, the fitted model $T_{avg}(n) = 92.1\cdot nH_n$}
    \label{fig:1DTimesFit1}
\end{figure}
Analogous to the procedure employed with the 2-dimensional system, multiple simulations are executed \cite{time_measurements_1D}, each utilizing a random generator initialized with a seed that produces a uniformly random sequence. For each system size, $n/2$ measurements are taken to ensure sufficient data points to detect where the highest density of runtimes accumulates, and thus infer the average complexity. In this case, since we know the complexity bound that the dataset should follow, it is fitted with a model based on such bound, parameterized with a constant that allows it to reach the height of the measurements without varying its asymptotic growth. Additionally, it is worth highlighting that the algorithm used for the measurements implements a clearing of the visited cells register in each iteration, employing a linear data structure such that the number of state changes needed in each clearing equals the number of visited cells, avoiding traversing the entire register each time it resets.
\[
\begin{array}{l|lll}
 \text{} & \text{DF} & \text{SS} & \text{MS} \\
\hline
 \text{Model} & 1 & 1.78911\times 10^{16} & 1.78911\times 10^{16} \\
 \text{Error} & 101788 & 1.76853\times 10^{14} & 1.73746\times 10^9 \\
 \text{Uncorrected Total} & 101789 & 1.8068\times 10^{16} & \text{} \\
 \text{Corrected Total} & 101788 & 2.69234\times 10^{15} & \text{} \\
\end{array}
\]

\[
\begin{array}{l|lllll}
\text{Parameter} & \text{Estimate} & \text{Standard Error} & \text{t-Statistic} & \text{P-Value} & \text{Confidence Interval} \\
\hline
x & 92.0549 & 0.0286871 & 3208.93 & 0. & \{91.9987,92.1111\} \\
\end{array}
\]

As a result of the fit, an adjusted $R^2$ of 0.990212 is attained. In conjunction with the above ANOVA table and the value of the statistical constant parameter $x$, it can be inferred that the model captures most of the variability in the data. Furthermore, graphically, it is observed that the model resides in the region of the dataset with the highest measurement density, aligning with the average case analyzed. This serves as a good indicator that the asymptotic bound of complexity is congruent with the empirical algorithm's growth.

\begin{figure}[H]
    \centering
    \includegraphics[width=10cm,clip]{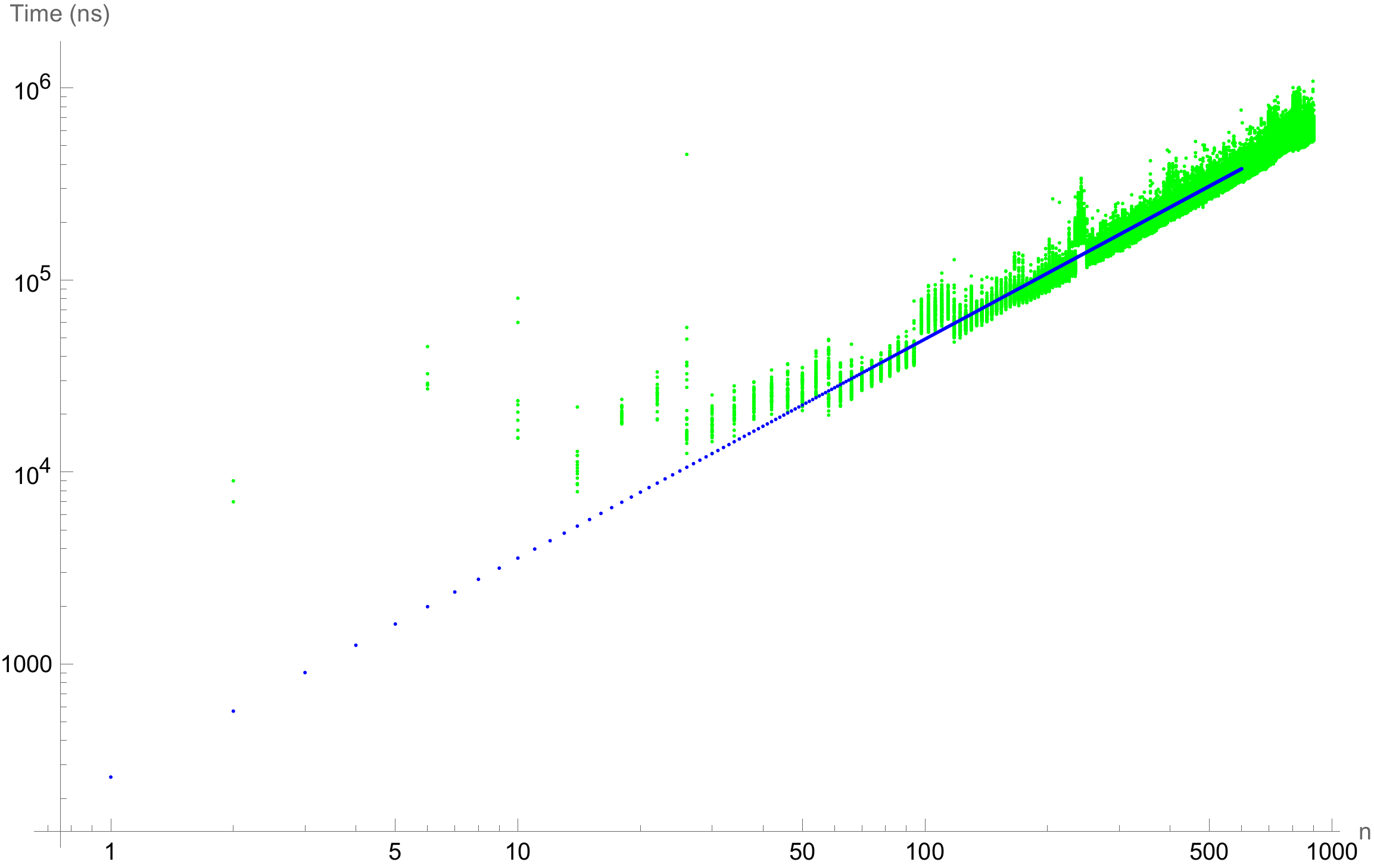}
    \caption{Runtime measurements form Figure 47 under a logarithmic transformation. In blue, the fitted model $T_{avg}(n) = e^{1.14\cdot ln(n) + 5.55}$}
    \label{fig:1DTimesFit2}
\end{figure}

Just in case, besides fitting the dataset with a parameterization of the resultant bound from the temporal analysis, a logarithmic transformation of the data points is performed to allow the use of a model that finds the optimal exponent for a model of the form $n^x$, with the addition of an offset to adjust its height. With this, it will be empirically verified that a model of this form is effective for bounded values of the system size; however, a discrepancy will exist with the fit that reaches the actual bound, because it is the latter that exactly models the average case of the algorithm. Formally, the reason for this phenomenon lies in the nature of the exponent $x$, which, being constant, cannot vary as it should to cause the ratio between it and the actual bound to be 1 when $n \to \infty$.

\begin{align}
    nH_n=n^x\implies x=\frac{\log(nH_n)+2i\pi c}{\log(n)}\colon c\in\mathbb{Z}
\end{align}
Specifically, for the exponential model to be asymptotically equivalent to $I(n)$, the exponent should depend on the size of the system as shown above, which is not considered in this this fitting procedure.

\[
\begin{array}{l|lllll}
 \text{} & \text{DF} & \text{SS} & \text{MS} & \text{F-Statistic}  \\
\hline
 x & 1 & 33045.8 & 33045.8 & 2.80996\times 10^6  \\
 \text{Error} & 101786 & 1197.03 & 0.0117602 & \text{}  \\
 \text{Total} & 101787 & 34242.8 & \text{} & \text{}  \\
\end{array}
\]

\[
\begin{array}{l|lllll}
\text{Parameter} & \text{Estimate} & \text{Standard Error} & \text{t-Statistic} & \text{P-Value} & \text{Confidence Interval} \\
\hline
ln(b) & 5.5533 & 0.00430183 & 1290.91 & 0. & \{5.54487,5.56173\} \\
x & 1.14012 & 0.000680146 & 1676.29 & 0. & \{1.13879,1.14146\} \\
\end{array}
\]
So, with a constant exponent, an adjusted $R^2$ of 0.965043 is achieved and the superior metrics for the model parameters, which despite exhibiting worse generalization in this case, still reach relatively acceptable values.

\begin{figure}[H]
    \centering
    \includegraphics[width=10cm,clip]{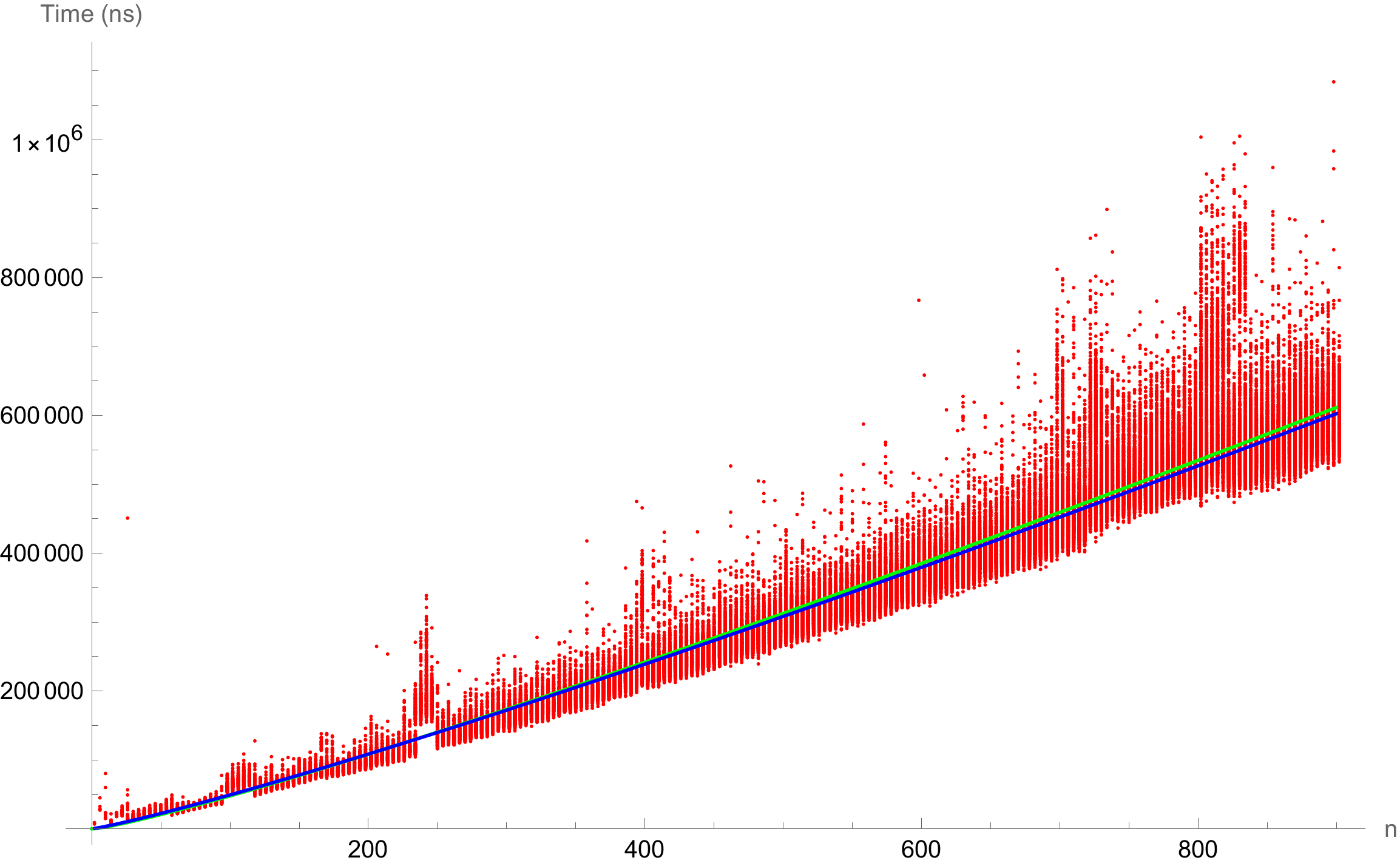}
    \caption{Runtime measurements of the simulations over a 1D system, along with both fitted models. In green, the one from Figure 46 and in green the Figure 47 one.}
    \label{fig:1DTimesFitComparison}
\end{figure}
Finally, when plotting both models on the dataset, a slight difference is discerned, which increases as the system size grows. This occurs because they fit the data well with a bounded $n$, meaning they are not asymptotically equivalent but show a similarity in small system simulations. Likewise, since the best-case bound of any algorithm coincides in this case with the average time complexity, it can be seen how the function that bounds the dataset from below grows similarly to $I(n)$, as they are asymptotically comparable. This arises mainly because, in 1 dimension, the only path that can emerge in the system has the same properties as the best-case path of higher dimensions, so the average of all possible insertion sequences according to their probability of occurrence results in the same growth. Additionally, above the graph of both models, it can be observed how there are data points that tend to follow the worst-case growth function, although the low probability of their occurrence causes the density of measurements to decline for growths close to the upper bound $n^2$.
\section{2-dimensional case}
For the moment, the time complexity of the algorithm has been analyzed in the edge case where the system is formed by a matrix of size $(n,1)$, which can also be denoted as a one-dimensional system due to its structure. This will serve for future analyses, both to validate the used methodology and to compare the results and understand the foundation of its origin. However, the aim is to analyze what happens in a 2-dimensional system, primarily in matrices with an aspect ratio of 1, although most metrics, and even results regarding time complexity with different aspect ratios, vary depending on the size of the system, so the analysis is adaptable to such systems depending on the product of the sizes of its sides. Therefore, for this purpose, the same methodology as in the previous section will be applied to analyze in detail the worst case in which the algorithm may fall and the average case, which will provide the necessary information for its contrast with other implementations, although at this point it is feasible to infer some of the variants that exhibit equivalent asymptotic growth.

\subsection{Average cluster size}
\label{subsec:AverageClusterSize2D}
To begin with, in 2-dimensional systems we do not have an exact formula for determining the average cluster size, although given the restrictions that this magnitude must meet, a very similar expression to the exact formula for 1-dimensional systems is obtained:
\begin{align}    
    c(n,k)=\frac{s(n,k)}{N(n,k)}\approx \frac{n^2}{n^2-k+1}
\end{align}
Presently, we only have the above formula, and the two estimators from the scalar field constructed earlier, which contain the probabilities that a traversal over a cluster has a certain length. Although none of these expressions have been proven correct in modeling the desired magnitude, their properties such as asymptotic behavior as $n \to \infty$ with respect to the asymptotic growth of $k$ are preserved, as we will see later, as well as restrictions or monotonicity. This will be more beneficial for analysis than for finding an exact expression of the average cluster size, as knowing it with certainty implies having enough information about the state space to generalize, from the extreme cases of the average size, an exact value for all valid $n$ and $k$, which is more complicated than in the case of one-dimensional systems. Notwithstanding the difficulty, we can proceed in the same manner as in the previous section, that is, starting from the definition based on the ratio between the total size accumulated by all the clusters of a state space restriction, with $k$ elements, and the total number of clusters that generate it. Therefore, first, the initial values of $c(n,k)$ are calculated for small quantities of the number of elements, which will be simpler and may lead to the identification of a possible pattern with which to generalize to larger quantities.

\begin{align}    
    c(n,1)&=1\\ 
    c(n,2)&=\frac{\displaystyle 2 \binom{n^2}{2}}{\displaystyle 2 \left(\binom{n^2}{2}-(2 (n-1) (n-1)+2 n (n-1))\right)+2 (n-1) (n-1)+2 n (n-1)}\\
    c(n,3)&=3 \binom{n^2}{3} (2 (\binom{n^2}{3}-\frac{1}{6} (n-2) \left(n^4+3 n^3-20 n^2-30 n+132\right) (n-1)-(2 (n-2) (n-2)  \\\notag 
    &+8 (n-1) (n-2)+4 (n-1) (n-2)+2 n (n-2)+4 (n-1) (n-1))) \\\notag 
    &+\frac{3}{6} (n-1) \left(n^4+3 n^3-20 n^2-30 n+132\right) (n-2)+2 (n-2) (n-2)\\\notag 
    &+8 (n-1) (n-2)+4 (n-1) (n-2)+2 n (n-2)+4 (n-1) (n-1))^{-1}
\end{align}

As can be seen, the term $s(n,k)$ in the numerator exhibits the same form as in 1 dimension, with the sole difference being that the number of elements in the system corresponds to the size of a matrix. With this, the total size of all clusters is accounted for, which in turn is divided by the $N(n,k)$ of each case. In the first 2 cases, it is easy to count the arrangements of a cluster of size 1, or 2, even 2 clusters of size 1. Yet, starting from 3 elements, the polyplet \cite{WeissteinPolyplet,MathSEAnswer2024,Karonen2018} that constitutes the largest cluster has a large number of variations, such as rotations, reflections, etc. Even considering the translations of a fixed polyplet of that size, we would still have the problem that there is no known formula, for now, for counting how many exist. Additionally, even knowing how many polyplets \cite{Kamenetsky2015,ReidRectifiablePolyominoes} exist for a given size, it would be necessary to know exactly how many of them occupy a certain space within the system, in order to infer their translations, which is the point of our inquiry. Therefore, under these conditions, the upper expressions are simplified in order to find a pattern, analogous to what was done for one-dimensional systems.

\begin{align}    
    c(n,1)&=1\\ 
    c(n,2)&=\frac{n^2 (n+1)}{n^3+n^2-4 n+2}=\frac{n^2 (n+1)}{n \left(n^2+n-4\right)+2}\\
    c(n,3)&=\frac{n^2 \left(n^3+n^2-2 n-2\right)}{n^5+n^4-10 n^3+2 n^2+24 n-16}=\frac{n^2 (n+1) \left(n^2-2\right)}{n \left(n \left(n \left(n^2+n-10\right)+2\right)+24\right)-16}\\\notag
    \vdots
\end{align}

In this case, there are 2 possible forms for each $c(n,k)$: one where the denominator is fully expanded, providing the opportunity to search for a generating function of its coefficients, signs, or degree with respect to $k$, and another form where both terms are factored, although not completely. The expanded form offers more options for generalization, as it allows the analysis of the parameters defining both polynomials, notwithstanding the persistent factor of $n^2$ in the numerator, which in turn indicates that it is divisible by that amount. Additionally, it is worth noting that the degrees of both polynomials are equal, assuming they exist for all valid $n$ and $k$, so if this property holds for larger numbers of elements, it would suggest that the average cluster size tends to 1, given that the coefficients of the term with the highest growth are 1. However, in light of these considerations, it is not possible to generalize $c(n,k)$ for 2-dimensional systems from intermediate evaluations, not even with the help of Wolfram's computational methods. Therefore, another way to obtain an exact expression for this magnitude involves formulating the recursion we used to derive the expression for one-dimensional systems, adapted for enumerating clusters in square matrices.

\begin{align}
    N(n,k)= \begin{cases} 
0 & \text{if } k \leq 0 \text{ or } k>n^2 \\
n^2 & \text{if } k=1
\end{cases}
\end{align}

First, since the only unknown is the function that counts the number of clusters, the constraints that it must satisfy are defined first, which we will use later in solving its recurrence relation. The foremost of these ensures that there are no clusters in invalid values of the number of elements, such as when it is less than 1 or greater than the size of the system. As for the other, it mainly constitutes the stopping condition with respect to the parameter $k$, since it will decrease in the recursive definition.

\begin{align}
    N(n,k)&=N(n-1,k)+N(n-1,k-1)+4\binom{(n-1)^2-2}{k-1}+\binom{(n-1)^2-1}{k-1}\\\notag
    &+(2n-1-5)\binom{(n-1)^2-3}{k-1}+\cdots+\left(N(n-1,k-(2n-1))+\binom{(n-2)^2}{k-(2n-1)}\right)
\end{align}

Once the base cases are formalized, the number of clusters $N(n,k)$ is constructed from its evaluation in the system of the immediately smaller size. Nonetheless, the maximum number of elements in both systems differs, specifically by $2n-1$ cells, in which from 0 to $2n-1$ elements can be placed. Hence, it is necessary to quantify, out of all these possibilities, how many clusters each one adds to the total count. In the event that there are no elements placed in the $2n-1$ additional cells of the system of size $(n-1)^2$, the number of clusters that this subsystem contains is directly the recursive evaluation $N(n-1,k)$. But if one of the $k$ elements is found outside the subsystem in the additional cells, it is not enough to count the clusters formed by the $k-1$ elements in the subsystem via the term $N(n-1,k-1)$; it is also essential to include the size-1 clusters formed by the leftover element when its neighboring cells in the subsystem are empty. Above, the inclusion of these additional clusters is shown, which with just one element is simple to enumerate. However, for larger numbers, it becomes incrementally more elaborate, because with 2 elements, special cases must be considered where the cluster reduces the number of neighboring cells when it is in one of the corners of the matrix. Moreover, finding a generating function for the amount of clusters added to each recursive term $N(n-1,k-i)$ is not as immediate as in the analysis of 1-dimensional systems. Therefore, in the superior definition, only the last case is shown where the $2n-1$ cells are occupied by elements, which is immediately evident to find. Since all intermediate summands in the previous plan cannot be generally constructed, the approach is approximated as closely as possible to the actual magnitude. Thus, by leveraging the exact expression for $c(n,k)$ in 1 dimension, we can find out what the exact expression of $N(n,k)$ will be for systems in that dimension, which facilitates the quantification of clusters with $k$ elements existing in the $2n-1$ additional cells that prevent us from formalizing the exact cluster count in two dimensions.

\begin{align}    
    c(n,k)=\frac{s(n,k)}{N(n,k)}= \frac{n}{n-k+1}\implies N(n,k)=\frac{s(n,k)}{c(n,k)}=\frac{\displaystyle k\binom{n}{k}(n-k+1)}{n}
\end{align}
In one dimension, the number of clusters that form $k$ elements in $n$ cells is derived from the expression above. In this way, an estimator for this magnitude can be constructed in 2-dimensional systems with an aspect ratio equal to 1.

\begin{align}
    \hat{N}_0(n,k)&=\hat{N}_0(n-1,k)+\frac{\displaystyle 0\binom{2n-1}{0}(2n-1-0+1)}{2n-1}\binom{(n-2)^2}{k}\\\notag
    &+\hat{N}_0(n-1,k-1)+\frac{\displaystyle 1\binom{2n-1}{1}(2n-1-1+1)}{2n-1}\binom{(n-2)^2}{k-1}\\\notag
    &+\hat{N}_0(n-1,k-2)+\frac{\displaystyle 2\binom{2n-1}{2}(2n-1-2+1)}{2n-1}\binom{(n-2)^2}{k-2}\\\notag
    &+\cdots+\hat{N}_0(n-1,k-(2n-1))+\frac{\displaystyle (2n-1)\binom{2n-1}{2n-1}(2n-1-(2n-1)+1)}{2n-1}\binom{(n-2)^2}{k-(2n-1)}\\\notag
\end{align}
In summary, since it is not feasible to account for the exact amount of clusters contributed by each amount of elements located in the $2n-1$ additional cells, an approximation can be derived under the assumption that, in all combinations of the elements their neighboring cells are constant. That is, in the edge cases where the clusters are located in the corners, their neighborhood with respect to the cells of the $n-1$ side subsystem is usually reduced, which is the main reason why this counting is complicated. Thus, by ignoring this variation and establishing a constant number of neighboring cells that must be empty for the cluster to be counted, we achieve an approximation to the actual magnitude. Additionally, as the constant number of neighboring cells coincides with the maximum number of them that exists when the cluster traverses one of the edges without reaching the corners, we can infer that the total amount $\hat{N}_0(n,k)$ is less than the true one. And, more notably, the number of edge cases, or corners of the matrix, is constant with respect to the increase in the size of the edges of the matrix, so that of the $2n-1$ additional cells, only 4 or 5 cells will influence the neighborhood, also depending on the size of the cluster that traverses them. With this, it could be hypothesized that the estimator has an asymptotic growth equivalent to the true magnitude, although the demonstration of this proposition could be even less feasible to solve than addressing the recursion itself.

\begin{align}
    \hat{N}_0(n,k)=\sum _{i=0}^{2n-1} \hat{N}_0(n-1,k-i)+\frac{\displaystyle i\binom{n}{i}(n-i+1)}{n}\binom{(n-2)^2}{k-i}
\end{align}
Therefore, to visually infer if it makes sense to assert that the asymptotic growth of the estimator and the actual magnitude are comparable, the number of clusters returned by the estimator for an arbitrary system size is evaluated:

\begin{figure}[H]
    \centering
    \includegraphics[width=10cm,clip]{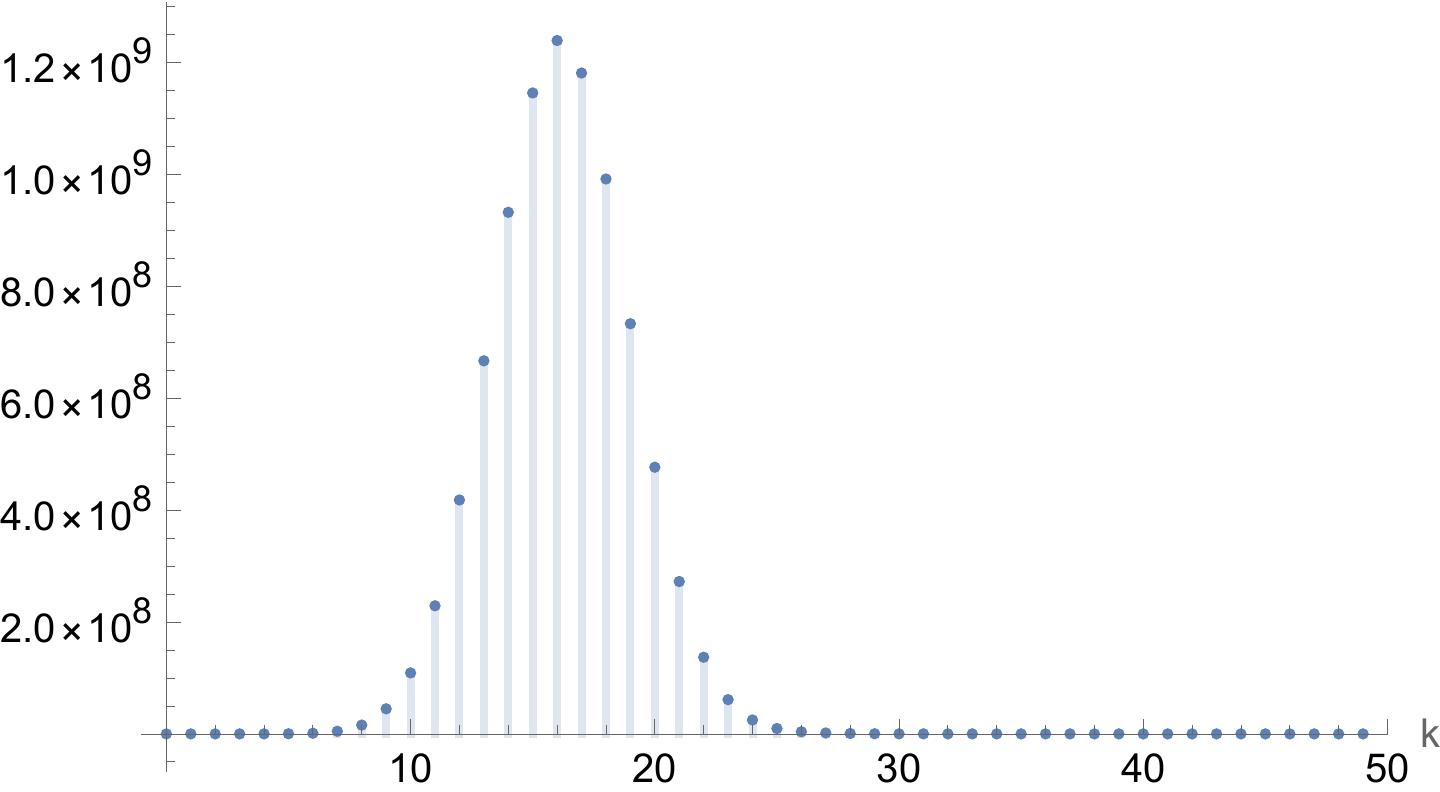}
    \caption{$\hat{N}_0(n,k)$ estimator plotted for a 2D system with a square matrix of side $n=7$ for $0\leq k\leq n^2$ }
    \label{fig:2DClusterNumberRecursion1}
\end{figure}
As noticed, the number of clusters, if normalized, would be similar to a density function that characterizes the probability of the presence of a cluster within the system given a certain $k$. Similarly, to determine the potential form of this density function, even if it may not be normalized, it is necessary to resolve the non-recursive term of the sum in $\hat{N}_0(n,k)$, as this determines the amount of clusters contributed by each recursive term.
\begin{align}
    &\sum _{i=0}^{2n-1} \frac{\displaystyle i\binom{n}{i}(n-i+1)}{n}\binom{(n-2)^2}{k-i}=\\\notag
    &=2 (n-1) \binom{n}{2 n} \binom{(n-2)^2}{k-2 n} \, _3F_2\left(1,n,2 n-k;2 n+1,n^2-2 n-k+5;1\right)\\\notag
    &-\binom{(n-2)^2}{k-1} \, _3F_2\left(2,1-k,1-n;1,n^2-4 n-k+6;1\right)\\\notag
    &+3 \binom{n}{2 n+1} \binom{(n-2)^2}{k-2 n-1} \, _3F_2\left(2,n+1,-k+2 n+1;2 n+2,n^2-2 n-k+6;1\right)\\\notag
    &+\frac{\binom{n}{2 n+1} \binom{(n-2)^2}{k-2 n-1} \, _4F_3\left(2,2,n+1,-k+2 n+1;1,2 n+2,n^2-2 n-k+6;1\right)}{n}\\\notag
    &-\frac{\binom{n}{2 n+1} \binom{(n-2)^2}{k-2 n-1} \, _3F_2\left(2,n+1,-k+2 n+1;2 n+2,n^2-2 n-k+6;1\right)}{n}\\\notag
    &+\frac{n \Gamma \left(n^2-3 n+4\right)}{\Gamma (k) \Gamma \left(n^2-3 n-k+5\right)}+\frac{\Gamma \left(n^2-3 n+4\right)}{\Gamma (k) \Gamma \left(n^2-3 n-k+5\right)}
\end{align}
Upon using Wolfram's $Sum[]$ function to computationally build a closed form of the sum, it can be visualized for different $n$, which will indicate the extent to which the recursive terms contribute to the final count compared to the sum of the non-recursive terms.
\begin{figure}[H]
    \centering
    \includegraphics[width=9cm,clip]{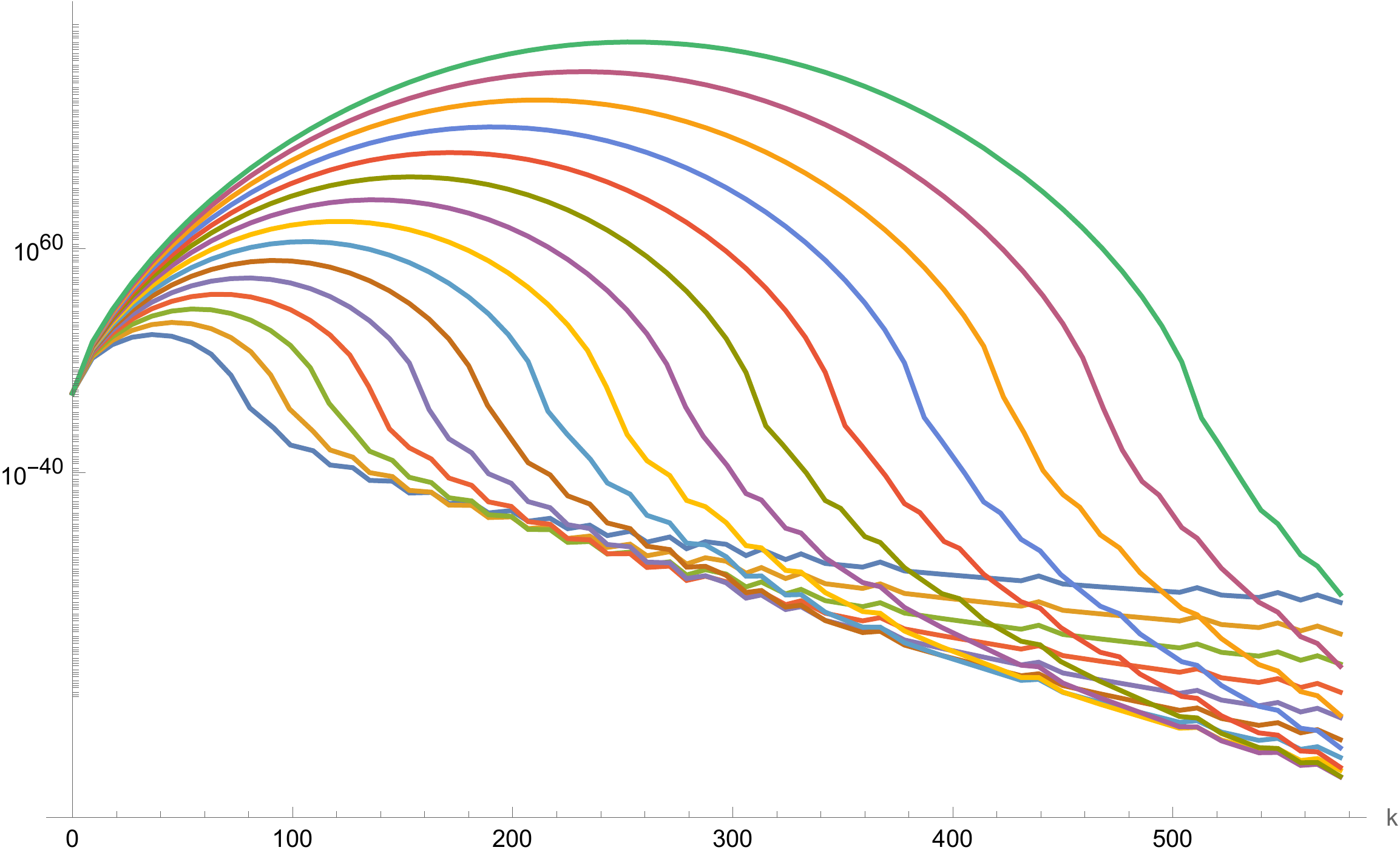}
    \caption{Non recursive term of the estimator $\hat{N}_0(n,k)$ plotted for multiple 2D systems with a square matrix of side $10\leq n\leq24$, from the left plot in blue to the right one in green, with $0\leq k\leq 24^2$}
    \label{fig:2DClusterSizePartialSum1}
\end{figure}
As can be seen, the difference between each pair of evaluations for contiguous system sizes prevents comparison between the different graphs, so they are visualized on a logarithmic scale. Thus, at first glance, they all have the same shape as the possible density function being sought, which is consistent with the recursive definition, since each evaluation of $\hat{N}_0(n,k)$ adds a linear amount of recursive terms, each of them contributing a number of clusters equal to that represented in the upper figure. With this, without exactly knowing the increase in asymptotic growth contributed by the recursive terms compared to those visualized in Figure 50, we can consider the sum of the non-recursive terms as an approximation to the estimator, given that they have a very similar form and asymptotic growth whose difference is mainly due to the contribution made by the sum of recursive terms. However, such a difference can be large enough that by ignoring it, the asymptotic growth of the estimator causes $c(n,k)$ not to return results consistent with the actual magnitude. Therefore, there is a possibility of considering other estimators with which we can infer or at least bound such magnitude.

\begin{align}
    \hat{N}_1(n,k)&=\hat{N}_1(n-1,k)+\frac{\displaystyle 0\binom{2n-1}{0}(2n-1-0+1)}{2n-1}\binom{(n-1)^2-0}{k-0}\\\notag
    &+\hat{N}_1(n-1,k-1)+\frac{\displaystyle 1\binom{2n-1}{1}(2n-1-1+1)}{2n-1}\binom{(n-1)^2-1}{k-1}\\\notag
    &+\hat{N}_1(n-1,k-2)+\frac{\displaystyle 2\binom{2n-1}{2}(2n-1-2+1)}{2n-1}\binom{(n-1)^2-2}{k-2}+\cdots\\\notag
    &+\hat{N}_1(n-1,k-(2n-1))+\frac{\displaystyle (2n-1)\binom{2n-1}{2n-1}(2n-1-(2n-1)+1)}{2n-1}\binom{(n-1)^2-(2n-1)}{k-(2n-1)}\\\notag
\end{align}
Returning to the estimator's approach, there exists a possibility to further streamline, conceptually, the counting of clusters that add the elements located in the $2n-1$ additional cells. As seen in the new upper estimator, for each combination of $k$ elements over the additional cells, regardless of the number of clusters they form among them, it is assumed that there is a quantity of empty cells contiguous to these elements located in the subsystem of the recursive case. For example, if there are 2 elements in the additional space to the subsystem, as many clusters are added to the count as the number of clusters of these 2 elements existing in a 1-dimensional system of size $2n-1$, multiplied by the combinations of the remaining $k-2$ elements of the recursive case over the subsystem of size $n-1$ with 2 empty cells in all combinations of the 2 elements.
\begin{align}
    \hat{N}_1(n,k)=\sum _{i=0}^{2n-1} \hat{N}_1(n-1,k-i)+\frac{\displaystyle i\binom{n}{i}(n-i+1)}{n}\binom{(n-1)^2-i}{k-i}
\end{align}
In this way, an overestimation of the actual number of clusters occurs since the assumption that there are always the same number of empty cells in the subsystem as elements in the additional space leads to the diagonal neighboring cells to the extreme elements of each cluster located in the $2n-1$ cells of such space not being counted. Therefore, by neglecting them, the space of the subsystem on which the combinations of the remaining $k-i$ elements are counted increases, resulting in an additional number of clusters that, fortunately, we can consider as an upper bound for this magnitude. That is, by ignoring the diagonal neighborhood presented by the extreme elements of each cluster in the additional space with the cells of the subsystem, we ensure that the number of combinations of the remaining elements over the valid space of the subsystem is maximal, always maintaining a dependency on the size of each cluster in the additional space, ensuring minimal deviation from the actual measure.
\begin{figure}[H]
    \centering
    \includegraphics[width=10cm,clip]{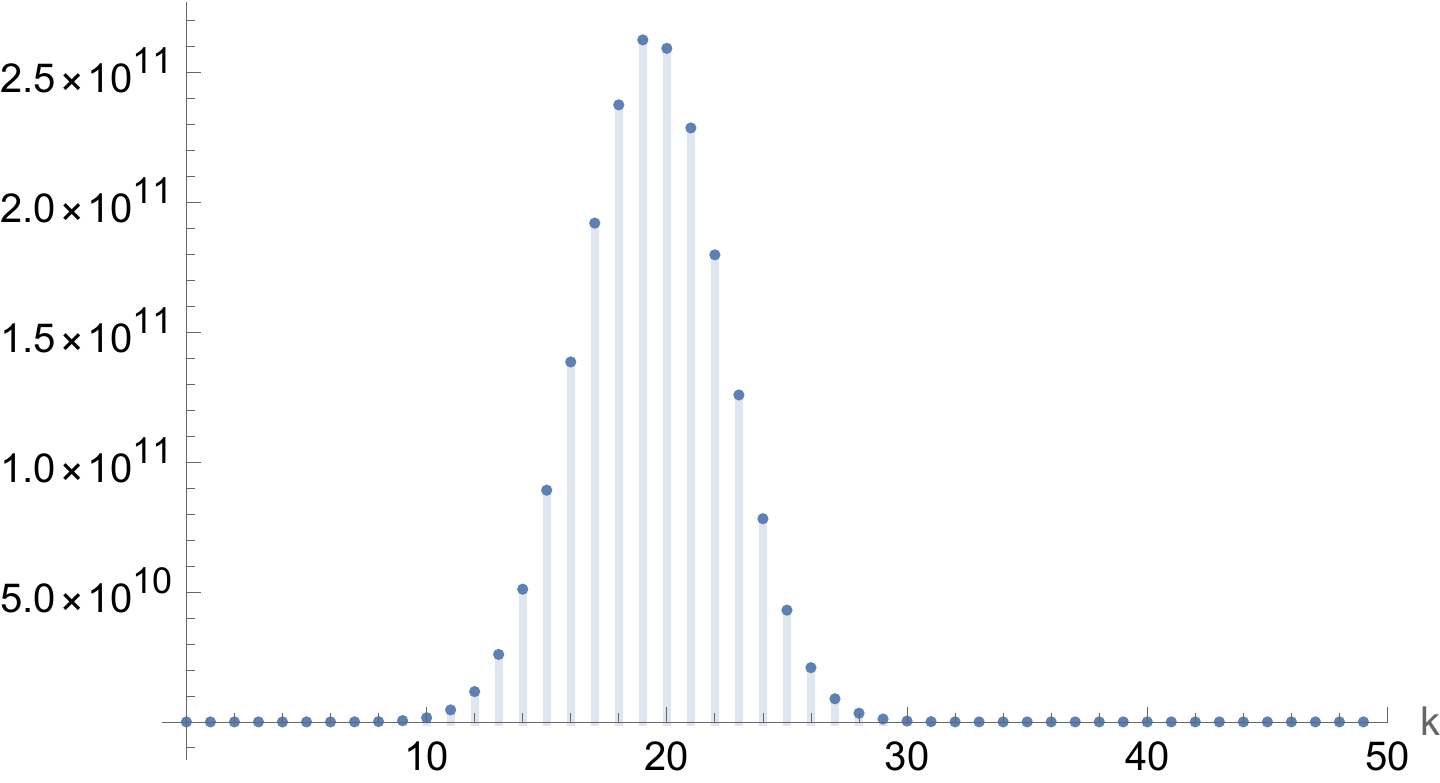}
    \caption{$\hat{N}_1(n,k)$ estimator plotted for a 2D system with a square matrix of side $n=7$ for $0\leq k\leq n^2$ }
    \label{fig:2DClusterNumberRecursion2}
\end{figure}
Through graphical analysis, it is verified that the new estimator provides a function with much larger values in the specific case where $n=7$, so it is consistent with what was assumed in its construction. Thus, after having its recursive definition posed, it is possible to proceed in the same way by solving the sum of its non-recursive terms:

\begin{align}
    &\sum _{i=0}^{2n-1} \frac{\displaystyle i\binom{n}{i}(n-i+1)}{n}\binom{(n-1)^2-i}{k-i}=\\\notag
    &=2 (n-1) \binom{n}{2 n} \binom{n^2-4 n+1}{k-2 n} \, _3F_2\left(1,n,2 n-k;2 n+1,-n^2+4 n-1;-1\right)\\\notag
    &-\binom{(n-2) n}{k-1} \, _3F_2\left(2,1-k,1-n;1,2 n-n^2;-1\right)\\\notag
    &+3 \binom{n}{2 n+1} \binom{(n-4) n}{k-2 n-1} \, _3F_2\left(2,n+1,-k+2 n+1;2 n+2,4 n-n^2;-1\right)\\\notag
    &+\frac{\binom{n}{2 n+1} \binom{(n-4) n}{k-2 n-1} \, _4F_3\left(2,2,n+1,-k+2 n+1;1,2 n+2,4 n-n^2;-1\right)}{n}\\\notag
    &-\frac{\binom{n}{2 n+1} \binom{(n-4) n}{k-2 n-1} \, _3F_2\left(2,n+1,-k+2 n+1;2 n+2,4 n-n^2;-1\right)}{n}\\\notag
    &+n \binom{(n-2) n}{k-1} \, _2F_1(1-k,1-n;-((n-2) n);-1)+\binom{(n-2) n}{k-1} \, _2F_1(1-k,1-n;-((n-2) n);-1)
\end{align}
Regarding its solution, it is not immediately clear which of its terms determines its asymptotic growth, which would be useful for simplifying the complete recurrence solution. Furthermore, given that we know how many terms are summed for each evaluation $\hat{N}_1(n,k)$, we could infer the asymptotic growth of the upper bound of the number of clusters existing in a system with $k$ elements, and therefore an approximation to $c(n,k)$. For this reason, we proceed to find an asymptotically equivalent bound to the non-recursive term of the recurrence, specifically the form that it takes in each summand using the function $Asymptotic[]$ from Wolfram \cite{WolframAsymptotic}, as it is a lengthy process to conduct:

\begin{align}
    &[n\to\infty]\enspace \frac{\displaystyle i\binom{n}{i}(n-i+1)}{n}\binom{(n-1)^2-i}{k-i}\sim \\\notag
    &\sim\frac{n^{-i+2 k-2} \left(3 i^4-22 i^3+3 i^2 (8 k-4 n+7)+i (-72 k+36 n+70)+36 k^2-12 k (4 n+5)+24 n (n+1)\right)}{24 \Gamma (i) \Gamma (-i+k+1)}\sim\\\notag
    &\sim n^{2 (2-i)+i-2 (2-k)-1} \left(\frac{i^2-3 i+4 k-2}{2 (i-k) \Gamma (i) \Gamma (k-i)}+\frac{-3 i^4+22 i^3-24 i^2 k-21 i^2+72 i k-70 i-36 k^2+60 k}{24 n (i-k) \Gamma (i) \Gamma (k-i)} \right.\\ \notag 
    & \left.+\frac{n}{(k-i) \Gamma (i) \Gamma (k-i)}\right)
\end{align}
Now, with an asymptotically equivalent bound to the form of each summand of the non-recursive part as $n \to \infty$, which is the accumulation point of interest in the subsequent analysis, it is possible to use its last shown formulation to decompose the sum of non-recursive terms into simpler ones that provide an asymptotic bound for their total sum. Therefore, since it is this sum that determines the growth of the estimator according to its recurrence equation, we proceed with its resolution:

\begin{align}
    &[n\to\infty]\enspace \sum _{i=0}^{2n-1}\frac{\displaystyle i\binom{n}{i}(n-i+1)}{n}\binom{(n-1)^2-i}{k-i}\sim\\\notag
    &\sim\sum _{i=0}^{2n-1} \left(n^{2 (2-i)+i-2 (2-k)-1} \frac{i^2-3 i+4 k-2}{2 (i-k) \Gamma (i) \Gamma (k-i)}\right) + \sum _{i=0}^{2n-1} \left(n^{2 (2-i)+i-2 (2-k)-1}\frac{n}{(k-i) \Gamma (i) \Gamma (k-i)}\right) \\\notag
    &+ \sum _{i=0}^{2n-1} \left(n^{2 (2-i)+i-2 (2-k)-1}\frac{-3 i^4+22 i^3-24 i^2 k-21 i^2+72 i k-70 i-36 k^2+60 k}{24 n (i-k) \Gamma (i) \Gamma (k-i)}\right)    
\end{align}
First, it is decomposed into simpler fractional forms that facilitate the summation of each of the asymptotic bound's terms, allowing them to be subsequently added to the total. Nevertheless, the only sum that has a closed form is the following:
\begin{align}
    \sum _{i=0}^{2n-1} n^{2 (2-i)+i-2 (2-k)-1}\frac{n}{(k-i) \Gamma (i) \Gamma (k-i)}=n^k \left(\frac{(n+1)^{k-1}}{(k-1) \Gamma (k-1)}-\frac{n^{k-2 n} \, _2F_1\left(1,2 n-k;2 n;-\frac{1}{n}\right)}{(k-2 n) \Gamma (2 n) \Gamma (k-2 n)}\right)
\end{align}
As for the others, a solution has not been identified in the same way as in the superior case, since they contain polynomials dependent on the index of the sum, as well as gamma functions in the denominator whose arguments depend on the index and the number of elements in the system, which despite being constant during the sum, can vary in value and asymptotic growth.

\begin{dmath}
    [n\to\infty]\enspace \sum _{i=0}^{2n-1}\frac{\displaystyle i\binom{n}{i}(n-i+1)}{n}\binom{(n-1)^2-i}{k-i}\sim \pi  \left(\frac{864 k^2-1800 k-1991}{2304 \sqrt{2} \pi ^2 n}-\frac{36 k-13}{48 \sqrt{2} \pi ^2}+\frac{n}{2 \sqrt{2} \pi ^2}\right) n^{-2 (2-k)} \sin (\pi  n) \sin (\pi  (k-1)-2 \pi  n) 2^{-\left\lfloor \frac{\arg (-k+2 n+2)}{2 \pi }+\frac{1}{2}\right\rfloor -\left\lfloor \frac{\arg (n+1)}{2 \pi }+\frac{1}{2}\right\rfloor +1} (-\csc (\pi  n))^{\left\lfloor \frac{\arg (n+1)}{2 \pi }+\frac{1}{2}\right\rfloor } (-\csc (\pi  (k-2)-2 \pi  n))^{\left\lfloor \frac{\arg (-k+2 n+2)}{2 \pi }+\frac{1}{2}\right\rfloor } \exp \left(\left(-k+2 n+\frac{3}{2}\right) \log \left(2 n (-1)^{\left\lfloor \frac{\arg (-k+2 n+2)}{2 \pi }+\frac{1}{2}\right\rfloor }\right)+\left(n+\frac{1}{2}\right) \log \left(n (-1)^{\left\lfloor \frac{\arg (n+1)}{2 \pi }+\frac{1}{2}\right\rfloor }\right)+\frac{6 k^2-18 k+239}{24 n}+n (-5 \log (n)-2-2 \log (2))+6\right) \, _3F_2\left(1,n+1,-k+2 n+1;2 n+2,4 n-n^2;-1\right)+n^{-2 (2-k)} \left(\frac{k^2-8 k+12}{3 n \Gamma (k-1)}-\frac{n^2}{(k-1) \Gamma (k-1)}+\frac{2 n}{\Gamma (k-1)}-\frac{3 (k-2)}{2 \Gamma (k-1)}\right) \, _3F_2\left(2,1-k,1-n;1,2 n-n^2;-1\right)+n^{-2 (2-k)} \, _2F_1\left(1-k,1-n;2 n-n^2;-1\right) \left(\frac{-k^2+8 k-12}{3 n \Gamma (k-1)}+\frac{n^2}{(k-1) \Gamma (k-1)}-\frac{2 n}{\Gamma (k-1)}+\frac{3 (k-2)}{2 \Gamma (k-1)}\right)+n^{-2 (2-k)} \, _2F_1\left(1-k,1-n;2 n-n^2;-1\right) \left(\frac{-k^2+8 k-12}{3 \Gamma (k-1)}-\frac{5 \left(k^3-k^2-14 k+24\right)}{24 n \Gamma (k-1)}+\frac{n^3}{(k-1) \Gamma (k-1)}-\frac{2 n^2}{\Gamma (k-1)}+\frac{3 (k-2) n}{2 \Gamma (k-1)}\right)
\end{dmath}
Alternatively, it is feasible to employ methods that, from the original definition of the sum of the recursive terms in the recurrence, obtain an asymptotic expansion with which we infer its growth as $n \to \infty$. Thus, as shown above, with the assistance of $AsymptoticSum[]$ \cite{WolframAsymptoticSum}, which uses techniques based on tools such as the Euler-Maclaurin formula \cite{Kac2002, Sarafyan1979}, Saddle-point methods \cite{Butler2019}, or singularity analysis \cite{Flajolet1990}, we manage to reach a closed form for the asymptotic bound that determines the growth of the sum. However, no simplification is achieved that easily infers such growth, so the only conclusion that can be drawn regarding its asymptotic behavior is that in view of some of the terms present, both in this last bound and in previous results, it is possible that at the accumulation point evaluated, the sum of non-recursive terms has a behavior similar to the numerator of the metric $c(n,k)$, which accounts for the accumulated size of all clusters, which makes sense, since the accumulated size must depend on their quantity.

\begin{figure}[H]
    \centering
    \includegraphics[width=10cm,clip]{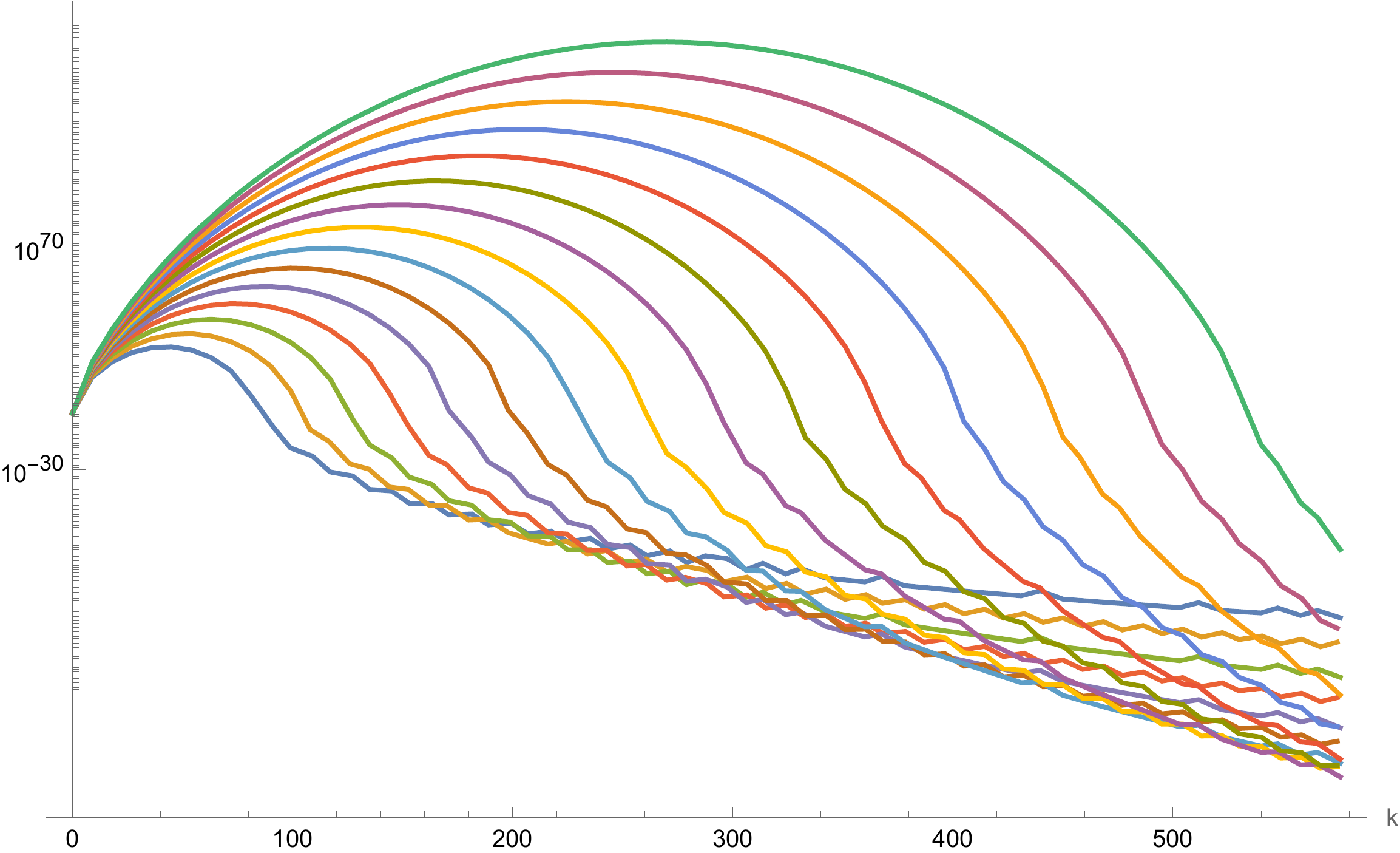}
    \caption{Non recursive term of the estimator $\hat{N}_1(n,k)$ plotted for multiple 2D systems with a square matrix of side $10\leq n\leq24$, from the left plot in blue to the right one in green, with $0\leq k\leq 24^2$}
    \label{fig:2DClusterSizePartialSum2}
\end{figure}

To verify this relationship between $s(n,k)$ and the number of existing clusters, the non-recursive terms of the estimator that provides the upper bound for $N(n,k)$ in systems of different sizes are first visualized. As shown above, it is necessary to plot it on a logarithmic scale, similar to what was performed with the first estimator. In this case, it is observed that the values returned are higher, which is consistent with the approach used to bound the actual magnitude of $N(n,k)$. Additionally, its shape is identical, resembling the possible density function associated with the intermediate evaluation of the complete recursion of the estimator. Regarding its interpretation, it is assumed that at each point $k$ the probability that some cluster exists in the system is obtained. Consequently, being this estimator an upper bound, it could be considered as a threshold that bounds this probability, which could be applied in constraints if it is necessary to reach the actual magnitude from a parametrization, or in its asymptotic analysis.
\begin{figure}[H]
    \centering
    \includegraphics[width=10cm,clip]{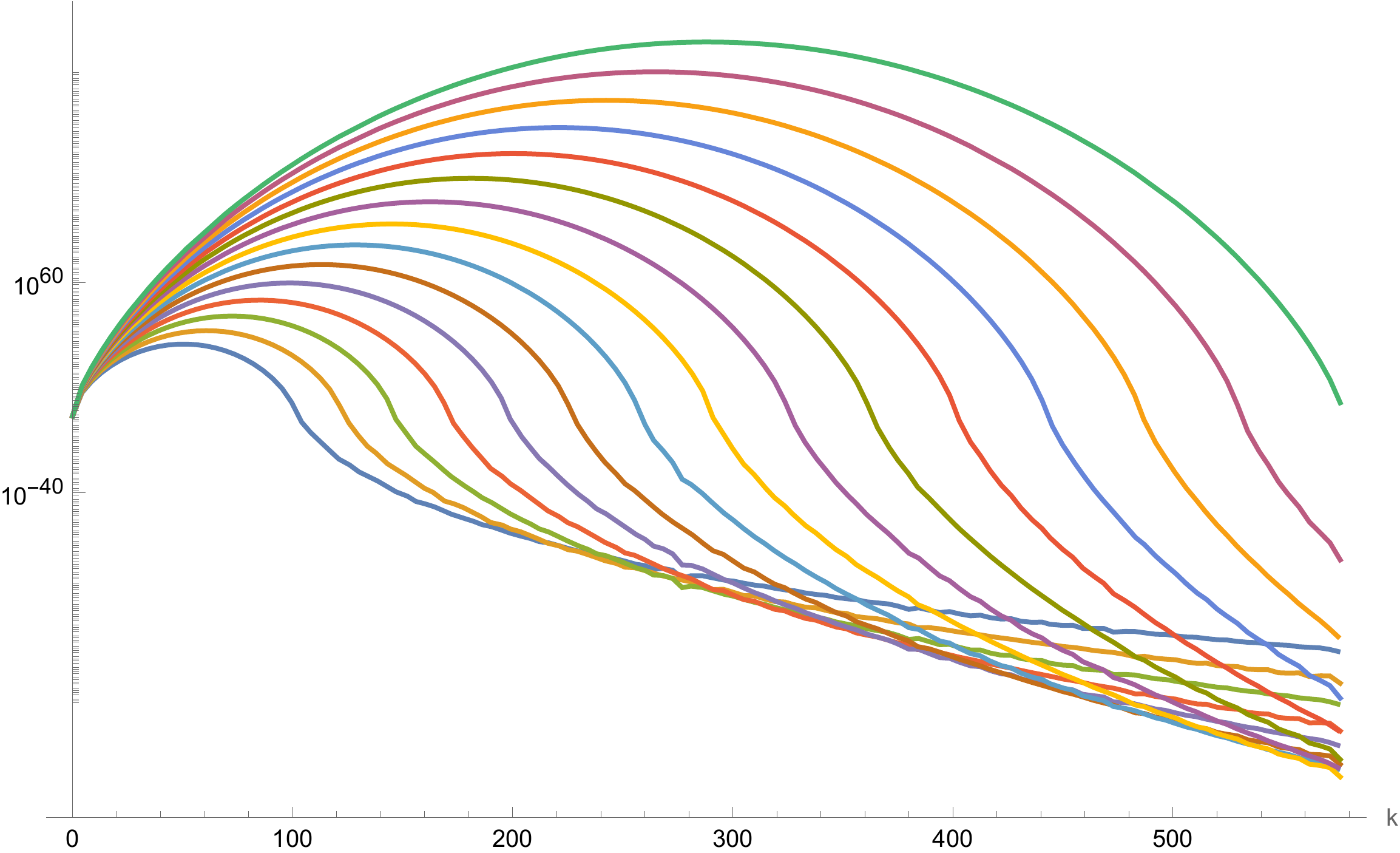}
    \caption{$s(n,k)$ plotted for multiple 2D systems with a square matrix of side $10\leq n\leq24$, from the left plot in blue to the right one in green, with $0\leq k\leq 24^2$}
    \label{fig:2DClusterSizes}
\end{figure}
Lastly, after visualizing the the aggregate sizes of the clusters, a strong similarity with respect to the denominator of $c(n,k)$ is observed, although it returns smaller values. With this, we can ensure that experimentally, the dependence of $s(n,k)$ on the number of clusters is fulfilled, which is useful for its subsequent asymptotic analysis. And, regarding the metric $c(n,k)$, which is the objective sought, for now, we only have the probabilistic estimators from section 3.1, although the use of the estimators for $N(n,k)$ can be considered as a possible alternative, especially when working with its asymptotic behavior.
\subsection{Expected iterations for process termination}
Besides the size of the clusters residing in the system at each iteration, it is necessary to know how many of them are executed before a process reaches its terminal state. Therefore, an expression is calculated that, similarly to the metric $c(n,k)$, returns the average number of iterations a process lasts when executed on a two-dimensional system with an aspect ratio of 1, although other systems will also be considered to better understand their source.
\begin{align}
    Pr[\mathbf{I}_n\leq i]=\sum _{j=0}^i \frac{p(n,j)}{\displaystyle\binom{n^2}{j}}H(i,j) \quad\quad\left[H(i,k)=\frac{(n^2)!}{(n^2)^i (n^2-k)!} \stirling{i}{k}\right]
\end{align}
To begin with, analogously to the analysis of systems in 1 dimension, we consider the probability that the process concludes within a maximum of $i$ iterations. Based on the general upper formulation, depending on the number of terminal states $p(n,j)$ for each quantity of elements, or the probability that in a sequence of $i$ insertions, $j$ elements are inserted, we can generalize for systems of different natures. In this specific case, the metric $H(i,j)$ adapts according to the maximum size of the analyzed system, and, regarding the exact number of terminal states, an attempt will be made to find a closed form or an expression that models that magnitude. However, due to the characteristics of the system, this task will be markedly more complex than in one dimension.
\begin{figure}[H]
    \centering
    \includegraphics[width=14cm,clip]{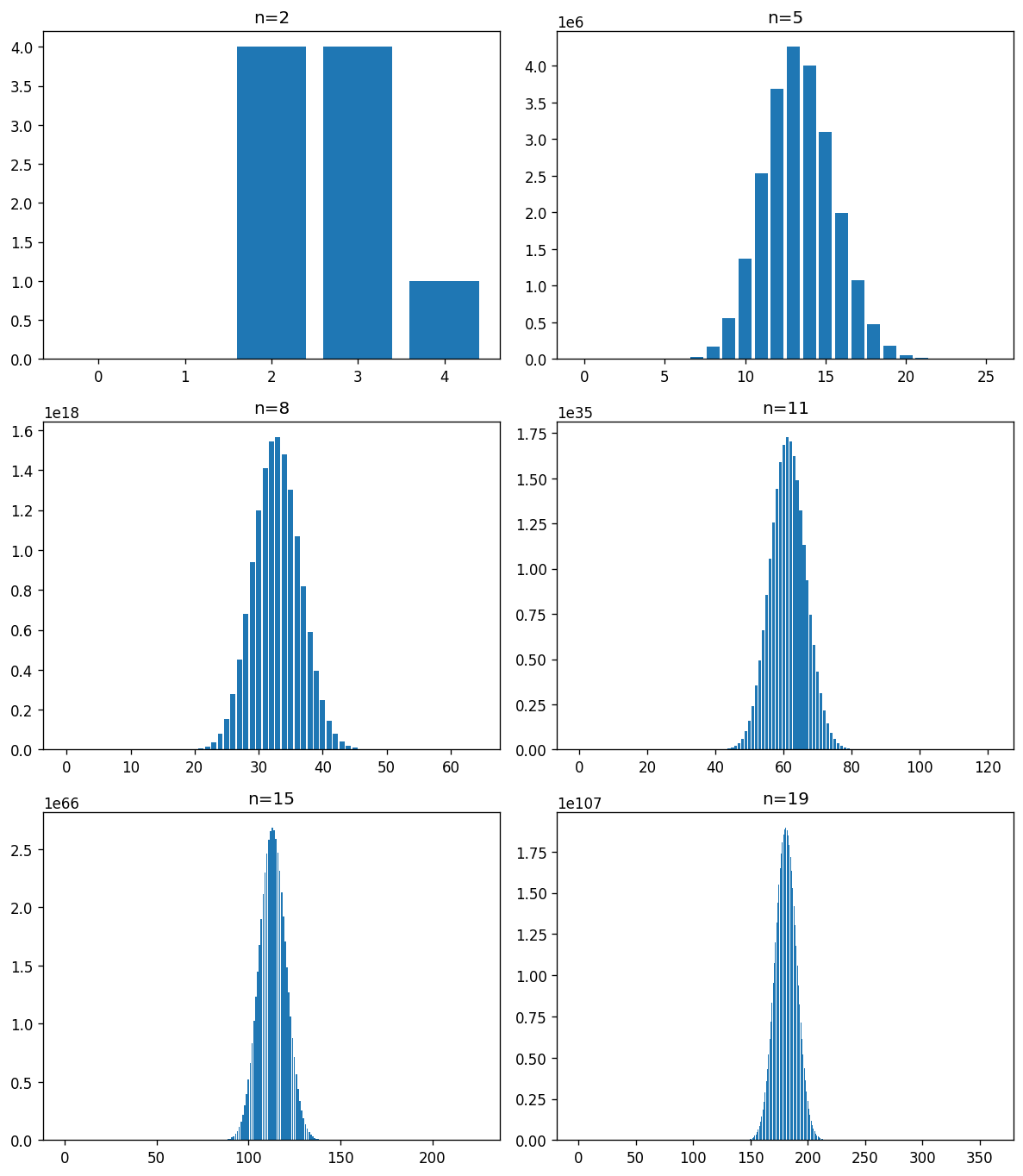}
    \caption{Exact amount of terminal states $p(n,k)$ with $0\leq k\leq n^2$ plotted for systems of different sizes. These values were computed empirically. \cite{MertensPercolation}}
    \label{fig:SamplePaths}
\end{figure}
As that the main issue when studying the duration of a process lies in $p(n,k)$, it is advisable to visualize the values it provides within the valid range of $k$. That is, the amount of terminal states for a system with exactly $k$ elements has a direct correlation with the combinations of those $k$ existing elements that satisfy the process termination condition. According to the algorithm's definition, the formation of a path that connects the extreme rows of the system matrix is enough to conclude its execution, so $p(n,k)$ can be interpreted as the number of combinations of $k$ elements in such a way that there is at least one path connecting the extremes of the system in all of them. Thus, as can be seen in Figure 54, certain constraints that this magnitude must meet can be inferred, which will limit the range of values of $k$ that have a more complex dependency on the system and the current iteration.

\begin{align}
    p(n,j)=0\enspace [j<n]
\end{align}

The first of all states that for any number of elements $j<n$, there is no terminal state in the state space $S_{n,j}$. This is the most evident, since the shortest path that can be formed in any two-dimensional system has a length equivalent to the side of the system $n$, and, in this way, until there are enough elements to form a path of such length, it can be guaranteed that the process does not terminate.

\begin{align}
    p(n,n)=(n + 2) 3^{(n-2)} + 2 \sum_{k=0}^{n-3} M_{k} (n-k-2) \cdot 3^{(n-k-3)}
\end{align}

Additionally, in the case where $j=n$, although it may not be visually apparent, an expression for $p(n,n)$ can be found as shown above \cite{Mertens2022,Mertens2019}. In summary, this number of elements constitutes a special case, coinciding with the minimum length of a path. Therefore, counting how many paths of length $n$ there are out of all $\binom{n^2}{n}$ possible combinations, assuming that the system is a square matrix, is possible by considering all the forms that these paths can exhibit according to the established neighborhood. In this algorithm, by enabling diagonal neighboring cells for each cell, the paths not only form a straight line between the extreme rows of the matrix but can also present "curves" caused by the translation of a number of elements that can vary throughout the range from 1 to $n$. Likewise, these "curves" may further subdivide the path into segments with differing directions, which together with the translations of the elements that compose the path lead to the emergence of Motzkin numbers \cite{Donaghey1977} in $p(n,n)$.

\begin{align}
    p(n,j)=\binom{n^2}{j}-n\enspace[j=n^2-n]
\end{align}

Beyond $j = n$ lies the challenge \cite{Krattenthaler2017} of counting terminal states, and although there are expressions that quantify the paths formed by $j > n$ elements with Von Neumann neighborhood, which we could leverage in our problem by subtracting that quantity from the total combinations $\binom{n^2}{j}$, currently, there is no general formula that from $n$ and $j$ returns the expected magnitude for the neighborhood considered in this problem, nor the one that appears to be complementary in this context. Therefore, to elucidate the relationship between the counting of terminal states with the neighborhood of our problem and that of Von Neumann, the case where $j = n^2 - n$ is shown above, meaning the system is full except for $n$ cells. Specifically, with Von Neumann neighborhood, there are only $n$ paths of size $n$ in a matrix of those dimensions, since without connectivity between diagonal cells, the only possible form of a path is a line of $n$ elements shifted through the columns of the matrix. Thus, if we subtract this number of paths from the total combinations of $n^2 - n$ elements, we obtain the number of terminal states with Moore neighborhood $p(n, j)$. The main reason lies in the complementarity of the paths formed by both neighborhoods. So, to enumerate the states $p(n, j)$, we can proceed by subtracting from the total combinations of $j$ elements the number of states that are not terminal, i.e., the number of 'complementary' paths that block the generation of paths in the system. This is, if we consider all existing paths that begin and end in the extreme columns of the matrix and originate through the Von Neumann neighborhood, their number will determine how many states do not contain valid paths according to the definition of our algorithm, since if there is a path formed by empty cells that connects the columns of the matrix, it will be impossible for a path of elements with Moore neighborhood to cross the matrix from its extreme rows.

\begin{align}
    p(n,j)=\binom{n^2}{j}\enspace[j> n^2-n]
\end{align}
Consequently, when $j > n^2 - n$, that is, when the number of empty cells in the system is less than $n$, it is not possible to form a path of empty cells with Moore neighborhood that connects the extreme columns of the system matrix. As a result, all states $p(n, j)$ will be terminal, matching the combinations of the existing elements. In summary, the restrictions for special cases regarding the number of elements solve part of the potential expression for $p(n, j)$, given that as the system size increases, the first and last $n$ quantities of elements have a closed form that we can work with in further analysis. Moreover, while this permits an alternative interpretation of the problem and includes the paths of empty cells with Von Neumann neighborhood to perform the desired counting of terminal states, there is no closed expression for the general case where a path of $j$ empty cells 'blocks' the existence of the paths considered in this context.
\\\\
A priori, the Von Neumann neighborhood appears simpler than that established in this algorithm. However, the impediment to achieving a closed expression for $p(n,j)$ lies not solely in the neighborhood but also in the variability of path lengths within this system and the absence of restrictions with which the connection between the extreme rows of the system is established. Therefore, given the difficulty of articulating $p(n,j)$ precisely, it is advisable to first solve certain cases in which the number of terminal states facilitates the characterization of the probability distribution of $\mathbf{I}_n$.
\begin{align}
    &Pr[\mathbf{I}_n\leq i]=\\\notag
    &=\sum _{j=0}^i \frac{\displaystyle\delta _{j,n^2}}{\displaystyle\binom{n^2}{j}}H(i,j)=\sum _{j=0}^i \frac{\displaystyle\delta _{j,n^2}}{\displaystyle\binom{n^2}{j}}\frac{(n^2)!}{(n^2)^i (n^2-j)!} \stirling{i}{j}=\frac{1}{\displaystyle\binom{n^2}{n^2}}\frac{(n^2)!}{(n^2)^i (n^2-n^2)!} \stirling{i}{n^2}=\frac{(n^2)!}{(n^2)^i} \stirling{i}{n^2}
\end{align}
In systems with 1 dimension, it was seen that the average number of iterations to fill the system was $nH_n$. Thus, it may be hypothesized that, in the case of a square matrix, approximately $n^2H_{n^2}$ iterations might be necessary to fully populate it; however, a two-dimensional system of size $n^2$ is not comparable to a one-dimensional one of length $n^2$. It is therefore necessary to propose a probability as shown above and check if the hypothetical number of iterations matches that which actually provides a bidimensional system. In the modeling of a scenario where a process needs to completely fill the system to finish, it can be ensured that $p(n,j)$ is zero for any number of elements $j$ except the one that meets the condition of filling the matrix. That is, all the terms of the upper sum are zero except where $j=n^2$, since it is the only quantity that meets the termination condition of the process.

\begin{align}
    I(n,n^2)=\sum _{i=0}^\infty 1-Pr[\mathbf{I}_n\leq i]=\sum _{i=0}^\infty 1-\frac{(n^2)!}{(n^2)^i} \stirling{i}{n^2}=\boxed{n^2H_{n^2}}
\end{align}

Once the distribution of $\mathbf{I}_n$ has been characterized under the previously described conditions, we proceed to calculate its expectation, which coincides with the metric $I(n,n^2)$ requisite for resolving this case.

\begin{figure}[H]
    \centering
    \includegraphics[width=10cm,clip]{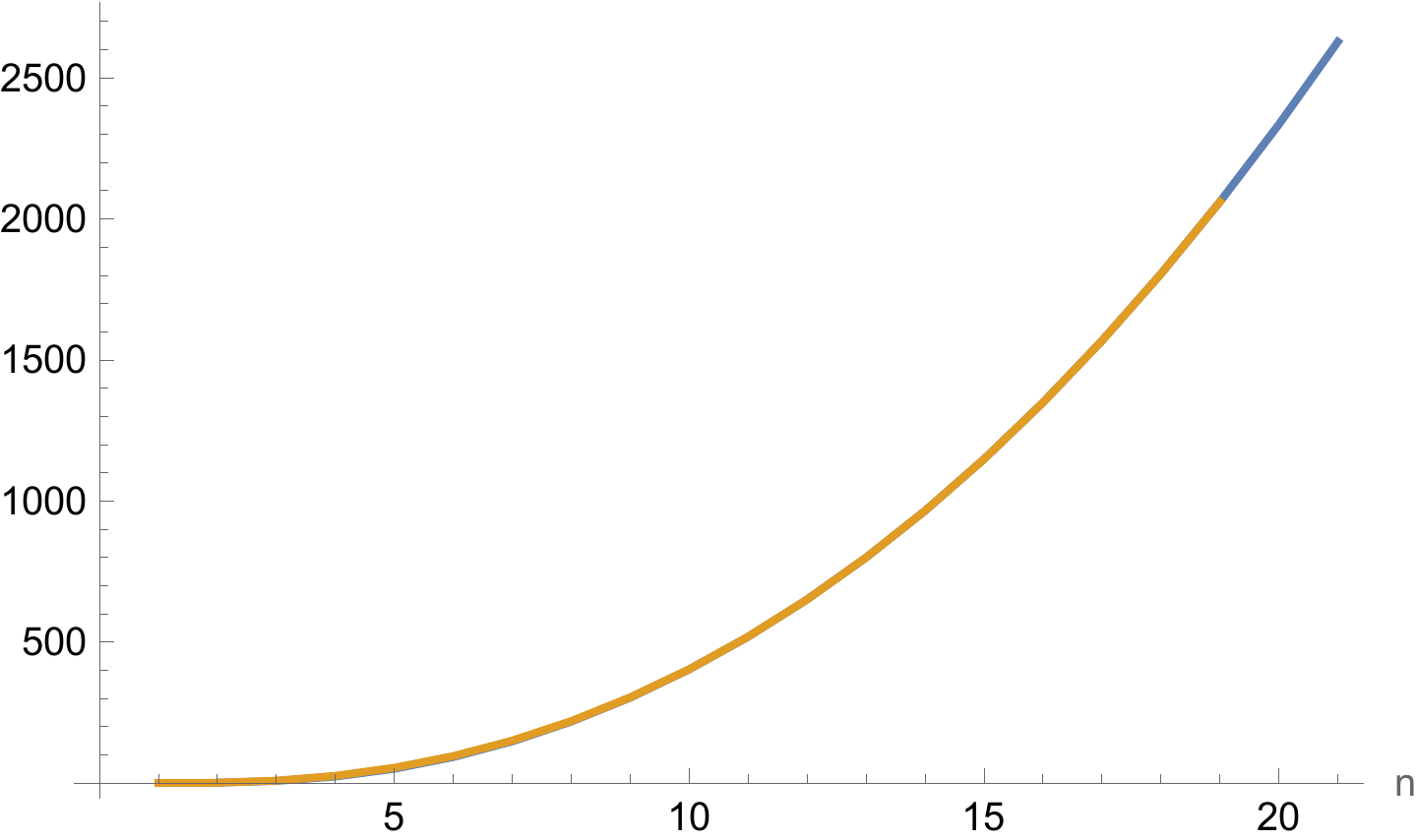}
    \caption{$I(n,n^2)$ evaluated numerically for the first values of $n$ plotted in blue, along with the function $n^2H_{n^2}$ plotted in orange with a small range of evaluation.}
    \label{fig:2DiterationsMaxFit}
\end{figure}

Furthermore, if we numerically evaluate the function $I(n,n^2)$, we observe that it aligns with its resulting closed form. This result may initially appear unhelpful, although later we will see similar metrics that are not so practical to solve exactly, so their numerical evaluation for the first values of $n$ will serve to infer a simpler expression with which to work in the analysis. Regarding the specific result, it is verified that the proposed function $n^2H_{n^2}$ is indeed the one that determines the duration of the percolation process when the 2-dimensional system must be completely filled. With this, if we compare it with a process of its immediately lower dimension, we can infer how in the case of needing to fill the system $I(n,n^2)$ has the same form, except for the size of the system. This is due to the factors involved in computing this metric, which are mainly the number of cells in which elements can be inserted and the amount of them necessary for the process to conclude. Thus, given that these factors depend solely on the geometry of the object that forms the system, and not on the neighborhood, it is plausible to adapt the metric $I(n,k)$  obtained in the 1-dimensional analysis derived from the coupon collector problem in order to estimate the duration of a bidimensional process that must insert exactly $k$ elements:
\begin{align}
    I(n,k)=n^2(H_{n^2}-H_{n^2-k})
\end{align}
Regarding the analysis, the aforementioned metric is suitable to measure the duration of the process if we knew the average number of elements that the system has upon completion, however, this quantity is unknown at the moment. Now, we will proceed to calculate $I(n)$ under special conditions. Specifically, since we know the upper bound of the number of terminal states $p(n,k)$, we will use it instead of the actual magnitude in order to verify that the approach of $I(n)$ is valid regardless of the conditions imposed on it, besides obtaining a measure corresponding to a possible best case of the algorithm in terms of iteration count.
\begin{figure}[H]
    \centering
    \begin{subfigure}[b]{0.49\textwidth}
        \centering
        \includegraphics[width=\textwidth,clip]{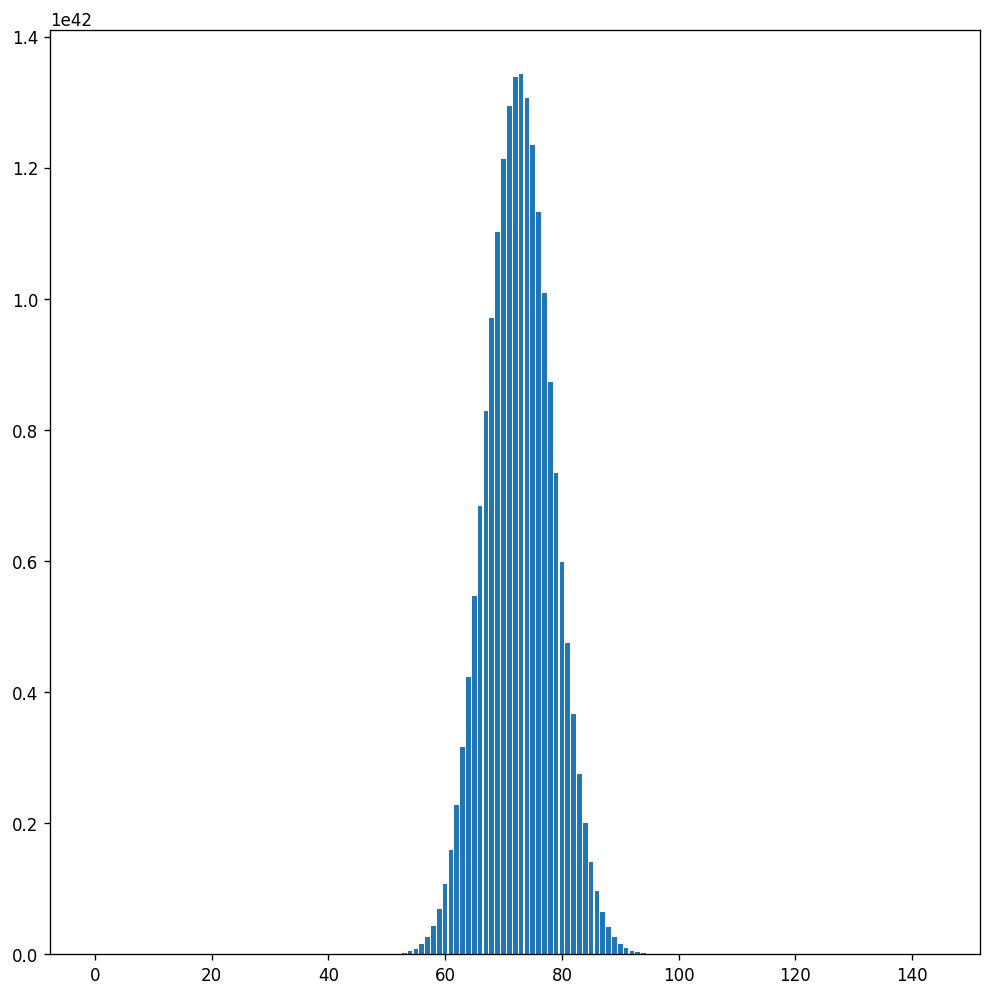}
        \caption{}
        \label{fig:SamplePaths2}
    \end{subfigure}
    \vfill
    \begin{subfigure}[b]{0.49\textwidth}
        \centering
        \includegraphics[width=\textwidth,clip]{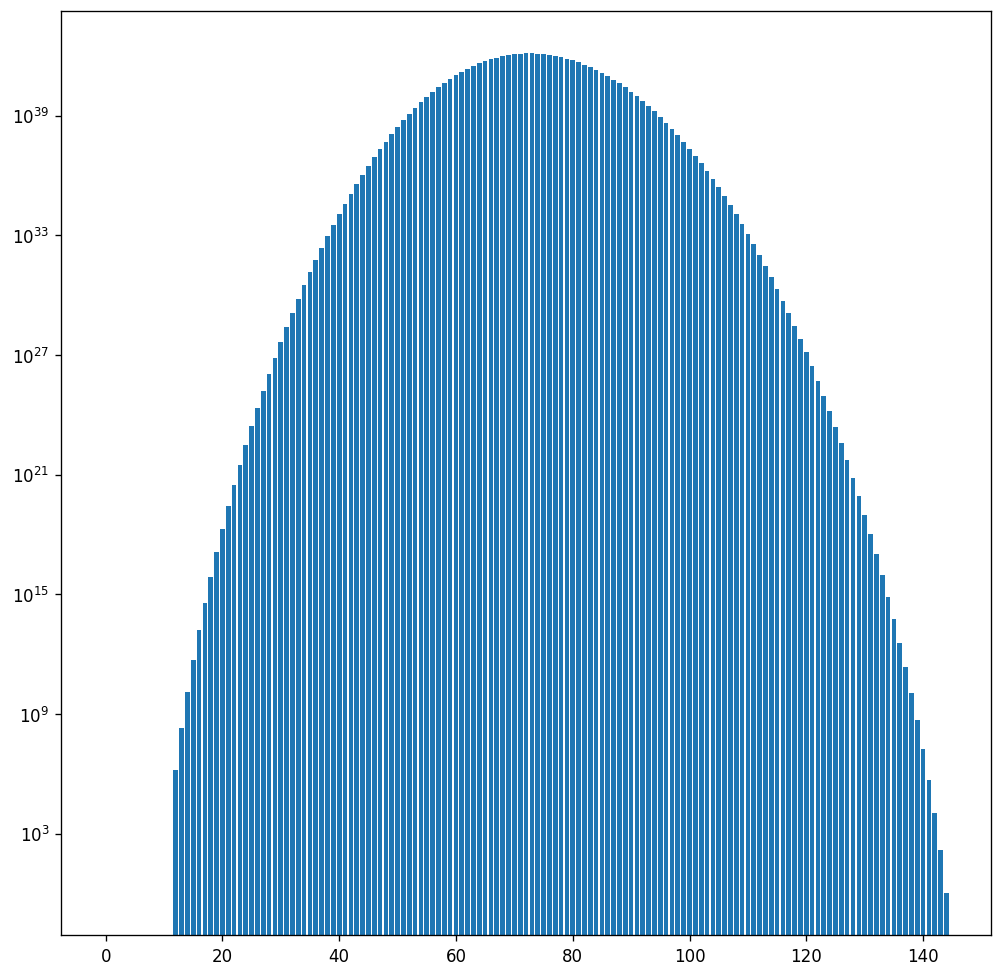}
        \caption{}
        \label{fig:SamplePaths2Log}
    \end{subfigure}
    \caption{(a) Number of terminal states of a system of size $n=12$ for a count of elements $0\leq k\leq n^2$. (b) Logarithmic scale plot of (a).}
    \label{fig:SamplePaths2dual}
\end{figure}

At the outset, to visualize the upper bound of terminal states, we will use a system of small size, for example $n=12$. Prior to overlaying this on the upper plot in logarithmic scale and examining the magnitude of the difference between the bound and the actual $p(n,k)$, it is possible to deduce what $I(n)$ should return in case the number of terminal states is replaced by its maximum value. Observations of the graph reveal that the shape of the actual magnitude is very similar to that shown by a function of the form $\binom{n^2}{k}$, which determines the total number of combinations of $k$ elements over the system, that is, the possible states and in turn the upper bound of $p(n,k)$. Thus, if we use this bound in the expression of the probability distribution of $\mathbf{I}_n$, we will be establishing the ratio of terminal states to existing states as 1, which represents a 100\% probability that the process will end in the first iteration with a sufficient number of elements to form the minimum length path. This interpretation stems from the analysis of the best case, in which it can be assumed that a process inserts elements in such a way as to form the minimum path in $n$ iterations, which implies that an effective insertion occurs in all of them. Therefore, using the upper bound in the probability distribution ensures that $I(n)$ returns a result consistent with the best case of the algorithm, due to the established ratio of terminal states.
\begin{figure}[H]
    \centering
    \includegraphics[width=10cm,clip]{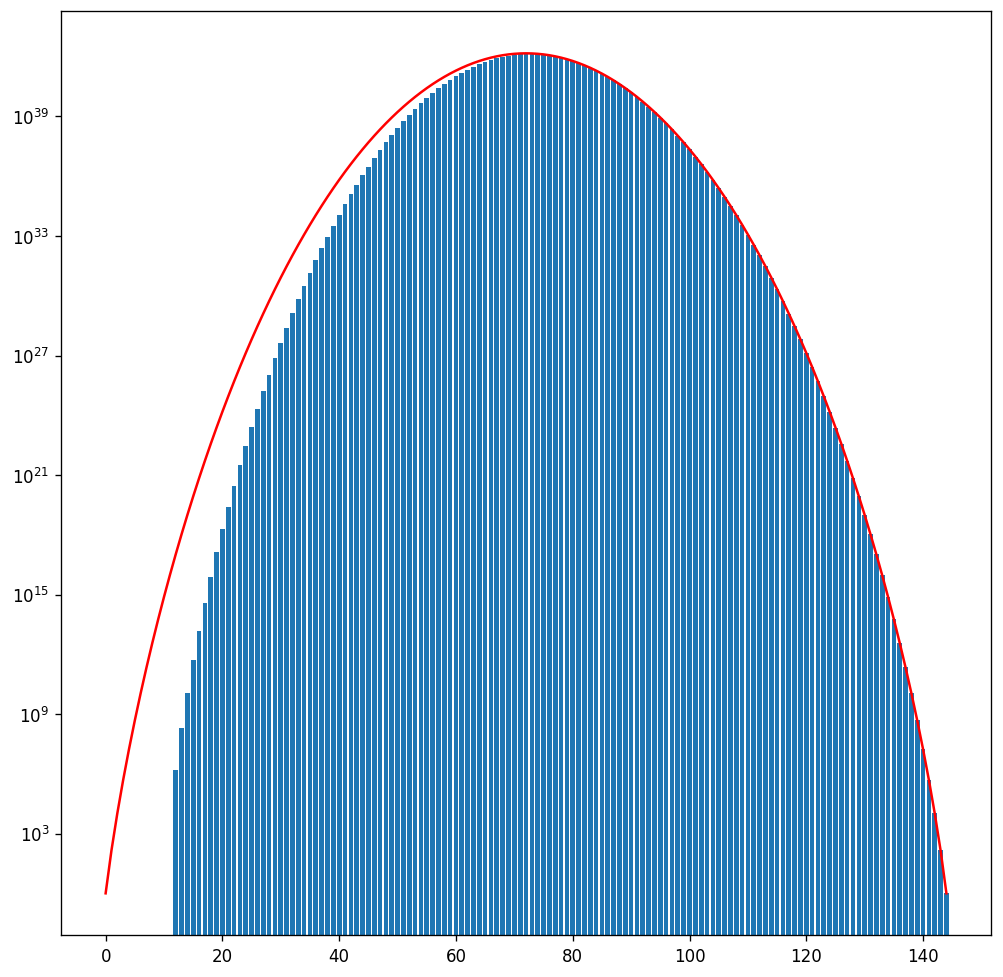}
    \caption{Logarithmic plot from Figure 56 (b) in blue and $\binom{n^2}{k}$ in log scale plotted in red for $0\leq k\leq n^2$}
    \label{fig:SamplePaths2LogFit}
\end{figure}
Regarding the terminal states, above it is observed how the difference between the magnitude and its upper bound tends to 0 as the number of elements in the system increases. Likewise, the disparity appears to be small enough to be negligible from $k=n^2/2$. This phenomenon may be attributed to the aforementioned complementarity, as the midpoint of the occupancy ratio establishes a division between the count of terminal states from the inserted elements, and its alternative by considering the empty cells that block the appearance of element paths. Nonetheless, with this, the true magnitude could be approximated through an interpolation of the midpoint and the case $k=n$, leaving the rest of the range equal to the upper bound. However, the expression for the case $k=n$ is not closed, which would complicate the process and possibly prevent a reliable asymptotic approximation.

\begin{align}
    Pr[\mathbf{I}_n\leq i]=\sum _{j=0}^i \frac{\displaystyle\binom{n^2}{j}}{\displaystyle\binom{n^2}{j}}H(i,j)=\sum _{j=n}^i H(i,j)
\end{align}
Then, we proceed to substitute the function $p(n,j)$ directly by the total number of combinations of $j$ elements, denoting its maximum bound. In this way, we characterize the probability distribution followed by the random variable in the case defined by the imposed restrictions.
\begin{align}
     I(n)=\sum _{i=0}^\infty Pr[\mathbf{I}_n> i] = \sum _{i=0}^\infty 1-Pr[\mathbf{I}_n\leq i]=\sum _{i=0}^\infty \left(1-\sum _{j=n}^i H(i,j)\right)
\end{align}
The metric of interest, as denoted by $I(n)$, can be derived utilizing the distribution function. Still, the lower limit of the summation in the expression of the distribution function hinders, contrary to what happened earlier, the calculation of the final closed-form of the metric, or even its asymptotic bound when $n \to \infty$.
\begin{figure}[H]
    \centering
    \includegraphics[width=10cm,clip]{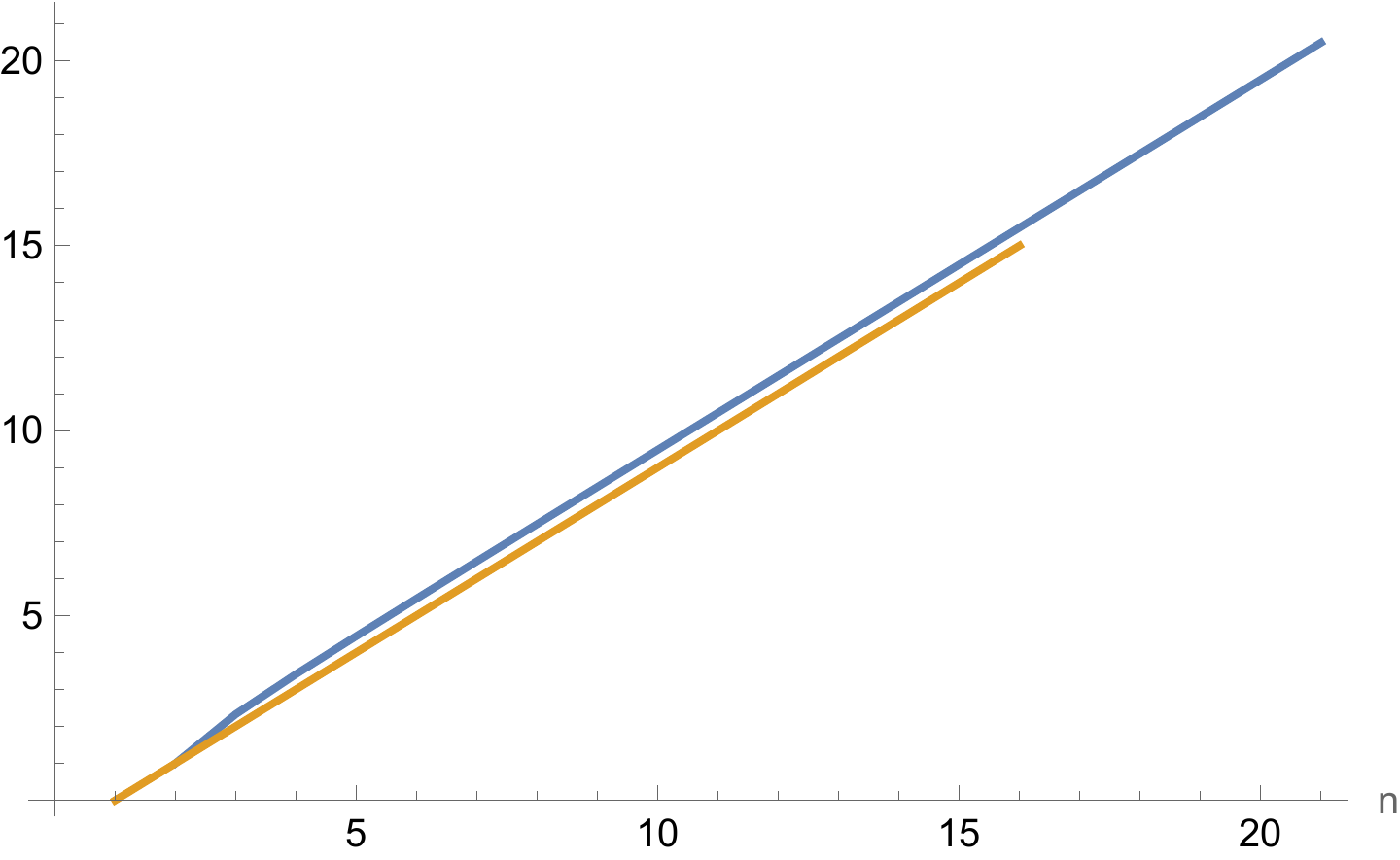}
    \caption{$I(n)$ evaluated numerically for multiples values of $n$ plotted in blue and $y=n$ function in orange.}
    \label{fig:2DiterationsWorstFit}
\end{figure}
Therefore, the metric $I(n)$ is evaluated numerically for certain values of $n$ along with the function that potentially might serve as solution. To ensure that the numerical evaluation is as close as possible to the actual value of the metric, an upper limit for the sum through the iterations $i$ is selected sufficiently large to guarantee adequate precision in all evaluations of $n$. Thus, considering that for all iterations from 0 to $n$ the probability that the process ends is null under the imposed restrictions, and $H(i,j)$ tends to 1 as the length of the insertion sequence $i$ increases, we can infer that a suitable upper bound for the sum would be of the order $n^3$. Similarly, with this information, it can also be seen how $I(n)$ will have a value proportional to $n$ plus the amount of probability that it takes $H(i,j)$ to converge to 1. Although it cannot be guaranteed that the result is exactly $n$, the above graph indicates a strong similarity to $n$, so in broad terms it can be concluded that the iterations that a process last whose insertion sequence follows the best case of the algorithm are of order $O(n)$.
\\\\
Additionally, besides analyzing edge cases for $I(n)$, it is convenient to try to find its general formulation for all $n$ based on the determinant function of the terminal states $p(n,k)$. That is, the main difficulty in calculating the duration of a process lies in the modeling of its terminal states, more specifically the quantity of those that have at least one valid path, or its complement. Consequently, recognizing that the search for different sequences of $p(n,k)$ for specific $n$ in OEIS and similar sites does not yield useful outcomes, or that for a certain range of $k$ the general expression is not immediately found, there is the possibility of applying a reversible transformation to these sequences in order to find general expressions for the magnitude under the effect of the transformation. With this in mind, the first way to proceed may involve interpolating the values of a specific sequence of $p(n,k)$, leading to polynomials whose coefficients and degrees are potentially easier to infer.
\begin{align}
    &P(2,j)=\frac{1}{24} \left(9 j^4-86 j^3+243 j^2-166 j\right)\\
    &P(3,j)=\frac{1}{362880}\left(502 j^9-18945 j^8+291552 j^7-2350026 j^6+10651494 j^5 -27496665 j^4 \right.\\\notag
    &\left. +40248308 j^3-31279644 j^2+9953424 j\right)\\
    &P(4,j)=\frac{1}{20922789888000}\left(177541 j^{16}-22889912 j^{15}+1342669060 j^{14}-47417636000 j^{13}\right.\\\notag
    &\left.+1124531200702 j^{12}-18915280062224 j^{11}+232422190140140 j^{10}-2117978597020000 j^9\right.\\\notag
    &\left.+14389512273652373 j^8-72689238663057016 j^7+270052102151435240 j^6\right.\\\notag
    &\left.-722263115740129600 j^5+1340234906635554384 j^4-1618739915026750848 j^3\right.\\\notag
    &\left.+1127409710876962560 j^2-336490569254246400 j\right)\\\notag
    &P(5,j)=\frac{1}{15511210043330985984000000}\left(-230218824 j^{25}+73266563875 j^{24}-11011110755100 j^{23}\right.\\\notag
    &\left.+1039173229994050 j^{22}-69080938164907680 j^{21}+3439419373674609325 j^{20}\right.\\\notag
    &\left.-133132204898001530100 j^{19}+4105134296351411573800 j^{18}-102505912314431497670040 j^{17}\right.\\\notag
    &\left.+2095919942184623633667325 j^{16}-35348224418771136389804100 j^{15}\right.\\\notag
    &\left.+493855634113556120475623050 j^{14}-5726224068277869336725620560 j^{13}\right.\\\notag
    &\left.+55080845465490791263039627075 j^{12}-438409595841983065299380261100 j^{11}\right.\\\notag
    &\left.+2873364603192255744268399816300 j^{10}-15391336265631458455674072223920 j^9\right.\\\notag
    &\left.+66663169677225062255000524037200 j^8-230012314720755036612958134369600 j^7\right.\\\notag
    &\left.+619268712039956935144120658932800 j^6-1263340003788588238905326578718976 j^5\right.\\\notag
    &\left.+1869964560876940761407758850995200 j^4-1874467230564580371406607777280000 j^3\right.\\\notag
    &\left.+1123137350768689633740585922560000 j^2-298307678156803870881095024640000 j\right)
\end{align}
Let $P(n,j)$ represent the interpolating polynomial constructed from a sequence of evaluations of $p(n,j)$, such as $p(2,j)=\{0,0,4,4,1\} \forall j\in[0,4]$, the results of the function $InterpolatingPolynomial[]$ \cite{Wolfram_InterpolatingPolynomial_1991} from Wolfram applied to each sequence of exact values of the number of terminal states for each $j$ are shown above.
Upon initial inspection, the only discernible patterns in the polynomials are the factor $\frac{1}{n^2!}$ and the alternation of signs in each monomial. However, the problem arises when trying to find a function that generates the coefficients of each monomial from an index dependent on its degree, whose maximum is equivalent to $n^2$. Since there is no apparent pattern or sequence for the coefficients, it is not possible to solve $p(n,j)$ from the polynomials in the form presented above. Therefore, we proceed by factoring them as much as possible, in case their roots follow a known sequence:
\begin{align}
    &P(2,j)=\frac{1}{24} (j-1) j \left(9 j^2-77 j+166\right)\\
    &P(3,j)=\frac{1}{362880}(j-2) (j-1) j (502 j^6-17439 j^5+238231 j^4-1600455 j^3+5373667 j^2\\\notag
    &-8174754 j+4976712)\\\notag
    &P(4,j)=\frac{1}{20922789888000}(j-3) (j-2) (j-1) j (177541 j^{12}-21824666 j^{11}+1209768113 j^{10}\\\notag
    &-39917890750 j^9+871585458963 j^8-13239411901518 j^7+143158771337939 j^6\\\notag
    &-1108162925321890 j^5+6086351765594596 j^4-23122383262921016 j^3\\\notag
    &+57718955600437248 j^2-85085055651807360 j+56081761542374400)\\\notag
    &P(5,j)=-\frac{1}{15511210043330985984000000}(j-4) (j-3) (j-2) (j-1) j (230218824 j^{20}\\\notag
    &-70964375635 j^{19}+10293409339910 j^{18}-933743872506525 j^{17}+59379676368912054 j^{16}\\\notag
    &-2812425200835749670 j^{15}+102882729981282627420 j^{14}-2974880720837948301850 j^{13}\\\notag
    &+69014163184432481319044 j^{12}-1296445850407086440549055 j^{11}\\\notag
    &+19817056981683686939996430 j^{10}-246787354235949490831245225 j^9\\\notag
    &+2498274899122664683934183454 j^8-20438571526511957823031210840 j^7\\\notag
    &+133769282028212340103249556240 j^6-689485138024417797453152816400 j^5\\\notag
    &+2732672840495306733962828646624 j^4-8025471623370468011596662124800 j^3\\\notag
    &+16429329399816634773156880320000 j^2-20902625886472843169651581440000 j\\\notag
    &+12429486589866827953378959360000)
\end{align}
After factoring, the only extracted roots are those corresponding to the null values of the first $j$. Nonetheless, the challenge of finding a generating function for the coefficients persists, although this time with one of lower degree. In summary, the few patterns seen in the degree, multiplier, or certain roots of the interpolators can, at most, constrain the original problem of characterizing $p(n,j)$. But, under no circumstances do they succeed in modeling the number of terminal states for all $j$. Subsequently, as a last resort that can be employed to try to characterize the interpolators, more specifically their coefficients, is the construction of a Hankel matrix \cite{widom1966hankel,Llovet1989} from its sequence:
\begin{dmath}
    \mathcal{H}_2=\left(
\begin{array}{ccccc}
 0 & -\frac{83}{12} & \frac{81}{8} & -\frac{43}{12} & \frac{3}{8} \\
 -\frac{83}{12} & \frac{81}{8} & -\frac{43}{12} & \frac{3}{8} & 0 \\
 \frac{81}{8} & -\frac{43}{12} & \frac{3}{8} & 0 & 0 \\
 -\frac{43}{12} & \frac{3}{8} & 0 & 0 & 0 \\
 \frac{3}{8} & 0 & 0 & 0 & 0 \\
\end{array}
\right)
\end{dmath}
\begin{dmath}
    \mathcal{H}_3=\\\left(
\begin{array}{cccccccccc}
 0 & \frac{69121}{2520} & -\frac{868879}{10080} & \frac{10062077}{90720} & -\frac{9699}{128} & \frac{253607}{8640} & -\frac{6217}{960} & \frac{3037}{3780} & -\frac{421}{8064} & \frac{251}{181440} \\
 \frac{69121}{2520} & -\frac{868879}{10080} & \frac{10062077}{90720} & -\frac{9699}{128} & \frac{253607}{8640} & -\frac{6217}{960} & \frac{3037}{3780} & -\frac{421}{8064} & \frac{251}{181440} & 0 \\
 -\frac{868879}{10080} & \frac{10062077}{90720} & -\frac{9699}{128} & \frac{253607}{8640} & -\frac{6217}{960} & \frac{3037}{3780} & -\frac{421}{8064} & \frac{251}{181440} & 0 & 0 \\
 \frac{10062077}{90720} & -\frac{9699}{128} & \frac{253607}{8640} & -\frac{6217}{960} & \frac{3037}{3780} & -\frac{421}{8064} & \frac{251}{181440} & 0 & 0 & 0 \\
 -\frac{9699}{128} & \frac{253607}{8640} & -\frac{6217}{960} & \frac{3037}{3780} & -\frac{421}{8064} & \frac{251}{181440} & 0 & 0 & 0 & 0 \\
 \frac{253607}{8640} & -\frac{6217}{960} & \frac{3037}{3780} & -\frac{421}{8064} & \frac{251}{181440} & 0 & 0 & 0 & 0 & 0 \\
 -\frac{6217}{960} & \frac{3037}{3780} & -\frac{421}{8064} & \frac{251}{181440} & 0 & 0 & 0 & 0 & 0 & 0 \\
 \frac{3037}{3780} & -\frac{421}{8064} & \frac{251}{181440} & 0 & 0 & 0 & 0 & 0 & 0 & 0 \\
 -\frac{421}{8064} & \frac{251}{181440} & 0 & 0 & 0 & 0 & 0 & 0 & 0 & 0 \\
 \frac{251}{181440} & 0 & 0 & 0 & 0 & 0 & 0 & 0 & 0 & 0 \\
\end{array}
\right)
\end{dmath}

\begin{align}
    |\mathcal{H}_2|\approx0.00741577,|\mathcal{H}_3|\approx-2.56689\cdot10^{-29},|\mathcal{H}_4|\approx6.13118\cdot10^{-138},|\mathcal{H}_5|\approx-2.87656\cdot10^{-438}
\end{align}
The idea behind the Hankel matrix, as presented above with coefficients multiplied by the common factor $\frac{1}{n^2!}$, lies in the properties exploitable for the generation of its entries. Lastly, although it is only posited here as a prospective line of research, we can observe the presence of some properties of interpolants such as the alternation of signs in the determinant's value.
\\\\
In conclusion, interpolating sequences of terminal state quantities is a method that could potentially result in a solution for $p(n,k)$. Nevertheless, as demonstrated previously, no concrete expression has been achieved for this quantity, leaving the method as future investigation. Therefore, in the absence of the magnitude for systems of 2 dimensions with an aspect ratio equal to 1, it is worth considering the analysis of systems whose shape simplifies the calculation of $I(n)$. The simplest of all, which has already been analyzed in section \textcolor{blue}{\ref{1-dimensional-case-analysis}}, consists of a matrix with only 1 column, forming a 1-dimensional system. In this edge case, the only possible path is generated when the system is completely filled, rendering the expression for $p(n,k)$ immediate to derive. However, for our purpose of extending this case to more complex 2-dimensional systems, matrices of general size $(n,m)$ can be considered, with the lowest value $m=2$ potentially providing useful results for the analysis of higher $m$ values. So, we proceed to characterize the distribution function of the random variable $\mathbf{I}_{n,2}$, which is defined as the exact number of iterations that a percolation process executed on an $(n,2)$ matrix lasts. Subsequently, if it is feasible to solve $I(n)$ completely, together with cases of higher $m$ values, alternative methods can be explored that, starting from specific cases like $(n,2)$ with $n=2$ or $(n,3)$ with $n=3$, allow a generalization for all valid $m$ values and, consequently, result in to the desired metric $I(n)$ for square matrix systems.
\begin{align}
    Pr[\mathbf{I}_{n,2}\leq i]=\sum _{j=0}^i \frac{p_{n,2}(n,j)}{\displaystyle\binom{2n}{j}}H(i,j) \quad\quad\left[H(i,k)=\frac{(2n)!}{(2n)^i (2n-k)!} \stirling{i}{k}\right]
\end{align}
To this end, first the definition of the probability that the process takes at most $i$ iterations is used, adjusting the metric $H(i,k)$ to the corresponding size of the system, which in this context becomes $2n$. Then, as was the case with square matrices, it is necessary to ascertain the precise number of terminal states present once the process has executed $j$ effective insertions, so an expression for $p_{n,2}(n,j)$ is obtained, which due to the nature of the system will be simpler than in other situations.
\begin{align}
    &p_{n,2}(n,2n)=\binom{2n}{2n}=1\\
    &p_{n,2}(n,2n-1)=\binom{2n}{2n-1}=2n\\
    &p_{n,2}(n,2n-2)=\binom{2 n}{2 n-2}-n \binom{2 n-2}{0}=2 \left(n^2-n\right)\\
    &p_{n,2}(n,2n-3)=\binom{2 n}{2 n-3}-n \binom{2 n-2}{1}=\frac{4}{3} \left(n^3-3 n^2+2 n\right)\\
    &p_{n,2}(n,2n-4)=\binom{2 n}{2 n-4}-n \binom{2 n-2}{2}+\binom{n}{2}=\frac{2}{3} \left(n^4-6 n^3+11 n^2-6 n\right)\\
    &p_{n,2}(n,2n-5)=\binom{2 n}{2 n-5}-n \binom{2 n-2}{3}+\binom{2 n-4}{1} \binom{n}{2}=\frac{4}{15} \left(n^5-10 n^4+35 n^3-50 n^2+24 n\right)\\
    &p_{n,2}(n,2n-6)=\frac{4}{45} \left(n^6-15 n^5+85 n^4-225 n^3+274 n^2-120 n\right)\\
    &p_{n,2}(n,2n-7)=\frac{8}{315} \left(n^7-21 n^6+175 n^5-735 n^4+1624 n^3-1764 n^2+720 n\right)\\
    &p_{n,2}(n,2n-8)=\frac{2}{315} \left(n^8-28 n^7+322 n^6-1960 n^5+6769 n^4-13132 n^3+13068 n^2-5040 n\right)
\end{align}
As previously discussed, the basic cases where the number of terminal states is zero are omitted, since the process will not change state until there are $n$ elements. Thus, $p_{n,2}(n,j)$ is solved for specific values of $j$ that start from the maximum number of elements. The advantage of this approach lies in the neighborhood used to count the paths that form the non-terminal states. That is, if counting started from $j=n$ onwards, the Moore neighborhood would have to be used, which, although a simple system, would complicate the calculation. However, by considering the non-terminal states to obtain their complement, which is the desired quantity, it is only necessary to block the formation of paths in the system by counting paths formed by empty and adjacent cells using the Von Neumann neighborhood. This is particularly advantageous for constructing the expressions of the higher metrics, which can be rewritten as they appear in the last cases in the form of a polynomial with a numerical factor extracted by expanding the binomial terms. As for the pattern followed by this factor, it can be seen that its numerator appears to be the Gould sequence \cite{Shunia2023}, recorded as \href{https://oeis.org/A001316}{A001316} in the OEIS.
\begin{align}
    G(k)=\sum _{i=0}^k \left(\binom{k}{i} \bmod 2\right)=\frac{2^k}{\gcd \left(2^k,k!\right)}
\end{align}
Apart from the original formulation for its sequence $G(k)$, it can also be rewritten \cite{Sil_4968430, OmerSimhi_4976362} in terms of the greatest common divisor of $2^k$ and $k!$, which will simplify the final form of the polynomials.

\begin{align}
    \mu(k)=\frac{k!}{\gcd \left(2^k,k!\right)}
\end{align}
Regarding the denominator, its values follow a sequence also recorded as \href{https://oeis.org/A049609}{A049606} in the OEIS, which has a relation with Gould's sequence that allows us to characterize the factor of all polynomials $p_{n,2}(n,2n-j)$. Thus, if we denote the denominator as $\mu(k)$, it is noted that the greatest common divisor portion coincides precisely with that identified in $G(k)$.
\begin{align}
    p_{n,2}(n,k)&=\frac{G(2n-k)}{\mu(2n-k)}\sum _{i=0}^{2 n-k} n^i \left[ \begin{matrix} 2n-k\\i \end{matrix} \right]\\\notag
    &=\frac{\displaystyle\frac{2^{2n-k}}{\gcd \left(2^{2n-k},(2n-k)!\right)}}{\displaystyle\frac{(2n-k)!}{\gcd \left(2^{2n-k},(2n-k)!\right)}}\sum _{i=0}^{2 n-k} n^i \left[ \begin{matrix} 2n-k\\i \end{matrix} \right]\\\notag
    &=\frac{2^{2n-k}}{(2n-k)!}\sum _{i=0}^{2 n-k} n^i \left[ \begin{matrix} 2n-k\\i \end{matrix} \right]=\frac{2^{2 n-k} n^{(2 n-k)}}{(2 n-k)!}=\frac{2^{2 n-k} \Gamma (n+1)}{\Gamma (k-n+1) \Gamma (-k+2 n+1)}
\end{align}
And, as for the rest of the polynomial, it is composed of monomials with a degree whose dependence on $j$ is linear. Its coefficients, when seeking the sequence they constitute, it is concluded that they are the Stirling numbers of the first kind \href{https://oeis.org/A008275}{A008275}, which in turn provide the appropriate sign for each monomial. Therefore, the final form of $p_{n,2}(n,k)$ for all valid values of $k$ is presented above.

\begin{align}
    Pr[\mathbf{I}_{n,2}\leq i]=\sum _{j=0}^i \frac{\displaystyle\frac{2^{2 n-j} n^{(2 n-j)}}{(2 n-j)!}}{\displaystyle\binom{2n}{j}}\frac{(2n)!}{(2n)^i (2n-j)!} \stirling{i}{j}
\end{align}

Once the metric that determines the number of terminal states for each $k$ is acquired, the probability distribution of $\mathbf{I}_{n,2}$ can be fully characterized by substituting its expression in the ratio against the total number of existing states $\binom{2n}{j}$.

\begin{figure}[H]
    \centering
    \begin{subfigure}[b]{0.49\textwidth}
        \centering
        \includegraphics[width=\textwidth,clip]{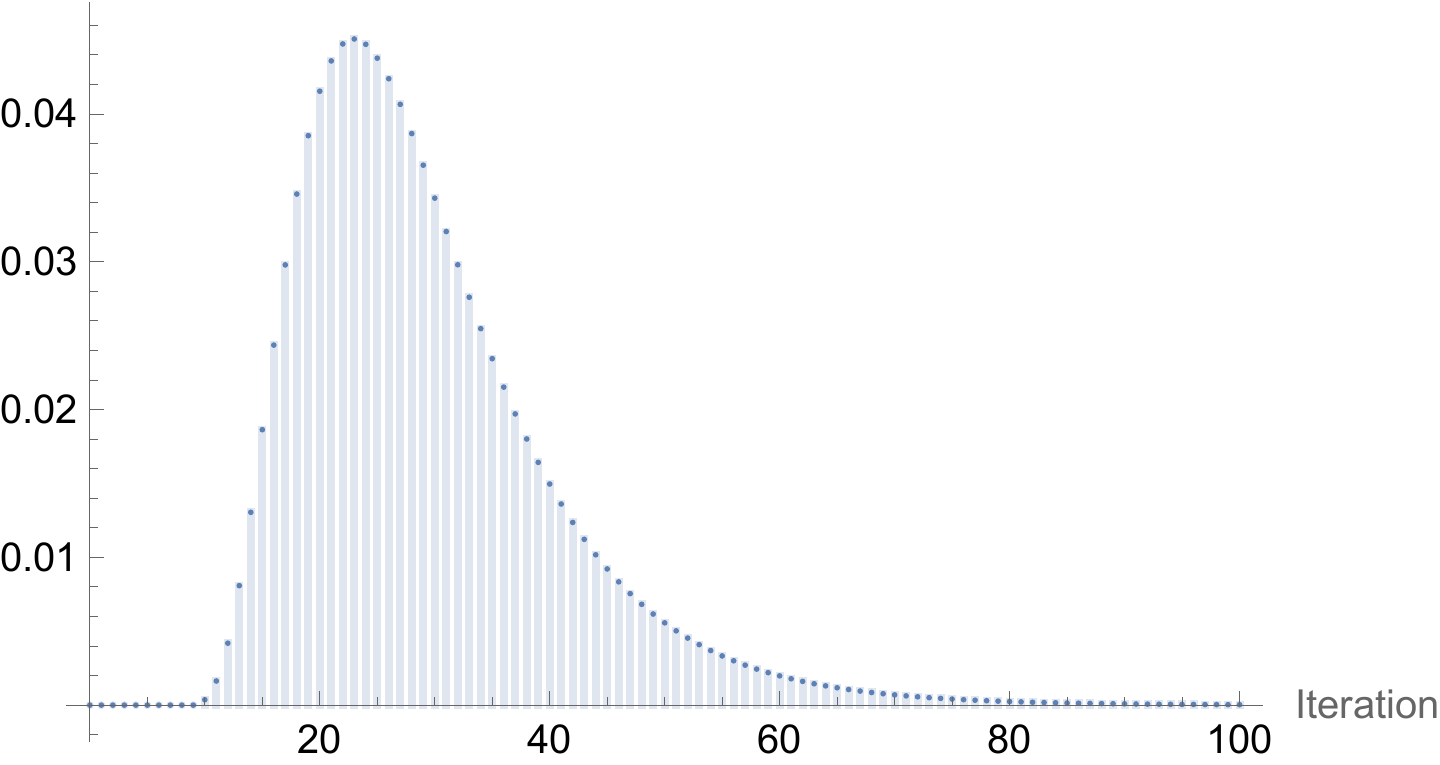}
        \caption{}
        \label{fig:2Dcase2density}
    \end{subfigure}
    \hfill
    \begin{subfigure}[b]{0.49\textwidth}
        \centering
        \includegraphics[width=\textwidth,clip]{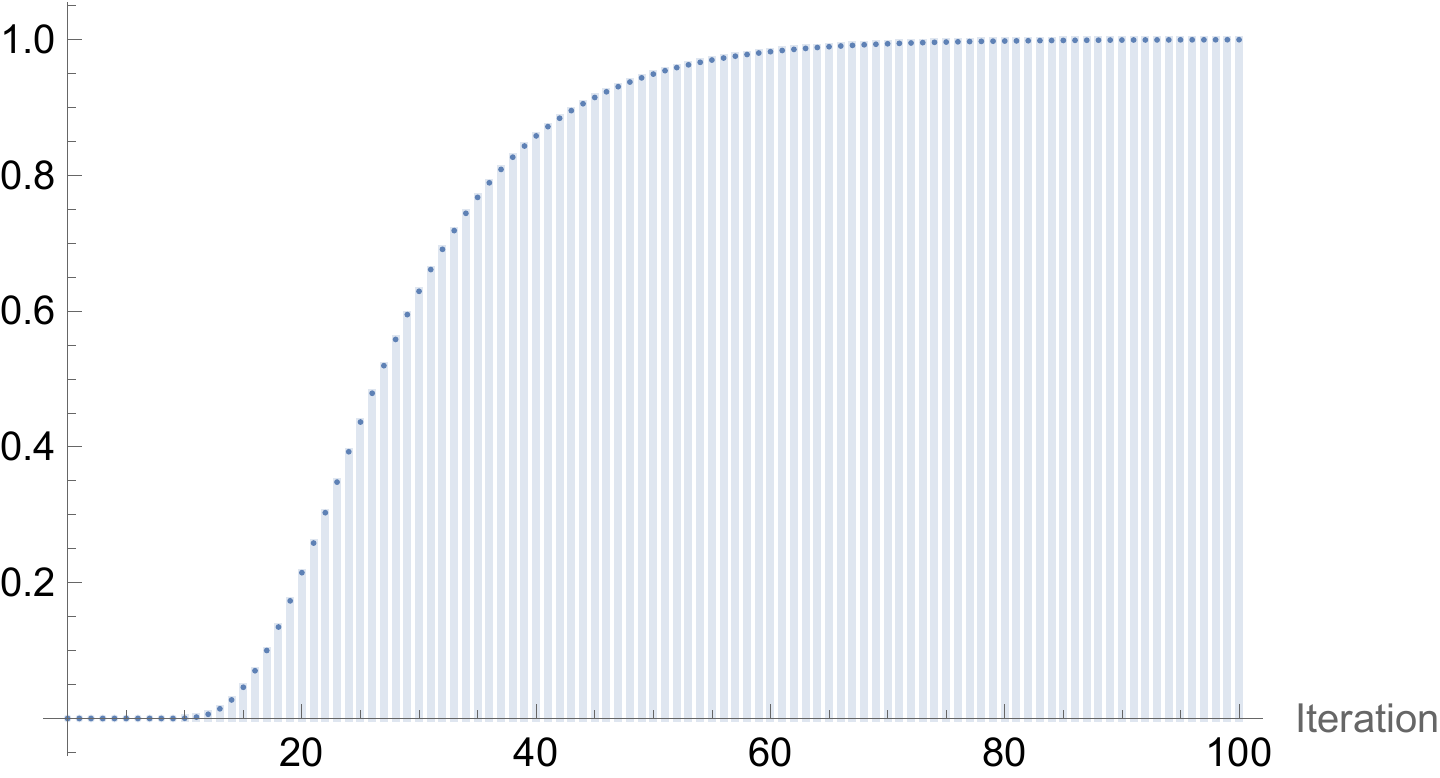}
        \caption{}
        \label{fig:2Dcase2CDF}
    \end{subfigure}
    \caption{(a) Probability density function of $\mathbf{I}_{n,2}$ along with its cumulative distribution function in (b), plotted with $n=10$}
    \label{fig:2Dcase2dual}
\end{figure}

With regard to the density function, it is noticeable that its shape bears a superficial resemblance to that obtained experimentally for square matrix systems. This similarity is not necessarily strong, as, despite being plotted with a small $n$, the $(10,2)$ sized system exhibits significantly less capacity for element insertion, so it will terminate sooner than the square matrix. Nonetheless, the value of this similarity lies in verifying that the shape does not vary drastically from the expected behavior. Additionally, since we have simulation data for $(n,2)$ sized systems, we can use them to verify the correctness of the superior density function, and therefore the probability distribution of $\mathbf{I}_{n,2}$.

\begin{figure}[H]
    \centering
    \includegraphics[width=11cm,clip]{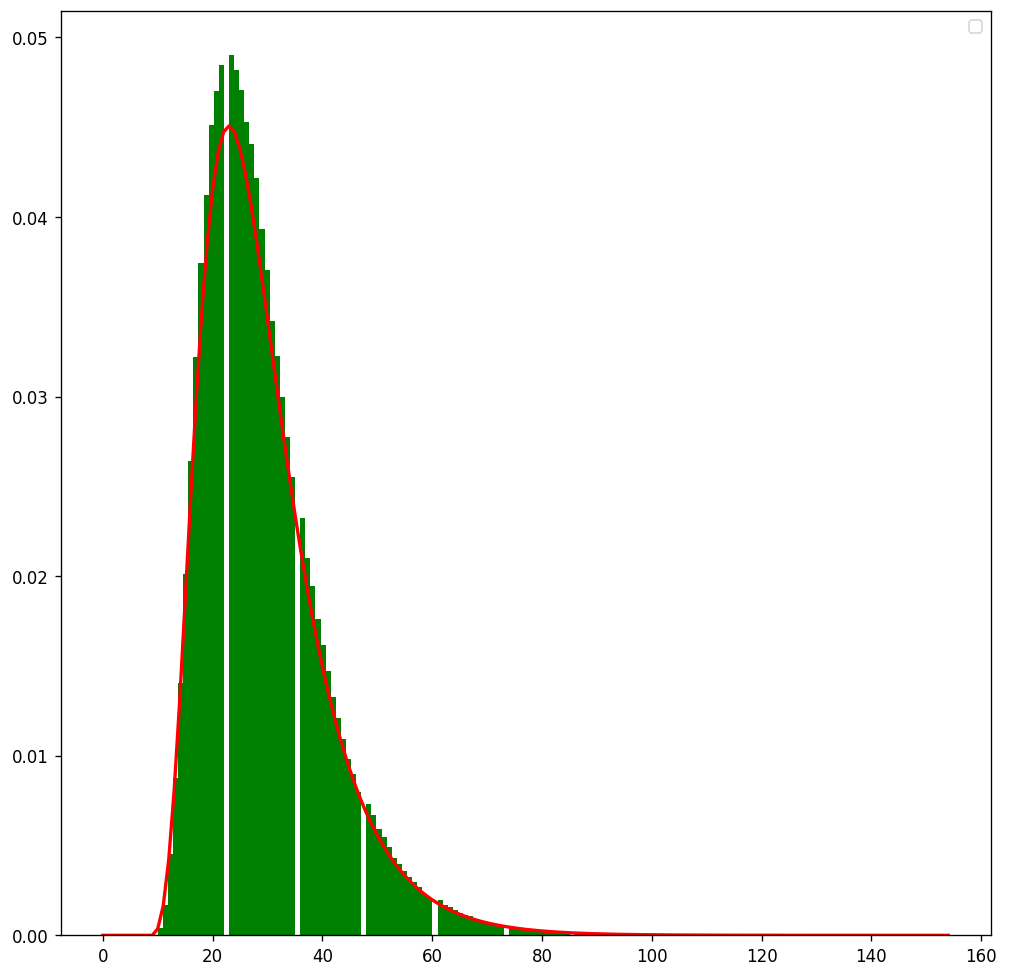}
    \caption{Histogram of numbers of iterations to reach the terminal state in a system of size $(10,2)$ plotted in green, along with the density function for $\mathbf{I}_{n,2}$ depicted in red.}
    \label{fig:2Dcase2fit}
\end{figure}

By overlaying experimental data from a sufficiently large number of system simulations with the desired size and density function, a clear similarity is observed which, although not perfect, indicates a reasonable resemblance whereby it is concluded that the previous probability distribution fits the modeled magnitude. However, it should be acknowledged that the evaluation of $Pr[\mathbf{I}_{n,2}=i]$ in the graph was conducted numerically, which contributes to increasing the difference between the histogram and the function it is supposed to follow. Additionally, it should also be noted that the small size of the system may not fully fit the exact density function. Nevertheless, asymptotically, the histogram, made with a very large number of simulations, and the function will coincide, observable by performing the experiment for higher $n$ values.

\begin{align}
     I(n)=\sum _{i=0}^\infty Pr[\mathbf{I}_{n,2}> i] = \sum _{i=0}^\infty \left(1-Pr[\mathbf{I}_{n,2}\leq i]\right) = \sum _{i=0}^\infty \left(1-\sum _{j=0}^i \frac{\displaystyle\frac{2^{2 n-j} n^{(2 n-j)}}{(2 n-j)!}}{\displaystyle\binom{2n}{j}}\frac{(2n)!}{(2n)^i (2n-j)!} \stirling{i}{j}\right)
\end{align}
Ultimately, through the complement of the cumulative distribution function \cite{NAIR2018429, Wolfram_SurvivalFunction_2010} probability, the expected number of iterations that a process lasts is determined. In this case, it does not have an easily inferable closed form, so it is evaluated numerically at certain values of $n$ and compared with the result it can potentially provide.

\begin{figure}[H]
    \centering
    \includegraphics[width=10cm,clip]{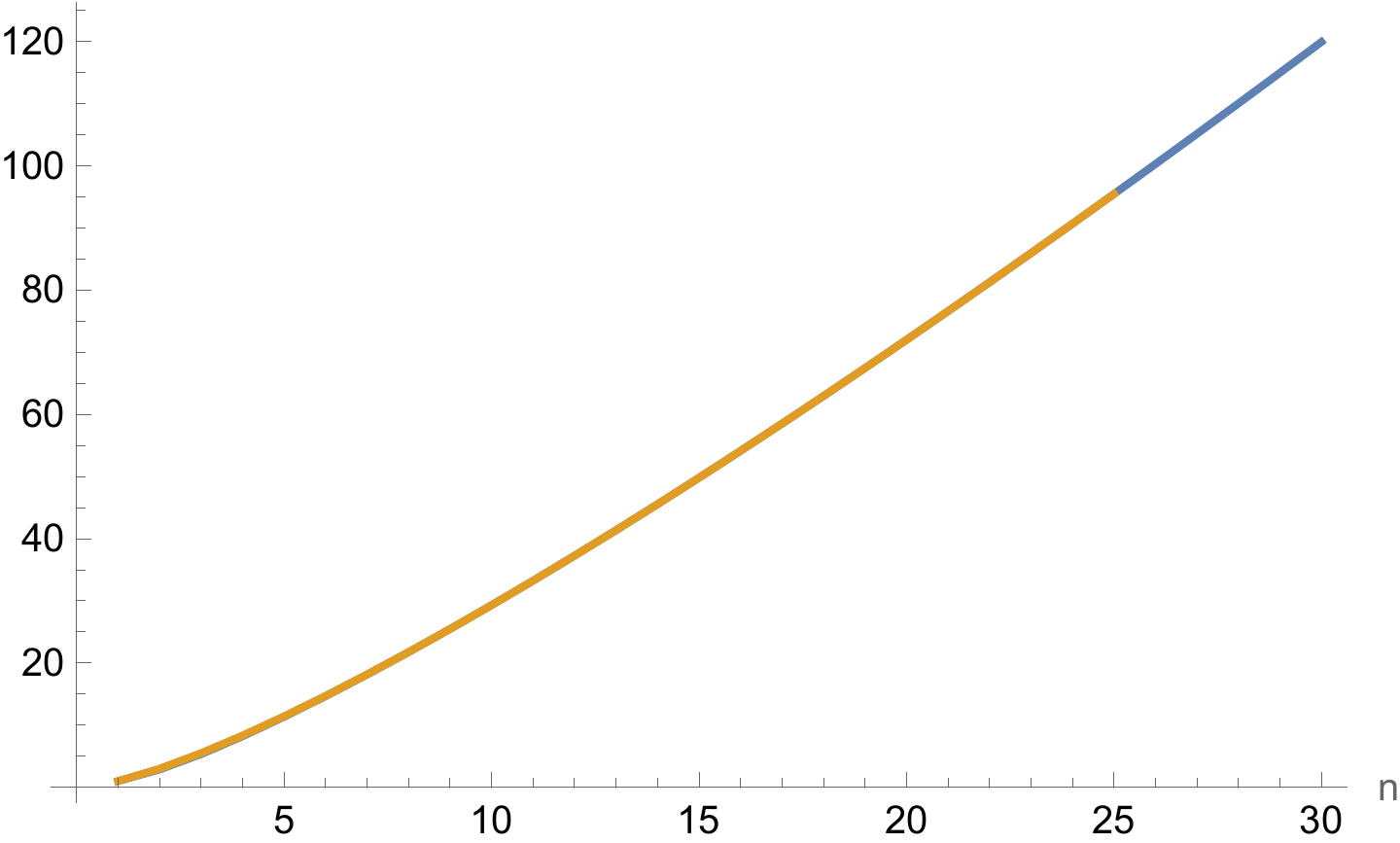}
    \caption{$I(n)$ evaluated numerically in blue and $nH_n$ function plotted in orange for a reduced range of $n$ values.}
    \label{fig:2Dcase2iterationsAverageFit}
\end{figure}

Given the similarity of a system of size $(n,2)$ with a one-dimensional system, the difference between the metric $I(n)$ in both cases should not modify its asymptotic growth as $n\to\infty$. That is, the system formed by a matrix of 2 columns behaves in the same way as one with a single column, although it has a slightly more space to insert elements and thus form paths. Similarly, most of the paths that can be generated in such a system are very similar to those that appear in the best case, joining both ends of the system with a straight line. Therefore, knowing that $I(n)=nH_n$ in one-dimensional systems, this same formula can be proposed as the approximate value that the metric reaches in the analyzed case. For this reason, Figure 61 shows the numerical evaluation of $I(n)$ for matrices with only 2 columns along with the function $nH_n$. At first glance, there is a strong similarity between the two, although it cannot be assured that they are equal due to the reduced range of $n$ in which they have been evaluated. However, given the variability in the analyzed system paths, it is most likely that the actual $I(n)$ is composed of a function asymptotically equivalent to $nH_n$, although with some multipliers or offsets that vary its growth for bounded $n$ given its difference with respect to one-dimensional systems.
\\\\
To summarize, a percolative process executed on a matrix $(n,2)$ lasts a similar number of iterations as that of a process in a one-dimensional system. Mainly, this occurs because the increase in space in the system is proportional to the difference in the parameter $m$ of the matrix size. Specifically, as the space increases with $m$, it is easily inferable that the duration of the process will be longer, considering that $n$ is fixed. Even without knowing how much longer it is, it is shown above that for bounded values of $n$ the difference is minimal between both systems. However, with a sufficiently large $m$, a more significant deviation would be appreciated, which in turn can be used to try to generalize the metric $I(n)$ for arbitrary values of $m$.

\begin{align}
    p_{n,3}(n,3n)&=\binom{3n}{3n}=1\\\notag
    p_{n,3}(n,3n-1)&=\binom{3n}{3n-1}=3n\\\notag
    p_{n,3}(n,3n-2)&=\binom{3n}{3n-2}=\frac{3}{2} \left(3 n^2-n\right)\\\notag
    p_{n,3}(n,3n-3)&=\binom{3 n}{3 n-3}-n \binom{3 n-3}{0}=\frac{9}{2} \left(n^3-n^2\right)\\\notag
    p_{n,3}(n,3n-4)&=\binom{3 n}{3 n-4}-n \binom{3 n-3}{1}-2 (n-1)=\frac{1}{8} \left(27 n^4-54 n^3+9 n^2+2 n+16\right)\\\notag
    p_{n,3}(n,3n-5)&=\binom{3 n}{3 n-5}-n \binom{3 n-3}{2}-6 (n-1) (n-2)-2 (n-2)=\\\notag
    &=\frac{1}{40} \left(81 n^5-270 n^4+135 n^3+30 n^2+424 n-320\right)
\end{align}

Consequently, despite not having derived the closed form of $I(n)$ for systems with 2 columns ($m=2$), we have an expression that we can evaluate numerically and compare with possible functions that have the same asymptotic growth, or even return all the exact values of the magnitude $I(n)$. In this way, if we solve the magnitude for cases where $m$ is higher, there is a possibility of obtaining a general expression for any $m$, which would imply that later substituting $m\to n$, the metric would return the duration of a percolative process in square matrices. To this end, the first expressions for the number of terminal states in systems with $m=3$, denoted as $p_{n,3}(n,j)$, have been presented above. As before, we begin with the maximum number of elements to simplify the calculation. However, after reducing the expressions to polynomials and attempting to model them, a clear sequence for their coefficients, nor for the numerical factor that multiplies the entire polynomial has not been found. Regarding their degree and signs, a general pattern can be inferred for any number of elements, although it is not enough to characterize the polynomial precisely.
\\\\
In conclusion, the magnitude that determines and in turn prevents the exact calculation of $I(n)$ for $m=3$ is the count of terminal states. It is conceivable that a general solution for all valid $j$ can be reached using the aforementioned method, perhaps with generating functions or more advanced techniques. But, for now, this will be proposed and the same process will be repeated for the case of $m=4$, given that its parity may influence the properties exhibited by the polynomials and that we can take advantage of for our purpose.

\begin{align}
    p_{n,4}(n,4n)&=\binom{4n}{4n}=1\\\notag
    p_{n,4}(n,4n-1)&=\binom{4n}{4n-1}=4n\\\notag
    p_{n,4}(n,4n-2)&=\binom{4n}{4n-2}=2 \left(4 n^2-n\right)\\\notag
    p_{n,4}(n,4n-3)&=\binom{4n}{4n-3}=\frac{4}{3} \left(8 n^3-6 n^2+n\right)\\\notag
    p_{n,4}(n,4n-4)&=\binom{4n}{4n-4}-n \binom{4 n-4}{0}=\frac{2}{3} \left(16 n^4-24 n^3+11 n^2-3 n\right)\\\notag
    p_{n,4}(n,4n-5)&=\binom{4n}{4n-5}-n \binom{4 n-4}{1}-4 (n-1)=\frac{4}{15} \left(32 n^5-80 n^4+70 n^3-40 n^2+3 n+15\right)\\\notag
    p_{n,4}(n,4n-6)&=\binom{4n}{4n-6}-n \binom{4 n-4}{2}-6 (n-2)-2 (n-1) ((4 n-7)+(4 n-6))=\\\notag
    &=\frac{2}{45} \left(128 n^6-480 n^5+680 n^4-630 n^3+182 n^2+570 n-315\right)
\end{align}
In $m=4$, the process yields polynomials for which, as in the previous case, no generating function has been found to model the coefficients of each monomial. Likewise, the sequence of the multiplier could not be characterized, although the degree and signs remain easy to extrapolate. Therefore, it is left as it appears above. And, in terms of subsequent analysis, we will use the results obtained in this section in special cases, such as when the system must be filled with elements or end in the smallest number of iterations.
\subsection{Worst case analysis}
First of all, the best case for this dimension is extrapolated from the one analyzed in section \textcolor{blue}{\ref{Methodology}}, as in any system the shortest path will always have the same shape and length. So, we proceed to analyze the worst case using a methodology analogous to that used in one-dimensional systems.

\begin{align}
    T_{worst}(n) = \sum _{i=0}^{I(n,n^2)} \left(1-\frac{E(i)}{n^2}\right)\cdot c(n,k)=\sum _{i=0}^{I(n)} \left(1-\frac{E(i)}{n^2}\right)\cdot E(i)
\end{align}

The total work formulation remains invariant; that is, it is defined by the sum of the work performed in each process iteration, which is expressed as the insertion probability multiplied by the average cluster size. Since this is the worst case, it is considered that the process lasts the maximum number of iterations, which coincides with the necessary ones to insert $n^2$ elements. As for the cluster size, it is considered to be the maximum possible on average, which is equivalent to the total number of elements inserted up to an iteration. This ensures that the work of traversing a cluster is always maximal in every iteration, since if all the inserted elements are together and form a cluster that does not cause a state change in the system, the algorithm will traverse it completely in order to reach the extreme rows.

\begin{align}
    T_{worst}(n) = \int _{0}^{I(n)} \left(1-\frac{E(i)}{n^2}\right)\cdot E(i) \, di
\end{align}
With this, the integral sum of the work is first posed, which is anticipated to be comparatively simpler to solve, and it will provide an equivalent result to that of the discrete formulation.
\begin{align}
    \int \left(1-\frac{E(i)}{n^2}\right)\cdot E(i) \, di&=\int \left(1-\frac{n^2\left(1-\left(1-\frac{1}{n^2}\right)^i\right)}{n^2}\right)\cdot n^2\left(1-\left(1-\frac{1}{n^2}\right)^i\right) \, di=\\\notag
    &=\int -n^2 \left(\left(1-\frac{1}{n^2}\right)^i-1\right) \left(1-\frac{1}{n^2}\right)^i \, di=\\\notag
    &=-n^2\int \left(\left(1-\frac{1}{n^2}\right)^i-1\right) \left(1-\frac{1}{n^2}\right)^i \, di=\\\notag
    &=-\frac{n^2}{\ln\left(1-\frac{1}{n^2}\right)}\int u-1 \, du=\quad\quad [u=\left(1-\frac{1}{n^2}\right)^i]\\\notag
    &=\frac{n^2 u}{\ln\left(1-\frac{1}{n^2}\right)}-\frac{n^2 u^2}{2\ln\left(1-\frac{1}{n^2}\right)}+C=-\frac{n^2 \left(\left(1-\frac{1}{n^2}\right)^i-2\right) \left(1-\frac{1}{n^2}\right)^i}{2 \ln \left(1-\frac{1}{n^2}\right)}+C\\\notag
\end{align}
Following the determination of its antiderivative, the limits of integration are evaluated to produce $T_{worst}(n)$:
\begin{align}
    T_{worst}(n) = \left . -\frac{n^2 \left(\left(1-\frac{1}{n^2}\right)^i-2\right) \left(1-\frac{1}{n^2}\right)^i}{2 \ln \left(1-\frac{1}{n^2}\right)} \right|_{i=0}^{i=I(n)} = -\frac{n^2 \left(\left(1-\frac{1}{n^2}\right)^{I(n)}-1\right)^2}{2 \ln \left(1-\frac{1}{n^2}\right)}
\end{align}
Subsequently, and as has happened on previous occasions, the asymptotic growth of $T_{worst}(n)$ cannot be easily deduced at first glance, despite the intuition that being the worst case of a 2-dimensional system, the bound would be of the order $O(n^4)$ due to the system size and the work done in each iteration. Consequently, the limit of the time complexity as $n \to \infty$ is evaluated by substituting the metric $I(n)$ with the specification $I(n,k)$, which provides a measure of the complexity when the system reaches a quantity $k$ of elements.
\allowdisplaybreaks
\begin{align}
    \lim_{n\to\infty} T_{worst}(n) &= \lim_{n\to\infty} -\frac{n^2 \left(\left(1-\frac{1}{n^2}\right)^{I(n,k)}-1\right)^2}{2 \ln \left(1-\frac{1}{n^2}\right)} =\\ \notag
    &= \lim_{n\to\infty} \frac{n^2 \left(\left(1-\frac{1}{n^2}\right)^{I(n,k)}-1\right)^2}{\frac{2}{n^2}}=\\ \notag
    &= \lim_{n\to\infty} \frac{1}{2}n^4 \left(\left(1-\frac{1}{n^2}\right)^{2I(n,k)}-2\left(1-\frac{1}{n^2}\right)^{I(n,k)}+1\right)=\\ \notag 
    &= \lim_{n\to\infty} \frac{1}{2}n^4 \left(e^{-2I(n,k)/n^2}-2e^{-I(n,k)/n^2}+1\right)=\\ \notag
    &= \lim_{n\to\infty} \frac{I(n,k)n^2\left(e^{I(n,k)/n^2}-1\right)}{2e^{2I(n,k)/n^2}}=\\ \notag
    &= \lim_{n\to\infty} \frac{I(n,k)n^2\left(e^{I(n,k)/n^2}-1\right)}{2}=\lim_{n\to\infty} \frac{I(n,k)^2e^{I(n,k)/n^2}}{2}=\boxed{\lim_{n\to\infty} \frac{I(n,k)^2}{2}}
\end{align}
According to this limit, the worst-case complexity of the algorithm grows asymptotically at the same rate as the square of $I(n,k)$, and since the system must be completely filled during the process, it could be inferred that the complexity is of the order $O(k^4)$ with $k=n^2$. However, when the elements reach their maximum, an indetermination occurs when substituting the metric $I(n)$ for the specification $I(n,k)$. Hence, it cannot be assured that the form of the complexity obtained above is the actual one. So, it becomes essential to verify its validity through the ratio between $T_{worst}(n)$ and the prospective bound.
\begin{align}
    \lim_{n\to\infty} \frac{T_{worst}(n)}{(n^2H_{n^2})^2} &= \lim_{n\to\infty} -\frac{n^2 \left(\left(1-\frac{1}{n^2}\right)^{n^2H_{n^2}}-1\right)^2}{2 \ln \left(1-\frac{1}{n^2}\right)n^4(H_{n^2})^2} =\\ \notag
    &= \lim_{n\to\infty} \frac{n^2 \left(\left(1-\frac{1}{n^2}\right)^{n^2H_{n^2}}-1\right)^2}{2n^2(H_{n^2})^2}=\\ \notag
    &= \lim_{n\to\infty} \frac{1}{2(H_{n^2})^2} \left(\left(1-\frac{1}{n^2}\right)^{2n^2H_{n^2}}-2\left(1-\frac{1}{n^2}\right)^{n^2H_{n^2}}+1\right)=\\ \notag
    &= \lim_{n\to\infty} \frac{1}{2(H_{n^2})^2} \left(e^{-2H_{n^2}}-2e^{-H_{n^2}}+1\right)=\\ \notag
    &= \lim_{n\to\infty} \frac{1}{2(H_{n^2})^2} \left(\frac{1}{e^{2H_{n^2}}}-\frac{2}{e^{H_{n^2}}}+1\right)=\\ \notag
    &= \lim_{n\to\infty} \frac{1}{2(H_{n^2})^2} \left(\frac{1}{e^{4\ln(n)+2\gamma}}-\frac{2}{e^{2\ln(n)+\gamma}}+1\right)=\\\notag
    &= \lim_{n\to\infty} \frac{1}{2(H_{n^2})^2} \left(\frac{e^{-2 \gamma }}{n^4}-\frac{2 e^{-\gamma }}{n^2}+1\right)=\lim_{n\to\infty} \frac{1}{2(H_{n^2})^2}=0
\end{align}
On the one hand, $I(n,n^2)$ can be expressed as $(n^2H_{n^2})^2$, leading to a possible bound that might have the same growth as the complexity when $n\to\infty$. Yet, when checking the ratio between the complexity and this function, it is observed that it converges to 0, indicating that the proposed bound reaches increasingly higher values with respect to the complexity as it approaches the accumulation point.
\begin{align}
    \lim_{n\to\infty} \frac{T_{worst}(n)}{n^4} &= \lim_{n\to\infty} -\frac{n^2 \left(\left(1-\frac{1}{n^2}\right)^{n^2H_{n^2}}-1\right)^2}{2 \ln \left(1-\frac{1}{n^2}\right)n^4} =\\ \notag
    &= \lim_{n\to\infty} \frac{n^2 \left(\left(1-\frac{1}{n^2}\right)^{n^2H_{n^2}}-1\right)^2}{2n^2}=\\ \notag
    &= \lim_{n\to\infty} \frac{1}{2} \left(\left(1-\frac{1}{n^2}\right)^{2n^2H_{n^2}}-2\left(1-\frac{1}{n^2}\right)^{n^2H_{n^2}}+1\right)=\\ \notag
    &= \lim_{n\to\infty} \frac{1}{2} \left(e^{-2H_{n^2}}-2e^{-H_{n^2}}+1\right)=\\ \notag
    &= \lim_{n\to\infty} \frac{1}{2} \left(\frac{1}{e^{2H_{n^2}}}-\frac{2}{e^{H_{n^2}}}+1\right)=\\ \notag
    &= \lim_{n\to\infty} \frac{1}{2} \left(\frac{1}{e^{4\ln(n)+2\gamma}}-\frac{2}{e^{2\ln(n)+\gamma}}+1\right)=\\\notag
    &= \lim_{n\to\infty} \frac{1}{2} \left(\frac{e^{-2 \gamma }}{n^4}-\frac{2 e^{-\gamma }}{n^2}+1\right)=\boxed{\frac{1}{2}}
\end{align}
Conversely, an alternative possibility is to consider $I(n,n^2)$ as the size of the system $n^2$, which in turn would be comparable to the asymptotic growth of $I(n,k)$ when $k\to n$. Although this trend holds, it does not in the limit situation when $k$ is directly substituted by $n^2$, which is the main reason for the indeterminacy that requires these validations in the analysis. Thus, by verifying whether the possible bound $(n^2)^2$ grows in the same way as the complexity of the algorithm, it can be seen above how the limit of its ratio converges to a real value of $\frac{1}{2}$. With this, it can be concluded that this bound is definitive, except for a factor of 2 that would be needed to ensure that the ratio converges to 1, and consequently imply that the bound is precise \cite{Clark925053}.

\begin{align}
    T_{worst}(n) &=\sum _{i=0}^{I(n)} \left(1-\frac{E(i)}{n^2}\right)\cdot E(i) =\\\notag
    &=\sum _{i=0}^{I(n)} -n^2 \left(\left(1-\frac{1}{n^2}\right)^i-1\right) \left(1-\frac{1}{n^2}\right)^i =\\ \notag
    &=\frac{n^2 \left(n^2-1\right) \left(\left(1-\frac{1}{n^2}\right)^{I(n)}-1\right) \left(\left(n^2-1\right) \left(1-\frac{1}{n^2}\right)^{I(n)}-n^2\right)}{2 n^2-1}
\end{align}
So far, through the analysis based on the integral formulation of $T_{worst}(n)$, a growth of $\Theta(n^4)$ is obtained in the temporal cost of the percolation process in its worst case. However, it is now essential to ascertain that through its formulation with the discrete sum, the same bound is reached, which, while not exactly the same for measuring the cost of specific cases with $n$, retains its asymptotic behavior as the system enlarges. For this, the $Sum[]$ function from Wolfram was used above to find a closed expression for the sum of the work performed by the algorithm over the iterations of its execution. And, as it remains true, no clear growth is distinguished, so the subsequent limit is computed:

\begin{align}
    \lim_{n\to\infty} T_{worst}(n) &= \lim_{n\to\infty} \frac{n^2 \left(n^2-1\right) \left(\left(1-\frac{1}{n^2}\right)^{I(n,k)}-1\right) \left(\left(n^2-1\right) \left(1-\frac{1}{n^2}\right)^{I(n,k)}-n^2\right)}{2 n^2-1} =\\ \notag
    &= \lim_{n\to\infty} \frac{n^2 \left(n^2-1\right) \left(e^{-I(n,k)/n^2}-1\right) \left(\left(n^2-1\right) e^{-I(n,k)/n^2}-n^2\right)}{2 n^2-1} =\\ \notag
    &= \lim_{n\to\infty} -\frac{n^2 e^{-\frac{I(n,k)}{n^2}}}{2 n^2-1}+\frac{n^2 e^{-\frac{2 I(n,k)}{n^2}}}{2 n^2-1}-\frac{2 n^6 e^{-\frac{I(n,k)}{n^2}}}{2 n^2-1}+\frac{n^6 e^{-\frac{2 I(n,k)}{n^2}}}{2 n^2-1}+\frac{3 n^4 e^{-\frac{I(n,k)}{n^2}}}{2 n^2-1}\\\notag
    &-\frac{2 n^4 e^{-\frac{2 I(n,k)}{n^2}}}{2 n^2-1}+\frac{n^6}{2 n^2-1}-\frac{n^4}{2 n^2-1} =\\ \notag
    &= \lim_{n\to\infty} -\left(n^2-1\right) n^2 e^{-I(n,k)/n^2}+\frac{\left(n^2-1\right)^2 n^2 e^{-2I(n,k)/n^2}+\left(n^2-1\right) n^4}{2 n^2-1} =\\ \notag
    &= \lim_{n\to\infty} \frac{1}{2} \left(I(n,k)^2+I(n,k)\right) \sim \boxed{\lim_{n\to\infty} \frac{I(n,k)^2}{2}}\quad\quad[k=o(n^2)]
\end{align}
Similarly, it is deduced that asymptotically the complexity depends on the square of the number of iterations needed by the process to insert $k$ elements. With this, we could conclude that the resulting bound will be the same as in the integral formulation, although the nature of the expression of $T_{worst}(n)$ for both formulations together with the indeterminacy occurring at $k=n^2$ may invalidate this guarantee. To resolve this issue, it is simply verified whether the possible bounds deriving from $\frac{I(n,k)^2}{2}$ have the same growth as the complexity using the limit of its ratio, which ensures independently of previous determinations whether the verified bounds match the growth of $T_{worst}(n)$ or not.
\begin{align}
    \lim_{n\to\infty} T_{worst}(n) &= \lim_{n\to\infty} \frac{n^2 \left(n^2-1\right) \left(\left(1-\frac{1}{n^2}\right)^{I(n,k)}-1\right) \left(\left(n^2-1\right) \left(1-\frac{1}{n^2}\right)^{I(n,k)}-n^2\right)}{2 n^2-1} =\\ \notag
    &=\lim_{n\to\infty} \frac{n^2 \left(n^2-1\right) \left(\left(1-\frac{I(n,k)}{n^2}\right)-1\right) \left(\left(n^2-1\right) \left(1-\frac{I(n,k)}{n^2}\right)-n^2\right)}{2 n^2-1}=\\ \notag
    &=\lim_{n\to\infty} \frac{n^2 \left(n^2-1\right) \left(-\frac{I(n,k)}{n^2}\right) \left(\left(n^2-1\right) \left(1-\frac{I(n,k)}{n^2}\right)-n^2\right)}{2 n^2-1}=\\ \notag
    &=\lim_{n\to\infty} \frac{I(n,k)^2 n^2}{2 n^2-1}-\frac{2 I(n,k)^2}{2 n^2-1}+\frac{I(n,k)^2}{n^2 \left(2 n^2-1\right)}+\frac{I(n,k) n^2}{2 n^2-1}-\frac{I(n,k)}{2 n^2-1}=\\ \notag
    &=\lim_{n\to\infty} \frac{I(n,k) \left(n^2-1\right) \left(I(n,k) \left(n^2-1\right)+n^2\right)}{n^2 \left(2 n^2-1\right)}=\\ \notag
    &=\lim_{n\to\infty} -\frac{I(n,k)^2}{n^2}+\frac{I(n,k)^2-I(n,k)}{2 \left(2 n^2-1\right)}+\frac{1}{2} (I(n,k)+1) I(n,k)=\\ \notag
    &= \lim_{n\to\infty} \frac{1}{2} \left(I(n,k)^2+I(n,k)\right) \sim \boxed{\lim_{n\to\infty} \frac{I(n,k)^2}{2}}
\end{align}
It is also worth noting that the previous limit has alternative resolutions depending on the substitution by infinitesimals applied at the beginning. That is, since only the range $k = o(n^2)$ is being considered, there are several possibilities to substitute the terms where $I(n,k)$ appears with equivalent expressions as $n \to \infty$.
\begin{align}
    \lim_{n\to\infty} \frac{T_{worst}(n)}{n^4} &= \lim_{n\to\infty} \frac{n^2 \left(n^2-1\right) \left(\left(1-\frac{1}{n^2}\right)^{n^2H_{n^2}}-1\right) \left(\left(n^2-1\right) \left(1-\frac{1}{n^2}\right)^{n^2H_{n^2}}-n^2\right)}{(2 n^2-1)n^4} =\\ \notag
    &=\lim_{n\to\infty} \frac{n^2 \left(n^2-1\right) \left(e^{-H_{n^2}}-1\right) \left(\left(n^2-1\right) e^{-H_{n^2}}-n^2\right)}{(2 n^2-1)n^4}=\\ \notag
    &=\lim_{n\to\infty} \frac{n^2 \left(n^2-1\right) \left(-1\right) \left(1-n^2\right)}{(2 n^2-1)n^4}=\lim_{n\to\infty} \frac{n^6}{2n^6}=\boxed{\frac{1}{2}}
\end{align}

Continuing with the verification of the bounds, the one resulting from the analysis of the integral formulation is verified first, which in this case also leads to the ratio with the complexity converging to the same value of $\frac{1}{2}$. This indicates that the bound has the same growth as the worst-case cost expression, except for a multiplier that is negligible in this context.

\begin{align}
    \lim_{n\to\infty} \frac{T_{worst}(n)}{n^4(H_{n^2})^2} &= \lim_{n\to\infty} \frac{n^2 \left(n^2-1\right) \left(\left(1-\frac{1}{n^2}\right)^{n^2H_{n^2}}-1\right) \left(\left(n^2-1\right) \left(1-\frac{1}{n^2}\right)^{n^2H_{n^2}}-n^2\right)}{(2 n^2-1)n^4(H_{n^2})^2} =\\ \notag
    &=\lim_{n\to\infty} \frac{n^2 \left(n^2-1\right) \left(e^{-H_{n^2}}-1\right) \left(\left(n^2-1\right) e^{-H_{n^2}}-n^2\right)}{(2 n^2-1)n^4(H_{n^2})^2}=\\ \notag
    &=\lim_{n\to\infty} \frac{n^2 \left(n^2-1\right) \left(-1\right) \left(1-n^2\right)}{(2 n^2-1)n^4(H_{n^2})^2}=\lim_{n\to\infty} \frac{n^6}{2n^6(H_{n^2})^2}=\lim_{n\to\infty} \frac{1}{2(H_{n^2})^2}=0
\end{align}

Lastly, when verifying the secondary bound $n^4(H_{n^2})^2$, a ratio with respect to the complexity of 0 is calculated, indicating a greater growth by the bound. This could be noticeable without performing this verification, because the inclusion of the squared harmonic number, which is a function whose limit tends to infinity at the evaluated accumulation point, increases the growth of the other product term $n^4$.

\begin{figure}[H]
    \centering
    \begin{subfigure}[b]{0.49\textwidth}
        \centering
        \includegraphics[width=\textwidth,clip]{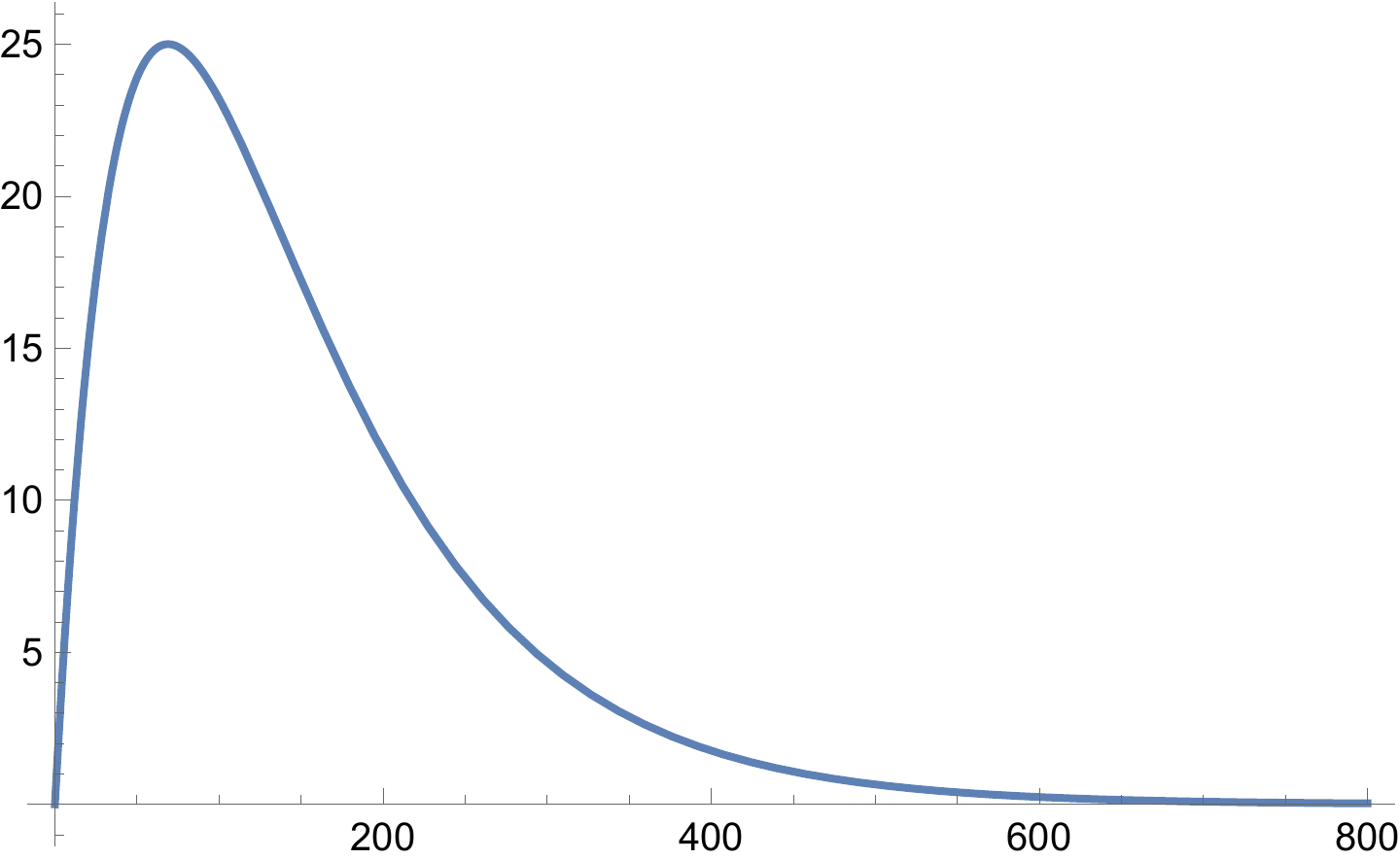}
        \caption{}
        \label{fig:2DWorstCaseInsertion}
    \end{subfigure}
    \hfill
    \begin{subfigure}[b]{0.49\textwidth}
        \centering
        \includegraphics[width=\textwidth,clip]{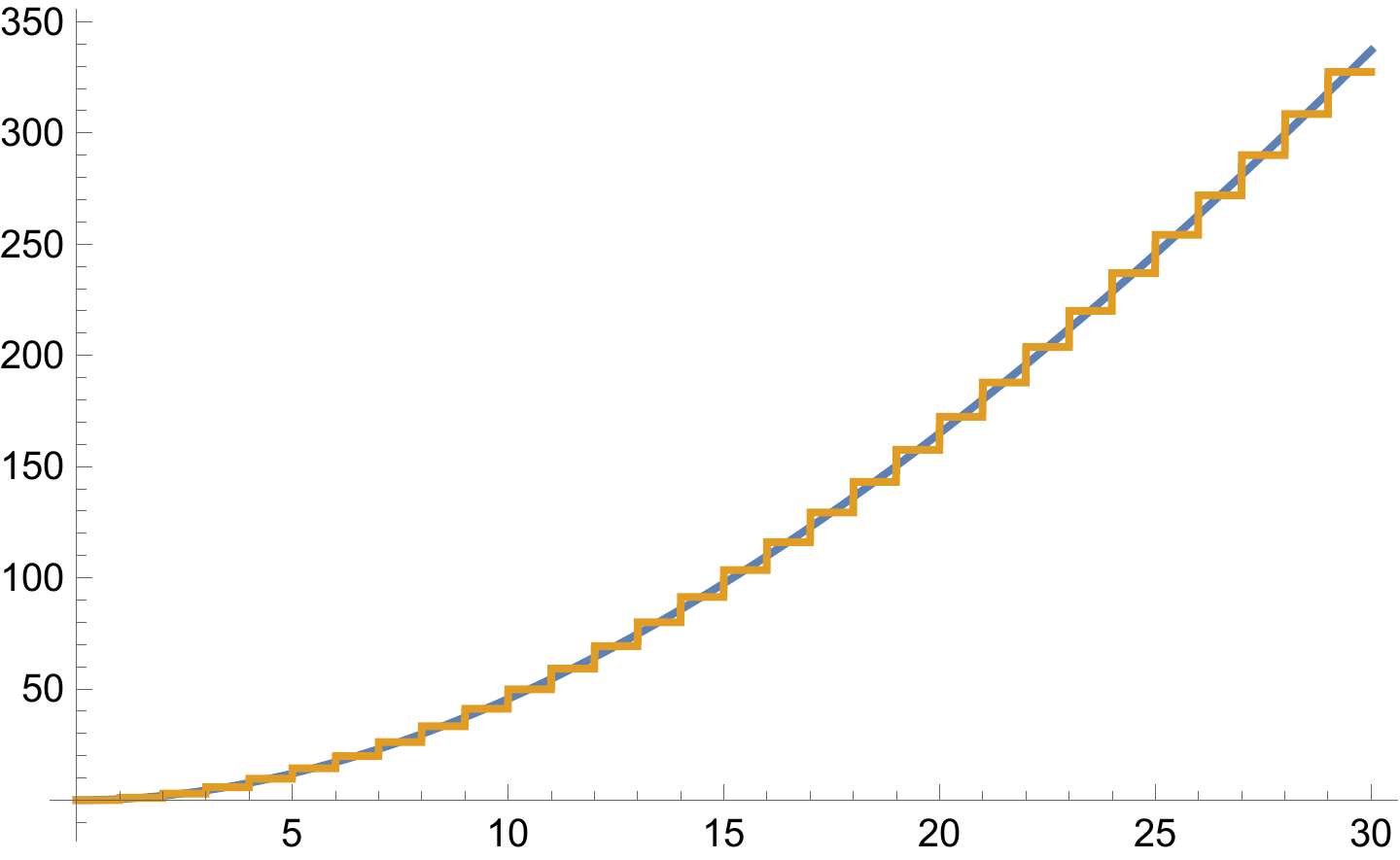}
        \caption{}
        \label{fig:2DWorstCaseSums}
    \end{subfigure}
    \caption{(a) $\left(1-\frac{E(i)}{n^2}\right)\cdot E(i)$ insertion complexity plotted for a 2D system of size $n=10$ (b) $T_{worst}(n)$ in its integral formulation plotted in blue for a 1D system of size $n=20$, and in orange its discrete variant.}
    \label{fig:2DWorstCaseDual}
\end{figure}
In the aforementioned figure, the cost associated with executing an insertion throughout iterations within a system of size $n=10$ is depicted, as well as the expressions of $T_{worst}(n)$ in integral and discrete forms. Regarding the latter, it is noted that they are not exactly the same, although their growth is equivalent. Therefore, after this procedure, it can be concluded that the worst-case complexity of the process executed in a 2-dimensional system is of the order $\Theta(n^4)$, which has a simple interpretation, since if in all the insertions a cell adjacent to the cluster formed by all the elements inserted up to that point is selected, the work of traversing the cluster will be maximum, and if the insertion sequence places the elements in such a way that it requires $\approx n^2-n=O(n^2)$ to form a valid path, the total workload will be $\sum _{i=0}^{n^2-n}i=O(n^4)$, coinciding with the attained bound.
\subsection{Average case analysis}
\label{subsec:AverageCaseAnalysis2D}
Next, knowing the algorithm's complexity in its best and worst-case scenarios, the only thing left to deduce is the growth of its average cost. Hence, the same methodology as in section \textcolor{blue}{\ref{1D-Average-case-analysis}} is followed to determine the work the algorithm performs on average over a system of a certain size.
\begin{align}
    T_{avg}(n) = \sum _{i=0}^{I(n)} \left(1-\frac{E(i)}{n^2}\right)\cdot c(n,E(i))=\sum _{i=0}^{I(n)} \left(1-\frac{E(i)}{n^2}\right)\cdot \frac{n^2}{n^2-E(i)+1}
\end{align}
Initially, the formulation is identical to the one-dimensional system analysis, considering the sum of the product between the insertion probability and the average cluster size. However, for 2-dimensional systems, we do not have an exact formula for the metric $c(n,k)$, which forces us to provisionally choose the same metric for systems of one lower dimension, considering its size equal to that of the analyzed 2-dimensional system. The advantage of this decision is based on the simplicity of the metric's expression, which provides an approximate idea of the cost growth. Thus, although it may seem that opting for this metric changes the final asymptotic bound, its expression meets the metric restrictions seen in section \textcolor{blue}{\ref{subsubsec:AverageInsertionRuntime}}, from which its form is deduced, reducing the likelihood of the final complexity differing significantly.

\begin{align}
    \lim_{n\to\infty} c(n,k)&=\lim_{n\to\infty} \frac{n^2}{n^2-k+1}=\begin{cases}
    1 & \text{if } k=o(n^2)\\
    \infty & \text{if } k= \Theta(n^2)
    \end{cases}
\end{align}

In this case, we could proceed as usual, solving the discrete sum or integral and then finding its asymptotic bound. Nevertheless, in the absence of the metric $I(n)$, we are presented with several options. The simplest way is to resolve the complexity in terms of the remaining metric and work as if that quantity were known, which has its pros and cons. On the other hand, the methodology we will ultimately use is based on studying the behavior of the insertion probability and the metric $c(n,k)$ at the accumulation point $n \to \infty$, subsequently bounding the asymptotic growth of $I(n)$, and using all the results to construct the bound of $T_{avg}(n)$. To this end, the result of the limit of the average number of clusters when the system increases in size is shown above. As can be seen, if the number of elements on the system is given by a function whose growth does not exceed the order of $n^2$, the average cluster size converges to 1. Conversely, if the system is full, that is, the number of elements adheres to the bound $\Theta(n^2)$, the limit diverges to infinity with a determined growth:

\begin{align}
    \lim_{n\to\infty} \frac{c(n,n^2)}{n^2}&=\lim_{n\to\infty} \frac{n^2}{n^2(n^2-n^2+1)}=1 \implies c(n,n^2)=\Theta(n^2)
\end{align}

In particular, if the denominator equals 1, which is the case when the system contains $n^2$ elements, the limit converges to a value that, if divided by a function of the same order as the denominator of the metric $c(n,k)$, results in 1, leading to the asymptotic growth of $c(n,n^2)$. With this result, the aim is to study the behavior of the metric $c(n,k)$ when $n\to\infty$ for all valid values of $k$, which can later be used to substitute the metric expression with the result of its convergence in $T_{avg}(n)$, if permitted by the insertion probability. However, the metric expression used to perform the limit does not exactly model the actual magnitude, so it will be necessary to perform tests with its estimators or even approximate its asymptotic growth from its definition.

\begin{align}
    \lim_{n\to\infty} I(n,k)&=\lim_{n\to\infty} n^2(H_{n^2}-H_{n^2-k})=\\ \notag
    &=\lim_{n\to\infty} n^2(\log(n^2)-\log(n^2-k))=\\ \notag
    &=\lim_{n\to\infty} \frac{\displaystyle\log(n^2)-\log(n^2-k)}{\frac{1}{n^2}}=\\ \notag
    &=\lim_{n\to\infty} \frac{\frac{1}{n^2}-\frac{1}{n^2-k}}{\frac{-1}{n^2}}=\lim_{n\to\infty} \frac{k n^2}{n^2-k} = k\lim_{n\to\infty} \frac{n^2}{n^2-k} = k \quad \colon \quad \boxed{n>0\enspace\land k=o(n^2)}
\end{align}

Before continuing, it is pertinent to consider the previously established asymptotic equivalence of $I(n,k)$ within two-dimensional systems. This is, as only the number of cells in the system influences this metric, we simply need to substitute the maximum size of the system into its original expression, derived from the coupon collector's problem, to arrive at the same conclusion as in the analysis of one-dimensional systems. However, in this case, the asymptotic equivalence with $k$ is only valid for $k=o(n^2)$. This will allow us, if necessary, to restrict the duration of the process under consideration by the terminal number of elements.

\begin{align}
    \lim_{n\to\infty} c(n,E(i))&=\lim_{n\to\infty} \frac{n^2}{n^2-E(i)+1}=\\\notag
    &=\lim_{n\to\infty} \frac{n^2}{n^2-n^2 \left(1-\left(1-\frac{1}{n^2}\right)^i\right)+1}=\\\notag
    &=\lim_{n\to\infty} \frac{n^2}{n^2 \left(1-\frac{1}{n^2}\right)^i+1}=\\\notag
    &=\lim_{n\to\infty} \frac{1}{e^{-i/n^2}+\frac{1}{n^2}}=\\\notag
    &=\lim_{n\to\infty} e^{\left(\frac{i}{n^2}-\frac{1}{n^2}\right)}=\lim_{n\to\infty} e^{\frac{i}{n^2}}=1 \quad [i=o(n^2)]
\end{align}

Likewise, in addition to parameterizing the metric $c(n,k)$ with a determined number of elements $k$, it is advisable to perform the same procedure with its estimation $E(i)$ for all iterations of a process. That is, to confirm the correctness of this estimation and ensure the tendency of the average cluster size previously achieved, the limit of $c(n,E(i))$ is computed. As observed, it converges to the same value as formerly calculated, at least for all iterations whose index is determined by a function of the set $o(n^2)$. Yet, given the previous asymptotic equivalence of $I(n,k)$, it can be inferred how the previous set can extend up to $o(n^2 H_{n^2})$ \cite{Rutanen2016}.

\begin{align}
    \lim_{n\to\infty} \frac{c(n,E(n^2H_{n^2}))}{n^2}&=\lim_{n\to\infty} \frac{n^2}{n^2(n^2-E(n^2H_{n^2})+1)}=\\\notag
    &=\lim_{n\to\infty} \frac{n^2}{n^2\left(n^2-n^2 \left(1-\left(1-\frac{1}{n^2}\right)^{n^2H_{n^2}}\right)+1\right)}=\\\notag
    &=\lim_{n\to\infty} \frac{n^2}{n^2\left(n^2 \left(1-\frac{1}{n^2}\right)^{n^2H_{n^2}}+1\right)}=\\\notag
    &=\lim_{n\to\infty} \frac{1}{n^2\left(e^{-n^2H_{n^2}/n^2}+\frac{1}{n^2}\right)}=\\\notag
    &=\lim_{n\to\infty} \frac{1}{n^2e^{-H_{n^2}}+1}=\\\notag
    &=\lim_{n\to\infty} \frac{1}{\frac{n^2}{\displaystyle e^{\ln(n^2)+e^\gamma}}+1}=\boxed{\frac{1}{1+e^{-\gamma}}} \enspace\implies c(n,E(n^2H_{n^2}))=\Theta(n^2)
\end{align}

Thus, in the limiting scenario where the system must be filled, which is characterized by the execution of $n^2H_{n^2}$ iterations, an asymptotic growth of $c(n,E(i))$ aligns with that obtained through parameterization based on the quantity of elements $k$, indicating that the estimation of elements $E(i)$ is consistent with the theoretical expectations observed during the algorithm's execution.

\begin{align}
    \lim_{n\to\infty} \hat{c}_0(n,k)&=\lim_{n\to\infty} \frac{2 n^2}{3 n^2 ln^2(p_{k,n})+2}=\\\notag
    &=\lim_{n\to\infty} \frac{2 n^2}{3 n^2 \ln^2\left(\frac{\displaystyle k}{\displaystyle n^2}\right)+2}=\\\notag
    &=\lim_{n\to\infty} \frac{2}{3\ln^2\left(\frac{\displaystyle k}{\displaystyle n^2}\right)}=0\quad [k=o(n^2)]
\end{align}

Likewise, in addition to verifying the trend of the cluster size metric through variations in its parameterization, it is advisable to calculate the limit using the estimators constructed in section \textcolor{blue}{\ref{subsubsec:ImageRangeCorrection}}.

\begin{align}
    \lim_{n\to\infty} \hat{c}_0(n,n^2)&=\lim_{n\to\infty} \frac{2 n^2}{3 n^2 ln^2(p_{n^2,n})+2}=\\\notag
    &=\lim_{n\to\infty} \frac{2 n^2}{3 n^2 \ln^2\left(\frac{\displaystyle n^2}{\displaystyle n^2}\right)+2}=\\\notag
    &=\lim_{n\to\infty} \frac{2 n^2}{2}=\boxed{\lim_{n\to\infty} n^2}\implies \hat{c}_0(n,n^2)=\Theta(n^2)
\end{align}

Considering the continuously derived measure, we can discern in the two previous limits how its convergence is equivalent to the result using the initial expression of $c(n,k)$. On the one hand, this verifies the validity of the estimator with respect to the asymptotic behavior  the magnitude should follow, and it also confirms its trend for different values of $k$ at the studied accumulation point.

\begin{align}
    &\lim_{n\to\infty} \hat{c}_1(n,k)=\\\notag
    &=\lim_{n\to\infty} \frac{n^2 \left(p_{k,n} \left(p_{k,n}^5-p_{k,n}^4+p_{k,n}^3+2 p_{k,n}^2+p_{k,n}-1\right)+1\right)}{n^2 \left(p_{k,n}^2+1\right) \left(p_{k,n}^2+p_{k,n}+1\right) (p_{k,n}-1)^2+p_{k,n} \left(p_{k,n}^5-p_{k,n}^4+p_{k,n}^3+2 p_{k,n}^2+p_{k,n}-1\right)+1} =\\\notag
    &=\lim_{n\to\infty} \frac{n^2 \left(\frac{k \left(\frac{k^5}{n^{10}}-\frac{k^4}{n^8}+\frac{k^3}{n^6}+\frac{2 k^2}{n^4}+\frac{k}{n^2}-1\right)}{n^2}+1\right)}{n^2 \left(\frac{k^2}{n^4}+\frac{k}{n^2}+1\right) \left(\frac{k^2}{n^4}+1\right) \left(\frac{k}{n^2}-1\right)^2+\frac{k \left(\frac{k^5}{n^{10}}-\frac{k^4}{n^8}+\frac{k^3}{n^6}+\frac{2 k^2}{n^4}+\frac{k}{n^2}-1\right)}{n^2}+1}=\\\notag
    &=\lim_{n\to\infty}\frac{n^2 \left(k^6-k^5 n^2+k^4 n^4+2 k^3 n^6+k^2 n^8-k n^{10}+n^{12}\right)}{k^6 n^2+k^6-k^5 n^4-k^5 n^2+k^4 n^6+k^4 n^4-2 k^3 n^8+2 k^3 n^6+k^2 n^{10}+k^2 n^8-k n^{12}-k n^{10}+n^{14}+n^{12}}=\\\notag
    &=\lim_{n\to\infty}\frac{n^{14}}{n^{14}}=1\quad [k=o(n^2)]
\end{align}

In the case of the discrete estimator, the process is relatively lengthier owing to its expression, although the same results are eventually reached. This occurs both when $k$ covers all the quantities of elements except the maximum, where the limit converges to 1 as seen above, and when the system is full and the average cluster size grows according to the system size.

\begin{align}
    &\lim_{n\to\infty} \hat{c}_1(n,n^2)=\\\notag
    &=\lim_{n\to\infty} \frac{n^2 \left(p_{n^2,n} \left(p_{n^2,n}^5-p_{n^2,n}^4+p_{n^2,n}^3+2 p_{n^2,n}^2+p_{n^2,n}-1\right)+1\right)}{n^2 \left(p_{n^2,n}^2+1\right) \left(p_{n^2,n}^2+p_{n^2,n}+1\right) (p_{n^2,n}-1)^2+p_{n^2,n} \left(p_{n^2,n}^5-p_{n^2,n}^4+p_{n^2,n}^3+2 p_{n^2,n}^2+p_{n^2,n}-1\right)+1} =\\\notag
    &=\boxed{\lim_{n\to\infty} n^2}\implies \hat{c}_1(n,n^2)=\Theta(n^2)
\end{align}

In light of the foregoing, it could be inferred that the average cluster size when the system size tends to infinity converges to 1 except when the system is full of elements. However, the previous checks have been carried out only with estimators and expressions that do not model the magnitude $c(n,k)$ exactly, which prevents ensuring that the studied convergence is met in practice. Therefore, the validity of the results of such convergence is demonstrated by defining the average cluster size. But first, it is worth noting that, according to the results in section \textcolor{blue}{\ref{subsubsec:AverageInsertionRuntime}}, by parameterizing the possible shape of the magnitude and calculating the average of all values achievable by such parameter, it is shown that for all quantities of elements in the system except the maximum, $c(n,k)$ converges to 1 as established. Likewise, $k=n^2$ obtained an identical convergence value, which does not coincide with that determined in this section because the parameterization was carried out with an arbitrary valid form, which may potentially not follow the true magnitude.

\begin{align}    
    c(n,k)=\frac{s(n,k)}{N(n,k)}=\frac{\displaystyle k\binom{n^2}{k}}{\displaystyle k\cdot t(n,k)+(k-1)\cdot t(n,k-1)+\cdots+t(n,1)}=\frac{\displaystyle k\binom{n^2}{k}}{\displaystyle \sum _{i=1}^{k}i\cdot t(n,i)}
\end{align}

Thus, to show what the average cluster size tends to as $n \to \infty$, we start from the aforementioned definition, in which the size of all clusters is divided by the number of existing ones. In this way, the denominator can be rewritten as the sum of a specific number of clusters multiplied by the combinations $t(n,k')$ that can be made of $k$ elements organized into exactly $k'$ clusters in a system of size $n$. Now, to prove that the magnitude tends to 1 at the desired accumulation point, it is enough to demonstrate that the asymptotic growth of the numerator and the denominator is equivalent.

\begin{align}    
    t(n,k)\sim\binom{n^2}{k}\enspace[n\to\infty]
\end{align}

In the case of the numerator, it is straightforward due to its existence in closed form. Conversely, for the denominator, it is necessary to find the asymptotic growth of the term $t(n,k')$ with the highest growth. Above, the bound followed by the term that models the combinations of $k$ elements when all of them form clusters of size 1 is shown, which is the same as counting combinations of $k$ kings on an $n\times n$ chessboard so that no pair of them attacks each other \cite{Kotesovec2013, caduk4946252, Brown2022, Bagno2020}. With this, and given that the maximum number of combinations of $k$ elements with any cluster configuration is $\binom{n^2}{k}$, it is concluded that the superior is the term with the highest growth in the sequence that constitutes $N(n,k)$, multiplied by its corresponding number of clusters contributed to the count $k$.

\begin{align}    
    t(n,k-1)\sim\binom{n^2}{k}\binom{k}{2}\frac{8}{n^2-1}\enspace[n\to\infty]
\end{align}

Subsequently \cite{Lavrov_4960032}, it is pertinent to determine the asymptotic growth for the scenario wherein $t(n,k')$ enumerates the combinations of $k-1$ clusters. Nonetheless, due to the complexity of this task, an upper bound is initially established.

\begin{align}
    \lim_{n\to\infty} c(n,k)&=\lim_{n\to\infty} \frac{\displaystyle k\binom{n^2}{k}}{\displaystyle k\cdot t(n,k)+(k-1)\cdot t(n,k-1)}=\\\notag
    &=\lim_{n\to\infty} \frac{\displaystyle k\binom{n^2}{k}}{\displaystyle k\cdot \binom{n^2}{k}+(k-1)\cdot \binom{n^2}{k}\binom{k}{2}\frac{8}{n^2-1}}=\\\notag
    &=\lim_{n\to\infty} \frac{\displaystyle k\binom{n^2}{k}}{\displaystyle \binom{n^2}{k}\left(k+(k-1)\binom{k}{2}\frac{8}{n^2-1}\right)}=\\\notag
    &=\lim_{n\to\infty} \frac{\displaystyle k\binom{n^2}{k}}{\displaystyle \binom{n^2}{k}k}=1\quad[k=o(n)]
\end{align}

By integrating it into the definition of $c(n,k)$, along with the asymptotic equivalence of the term with the largest growth in the numerator, convergence to 1 is maintained for values of $k$ that grow more slowly than $n$. This yields insights about the behavior of the metric in a range of $k$ that is not traversed during a complete percolation process, meaning there are quantities of elements that reach $n^2$, so it is necessary to perform a substitution like the one above that provides enough information to verify the convergence of the limit over the entire valid range of $k$.

\begin{align}
    \lim_{n\to\infty} c(n,k)&=\lim_{n\to\infty} \frac{\displaystyle k\binom{n^2}{k}}{\displaystyle k\cdot \binom{n^2}{k}+\sum _{i=2}^{k} \left((i-1)\cdot \binom{n^2}{k}\binom{k}{2}\frac{8}{n^2-1}\right)}=\\\notag
    &=\lim_{n\to\infty} \frac{\displaystyle k\binom{n^2}{k}}{\displaystyle k\cdot \binom{n^2}{k}+\binom{n^2}{k}\binom{k}{2}\frac{8}{n^2-1}\sum _{i=2}^{k} (i-1)}=\\\notag
    &=\lim_{n\to\infty} \frac{\displaystyle k\binom{n^2}{k}}{\displaystyle k\cdot \binom{n^2}{k}+\binom{n^2}{k}\binom{k}{2}\frac{4k(k-1)}{n^2-1}}=\\\notag
    &=\lim_{n\to\infty} \frac{\displaystyle k\binom{n^2}{k}}{\displaystyle \binom{n^2}{k}\left(k+\binom{k}{2}\frac{4k(k-1)}{n^2-1}\right)}=\lim_{n\to\infty} \frac{\displaystyle k\binom{n^2}{k}}{\displaystyle \binom{n^2}{k}k}=1\quad[k=o(n^{2/3})]
\end{align}
As a first approximation, we could assume that all the $t(n,k')$ entries following $t(n,k-1)$ in the denominator sequence grow at a rate at most equal to that of $t(n,k-1)$, therefore, the upper sum can be formulated. However, this assumption reduces the range of $k$ within which we can ensure that the metric converges.

\begin{align}
    \lim_{n\to\infty} c(n,k)&=\lim_{n\to\infty} \frac{\displaystyle k\binom{n^2}{k}}{\displaystyle \sum _{i=1}^{k} i\cdot \binom{n^2}{k}}=\\\notag
    &=\lim_{n\to\infty} \frac{\displaystyle k\binom{n^2}{k}}{\displaystyle \binom{n^2}{k}\sum _{i=1}^{k} i}=\lim_{n\to\infty} \frac{\displaystyle k}{\displaystyle \frac{1}{2} k (k+1)}=\frac{2}{k+1}
\end{align}
Ultimately, each term in the sequence located in the denominator can be regarded as asymptotically equivalent to the maximum conceivable value $t(n,k)$, which in turn coincides with the numerator. In this case, the limit converges depending on the value of $k$, which, as can be observed, tends to 0 when the system is sufficiently large and inserts a correspondingly large number of elements. Accordingly, in order to study the convergence of the metric $c(n,k)$ as accurately as possible, it is crucial to find sufficiently tight asymptotic bounds for the terms $t(n,k')$ of $N(n,k)$, or at least for the first one that has a growth lower than the maximum possible, whose exact growth we already know.

\begin{align}    
    t(n,k-1)\sim\binom{n^2}{k-1}\enspace[n\to\infty]
\end{align}

Consequently, if we consider that the combinations of $k$ clusters of size 1 have an asymptotic growth equivalent to $\binom{n^2}{k}$, it is possible to suggest that the number of combinations for configurations with one fewer cluster grows in the form $\binom{n^2}{k-1}$. That is, if the cluster of 2 elements in the configuration is considered as a cluster of size 1 with an extended neighborhood according to the dimensions of the cluster of size 2, the possible combinations with that configuration will adhere to the upper bound, given that asymptotically it is equivalent to counting the combinations of $k-1$ elements over the system.

\begin{align}
    \lim_{n\to\infty} \frac{\displaystyle\binom{n^2}{k-1}}{\displaystyle\binom{n^2}{k}\binom{k}{2}\frac{8}{n^2-1}}&=\lim_{n\to\infty} \frac{\displaystyle(n^2-1)\binom{n^2}{k-1}}{\displaystyle8\binom{n^2}{k}\binom{k}{2}}=\\\notag
    &=\lim_{n\to\infty} \frac{\displaystyle(n^2-1)\binom{n^2}{k-1}}{\displaystyle8\binom{n^2}{k}\binom{k}{2}}=\\\notag
    &=\lim_{n\to\infty} \frac{n^2-1}{4 (k-1) \left(-k+n^2+1\right)}=\\\notag
    &=\lim_{n\to\infty} \frac{n^2-1}{n^2-k+1}\frac{1}{4 (k-1)}=\\\notag
    &=\frac{1}{4 (k-1)} \implies \boxed{t(n,k-1)\sim4 (k-1)\binom{n^2}{k-1}}
\end{align}

Upon identifying a potential asymptotic bound that is tighter than the original for $t(n,k-1)$, we proceed to calculate its ratio at the accumulation point where the convergence of $c(n,k)$ will be studied. From this, it is deduced that there is a $k$-dependent difference in the growth rates of both bounds, leading to the bound that will finally be used in the substitution. In summary, the bound consists of the previously proposed term $\binom{n^2}{k-1}$ multiplied by its difference with respect to the one originally substituted for $t(n,k-1)$, which, depending solely on $k$, simplifies the evaluation at the accumulation point and ensures that the infinitesimal provides sufficient information to guarantee the convergence across the entire applicable range of $k$.

\begin{align}
    \lim_{n\to\infty} c(n,k)&=\lim_{n\to\infty} \frac{\displaystyle k\binom{n^2}{k}}{\displaystyle k\binom{n^2}{k}+(k-1)4 (k-1)\binom{n^2}{k-1}}=\\\notag
    &=\lim_{n\to\infty} \frac{\displaystyle k\binom{n^2}{k}}{\displaystyle k\binom{n^2}{k}+k^2 4\binom{n^2}{k-1}}=\\\notag
    &=\lim_{n\to\infty} \frac{\displaystyle k\binom{n^2}{k}}{\displaystyle k\left(\binom{n^2}{k}+4k\binom{n^2}{k-1}\right)}=\\\notag
    &=\lim_{n\to\infty} \frac{\displaystyle k\binom{n^2}{k}}{\displaystyle k\left(\frac{n^{2k}}{k!}+4k\frac{n^{2(k-1)}}{(k-1)!}\right)}=\\\notag
    &=\lim_{n\to\infty} \frac{\displaystyle k\binom{n^2}{k}}{\displaystyle k\left(\frac{n^{2k}}{k!}\right)}=1\quad\left[N(n,k)\sim k\binom{n^2}{k}+O\left(k^2\binom{n^2}{k-1}\right)\right]
\end{align}

By substituting the initial terms of $N(n,k)$ with their corresponding bound, this time a convergence to 1 is achieved for all values of $k$ the algorithm can reach. Before concluding, it should be noted that the remaining terms in the denominator can be neglected for the same reasoning with which the final bound of $t(n,k-1)$ was computed, which demonstrates an asymptotic growth as indicated above. Therefore, with this procedure, the tendency to 1 of the average cluster size is confirmed as $n\to\infty$, at least in $o(n^2)$, where the majority of algorithm iterations occur. That is, the asymptotic growth shown in $k=n^2$ is negligible, given that this situation occurs a constant and equal to 1 number of times in the algorithm. Finally, the intuition of this result is based on the amount of size 1 clusters that exist for each configuration of $k'$ clusters. In summary, since the combinations of size 1 cluster configurations asymptotically surpass the others, as well as the number of combinations of different $k'$ whose largest part of clusters have that size, they contribute more significantly on average compared to other sizes. So, in the limit situation where the system is an 'infinite' matrix, their contribution causes the metric to converge to 1, except when $k=n^2$, since there is only one cluster, the metric increases, by definition through an order of $\Theta(n^2)$, as shown below:

\begin{align}
    \lim_{n\to\infty} c(n,n^2)&=\lim_{n\to\infty} \frac{\displaystyle n^2\binom{n^2}{n^2}}{\displaystyle 1}=\lim_{n\to\infty} n^2\implies \boxed{c(n,n^2)=\Theta(n^2)}
\end{align}

After studying the trend of the average cluster size, we now proceed to repeat the same procedure for the insertion probability, with the difference that this expression is closed, providing results that do not require additional verifications or modifications in the limit approach.
\begin{align}
    \lim_{n\to\infty} \left(1-\frac{k}{n^2}\right)=\begin{cases}
    1 & \text{if } k=o(n^2)\\
    0 & \text{if } k= \Theta(n^2)
    \end{cases}
\end{align}
Initially, if the probability of an insertion occurring in a system with exactly $k$ elements is considered, it is verifiable that for all valid $k$ the probability converges to 1, except in the edge case $k=\Theta(n^2)$ where it becomes impossible to perform an insertion because the system is full.
\begin{align}
    \lim_{n\to\infty} \left(1-\frac{E(i)}{n^2}\right)&=\lim_{n\to\infty} \left(1-\frac{n^2\left(1-\left(1-\frac{1}{n^2}\right)^i\right)}{n^2}\right)=\lim_{n\to\infty} \left(1-\frac{1}{n^2}\right)^i
\end{align}
Likewise, if we use the metric $E(i)$ as an estimate of the number of elements in a certain iteration $i$, we arrive at a form in which a function will later be included to traverse all values of $i$ corresponding to states where the system has made $k$ successful insertions, or in other words, the index of each iteration that a process lasts.
\begin{figure}[H]
    \centering
    \includegraphics[width=10cm,clip]{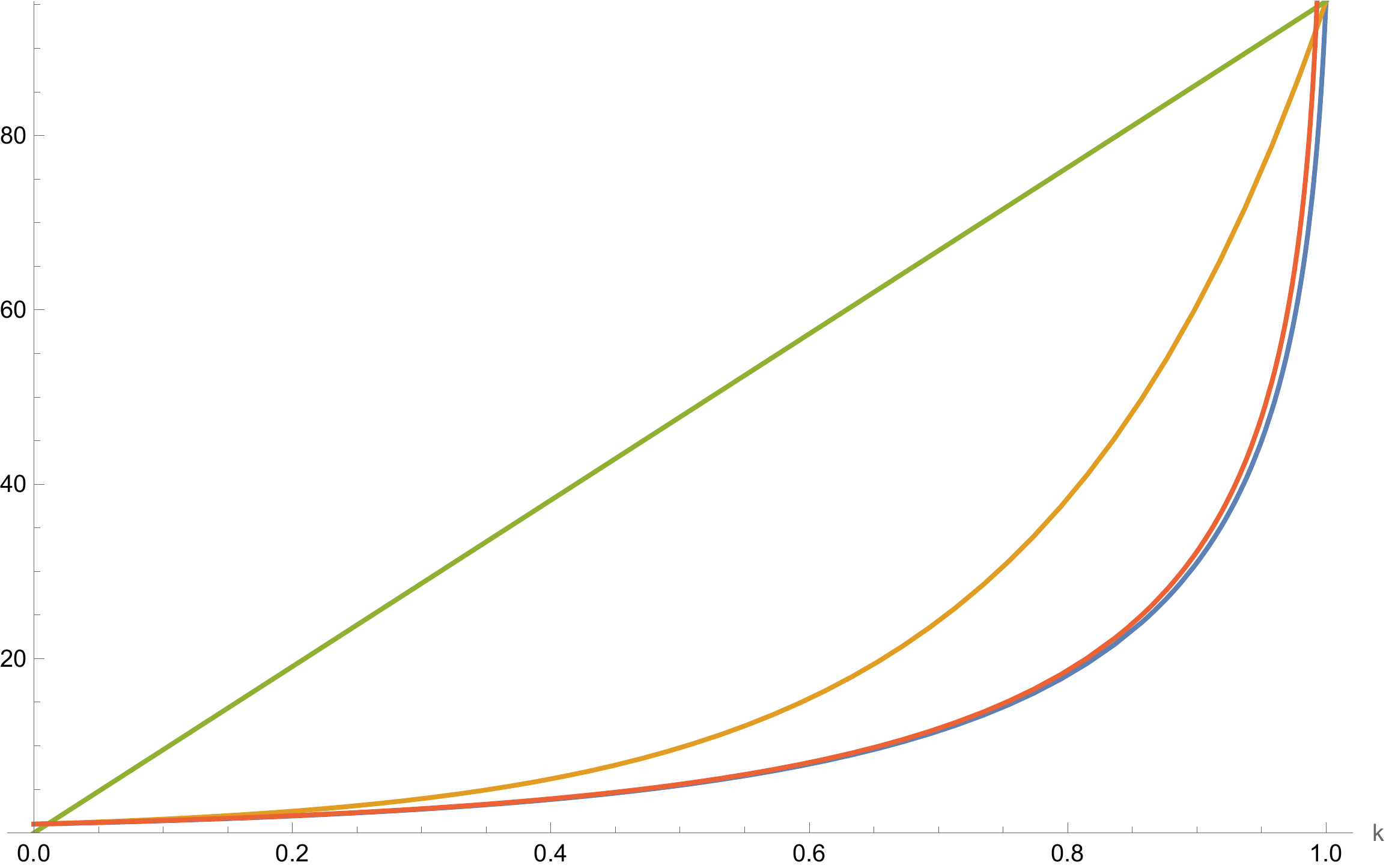}
    \caption{$(n^2H_{n^2})k$ function plotted in green, $(n^2H_{n^2})^k$ in orange, $n^2(\ln(n^2)-\ln(n^2-(n^2)^k))$ in red, and $n^2(H_{n^2}-H_{n^2-(n^2)^k})$ in blue with $n=5$ and $0\leq k\leq 1$}
    \label{fig:2DAverageCaseInterpolators}
\end{figure}
Here are shown different parameterizations with which the iteration steps can be modeled throughout the execution of a process. Specifically, to fill the system, $I(n,n^2)=n^2H_{n^2}$ iterations are necessary, so by multiplying its value by a constant between 0 and 1, or by using $I(n,k)$, it is possible to characterize the behavior of the insertion probability in a specific iteration. And, since the parameter that controls the iteration step is variable, we can express the limit of the insertion probability as a function of it.
\begin{align}
    \lim_{n\to\infty} \left(1-\frac{E(n^2(H_{n^2}-H_{n^2-k}))}{n^2}\right)&=\lim_{n\to\infty} \left(1-\frac{1}{n^2}\right)^{n^2(H_{n^2}-H_{n^2-k})}=\\\notag
    &=\lim_{n\to\infty} e^{-(H_{n^2}-H_{n^2-k})}=\\\notag
    &=\lim_{n\to\infty} \frac{\displaystyle e^{H_{n^2-k}}}{\displaystyle e^{H_{n^2}}}=\\\notag
    &=\lim_{n\to\infty} \frac{\displaystyle e^{\ln(n^2-k)+\gamma}}{\displaystyle e^{\ln(n^2)+\gamma}}=\\\notag
    &=\lim_{n\to\infty} \frac{\displaystyle n^2-k}{\displaystyle n^2}=\lim_{n\to\infty} 1-\frac{k}{n^2}=\begin{cases}
    1 & \text{if } k=o(n^2)\\
    0 & \text{if } k= \Theta(n^2)
    \end{cases}
\end{align}

For instance, if we use $I(n,k)$ instead of $i$ when studying the probability convergence, we obtain the same result as before, which implies that both the probability expression and the metric $I(n,k)$ are consistent with what should happen in the execution of the algorithm, especially in the edge case when the system is full of elements.

\begin{align}
    \lim_{n\to\infty} \left(1-\frac{E((n^2H_{n^2})^\xi)}{n^2}\right)&=\lim_{n\to\infty} \left(1-\frac{1}{n^2}\right)^{(n^2H_{n^2})^\xi}=\\\notag
    &=\lim_{n\to\infty} e^{-n^{2\xi-2}(H_{n^2})^\xi}=\\\notag
    &=\lim_{n\to\infty} \frac{1}{e^{n^{2\xi-2}(H_{n^2})^\xi}}=\\\notag
    &=\lim_{n\to\infty} \frac{1}{e^{n^{2\xi-2}(\ln(n^2)+\gamma)^\xi}}=\\\notag
    &=\lim_{n\to\infty} \frac{1}{e^{n^{2\xi-2}(2\ln(n)+\gamma)^\xi}}=\begin{cases}
    \lim_{n\to\infty} \frac{1}{e^0}=1 & \text{if } 0\leq\xi<1\\
    \lim_{n\to\infty} \frac{1}{e^\infty}=0 & \text{if } \xi=1
    \end{cases}
\end{align}

On the other hand, if we use a different parameterization than $I(n,k)$ as shown above, it is verifiable that convergence is maintained, so it can be concluded that the insertion probability tends to 1 as $n\to\infty$. In this case, the interpretation of this result is based on the available space that a system has at infinity compared to the number of elements it contains. Thus, in an infinite matrix, any number of elements will be negligible compared to the total size of the system, except when it is an amount equivalent to the total system, in which case it will be impossible to insert new elements. But, for this to happen it is necessary to 'reach' the infinity at which the size of the system resides, so the previous convergence is only taken into account to asymptotically analyze the cost of the algorithm.

\begin{align}
    \lim_{n\to\infty} T_{avg}(n) &= \lim_{n\to\infty} \sum _{i=0}^{I(n)} \left(1-\frac{E(i)}{n^2}\right)\cdot c(n,E(i))\\\notag&=\lim_{n\to\infty} \sum _{i=0}^{I(n)} 1=\lim_{n\to\infty} I(n) \implies \boxed{\boxed{T_{avg}(n)=\Theta(I(n))}}
\end{align}

Thus, with the results of the study in $n\to\infty$ of the metric $c(n,k)$ and the insertion probability, which is also interpretable as the probability that in an iteration $i$ the algorithm performs a work proportional to the average cluster size in it, its expressions can be replaced by the value to which they converge at the studied accumulation point. Therefore, as shown above, the sum of the work of all insertions grows asymptotically with the upper limit of the sum $I(n)$, since all the summand terms have the same constant value when the system size is large enough. Therefore, before concluding that the work of the algorithm is determined by the duration of a process, at least in the average case, it is advisable to visualize why it occurs. Thus, it proceeds to study the form of both functions that compose the work of each insertion throughout the execution of a process.
\begin{figure}[H]
    \centering
    \includegraphics[width=10cm,clip]{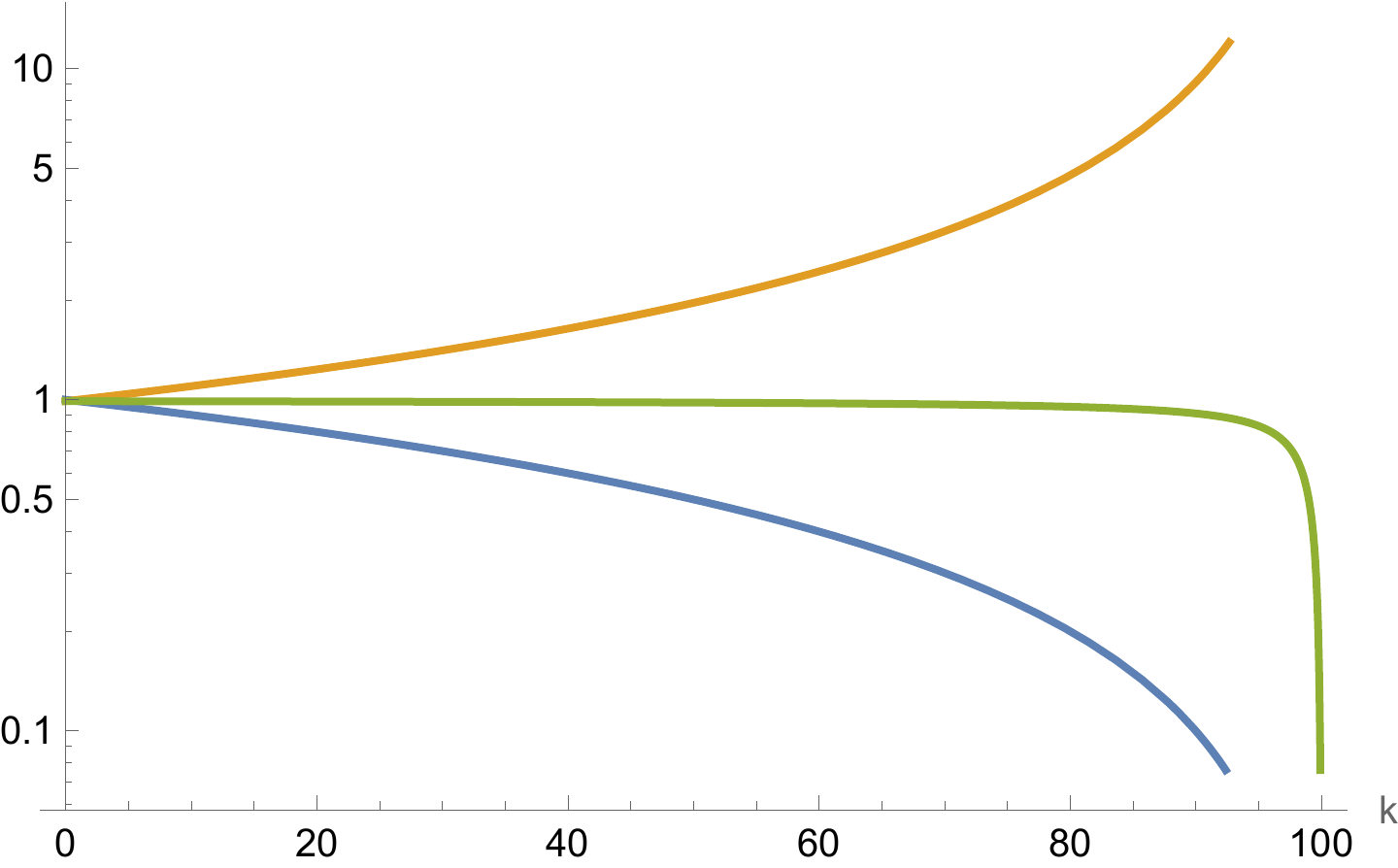}
    \caption{Average cluster size metric $\frac{n^2}{-k+n^2+1}$ plotted in orange, insertion probability $1-\frac{k}{n^2}$ in blue, and its product in green for a system with size $n=10$, in logarithmic scale.}
    \label{fig:2DAverageInsertionComplexity}
\end{figure}
On the one hand, the behavior of $c(n,k)$ was previously analyzed in section \textcolor{blue}{\ref{subsubsec:AverageInsertionRuntime}}, so assuming that the above graph faithfully represents the evolution of the metric in the range of $k$ during which the algorithm is executed, and based on the checks performed in this section regarding its asymptotic behavior, it is observed that it maintains the same properties and meets the same fundamental constraints. On the other hand, the probability of an insertion occurring in a specific iteration where the system already has $k$ elements starts at 1 and gradually tends to 0 as the process advances. That is, the insertion of new elements causes a change in the probability of an insertion occurring in the next iteration, specifically a decrease. Therefore, the only magnitude that determines the value of $c(n,k)$ and the insertion probability is the number of elements, which we estimate through $E(i)$. Thus, as metrics tending to complementary values, it can be inferred that their product tends to a constant value in $0 \leq k \leq n^2$, which in this case is 1 since the average cluster size also starts at 1 when the process begins.
\begin{align}
    \lim_{n\to\infty} \left(1-\frac{E(n^2(H_{n^2}-H_{n^2-\xi n^2}))}{n^2}\right)&=\lim_{n\to\infty} \left(1-\frac{1}{n^2}\right)^{n^2(H_{n^2}-H_{n^2-\xi n^2})}=\\\notag
    &=\lim_{n\to\infty} e^{-(H_{n^2}-H_{n^2-\xi n^2}))}=\\\notag
    &=\lim_{n\to\infty} \frac{\displaystyle e^{H_{n^2-\xi n^2}}}{\displaystyle e^{H_{n^2}}}=\\\notag
    &=\lim_{n\to\infty} \frac{\displaystyle e^{\ln(n^2-\xi n^2)+\gamma}}{\displaystyle e^{\ln(n^2)+\gamma}}=\\\notag
    &=\lim_{n\to\infty} \frac{\displaystyle e^{\ln(n^2(1-\xi))}}{\displaystyle e^{\ln(n^2)}}=\lim_{n\to\infty} e^{\ln(n^2(1-\xi))-\ln(n^2)}=1-\xi
\end{align}
Formally, this is evidenced through a specific interpolation of $i$ as shown above. In summary, the number of elements of $I(n,k)$ is parameterized by the product of $n^2$ and a constant $0 \leq \xi \leq 1$. So, when solving the limit as $n \to \infty$, the value to which the insertion probability tends depends on the introduced parameter $\xi$, which is beneficial because the limit of the metric $c(n,k)$ characterized with the same parametrization also depends on the new parameter, subsequently allowing the product of both results to be found and the convergence of the product limit to be analyzed.
\begin{align}
    \lim_{n\to\infty} \frac{n^2}{n^2-E(n^2(H_{n^2}-H_{n^2-\xi n^2}))+1}&=\lim_{n\to\infty} \frac{n^2}{n^2-n^2\left(1-\left(1-\frac{1}{n^2}\right)^{n^2(H_{n^2}-H_{n^2-\xi n^2})}\right)+1}=\\\notag
    &=\lim_{n\to\infty} \frac{1}{1-\left(1-\left(1-\frac{1}{n^2}\right)^{n^2(H_{n^2}-H_{n^2-\xi n^2})}\right)}=\\\notag
    &=\lim_{n\to\infty} \frac{1}{\left(1-\frac{1}{n^2}\right)^{n^2(H_{n^2}-H_{n^2-\xi n^2})}}=\\\notag
    &=\lim_{n\to\infty} e^{\ln(n^2)-\ln(n^2(1-\xi))}=\frac{1}{1-\xi}
\end{align}
Next, after solving the limit of the metric $c(n,k)$, it is observed that its convergence is guaranteed for all values of $\xi$ except 1, which is easily interpretable according to the previous results of this average analysis. That is, as long as the system contains a number of elements fewer than the maximum, the average cluster size will converge to a constant, which in this instance varies with the parameter $\xi$. And, in the case that it fills up, the limit should diverge to infinity with a growth of $\Theta(n^2)$, or as determined by the overall size of the system.
\begin{align}
    &\lim_{n\to\infty} \left(1-\frac{E(n^2(H_{n^2}-H_{n^2-\xi n^2}))}{n^2}\right)\frac{n^2}{n^2-E(n^2(H_{n^2}-H_{n^2-\xi n^2}))+1}=\\\notag
    &=\lim_{n\to\infty} \left(1-\frac{E(n^2(H_{n^2}-H_{n^2-\xi n^2}))}{n^2}\right)\cdot\lim_{n\to\infty} \left(\frac{n^2}{n^2-E(n^2(H_{n^2}-H_{n^2-\xi n^2}))+1}\right)\\\notag
    &=\frac{1-\xi}{1-\xi}=1\quad[0\leq\xi<1]
\end{align}
With this, we can ensure that the limit of the product of both magnitudes tends to 1 in the corresponding range of $k$, which in this case is given by the value of the parameter, also interpreted as the proportion of occupied cells in the system. In conclusion, the work performed by the algorithm in each iteration tends to be constant throughout the process for large system sizes, implying that the time complexity of its execution depends directly on the number of iterations it lasts, determined by $I(n)$, which, even without a closed form for its formulation, can be asymptotically bounded. However, before proceeding with such bounding, it is convenient to visualize the similarity between the functions of Figure 64 and themselves when the number of elements is replaced by its estimate $E(i)$. With this, we not only confirm the shape they should follow, but also reiterate the correctness of the expression of $E(i)$, which is fundamental to validate the procedure up to this point. Then, as depicted in the lower graph, the average cluster size behaves in a very similar way, although with visually reduced growth since the inclusion of $E(i)$ increases the distance on the abscissa axis for the metric to reach the same values. Likewise, the insertion probability also maintains its shape, although undergoing the same transformation, it manifests a decrease that in logarithmic scale appears to be linear. And, finally, the product tends to 1, which is coherent since the transformation performed on the metric $c(n,k)$ and the respective probability affects the number of elements on the system, determining the growth of both functions and their values in a bijective manner over all $k\in[0,n^2)$.
\begin{figure}[H]
    \centering
    \includegraphics[width=10cm,clip]{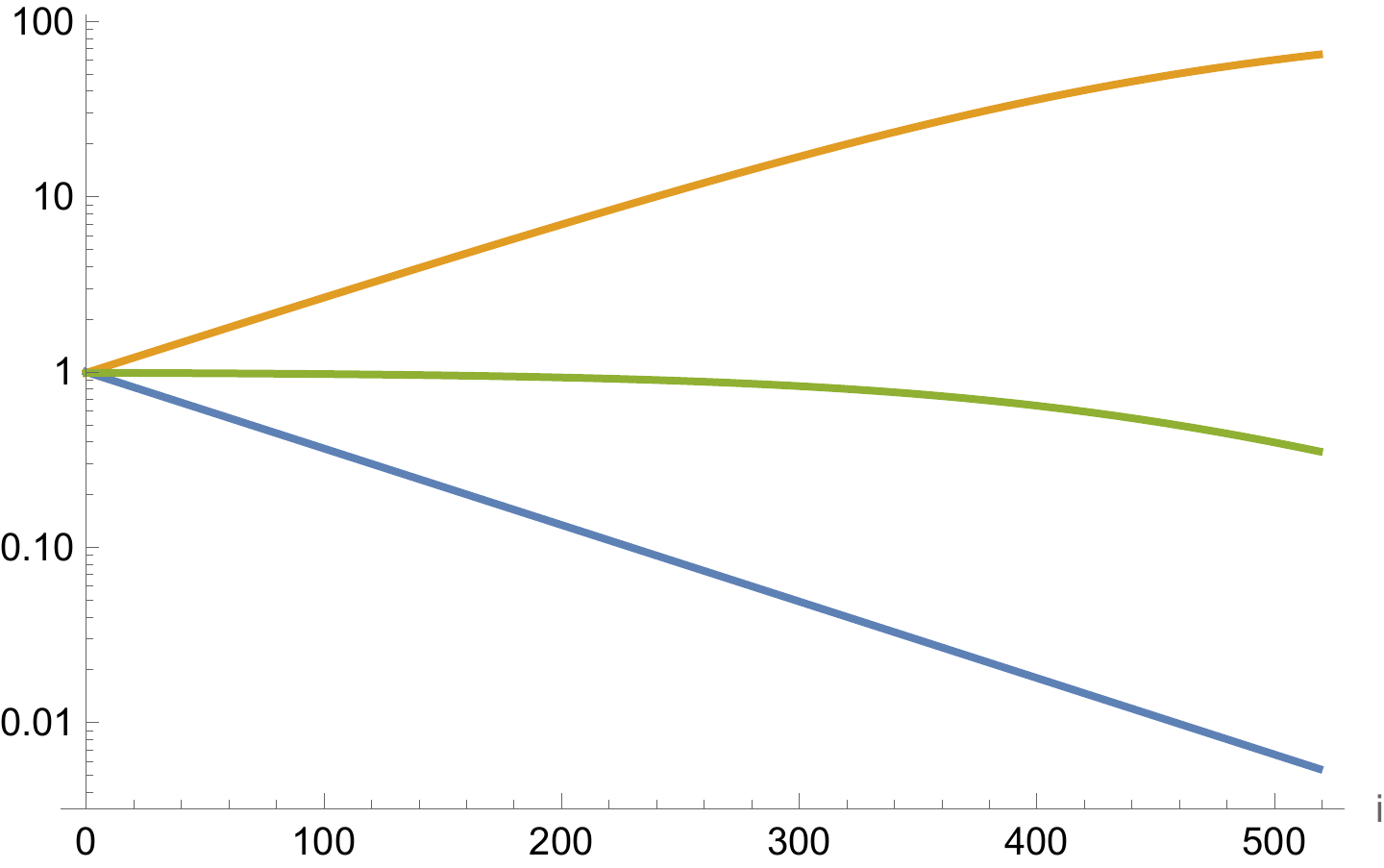}
    \caption{Average cluster size metric $\frac{n^2}{n^2-\left(n^2 \left(1-\left(1-\frac{1}{n^2}\right)^i\right)\right)+1}$ plotted in orange, insertion probability $1-\frac{n^2 \left(1-\left(1-\frac{1}{n^2}\right)^i\right)}{n^2}$ in blue, and its product in green for a system with size $n=10$, in logarithmic scale until $i$ reaches the upper bound $n^2H_{n^2}$.}
    \label{fig:2DAverageInsertionComplexity2}
\end{figure}
Now, knowing that the time complexity of the average case is given by $I(n)$, for which there is no exact expression that models its magnitude for all $n \in \mathbb{N}$, it is necessary to bound its asymptotic growth, which will provide a reliable measure for $T_{avg}(n)$.
\begin{align}
    \lim_{n\to\infty} \frac{E(i)}{E(n^2H_{n^2})}=L\quad\colon L\in[0,1]
\end{align}
To this end, it is assumed that the ratio between the number of elements existing at the end of the process and the size of the system must be a real number between 0 and 1. This is equivalent to assuming the existence of the percolation threshold \cite{Wierman2009, Uzzell4336421, Bauer2001, Milovanov2012, Easo2023, Galam1996}, which denotes the occupation ratio $p_c$ at which the critical point is reached that causes a change of state in the system. Hence, the superior equation denotes the limit of the ratio between $E(i)$, where $i$ is the index of the iteration at which the process ends, on average, and $E(n^2H_{n^2})$, which represents the total number of cells in the system, equivalent to the maximum number of insertable elements.

\begin{align}
    \lim_{n\to\infty} \frac{E(i)}{E(n^2H_{n^2})}&=\lim_{n\to\infty} \frac{n^2 \left(1-\left(1-\frac{1}{n^2}\right)^i\right)}{n^2 \left(1-\left(1-\frac{1}{n^2}\right)^{n^2H_{n^2}}\right)}=\\\notag
    &=\lim_{n\to\infty} \frac{\displaystyle1-\left(1-\frac{1}{n^2}\right)^i}{\displaystyle1-\left(1-\frac{1}{n^2}\right)^{n^2H_{n^2}}}=\\\notag
    &=\lim_{n\to\infty} \frac{\displaystyle1-\left(1-\frac{1}{n^2}\right)^i}{\displaystyle1-e^{-H_{n^2}}}=\\\notag
    &=\lim_{n\to\infty} 1-\left(1-\frac{1}{n^2}\right)^i=\\\notag
    &=\lim_{n\to\infty} 1-\frac{1}{e^{i/n^2}}=\begin{cases}
    \lim_{n\to\infty} 1-\frac{1}{e^0}=0 & \text{if } i=o(n^2)\\
    \lim_{n\to\infty} 1-\frac{1}{e}\approx 0.6321 & \text{if } i=\Theta(n^2)
    \end{cases}
\end{align}
After evaluating the limit, it is concluded that the only asymptotic growth of $i$, and therefore of the metric $I(n)$, for which the ratio converges to a value that meets the same conditions as $p_c$ is $\Theta(n^2)$.
\begin{align}
    \lim_{n\to\infty} \frac{E(n^2(H_{n^2})^\xi)}{E(n^2H_{n^2})}&=\lim_{n\to\infty} 1-\left(1-\frac{1}{n^2}\right)^{n^2(H_{n^2})^\xi}=\\\notag
    &=\lim_{n\to\infty} 1-\frac{1}{e^{(H_{n^2})^\xi}}=\\\notag
    &=\lim_{n\to\infty} 1-\frac{1}{e^{(\ln(n^2)+\gamma)^\xi}}=\\\notag
    &=\lim_{n\to\infty} 1-\frac{1}{e^{2^{\xi}\ln^\xi(n)}}=\begin{cases}
    1-\frac{1}{e} & \text{if } \xi=0\\
    \lim_{n\to\infty} 1-\frac{1}{e^\infty}=1 & \text{if } \xi>0
    \end{cases}
\end{align}
Correspondingly, when studying convergence in functions whose growth exceeds the bound $n^2$ but does not reach the maximum number of iterations $n^2H_{n^2}$, it is inferred that the ratio $p_c$ converges to 1 for all values of $\xi$ that place $i$ in the set $\omega(n^2)$. In summary, if the percolation threshold exists for growths of $I(n)$ of order $n^2$, then for any greater growth, the system will be filled with elements after the corresponding iterations. Another way to view the result of the upper limit is to write $I(n)$ as the product of $n^2$ and a function $h(n)$ that must tend to infinity as $n\to\infty$. In this way, for any function that meets this condition, regardless of its growth, the limit will converge to 1 in the same manner. Therefore, it can be concluded that, asymptotically, the average number of iterations a process lasts is of order $n^2$, which implies that the complexity $T_{avg}(n)$ will also conform to that order.
\begin{align}
    \boxed{\boxed{T_{avg}(n)=O(n^2)}}
\end{align}

And, prior to concluding this analysis, it is worthwhile to highlight a possible way to approximate $p_c$ in two-dimensional systems with aspect ratio 1, or in other words, the site percolation threshold for systems with Moore neighborhood.
\begin{align}
    \lim_{n\to\infty} \frac{E(\xi n^2)}{E(n^2H_{n^2})}&=\lim_{n\to\infty} 1-\left(1-\frac{1}{n^2}\right)^{\xi n^2}=\lim_{n\to\infty} 1-\frac{1}{e^{\xi n^2/n^2}}=1-\frac{1}{e^{\xi}}
\end{align}
First, knowing that the threshold is determined by the ratio of elements in the final state of the system to its size, and given that $E(I(n))$ is equivalent to this amount of elements, we can propose the superior parametrization in which $\xi$ is any real value. Hence, its resolution indicates a dependency between $\xi$ and the threshold $p_c$, fixed for each system of size $n$.

\begin{align}
    \lim_{n\to\infty} \frac{E(n^2(H_{n^2}-H_{n^2-\xi n^2}))}{E(n^2H_{n^2})}&=\lim_{n\to\infty} 1-\left(1-\frac{1}{n^2}\right)^{n^2(H_{n^2}-H_{n^2-\xi n^2})}=\\\notag
    &=\lim_{n\to\infty} 1-\frac{1}{e^{(H_{n^2}-H_{n^2-\xi n^2})}}=\\\notag
    &=\lim_{n\to\infty} 1-\frac{e^{H_{n^2-\xi n^2}}}{e^{H_{n^2}}}=\\\notag
    &=\lim_{n\to\infty} 1-\frac{e^{\ln(n^2-\xi n^2)+\gamma}}{e^{\ln(n^2)+\gamma}}=\\\notag
    &=\lim_{n\to\infty} 1-\frac{n^2(1-\xi)}{n^2}=1-(1-\xi)=\xi\quad\colon\xi\in[0,1]
\end{align}
On the other hand, we have the function $I(n,k)$ which denotes the iterations required to insert $k$ elements, which in turn can be substituted by the previous parametrization $\xi n^2$. Though, this is possible because asymptotically $I(n,k) \sim k$ holds as long as the system is not intended to be entirely filled.

\begin{figure}[H]
    \centering
    \includegraphics[width=10cm,clip]{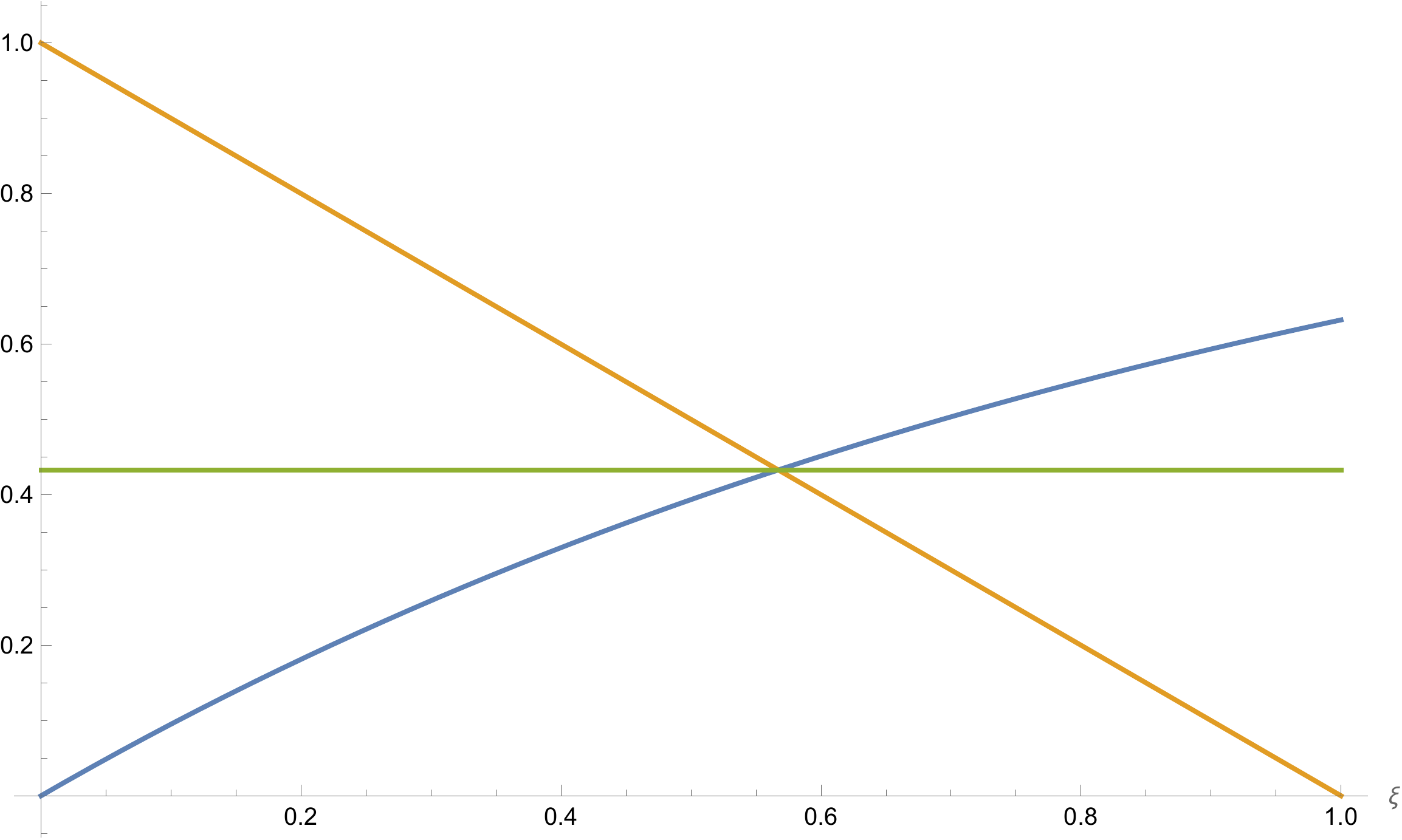}
    \caption{$1-\frac{\displaystyle1}{\displaystyle e^{\xi}}$ function plotted in blue, $1-\xi$ in orange, and its intersection ordinate $1-W(1)$ in green.}
    \label{fig:2Dthreshold1}
\end{figure}
As illustrated in Figure 66, when plotting the limit of the first parametrization and the complementary of the second, both intersect at a very specific point derived in the following way:

\begin{align}
    1-\frac{1}{e^{\xi}}=1-\xi\implies \boxed{\xi=W(1)\approx0.567143}
\end{align}
The range of interest for the parameter, especially in the second parameterization wherein it is included within $I(n,k)$, are the real values in the interval $[0,1]$. Therefore, since it coincides with the interval in which $p_c$ resides, it is pertinent to consider whether the point at which both dependencies are equal is the system's threshold.

\begin{align}
    p_c\approx1-\frac{1}{e^{W(1)}}=1-W(1)\approx\boxed{0.432856}
\end{align}
At present, the only approximately known value for the actual threshold is $p_c \approx 0.407$ \cite{Malarz2005, Majewski2007, Feng2008, Ouyang2018, Xu2021, Akhunzhanov2022}, derivable through the Monte Carlo method  \cite{Pawlowska2013, Deng2005, Zukowski2023} with the algorithm under analysis. And, if we calculate the height at which the previous intersection occurs, we obtain a value very close to $p_c$; although given the difference, it can be assured that it is not the desired value. So, to reduce this difference and approach the threshold, it would be convenient to acquire further information regarding the parameterizations and their impact on the studied intersection, since starting from the ratio that characterizes $p_c$, it is possible that through a specific approach to $I(n)$, in combination with another function of the style of $I(n, k)$, a closed form for the threshold can be found, at least in 2-dimensional systems with Moore neighborhood and all defined properties.

\begin{align}
    \lim_{n\to\infty} \frac{E(n^2(H_{n^2}-\ln(n^2-\xi n^2)))}{E(n^2H_{n^2})}&=\lim_{n\to\infty} 1-\left(1-\frac{1}{n^2}\right)^{n^2(H_{n^2}-\ln(n^2-\xi n^2))}=\\\notag
    &=\lim_{n\to\infty} 1-\frac{1}{e^{(H_{n^2}-\ln(n^2-\xi n^2))}}=\\\notag
    &=\lim_{n\to\infty} 1-\frac{e^{\ln(n^2-\xi n^2)}}{e^{H_{n^2}}}=\\\notag
    &=\lim_{n\to\infty} 1-\frac{n^2-\xi n^2}{e^{\ln(n^2)+\gamma}}=\\\notag
    &=\lim_{n\to\infty} 1-\frac{n^2(1-\xi)}{n^2 e^{\gamma}}=1-\frac{1-\xi}{e^{\gamma}}\quad\colon\xi\in[0,1]
\end{align}
To ascertain the proximity with which the threshold can be approximated, a variation of the last parameterization is chosen above, in which $I(n,k)$ is introduced into the ratio expression. Specifically, one of the terms involving the harmonic number is replaced by its asymptotic equivalent, the natural logarithm, resulting in the limit depending on $\xi$ to vary subtly as a result of the Euler-Mascheroni constant $\gamma$ \cite{Produit629630, Feldman323094}, introduced through the infinitesimal substitution.
\begin{figure}[H]
    \centering
    \includegraphics[width=10cm,clip]{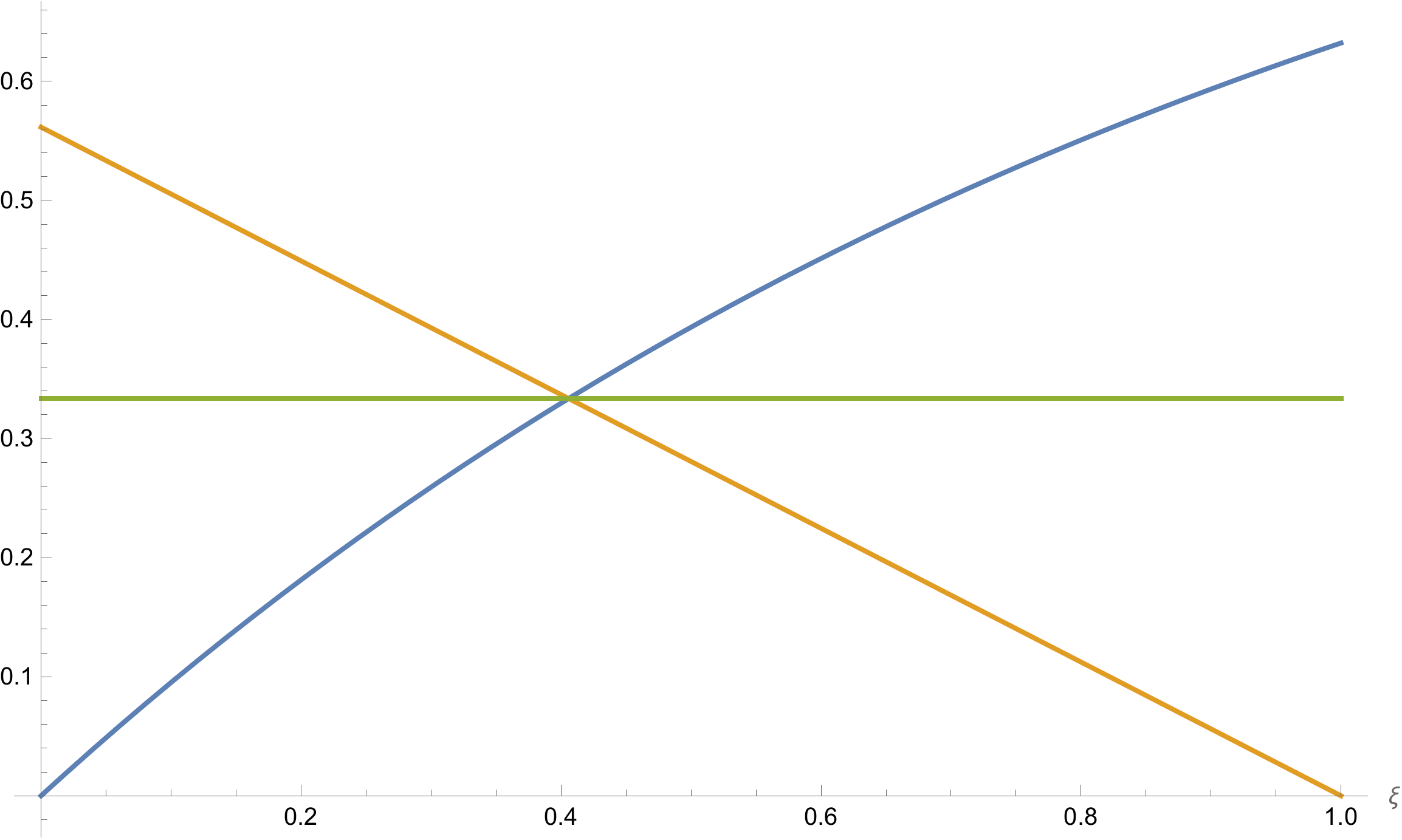}
    \caption{$1-e^{-\xi}$ function plotted in blue, $-e^{-\gamma } (\xi-1)$ in orange, and the intersection ordinate $1-e^{-\gamma } W\left(e^{-1+e^{\gamma }+\gamma }\right)$ in green.}
    \label{fig:2Dthreshold2}
\end{figure}
Graphically, this scenario closely resembles the original, with the primary difference being the closeness of the approximation to the actual threshold value.

\begin{align}
    1-\frac{1}{e^{\xi}}=1-\left(1-\frac{1-\xi}{e^{\gamma}}\right)
\end{align}

\begin{align}
    1-\frac{1}{e^{\xi}}=\frac{1-\xi}{e^{\gamma}}\implies \boxed{p_c\approx\xi=W\left(e^{-1+e^{\gamma }+\gamma }\right)-e^{\gamma }+1\approx0.405851}
\end{align}
Consequently, by imposing the condition that the result of both limits is equivalent, a value for $\xi$ much closer to $p_c\approx0.407$ is achieved. Nevertheless, this time the threshold value has been considered as the $\xi$ of the intersection, not its height, since it deviates from the expected value, as demonstrated below:

\begin{align}
    1-\frac{1}{e^{W\left(e^{-1+e^{\gamma }+\gamma }\right)-e^{\gamma }+1}}\approx0.333591
\end{align}

\subsection{Empirical time measurements}
Analogous to the analysis of one-dimensional systems, it is advisable to contrast the conclusions of this section with the runtime measurements of the algorithm. For this case, measurements were originally taken in section \textcolor{blue}{\ref{subsec:Average-time-complexity-estimation}} on which several adjustments of different models that seemed to fit their growth adequately were made. However, now these measurements will be repeated with a specific implementation of the algorithm that mitigates the impact of resetting the visited register on the resulting complexity. That is, an implementation will be used in which the cells visited in each execution of $helper()$ are stored in a specific data structure, to later return only those cells from the register to their original state.

\begin{figure}[H]
    \centering
    \includegraphics[width=10cm,clip]{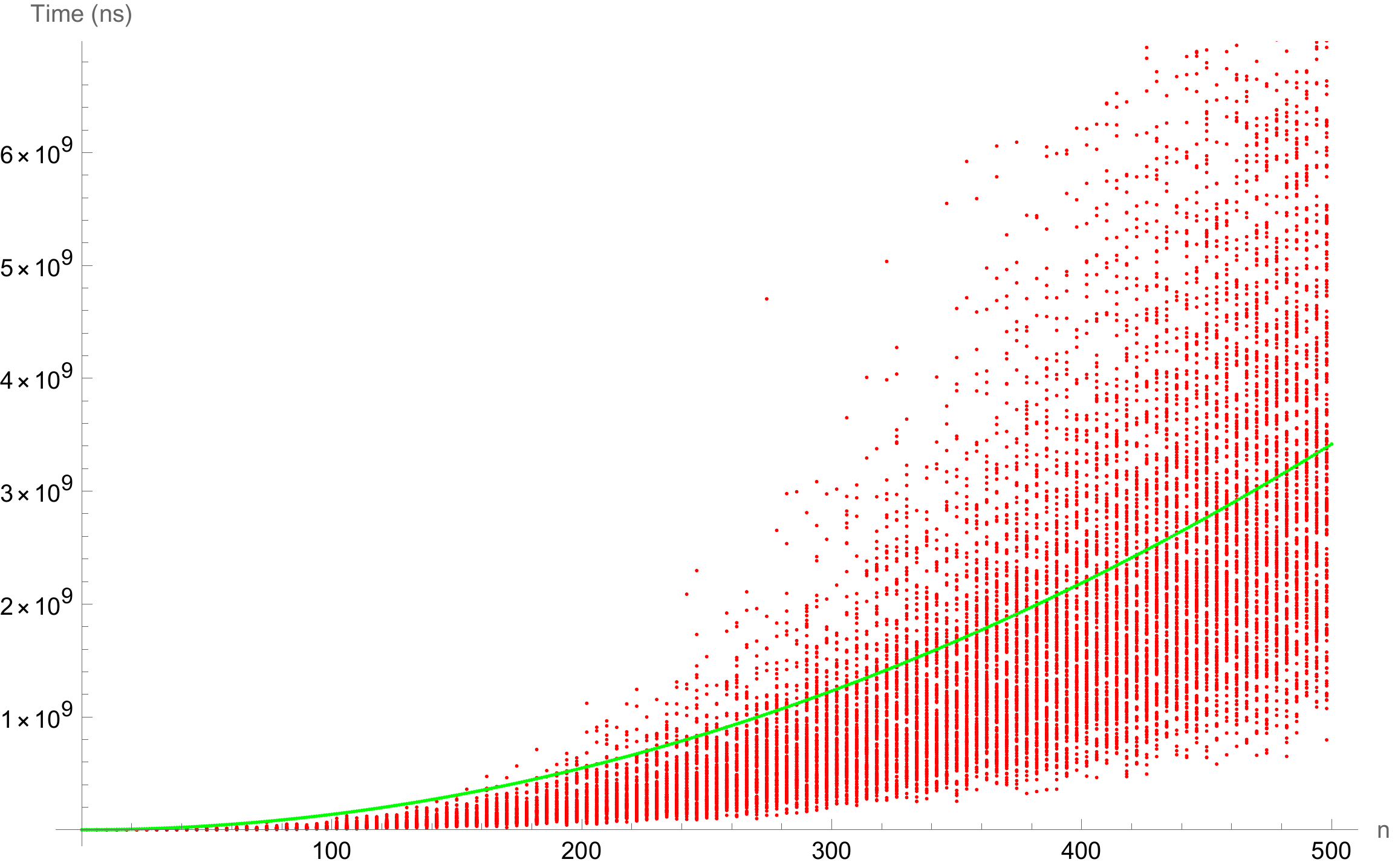}
    \caption{Runtime measurements for a 2D system with aspect ratio 1 plotted in red and the fitted model $T_{avg}(n)\approx1.37\cdot 10^4 n^2$ in green.}
    \label{fig:2DtimesFitAverage}
\end{figure}
The runtimes dataset is shown above \cite{time_measurements_2D2}, which, as noted, is very similar to that of section \textcolor{blue}{\ref{subsec:Average-time-complexity-estimation}}. Thus, knowing now that the average time complexity is of the order $n^2$, the model used to fit the growth will have a fixed exponent and a factor $x$ that will vary during the fitting process, in order to capture the appropriate height of the data points.
\[
\begin{array}{l|lll}
 \text{} & \text{DF} & \text{SS} & \text{MS} \\
\hline
 \text{Model} & 1 & 6.06135\times 10^{22} & 6.06135\times 10^{22} \\
 \text{Error} & 15608 & 2.34629\times 10^{22} & 1.50326\times 10^{18} \\
 \text{Uncorrected Total} & 15609 & 8.40763\times 10^{22} & \text{} \\
 \text{Corrected Total} & 15608 & 4.49827\times 10^{22} & \text{} \\
\end{array}
\]

\[
\begin{array}{l|lllll}
 \text{Parameter} & \text{Estimate} & \text{Standard Error} & \text{t-Statistic} & \text{P-Value} & \text{Confidence Interval} \\
\hline
 x & 13666.9 & 68.0617 & 200.802 & 0. & \{13533.5,\ 13800.3\} \\
\end{array}
\]
Therefore, after fitting the model $T_{avg}(n)=x\cdot n^2$ to the dataset, the above results are obtained, along with an adjusted $R^2$ of 0.720916. Overall, the model appears to fit the line where most measurements lie, which is a good indication that the proposed bound is consistent with the measurements. However, with a range of $n$ limited to $\approx500$ and varying only the constant, the model is close to the upper bound of the worst case for small values of $n$, although it descends to the region where the average case lies in the rest of the range.

\begin{figure}[H]
    \centering
    \includegraphics[width=10cm,clip]{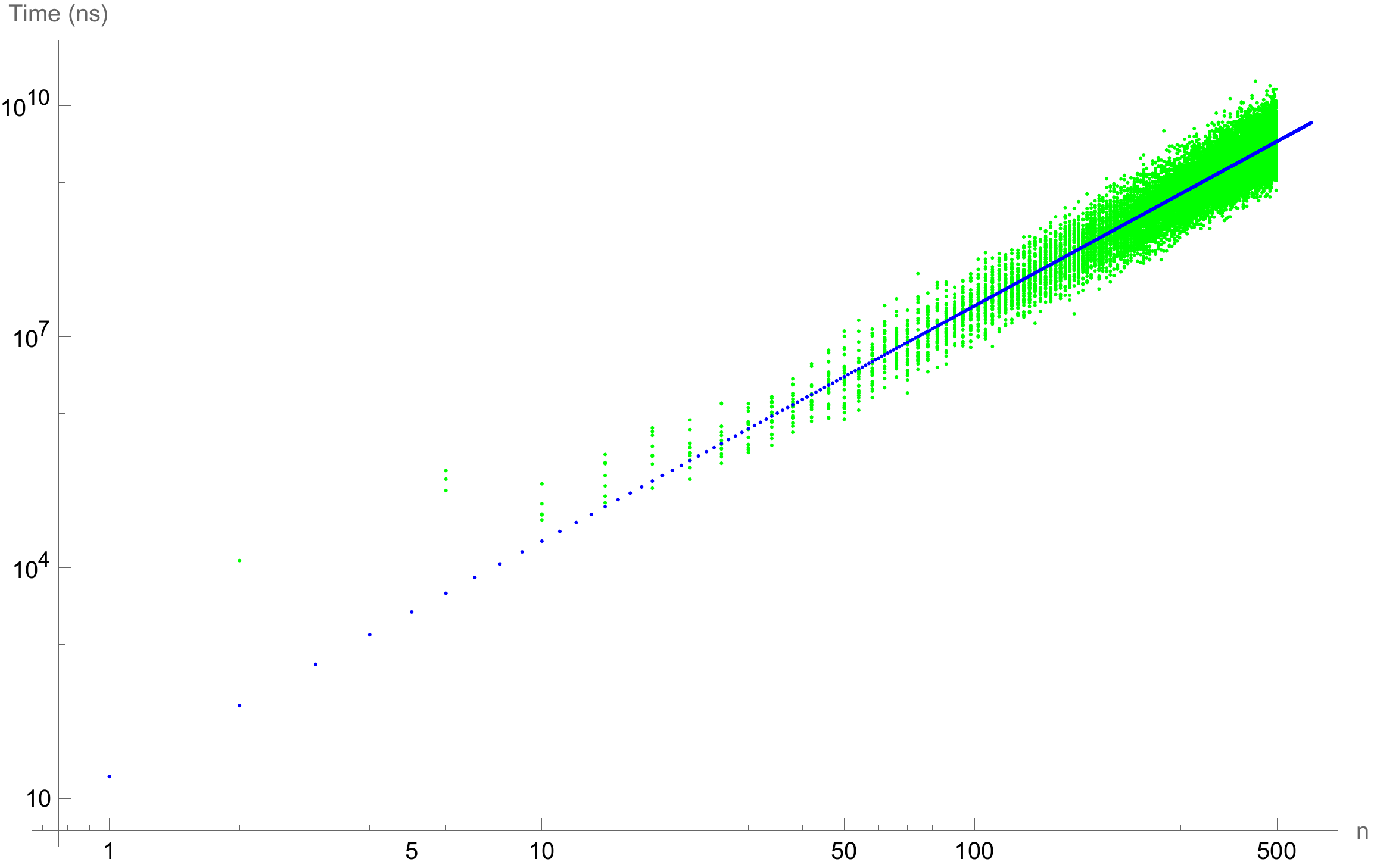}
    \caption{Runtime measurements from Figure 68 plotted in green in logarithmic scale, and the fitted model $T_{avg}(n)\approx e^{2.96 + 3.06 \ln(n)}$ in blue.}
    \label{fig:2DtimesFitLog}
\end{figure}
On the other hand, we could provide enough freedom to the model so that it is capable of inferring the exact exponent of $n$ that best fits the dataset. Accordingly, similar to the original fitting in section \textcolor{blue}{\ref{subsec:Average-time-complexity-estimation}}, a logarithmic transformation of the measurements is performed and fitted with a linear model.
\[
\begin{array}{l|lllll}
 \text{} & \text{DF} & \text{SS} & \text{MS} & \text{F-Statistic} & \text{P-Value} \\
\hline
 x & 1 & 36218.3 & 36218.3 & 123087. & 0. \\
 \text{Error} & 15606 & 4592.04 & 0.294248 & \text{} & \text{} \\
 \text{Total} & 15607 & 40810.3 & \text{} & \text{} & \text{} \\
\end{array}
\]

\[
\begin{array}{l|lllll}
 \text{Parameter} & \text{Estimate} & \text{Standard Error} & \text{t-Statistic} & \text{P-Value} & \text{Confidence Interval} \\
\hline
 \ln(b) & 2.96457 & 0.0499559 & 59.3437 & 0. & \{2.86665,\ 3.06249\} \\
 x & 3.05547 & 0.00870905 & 350.838 & 0. & \{3.0384,\ 3.07254\} \\
\end{array}
\]
In this case, an adjusted $R^2$ of 0.887471 and an order of $\approx n^3$ in the final bound is attained. With these results, a significant difference can be observed between the theoretically obtained bound of $n^2$ and the one provided by this model. Mainly, the reasons why this phenomenon occurs can be appreciated by comparing the 2 fitted models in this context.
\begin{figure}[H]
    \centering
    \includegraphics[width=10cm,clip]{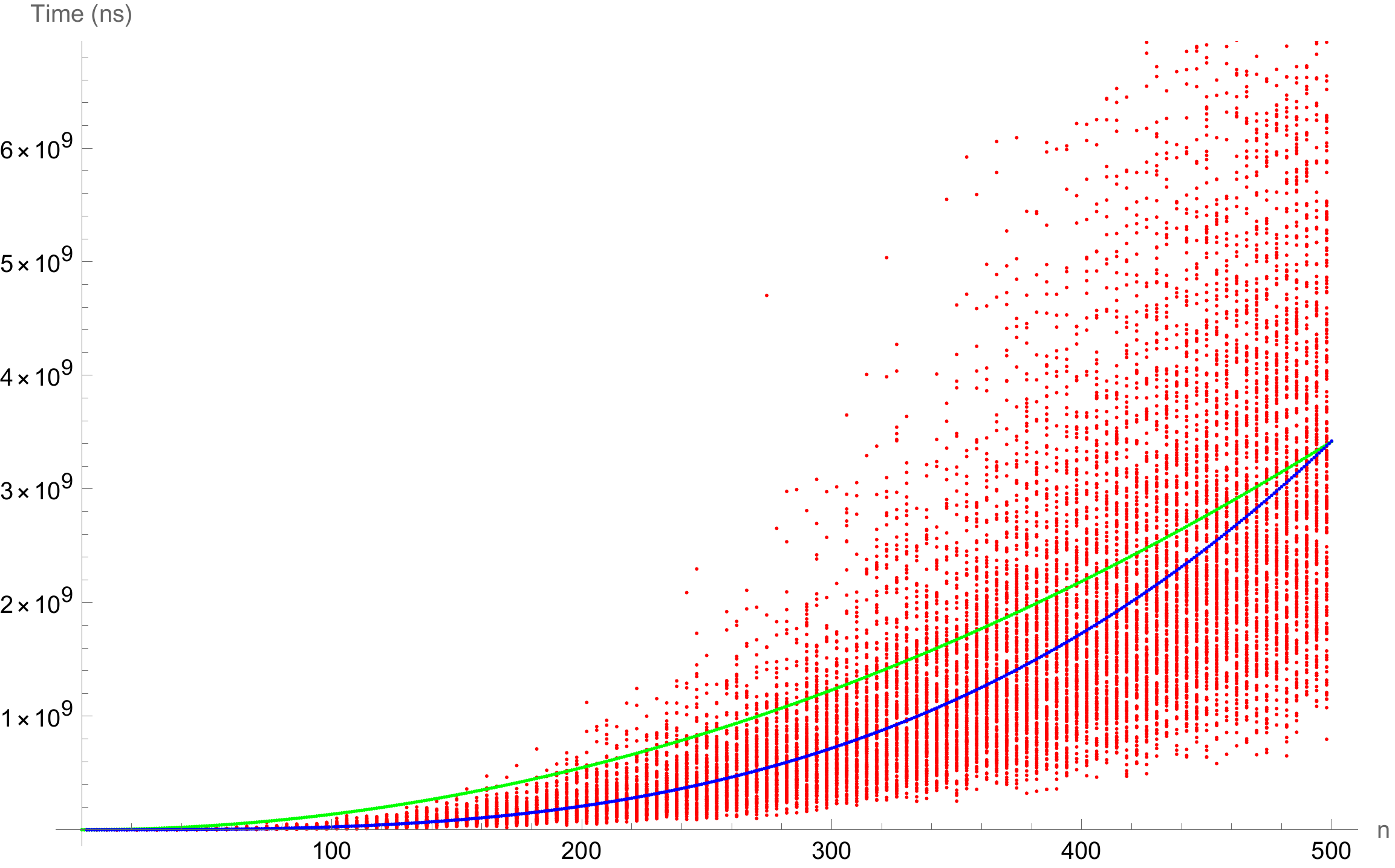}
    \caption{Runtime measurements for a 2D system with aspect ratio 1 plotted in red, the original fitted model $T_{avg}(n)\approx1.37\cdot 10^4 n^2$ in green, and the alternative one $T_{avg}(n)\approx e^{2.96 + 3.06 \ln(n)}$ in blue.}
    \label{fig:2DtimesFitComparisonAverage}
\end{figure}

On one hand, the model suggesting that the average temporal growth is on the order of $\approx n^3$ may overfit the concrete results of the dataset more than the trend for large values of $n$. Although it is not entirely appreciable, given that the overfitting is not so significant, we can see how the model tends to align increasingly closer to the worst-case with respect to areas with higher density of measurements. This model could serve as a precise approximation for the average case of small system sizes. However, to achieve the exact average bound, it is necessary to use the information contained in the theoretically calculated bound. That is, by adjusting the other model in the form $x \cdot n^2$, it may seem imprecise due to the adjusted $R^2$ and the error with respect to the measurements. But, unlike the other model, this one approaches considerably faster towards the area with the highest density of measurements than the progressive divergence shown by its counterpart. Additionally, if we try to infer what will happen for values of $n$ greater than 500, the model that overfits the data appears to have a growth closer to the worst-case than the average, while the other, although growing slower as its only variability lies in its factor $x$, appears to be closer to achieving its goal.
\section{3-dimensional case}
To conclude, after analyzing the complexity of the algorithm proposed in one- and two-dimensional systems, the same procedure will be applied to three-dimensional systems \cite{Wang2013, Kim2016, Sykes1976, Grassberger1992, BundeHavlin2012}. The primary significance of this section lies in the potential generalization for higher dimensions that can be considered with the results obtained up to this point and those forthcoming. For this purpose, the same methodology seen so far will be applied for several reasons. First, its coherence is verified, since if a contradiction occurs in this dimension, it is possible to study whether the failure comes from the nature of the system in this dimension or is something that transcends this property. On the other hand, the process of analysis is simplified and shortened, which allows focusing on the results and their comparison with those of previous sections rather than on the elaboration of the procedures.
\subsection{Average cluster size}
To begin with, the recursive approach from sections \textcolor{blue}{\ref{subsubsec:ActualClusterSize1D}} and \textcolor{blue}{\ref{subsec:AverageClusterSize2D}} could be used to obtain an expression for the number of clusters across all combinations of $k$ elements in the system $N(n,k)$. Nevertheless, as demonstrated in \textcolor{blue}{\ref{subsec:AverageClusterSize2D}}, no conclusive result was achieved for 2 dimensions, which suggests that its application in higher dimensions may not be viable.
\begin{align}
   f(x,y,z,p_{k,n}) = p_{k,n}^{\left(| x| +| y| +| z|+| x-y| +| x+y|\right)}
\end{align}
Accordingly, we proceed from the scalar field from section \textcolor{blue}{\ref{subsubsec:traversalReachProbabilityScalarField}} adapted to the geometry of the three-dimensional system, whose formulation is presented above.
\begin{figure}[H]
    \centering
    \includegraphics[width=10cm,clip]{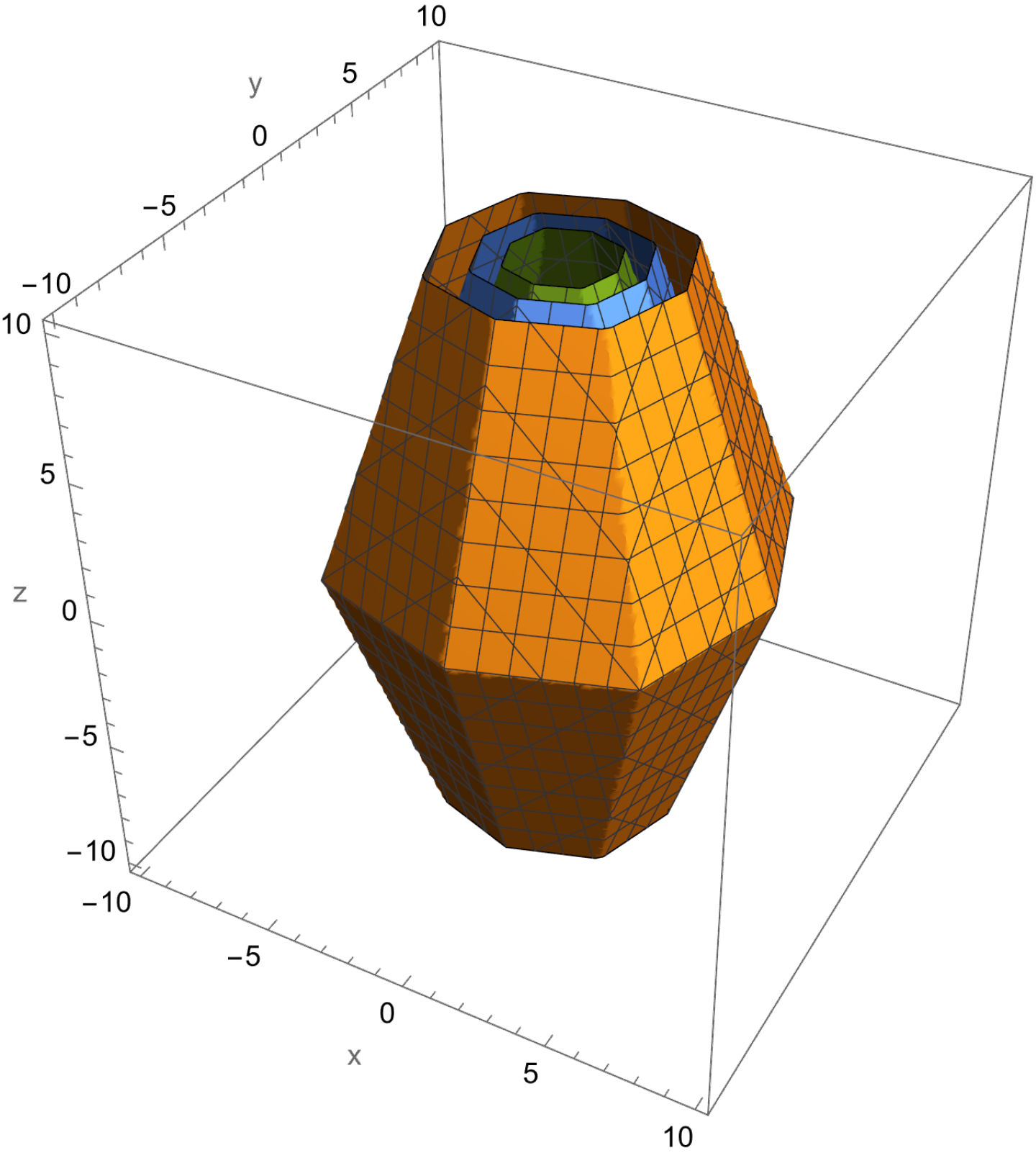}
    \caption{Contour plot of $f(x,y,z,p_{k,n})$ for $f(x,y,z,p_{k,n})=\{0.1,0.15,0.2\}$ plotted in orange, blue, and green, respectively and $p_{k,n}=0.9$.}
    \label{fig:3Dcontours}
\end{figure}

Regarding the field, the primary distinction from that proposed in two-dimensional systems lies in the number of variables. Additionally, in this case, the exponent of the occupancy ratio is formed by the planes that determine the axes $x=0$, $y=0$, and $z=0$, as well as those containing the diagonal advancement directions with respect to these axes, which are $x=y$ and $x=-y$. This is evident in the contour plot above where the field intersects specific fixed probability values.
\begin{align}
   f(x,y,z,p_{k,n})=f(-x,y,z,p_{k,n})\\\notag
   f(x,y,z,p_{k,n})=f(x,-y,z,p_{k,n})\\\notag
   f(x,y,z,p_{k,n})=f(x,y,-z,p_{k,n})\\\notag
   f(-x,y,z,p_{k,n})=f(x,-y,z,p_{k,n})\\\notag
   f(x,-y,z,p_{k,n})=f(-x,y,z,p_{k,n})\\\notag
\end{align}
Before proceeding with its summation, which will lead to an estimation of the metric $c(n,k)$, it is necessary to demonstrate its symmetry with respect to the planes encoded in its exponent. Specifically, the equalities that the field must satisfy for it to be summed easily are shown above. First, an equivalence is established between the points that are on both sides of the three-dimensional coordinate planes, and then the process is repeated for the planes characterizing the Moore neighborhood by including the corresponding advancement directions.
\begin{align}
   p_{k,n}^{| x| +| y| +| z|+| x-y| +| x+y|}=p_{k,n}^{|\pm x| +|\pm y| +|\pm z|+| x-y| +| x+y|}
\end{align}
In the first 3, the equivalences are met because the only term that differs in the exponents is one of the variables of the field function that appears with a complementary sign on both sides of the equality. And, being an input of the absolute value function, their values coincide, so it is concluded that the field is symmetrical with respect to the 3 planes of those equations. As for the others, the justification can be extracted from the procedure performed in section \textcolor{blue}{\ref{subsubsec:ContinuousSumScalarFieldProbabilities}} for the field of 2-dimensional systems. In summary, the terms $|x-y|$ of the exponent always have complementary signs for both variables on both sides of the equation, so the other planes also generate a symmetry in the field. After establishing its symmetry, the probability is summed across the entire domain of the field, which in this case is the space occupied by $\mathbb{R}^3$. First, for the continuous estimator, the $Integrate[]$ function of Wolfram is used to find the closed form of its integral sum, as illustrated below:
\begin{align}
   \hat{c}_0(n,k) &= \int_{-\infty}^{\infty}\int_{-\infty}^{\infty} \int_{-\infty}^{\infty} f(x,y,z,p_{k,n}) \, dx \, dy \, dz =\\\notag
   &= \int_{-\infty}^{\infty}\int_{-\infty}^{\infty} \int_{-\infty}^{\infty} p_{k,n}^{| x| +| y| +| z|+| x-y| +| x+y|} \, dx \, dy \, dz=\\\notag
   &=-\frac{4}{3 \ln ^3(p_{k,n})}\quad [\Re(\ln (p_{k,n}))<0]
\end{align}
And, with the formulation of the sum as a function of the occupation ratio, the transformation proposed in section \textcolor{blue}{\ref{subsubsec:ImageRangeCorrection}} is applied in order to correct the image of the sum function and reach the ultimate expression of the estimator $\hat{c}_0(n,k)$:

\begin{align}
   \hat{c}_0(n,k) = n^3\left(\frac{\displaystyle-\frac{4}{3 \ln ^3(p_{k,n})}}{\displaystyle -\frac{4}{3 \ln ^3(p_{k,n})}+n^3}\right)=\frac{(-4) n^3}{3 \ln ^3(p_{k,n}) \left(n^3-\frac{4}{3 \ln ^3(p_{k,n})}\right)}=\frac{4 n^3}{4-3 n^3 \ln ^3(p_{k,n})}
\end{align}
Similarly, to derive the alternative estimator $\hat{c}_1(n,k)$ characterized by the discrete sum of the scalar field of probabilities, the previous sum of $f(x,y,z,p_{k,n})$ is performed discretely:
\begin{align}
   \hat{c}_1(n,k) = \sum _{x=-\infty }^{\infty } \sum _{y=-\infty }^{\infty } \sum _{z=-\infty }^{\infty } f(x,y,z,p_{k,n}) = \sum _{x=-\infty }^{\infty } \sum _{y=-\infty }^{\infty } \sum _{z=-\infty }^{\infty } p_{k,n}^{| x| +| y| +| z|+| x-y| +| x+y|}
\end{align}
Thus, this time the sum can be decomposed into different regions according to the axes of symmetry. Specifically, the region in which the absolute values of the exponent can be ignored to obtain its sum is considered, and it is subsequently multiplied by the number of symmetric regions constituting the plane. Likewise, the planes composing the symmetries and the lines on which the coordinate axes lie are summed separately, as well as the point $(0,0,0)$ for simplicity:
\begin{align}
   \hat{c}_1(n,k) &= \sum _{x=-\infty }^{\infty } \sum _{y=-\infty }^{\infty } \sum _{z=-\infty }^{\infty } p_{k,n}^{| x| +| y| +| z|+| x-y| +| x+y|}=\\\notag
   &=32 \sum _{x=2}^{\infty } \sum _{y=1}^{x-1} \sum _{z=1}^{x-1} p_{k,n}^{x+x+x+y+y-y+z}+16 \sum _{x=1}^{\infty } \sum _{y=1}^{x-1} p_{k,n}^{x+x+x+x+y+y-y}\\\notag
   &+24 \sum _{x=1}^{\infty } \sum _{z=1}^{x-1} p_{k,n}^{x+x+x+z}+16 \sum _{x=1}^{\infty } \sum _{z=1}^{x-1} p_{k,n}^{x+x+x+x+z}\\\notag
   &+6 \sum _{x=1}^{\infty } p_{k,n}^{x+x+x}+12 \sum _{x=1}^{\infty } p_{k,n}^{x+x+x+x}+8 \sum _{x=1}^{\infty } p_{k,n}^{x+x+x+x+x}+1=\\\notag
   &=-\frac{p_{k,n}^{11}-p_{k,n}^{10}+p_{k,n}^9+12 p_{k,n}^8+7 p_{k,n}^7+7 p_{k,n}^4+4 p_{k,n}^3+p_{k,n}^2-p_{k,n}+1}{(p_{k,n}-1)^3 \left(p_{k,n}^2+1\right) \left(p_{k,n}^2+p_{k,n}+1\right) \left(p_{k,n}^4+p_{k,n}^3+p_{k,n}^2+p_{k,n}+1\right)}
\end{align}
After solving the sum, the subsequent transformation is performed to correct the range attained by its image, resulting in the final expression of the discrete estimator $\hat{c}_1(n,k)$:
\begin{align}
   \hat{c}_1(n,k) &= n^3\left(\frac{\displaystyle-\frac{p_{k,n}^{11}-p_{k,n}^{10}+p_{k,n}^9+12 p_{k,n}^8+7 p_{k,n}^7+7 p_{k,n}^4+4 p_{k,n}^3+p_{k,n}^2-p_{k,n}+1}{(p_{k,n}-1)^3 \left(p_{k,n}^2+1\right) \left(p_{k,n}^2+p_{k,n}+1\right) \left(p_{k,n}^4+p_{k,n}^3+p_{k,n}^2+p_{k,n}+1\right)}}{\displaystyle -\frac{p_{k,n}^{11}-p_{k,n}^{10}+p_{k,n}^9+12 p_{k,n}^8+7 p_{k,n}^7+7 p_{k,n}^4+4 p_{k,n}^3+p_{k,n}^2-p_{k,n}+1}{(p_{k,n}-1)^3 \left(p_{k,n}^2+1\right) \left(p_{k,n}^2+p_{k,n}+1\right) \left(p_{k,n}^4+p_{k,n}^3+p_{k,n}^2+p_{k,n}+1\right)}+n^3}\right)=\\\notag
   &=-n^3 \left(p_{k,n}^{11}-p_{k,n}^{10}+p_{k,n}^9+12 p_{k,n}^8+7 p_{k,n}^7+7 p_{k,n}^4+4 p_{k,n}^3+p_{k,n}^2-p_{k,n}+1\right)\left(n^3 p_{k,n}^{11}\right .\\\notag
   & \left .-n^3 p_{k,n}^{10}+n^3 p_{k,n}^9-2 n^3 p_{k,n}^8+n^3 p_{k,n}^7-2 n^3 p_{k,n}^6+2 n^3 p_{k,n}^5-n^3 p_{k,n}^4+2 n^3 p_{k,n}^3-n^3 p_{k,n}^2\right .\\\notag
   & \left .n^3 p_{k,n}-n^3-p_{k,n}^{11}+p_{k,n}^{10}-p_{k,n}^9-12 p_{k,n}^8-7 p_{k,n}^7-7 p_{k,n}^4-4 p_{k,n}^3-p_{k,n}^2+p_{k,n}-1\right)^{-1}
\end{align}

\begin{figure}[H]
    \centering
    \begin{subfigure}[b]{0.49\textwidth}
        \centering
        \includegraphics[width=\textwidth,clip]{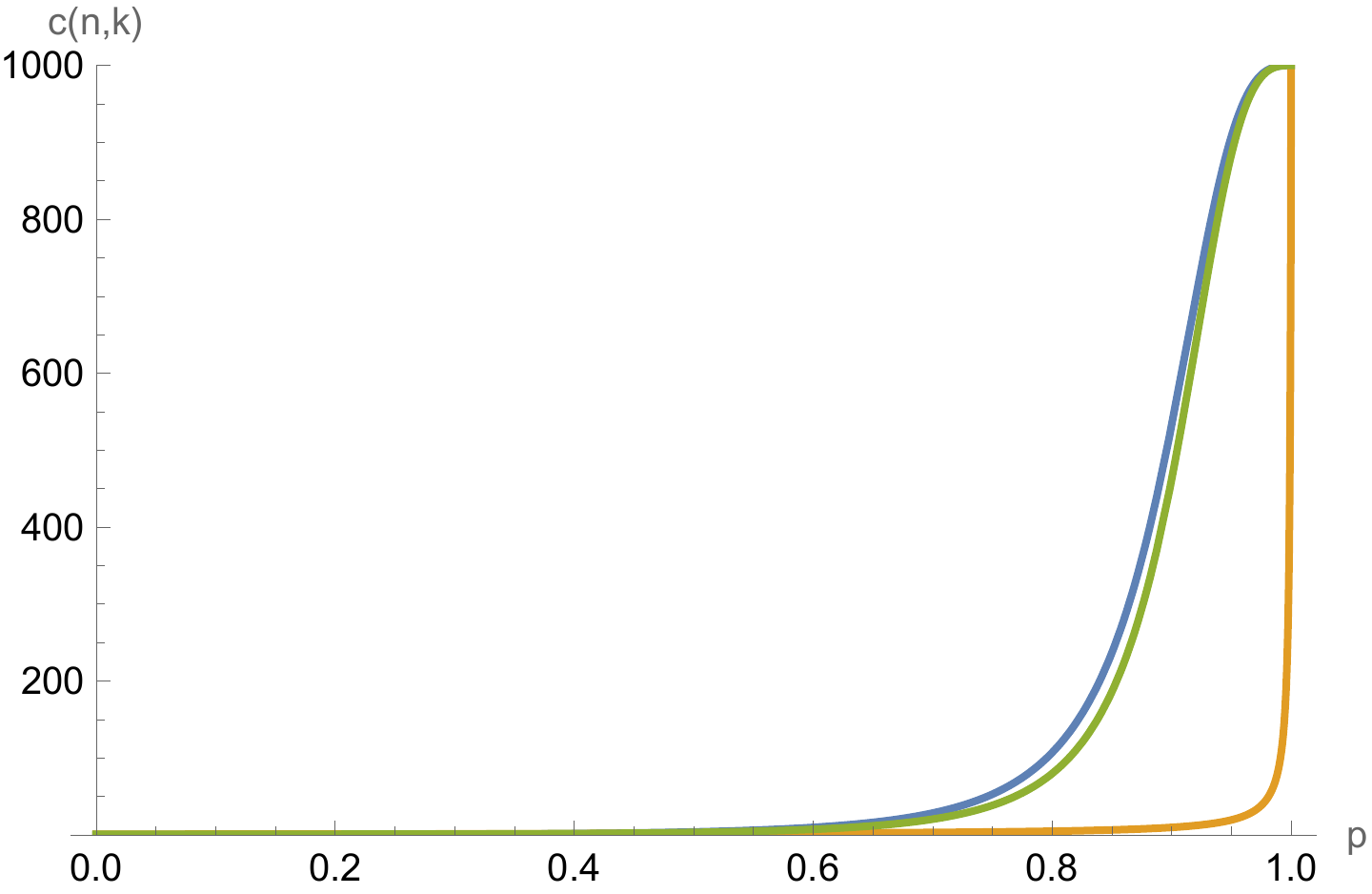}
        \caption{}
        \label{fig:3DclusterEstimators}
    \end{subfigure}
    \hfill
    \begin{subfigure}[b]{0.49\textwidth}
        \centering
        \includegraphics[width=\textwidth,clip]{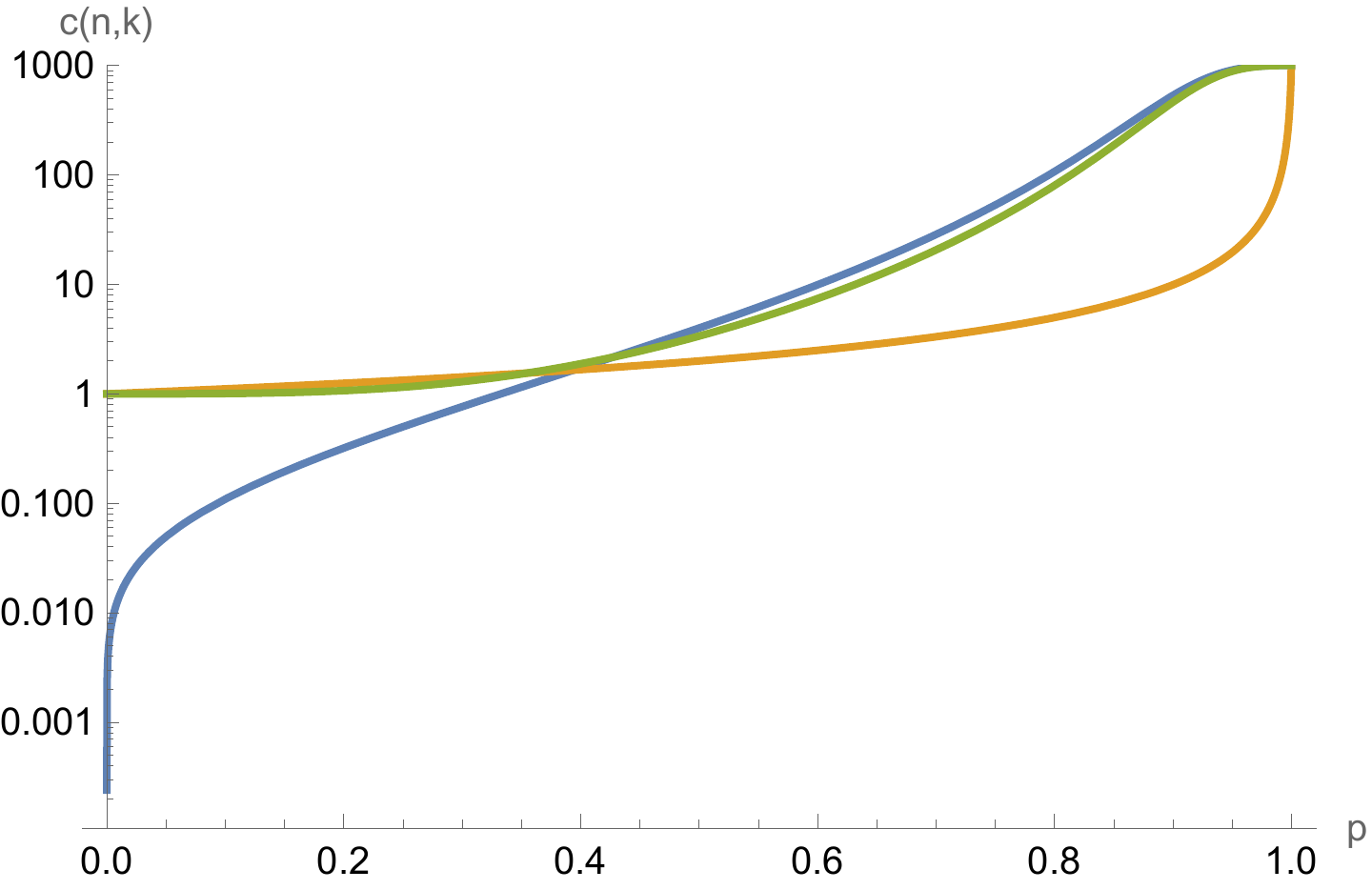}
        \caption{}
        \label{fig:3DclusterEstimatorsLog}
    \end{subfigure}
    \caption{(a) $\frac{n^3}{n^3-p_{k,n}n^3+1}$ cluster size approximation plotted in orange, $\hat{c}_0(n,k)$ and $\hat{c}_1(n,k)$ estimators plotted in blue and green respectively for a system size of $n=10$. (b) Logarithmic scale plot of (a)}
    \label{fig:3DclusterEstimatorsDual}
\end{figure}
Finally, in the figures presented above, the estimators of $c(n,k)$ calculated for 3-dimensional systems are depicted, along with the formula $\frac{n^3}{n^3-k+1}$ appreciated for its asymptotic properties. Although it is not guaranteed that its value is equivalent to the actual magnitude of $c(n,k)$, since it can be interpreted as the average cluster size of a one-dimensional system of length $n^3$, which implies that the neighborhood modifies the growth of the magnitude in the range of $p_{k,n}$, it will be used together with the estimators to infer the trend of the metric in special situations such as $n\to\infty$. Graphically, it is worth noting the resemblance of both estimators with those constructed in two-dimensional systems, especially the tendency of the continuous estimator toward zero, which causes a convergence to 0 in the subsequent analysis for a certain range of the occupation ratio values. As for its alternative, a clear similarity to $\frac{n^3}{n^3-k+1}$ is observed for small quantities of $k$ in the system, contrasting with the remaining quantities where it exhibits similar growth as the continuous estimator.

\begin{align}
   \lim_{n\to\infty} \hat{c}_0(n,k) = \lim_{n\to\infty}\frac{4 n^3}{4-3 n^3 \ln ^3(p_{k,n})} \approx -\frac{4}{3\ln ^3(p_{k,n})}
\end{align}
Likewise, identifying the intersection point where the estimators and $\frac{n^3}{n^3-k+1}$ converge may facilitate the determination of the percolation threshold $p_c$. Consequently, a possible approximation to this threshold is made by considering the points at which some of the estimators of $c(n,k)$ are equivalent to each other, which begin to be computed with the limit of the involved expressions as $n \to \infty$, as shown above with the continuum.
\begin{align}
   \lim_{n\to\infty} \hat{c}_1(n,k) \approx -\frac{1 - p_{k,n} + p_{k,n}^2 + 4p_{k,n}^3 + 7p_{k,n}^4 + 7p_{k,n}^7 + 12p_{k,n}^8 + p_{k,n}^9 - p_{k,n}^{10} + p_{k,n}^{11}}{(-1 + p_{k,n})^3 (1 + p_{k,n}^2) (1 + p_{k,n} + p_{k,n}^2) (1 + p_{k,n} + p_{k,n}^2 + p_{k,n}^3 + p_{k,n}^4)}
\end{align}
It can be noted that, at infinity the value of the estimators depends solely on the occupancy ratio:
\begin{align}
   \lim_{n\to\infty} \frac{n^3}{n^3-p_{k,n}\cdot n^3+1}\approx \frac{1}{1-p_{k,n}}
\end{align}
Thus, when establishing the equivalence between both estimators, several roots are procured, among which only one resides within the valid range for the percolation threshold, as well as the occupancy ratio in the terminal state:
\begin{align}
   \lim_{n\to\infty} \hat{c}_0(n,k)=\lim_{n\to\infty} \hat{c}_1(n,k) \implies \boxed{p_{k,n}\approx\begin{cases}
    0.426651120880672\\
    1.36390108983691
    \end{cases}}
\end{align}
In the same way, each of the estimators intersects the approximation $\frac{n^3}{n^3-p_{k,n} \cdot n^3+1}$ of the average cluster size, so from each of the produced roots, only those within the valid range where the percolation threshold resides are selected.
\begin{align}
   \lim_{n\to\infty} \hat{c}_0(n,k)=\frac{1}{1-p_{k,n}} \implies \boxed{p_{k,n}\approx0.394001218378434}
\end{align}

\begin{align}
   \lim_{n\to\infty} \hat{c}_1(n,k)=\frac{1}{1-p_{k,n}} \implies \boxed{p_{k,n}\approx0.355144}
\end{align}

Through this approach, approximations are derived to values that could potentially represent thresholds of systems characterized by specific properties \cite{Kaneko1999, Xu2013}. However, it will not be demonstrated here if the returned values are the actual thresholds, so it is proposed as further investigation.
\subsection{Worst case analysis}
To begin with, before addressing the average-case time complexity, the worst case is analyzed. So, similar to what was conducted in lower dimensions, the sum of the work executed by the algorithm across all iterations is examined:
\begin{align}
    T_{worst}(n) = \sum _{i=0}^{I(n,n^3)} \left(1-\frac{E(i)}{n^3}\right)\cdot c(n,k)=\sum _{i=0}^{I(n)} \left(1-\frac{E(i)}{n^3}\right)\cdot E(i)
\end{align}
As it is the worst case, the average cluster size is estimated as the number of elements present in each iteration, given by the metric $E(i)$, which in 3-dimensional systems is defined as:
\begin{align}
    E(i)=n^3\left(1-\left(1-\frac{1}{n^3}\right)^i\right)
\end{align}
Likewise, the work of all iterations required for the system to fill up is summed, so $I(n,n^3)$ is used as the upper limit of the sum. Therefore, for simplicity, the integral sum of $T_{worst}(n)$ is proposed, which will provide a comparatively simpler solution from which the asymptotic bound of the complexity can be inferred and subsequently contrasted with the one from the discrete approach.
\begin{align}
    T_{worst}(n) = \int _{0}^{I(n)} \left(1-\frac{E(i)}{n^3}\right)\cdot E(i) \, di
\end{align}

\begin{align}
    \int \left(1-\frac{E(i)}{n^3}\right)\cdot E(i) \, di&=\int \left(1-\frac{n^3\left(1-\left(1-\frac{1}{n^3}\right)^i\right)}{n^3}\right)\cdot n^3\left(1-\left(1-\frac{1}{n^3}\right)^i\right) \, di=\\\notag
    &=\int n^3 \left(1-\left(1-\frac{1}{n^3}\right)^i\right) \left(1-\frac{1}{n^3}\right)^i \, di=\\\notag
    &=\frac{-n^3}{\ln(1-\frac{1}{n^3})}\int u \, du=\quad\left[u=1-\left(1-\frac{1}{n^3}\right)^i\right]\\\notag
    &=\frac{-n^3 u^2}{2\ln(1-\frac{1}{n^3})}=\\\notag
    &=\frac{-n^3 \left(1-\left(1-\frac{1}{n^3}\right)^i\right)^2}{2\ln(1-\frac{1}{n^3})}=-\frac{n^3 \left(\left(1-\frac{1}{n^3}\right)^i-2\right) \left(1-\frac{1}{n^3}\right)^i}{2 \ln \left(1-\frac{1}{n^3}\right)}+C
\end{align}
After finding an expression for its antiderivative, it is evaluated between the limits 0 and $I(n)$, which in this context corresponds to $I(n,n^3)$:
\begin{align}
    T_{worst}(n) = \left . -\frac{n^3 \left(\left(1-\frac{1}{n^3}\right)^i-2\right) \left(1-\frac{1}{n^3}\right)^i}{2 \ln \left(1-\frac{1}{n^3}\right)} \right|_{i=0}^{i=I(n)} = -\frac{n^3 \left(\left(1-\frac{1}{n^3}\right)^{I(n)}-1\right)^2}{2 \ln \left(1-\frac{1}{n^3}\right)}
\end{align}
Finally, the formulation of the worst-case complexity lacks a bound that determines its growth as $n \to \infty$ and that can be easily inferred by simple observation. Hence, the same approach is employed as in the rest of the worst-case analysis of other sections, evaluating the limit of the complexity as the system size increases to infinity.
\begin{align}
    \lim_{n\to\infty} T_{worst}(n)&=\lim_{n\to\infty}-\frac{n^3 \left(\left(1-\frac{1}{n^3}\right)^{I(n,k)}-1\right)^2}{2 \ln \left(1-\frac{1}{n^3}\right)}=\\\notag
    &=\lim_{n\to\infty} -\frac{n^3 \left(\left(1-\frac{I(n,k)}{n^3}\right)-1\right)^2}{2 \ln \left(1-\frac{1}{n^3}\right)}=\\\notag
    &=\lim_{n\to\infty} \frac{1}{2}n^6 \left(\left(1-\frac{I(n,k)}{n^3}\right)-1\right)^2=\\\notag
    &=\lim_{n\to\infty} \frac{1}{2}n^6 \left(\left(1-\frac{I(n,k)}{n^3}\right)^2-2\left(1-\frac{I(n,k)}{n^3}\right)+1\right)=\\\notag
    &=\boxed{\lim_{n\to\infty} \frac{I(n,k)^2}{2}}
\end{align}
Here, a result equivalent to previous analyses is derived, in which the complexity depended on the square of the number of elements present in the terminal state of the process, which asymptotically aligns with the metric $I(n,k)$ as demonstrated above. Therefore, there are several possibilities that can determine the ultimate form of the bound to which $T_{worst}(n)$ asymptotically converges. On the one hand, one can consider a bound of order $(n^3)^2$ corresponding to the size of the system.
\begin{align}
    \lim_{n\to\infty} \frac{T_{worst}(n)}{n^6}&=\lim_{n\to\infty}-\frac{n^3 \left(\left(1-\frac{1}{n^3}\right)^{n^3H_{n^3}}-1\right)^2}{2 \ln \left(1-\frac{1}{n^3}\right)n^6}=\\\notag
    &=\lim_{n\to\infty}\frac{\left(e^{-H_{n^3}}-1\right)^2}{2}=\\\notag
    &=\lim_{n\to\infty}\frac{\displaystyle\left(\frac{1}{e^{H_{n^3}}}-1\right)^2}{2}=\lim_{n\to\infty}\frac{\displaystyle\left(-1\right)^2}{2}=\frac{1}{2}
\end{align}
And, by verifying its ratio with the complexity expression, it converges to $\frac{1}{2}$, which leads to the conclusion that the worst-case bound is of the order $\Theta(n^6)$. However, it remains necessary to discard the alternative candidate bound $n^6(H_{n^3})^2$ arising from the number of iterations lasting in the worst-case scenario. Thus, given the previous limit, it is easily inferable that with this bound the ratio will converge to 0 as the harmonic number's growth increases.
\begin{align}
    \lim_{n\to\infty} \frac{T_{worst}(n)}{n^6(H_{n^3})^2}&=\lim_{n\to\infty}-\frac{n^3 \left(\left(1-\frac{1}{n^3}\right)^{n^3H_{n^3}}-1\right)^2}{2 \ln \left(1-\frac{1}{n^3}\right)n^6H_{n^3}}=\\\notag
    &=\lim_{n\to\infty}\frac{\left(e^{-H_{n^3}}-1\right)^2}{2H_{n^3}}=\\\notag
    &=\lim_{n\to\infty}\frac{\displaystyle\left(\frac{1}{e^{H_{n^3}}}-1\right)^2}{2H_{n^3}}=\lim_{n\to\infty}\frac{\displaystyle\left(-1\right)^2}{2H_{n^3}}=0
\end{align}
Ultimately, after discarding the upper bound due to its higher growth compared to the actual complexity, wen can conclude with the bound of order $n^6$, although since it stems from the approach where the cost of each iteration is summed integrally, it is essential to verify that through the discrete sum the same growth is reached.

\begin{align}
    T_{worst}(n)&=\sum _{i=0}^{I(n)} \left(1-\frac{E(i)}{n^3}\right)\cdot E(i)=\\\notag
    &=\sum _{i=0}^{I(n)} n^3 \left(1-\left(1-\frac{1}{n^3}\right)^i\right) \left(1-\frac{1}{n^3}\right)^i=\\\notag
    &=\frac{n^3 \left(n^3-1\right) \left(\left(1-\frac{1}{n^3}\right)^{I(n)}-1\right) \left(\left(n^3-1\right) \left(1-\frac{1}{n^3}\right)^{I(n)}-n^3\right)}{2 n^3-1}
\end{align}
To this end, an expression of the sum that determines the total cost $T_{worst}(n)$ of the process as a function of $I(n)$ and the size of the system is first obtained. However, as it does not exhibit a bound that unambiguously determines its growth, as in previous occasions, the procedure of finding the limit of $T_{worst}(n)$ and subsequently checking the potential candidate bounds is reiterated.

\begin{align}
    \lim_{n\to\infty} T_{worst}(n)&=\lim_{n\to\infty} \frac{n^3 \left(n^3-1\right) \left(\left(1-\frac{1}{n^3}\right)^{I(n,k)}-1\right) \left(\left(n^3-1\right) \left(1-\frac{1}{n^3}\right)^{I(n,k)}-n^3\right)}{2 n^3-1}=\\\notag
    &=\lim_{n\to\infty} \frac{n^3 \left(n^3-1\right) \left(\left(1-\frac{I(n,k)}{n^3}\right)-1\right) \left(\left(n^3-1\right) \left(1-\frac{I(n,k)}{n^3}\right)-n^3\right)}{2 n^3-1}=\\\notag
    &=\lim_{n\to\infty} -\frac{\left(n^3-1\right) I(n,k) \left(\left(n^3-1\right) \left(1-\frac{I(n,k)}{n^3}\right)-n^3\right)}{2 n^3-1}=\\\notag
    &=\lim_{n\to\infty} -\frac{\displaystyle n^3 \left(-I(n,k)^2\right)-\frac{I(n,k)^2}{n^3}-n^3 I(n,k)+2 I(n,k)^2+I(n,k)}{2 n^3-1}=\\\notag
    &=\lim_{n\to\infty}  \frac{I(n,k)^2}{2}+\frac{I(n,k)^2}{2n^6}+\frac{I(n,k)}{2}-\frac{I(n,k)^2}{n^3}-\frac{I(n,k)}{n^3}=\\\notag
    &=\boxed{\lim_{n\to\infty}  \frac{I(n,k)^2}{2}+\frac{I(n,k)}{2}}
\end{align}
Regarding the complexity limit, it provides a result closely aligned to that derived from the integral approach, so it can be inferred that the final bound will be equally similar. Thus, since the square of $I(n,k)$ will always exhibit a greater growth than the metric itself, a possible candidate for the bound would be $(n^3)^2$, since the alternative term would be of order $O(n^3)$, which is negligible as $n\to\infty$.
\begin{align}
    \lim_{n\to\infty} \frac{T_{worst}(n)}{n^6}&=\lim_{n\to\infty} \frac{n^3 \left(n^3-1\right) \left(\left(1-\frac{1}{n^3}\right)^{n^3H_{n^3}}-1\right) \left(\left(n^3-1\right) \left(1-\frac{1}{n^3}\right)^{n^3H_{n^3}}-n^3\right)}{(2 n^3-1)n^6}=\\\notag
    &=\lim_{n\to\infty} \frac{n^3 \left(n^3-1\right) \left(e^{-H_{n^3}}-1\right) \left(\left(n^3-1\right) e^{-H_{n^3}}-n^3\right)}{(2 n^3-1)n^6}=\\\notag
    &=\lim_{n\to\infty} \frac{n^3 \left(n^3-1\right) (-1) \left(-n^3\right)}{(2 n^3-1)n^6}=\\\notag
    &=\lim_{n\to\infty} \frac{n^9}{2 n^9}=\frac{1}{2}
\end{align}
Upon checking if the time complexity is equivalent to the bound proposed at the specified accumulation point, its ratio converges to the same constant $\frac{1}{2}$ as in the integral approach. Therefore, its alternative $n^6(H_{n^3})^2$ is discarded as before, since it includes a harmonic number as a factor.
\begin{align}
    \lim_{n\to\infty} \frac{T_{worst}(n)}{n^6(H_{n^3})^2}&=\lim_{n\to\infty} \frac{n^3 \left(n^3-1\right) \left(\left(1-\frac{1}{n^3}\right)^{n^3H_{n^3}}-1\right) \left(\left(n^3-1\right) \left(1-\frac{1}{n^3}\right)^{n^3H_{n^3}}-n^3\right)}{(2 n^3-1)n^6(H_{n^3})^2}=\\\notag
    &=\lim_{n\to\infty} \frac{n^3 \left(n^3-1\right) \left(e^{-H_{n^3}}-1\right) \left(\left(n^3-1\right) e^{-H_{n^3}}-n^3\right)}{(2 n^3-1)n^6(H_{n^3})^2}=\\\notag
    &=\lim_{n\to\infty} \frac{n^3 \left(n^3-1\right) (-1) \left(-n^3\right)}{(2 n^3-1)n^6(H_{n^3})^2}=\\\notag
    &=\lim_{n\to\infty} \frac{n^9}{2 n^9(H_{n^3})^2}=\lim_{n\to\infty} \frac{1}{2 (H_{n^3})^2}=0
\end{align}

\begin{figure}[H]
    \centering
    \begin{subfigure}[b]{0.49\textwidth}
        \centering
        \includegraphics[width=\textwidth,clip]{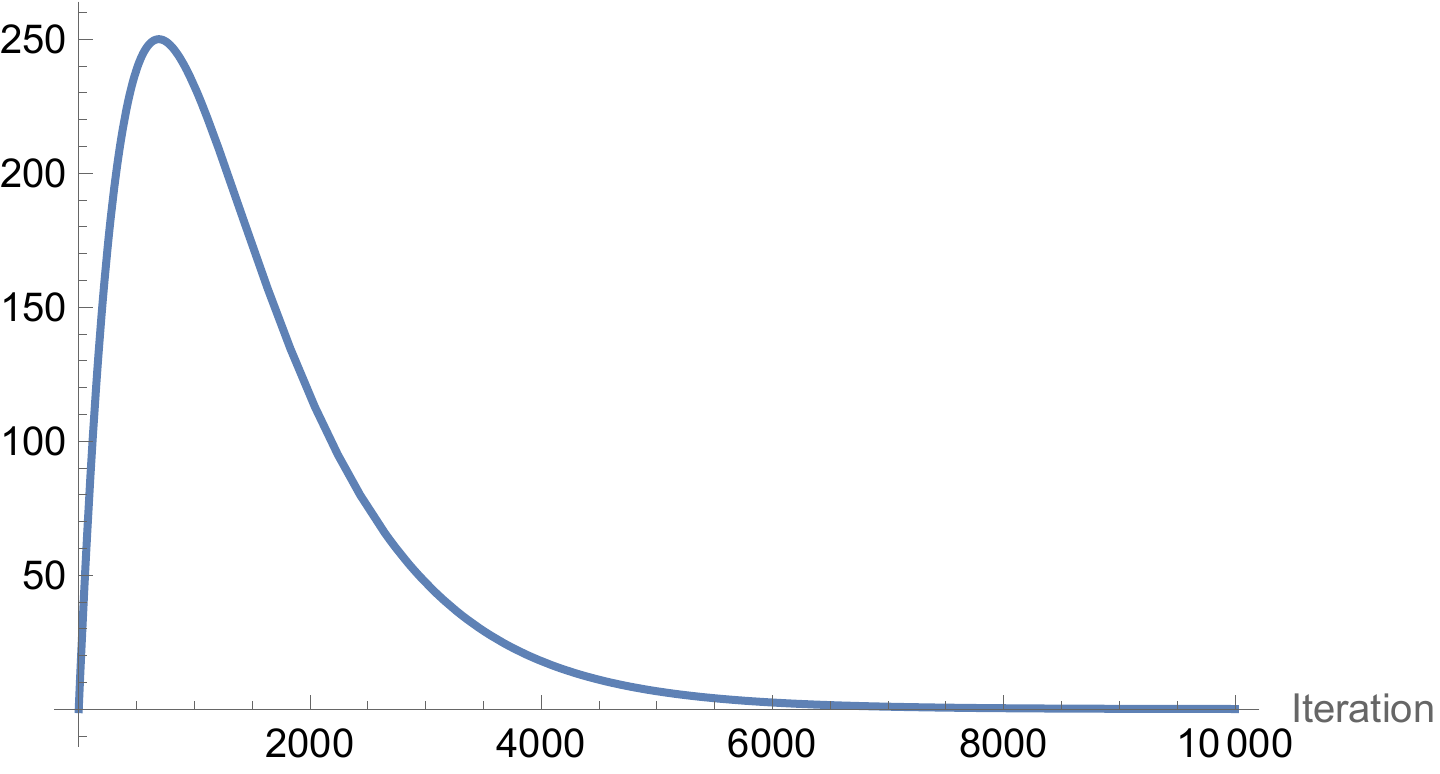}
        \caption{}
        \label{fig:3DWorstCase}
    \end{subfigure}
    \hfill
    \begin{subfigure}[b]{0.49\textwidth}
        \centering
        \includegraphics[width=\textwidth,clip]{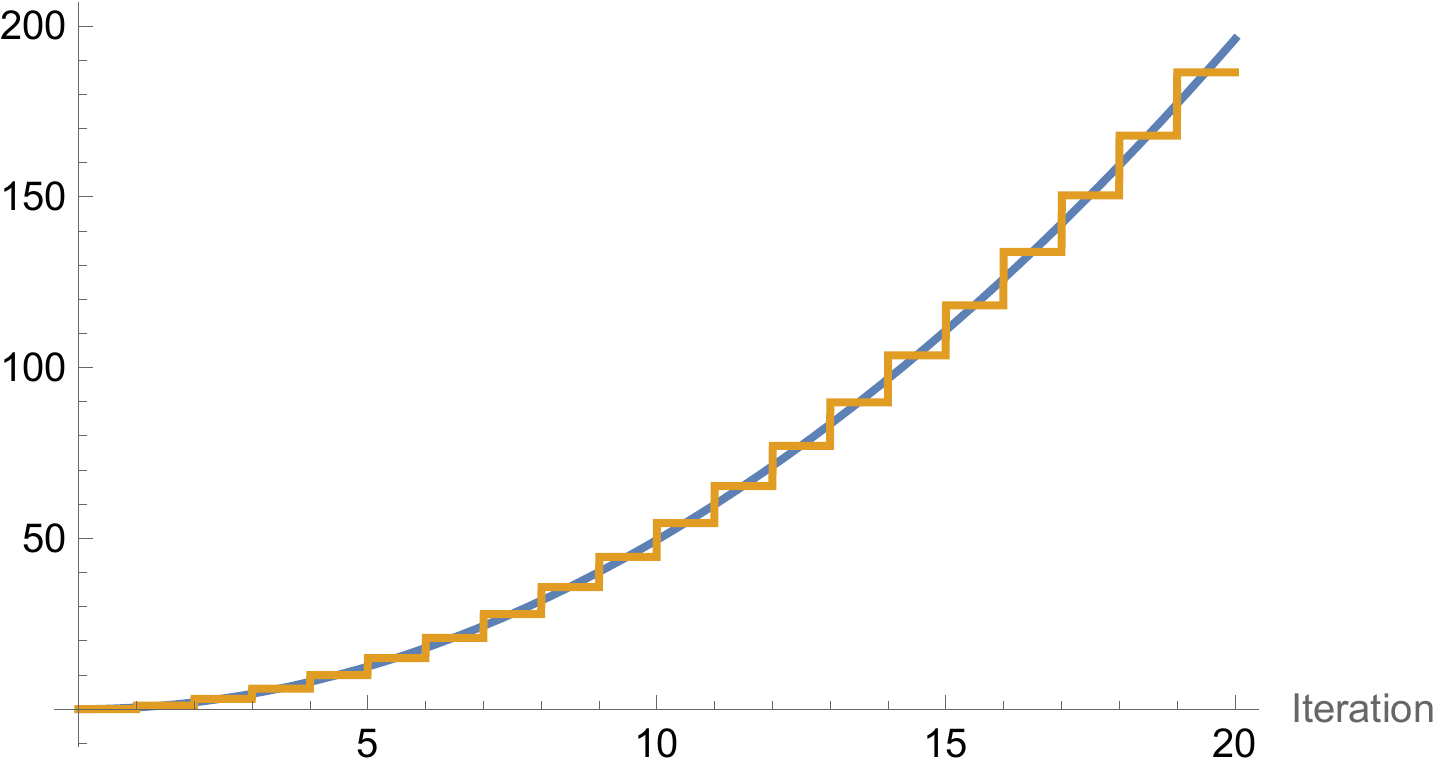}
        \caption{}
        \label{fig:3DWorstCaseSums}
    \end{subfigure}
    \caption{(a) $\left(1-\frac{E(i)}{n^3}\right)\cdot E(i)$ iteration complexity for a system of size $n=10$. (b) Integral formulation of $T_{worst}(n)$ plotted in blue and discrete variant in orange for a system of size $n=10$.}
    \label{fig:3DWorstCaseDual}
\end{figure}
Ultimately, the cost of executing an iteration as a process evolves is illustrated above, along with the time complexity of the entire process $T_{worst}(n)$, both in its integral and discrete formulation. In general, they are very similar to those of other systems with lower dimensions, because the worst case of this algorithm is mainly determined by the size of the system. In this case, the obtained bound can be interpreted such that in each iteration, a work proportional to $E(i)$ is executed, which is equivalent to the sequence $1+2+3+\cdots$ whose increase by 1 unit occurs simultaneously with an insertion. Thus, by having to fill the system with elements, the cost of the process can also be expressed as $\sum_{i=0}^{n^3} i = O(n^6)$.
\subsection{Average case analysis}
Subsequently, after understanding the time complexity growth in the worst-case scenario of the process for 3-dimensional systems, the analysis of the average case is conducted. This will lead to a bound that determines the average work required to complete the execution of the algorithm, considering all conceivable situations that may arise during its execution, and the probability of them occurring.
\begin{align}
    T_{avg}(n) = \sum _{i=0}^{I(n)} \left(1-\frac{E(i)}{n^3}\right)\cdot c(n,E(i))=\sum _{i=0}^{I(n)} \left(1-\frac{E(i)}{n^3}\right)\cdot \frac{n^3}{n^3-E(i)+1}
\end{align}
Initially, the average cost is formulated as the sum of the cost across all iterations, in which insertion probability is used along with an approximation to the metric $c(n, k)$. Specifically, the aforementioned formulation presents certain properties that asymptotically serve to determine the trend of the average cluster size in relation to the probability that a cost proportional to that size occurs during an insertion. Although the approximation does not return the exact values of this magnitude, it complies with the restrictions seen in section \textcolor{blue}{\ref{subsubsec:AverageInsertionRuntime}}, ensuring a degree of correctness, especially when varying the system size towards extreme accumulation points as we are interested in here, $n \to \infty$.

\begin{align}
    \lim_{n\to\infty} T_{avg}(n)&=\sum _{i=0}^{I(n)} \left(1-\frac{E(i)}{n^3}\right)\cdot \frac{n^3}{n^3-E(i)+1}=\\\notag
    &=\sum _{i=0}^{I(n)} \frac{n^3 \left(1-\frac{1}{n^3}\right)^i}{n^3-\left(n^3 \left(1-\left(1-\frac{1}{n^3}\right)^i\right)\right)+1}=\\\notag
    &=\frac{\psi _{\frac{n^3-1}{n^3}}^{(0)}\left(I(n)-\frac{\ln \left(-\frac{1}{n^3}\right)}{\ln \left(\frac{n^3-1}{n^3}\right)}+1\right)}{\ln \left(\frac{n^3-1}{n^3}\right)}-\frac{\psi _{\frac{n^3-1}{n^3}}^{(0)}\left(-\frac{\ln \left(-\frac{1}{n^3}\right)}{\ln \left(\frac{n^3-1}{n^3}\right)}\right)}{\ln \left(\frac{n^3-1}{n^3}\right)}
\end{align}
Hence, as an initial attempt, the sum of the total cost of the algorithm is solved, resulting in an expression like the one above. In it, since it is composed by Digamma functions, the growth bound as the system size escalates remains indiscernible. So, to address this concern, there are several alternatives. On one hand, the properties of the superior function could be studied and a conclusion drawn regarding the bound that asymptotically equals $T_{avg}(n)$. However, since this can be costly and long depending on the expression to analyze, we proceed through the integral sum of the complexity:
\begin{align}
    T_{avg}(n)\approx&\int_{0}^{I(n)} \left(1-\frac{E(i)}{n^3}\right)\cdot \frac{n^3}{n^3-E(i)+1} \, di=\\\notag
    &=\int_{0}^{I(n)} \frac{n^3 \left(1-\frac{1}{n^3}\right)^i}{n^3-\left(n^3 \left(1-\left(1-\frac{1}{n^3}\right)^i\right)\right)+1} \, di=\\\notag
    &=\frac{1}{\ln\left(1-\frac{1}{n^3}\right)}\int_{0}^{I(n)} \frac{1}{u} \, du=\quad\left[u=n^3-\left(n^3 \left(1-\left(1-\frac{1}{n^3}\right)^i\right)\right)+1\right]\\\notag
    &=\left . \frac{\ln(u)}{\ln\left(1-\frac{1}{n^3}\right)} \right|_{i=0}^{i=I(n)}=\\\notag
    &=\left . \frac{\ln(n^3-\left(n^3 \left(1-\left(1-\frac{1}{n^3}\right)^i\right)\right)+1)}{\ln\left(1-\frac{1}{n^3}\right)} \right|_{i=0}^{i=I(n)}=\\\notag
    &=\frac{\ln \left(n^3 \left(1-\frac{1}{n^3}\right)^{I(n)}+1\right)-\ln \left(n^3+1\right)}{\ln \left(1-\frac{1}{n^3}\right)}
\end{align}
The aim of altering the nature of the sum is to simplify the resulting complexity expression in order to subsequently find the bound that grows at the same rate, which leads to the growth of the average cost. In this case, through the integral, we obtain a formulation devoid of Digamma functions. Specifically, it is an expression reminiscent of those previously derived, from which we finally manage to find its asymptotic bound. Thus, since the same methodology is to be applied, it is convenient to try to ascertain the shape of this bound, or at least understand the factors upon which the complexity growth depends on as the system size increases, which is why the functions involved in this analysis are first plotted.

\begin{figure}[H]
    \centering
    \includegraphics[width=10cm,clip]{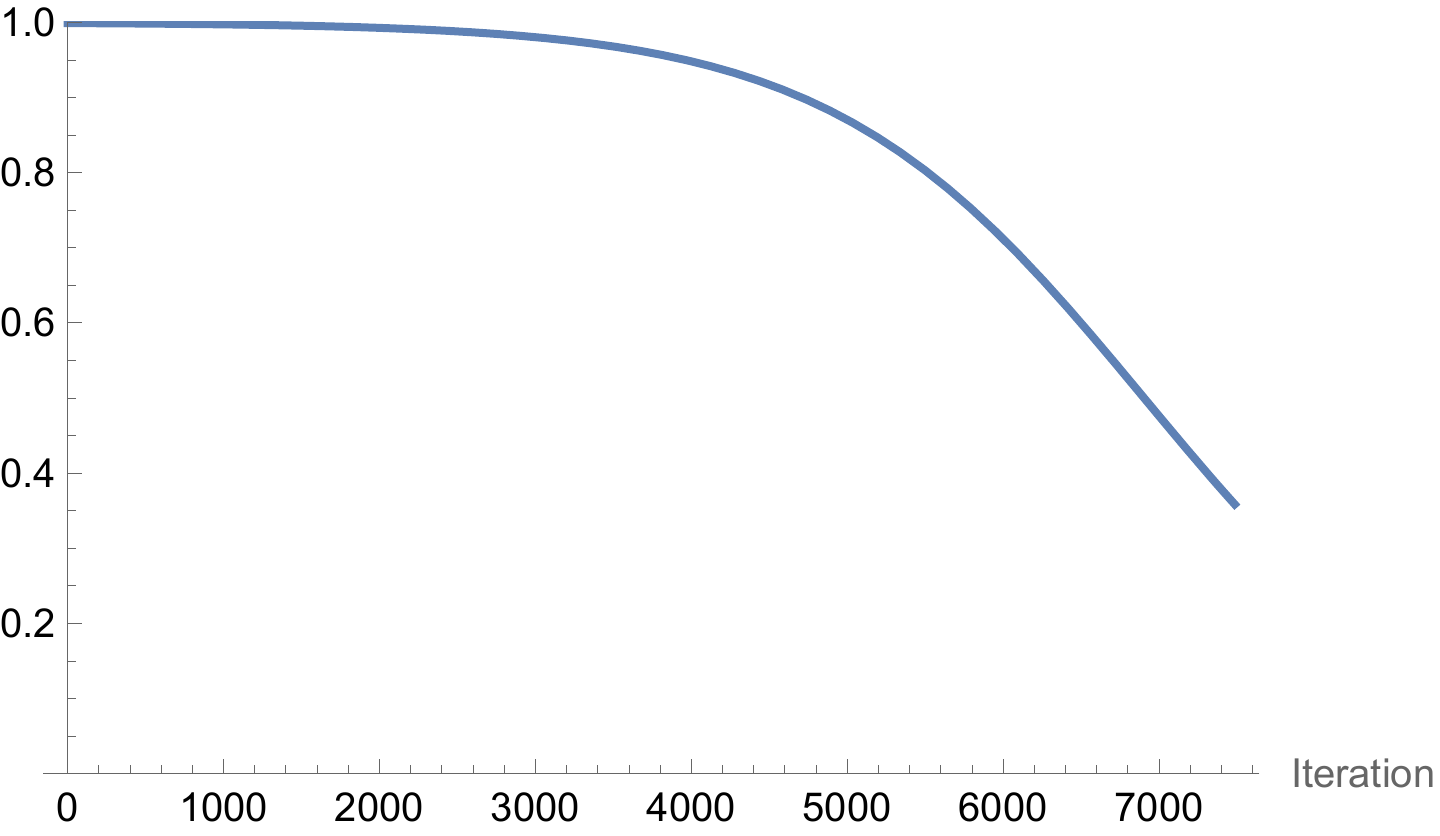}
    \caption{Iteration complexity $\left(1-\frac{E(i)}{n^3}\right)\cdot \frac{n^3}{n^3-E(i)+1}$ for a system of size $n=10$ from iteration 0 to $n^3H_{n^3}$}
    \label{fig:3DaverageInsertion}
\end{figure}
On one hand, there is the cost of an iteration, which, as observed, is close to 1 until it surpasses a threshold from which it decreases to 0 due to the decay of the insertion probability. Additionally, as $n$ increases, the threshold approaches infinity, so the cost will approach 1 over a broader range of iterations.
\begin{figure}[H]
    \centering
    \includegraphics[width=10cm,clip]{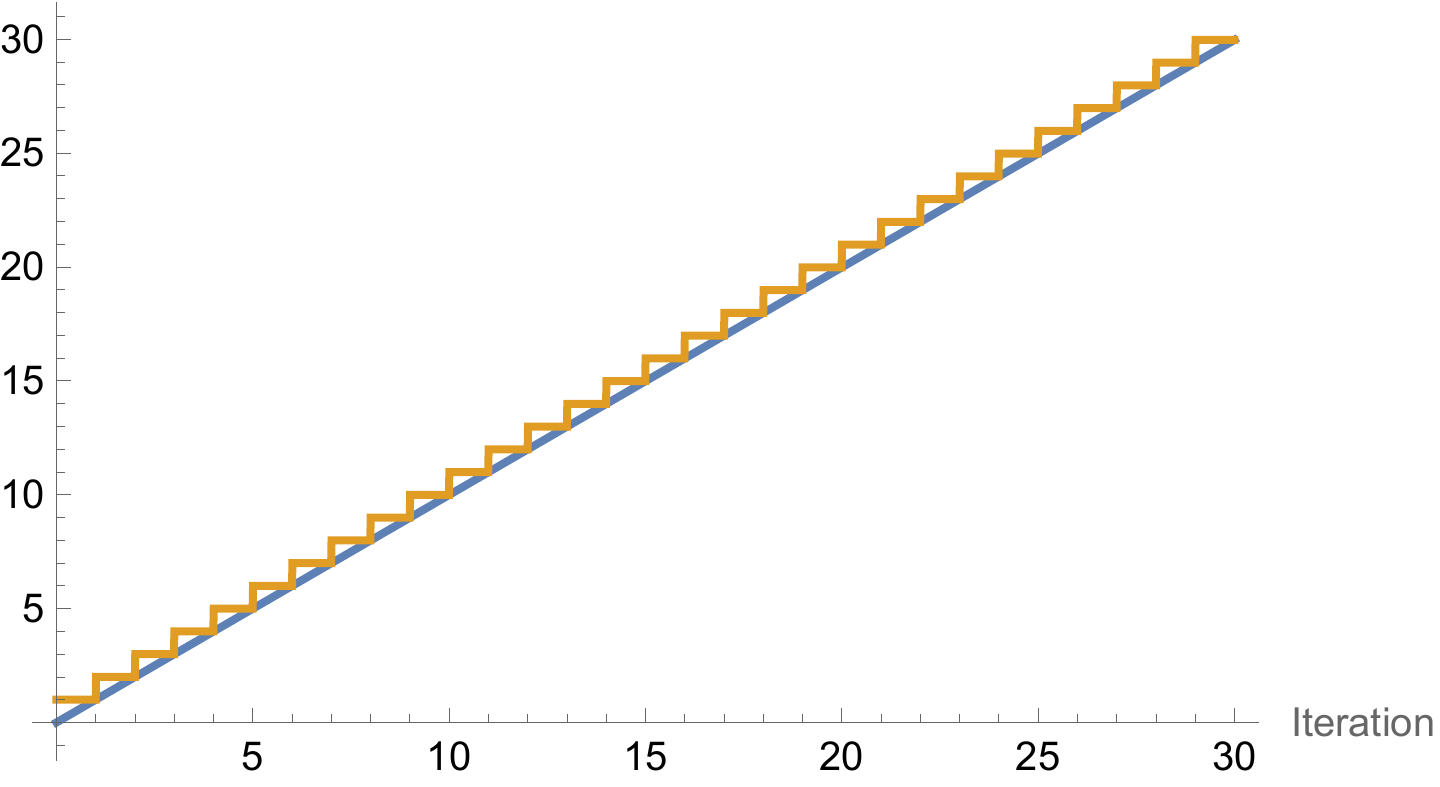}
    \caption{$T_{avg}(n)$ in its integral sum formulation plotted in blue for a system of size $n=10$, and in orange its discrete sum variant.}
    \label{fig:3DaverageComplexitySums}
\end{figure}
Regarding the total cost of the algorithm during the process, since it is defined as the sum over the cost of each iteration, approaching 1 as $n$ increases, it exhibits a linear growth with respect to the iterations. That is, if the process lasts for a certain number of iterations $I(n)$, the upper graph indicates that the cost of the algorithm grows linearly with respect to $I(n)$. Also, the difference between the integral and discrete formulations is not significant asymptotically, although for precise cost measurements it should be heeded.
\begin{figure}[H]
    \centering
    \includegraphics[width=10cm,clip]{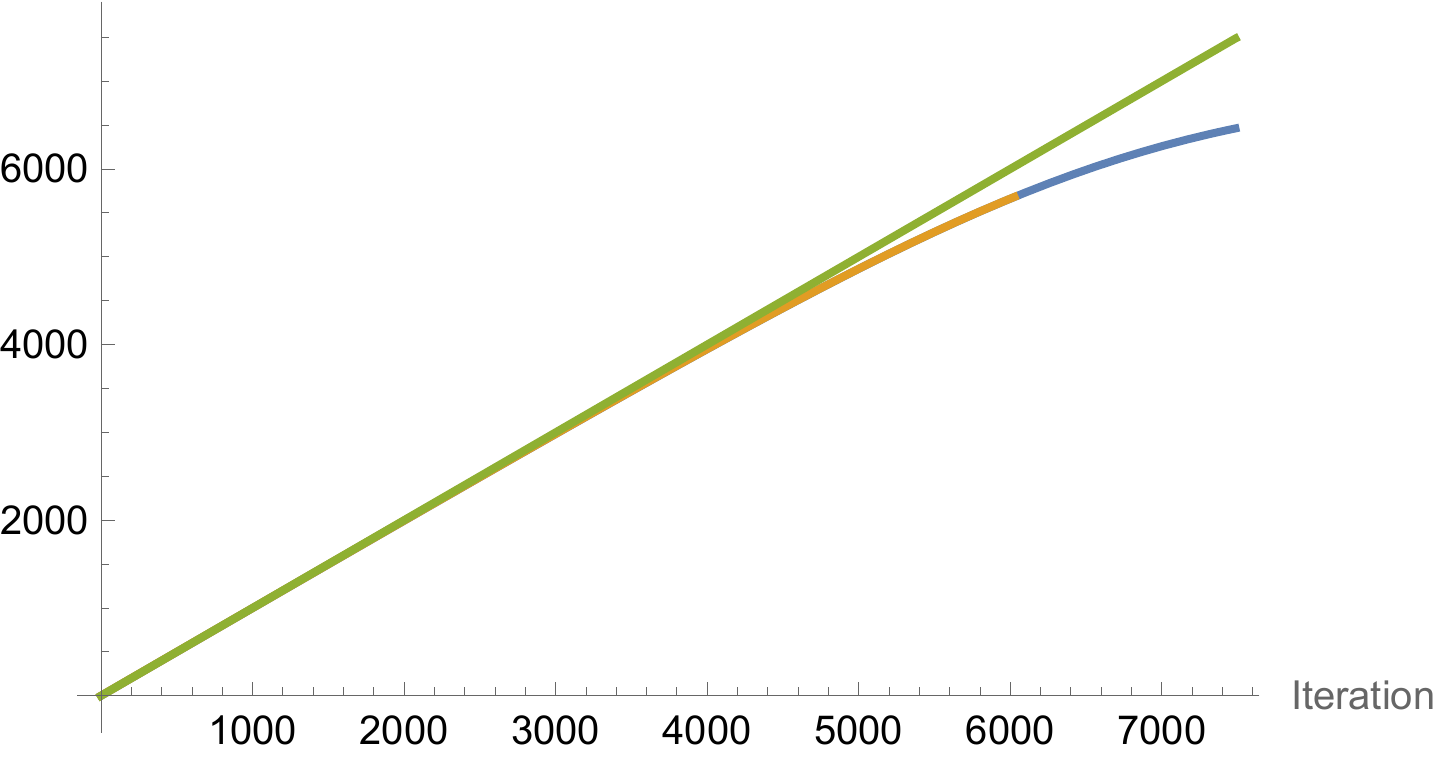}
    \caption{$T_{avg}(n)$ in its integral sum formulation plotted in blue for a system of size $n=10$, in orange its discrete sum variant, and the function $i$ in green, from iteration 0 to $n^3H_{n^3}$. }
    \label{fig:3DaverageComplexitySums2}
\end{figure}
Considering the previous plot in Figure 75, which presents an insufficiently small range of iterations, it is necessary to extend it to the upper limit of the magnitude $n^3H_{n^3}$ to confirm its correlation with $I(n)$. Thus, in Figure 76 the complexity of the process is plotted for different values of $I(n)$, from 0 to the maximum iteration the algorithm can last on average. As can be seen, for a reduced system size in most quantities of $I(n)$, the complexity is proportional to that magnitude, as observed in the original plot. Nevertheless, there is a threshold from which the complexity deviates from the line that represents this correlation. This could suggest that the linear relationship between complexity and process duration does not hold over the entire range of iterations it may require to reach the terminal state. Therefore, to verify if this occurs, even at infinity, the limit is examined as the system increases in size.

\begin{align}
    \lim_{n\to\infty} T_{avg}(n)&=\lim_{n\to\infty} \frac{\ln \left(n^3 \left(1-\frac{1}{n^3}\right)^{I(n)}+1\right)-\ln \left(n^3+1\right)}{\ln \left(1-\frac{1}{n^3}\right)}=\\\notag
    &=\lim_{n\to\infty} -n^3\left(\ln \left(n^3 \left(1-\frac{I(n)}{n^3}\right)+1\right)-\ln \left(n^3+1\right)\right)=\\\notag
    &=\lim_{n\to\infty} -n^3\left(\ln \left(n^3-I(n)+1\right)-\ln \left(n^3+1\right)\right)=\\\notag
    &=\lim_{n\to\infty} -n^3\left(\ln \left(\frac{n^3}{n^3+1}-\frac{I(n)}{n^3+1}+\frac{1}{n^3+1}\right)\right)=\\\notag
    &=\lim_{n\to\infty} -n^3\left(\ln \left(1-\frac{I(n)}{n^3+1}\right)\right)=\\\notag
    &=\lim_{n\to\infty} -n^3\left(-\frac{I(n)}{n^3+1}\right)=\boxed{\lim_{n\to\infty} I(n)}
\end{align}
Consequently, based on the outcome of the limit, it can be concluded that, asymptotically, the complexity is equivalent to $I(n)$, that is, to the duration of the process. Similarly, we can proceed using the Laurent series expansion \cite{GoodmansonWeissteinLaurentSeries} of complexity, yielding the same result:

\begin{align}
    \lim_{n\to\infty} T_{avg}(n)&=\lim_{n\to\infty} \frac{\ln \left(n^3 \left(1-\frac{1}{n^3}\right)^{I(n)}+1\right)-\ln \left(n^3+1\right)}{\ln \left(1-\frac{1}{n^3}\right)}=\\\notag
    &=\lim_{n\to\infty} I(n) - \frac{I(n)}{n^{3}} - \frac{(I(n) - 2)I(n)}{2\,n^{6}} - \frac{(2I(n)^{2} - 9I(n) + 12)I(n)}{12\,n^{9}} \\\notag
    &- \frac{(I(n)^{3} - 12I(n)^{2} + 28I(n) - 24)I(n)}{24\,n^{12}} + O\left(\frac{1}{n^{13}}\right)=\\\notag
    &=\boxed{\lim_{n\to\infty} I(n)}
\end{align}
After formally establishing the dependence between the complexity and the process duration, it is convenient to ensure that it is satisfied by applying the analysis procedure from section \textcolor{blue}{\ref{subsec:AverageCaseAnalysis2D}}, in which the limit of the cost of an iteration was calculated to simplify the sum of the total work. Therefore, by proceeding in the same manner, the limit of the expression with which the average cluster size has been approximated is first computed:
\begin{align}
    \lim_{n\to\infty} c(n,k)&=\lim_{n\to\infty} \frac{n^3}{n^3-k+1}=\begin{cases}
    1 & \text{if } k=o(n^3)\\
    \infty & \text{if } k= \Theta(n^3)
    \end{cases}
\end{align}
As a result, for any number of elements except the maximum, the metric $c(n,k)$ tends to 1 in the same way as it did in lower-dimensional systems. And, when $k=n^3$, since the only existing cluster is of a size equal to that of the system, its asymptotic growth exhibits an order of $n^3$.
\begin{align}
    \lim_{n\to\infty} \hat{c}_0(n,k) &=\lim_{n\to\infty} \frac{4 n^3}{4-3 n^3 \ln ^3(p_{k,n})}=\\\notag
    &=\lim_{n\to\infty} \frac{4 n^3}{\displaystyle4-3 n^3 \ln ^3\left(\frac{k}{n^3}\right)}=\\\notag
    &=\lim_{n\to\infty} \frac{4}{\displaystyle\frac{4}{n^3}-3 \ln ^3\left(\frac{k}{n^3}\right)}=\\\notag
    &=\lim_{n\to\infty} \frac{4}{\displaystyle-3 \ln ^3\left(\frac{k}{n^3}\right)}=0\quad[k=o(n^3)]
\end{align}
An analogous result is derived from the limit of the continuous estimator, where the average cluster size tends to 0 instead of 1 for all quantities of elements, except for the maximum one, which remains invariant with respect to the preceding approximation:
\begin{align}
    \lim_{n\to\infty} \frac{\hat{c}_0(n,n^3)}{n^3} &=\lim_{n\to\infty} \frac{4 n^3}{n^3(4-3 n^3 \ln ^3(p_{n^3,n}))}=\\\notag
    &=\lim_{n\to\infty} \frac{4}{\displaystyle4-3 n^3\ln ^3\left(\frac{n^3}{n^3}\right)}=\\\notag
    &=\lim_{n\to\infty} \frac{4}{\displaystyle4}=1\quad[k=\Theta(n^3)]
\end{align}
Nonetheless, by repeating the process with the discrete estimator, since it constitutes a more accurate approximation to the actual magnitude, it tends to 1 just like the original approximation of $c(n,k)$.
\begin{align}
    &\lim_{n\to\infty} \hat{c}_1(n,k)=\\\notag
    &=\lim_{n\to\infty} n^3\left(\frac{\displaystyle-\frac{p_{k,n}^{11}-p_{k,n}^{10}+p_{k,n}^9+12 p_{k,n}^8+7 p_{k,n}^7+7 p_{k,n}^4+4 p_{k,n}^3+p_{k,n}^2-p_{k,n}+1}{(p_{k,n}-1)^3 \left(p_{k,n}^2+1\right) \left(p_{k,n}^2+p_{k,n}+1\right) \left(p_{k,n}^4+p_{k,n}^3+p_{k,n}^2+p_{k,n}+1\right)}}{\displaystyle -\frac{p_{k,n}^{11}-p_{k,n}^{10}+p_{k,n}^9+12 p_{k,n}^8+7 p_{k,n}^7+7 p_{k,n}^4+4 p_{k,n}^3+p_{k,n}^2-p_{k,n}+1}{(p_{k,n}-1)^3 \left(p_{k,n}^2+1\right) \left(p_{k,n}^2+p_{k,n}+1\right) \left(p_{k,n}^4+p_{k,n}^3+p_{k,n}^2+p_{k,n}+1\right)}+n^3}\right)=\\\notag
   &=\lim_{n\to\infty} n^3 \left(k^{11}-k^{10} n^3+k^9 n^6+12 k^8 n^9+7 k^7 n^{12}+7 k^4 n^{21}+4 k^3 n^{24}+k^2 n^{27}-k n^{30}\right .\\\notag
   & \left .+n^{33}\right)\left(k^{11}-k^{10} (k+1) n^3+k^9 (k+1) n^6-(k-12) k^8 n^9-k^7 n^{15}+k^7 (2 k+7) n^{12}\right .\\\notag
   & \left .+2 k^6 n^{18}+k^4 (7-2 k) n^{21}+k^3 (k+4) n^{24}+k^2 (1-2 k) n^{27}-(k-1) n^{33}+(k-1) k n^{30}+n^{36}\right)^{-1}=\\\notag
   &=\lim_{n\to\infty} \frac{n^{36}}{n^{36}}=1\quad[k=o(n^3)]
\end{align}
And, in the edge case where the system is filled with elements, its asymptotic equivalence with the size of the system persists:
\begin{align}
    &\lim_{n\to\infty} \hat{c}_1(n,n^3)=\\\notag
    &=\lim_{n\to\infty} n^3\left(\frac{\displaystyle-\frac{p_{n^3,n}^{11}-p_{n^3,n}^{10}+p_{n^3,n}^9+12 p_{n^3,n}^8+7 p_{n^3,n}^7+7 p_{n^3,n}^4+4 p_{n^3,n}^3+p_{n^3,n}^2-p_{n^3,n}+1}{(p_{n^3,n}-1)^3 \left(p_{n^3,n}^2+1\right) \left(p_{n^3,n}^2+p_{n^3,n}+1\right) \left(p_{n^3,n}^4+p_{n^3,n}^3+p_{n^3,n}^2+p_{n^3,n}+1\right)}}{\displaystyle -\frac{p_{n^3,n}^{11}-p_{n^3,n}^{10}+p_{n^3,n}^9+12 p_{n^3,n}^8+7 p_{n^3,n}^7+7 p_{n^3,n}^4+4 p_{n^3,n}^3+p_{n^3,n}^2-p_{n^3,n}+1}{(p_{n^3,n}-1)^3 \left(p_{n^3,n}^2+1\right) \left(p_{n^3,n}^2+p_{n^3,n}+1\right) \left(p_{n^3,n}^4+p_{n^3,n}^3+p_{n^3,n}^2+p_{n^3,n}+1\right)}+n^3}\right)=\\\notag
   &=\lim_{n\to\infty} n^3\implies \boxed{\hat{c}_1(n,n^3)=\Theta(n^3)}\quad[k=\Theta(n^3)]
\end{align}
In 3-dimensional systems, there is no method to demonstrate the convergence to 1 of $c(n,k)$ in the same way as in 2-dimensional systems, since there are no exact expressions of magnitudes such as combinations of all elements in the system where each configuration has a specific number of clusters between 0 and $k$. However, given the previous tests for different estimates of $c(n,k)$, the tendency to 1 at infinity can be assumed, mainly because the expression $\frac{n^3}{n^3-k+1}$ used at the beginning derives from section \textcolor{blue}{\ref{subsubsec:AverageInsertionRuntime}}, where the restrictions of such magnitude were used to infer a sufficiently reliable expression to consider its asymptotic properties in this analysis.
\begin{align}
    \lim_{n\to\infty} \left(1-\frac{k}{n^3}\right)=\begin{cases}
    1 & \text{if } k=o(n^3)\\
    0 & \text{if } k= \Theta(n^3)
    \end{cases}
\end{align}
Subsequently, in addition to the average cluster size, it is necessary to evaluate the limit of the insertion probability, which is crucial to determine the amount of work proportional to $c(n,k)$ accounted for in each iteration based on the probability of inserting a new element. Therefore, following the limit computation, a result similar to that offered by the metric $c(n,k)$ is acquired, ensuring convergence to 1 throughout the process, except when the system is full, wherein the insertion probability diminishes to zero.

\begin{align}
    \lim_{n\to\infty} \left(1-\frac{E(i)}{n^3}\right)=\lim_{n\to\infty} \left(1-\frac{n^3\left(1-\left(1-\frac{1}{n^3}\right)^i\right)}{n^3}\right)=\lim_{n\to\infty} \left(1-\frac{1}{n^3}\right)^i
\end{align}
In a similar manner, if instead of $k$ we use $E(i)$ to denote the number of elements, it is possible to reach the same result. First, the insertion probability is simplified as seen above, and subsequently the index $i$ of the iteration is replaced by $I(n,k)$, determining the necessary iterations to reach each quantity $k$ of elements.
\begin{align}
    \lim_{n\to\infty} \left(1-\frac{E(n^3(H_{n^3}-H_{n^3-k}))}{n^3}\right)&=\lim_{n\to\infty} \left(1-\frac{1}{n^3}\right)^{n^3(H_{n^3}-H_{n^3-k})}=\\\notag
    &=\lim_{n\to\infty} e^{-(H_{n^3}-H_{n^3-k})}=\\\notag
    &=\lim_{n\to\infty} \frac{\displaystyle e^{H_{n^3-k}}}{\displaystyle e^{H_{n^3}}}=\\\notag
    &=\lim_{n\to\infty} \frac{\displaystyle e^{\ln(n^3-k)+\gamma}}{\displaystyle e^{\ln(n^3)+\gamma}}=\\\notag
    &=\lim_{n\to\infty} \frac{\displaystyle n^3-k}{\displaystyle n^3}=\lim_{n\to\infty} 1-\frac{k}{n^3}=\begin{cases}
    1 & \text{if } k=o(n^3)\\
    0 & \text{if } k= \Theta(n^3)
    \end{cases}
\end{align}
Through the previous substitution, it is inferred that the insertion probability converges to 1 except when $k=n^3$, which is consistent with the previous analysis.

\begin{align}
    \lim_{n\to\infty} T_{avg}(n) &= \lim_{n\to\infty} \sum _{i=0}^{I(n)} \left(1-\frac{E(i)}{n^3}\right)\cdot c(n,E(i))\\\notag&=\lim_{n\to\infty} \sum _{i=0}^{I(n)} 1=\lim_{n\to\infty} I(n) \implies \boxed{\boxed{T_{avg}(n)=\Theta(I(n))}}
\end{align}
Ultimately, given the convergence of $c(n,k)$ and the insertion probability, their values can supplant the original expressions in the sum of the total work of the percolation process, engendering the aforementioned dependency between the complexity $T_{avg}(n)$ and the average duration $I(n)$ of the process. After ensuring this dependency, the asymptotic growth of $I(n)$ is deduced, since in this dimension there is no expression for the number of terminal states $p(n,k)$, which impedes knowing the exact average duration of the process.

\begin{align}
    \lim_{n\to\infty} \frac{E(i)}{E(n^3H_{n^3})}=L\quad\colon L\in[0,1]
\end{align}
First, as was considered with 2-dimensional systems, the ratio between the expected number of elements in the terminal state and the size of the system is regarded as the site percolation threshold of the system \cite{Yoo2014, Novak2011, Schrenk2016, Grassberger2017}, which should lie between 0 and 1.

\begin{align}
    \lim_{n\to\infty} \frac{E(i)}{E(n^3H_{n^3})}&=\lim_{n\to\infty} \frac{n^3 \left(1-\left(1-\frac{1}{n^3}\right)^i\right)}{n^3 \left(1-\left(1-\frac{1}{n^3}\right)^{n^2H_{n^3}}\right)}=\\\notag
    &=\lim_{n\to\infty} \frac{\displaystyle1-\left(1-\frac{1}{n^3}\right)^i}{\displaystyle1-\left(1-\frac{1}{n^3}\right)^{n^2H_{n^3}}}=\\\notag
    &=\lim_{n\to\infty} \frac{\displaystyle1-\left(1-\frac{1}{n^3}\right)^i}{\displaystyle1-e^{-H_{n^3}}}=\\\notag
    &=\lim_{n\to\infty} 1-\left(1-\frac{1}{n^3}\right)^i=\\\notag
    &=\lim_{n\to\infty} 1-\frac{1}{e^{i/n^3}}=\begin{cases}
    \lim_{n\to\infty} 1-\frac{1}{e^0}=0 & \text{if } i=o(n^3)\\
    \lim_{n\to\infty} 1-\frac{1}{e}\approx 0.6321 & \text{if } i=\Theta(n^3)
    \end{cases}
\end{align}
Through this approach, by elaborating the limit, it is determined that for the ratio to converge to a constant within $[0,1]$, the process must last a number of iterations asymptotically equivalent to $\Theta(n^3)$, which closely resembles the bound $\Theta(n^2)$ from the analysis in section \textcolor{blue}{\ref{subsec:AverageCaseAnalysis2D}}.
\begin{align}
    \lim_{n\to\infty} \frac{E(n^3(H_{n^3})^\xi)}{E(n^3H_{n^3})}&=\lim_{n\to\infty} 1-\left(1-\frac{1}{n^3}\right)^{n^3(H_{n^3})^\xi}=\\\notag
    &=\lim_{n\to\infty} 1-\frac{1}{e^{(H_{n^3})^\xi}}=\\\notag
    &=\lim_{n\to\infty} 1-\frac{1}{e^{(\ln(n^3)+\gamma)^\xi}}=\\\notag
    &=\lim_{n\to\infty} 1-\frac{1}{e^{3^{\xi}\ln^\xi(n)}}=\begin{cases}
    1-\frac{1}{e} & \text{if } \xi=0\\
    \lim_{n\to\infty} 1-\frac{1}{e^\infty}=1 & \text{if } \xi>0
    \end{cases}
\end{align}
Likewise, when evaluating the convergence for iterations with a growth greater than $O(n^3)$ and less than the maximum feasible $O(n^3H_{n^3})$, it is evidenced that the ratio converges to 1, coinciding with the maximum value achievable by the threshold, which in this case must be less than this amount.
\begin{align}
    \boxed{\boxed{T_{avg}(n)=O(n^3)}}
\end{align}
To conclude, given that the complexity depends on $I(n)$, and its asymptotic growth is of the order $\Theta(n^3)$, it can be inferred that the average cost of executing the algorithm on a 3-dimensional system has an asymptotic growth as presented above. It is also worth noting that this bound originates from the average of all cases reachable by the algorithm, accounting for the probability of each occurring. Furthermore, although there are situations where the sequence of insertions can be extended indefinitely, the probability of it appearing in an execution is remarkably low, exerting a minimal influence on the overall average complexity.
\subsection{Experimental time measurements}
Ultimately, to conclude the analysis of the process on 3-dimensional systems, execution time measurements will be taken following the same procedure used for the previously analyzed systems \cite{time_measurements_3D1}.
\begin{figure}[H]
    \centering
    \includegraphics[width=10cm,clip]{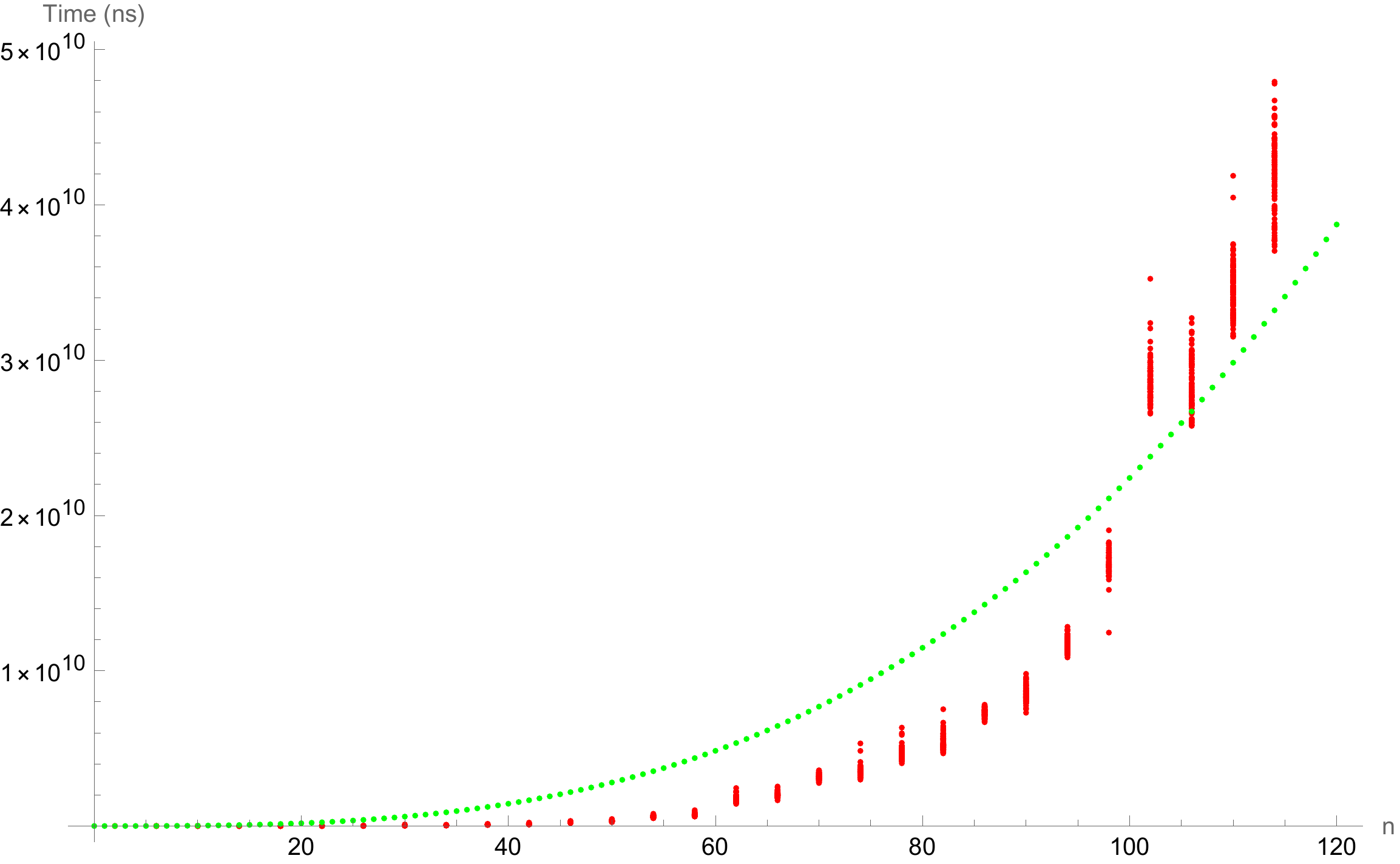}
    \caption{Runtime measurements for a 3D system plotted in red and the fitted model $T_{avg}(n)\approx2.24\cdot10^4 n^3$ in green.}
    \label{fig:3DtimesFit}
\end{figure}
In this case, the model used to fit the asymptotic growth of the measurements will be of the same order inferred in the theoretical analysis, specifically $x \cdot n^3$ where $x$ is a real parameter of the model.
\[
\begin{array}{l|lll}
 \text{} & \text{DF} & \text{SS} & \text{MS} \\
\hline
 \text{Model} & 1 & 4.83415\times 10^{23} & 4.83415\times 10^{23} \\
 \text{Error} & 1636 & 4.44247\times 10^{22} & 2.71544\times 10^{19} \\
 \text{Uncorrected Total} & 1637 & 5.2784\times 10^{23} & \text{} \\
 \text{Corrected Total} & 1636 & 3.02644\times 10^{23} & \text{} \\
\end{array}
\]

\[
\begin{array}{l|llllll}
\text{Parameter} & \text{Estimate} & \text{Standard Error} & \text{t-Statistic} & \text{P-Value} & \text{Confidence Interval} \\
\hline
x & 22,\!414.8 & 167.995 & 133.426 & 0.000 & \{22,\!085.3,\ 22,\!744.3\} \\
\end{array}
\]

Considering the preceding results and an adjusted $R^2$ of 0.915785, it becomes more challenging to ascertain that the bound $n^3$ matches the average cost growth of the algorithm. This arises due to several reasons, among the most important is the significantly higher cost of the process at this dimension, leading to a reduction in the number of measurements, an increase in the step of $n$, and a lower density of measurements for each of its values. In summary, the model shows similar growth, although with a clear difference concerning the points where the measurements are concentrated. Thus, it is also important to consider the difference between the worst and best case, which in this dataset is not as clear as in the previous ones, escalating the uncertainty in the parameter $x$. However, for values of $n$ higher than those shown in this dataset, it can be inferred that the model falls within those limit bounds, even in the last measured $n$ it approaches considerably the areas of higher data point density, suggesting a good fit for large values of the system size, which is valuable for this analysis. And, besides adjusting the constant that multiplies the inferred growth order in the average analysis, it is advisable to also adjust the model of the form $T_{avg}(n) \approx e^{\ln(b) + x \ln(n)}$ to the dataset to find the exponent $x$ that best models the growth. Therefore, after performing the corresponding logarithmic transformation of the data and proceeding with the linear model fit in the logarithmic space, the following results are presented:

\[
\begin{array}{l|lllll}
 \text{} & \text{DF} & \text{SS} & \text{MS} & \text{F-Statistic} & \text{P-Value} \\
\hline
 x & 1 & 10515.3 & 10515.3 & 64910.1 & 0. \\
 \text{Error} & 1634 & 264.704 & 0.161997 & \text{} & \text{} \\
 \text{Total} & 1635 & 10780. & \text{} & \text{} & \text{} \\
\end{array}
\]

\[
\begin{array}{l|llllll}
\text{Parameter} & \text{Estimate} & \text{Standard Error} & \text{t-Statistic} & \text{P-Value} & \text{Confidence Interval} \\
\hline
\ln(b) & 0.0441481 & 0.0856556 & 0.515414 & 0.606333 & \{-0.123858,\ 0.212154\} \\
x & 5.10641 & 0.0200428 & 254.775 & 0.000 & \{5.0671,\ 5.14572\} \\
\end{array}
\]

With this, and an adjusted $R^2$ of 0.97543, considerably higher than in the other model, it could be deduced that the resulting exponent of the order $n^{5.1}$ fits better with the average growth of the cost.

\begin{figure}[H]
    \centering
    \includegraphics[width=10cm,clip]{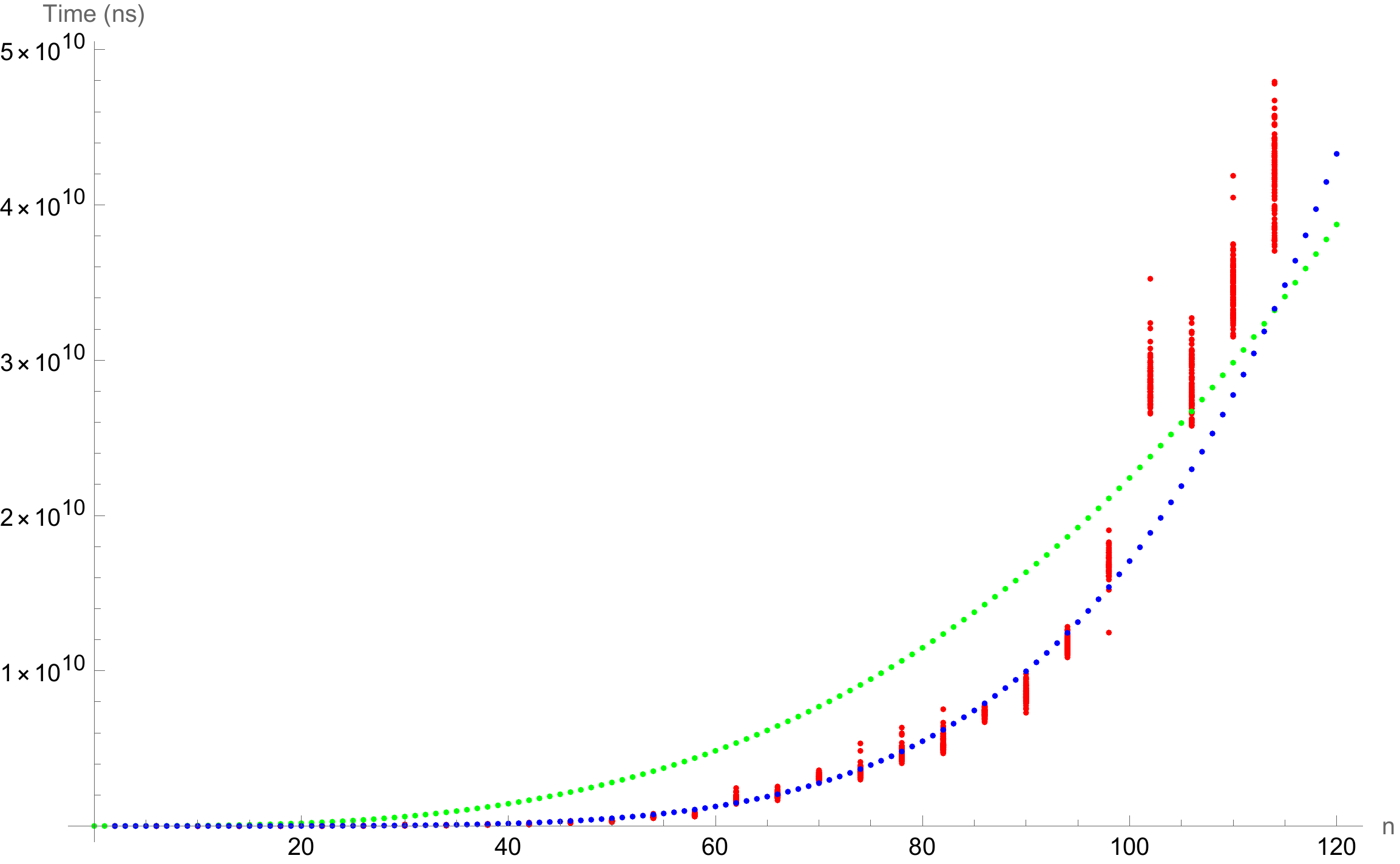}
    \caption{Runtime measurements for a 3D system plotted in red, the original fitted model $T_{avg}(n)\approx2.24\cdot10^4 n^3$ in green, and the alternative one $T_{avg}(n)\approx e^{0.04 + 5.11 \ln(n)}$ in blue.}
    \label{fig:3DtimesFitComparison}
\end{figure}
Graphical representations indicate that the linear model is closer to the regions with higher measurement density only for small values of $n$. Beyond this range, it grows faster than the $n^3$ bound, and its proximity to the worst case leads to dismissing its growth as an adequate average. Mainly, we know that the maximum number of iterations the algorithm can last is $n^3H_{n^3}$, and given that in the average case the cost of the algorithm is proportional to the duration $I(n)$ of the process, it is concluded that the $O(n^{5.1})$ growth derived from the linear model is inconsistent with the maximum order bound $n^3H_{n^3}$ achievable in the average case. Consequently, the original model fit is established as the most accurate asymptotic bound for the average case, even with the inaccuracies caused by the low number of measurements.
\section{Conclusion}
Finally, the runtime growth of the Monte Carlo algorithm has been characterized in every dimension up to the third, in which there isn't the same guarantee that the provided outcomes are precise. Also, the analysis of the space occupied by the algorithm can be deduced from the first sections, since it is highly dependent on the size of the system. Hence, as a conclusion, it is proposed as future work the study of the expressions that have not yet been determined, being mainly the number of terminal states for systems of 2 or more dimensions, and the average cluster size in systems of the same characteristics. Likewise, the possibility of finding $p_c$ in an exact way by means of some of the proposed methods is suggested, although from this point it cannot be assured whether they will prove successful in their purpose.

\end{document}